\documentclass[12pt,twoside,usenatbib]{book}
\usepackage{amsmath}
\usepackage{setspace}
\usepackage{a4wide,amsthm, amsfonts, amssymb, latexsym, epsfig,natbib, fancyhdr, mathrsfs}
\usepackage{multicol}
\usepackage{graphicx}
\usepackage{textcomp}
\usepackage{float}
\usepackage{amsxtra}
\usepackage[nottoc]{tocbibind}%this is to link the "Bibliography" to the ToC
\usepackage{hyperref}
\usepackage{tabularx}
\usepackage{threeparttable} % for table notes
\usepackage{xcolor} % for colored text

\hypersetup{colorlinks, citecolor=blue}

\usepackage{epigraph}
\usepackage[paper=a4paper]{geometry}
\usepackage{xcolor}
\usepackage{sectsty}

\usepackage{titlesec}
\titleformat{\chapter}{\normalfont\Huge}{\thechapter.}{20pt}{\huge\it\bf\color{black!70}}

\usepackage{stackengine}
\usepackage{scalerel}
\usepackage{xcolor}

\usepackage{graphicx}
\usepackage{amscd}
\usepackage[titles]{tocloft}
\usepackage{calligra}
\usepackage{auncial}
\usepackage[B1]{fontenc}
\usepackage{sqrcaps}
\usepackage{LobsterTwo}
\usepackage[T1]{fontenc}
\usepackage{ifthen}
\usepackage{enumerate}
\newcommand{\overbar}[1]{\mkern 1.5mu\overline{\mkern-1.5mu#1\mkern-1.5mu}\mkern 1.5mu}

\newcommand{\HI}{H\protect\scaleto{$I$}{1.2ex}}
\newcommand{\HII}{H\protect\scaleto{$II$}{1.2ex}}
\newcommand{\farcs}{.\!\!^{\prime\prime}}
\newcommand{\farcm}{.\mkern-4mu^\prime}
\newcommand{\kms}{\mbox{km\,s$^{-1}$}}

\usepackage[symbol]{footmisc}

\usepackage{microtype} 
\usepackage{amsmath}

\newenvironment{uprightmath}
 {\changecodes\ignorespaces}
 {\ignorespacesafterend}

\newcommand{\changecodes}{%
  \count255=`A
  \loop
  \mathcode\count255=\numexpr\mathcode\count255-\string"100\relax
  \ifnum\count255<`Z
    \advance\count255 1
  \repeat
  \count255=`a
  \loop
  \mathcode\count255=\numexpr\mathcode\count255-\string"100\relax
  \ifnum\count255<`z
    \advance\count255 1
  \repeat
} 
 
\numberwithin{equation}{chapter}
\RequirePackage{filecontents}        
\usepackage{mathrsfs}
\usepackage{fourier}
\begin{document}
%-------------------------------------------------

\thispagestyle{empty}

\begin{center}
{ \Huge{\bf  \HI{} 21 cm observations and dynamical models of superthin galaxies}}\\
\end{center}
\vspace{0.8in}
\begin{center}
{\large  A THESIS \\
 SUBMITTED FOR THE DEGREE OF\\
\vspace{0.06in}
%{\aunclfamily \Large Doctor of Philosophy} \\
%{\calligra \Large Doctor of Philosophy} \\
{\LobsterTwo{\LARGE Doctor of Philosophy}\\}
\vspace{0.2in}
IN THE FACULTY OF SCIENCE}\\
\vspace{0.75in}
{ \large by}\\
\vspace{0.5cm}
{\Large \bf K Aditya}\\
\vspace{0.8in}
\includegraphics[height=1.5in]{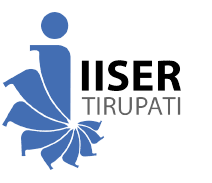} \\
\vspace{0.2in}
{\large{DEPARTMENT OF PHYSICS}}\\
\vspace{0.1cm}
{\large{Indian Institute of Science Education and Research}} \\
\vspace{0.1cm}
{\large{Tirupati \,-\, 517507}}\\
%\vspace{0.1cm}
{August 2022}
\end{center}

%\newpage
\let\cleardoublepage=\clearpage

\cleardoublepage
\pagenumbering{roman}

%-------------------------------------------------
%\include{copyright}
%\thispagestyle{empty}
\cleardoublepage
\let\cleardoublepage=\clearpage
%-------------------------------------------------------
\onehalfspacing
%\linespread{2.1}% use \doublespacing in order to creat double-spacing

%\addcontentsline{toc}{chapter}{Declaration}
\chapter*{}
\begin{center}
\textcopyright {\bf K Aditya\\
August 2022\\
All rights reserved\\}
\end{center}

\cleardoublepage

\addcontentsline{toc}{chapter}{Declaration}
\chapter*{Declaration}

\vspace*{.80in}
I hereby declare that the work reported in this doctoral thesis titled ``\HI{} 21 cm observations and dynamical models of superthin galaxies'' 
is entirely original and is the result of investigation carried out by me in the 
Department of Physics, Indian Institute of Science Education and Research, Tirupati -- 517507, under the supervision of Dr. Arunima Banerjee.
I further declare that this work has not formed the basis for the award of any degree, diploma, fellowship, associateship or 
similar title of any University or Institution.

\vspace*{0.5 in}

\begin{flushright}
K Aditya\\
\end{flushright}

\begin{flushleft}
Dr Arunima Banerjee\\
Thesis Supervisor\\
\end{flushleft}

\vspace {1cm}
August 2022\\
Department of Physics\\
Indian Institute of Science Education and Research\\
Tirupati -- 517507, India\\

\newpage

\cleardoublepage

\addcontentsline{toc}{chapter}{Dedication}
\chapter*{}
\begin{center}
%\centering
\vspace {-3.5cm}
{\Huge \it{Sarvam Brahmarpanam Astu}}\\
\end{center}

\cleardoublepage

\addcontentsline{toc}{chapter}{Acknowledgement}   
\chapter*{Acknowledgements}

Although I claim authorship to this thesis, this thesis would not have been possible without immense contribution and help from many people. 
This is a humble effort to express my gratitude to everyone who has been part of this journey, and I apologize forehand if I have missed acknowledging anyone. 
I would like to express my utmost gratitude to my thesis supervisor Dr. Arunima Banerjee, for her methodical guidance, support, and constant encouragement. 
I thank her for showing faith in me, even when I had little experience to start with the research, and for her continuous encouragement and support in times 
of my despair when I was not getting the results in my research. I am grateful to you for ensuring that we consistently made progress even in the middle of 
the pandemic. I wish to thank her for patiently and enthusiastically teaching me the nuances of research in galactic dynamics and introducing me to the exciting 
field of radio astronomy. Your guidance and support in the last five years have helped me grow as a researcher and as an individual. I always remember the advice you 
gave me when we visited IUCAA " you should believe in your results before you can convince others ``. I again sincerely thank you for all the help, support, and guidance, 
without which the thesis could not have been put together.

I am immensely grateful to Dr. Peter Kamphuis (Ruhr University Bochum, Germany) for teaching me radio data reduction, showing interest in my work, 
and patiently and tirelessly answering my questions. I would like to thank Dr Eugene Vasiliev, for his help with AGAMA. I would also like to thank Professor Soumitra Sengupta, 
Indian Association for the Cultivation of Science (IACS), Kolkata and Dr Indrani Banerjee, National Institute of Technology (NIT), Rourkela for a collaborative visit to 
IISER, Tirupati, which culminated in a paper on superthin galaxies in braneworld model. I am thankful to them for answering 
my innumerable questions patiently. I would like to express my thankfulness to Dr. Dmitry Makarov and his team at Special Astrophysical Observatory (SAO), Russia: Dr. Sviatoslav Borisov, Dr. Aleksandr Mosenkov, 
Dr. Aleksandra Antipova, for providing us with the optical observation of FGC 1440 and FGC 2366, which form a crucial part of the thesis.

I am grateful to my Research Advisory committee members, Professor Surhud More (Inter University Center for Astronomy and Astrophysics, Pune), 
Professor Nirupam Roy (Indian Institute of Science, Bangalore), and Dr. Jessy Jose (Indian Institute of Science Education and Research, Tirupati), for 
their comments and questions that have given us new ideas and helped improve the quality work. I also take this opportunity to thank the Chair, Department of Physics 
Professor G. Ambika for her constant support and encouragement. I am grateful to all the faculty members in the Department of Physics for their support
and encouragement.

I acknowledge Inter-University Center for Astronomy and Astrophysics, Pune, for hosting me for 
meeting with Dr. Peter Kamphuis, National Center for Radio Astrophysics (NCRA), Pune for providing computational resources and introducing 
me to radio astronomy through its summer school. Besides, I would like to thank the staff at the Giant Meterwave Radio Telescope, 
Khodad, for helping me with the radio observations, which form a crucial part of the thesis. Further, I would like to thank Professor 
Nissim Kanekar and Dr. Avishek Basu for their help and support during my visit to NCRA. Finally, I would like to thank Dr. Sandeep Kataria for 
introducing me to N-body simulations using GADGET at the ASI workshop 2018.

I would like to express my heartfelt gratitude to Professor K.N. Ganesh, Director IISER, Tirupati, Academic $\&$ Administrative staff, and the day-to-day working staff
who kept the institute running amidst the pandemic to continue research activities. I am particularly indebted to the IT department at IISER, Tirupati, 
who kept the computers working. Thank you very much, Mr. Satish Jadhav and Mr. Arunsairam Sekaran, for switching on my computer/ Anydesk many times in the middle 
of the pandemic. I could complete at least two chapters in the thesis for your help. I would like to thank the library for providing a congenial environment for writing the thesis and Mr. K. Murugaraj for helping me patiently with plagiarism-checking the thesis (multiple times).
Finally, I would like to thank the thesis examiners for taking their time to provide  detailed comments on the thesis.

My time at IISER, Tirupati would have become monotonous and morose if not for wonderful friends: Ganesh Narayanan, Krishan Gopal, Souren Adhikary, Arka Bhattacharya and Rashi Soni, 
thank you very much for being there for chai, humor, politics, and everything else. My special thanks to Souren Adhikary, who insisted that I should write the acknowledgment 
first, even before I started writing the thesis.

I would like to thank Krishna Tataji, Ambuja Amamma, Sheshu Mama, and Radhika Atta for their love, blessings and for taking care of me during 
the pandemic. Sagara, for your jokes and teaching me to ride a bike as you would like to think.

Amma and Tataji, thank you very much for your love and blessings. It is what keeps me going forward every day.

Dad, I wish you were here; you only saw me failing in the fourth class !

\cleardoublepage
\let\cleardoublepage=\clearpage
\addcontentsline{toc}{chapter}{Preface}

\chapter*{Preface}
\vspace {2.5cm}

\textbf{Superthin galaxies:} Superthin galaxies are a class of edge-on disc galaxies exhibiting extraordinarily high values of vertical-to-planar 
axes ratio $\rm a/b \sim$ 10-21, with no discernible bulge component. They are generally characterized by low values of central B-band surface brightness 
$\rm (\mu_{B}$= 23-26 $\rm{mag/arcsec}^{2})$, gas richness as given by high values of the ratio of the total mass in neutral hydrogen gas to the blue band 
luminosity ratio $\rm \frac{M_{HI}}{L_{B}} \sim1$ \citep{matthews2003high} and dynamical dominance of dark matter at all galactocentric radii \citep{banerjee2010dark}. 
Being gas-rich and dark matter-dominated, they constitute the proxies of a primeval galaxy population, hence the ideal test-beds for studying galaxy formation and 
evolution processes in the local universe. The term \emph{superthin} was first introduced by \cite{goad1981spectroscopic} who conducted a spectroscopic study of four 
edge-on galaxies: UGC 7321, UGC 7170, UGC 9242, and UGC 4278 (IC 2233). The Revised Flat Galaxy Catalogue (RFGC) sourced from the Palomar Observatory Sky Survey (POSS-II) 
survey is the primary catalog of edge-on disc galaxies \citep{ 1999BSAO...47....5K}. It contains 4,444 edge-on galaxies with $\rm a/b$  greater than $\rm 7$,  1150 galaxies 
with an $\rm a/b$ $>$ 10, and only 6 extremely thin galaxies with  $\rm a/b$ $>$ 20, indicating the paucity of extremely thin galaxies. \\

\textbf{Earlier work:} The origin of the superthin stellar discs in these galaxies is still not well understood. Interestingly, cosmological  
and cosmological hydrodynamical simulations have indicated the sparsity of thin, bulgeless discs \citep{pillepich2018simulating} though 
there is ample observational evidence that they are ubiquitous. This challenges our current understanding of galaxy formation and evolution under 
the $\rm \Lambda$ -CDM paradigm. Our current understanding of superthin galaxies originates from the dynamical studies of a small sample of superthin galaxies. 
The razor-thin appearance of the stellar discs indicates that these galaxies have not evolved dynamically in the vertical direction. However, such 
thin discs are generally prone to vertical bending instabilities \citep{khoperskov2017disk} which results in disc thickening. This implies that the superthin 
galaxies are intrinsically stable against bending instabilities. In fact, \cite{zasov1991thickness} argued that a massive 
dark matter halo stabilizes the superthin discs, and later studies showed that the disc is stable against local, dynamical instabilities. Dynamical modeling of 
the prototypical superthin galaxy, UGC 7321, indicated that it has a  compact dark matter halo \citep{o2010dark}, and the same was found for a sample of superthin galaxies in later studies. 
\cite{banerjee2013some} showed that the compact dark matter halo plays a decisive role in regulating its thin vertical structure. Further, \cite{jadhav2019specific} 
showed that superthin stellar discs may have a higher value of specific angular momentum than an ordinary disc galaxy for a given stellar mass, which may hold the clue 
to the large values of their planar-to-vertical axes ratio. \\

\textbf{This thesis:} The primary objective of this thesis is to identify the key dynamical mechanisms responsible for the superthin stellar discs. 
We use \HI{} 21cm radio-synthesis observations and stellar photometry to construct detailed dynamical models of a sample of superthin galaxies to determine
the primary mechanism responsible for the existence of superthin stellar discs in these galaxies. The size of the interstellar atomic hydrogen (\HI{}) disc 
is 3-4 times the size of the stellar disc, and hence \HI{} serves as an effective diagnostic tracer of the underlying gravitational potential. For a disc galaxy 
that is almost edge-on like the superthins, it is possible to determine the rotation curve traced by \HI{} as well as the thickness of the HI disc. While the 
rotation curve constrains the radial derivative of the potential, the vertical thickness imposes a constraint on the velocity dispersion of \HI{}. \HI{} 21cm 
radio-synthesis observations, in conjunction with stellar photometry, help construct mass models and hence determine the dark matter distribution 
in these galaxies. Further, the observed thickness of the stellar disc constrains the vertical velocity dispersion of the stars. Finally, superthin galaxies are 
dominated by dark matter at all radii, and hence their disc structure and kinematics are strongly regulated by dark matter. This is unlike the case for ordinary 
disc galaxies like the Milky Way, where dark matter governs the disc dynamics only in the outer galaxy. Therefore superthin galaxies also serve as laboratories in 
the local universe to test various models of dark matter, including $\rm \lambda-$CDM, and those inspired by alternative theories of gravity. Our study is based on 
a sample of superthin galaxies with $\rm 10< a/b < 16$ for which H1 21cm radio-synthesis data were already available in the literature. In addition, we had the two thinnest 
galaxies in our sample with $\rm a/b \sim 21$, for which we carried out Giant Meterwave Radio Telescope (GMRT) 21cm radio-synthesis observations. The stellar photometry for 
all the galaxies were already available in the literature. We then construct mass models as well as dynamical models based on Jeans Modelling and Distribution Function 
Modelling as constrained by the observations for our sample galaxies. For one of our sample galaxies, we construct dynamical models using the dark mass inspired by an 
alternative theory of gravity, namely the braneworld paradigm. We choose the best-fitting model employing the Markov Chain Monte Carlo (MCMC) technique, which constitutes 
an efficient method of scanning a multi-dimensional grid of model parameters. 

\section*{Organization of the chapters}
In \textbf{Chapter 1}, we present a general introduction to the morphological classification of galaxies, followed by the vertical structure of
disc galaxies and a description of the different dynamical models to be used in the thesis. We next introduce the \emph{dark matter models} or 
\emph{dark mass} in an alternative theory of gravity, namely the braneworld gravity, and present the theoretical model of a galactic disc consisting 
of gravitationally-coupled stars and gas responding to the external potential of the \emph{dark mass}. We then move on to describe \HI{} 21cm radio-synthesis
observations in disc galaxies, methods of data analysis, and modeling the structure and kinematics. We end the chapter with a description of the Markov Chain 
Monte Carlo technique, which is a method for efficiently scanning the multi-dimensional grid of dynamical parameters to choose the best-fitting dynamical model and 
determine the confidence limits on the same. \\

In \textbf{Chapter 2}, we use the multi-component galactic disc model of gravitationally-coupled stars and gas in the force field of the dark matter 
halo as well as the stellar dynamical code AGAMA (Action-based Galaxy Modelling Architecture), and determine the vertical velocity dispersion of stars 
and gas as a function of galactocentric radius for five superthin galaxies (UGC 7321, IC 5249, FGC 1540, IC2233, and UGC00711) using observed stellar 
and atomic hydrogen (\HI{}) scale heights as constraints, using a Markov chain Monte Carlo (MCMC) method. We find that the central vertical velocity dispersion 
for the stellar disc in the optical band varies between $\sim$ 10.2-18.4  \kms  and falls off with an exponential scale length of 2.6-3.2 $\rm R_{d}$ where $\rm R_{d}$ 
is the exponential stellar disc scale length. Besides, in the 3.6 $\rm \mu$m, the same, averaged over the two components of the stellar disc, varies between 5.9 
and 11.8 \kms, both of which confirm the presence of 'ultra-cold' stellar discs in superthin galaxies. Further, we find an average value of the 
vertical-to-radial velocity dispersion to be 0.3 compared to the value of 0.5 at the solar radius in the Milky Way. Interestingly, however, the 
average stellar vertical velocity dispersion as normalized by the asymptotic rotational velocity is comparable to the Milky Way's thin disc stars lying 
between $\rm \sim$ 200 pc to 200 pc. Finally, the global median of the multi-component disc dynamical stability parameter $\rm Q_{N}$ of our sample superthins is 
found to be $\rm 5 \pm 1.5$, which higher than the global median value of $\rm 2.2 \pm 0.6$ for a sample of spiral galaxies.\\

In \textbf{Chapter 3}, we consider a gravitational origin of dark matter in the braneworld scenario, where the higher dimensional Weyl stress term projected on 
to the three-brane acts as the source of dark matter. In the context of the braneworld model, this dark matter is referred to as the 'dark mass'. This model 
successfully reproduces the rotation curves of several LSB and high surface brightness galaxies. Therefore, it is interesting to study the prospect of 
this model in explaining the vertical structure of galaxies which has not been explored in the literature so far. Using our two-component model of gravitationally 
coupled stars and gas in the external force field of this dark mass, we fit both the observed rotation curve and the scale heights of stellar and atomic hydrogen (HI) 
gas of superthin galaxy 'UGC7321' using the Markov Chain Monte Carlo approach. We find that the observed scale heights of 'UGC7321' can be successfully modeled in 
the context of the braneworld scenario. In addition, the model predicted rotation curve also matches the observed one. We thereby estimate the posterior probability 
distribution of the vertical velocity dispersion of the stars and gas and the parameters corresponding to the density profile of the five-dimensional bulk 
We find that the values of vertical velocity dispersion obtained in the braneworld paradigm agree with those obtained in the first chapter using Newtonian gravity. \\

In \textbf{Chapter 4}, we present observations and models of the kinematics and distribution of neutral hydrogen (H I) in the superthin galaxy FGC 1440 
with an optical axial ratio a/b = 20.4, one of the thinnest galaxies known. Using the Giant Meterwave Radio telescope (GMRT), we image the galaxy with 
a spectral resolution of 1.7 \kms and a spatial resolution of $\rm 15.9^{\farcs} \times 13.5^{\farcs}$. We find that FGC 1440 has an asymptotic rotational velocity of 
141.8\kms. The structure of the \HI{} disc in FGC 1440 is that of a typical thin disc warped along the line of sight, but one cannot rule out the presence of a central thick 
\HI{} disc. We find that the dark matter halo in FGC 1440 could be modeled by a pseudo-isothermal (PIS) profile with  $\rm R_{c}/R_{d}<2$ , where $\rm R_{c}$ is the core 
radius of the PIS halo and $\rm R_{d}$ the exponential stellar disc scale length. We note that despite the unusually large axial ratio of FGC 1440, the ratio of the 
stellar vertical velocity dispersion to the rotation velocity is $\rm \sim$ 0.125 - 0.2, which is comparable to other superthins. Interestingly, unlike previously studied superthin 
galaxies, which are outliers in the log10(j*) - log10(M*) relation for ordinary bulgeless disc galaxies, FGC 1440 is found to comply with the same. The values of j for the stars, 
gas, and baryons in FGC 1440 are consistent with those of normal spiral galaxies with similar mass. \\

In \textbf{Chapter 5}, we present  GMRT \HI{} 21cm radio-synthesis observations of FGC2366, a superthin galaxy with $\rm a/b=21.6$ and also the thinnest galaxy known.
Employing the 3-D tilted-ring modelling, we determine the structure and kinematics of the \HI{}  disc, obtaining an asymptotic rotational velocity $\rm \sim$ 100 \kms and 
a total \HI{} mass $\rm \sim$ 10$^9 M_{\odot}$. Using available stellar photometry, we construct mass and dynamical models for FGC2366. We found that  FGC 2366 
hosts a compact dark matter halo, i.e the ratio of the core radius  of the dark matter halo to the disc scalelength $\rm ({R_{c}/R_{d}})$ $=0.35 \pm 0.03$, has a 
minimum of vertical-to-radial stellar velocity dispersion  $\rm (\sigma_{z}/\sigma_{R}) = 0.42 \pm 0.04$, a 2-component (star + gas) disc dynamical stability parameter 
$\rm Q_{RW} = 7.4 \pm 1.8 $ at 1.5$R_{d}$  and finally a specific angular momentum $\rm \sim$ $\rm{log}_{10}(j_{*})$  $\rm 2.67 \pm 0.02 $ for a stellar disc mass 
$\rm \sim$ $\rm{log}10(M_{*}/M_{\odot})=9.0$. To identify the physical mechanism primarily responsible for the superthin vertical structure, we carry out a 
Principal Component Analysis of the above dynamical parameters along with $\rm a/b$ for all superthins studied so far. We note that the first two principal 
components explain $\rm \sim$ 80$\rm \%$ of the variation in the data, and the major contributions are from $\rm a/b$, $\rm Q_{RW}$ and $\rm V_{\rm{rot}}/{(R_{c}/R_{d})}$. 
This possibly indicates that high values of the disc dynamical stability and dark matter dominance at inner galactocentric radii are fundamentally responsible 
for the superthin stellar discs.

\cleardoublepage

\addcontentsline{toc}{chapter}{List of Publications}    
\chapter*{List of Publications}

\begin{itemize}
\item \textbf{K Aditya}, Arunima Banerjee, How $\rm 'cold'$ are the stellar discs of superthin galaxies?, \textit{Monthly Notices of the Royal Astronomical Society, 
  Volume 502, Issue 4, April 2021, Pages 5049\textendash5064}
  
\item \textbf{Aditya Komanduri}, Indrani Banerjee, Arunima Banerjee, Soumitra Sengupta,
Dynamical modelling of disc vertical structure in superthin galaxy UGC 7321 in braneworld gravity: an MCMC study, \textit{Monthly Notices of the Royal Astronomical Society, 
Volume 499, Issue 4, December 2020, Pages 5690\textendash5701}
  
\item \textbf{K Aditya}, Peter Kamphuis, Arunima Banerjee, Sviatoslav Borisov, Aleksandr Mosenkov, Aleksandra Antipova, Dmitry Makarov
\HI{} 21 cm observation and mass models of the extremely thin galaxy FGC 1440, \textit{Monthly Notices of the Royal Astronomical Society, 
Volume 509, Issue 3, January 2022, Pages 4071\textendash4093}

\item \textbf{K Aditya}, Arunima Banerjee, Peter Kamphuis, Aleksandr Mosenkov, Dmitry Makarov, Sviatoslav Borisov
 \HI{} 21cm observations and dynamical modelling of the flattest/thinnest galaxy known: FGC 2366 \textbf{[Under Revision]}
\end{itemize}

\cleardoublepage

%\addcontentsline{toc}{chapter}{Contents}
\tableofcontents \cleardoublepage

\pagestyle{fancy}
\renewcommand{\chaptermark}[1]{\markboth{\textsl{\thechapter.\ #1}}{}}
\renewcommand{\sectionmark}[1]{\markright{\textsl{\thesection.\ #1}}}
\fancyhf{}
\renewcommand{\headrulewidth}{.05pt}
\fancyhead[LE]{\thepage} \fancyhead[RO]{\thepage}
\fancyhead[RE]{\leftmark} \fancyhead[LO]{\rightmark}

\fancypagestyle{plain}

\pagenumbering{arabic} \setcounter{chapter}{0}

\thispagestyle{empty}
\chapter[Introduction]{\fontsize{50}{50}\selectfont Chapter 1}
\chaptermark{\it Introduction}
%\vspace {2.5cm}
\textbf{\Huge{Introduction}}

Astronomy is as old as the human civilization itself. Since ancient times nomadic people have been using constellations to help them find directions. 
The early civilizations associated various stories and fables for remembering the various constellations and their relative positions in the sky. 
Away from the gaze of the city light, one can observe a band of diffuse white light strewn across the sky from horizon to horizon. This majestic sight 
has caught the imagination of humankind since time immemorial and has been subjected to various myths and legends. The ancient Greeks described this streak of 
diffuse white light as a river of milk flowing from the breast of Hera, wife of Zeus, whereas Romans baptized this as \textit{Via Lactea} or Milky Way. The early 
Indic civilizations called this streak of bright white light \textit{AkasaGanga}, which translates as \textit{The Ganges river of the Sky}. The word \textit{galaxy} 
itself comes from the Greek word for milk.

In 1610, Galileo Galilei built his telescope using the \textit{ Dutch Perspective Glass} patented by a Dutch spectacles maker Hans Lippershey in 1608 as 
\textit{a device to observe things at a distance}. Galileo could resolve the white band of light into a large number of faint stars using this 
telescope which could not be resolved with the naked eye. Thus, Galileo concluded that the Milky Way was not a luminous celestial fluid but a stellar system.  
\begin{figure*}
\hspace*{-7.0mm}
\resizebox{160mm}{90mm}{\includegraphics{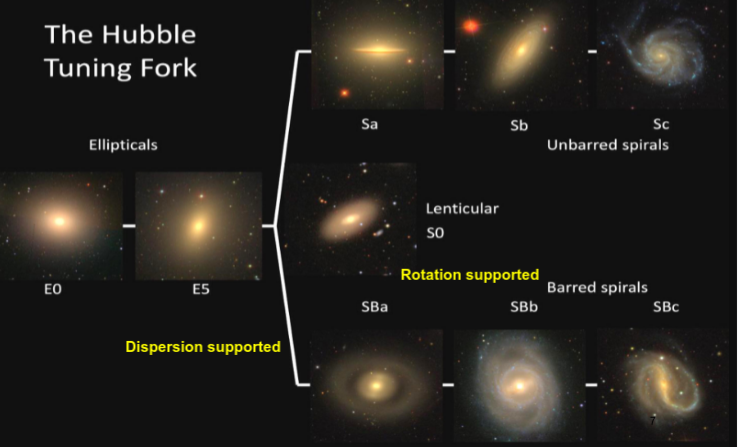}} 
\caption{The classification scheme proposed by Edwin Hubble. The galaxies on the left are the early-type elliptical galaxies, and on the right are the late-type spirals, which are further classified as \textit{barred spirals} and \textit{unbarred spiral}. [\textit{Image courtesy: www.sdss.org}] }	
\end{figure*}

In 1755, Immanuel Kant, in his seminal treatise \textit{General Natural History and Theory of the  Heavens}, 
showed that the stellar structure of the Milky Way galaxy is similar to the arrangement of the planets in the solar system. Kant hypothesized that 
the stellar system in the Milky Way has a disc-like structure, wherein the rotation velocity balances the inward gravitation pull. This disc-shaped 
stellar distribution is seen as a band of white light from the earth. Further, Kant also suggested that the faint elliptical patches (nebulae) 
in the sky might not be part of the Milky Way but might be \textit{island universes}, similar in structure to the Milky Way.

Willian Parsons (Lord Rosse), in 1845, resolved the faint nebulae in greater detail using the 72 inches telescope. Lord Rosse observed that the faint nebulae 
could be cataloged into two distinct categories; one with a completely featureless elliptical distribution of light, and the second class was less symmetric and 
had a distinct spiral structure. Further, the presence of the spiral structure in the faint nebulae reaffirmed the idea propounded by Lord Rosse that the nebulae, 
like Milky Way, rotated about their axis perpendicular to their planes. 

Harlow Shapley, in 1918, showed that the globular clusters in the Milky Way are not distributed uniformly along the plane. Shapely argued that since the 
globular clusters are a major structural component of the Milky Way, they should be distributed symmetrically about the system's center. Thus, from 
the asymmetry in the distribution of the globular clusters, Shapely concluded that our solar system does not lie at the center of the Milky Way. Further, upon measuring 
the distances to the globular clusters, Shapely found that the globular system extended 100 kpc across the disc. Thus, Shapely's model of the Milky Way was extended 10 times 
more than the heliocentric model of the Galaxy presented by Kapteyn. In Shapley's model of Milky Way, it was hard to believe if independent island universes 
like Milky Way, as suggested by Lord Rosse, could exist, as Shapely's model of Milk Way encompassed the then-known universe. 

Understanding the nature of the observed nebulae, i.e., whether the island universes proposed by Lord Rosse are independent entities like Milky Way or does the Milky Way 
by itself constitutes in its entirety the observable universe as suggested by Shapley culminated in a public debate now famously known as \textit{The Great Debate}
between Harlow Shapely and Heber Curtis at the National Academy of Sciences. The debate was finally resolved by Edwin Hubble in 1922 who obtained the distance to 
M31 or the Andromeda nebula by resolving the cepheid variable stars in the M31 using the 100-inch telescope at Mt. Wilson. This firmly established the fact that 
Milky Way is one amongst many nebulae and is not a universe by itself, as propounded by Shapely. Thus, the two distinct classes of nebulae observed by 
Lord Rosse are, in fact, spiral galaxies and elliptical galaxies. The ellipticals, spirals, and the subclasses within the spiral galaxies are organized 
into a classification scheme by Hubble in 1926, now known as \textit{Hubble's Tuning Fork Diagram} (Figure 1.1). 

Elliptical galaxies are generally spherical stellar systems, and the random motion of the stars stabilizes the system against their gravitation collapse. On the other hand, 
the spiral galaxies are a disc-like stellar system where the rotation of the disc 
counteracts the gravitation force and prevents the system from gravitation collapse.

\subsection{Building blocks of disc galaxies}

\begin{figure*}
\hspace{-8.0mm}
\resizebox{160mm}{90mm}{\includegraphics{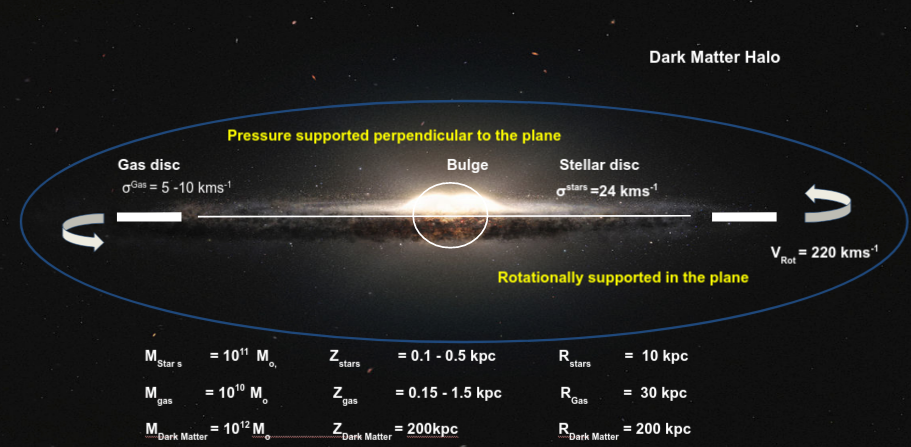}} 
%\vspace{1pt}
\caption{Structural and kinematical properties of a proto-typical galaxy like the Milky Way. [\textit{Image courtsey: www.eso.org/public/}] }	
\end{figure*}

A spiral galaxy consists of a disc of stars and gas embedded in the dark matter halo. The gas disc comprises molecular hydrogen concentrated close to the Galaxy's 
center and atomic hydrogen extending much beyond the optical radius. The stellar and the gas disc are rotation supported in the plane and are supported by the
random motion in the direction perpendicular to the plane. The observed rotation curve is a plot depicting the azimuthal velocity as 
a function of radius. The rotation curve due to the stars shows a keplerian decline. On the other hand, ${H\protect\scaleto{$I$}{1.2ex}}$ extends beyond 
the optical radius of the galaxy  and thereby traces the total potential of the galaxy. The total rotation velocity derived using interferometric observation
of ${H\protect\scaleto{$I$}{1.2ex}}$ 21 cm line remains remarkably constant in the outer regions of the galaxy, indicating the presence of excess mass called \textit{dark matter}.
Dark matter is the dominant mass component of the galaxy, typically 10 times more massive than the stellar disc and 100 times more massive than the gas disc. We show an 
artist's illustration of the Milky Way in Figure 1.2. In Figure 1.2, the central stellar light concentration in the galaxy is called the bulge, 
which consists of old metal-poor stars, which are pressure supported. The stars confined in the plane of the galaxy constitute the stellar disc, which are rotationally 
supported in the plane and supported by the random motion of the stars in the vertical direction and extends up to 10 kpc. The cold gaseous disc, which is predominantly 
atomic \HI{} gas, lies in the plane and extends further beyond the optical disc (30 kpc). The baryonic components are embedded in a dark matter halo, for which we only have gravitational evidence from the observed rotation
curve. This dark matter has a mass equal to $\rm 10^{12}M_{\odot}$, which is roughly 10 times the stellar mass and 100 times more than the 
mass of the gas disc.\\

\subsection*{The bulge}
The bulge is a spherical central light concentration at the center of a disc galaxy. The bulge typically consists of old stellar populations and is dynamically 
supported by the random motion of the stars. \textcolor{red}{However, it has been found that the bulge in the Milky Way consists of young stars as well; also the bulge is 
is slightly flattened with a cylindrical rotation.} Further, the bulge stars contain fewer heavier metals than the disc stars, as the bulge stars had 
formed during the initial formation of the galaxy. The light profile of the bulge \citep{1948AnAp...11..247D,de1956survey, sersic1968atlas} is described by

\begin{uprightmath}
\begin{equation}
I (r)=I_{e}exp\left(  b_{n}\bigg\lbrack(-R/R_{e})^{\frac{1}{n}} -1 \bigg\rbrack  \right)
\end{equation}
\end{uprightmath}
where $\rm I_{e}$ is the intensity at the effective radius $\rm R_{e}$ that encloses half of the total light from the model, $\rm n$ is the Sersic index, 
and describes the concentration of the light profile, and $\rm b_{n}$ is a number specified by $\rm n$.

\subsection*{The stellar disc}
The observed light profile of many external galaxies is described by a thick and thin stellar disc. Further, photometric studies of nearby galaxies 
\cite{dalcanton2000structural, dalcanton2002structural} show that the thick and thin disc discs are ubiquitous in external galaxies
(also see \cite{yoachim2005kinematics, yoachim2008lick}). Similarly, \cite{juric2008milky} analyzed data from Sloan Digital Sky Survey (SDSS) and found that the number 
density distribution of the Milky Way stars was well fitted by a thick $\rm \sim 0.9 kpc$ and a thin disc $\rm \sim 0.3 kpc$ components called geometric thick and thin disc. 
The surface density of the external disc galaxies is described by an exponential profile that falls off with radius. And in the vertical direction described by either an 
exponential profile $\rm \bigg(exp(z/h_{z})\bigg)$ or $\rm \bigg(sech^{2}(z/z_{z})\bigg)$, where $\rm h_{z}$ and $\rm z_{0}$ are the
half-width at the half maximum corresponding to the respective profiles.  

\begin{uprightmath}
\begin{equation}
\Sigma (R)= \Sigma_{0}exp(-R/R_{d})
\end{equation}
\end{uprightmath}
$\rm \Sigma_{0}$ is the central surface density, and the $\rm R_{d}$ is the disc scalelength of the exponential profile.

\subsection*{The gas disc}
The gaseous disc in the galaxies typically consists of atomic hydrogen gas (\HI{}) and molecular hydrogen gas (\HII{}). The neutral \HI{} is further 
classified as cold neutral medium (CNM)  and warm neutral medium (WNM), with a characteristic temperature equal to 100K and 8000 K, respectively.
CNM lies close to the plane of the galaxy with a scalehight equal to 100 pc reaching close to 220 pc in the outer region of the Milky Way, whereas 
the WNM has a scalehight $\sim$ 1kpc everywhere. Also, it is common for the  \HI{} surface densities to peak away from the center in many galaxies, indicating the presence 
of \HI{} holes in the center. On the other hand, the molecular gas is localized in the giant molecular clouds, which are active regions of star formation and 
characterized by turbulent pressures. The surface density of the  gas disc can be fitted with an off-centered double gaussian 
\citep{begum2004kinematics, patra2014modelling} as given below

\begin{uprightmath}
\begin{equation}
 \Sigma(r) =\Sigma_{01}e^-\frac{(R-a_{1})^{2}}{2b^{2}_{1}}  + \Sigma_{02}e^-\frac{(R-a_{2})^{2}}{2b^{2}_{2}}
\end{equation}
\end{uprightmath}
where $\rm \Sigma_{01}$ is the central gas surface density, $\rm a_{1}$ the center, and $\rm b_{1}$ scale length of the gas disc 1, and so forth.
\subsection*{The dark matter halo}
The dark matter halo is the dominant mass component of the galaxy, almost 10 times more massive than the stars and gas put together. Although 
direct searches for dark matter particles have not been fruitful \citep{agnese2018results}, the existence of dark matter is necessary explaining 
the large-scale structure formation \citep{springel2005simulations}, the observed rotation curve \citep{rubin1983dark}, the dynamical mass of galaxy clusters \citep{zwicky1937masses} 
and many other dynamical and structural constraints. The total rotation curve derived from the interferometric \HI{} 21 cm line is the primary diagnostic tracer of 
the dark matter in the disc galaxies\citep{de2008high, oh2015high, lelli2016sparc}. The rotation curve modeled based on the observed baryonic distribution
shows a Keplerian decline in the outer regions of the galaxy. However, it remains constant as obtained from \HI{} 21 cm observations, implying that there is excess mass in the galaxies over and above the observed baryonic mass. This excess mass (dark matter mass) can be obtained by subtracting the contribution of the stellar and the gas mass from the total mass given by \HI{} 21cm rotation curve. 
The dwarf galaxies and low surface brightness (LSB) galaxies have a slowly rising rotation curve, and the high surface brightness (HSB) galaxies typically have a steeply 
rising rotation curve. The observed rotation curve of 
the high surface brightness galaxies can be fitted using the cuspy Navarro-Frenk- White (NFW) dark matter halo profile first proposed by
\cite{navarro1996structure} using N-body simulations of structure formation in the Lambda - Cold Dark Matter paradigm $(\Lambda-CDM)$. 
On the other hand, the cuspy NFW profile fails to reproduce the inner slope of the slowly rising rotation curves of the dwarfs and LSBs. 
It has been shown that the inner slope of the dwarfs and LSBs is fitted well 
by cored pseudo-isothermal (PIS) dark matter halo profile \citep{de2008high, oh2015high}. This mismatch between the cuspy dark matter halo predicted 
by N-body simulations of structure formation in a CDM cosmology and the dark matter density profile predicted by the observed rotation curves of 
the dwarf galaxies is known as the $\rm core-cusp$ problem.
The NFW dark matter halo profile is given as 
\begin{uprightmath}
\begin{equation}
 \rho(r) = \frac{\rho_{0}}{\frac{R}{R_{s}}\big(   1+\frac{R}{R_{s}}     \big)^{2}  }
\end{equation}
\end{uprightmath}
where $\rm \rho_{0}$ is the central density and $\rm R_{s}$ is the characteristic scale of the NFW dark matter density profile.
Similarly, the cored PIS dark matter halo is given as

\begin{uprightmath}
\begin{equation}
 \rho(r) =\frac{\rho_{0}}{1+ \bigg(\frac{R}{R_{c}}\bigg)^{2} }
\end{equation}
\end{uprightmath}
Where $\rm \rho_{0}$ is central density and $\rm R_{c}$ is the core radius of the profile.

However, we may note here that it is possible to explain the observed properties of the galaxies without nvoking the dark matter hypothesis. In Section 1.2, we derive the vertical
dynamics of a prototypical superthin galaxy UGC 7321 using a novel higher dimensional gravity model called the braneworld model.

\section{Vertical structure of disc galaxies}

\begin{figure*}
\hspace{0.5cm}
\resizebox{140mm}{60mm}{\includegraphics{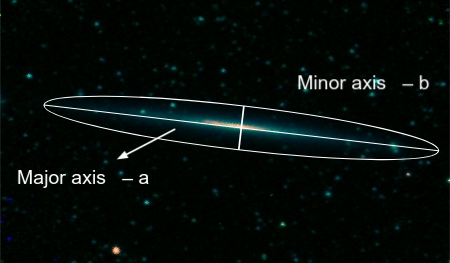}} 
%\vspace{1pt}
\caption{An edge-on disc galaxy inclined at $\rm 90^{\circ}$ to the line of sight. We have schematically indicated the major axis (a) to the minor axis (b) ratio. 
The superthin galaxy appears like \textit{edge of razor blade}.[\textit{Image courtsey: leda.univ-lyon1.fr}] }	
\end{figure*}
The vertical distribution of the stars in the Milky Way can be fitted with two exponentials, one with a larger and another with a smaller scaleheight. This separates 
Milky Way geometrically into a thick and a thin disc \citep{gilmore1983new, juric2008milky, bovy2012spatial}. The thick disc stars are kinematically hot, 
have \textcolor{red}{lower} metallicities, and show a rotational lag with respect to the young disc stars. Structural thick and thin discs are also ubiquitous in the external galaxies: See for 
example, \cite{dalcanton2000structural, dalcanton2002structural,yoachim2005kinematics, yoachim2008lick}. The vertical structure of disc galaxies contains important information 
about the formation and evolution: the oldest stars born during the earliest time of disc formation populate the thick discs and young stars born during recent starburst 
events constitute the thin discs. Understanding how the stars attain their present-day distribution into thick and thin discs and the different mechanisms that 
lead to such a distribution is a pertinent question for understanding the structural evolution of disc galaxies. 

Different processes have been suggested to explain the evolution of the vertical structure of the galaxies. \cite{quinn1993heating} suggest that the accretion 
of satellite galaxies by a pre-existing thin disc can lead to heating up of the disc and also explain the observed vertical profiles 
\citep{villalobos2008simulations, bekki2011origin}. Other mechanisms for disc heating include the massive clump formation in unstable gas-rich discs 
in the early universe \citep{agertz2009disc, ceverino2010high} or the star formation might have occurred away from the plane of the disc during the 
early phase of gas-rich mergers\citep{brook2004emergence}. Other mechanisms include dynamical heating of the thin disc due to the molecular clouds, 
spiral arms, galaxy interactions \citep{spitzer1951possible, lacey1991tidally} and the slow collapse of the proto-galaxy forming the thick and the thin disc 
\citep{eggen1962evidence}.
\begin{figure*}
\hspace{-15.0pt}
\resizebox{150mm}{90mm}{\includegraphics{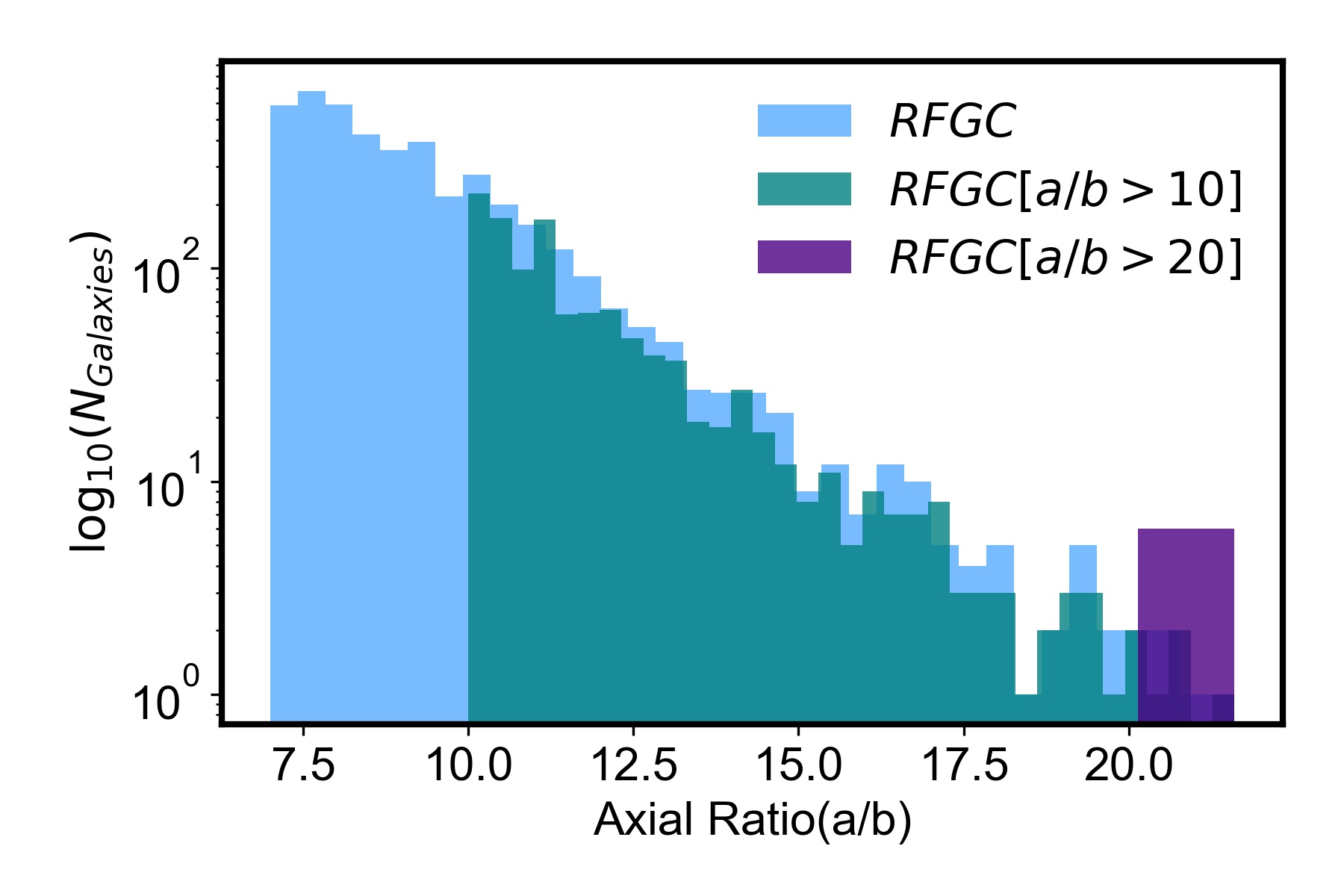}} 
%\vspace{1pt}
\caption{The distribution of the a/b ratio for the galaxies in the RFGC sample studied by \cite{karachentsev2003revised}.  }	
\end{figure*}
Typically galaxies inclined above 90$^{\circ}$ with respect to the line of sight or, in other words, galaxies with almost edge-on orientation are the best test beds 
to study the vertical structure of the galaxies, see Figure 1.3. Due to the edge-on orientation, one can directly observe the vertical features in the galaxy. 
Thus, edge-on galaxies are indispensable for studying the vertical structure of the galaxies. However, one of the possible disadvantages of the edge-on galaxies is that the dust lanes along the line of sight
can obscure the entire plane of the galaxy, seen as dark lanes in the optical images of edge-on disc galaxies.
The edge-on disc galaxies thus are perfectly suitable for deriving the radial variation of the scaleheight of stars and gas and studying the extended diffuse \HI{} 
\citep{swaters1997hi,heald2011westerbork} and the stellar haloes\citep{zibetti2004haloes, monachesi2016ghosts}.  
The thickness of gas layers starts to increase as we move outward along the radius and is called flaring. The flaring gas layer traces the mid-plane dark matter potential
in the galaxy's outer regions. The gas scaleheight at a given radius is determined by the balance between the pressure of the gas layers and the total gravitational 
potential at that radius. The flaring of the gas layers can be used to measure the potential of the dark matter halo. The dark matter halo dominates the total 
potential in outer regions. This flaring of gas layers has been used extensively by \cite{olling1995usage, olling1996highly, banerjee2011progressively,banerjee2008flattened} 
to derive the parameters corresponding to the shape of the dark matter halo in nearby galaxies, where the radial variation of the \HI{} scaleheight can be resolved.
\subsection{Superthin galaxies}
\begin{figure*}
\hspace{1.5cm}
\resizebox{120mm}{30mm}{\includegraphics{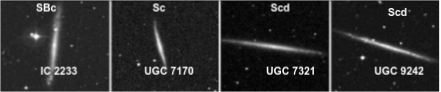}} 
%\vspace{1pt}
\caption{The sample of superthin galaxies studied by \cite{goad1981spectroscopic}. }	
\end{figure*}

This thesis concerns a particular class of edge-on disc galaxies known as superthin galaxies and extremely thin galaxies. Superthin galaxies are 
bulge-less razor-thin systems with major-to- minor axes ratios $\rm (10<a/b<20)$, i.e., the major axis diameter of the stellar disc is 10 times more than the minor axis diameter. 
Galaxies with a/b ratios greater than 20 have been referred to as extremely thin galaxies, as in this thesis. 
These galaxies are low surface brightness galaxies in 
B-band indicating that their deprojected central surface brightness in B-band is greater than $\rm 23.5 mag/arcsec^{2}$, so the superthin galaxies would hardly be visible, 
if they were oriented face-on. A large number of superthin galaxies were discovered in the Palomar Optical Sky Survey (POSS) and have been cataloged in the Revised Flat 
Galaxy Catalog (RFGC) \citep{karachentsev2003revised}, which has about 4444 galaxies with an a/b
ratio greater than 7, about 1100 galaxies with an a/b value greater than 10, and only 6 galaxies with an axis ratio greater than 20 (see Figure 1.4). Although the first image of a 
prototypical superthin galaxy IC 2233 was obtained as early as 1897 by Isaac Roberts, the first systematic spectroscopic study of the superthin galaxies was undertaken 
by \cite{goad1981spectroscopic} who also coined the term $\rm 'superthin'$ for these highly flattened disc galaxies. The galaxies studied by \cite{goad1981spectroscopic} are
shown in Figure 1.5. \cite{goad1981spectroscopic} conclude that the superthin galaxies have a modest central light concentration and small central velocity gradients. Further, 
superthin galaxies have been observed and catalogued as part of \HI{} observations \citep{matthews2000h,huchtmeier2005hi}, optical observations \citep{kautsch2006catalog, bizyaev2017very, 
kautsch2021spectroscopic, bizyaev2021spectral} and Near Infrared (NIR) observations \citep{bizyaev2020near,antipova2021database}. Also, \HI{} 21 cm interferometric observations have been 
used to map the distribution of neutral \HI{} and derive the kinematic properties of the superthin galaxies, see for example 
\citep{abe1999observation,van2001kinematics,uson2003hi,o2010dark,matthews2007h,kurapati2018mass}. In conjunction with optical observations, the \HI{} observation has 
been critical to our current understanding of the existence and origin of these superthin galaxies. The 
\HI{} observations and optical photometry of superthin galaxies have been used extensively to model the dark matter haloes either using the rotation curve 
\citep{van1986dark, kurapati2018mass,banerjee2017mass} constraint or by using both the observed gas scaleheight and the observed rotation curve as constraint  
\citep{banerjee2010dark,peters2017shape}. Thus, the \HI{} observations of the gas scaleheight provides an alternative to using the rotation curve to derive 
the density and the shape of dark matter haloes.. Studies by \cite{zasov1991thickness} and \cite{banerjee2013some} have shown that the
dark matter halo plays an important role in regulating the superthin disc structure. In the following section will present the formulation of the dynamical equations to model
the vertical structure of the disc galaxies.

The existence of the superthin stellar disc in these low surface brightness galaxies remains a mystery. The scaleheight of the superthin galaxies is determined by 
the balance between the vertical gradient of stellar velocity dispersion and the net vertical gravitational potential. \cite{banerjee2010dark} have found 
that the superthin galaxy UGC7321 has a dense, compact dark matter halo by modeling the superthin galaxies using a multi-component model (See Section 1.1.3 ). 
The multi-component model is constrained by using the observed rotation curve and the scale height data. 

UGC7321,IC5249, IC2233 \citep{banerjee2016mass}, and FGC1540 \citep{kurapati2018mass} are superthin galaxies for which the mass models 
were obtained using the \HI{} rotation curves using \HI{} 21cm radio-synthesis observations and the stellar photometry in 3.6$\mu$m.
These results suggest that the dense and compact dark matter halos in superthin galaxies significantly impact the structure and dynamics of the stellar disc. 
\cite{zasov1991thickness} showed that a massive dark matter halo is necessary to counteract bending instabilities in ultra-thin galaxies. It has also been shown by \cite{garg2017origin} that the 
dark matter halo is responsible for regulating the stability of low surface brightness galaxies against local, axisymmetric perturbations using the multi-component 
disc dynamical stability parameter $\rm Q_{RW}$ developed by \citet{romeo2011effective}.

Further, as the the shape of disc galaxies is primarily driven by 
the angular momentum of their stellar discs, it is possible that the high planar-to-vertical axis ratios of stellar discs in superthin galaxies may be 
an outcome of high specific angular momentum. 

\subsection{Dynamical equations governing the vertical disc structure}
Studies of the stars in the solar neighborhood have shown that their vertical velocity dispersion increases with age. This age dependence points out that 
the dispersion of the stars born close to the midplane increase with time due to the effect of the various disc heating mechanisms. \cite{spitzer1953possible} suggested that 
the disc stars can be heated by giant molecular clouds (GMCs) orbiting the disc. Other methods for disc heating include heating of the stars due to the transient spiral waves
\citep{goldreich1965gravitational, barbanis1967orbits, binney2011galactic}, minor mergers with satellite galaxies \citep{toth1992galactic} 
and heating by stellar-mass black holes \citep{lacey1985massive}. Thus, the observed vertical structure of the stellar disc results from heating the disc stars that originally formed near the midplane with small velocity dispersion. 

In this section, we will present the mathematical tools necessary 
to model the observed vertical structure of the superthin galaxies using Boltzmann's equations along with Poisson's equations. 
Since it is not feasible to follow the orbit of each star, we model the galaxy as a distribution of stars, where the probability of finding the star 
at a given point in phase space (q,p) at a given time instance is given as $\rm f(q,p,t)d^{3}qd^{3}p$. f(q,p,t) is defined as the distribution function, and q and p are a pair of 
canonical coordinates. The distribution function is normalized by its definition such that, 

\begin{uprightmath}
\begin{equation}
 \int f(q,p,t)d^{3}q d^{3}p =1
\end{equation}
\end{uprightmath}
where the integration is over the entire phase space.

With the above definition of the distribution function in place, the total derivative of the distribution function is written as;

\begin{uprightmath}
\begin{equation}
 \frac{df}{dt} =\frac{\partial f}{\partial t} + \frac{\partial f}{\partial q} \dot{q}  +  \frac{\partial f}{\partial p} \dot{p}
\end{equation}
\end{uprightmath}

Using the notation w=(q,p), the above equation can be condensed to read 

\begin{uprightmath}
\begin{equation}
 \frac{df}{dt} =\frac{\partial f}{\partial t} + \sum^{6}_{i=1} \frac{\partial f}{\partial w_{i}} \dot{w_{i}}  
\end{equation}
\end{uprightmath}

$\rm \frac{df}{dt}$ in the above equation represents the rate of change of the local probability density as seen by an observer who moves through the phase space with the star.
Now, imposing the condition that the local phase space volume is incompressible or, in other words, the phase space density around the given star remains constant, 
the collisionless Boltzmann equation reads 

\begin{uprightmath}
\begin{equation}
 \frac{df}{dt}=0
\end{equation}
\end{uprightmath}

The collisionless Boltzmann equation in the  galactic cylindrical coordinates $\rm (R,\phi,z)$ is given by

\begin{uprightmath}
\begin{equation}
 \frac{\partial f}{\partial t} + v_{R}\frac{\partial f}{\partial R}  + \frac{v_{\phi}}{R}\frac{\partial f}{\partial \phi} + v_{z}\frac{\partial f}{\partial z} +
 \frac{\partial f}{\partial v_{R}}\bigg( \frac{v^{2}_{\phi}}{R} - \frac{\partial \Phi}{\partial R}  \bigg) - \frac{1}{R} \frac{\partial f}{\partial v_{\phi}} 
 \bigg( v_{R}v_{\phi} + \frac{\partial \Phi}{\partial \phi}\bigg) -\frac{\partial f}{\partial v_{z}} \frac{\partial \Phi}{\partial z}=0
\end{equation}
\end{uprightmath}

Integrating the above equation over the velocities and using $\rm \rho= \int fd^{3}v$, which is the density at any point in the coordinate space 
and defining the mean stellar velocity of the $\rm i^{th}$ component as $\rm \overbar{v_{i}}=\int fv_{i}d^{3}v$, we can write 

\begin{uprightmath}
\begin{equation}
\frac{\partial \rho}{\partial t} + \frac{1}{R} \frac{\partial( R \rho \overbar{v_{R}}) }{\partial R} + \frac{\partial ( \rho \overbar{v_{z}}) }{\partial z} =0
\end{equation}
\end{uprightmath}

Further, by multiplying Equation 1.11 by $\rm v_{R}$ and $\rm v_{z}$ and integrating over the velocity space, we obtain

\begin{uprightmath}
\begin{equation}
 \frac{\partial ( \rho \overbar{v_{R}} )}{\partial t} + \frac{\partial (\rho \overbar {v^{2}_{R}} )}{\partial R} + \frac{\partial (\rho \overbar{v_{R}v_{z}}) }{\partial z}+
 \rho\bigg( \frac{(\overbar{v^{2}_{R}} -  \overbar{v^{2}_{\phi}}  )  }{R}  + \frac{\partial \Phi}{\partial R}       \bigg)=0
\end{equation}
\end{uprightmath}

\begin{uprightmath}
\begin{equation}
 \frac{\partial (\rho \overbar{v_{\phi}} )}{\partial t} + \frac{\partial (\rho \overbar{v_{R}v_{\phi}}) }{\partial R} +\frac{\partial (\rho \overbar{v_{\phi} v_{z} })  }{\partial z}
 +\frac{2\rho \overbar {v_{\phi}v_{R} }}{R}=0
\end{equation}
\end{uprightmath}

\begin{uprightmath}
\begin{equation}
 \frac{\partial (\rho \overbar{v_{z}}) }{\partial t} + \frac{\partial (\rho \overbar{v_{R}v_{z} }) }{\partial R} +  \frac{\partial (\rho \overbar{v^{2}_{z}} ) }{\partial z}
 +\frac{ (\rho \overbar{v_{R} v_{z}} ) }{R} + \rho \frac{\partial \Phi}{\partial z}=0
\end{equation}
\end{uprightmath}

Together, the set of equations  1.12, 1.13, and 1.14 are known as Jean's equations. Equation 1.14 under suitable assumptions can be recast \citep{binney2011galactic} as 

\begin{uprightmath}
\begin{equation}
\frac{1}{\rho} \frac{\partial (\rho \overbar{v^{2}_{z}} ) }{\partial z} = -\frac{\partial \Phi}{\partial z}
\end{equation}
\end{uprightmath}

Equation 1.15, in conjunction with the Poisson equation, constitutes the cornerstone for studying the vertical structure of the disc galaxies. The isothermal gas sheet also 
obeys an equation similar to Equation 1.15 i.e. 
\begin{uprightmath}
\begin{equation}
\frac{1}{\rho} \frac{\partial P }{\partial z} = -\frac{\partial \Phi}{\partial z} 
\end{equation}
\end{uprightmath}
where  $\rm P=\rho v_{z}^{2}$. Equation 1.16 is known as the equation of vertical hydrostatic equilibrium and represents the balance between gravitational force and the 
pressure gradient arising from 
the random motion of the gas molecules in a vertical column.

\subsection{Dynamical modeling using Jean's equations}
The galactic disc is modeled as a system of co-planar and concentric axisymmetric discs of stars and gas, each characterized by its velocity dispersion.
The stellar and the gas components are gravitationally coupled and are in the external force field of a rigid dark matter halo potential. 
The stellar and the gas disc are in vertical hydrostatic equilibrium and are characterized by a constant velocity dispersion at a given value of z 
. The joint Poisson distribution in terms of galactic cylindrical coordinates $\rm (R,\phi,z)$ is;

\begin{uprightmath}
\begin{equation}
\frac{\partial ^{2} \Phi _{\rm{total}}}{\partial z^{2}} +\frac{1}{R} \frac{\partial}{\partial R}(\frac{R \partial \Phi _{\rm{total}}}{\partial R}) =
4 \pi G(\sum_{i=1}^{n}\rho _{i} + \rho_{DM})
\end{equation}
\end{uprightmath}
    
The azimuthal and the radial term vanish due to the galaxy's axisymmetry and flat rotation curve, respectively. Thus Poisson's equation reduces to

\begin{uprightmath}
\begin{equation}
\frac{\partial ^{2} \Phi _{\rm{total}}}{\partial z^{2}}  =4 \pi G(\sum_{i=1}^{n}\rho _{i} + \rho_{DM})
\end{equation}
\end{uprightmath}
The equations governing the dynamics of the $\rm j^{th}$ component of the disc (j=stars, gas) \citep{narayan2002vertical} are obtained by 
combining the joint Poisson's equation and the equation for vertical hydrostatic equilibrium, we get:

\begin{uprightmath}
\begin{equation}
\frac{\partial^{2}\rho_{j}}{\partial z^{2}} = \sum_{i=1}^{n} -4\pi G
\frac{\rho_{j}}{\langle(\sigma_z)_{j}^2\rangle} (\rho_{i} +  \rho_{DM}) +
(\frac{\partial\rho_j}{\partial z})^2 \frac{1}{\rho_{j}};
\end{equation}
\end{uprightmath}
where $\rm \rho_{j}$, $\rm j$ stands for the density and $\rm \langle(\sigma_z)_{j}^2\rangle$ is the vertical velocity dispersion of the $\rm j^{th}$ component 
with $\rm j = 1$ to $n$. In the current work, the gas component can be modeled as a single-component disc i.e the gas component is always characterized by a single 
velocity dispersion value. 
The stellar component is modeled either as a single component or, at the most, as a multi-component disc. 
The dark matter is modeled as a pseudo-isothermal profile \citep{de1988potential} given by

\begin{uprightmath}
\begin{equation}
\rho_{DM}=\frac{\rho_{0}}{(1+\frac{m^{2}}{R^{2}})}
\end{equation}
\end{uprightmath}
where    
\begin{uprightmath}
\begin{equation}
m^2= R^2+\frac{z^2}{q^2}
\end{equation}
\end{uprightmath}
$\rm \rho_0$ is the central core density, $\rm R_{c}$ the core radius, and q the vertical-to-planar axes ratio of the spheroidal the halo.
For a spherical halo $\rm q=1$, oblate $\rm q<1$, prolate $\rm q>1$. We assume a spherical halo in this work.
$\rm \rho_{DM}$ is an input parameter, having been determined by earlier mass modeling studies. The radial dependence of vertical velocity 
dispersion of the stars is parametrized as :

\begin{uprightmath}
\begin{equation}
\sigma_{z}(R)=\sigma_{0s} \exp(-R/{\alpha_{s}}R_{d})
\end{equation} 
\end{uprightmath}

Here $\rm \sigma_{0s}$ and $\rm \alpha_{s}$ are the free parameters. This is closely following \cite{van1989photometry} who modeled the observed near-constant
stellar scale height for a sample of edge-on disc galaxies. For a self-gravitating, isothermal stellar disc in vertical hydrostatic equilibrium, \cite{van1989photometry} 
found $\rm \alpha_{s} = 2$. However, modeling the galactic disc as a self-gravitating stellar disc is reasonable for ordinary galaxies like our Galaxy for which the stellar 
disc dominates the disc dynamics in the inner regions. For low surface brightness galaxies like the superthins, which are gas-rich and dark matter dominated, the multi-component 
model described above is an appropriate choice. Consequently, $\rm \alpha_{S}$ need not be necessarily equal to $\rm2$ to comply with near-constant stellar scale height. 
Finally, the \HI{} vertical velocity dispersion is either parametrized as a polynomial 
\begin{uprightmath}
\begin{equation}
\sigma_{z}(R) = \sigma_{\rm{0HI}} + \alpha_{HI} R +\beta_{HI} R^{2} 
\end{equation}
\end{uprightmath}

with $\rm \sigma_{\rm{0HI}}$, $\rm \alpha_{HI}$ and $\rm \beta_{HI}$ as free parameters
or as an exponential 
\begin{uprightmath}
\begin{equation}
 \sigma_{z}(R)=\sigma_{0HI}e^{-R/{\alpha_{HI}}}
\end{equation}
\end{uprightmath}

with $\rm \sigma_{0HI}$ and $\rm \alpha_{HI}$ as free parameters
to comply with the observed \HI{} scaleheight data. Equation (1.19) thus represents $\rm 2(3)$ coupled non-linear ordinary differential equation in the variables $\rm \rho_{i}$ 
where j = 1 to 2(3).  For a given set of values of the free parameters, the above equations determines $\rm \rho_{j}$ as a function of $\rm z$ and hence scaleheight 
for all $\rm j$ at a given $\rm R$. The observed scaleheight values constrain the velocity dispersion profiles. 
The above equation is solved iteratively using the Runge-Kutta method with initial conditions at midplane $\rm z=0$ given by;
\begin{uprightmath}
\begin{equation}
\frac{d\rho_j}{dz}
\end{equation}
\end{uprightmath}

and    
\begin{uprightmath}
\begin{equation}
\rho_{j}=(\rho_{0})_{j}    
\end{equation}
\end{uprightmath}

\subsection{Distribution function based modeling: Action-Angle based Galaxy Modeling Architecture (AGAMA)}
The multi-component model is immensely useful for deriving the vertical velocity dispersion when the scaleheight of the stars and gas are known as in the case of the 
edge-on galaxies or can be used to derive the scaleheight of the stellar and the gas disc when the vertical velocity dispersion is known from the observation. The 
multi-component model, on the other hand, can not be used to model the radial velocity dispersion as the $\rm \frac{\partial \Phi}{\partial R} =0$ 
due to a flat rotation curve; thus, the radial Jean's equation does not enter into the multi-component model. The ratio of $\rm \frac{\sigma_{z}}{\sigma_{R}}$
is an important diagnostic for measuring the relative importance of the disc heating agents in the vertical as compared to the radial direction in the galaxy; see, 
for example \cite{2001ASPC..230..221M}. Thus, in order to derive the radial and vertical velocity dispersion of the stars, we use the distribution function-based 
action-angle modeling method implemented in the publicly available toolkit  Action-based Galactic Modelling Architecture (AGAMA) \citep{vasiliev2018agama}. 
If the distribution function (Equation 1.6) is a solution of the steady-state collisionless Boltzmann equation, then it is a function of the integrals of the motion (Jeans Theorem). The integrals of motion are defined such that

\begin{uprightmath}
\begin{equation}
 I[q(t1),p(t1)] = I[q(t2),p(t2)]
\end{equation}
\end{uprightmath}

Thus, $\rm I[q(t),p(t)]$ is an integral of motion if
\begin{uprightmath}
\begin{equation}
 \frac{d I[q(t),p(t)]}{dt} =0
\end{equation}
\end{uprightmath}

The core idea of the distribution function-based modeling of the galaxies is based on the Jeans theorem, which states that
\begin{itemize}
 \item  Any steady-state solution of the collisionless Boltzmann equation depends on the phase space coordinates only through the integrals of motion in the given potential.
 Moreover, any function of the integrals yields a steady-state solution of the Boltzmann equation.
\end{itemize}

Jeans theorem can be proved as follows:

\begin{uprightmath}
\begin{equation}
 \frac{d I[q(t),p(t)]}{dt} =\frac{\partial I[q(t),p(t)] }{\partial t} + \frac{\partial I[q(t),p(t)] }{\partial q} \dot{q}  +  \frac{\partial I[q(t),p(t)] }{\partial p} \dot{p} =0
\end{equation}
\end{uprightmath}

Using Hamilton's equation $\rm \dot{q}=\frac{dH}{dp}$  and $\rm \dot{p}=-\frac{dH}{dq} = -\frac{d\Phi}{dq}$

\begin{uprightmath}
\begin{equation}
 \frac{\partial I[q(t),p(t)] }{\partial q} \dot{q}  -  \frac{\partial I[q(t),p(t)] }{\partial p} \frac{d\Phi}{dq} =0
\end{equation}
\end{uprightmath}

Thus, the condition for $\rm I[q(t),p(t)]$ to be the integral of motion is same as $\rm I[q(t),p(t)]$ to be solution of steady-state collisionless Boltzman equation. Further, if 
we consider the distribution function to be a function of the integral of motion such that $\rm f(I[q(t),p(t)],..........I_{n}[q(t),p(t)] )$ then 
\begin{uprightmath}
\begin{equation}
 \frac{d f(I[q(t),p(t)],..........I_{n}[q(t),p(t)] ) }{dt}= \sum^{n}_{m=1} \frac{df}{dI_{m}} \frac{dI_{m}}{dt} =0
\end{equation}
\end{uprightmath}

Thus, the above proves the Jeans theorem, i.e., finding the solution to Jean's equations is similar to computing the distribution functions as functions of the 
integrals of motion. 

The distribution functions implemented in AGAMA \citep{2001ASPC..230..221M} are functions of actions and angles instead of the integrals of motion. As the action themselves are integrals of motion and 
can be expressed in terms of more familiar integrals such as energy $\rm E$, angular momentum $\rm L$ or $\rm L_{z}$, 
and the third integral $\rm I_{3}$ if it exists.

The stellar disc and the dark matter halo are defined by their 
distribution functions in terms of actions, whereas the \HI{} is included as a static density component serving as a source of a rigid external potential. 
We model the stellar disc with an exponential surface 
density consistent with the parameters obtained from the optical photometry and model the stellar distribution function using the quasi-isothermal distribution function. 
The stellar surface density is given by

\begin{uprightmath}
\begin{equation}
 \Sigma_{s} (R) =\Sigma_{0,s}e^{\frac{-R}{R_{d,s}}}e^{\frac{-z}{h_{z,s}}}
\end{equation}
\end{uprightmath}

where, $\rm \Sigma_{0,s}$ is the central stellar density, $\rm R_{d,s}$ is the disc scalelength and $\rm h_{z,s}$ is the exponential scaleheight. 
The quasi-isothermal distribution function is given as follows:
\begin{uprightmath}
\begin{equation}
 f(J)=f_{0}(J_{\phi}) \frac{\kappa}{\sigma_{R}^{2}} e^{- \frac{-\kappa J_{R}}{\sigma_{R}^{2}}}\frac{\nu}{\sigma_{z}^{2}}e^{\frac{-\nu J_{z}}{\sigma_{z}^{2}}}
\end{equation}
\end{uprightmath}

Where $\rm \kappa$ and $\rm \nu$ are the radial and vertical epicyclic frequencies, respectively. $\rm \sigma_{R}$ and $\rm \sigma_{z}$ are the stellar velocity dispersion in the 
$\rm R$ and $\rm z$ directions, respectively. The radial and the vertical epicyclic frequencies $\rm \kappa$ and $\rm \nu$ can be derived from the total potential and are given as
\begin{uprightmath}
\begin{equation}
\!
\begin{aligned}
 \kappa =\sqrt{    \frac{\partial^{2} \Phi }{\partial R^{2}} +\frac{3}{R} \frac{\partial \Phi }{\partial R}             }\\
 \nu =  \sqrt{    \frac{\partial^{2} \Phi }{\partial z^{2}}}
\end{aligned} 
\end{equation}
\end{uprightmath}

The actions $\rm J_{R}$ and $\rm J_{z}$ are related to the integrals of motion as $\rm E_{R}=\kappa J_{R}$ and $\rm E_{z}=\nu J_{z}$
The radial dependence of the velocity dispersion in the radial and the vertical directions is modeled as \citep{2001ASPC..230..221M}

\begin{uprightmath}
\begin{equation}
\!
\begin{aligned}
\sigma_{R}(R)&= \sigma_{R,0}e^{\frac{-R}{R_{\sigma_{R}}}}\\
\sigma_{z}(R)&= \sigma_{z,0}e^{\frac{-R}{R_{\sigma_{z}}}} 
\end{aligned}
\end{equation}
\end{uprightmath}

$\rm J_{R}$, $\rm J_{z}$ and $\rm J_{\phi}$ are the actions of the stellar discs in the $\rm R$, $\rm z$ and $\rm \phi$ directions, respectively. 

Here, $\rm J^{2}_{\phi}=R^{3} \frac{\partial \Phi }{\partial R}$, $J^2 = {J_R}^2 + {J_{\phi}}^2 + {J_{z}}^2$ and  
$\rm f_{0}(J_{\phi}) = \frac{\Sigma(R) \Omega (R)}{2 \pi^{2} \kappa^{2}(R) }$, where $\rm \Phi$ and $\rm \Omega$ are the total gravitational potential and angular velocity, respectively. The \HI{} density is modeled as a static potential as follows:

\begin{uprightmath}
\begin{equation}
  \Sigma_{g} (R) =\Sigma_{0,g}e^{-(\frac{R}{a_{g}})^{2}}
\end{equation} 
\end{uprightmath}

The dark matter density is modeled as 3-parameter spherically symmetric function given by:

\begin{uprightmath}
\begin{equation}
 \rho_{r}= \rho_{0} (r/a)^{-\gamma}\bigg(1+\bigg(\frac{r}{a}\bigg)^{\alpha} \bigg)^{(\frac{\gamma - \beta}{\alpha})}
\end{equation}
\end{uprightmath}

we use, $\rm \alpha=1$, $\rm \beta=3$, $\rm \gamma =1$, in the above to mimic the cuspy NFW profile (See Equation 1.4), where $\rm \rho_{0}$ is the central density and $\rm a$ 
is the characteristic scale of the NFW halo given by $\rm a=R_{200}/c$. $\rm R_{200}$ and $\rm c$ are the virial radius and the concentration parameter of the NFW profile, respectively. 
Virial radius is defined as the radius at which the density of the dark matter halo equals the critical density of the universe.
Similarly, we set $\rm \alpha=2$, $\rm \beta=2$, $\rm \gamma =0$, for the cored PIS halo (Equation 1.5), then $\rm a$ mimics 
the core radius of the density profile and $\rm \rho_{0}$ the central density. 
We derive the distribution function for the dark matter halo corresponding to the input density profile using the $\rm Quasi-Spherical$ distribution function implemented in AGAMA,
with the help of the generalized Eddington inversion formula. The general method for setting up equilibrium models of the galaxies using AGAMA consists of manually choosing 
the parameters of the density profiles and the distribution functions self-consistently to match the observations \citep{10.1093/mnras/stab155}.
AGAMA implements a more sophisticated method known as the self-consistent iterative method for constructing realistic models of the galaxy to match the observed properties. It consists 
of computing the potential and evaluating the actions and the distribution function as a function of the action and angles, then computing the density from 
the distribution function. It iteratively repeats the procedure until the distribution function's density profile matches each component's observed density profile. The self-consistent iterative method in AGAMA consists of the following steps:

\begin{itemize}
 \item Input the density of the stellar disc, gas disc, and the dark matter halo. We get the stellar density from the optical or 3.6$\rm \mu$m photometry, \HI{} density from the 
        \HI{} 21 cm synthesis observations, and the dark matter density from the mass modeling.
        
 \item Combine these densities and integrate the Poison's equation to obtain the system's total potential.
 
 \item Compute the actions and the angles from the total potential using the action finder implemented in AGAMA, which is based on axisymmetric stackel fudge 
       \citep{binney2012actions}.
       
 \item Set up the distribution function, which are functions of actions and angles, and create the galaxy model object.
 
 \item Iteratively samples the densities from the galaxy model object till the densities match with the observed densities.
\end{itemize}

One of the significant differences between AGAMA and the multi-component model is that we use the density of the gas disc as a static background potential in AGAMA, whereas the 
gas density is taken to be gravitationally coupled to the stars. In the multi-component model, we 
explicitly solve the system of Jean's+ Poisson equation, whereas in AGAMA, we construct solutions to the Jeans equations by setting up distribution functions which are
a function of the integrals of motion. Jean's theorem guarantees that these distribution functions will be solutions to Jean's equations. Another critical difference is 
that AGAMA allows for simultaneous computation of the $\rm \sigma_{z}$ and $\rm \sigma_{R}$; on the other hand, we can only compute the $\rm \sigma_{z}$ using the 
multi-component model, which we use as an input parameter in the AGAMA.

\subsection{Dynamical stability of disc galaxies: Multi-component stability parameter}
The Toomre stability criterion \cite{toomre1964gravitational} $\rm Q= \frac{\kappa \sigma  }{\pi G \Sigma}$ shows the equilibrium condition between the epicyclic frequency 
of self-gravitating matter $\rm \kappa$, the radial velocity dispersion $\rm \sigma$ and the surface density $\rm \Sigma$. 
Where, $\rm \kappa$ at a radius R is defined as $\rm \kappa^2(R)= \big( R\frac{d\Omega (R)^{2}}{dR} + 4\Omega (R)^{2} \big)$, and $\rm \Omega$
is the angular frequency defined as $\rm \Omega (R)^{2}=\frac{1}{R}\frac{d\Phi_{Total}}{dR}= \frac{V^{2}_{rot}}{R^{2}}$. $\rm \Phi_{Total}$ is the total gravitational 
potential, and $\rm V_{Rot}$ is the total rotation velocity. 

The multi-component disc stability parameter $\rm Q_{RW}$ \citep{romeo2011effective} appraising the stability of the composite $\rm star+ gas$ disc is given by

\begin{uprightmath}
\begin{equation}
\frac{1}{Q_{RW}} = \left\{
                \begin{array}{ll} \frac{ W_{\sigma} }{T_{s}Q_{s}} + \frac{1}{T_{g}Q_{g}}  \hspace*{0.5cm} if \hspace*{0.5cm}  T_{s}Q_{s} > T_{g}Q_{g}\\  
         \frac{1}{T_{s}Q_{s}} +\frac{W_{\sigma}}{T_{s}Q_{s}}                            \hspace*{0.5cm} if \hspace*{0.5cm} T_{s}Q_{s} < T_{g}Q_{g}  
         \end{array}
              \right.
\end{equation}
\end{uprightmath}

\noindent where the weight function $\rm W$ is given by
\begin{uprightmath}
\begin{equation}
 W_{\sigma} =\frac{2\sigma_{s} \sigma_{g}}{\sigma_{s} ^{2} + \sigma_{g}^ {2}}
\end{equation}
\end{uprightmath}

The thickness correction is defined as;
\begin{uprightmath}
\begin{equation}
 T \approx 0.8 + 0.7 \frac{\sigma_{z}}{\sigma_{R}}
\end{equation}
\end{uprightmath}

$\rm\sigma_{s}$ and $\rm \sigma_{g}$ are the stellar and gas velocity dispersion. The value of radial velocity dispersion $\rm \sigma_{s}$ is needed for computing the stability
from the action-angle modeling. $\rm Q_{s}$ and $\rm Q_{g}$ are the stellar and gas Toomre criterion. 
$\rm Q_{RW} >1$ means that the $\rm stars+gas$ disc is stable against axisymmetric perturbations.

\subsection{Specific angular momentum of galactic discs}
The Fall relation connects the mass of the disc galaxies to their specific angular momentum. The Fall relation is well established for disc galaxies of varied morphologies 
\citep{posti2019galaxy, 2021A&A...647A..76M, marasco2019angular}. Studies of superthin galaxies and low surface brightness galaxies 
indicated that low surface brightness galaxies had higher specific angular momentum than typical 
disc galaxies, suggesting that high specific angular momentum drives the superthin disc structure. 
Given a rotation curve $\rm V(R)$ and a surface density profile $\rm \Sigma(R)$, the specific angular momentum $\rm j$ is given by
\begin{uprightmath}
\begin{equation}
 j_{i}(<R)= \frac{2 \pi \int^{R} _{0} R^{'2} \Sigma_{i}(R^{'}) V_{i}(R^{'})dR^{'} }{ 2 \pi \int^{R} _{0} R^{'} \Sigma_{i}(R^{'}) dR^{'}}
\end{equation}
\end{uprightmath}

In the above equation $\rm i$ indexes over stars$\rm (*)$, gas $\rm (g)$ and baryons ($\rm b$, defined as the sum of stars and gas).

\section{'Dark Mass' from an alternative theory of gravity: The braneworld paradigm}
Our current understanding of observational cosmology strongly depends on the dark matter and dark matter hypothesis. Galaxies act as valuable probes for understanding the
gravitational properties of dark matter. The first observational evidence of dark matter came from studies of galaxy clusters. \citep{zwicky1937masses} pointed out 
that the dynamical mass of galaxy clusters obtained by the spectroscopic method far exceeds the mass estimated by the photometric method. 
The spectroscopic method measures the radial velocity dispersion of the galaxies in the cluster and then uses the virial theorem to estimate the mass enclosed inside the radius 
at which the galaxy clusters velocity dispersion was measured. On the other hand, the photometric method consists of measuring the luminosity of the galaxy cluster and 
then converting it into the mass using a suitable choice of mass-to-light ratio. It was found that the mass of a galaxy cluster 
estimated by the spectroscopic method is 100 times more than the mass estimated by the photometric method. This indicates that there is more mass in the cluster than is traced by
the luminous component. Further, for the mass estimated from the photometry to match that from the spectroscopy, one would need a mass to light $\rm (M/L)$ ratio close to 400, 
which exceeds the value observed in the solar neighborhood (M/L = 2 - 3). The second substantial evidence for the presence of dark matter comes from studying the 
rotation curve of spiral galaxies. If we derive the rotation curve of a spiral galaxy corresponding to the surface density of stars and gas obtained from optical photometry and \HI{} 
21 cm interferometric observations, we find that their theoretical
rotation curve shows a Keplerian decline. However, \cite{rubin1970rotation} from the spectroscopic studies of the stellar emissions show that the rotation curve, 
instead of showing a Keplerian decline, remained flat. 

The final convincing evidence for the presence of dark matter came from spectroscopic studies of the \HI{} rotation curves of 20 spiral galaxies by \cite{bosma1978distribution}.
\cite{bosma1978distribution}, showed that the rotation curve due to \HI{} which extends further beyond the optical radius, shows no signs of
decline but remains flat, indicating that the mass in the outer regions should increase, which is ascribed to the dark matter halo. 
Another important piece of the puzzle about the origin of dark matter comes from structure formation in the universe. Gravitational potential wells due to 
dark matter overdensities are the drivers of the gravitational instabilities, which lead to the structure formation in the universe. Although 
the dark matter hypothesis resolves a host of astrophysical phenomena, the fundamental particle constituting the dark matter has evaded the dark matter search experiments, 
See, for example, Cryogenic Dark Matter Search (CDMS) \citep{agnese2013silicon}. It opens up the possibility of the gravitational origin of the dark matter, where 
Newtonian or Einstein's gravity is modified to explain the missing mass problem and other astrophysical phenomena that are conventionally 
(\citep{pawlowski2015persistence, Kroupa:2014ria, Peebles:2010di}) explained using dark matter.

\cite{milgrom1983modification} made the first attempt at modifications of
Newtonian gravity, known as modified Newtonian Dynamics (MOND), has successfully explained the observed rotation curves of the spiral galaxies 
\citep{sanders1998rotation,de1998testing, sanders2007confrontation}. Other than the modifications of the Newtonian gravity, there have been attempts at modifications of 
the Einsteins equations (\citep{ Csaki:1999mp,binetruy2000brane, Maartens:2001jx}) in order to explain the various astrophysical phenomenon which
warrants the presence of dark matter. The braneworld gravity model was developed as a fundamental theory that can unify the standard model with gravity
and provide an alternative to the dark matter hypothesis.
In the braneworld model the standard model particles and field are confined in the brane whereas the the gravity reside in the bulk 
\citep{Antoniadis:1990ew,Antoniadis:1998ig,ArkaniHamed:1998rs,Randall:1999vf,Randall:1999ee,Csaki:1999mp,Garriga:1999yh}. \cite{maartens2004brane} 
consider a 3-brane embedded in a five-dimensional bulk, and the Einsteins equations are modified due to the non-local effects of the bulk Weyl tensor, the Weyl stress 
tensor acts as a fluid with its energy and density. \cite{mak2004can,harko2006galactic, boehmer2007galactic, rahaman2008galactic, gergely2011galactic} 
showed that such a model successfully complies with observed rotation curves of galaxies. 

\subsection{Motion of test particles in the braneworld model}
In braneworld model, we consider a 3 brane to be embedded in a 5-D bulk, where the standard model particles are embedded in the brane and the gravity is confined to the bulk.
The extra dimensions bestow an effective energy-momentum tensor on the brane due to the non-local effects of the bulk Weyl tensor, which plays the role of the dark matter.
Cosmological simulations show that the mass distribution of the dark matter is isotropic \citep{de1988potential, navarro1997universal}, so we use a static spherically symmetric 
metric given by 

\begin{uprightmath}
\begin{align}
ds^2=-e^{\nu(r)}dt^2 + e^{\lambda(r)}dr^2 + r^2(d\theta^2 +\rm{sin}^2 \theta d\phi^2 )\label{15}
\end{align}
\end{uprightmath}

The circular velocity of the test particles in the above spacetime is given by \citep{gergely2011galactic} 
\begin{uprightmath}
\begin{align}
v_{c}^2=\frac{r\nu^\prime}{2} \label{16}
\end{align}
\end{uprightmath}

The gravitational field equations and the energy-momentum tensor conservation are given by

\begin{uprightmath}
\begin{align}
&-e^{-\lambda}\bigg(\frac{1}{r^2}-\frac{\lambda^\prime}{r}\bigg)+\frac{1}{r^2}=3\alpha U   \\
&e^{-\lambda}\bigg(\frac{\nu^\prime}{r}+\frac{1}{r^2}\bigg)-\frac{1}{r^2}=\alpha (U+2P)    \\
& \frac{e^{-\lambda}}{2}\bigg(\nu^{\prime \prime}+  \frac{\nu^{\prime 2}}{2}+\frac{\nu^{\prime}-\lambda{^\prime}}{r}-\frac{\nu^{\prime}\lambda^{\prime}}{2}\bigg)                  =\alpha (U-P)\\
&\nu^{\prime} = - \frac{U^{\prime} + 2P^{\prime}}{2U + P} - \frac{6P}{(2U + P)r}
\end{align}
\end{uprightmath}

where prime denotes derivative with respect to $\rm r$ and $\rm \alpha = \frac{1}{4\pi G_{4}\lambda_{T}}$.
One can show that the solution of these equations leads to the following form for $\rm e^{-\lambda}$,
\begin{uprightmath}
\begin{align}
e^{-\lambda} &= 1 - \frac{\Lambda_{4}}{3}r^{2} - \frac{Q(r)}{r} - \frac{C}{r} \label{16-1}
\end{align}
\end{uprightmath}

where $\rm C$ is an arbitrary integration constant and $\rm Q(r)$ is defined as \citep{gergely2011galactic},

\begin{uprightmath}
\begin{align}
Q(r) = \frac{3}{4\pi G_{4}\lambda_{T}}\int r^{2}U(r)dr \label{16-2}
\end{align}
\end{uprightmath}
 
From the form of $\rm e^{-\lambda}$, it can be inferred that $\rm Q(r)$ is the gravitational mass originating from the dark radiation and can be 
interpreted as the ``dark mass" term.

Further, one can show that for a static, spherically symmetric spacetime, the dark radiation $\rm U(r)$ and dark 
pressure $\rm P(r)$ satisfy,
\begin{uprightmath}
\begin{align}
\frac{dU}{dr} = -2\frac{dP}{dr} - 6\frac{P}{r} - \frac{(2U + P)[2G_{4}M + Q + \{ \alpha(U + 2P+\frac{2}{3} \chi) \}r^{3}]}{r^{2}\big{(} 1 - \frac{2G_{4}M}{r} - \frac{Q(r)}{r} - \frac{\chi }{3}r^{2}\big{)}} \label{17}
\end{align}
\end{uprightmath}

and
\begin{uprightmath}
\begin{align}
\frac{dQ}{dr} = 3\alpha r^{2}U. \label{18}
\end{align}
\end{uprightmath}

where $\rm \alpha = \frac{1}{4\pi G_{4}\lambda_{T}}$ and $\rm \chi = -\Lambda_{4}$ \citep{gergely2011galactic}. 
In the subsequent calculations, we will neglect the effect of the cosmological constant $\rm \Lambda_{4}$ \citep{gergely2011galactic} on the vertical 
scale height of the galaxies, i.e. we will take $\rm \chi = -\Lambda_{4}=0$.  Since the observed cosmological constant required to explain the 
accelerated expansion of the universe is extremely small ($\rm \Lambda_{4}\approx 10^{-52} \rm m^{-2}~ or~ 10^{-122} $ in Planckian units), its effect on the 
the mass-energy of the galaxy can be ignored as being several orders of magnitude smaller than the observed masses. 
Equations (1.50) and (1.51) can be recast into a more convenient form, namely,
\begin{uprightmath}
\begin{align}
\frac{d\mu}{d\theta} &= -(2\mu + p)\frac{\tilde{q} + \frac{1}{3}(\mu + 2p) }{1 - \tilde{q} } - 2\frac{dp}{d\theta} + 2\mu - 2p \label{19}\\
&=-2v^2_{tg}(2\mu+p)- 2\frac{dp}{d\theta} + 2\mu - 2p \nonumber \\
\frac{d\tilde{q}}{d\theta} &= \mu - \tilde{q} \label{20} 
\end{align} 
\end{uprightmath}

by defining the variables, 

\begin{uprightmath}
\begin{align}
\tilde{q} = \frac{2G_{4}M + Q}{r} ;~~~\mu = 3\alpha r^{2}U ; ~~~ p= 3\alpha r^{2}P ; ~ ~~\theta = ln ~ r ; \label{20-1}
\end{align}
\end{uprightmath}

Equations (1.52) and (1.53) can be referred to as the differential equations governing the source terms on the brane, while the circular velocity 
of the test particle, $\rm v_c$ assumes the form,

\begin{uprightmath}
  \begin{align}
 v^2_{c}=\frac{1}{2}\frac{\tilde{q}+\frac{1}{3}(\mu+2p)}{1-\tilde{q}} \label{20-2}
  \end{align}
\end{uprightmath}

The equation of state for the Weyl fluid can be rewritten as \citep{gergely2011galactic}, 
\begin{uprightmath}
\begin{align}
p(\mu)=(a-2)\mu-B 
\label{26}
\end{align}
\end{uprightmath}

The equation of state is similar to the boundary condition imposed on the brane, which is preserved during the extra-dimensional evolution \citep{Keresztes:2009hy,Keresztes:2009wb} due to the static character of the problem. The above equation of state reduces to 
that of Schwarzschild limit for the case of a = 3 and B = 0.

\subsection{Density profile for the Weyl fluid}
Equation  1.52 can be simplified as

\begin{uprightmath}
\begin{align}
(2a-3)\frac{d\mu}{d\theta}=-\frac{(a\mu-B)(\tilde{q}+(2a-3)\mu/3-2B/3)}{1-\tilde{q}}+2\mu(3-a)+2B \label{22}
\end{align}
\end{uprightmath}

whereas the equation for the reduced dark radiation assumes the form,
\begin{uprightmath}
\begin{align}
&\mu(\theta)=\theta^{2(3-a)/(2a-3)}\rm{exp}\bigg[-\frac{2a}{2a-3}\int v^2_{c}(\theta)d\theta   \bigg]\times \nonumber \\
&\bigg \lbrace C -\frac{3B}{2a-3}\int [1+v_c^2(\theta)]\theta^{-2(3-a)/(2a-3)}\times \nonumber \\
& \rm{exp}\bigg[\frac{2a}{2a-3}\int v_c^2 (\theta) d\theta\bigg]   \bigg \rbrace \label{23}
\end{align}
\end{uprightmath}

where $\rm C$ is an arbitrary integration constant \citep{gergely2011galactic}.
We consider the case when $\rm a\neq \frac{3}{2}$ and $\rm \tilde{q}<<1$.The ratio $\rm \tilde{q}\approx \frac{GM_U}{R}\approx 10^{-7}<< 1$,given a dark matter halo mass equal 
to $\rm M  \approx 10^{12}M_\odot$ and radius $\rm R  \approx 100$ kpc, remains roughly constant at all radius as mass of galaxy increases with the radius.
So we neglect the higher order terms in $\rm \tilde{q}$ owing to the smallness of the $\rm \tilde{q}$, such that equation 1.53 assumes the form,

\begin{uprightmath}
\begin{align}
\frac{d^2\tilde{q}}{d\theta^2}+m\frac{d\tilde{q}}{d\theta}-n\tilde{q}=b  
\label{24}
\end{align}
\end{uprightmath}

\begin{uprightmath}
\begin{align} 
 m=1-\frac{B}{3}-\frac{2}{3}\frac{a(B-3)+9}{2a-3}~~~~a\neq \frac{3}{2}
\label{25}
\end{align}
\end{uprightmath}

\begin{uprightmath}
\begin{align} 
n=\frac{2}{3}\frac{a(2B-3)+9}{2a-3} ~~~~a\neq \frac{3}{2}
\label{26}
\end{align}
\end{uprightmath}

\begin{uprightmath}
\begin{align}  
 b=\frac{2}{3}\frac{B(B-3)}{3-2a}   ~~~~a\neq \frac{3}{2} 
\label{27}
\end{align}
\end{uprightmath}

The general solution of Equation (1.59) is,
\begin{uprightmath}
\begin{align} 
\tilde{q}(r)=q_0+ C_1 r^{l_1} + C_2 r^{l_2}
\label{28}
\end{align}
\end{uprightmath}

where $\rm C_1$ and $\rm C_2$ are constants of integration and $\rm q_0$ is given by,
\begin{uprightmath}
\begin{align} 
q_0=-\frac{b}{n}=\frac{B(B-3)}{a(2B-3)+9}
\label{29}
\end{align}
\end{uprightmath}

while 
\begin{uprightmath}
\begin{align}
 l_{1,2}=\frac{-m\pm\sqrt{m^2+4n}}{2}
\label{30}
\end{align}
\end{uprightmath}

The solution for reduced dark radiation is given by,
\begin{uprightmath}
\begin{align}
\mu(r)=q_0+C_1(1+l_1)r^{l_1}+C_2(1+l_2)r^{l_2} 
\label{31}
\end{align}
\end{uprightmath}

In the original radial coordinate $\rm r$, the solution for dark radiation $\rm U(r)$ is,

\begin{uprightmath}
\begin{align}
 \rho_h(r)=3\alpha U(r)=\frac{q_0}{r^2}+C_1(1+l_1)r^{l_1-2}+C_2(1+l_2)r^{l_2-2} \label{32}
\end{align}
\end{uprightmath}

The profile of the dark mass is given by,
\begin{uprightmath}
\begin{align}
Q(r)=r(q_0+C_1r^{l_1}+C_2r^{l_2})-2GM
\label{33}
\end{align}
\end{uprightmath}

where $\rm M$ is the baryonic mass. The tangential velocity of a test particle in the dark mass-dominated region is given by,

\begin{uprightmath}
\begin{align}
v_c^2\approx v_{c_\infty}^2+\gamma r^{l_1} +\eta r^{l_2} ~~~\rm{where,}
\label{34}
\end{align}
\end{uprightmath}

\begin{uprightmath}
\begin{align}
v_{c_\infty}^2=\frac{1}{3} (a q_0-B) 
\label{35}
\end{align}
\end{uprightmath}

\begin{uprightmath}
\begin{align}
\gamma=\frac{C_1}{2}\bigg[1+\frac{(2a-3)}{3}(1+l_1)   \bigg]
\label{36}
\end{align}
\end{uprightmath}

\begin{uprightmath}
\begin{align}
\eta=\frac{C_2}{2}\bigg[1+\frac{(2a-3)}{3}(1+l_2)\bigg]
\label{37}
\end{align}
\end{uprightmath}

\citep{gergely2011galactic}.
A negative value of $\rm l_1$ and $\rm l_2$ guarantees a flat rotation curve at a large radius. The constrain on the 
Weyl parameters from the rotation curve have been derived in \cite{gergely2011galactic}. When $\rm \tilde{q}<<1$ and $\rm a\ne3/2$, we can further 
simplify the parameters $\rm m$, $\rm n$, $\rm q_0$ and $\rm v^2_{c_\infty}$ such that,

\begin{uprightmath}
\begin{align}
m\approx \frac{4a-9}{2a-3},
\label{38}
\end{align}
\end{uprightmath}

\begin{uprightmath}
\begin{align}
n\approx -2\frac{a-3}{2a-3},
\label{39}
\end{align}
\end{uprightmath}

\begin{uprightmath}
\begin{align}
q_0\approx \frac{B}{a-3}~~~\rm{and}
\label{40}
\end{align}
\end{uprightmath}

\begin{uprightmath}
\begin{align}
v^2_{c_\infty}\approx \frac{a}{3} \bigg(q_0-\frac{B}{a}	\bigg)
\label{41}
\end{align}
\end{uprightmath}

Equations (1.73) and (1.74) imply that
\begin{uprightmath}
\begin{align}
l_1\approx -1 ~~~\mathrm{and} ~~~l_2\approx -1+\frac{3}{2a-3}
\label{42}
\end{align}
\end{uprightmath}

For $\rm v^2_{c_\infty}$ to be positive,  $\rm a<3/2$, $\rm B\leq 0$ while when $\rm a>3$, $\rm B>0$, this ensures that  $\rm a$ cannot assume values 
between $\rm 3/2$ to $\rm 3$. The density profile for the Weyl fluid assumes the form,

\begin{uprightmath}
\begin{align}
\rho_h(r)\approx \frac{q_0}{r^2}+\frac{3C_2}{2a-3}r^{-3(1-\frac{1}{2a-3})}
\label{43}
\end{align} 
\end{uprightmath}

while the rotation curve is given by,
\begin{uprightmath}
\begin{align}
v_c^2\approx \frac{B}{a-3}+ \frac{C_1}{2}r^{-1}+C_2 r^{-1+\frac{3}{2a-3}}
\label{44}
\end{align}
\end{uprightmath}

By taking $\rm C_1=\frac{2G(M_b+M_U)}{c^2}$, $\rm C_2=C c^2 R_{c(DM)}^{1-\alpha_{DM}}=-\beta_{DM} $ and by defining $\rm \alpha_{DM}=3/(2a-3)$ and $\rm\beta_{DM}=B/(a-3)$, 
the final expressions for the rotation curve and the density profile are given by,

\begin{uprightmath}
\begin{align}
\bigg(\frac{v_c(r)}{c}\bigg)^2\approx \frac{G(M_b+M_U)}{c^2 r}+\beta_{DM}\bigg[1 -\bigg(\frac{R_{c(DM)}}{r}\bigg)^{1-\alpha_{DM}}\bigg]~~~\rm{and}
\label{45}
\end{align}
\end{uprightmath}

\begin{uprightmath}
\begin{align}
\rho_h(r)\approx \frac{c^2\beta_{DM}}{Gr^2}\bigg[1-\alpha_{DM}\bigg( \frac{R_{c(DM)}}{r}\bigg)^{1-\alpha_{DM}}\bigg] 
\label{46}
\end{align}
\end{uprightmath}

Further constraints on the parameters $\rm \alpha$ and $\rm \beta$ are obtained from constraints on the parameters $\rm a$ and $\rm B$, which give  
$\rm \alpha_{DM}<0$ or $\rm 0<\alpha_{DM}<1$ and $\rm 0<\beta_{DM}<<1$ \citep{gergely2011galactic}. 

\subsection{Weyl fluid as a proxy for dark matter in galaxies}
The low surface brightness galaxies (LSBs) are dark matter-dominated systems with a constant dark matter core with a core radius of a few kpc \citep{de2005halo}.
The braneworld model provides a universal rotation curve solution to the core-cusp problem, \cite{gergely2011galactic} show that the 
braneworld model can reproduce the observed rotation curve of 9 low surface brightness galaxies and 9 high surface brightness galaxies.
In the braneworld model, the origin of dark matter is attributed to higher-dimensional gravity
Assuming the core radius to be $\rm R_{c(DM)}$ and the mass of the core to be $\rm M_{DM}$, the density profile describing the dark matter in 
low surface brightness galaxies, which follows from equation 1.80 \citep{gergely2011galactic, komanduri2020dynamical}, is given by     

\begin{uprightmath}
\begin{align}
 \rho_{_{DM}}(r)=\frac{3M_{DM}}{4 \pi  R^{3}_{c(DM)}}(1-H_{k_{DM}}(r)) + 
H_{k_{DM}}(r)\left \lbrace \frac{c^{2} \beta_{DM}}{Gr^{2}}\left(1-\alpha_{DM} \bigg(\frac{R_{c(DM)}}{r}\bigg)^{1-\alpha_{DM}}\right )\right\rbrace
\label{47}
\end{align}
\end{uprightmath}
The Equation (1.82) can be derived by integrating Poisson's equation using the expression for tangential velocity given by 
Equation (1.80) in post-Newtonian approximation \citep{gergely2011galactic, komanduri2020dynamical}.

where, $\rm H(k)$ is a smoothening function that ensures smooth transition from the region of constant density core to the constant 
rotation curve regime where the mass is proportional to the radius. The smoothening function is given by,
\begin{uprightmath}
\begin{equation}
H_{k_{DM}}(r)=\frac{1}{1+\rm{exp}(-2k_{DM}(r-R_{c(DM)}))} 
\label{48}
\end{equation}
\end{uprightmath}

such that it smoothly approaches the Heaviside step function as $\rm k_{DM}$ tends to infinity, i.e.,

 \[    H(R_{c(DM)})= \lim_{k_{DM} \to \infty} H_{k_{DM}}(R_{c(DM)})=\left\{
                \begin{array}{ll}
                  0 ~~~~~r<R_{c(DM)}\\
                  1 ~~~~~r\geq R_{c(DM)}\\
                \end{array}
              \right.
  \]
 
\cite{gergely2011galactic}, show that the braneworld model consistently explains the observed rotation curves of both the 
high surface brightness galaxies and the low surface brightness galaxies. Thus, effectively \cite{gergely2011galactic} show that the braneworld model presents 
a universal rotation curve proving a solution to the core-cusp paradox.

\section{Best-fitting dynamical models using Markov Chain Monte Carlo (MCMC) method}
Markov Chain Monte Carlo are methods for sampling from a probability distribution and is extensively used for probabilistic inference and parameter estimation problems.
The details about implementation and application of MCMC are given in \cite{foreman2013emcee, hogg2018data}
In the context of bayesian probabilistic inference, the quantity of interest is the posterior distribution function 
$\rm p(\theta|D)$, which is the probability of parameters $\rm \theta$ given the data $\rm D$.  The posterior probability distribution function $\rm p(\theta|D)$ from conditional 
probability is given by
\begin{uprightmath}
\begin{equation}
 p(\theta|D) =\frac{P(D|\theta) p(\theta)}{p(D)}
\end{equation}
\end{uprightmath}

where, $\rm p(D|\theta)$ is the probability distribution function for the data $\rm D$ given the parameter $\rm \theta$, $\rm p(\theta)$ and $\rm p(d)$
being the priors and the evidence, respectively. Priors express one's belief about a quantity before taking into account evidence. 
The evidence $\rm p(D)$, usually cancels out as, one is interested in computing the ratio of $\rm p(\theta_{i+1}|D)$ to $\rm p(\theta_{i}|D)$. The probability distribution function $\rm p(D|\theta)$ is constructed from the likelihood function and, in the case of the uniform priors, is proportional to the posterior. 
In this work, we use the gaussian log-likelihood function defined as following

\begin{uprightmath}
\begin{equation}
  \mathcal{L} = -\sum_{n} \frac{ (Y_{n} - M_{n})^{2} }{\sigma^{2}_{n}}
\end{equation}
\end{uprightmath}

where $\rm Y_{n}$ is the data, and $\rm \sigma_{n}$ is the error on the observed data points, respectively. $M_{n}$ is the value of the model 
corresponding to the $\rm n^{th}$ data point.

We use Metropolis Hasting algorithm \citep{hastings1970monte}to calculate the posterior distribution for our model by sampling from the log-likelihood distribution.
The Metropolis-Hasting algorithm consists of the following steps:
\begin{itemize}
 \item Chose a good starting guess value for the parameters $\rm (\theta_{0})$ and estimate the log-likelihood function $\rm p(\theta_{0}|D)$.
 \item Next, we choose value of $\rm \theta_{1}$ such that $\rm \theta_{1}$ is correlated to $\rm \theta_{0}$. This is achieved by sampling a random number X from a gaussian distribution with a mean 0 and variance 1, such that $\rm \theta_{1}=\theta_{0} + X$. We then compute the log-likelihood function at 
 $\rm \theta_{1}$ i.e. $\rm p(\theta_{1}|D)$.
 
\item We define r as  

\begin{uprightmath}
\begin{equation}
r= \frac{\rm p(\theta_{i+1}|D)}{ p(\theta_{i}|D)}
\end{equation}
\end{uprightmath} 

\item If r is larger than 1, then the new values are more likely than the old ones. In that case, we accept the proposed value $\rm \theta_{1}$ and add it to our chain.

\item If r is less than 1, we draw another random number, Y, from a uniform distribution in the interval 0 to 1. We then compare the value of r with the variable Y . If $\rm r > Y$, we accept the new values of $\rm \theta$.
However, if $\rm r < Y$, we throw away our proposed new value of $\rm \theta$ and start again by drawing another set of p parameters.
\end{itemize}

The multi-component model of the vertical structure of a galaxy consists of 3 coupled non-linear differential equations in case the galaxy has a thick + thin stellar disc and a 
gas disc, or two-coupled differential equations in the case when the galaxy has only one stellar disc and a gas disc. An exponential parameterizes the radial profile of the 
velocity dispersion profile of the stars with two parameters $\rm \sigma_{0s}$ and $\rm \alpha_{s}$ (Equation 1.22), and the gas dispersion by three parameters;
$\rm \alpha_{\HI{}}$, $\rm \beta_{\HI{}}$ and $\rm \sigma_{0\HI}$ (Equation 1.23). In the case of galaxies with thick and thin stellar discs, the multi-component 
model has seven free parameters; $\rm \sigma_{0sI}$ and  $\rm \alpha_{0sI}$ for the thick stellar disc, $\rm \sigma_{0sII}$ and $\rm \alpha_{0sII}$ for the thin
stellar disc and $\rm \alpha_{\HI{}}$, $\rm \beta_{\HI{}}$ and $\rm \sigma_{0\HI}$ for the gas disc. On the other hand, in galaxies with only a thin stellar disc, we have 5 free parameters; $\rm \sigma_{0s}$ and $\rm \alpha_{s}$ for the stellar disc and  
$\rm \alpha_{\HI{}}$, $\rm \beta_{\HI{}}$ and $\rm \sigma_{0\HI}$ for the gas disc. For estimating the posterior distribution of 
the free parameters describing the multi-component model, we use the $\rm modMCMC$ task from the publicly available R package Flexible Modeling Environment (FME) 
\citep{soetaert2010inverse}, which implements MCMC using adaptive Metropolis procedure \citep{haario2006dram}. We again use the MCMC method to find optimum parameters 
describing the braneworld model. The braneworld model has 5 free parameters for the vertical velocity dispersion of the stars and gas and 5 free parameters describing 
the braneworld gravity.

%%%%%%%%%%%%%%%%%%%%%%%%%%%%%%%%%%%%%%%%%%%%%%%%%%%%%%%%%%%%%%%%%%%%%%%%%%%%%%%%%%%%%%%%%%%%%%%%%%%%%%%%%%%%%%%%%%%%%%%%%%%%%%%%%%%%%%%%
%%%%%%%%%%%%%%%%%%%%%%%%%%%%%%%%%%%%%%%%%%%%%%%%%%%%%%%%%%%%%%%%%%%%%%%%%%%%%%%%%%%%%%%%%%%%%%%%%%%%%%%%%%%%%%%%%%%%%%%%%%%%%%%%%%%%%%%%
\section{\HI{} 21cm radio-synthesis observations of edge-on galaxies}
Hydrogen is the most abundant element in the universe, permeating the intergalactic medium (IGM) and interstellar medium (ISM), 
and is the primary fuel for forming a star. Based on the abundance of the neutral hydrogen gas, the galaxies can be divided into two major groups, one being 
the gas-rich late-type galaxies, with an abundance of neutral \HI{} gas for sustaining star formation and early-type galaxies, in which the star formation activity has 
depleted the gas reservoirs \citep{draine2010physics}. The gas in the interstellar medium exists in the following thermal phases with their characteristic temperature and densities:
\begin{itemize}
 \item \textbf{Cold neutral medium}\\ In a typical Milky Way-like galaxy, cold neutral medium exists at temperature 10 K - 100 K with a density close to 20 -50 $\rm particle/cm^{3}$. The gas lies close to the 
 plane with scalehights in the range of 100 - 300 pc.
 
 \item \textbf{Molecular Gas}\\ At temperatures less than 50 K and density in range $\rm 10^{2} -10^{6} particles/cm^{3}$, the hydrogen gas is found in molecular phase. 
  The molecular gas disc is the active site of star formation and is confined in a very thin disc close to the center with scaleheight equal to 80 pc.
 
 \item \textbf{Warm neutral and Warm ionized medium}\\ The hydrogen gas exists at a temperatures 6000 - 10000 K and densities  $\rm 0.2 - 0.5 particles/cm^{3}$. The warm medium 
 can exist in both ionized (WIM) and neutral states (WNM). The warm gas becomes ionized by radiation from the young stars; on the other hand, the background radiation fields 
only have sufficient energy to heat the gas. Thus the gas remains neutral. WNM extends from 300 pc to 400 pc above the plane, whereas the WIM lies close to 1000 pc.
 
 \item \textbf{Hot ionized medium}\\ The hot ionized medium consists mainly of the coronal gas at temperatures $\rm 10^{6}\, K$ and $\rm 10^{7} \, K$. The gas is completely 
 ionized and has a density equal to $\rm 10^{-4} particles/cm^{3}$ to $\rm 10^{-2}particles/cm^{3}$ with a scaleheight in between 1000 pc to 3000 pc. The high temperatures result from supernova explosions or other high energetic events.
  
\end{itemize}

The neutral hydrogen gas is uniformly distributed over the entire galaxy and extends beyond the optical radius, making it a good trace of the kinematics and the dynamics
at large radii. The neutral \HI{} is traced by the 21 cm line in both absorption and the emission. The 21 cm line originates from the interaction of the magnetic moment 
of their proton with the magnetic moment of the electron. The ground state of the hydrogen atom is described by the four quantum numbers n=1,l=0,$\rm m_{l}=0$, and 
$\rm m_{s}= \pm 1/2$, and the energy of this state is determined by the principal quantum number n=1. But the n=1 state is four-fold degenerate, as $\rm s=1/2$ for 
electrons give $\rm m_{s}=\pm1/2$ and the  spin of proton $\rm i=1/2$ give $\rm m_{i}=\pm 1/2$. This degeneracy is lifted by hyperfine splitting, which arises 
due to the interaction between the proton's magnetic moment and the electron's magnetic moment. When the spin of the proton and electron are anti-parallel, then 
$\rm f_{0}=|s-i|=0$, and $\rm m_{f}=0$, and on the other hand when the spin of the proton and the electron are parallel then $\rm f_{1}=s+i=1$, and $\rm m_{f}=0,\pm1$. The 
state with the spin parallel has slightly higher energy than the spin anti-parallel. Thus it is the de-excitation of the atom from a state with a parallel 
spin to a state with an anti-parallel spin that emits a 21 cm photon. In other words, absorption of a 21 cm photon will excite the atom from a spin anti-parallel state 
to a parallel spin state. The interaction energy between the magnetic moment is of order $\rm \frac{hc}{\lambda} =5.9\mu eV$. The Einsteins coefficient for the spontaneous transition for hyperfine 
splitting is equal to  $\rm 10^{15}s$, indicating that an \HI{} atom spends close to $\rm 10^{7}$ years in the excited state, but the weak emission is compensated by the 
presence of a huge quantity of hydrogen in the interstellar medium. 

The earth's atmosphere is opaque to the electromagnetic radiation in the near-infrared (NIR), ultra-violet (UV), X-rays, and gamma rays. 
Thus the ground-based observations are amenable only in the optical and the radio regimes \citep{thompson2017interferometry}. Radio observation from the earth is only possible in the frequency range $\rm \nu= 10Mhz$ 
$\rm (\lambda= 30m)$ to  $\rm \nu=1.5THz$ $(\lambda=0.2mm)$. Thus, the radio window is highly extended, as it spans five decades between 10 MHz to 10 THz.
The advent of radio astronomy began with the serendipitous discovery of radio emissions from the center of the Galaxy by Karl Jansky in 1920. 
Karl Jansky, at that time, was trying to understand the origin of the static noise in short wave $\rm (\lambda=15m)$ radio transmission-based communication systems.
Jansky found that the antennae detected steady radiation, which peaked at the same sidereal time every day, and deduced that 
the origins of the radiations lie outside the solar systems and emanated from the center of the Galaxy. Further, the initial advancement in radio astronomy was 
made by Grote Reber, who mapped the Galaxy at 160 MHz and showed that the radio emissions had a distinct nonthermal spectrum. Also, Grote Reber was the first to build 
a radio telescope with a parabolic reflector. The field of radio astronomy advanced rapidly after the end of the second world war with the help of 
the technological advancement made in radar technology during the war. Despite all the advancements, the size of the radio telescope remained a significant 
hindrance for the astronomers to achieve the sensitivity and resolution required for imaging studies of gas distribution in the Milky Way and external galaxies. 
A large aperture of the radio telescope increases the collecting area, thus effectively increasing the instrument's sensitivity. 
Besides, a larger aperture also increases the maximum resolution of the instrument. Rayleigh's criterion for the angular resolution is given by 
\begin{uprightmath}
\begin{equation}
 \Delta \theta \propto \frac{\lambda}{D}
\end{equation}
\end{uprightmath}

where, $\rm \Delta \theta$, is the angular resolution, $\lambda$ is the observing wavelength, and $\rm D$ is the diameter of the dish. The above immediately shows that 
for a given diameter, a radio telescope will have a lower angular resolution than the optical telescope since photons in the radio regime have a longer 
wavelength than the optical photons. For example, in order to achieve an 
angular resolution equal to $\rm 0.8^{\farcs}$, one would need a 300km dish to observe radio waves at 300MHz, whereas an optical telescope can obtain a similar angular 
resolution with a diameter equal to 15 cm at 500 nm. 

The limits on the sensitivity and resolution arising from 
the limited size of the radio telescope were solved using Earth Aperture Synthesis Techniques developed by Martin Ryle in 1960. It relies on a simple idea that the baseline vector 
between two antennae on the earth is continuously changing because of the earth's rotation, as seen from the source of interest. This can be achieved with atleast 
two telescopes, one of which is mobile. Measuring the correlation of the voltages between the two antennae allows measurement of a single Fourier component of the source
brightness distribution since the measured correlator outputs (visibilities) are related to the source brightness distribution through a Fourier transform. 
If one of the telescopes is mobile, then moving it across a distance of D and measuring all the 
Fourier components is equivalent to measuring the signals with a single telescope of diameter equal to D, i.e., one will obtain an image of the sky with the same resolution as 
with a telescope of diameter equal to D. In other words, one has synthesized an aperture size of D. Even for a pair of fixed antenna, the Fourier components measured by 
the antennae continuously change as the source rises and sets. Thus, instead of using a single dish, we combine the signals from multiple telescopes to 
synthesize an aperture much larger telescope than a single dish. For an array containing N antennae, one measures $\rm {}^{N}C_{2}$ Fourier components. 
Thus, the spatial correlation functions $\rm (V(r1,r2))$ measured by the antennae are the Fourier transform of the source brightness distribution I(s)\citep{chengalur2007low}:
\begin{uprightmath}
\begin{equation}
 V(r1,r2) =\langle E(r1) E(r2) \rangle = \mathcal{F}[ I(s)]
\end{equation}
\end{uprightmath}

The visibility function $\rm V(r1,r2)$  at antennae at point $\rm P_{1},P_{2}$ is given by the "Cittert-Zernike'' theorem \citep{born2013principles}
\begin{uprightmath}
\begin{equation}
\langle E(r1) E(r2) \rangle = \frac{1}{R^{2}}\int \mathcal{I}(l,m) e^{-ik\bigg( l(x_{2}-x_{1}) + m(y_{2}-y_{1}) +n(z_{2}-z_{1})    \bigg)  }  \frac{dldm}{\sqrt{1-l^{2}-m^{2} }}
\end{equation}
\end{uprightmath}

In the above equation, $\rm \mathcal{I}(l,m)$ is the intensity at the source. Assuming that the source lies on the celestial sphere of radius R, one 
can write $\rm x^{'}_{1}=Rcos \theta_{x}=Rl$, $\rm y^{'}_{1}=Rcos \theta_{y}=Rm$, $\rm z^{'}_{1}=Rcos \theta_{z}=Rn$, where $\rm x^{'}_{1}, y^{'}_{1}$, $z^{'}_{1}$ and 
are points on the source and $\rm x_{1}, y_{1}$, $z_{1}$ are the observation points. $\rm \theta_{x},\theta_{y}, \theta_{z}$ are the angles between the radius vector drawn from 
the source to the origin and the unit basis vectors. l, m, and n are the direction cosines and satisfy the relation $\rm l^{2} + m^{2} + n^{2} =1$, 
thus specifying two direction cosines (l,m) are sufficient to describe the intensity distribution of the source on the celestial sphere.
Further defining the baseline coordinates as $\rm u=\frac{x_{2} -x_{1}} {\lambda}$, $v=\frac{y_{2} -y_{1}} {\lambda}$, $\rm w=\frac{z_{2} -z_{1}} {\lambda}$, and 
and the spatial correlation function $\rm \langle E(r1) E(r2) \rangle$ as the visibility function $\rm \mathcal{V}(u,v,w)$, the relation between the visibilities measured by 
the interferometer and the sky brightness distribution is given by,

\begin{uprightmath}
\begin{equation}
\mathcal{V}(u,v,w)= \int \mathcal{I}(l,m) e^{-i2\pi\bigg( lu + mv +nw)    \bigg)  }  \frac{dldm}{\sqrt{1-l^{2}-m^{2} }}
\end{equation}
\end{uprightmath}

In the above equation we have neglected the constant term $\frac{1}{R^{2}}$. The above equation reduces to a Fourier transform when

\begin{itemize}
\item  when the observations are confined in the u-v plane, i.e. w=0; then 

\begin{uprightmath}
\begin{equation}
\mathcal{V}(u,v,w)= \int \mathcal{I}(l,m) e^{-i2\pi\bigg( lu + mv )    \bigg)  }  \frac{dldm}{\sqrt{1-l^{2}-m^{2} }} 
\end{equation}
\end{uprightmath}

 \item when the source distribution is confined to a very small region of sky i.e. $\rm n=\sqrt{1- l^{2} -m^{2}} =1$
\begin{uprightmath} 
 \begin{equation}
 \mathcal{V}(u,v,w)= e^{-i2\pi w} \int \mathcal{I}(l,m) e^{-i2\pi\bigg( lu + mv     \bigg)  } dldm 
 \end{equation}
\end{uprightmath}

\end{itemize}

So even by using a simple two-element interferometer and measuring the visibility function $\rm \mathcal{V}(u,v,w)$, 
by moving one of the interferometers across the source, one can reconstruct the sky brightness distribution  $\rm \mathcal{I}(l,m)$ by taking an 
inverse Fourier transform of $\rm \mathcal{V}(u,v,w)$.

The sensitivity of an N-element interferometer is given by

\begin{uprightmath}
\begin{equation}
 \Delta S [Jy]= \frac{SEFD}{\sqrt{N(N-1) \times N_{pol} \times \Delta \nu \times \tau  }}
\end{equation}
\end{uprightmath}

SEFD is the system equivalent flux density, N is the  number of antennae, $\rm N_{pol}$ is the number of polarizations, $\rm \Delta \nu$ is the channel width or the 
the spectral resolution of the observation, $\rm \tau$ is the on-source time.

We define the brightness temperature as 
\begin{uprightmath}
\begin{equation}
 T_{B}= \frac{1.36 \lambda^{2}\Delta S}{\theta^{2}}
\end{equation}
\end{uprightmath}

In the above equation, $\rm \lambda$ is the wavelength of the spectral line one intends to detect. In the case of \HI{} observations $\rm \lambda$  is equal to 21.1 cm. 
$\rm \theta$ is the synthesized beam size or the spatial resolution of the observations measured, usually as a second of arc.

The column density of the $\rm \HI{}$ detectable with given the linewidth $\rm dV(km/s)$, system temperature $\rm (T_{B})$  is given by
\begin{uprightmath}
\begin{equation}
 N_{\HI{}}(atoms/cm^{2})= 1.823 \times 10^{18} \int^{\infty}_{-\infty} T_{B}[K] dV[km/s]
\end{equation}
\end{uprightmath}

Thus the limiting column density or the sensitivity of the observation $\rm (N_{HI})$ achievable for an on source time $\rm (\tau)$ is a fine balance between 
the spectral resolution $\rm(\Delta \nu)$, the spatial resolution $\rm(\theta)$ at a given SEFD. 

\subsection{The Giant Meter Wave Radio Telescope (GMRT)}
We use the Giant Meter Wave Radio Telescope (GMRT) located in Pune, India, to map the distribution of neutral hydrogen in two of the thinnest known galaxies (Figure 1.6)
\citep{chengalur2007low, lal2013gmrt}.
GMRT, unlike Very Large Array (VLA), located in New Mexico in which the antennae are mounted on rails,  is a fixed interferometer. 
GMRT consists of 30 fully steerable antennae, each with a diameter equal to 45 meters, spread over a distance of 25 km. Of the 25 antennae, 14 are arranged randomly in a 
compact array over a region of 1 sq km, whereas the remaining 16 are arranged to form 3 arms resembling $\rm 'Y-shape'$, with the longest baseline equal to 25 km. The compact, 
dense central array offers high sensitivity and reduces the gaps in the UV plane, and the long baseline allows high-resolution observations. GMRT offers observation in 
five spectral bands:   153MHz, 233MHz, 325MHz, 610 MHz, and 1420 MHz in dual-polarization mode. We use GMRT at 1420 MHz, as it corresponds to the rest frequency 
of \HI{} 21 cm line. At 1420 MHz,  the approximate synthesized beam size for a full synthesis observation is $\rm 2^{``}$ and the best rms sensitivity achievable is 0.03 mJy.
The backend GSB allows a maximum bandwidth of 32 MHz spread over 512 spectral channels. GMRT uses three standard flux density calibrators, 3C286, 3C48, and 3C147, for 
amplitude and bandpass calibration (see the following section for details). The phase calibrators are chosen from the VLA calibrators list. The observations usually are affected by interference from 
mobile telecommunication devices, passing satellites, and solar interference. The RFI environment is slightly congenial at night than during the daytime. Usual GMRT 
observations consist of observations of the phase and flux calibrators interspersed in between the observation of the target for bandpass and the phase calibration. The 
flux calibrators are usually observed at the beginning and the end of the observations.
\begin{figure*}
\hspace{-15.0pt}
\resizebox{150mm}{60mm}{\includegraphics{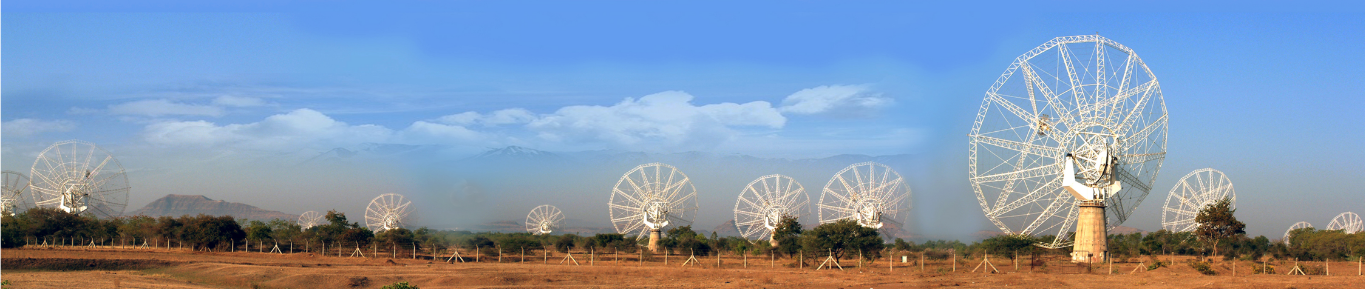}} 
%\vspace{1pt}
\caption{The Giant Meterwave Radio Telescope (GMRT) is located in Narayangaon, Pune, India. We have used GMRT to map the neutral \HI{} distribution in two of 
the thinnest known galaxies.[\textit{Image courtsey: www.skatelescope.org/ }]}	
\end{figure*}
\subsection{Data Analysis}
The data at the GMRT Observatory is recorded in LTA format, which is raw telescope format, or Flexible Image Transport System (FITS) format, which needs to be converted into a 
measurement set for analysis using the Common Astronomy Software (CASA). The standard steps for data analysis of spectral line observations follow the 
following sequence: 1) Flagging  2) Calibration  3)Imaging + Self-calibration 4) continuum subtraction and Imaging of the channels containing the spectral line.

\begin{itemize}
 \item \textbf{Flagging}\\ The interferometric observations, in general, are affected by spurious noise due to solar interference, thermal emission from the ground, 
 satellites transiting through the field of view of the observation, and other means like telecommunication channels. Thus, it is imperative to remove the  
 radio interference from the data set, and this process constitutes flagging. One of the first steps is to identify the bad or non-working antennae in the data set 
 and flag them all together. Another source of error comes if the antennae are very close to one another; the antennae block or shadow each other. Also, there 
 are times when the correlator writes out zero data to the measurement set. Besides, arrays take a certain time to settle before each scan, so it is important to clip 
 data at their scan boundaries. Finally, manual flagging is an iterative process by plotting the amplitude of the visibilities as a function of the channel to identify bad channels 
 and visibility as a function of time to know if the observations are affected by RFI during a particular interval. Similarly, one can identify non-working baseline, scans, and 
 so forth by plotting the amplitude of the visibilities as a function of each quantity. Alternative one may also use automated flagging  programs like AOFlagger \citep{offringa2010aoflagger}.

 \item \textbf{Calibration}\\ The observations carried out with a radio interferometer are not free from errors. The observations are usually affected by environmental effects 
 like ionospheric attenuation and faults with systems like the error in the pointing of the antenna. The instrumental factors that affect the observation can be divided into 
 long-term and short-term effects. Long-term effects include antenna position coordinates and antenna pointing corrections resulting from axis-misalignment, short-term 
 problems consist of atmospheric attenuation, variation of antenna gain, and variation in the local oscillator phase, which affect the measurements during the observation itself.
 The primary purpose of calibration is to remove the effects of instrumental factors and the atmosphere factors in the measurement so that the observed visibilities are close 
 to the true visibility. Calibration is achieved by observing sources with known flux and spectral response. The observation of the target is interspersed by observation of the 
 calibrators to track changes in the observational parameters.

\begin{itemize}
 \item \textbf{Absolute flux calibration}\\
 It is used to determine the true flux of the source by observing a bright invariant calibrator at the beginning and at the end of the observation. Flux calibration ensures 
 that the flux of the observations is scaled correctly.
 
 \item \textbf{Bandpass calibration}\\
 Bandpass calibration is carried out to correct the spectral response of the antennae during observation. The errors can be introduced due to system as well 
 as atmospheric attenuation. Bandpass calibration requires a bright invariant point source or a source whose spectrum is well modeled. 
 
 \item \textbf{Delay calibration}\\
 Delay calibration is done to correct for phase delay errors or the residual path length for a particular frequency and the correlator settings. It fixes the phase slope 
 across the band.
 It is usually done before the bandpass calibration using the same calibrator source used for the bandpass calibration.
 
 \item \textbf{Gain calibration}\\
 Gain calibration is done to determine the complex value gains to correct the system gain variations, using a relatively bright calibration source close to the target. 
 Gain calibration is performed in order to track the variation of the local effects, so a calibrator source close to 
 the target will be subjected to the same variation.
 
\end{itemize}
 
The flux, delay, bandpass, and gain calibration can all be done with the same calibrator if the calibration source is close to the target.

\item \textbf{Imaging + Self Calibration}\\
Imaging is reconstructing the source intensity distribution from the visibilities measured by the interferometer. In essence, from the visibility measured by
the interferometer, one intends to deduce the model of the sky, which is as close to the true sky as possible. When the w-term is equal to zero (see Equation 1.90),
(or all the antennae are in plane) the Cittert-Zernike theorem can be written as:

\begin{uprightmath}
 \begin{equation}
    \mathcal{V}(u,v)= \int \mathcal{I}(l,m) e^{-i2\pi\bigg( lu + mv     \bigg)  } dldm 
 \end{equation}
\end{uprightmath}

In the above equation, we use the assumption that all the antennae are located in the u-v plane and that the w-term is equal to zero. Since, 
the interferometer can make observations only at a finite number of points; we multiply the above equation with a sampling function 
$\rm \mathcal{S}(u,v)$, such that $\rm \mathcal{S}(u,v)$ =1, where there are measurements and 0 everywhere else.

\begin{uprightmath}
 \begin{equation}
    \mathcal{V}(u,v)=  \mathcal{S}(u,v) \int \mathcal{I}(l,m) e^{-i2\pi\bigg( lu + mv     \bigg)  } dldm  = \mathcal{S} \mathcal{F} I
 \end{equation}
\end{uprightmath}

 In the above equation $\rm \mathcal{F}$  is the Fourier transform. Besides, all the measurements have equal weights and are called naturally weighed. 
Taking the inverse Fourier transform of the above equation:

 \begin{uprightmath}
 \begin{equation}
  \mathcal{F}^{-1} \bigg(V \bigg) = \mathcal{F}^{-1} \bigg( \mathcal{S} \mathcal{F} I \bigg)
 \end{equation}
\end{uprightmath}

 which gives
 
\begin{uprightmath}
\begin{equation}
  I^{D}= I^{PSF} * I
\end{equation}
\end{uprightmath}

 $\rm I^{D}$ is called the dirty beam and is constructed by taking the inverse Fourier transform of the measured visibilities. Following the above equation, 
 the dirty beam is defined as the convolution of the true image with the point spread function of the instrument $\rm I^{PSF}$. $I^{PSF}$ is defined as the inverse 
 Fourier transform of the sampling function $\rm \mathcal{S}$. Finally, to reconstruct the true image of the sky, one carries out deconvolution by eliminating the corrupting 
 instrumental effects. The clean algorithm in CASA is implemented in the task $\rm 'tclean'$. The $\rm 'hogbom'$ algorithm \citep{hogbom1974aperture} implemented in 
 $\rm 'tclean'$ follows these steps:
 \begin{enumerate}
  \item Hogbom algorithm first finds the position and the strength of the peak flux in the dirt image.
  
  \item Multiply the peak flux with the dirty beam and subtract it from the  $\rm 'dirty image'$.
  
  \item Save the position and the strength of the peak flux in the $\rm 'model image'$
  
  \item Then, one iterates the above steps till there is no brighter flux than the user-specified threshold.
  
  \item Once all the flux sources above the user-specified threshold are subtracted, one will be left with a $\rm 'residual image'$.
  
  \item Form a clean image by convolving the model image with an idealized clean beam, i.e., a Gaussian with a central peak of the dirty beam.
  
  \item Finally, add the residuals to the $\rm 'clean-image'$.
  
 \end{enumerate}
 
For GMRT observation, one can not neglect the w term as the antennae are arranged in $\rm Y-shape$ to maximize the UV coverage. Thus the relation between the visibility and 
the source brightness distribution is no longer a straightforward Fourier transform. Self-calibration is an iterative process in which the source is used to calibrate 
the phases and the amplitudes of visibilities, given the source has a high signal-to-noise ratio. One starts by generating the 
model image of the source, and then this model is used to determine the gains as a function of time. The process is iterated until the signal-to-noise ratio of the 
continuum image saturates.

\item \textbf{Imaging the spectral line}\\
The spectral line observation, even in the case of narrow bandwidth, has continuum flux present in it. Thus, it becomes imperative to subtract the continuum flux density 
from the spectral line data. This is achieved by performing a linear fit to the visibilities as a function of the frequencies. The best-fit continuum is  
subtracted from the original visibilities to get continuum-free data. One then makes a dirty cube by specifying the central frequency of observation in order to identify 
the channels containing the spectral line. After that, one subtracts the channel containing the spectral line from the measurement set 
and iteratively cleans the data cube until one reaches the desired rms. The result of the cleaning process is a datacube that contains information about the flux 
of the source along the spatial and frequency axes.

\end{itemize}

\subsection{Analysis of the data cube}

\begin{figure*}
\hspace{-15.0pt}
\resizebox{180mm}{80mm}{\includegraphics{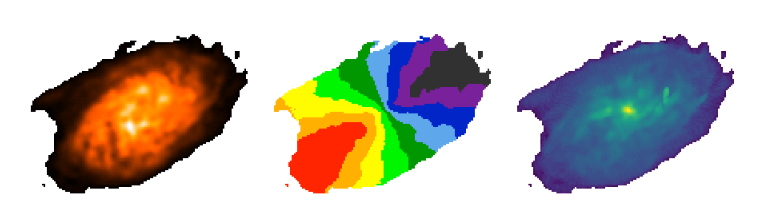}} 
%\vspace{1pt}
\caption{[Left] Moment 0 map showing the line of sight column density, [Middle] Moment 1 map indicating the velocity field and [Right] the Moment 2 indicating the 
disordered motion for a spiral galaxy NGC 2403. [\textit{Image courtesy: https://bbarolo.readthedocs.io/}]}	
\end{figure*}

The data cubes are the final products of the single-dish or interferometer observations. A data cube contains two spatial axes indicating the coordinates of the emission
and a frequency or the velocity axis indicating the frequency/velocity at which the \HI{} line is emitted due to Doppler shift. Each pixel of the data cube is called a 
$\rm 'voxel'$. Each voxel stores the brightness of the \HI{} emission at a given sky position and at the frequency at which \HI{} was emitted. The \HI{} data cube can also be understood 
as a collection of two-dimensional images containing the spatial position of the \HI{} emitted at the corresponding Doppler shifted frequency.
The data cube can be used to derive various diagnostics for inferring the physical properties of the structure and kinematics of galaxies:

\begin{itemize}
 \item The Moment 0 map gives the total intensity along the line of sight. The first image in the Figure 1.7 shows the Moment 0 map.
 
 \begin{uprightmath}
 \begin{equation}
  I(Jy beam^{-1} kms^{-1}) = \sum^{i=n}_{i=1} I_{i} (x,y)\Delta \nu
 \end{equation}
\end{uprightmath}

$\rm I_{i}$ is the emission at a specific channel, and $\Delta \nu$ is the channel separation. The total intensity map is proportional to the \HI{} column density.
 
\item The Moment 1 map or the velocity field is given by.  
\begin{uprightmath}
\begin{equation}
 \langle V \rangle =\frac{ \sum^{i=n}_{i=1}I_{i} (x,y)  V_{i}   }{  \sum^{i=n}_{i=1} I_{i} (x,y) }
\end{equation}
\end{uprightmath}

The Moment 1 map can be used to extract the galaxy's rotation curve by performing tilted ring modeling (TRM) for galaxies with intermediate inclination. The second image in Figure 1.7 
indicates the moment 1 map.

\item The Moment 2 map or the line width map is given by

\begin{uprightmath}
\begin{equation}
 \sigma =  \sqrt{  \frac{ \sum^{i=n}_{i=1}I_{i} (x,y)  (V_{i} - \langle V \rangle)^{2}    }{  \sum^{i=n}_{i=1} I_{i} (x,y) }  }
\end{equation}
\end{uprightmath}

The Moment 2 map estimates the disordered motion in the galaxy. The third image in Figure 1.7 depicts the Moment 2 map.

\item The \HI{} column density along the line of sight is given by 

\begin{uprightmath}
\begin{equation}
 N_{\HI{}}(atoms/cm^{2})  =  \frac{1.106 \times 10^{24}}{ \frac{B_{maj}}{arcsec} \times \frac{B_{min}}{arcsec} } \int \frac{S} {Jy beam ^{-1}} \frac{V}{kms^{-1}}
\end{equation}
\end{uprightmath}

The above equation neglects the \HI{} self-absorption and dust extinction along the line of sight thus the observed intensity is directly proportional to the column density
$\rm S$ is the brightness of emissions integrated through the spectral channels, and $\rm B_{maj}$ and $\rm B_{min}$ are the synthesized beam size or the full width at the 
half maximum of the instrumental beam.

\item The surface brightness is given by

\begin{uprightmath}
\begin{equation}
 \Sigma_{\HI{}}(M_{\odot}pc^{-2})  =  \frac{8.794 \times 10^{3}}{ \frac{B_{maj}}{arcsec} \times \frac{B_{min}}{arcsec} } \int \frac{S} {Jy beam ^{-1}} \frac{V}{kms^{-1}}
\end{equation}
\end{uprightmath}

\item Integrating the above equation over the surface of the source, one obtains the \HI{} mass given by 

\begin{uprightmath}
\begin{equation}
 M_{\HI}(M_{\odot}) =2.356 \times 10^{5} \frac{F_{\HI}}{[Jy kms^{-1}]} \frac{D}{[Mpc]} 
\end{equation}
\end{uprightmath}

In the above equation, $\rm F_{\HI{}}$ is the total flux of the source multiplied by the channel width, and $\rm D$ is the distance to the source.

\end{itemize}

\subsection{Structural and kinematic properties of gas distribution: 2D methods}
The \HI{} data cubes can be used to derive the structural and kinematic properties of the gas distribution in the galaxies. Tilted ring modeling (TRM) is one such method 
used to extract the rotation velocity, position velocity, and inclination of the gas distribution. TRM was introduced by \cite{rogstad1974aperture}
in order to derive the one-dimensional representation of the gas rotation as a function of the radius. The galaxy is modeled as a sequence of independent rings with
its own kinematic and geometric properties. Each point in the velocity field map of the Moment 1 map can be defined as the line of sight velocity $\rm V_{los}$ at the position 
$\rm (x,y)$. Ignoring the non-rotational motion, the  velocity field to the first approximation can be written as

\begin{uprightmath}
\begin{equation}
 V_{los}(x,y)=  V_{sys} + V_{rot}(R)cos(\theta) sin(i)
\end{equation}
\end{uprightmath}

$\rm V_{sys}$ is the systemic velocity of the gas distribution, $V_{rot}$ is the rotation velocity of the gas disc at the radius R, $\rm i$ the inclination angle and 
$\rm \theta$ the azimuthal angle of the rings in the plane of the galaxy. $\rm \theta$ is related to the inclination angle $\rm i$, the center of the galaxy 
$\rm (x_{0},y_{0})$ and the position angle of the galaxy through the following equation:

\begin{uprightmath}
\begin{equation}
 Rcos\theta =-(x-x_{0})sinPA + (y-y_{0})cosPA  
\end{equation}
\end{uprightmath}

\begin{uprightmath}
\begin{equation}
 Rsin\theta =-(x-x_{0})cosPA + (y-y_{0})sinPA  
\end{equation}
\end{uprightmath}

PA is the position angle and is measured from the north to the major axis in the galaxy, and $\rm R$ is the mean radius of the ring in 
the plane of the galaxy. The above equations give transformation between the pixel coordinates $\rm (x,y)$ to the polar coordinates \citep{begeman1989hi}. In the above model 
\citep{begeman1989hi}  the velocity field is parameterized  by the following parameters
\begin{itemize}
 \item The rotation center of the galaxy $\rm (x_{0},y_{0})$.
 
 \item The velocity center of the galaxy called the systemic velocity $\rm V_{sys}$
 
 \item  The circular velocity $\rm V_{rot}$ at a distance $\rm R$ from the center.
 
 \item The position angle $\rm PA$ of the galaxy.
 
 \item The inclination angle $\rm i$ of the galaxy.
 
\end{itemize}

The aforesaid parameters are determined by making an initial guess and fitting the model to the $\rm V_{los}$ determined from the velocity field. 
The tilted ring model assumes that the gas lies in a thin disc and follows circular orbits, but the model can be modified to include the deviations from the circular motion, 
to include spiral perturbations and warps \citep{sellwood2015diskfit}. The tilted ring method is an example of a two-dimensional fitting method in which the parameters are fitted to the velocity field. Another example of a two-dimensional method is the envelope tracing method \citep{mathewson1992southern, takamiya2002iteration}, which
use the major axis position velocity diagram to derive the rotation curve. The two-dimensional methods give very good results for nearly face-on galaxies 
with high spatial resolution.

\subsection{Structural and kinematic properties of gas distribution: 3D methods}

\begin{figure*}
\hspace{-11.0pt}
\resizebox{140mm}{50mm}{\includegraphics{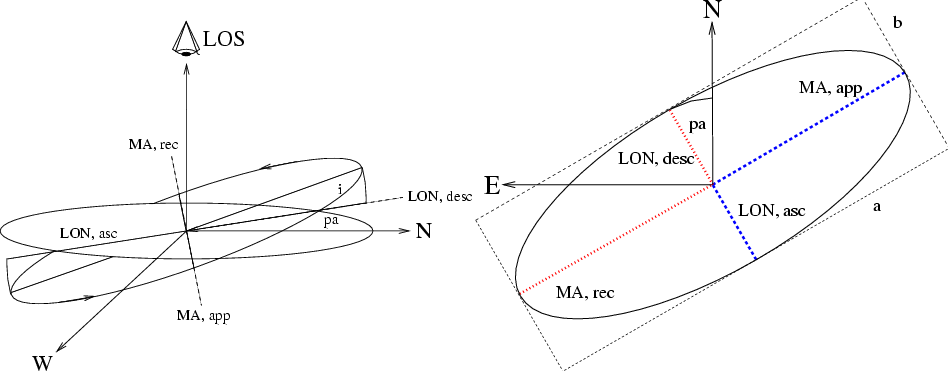}} 
%\vspace{1pt}
\caption{The left panel shows the orientation parameters, inclination and the position angle for TiRiFiC. The right panel shows the orbit of tracer material at an arbitrary position. 
[\textit{Image courtsey: \cite{jozsa2012tirific} }] }	
\end{figure*}

The two-dimensional fitting methods do not derive the model parameters directly from the data cube but from the secondary products from the data cube. So the 
errors and assumptions inherent in the derivation of velocity fields and major axis PV diagram also affect the derivation of the final structural parameters \citep{jozsa2012tirific}. 
For example, in the case of galaxies with warps, or edge-on galaxies, one can not derive a unique velocity field as the line of sight intersects the 
disc multiple times, which implies multiple values of velocities for the same point in the sky. Further, problems arise if the observation has a very small number of 
resolution elements, then the rotation curve derived in the inner region becomes highly inaccurate due to beam smearing. An accurate determination of the rotation curve 
is of paramount importance for inferring the shape of the dark matter halo. Thus, to explore the degeneracy in the parameters fitted to the tilted ring models, 
one derives a model data cube \citep{van1992groningen} and makes a channel-by-channel comparison of the model and the data.
These problems can be circumvented by fitting the data cube directly instead of fitting the secondary 
data products. 

The 3D fitting method was first introduced by \cite{swaters1999kinematically}, who showed that the kinematic and structural properties could be directly 
derived from the 3D datacube without fitting the velocity field or the PV diagrams. The 3D methods eliminate the beam smearing effects as the final model data cubes 
are convolved with the instrumental response. The 3D modeling method consists of $\chi^{2}$ minimization of the channel maps between the data and model instead of 
using a single representation like a velocity field or a major axis PV diagram. The current state of the art 3D tilted ring modeling software are TiRiFiC 
\citep{jozsa2012tirific}, an automated version of the TiRiFiC called Fully Automated TiRiFiC (FAT) \citep{kamphuis2015fat} and 3DBarolo \citep{teodoro20153d}.
One of the main disadvantages of the 3D method is that it is computationally expensive as it generates a model data cube at each iteration of the 
$\rm \chi^{2}$ minimization for different parameter combinations. Further, the models are highly sensitive to the inhomogeneity in the gas distribution.
TiRiFiC performs a tilted ring modeling over the entire data cube. It constructs a tilted ring model at each sampling radius and computes the pixel by pixel 
residuals between the model data cube and the observed data cube at each ring. The model data cube is smoothed to the spatial scale of the instrumental resolution 
to take into account the observational effects. The geometry of the galaxy model constructed with TiRiFiC is indicated in Figure 1.8.
The following free parameters parametrize the 3D models.

\begin{itemize}
 \item  The radius, which determines the ring size
 \item  The circular velocity at each radius
 \item  Scaleheight of the vertical gas density distribution
 \item  Surface brightness
 \item  Inclination angle
 \item  Position angle
 \item Coordinates specifying the center, 1)Right ascension, 2)Declination  3) Systemic velocity
 \end{itemize}
The above parameters constitute the local parameters, which can vary from ring to ring. This approach is helpful in describing the inhomogeneities in the gas distribution. 
Global parameters for TiRiFiC are

\begin{itemize}
 \item The velocity dispersion, which also takes into account the instrumental dispersion.
 \item The density distribution of the gas disc, which can be modeled as $\rm gaussian$, $\rm lorentzian$ or $\rm sech^{2}$ profile
 \item The constant total flux of the single point source, which determines the number of point sources for generating the model data cube.
\end{itemize}

Once the above parameters are specified, TiRiFiC generates a model data cube and carries the $\rm \chi^{2}$
minimization of the data and the model. The kinematic and structural parameters obtained by TiRiFiC, unlike those obtained with the 2D method, are not affected by
beam smearing and the projection effects. The total rotation curve and gas surface density obtained from TiRiFiC constitute important parameters for deriving 
the density distribution of the dark matter.

\subsection{Mass Modeling}
The total rotation curve can be obtained by adding the circular velocity of the baryons and the dark matter components in the quadrature.

\begin{uprightmath}
\begin{equation}
V^{2}_{Total}=\gamma V^{2}_{*} + V^{2}_{gas} + V^{2}_{DM}
\end{equation}
\end{uprightmath}

In the above equation, $\rm V_{gas}$, $V_{*}$, $V_{DM}$ are the circular velocity profiles required to balance the gravitational potential due to the stars, gas, and 
the dark matter distribution respectively, and $\rm \gamma$ is the mass to light ratio.
The circular velocity of the gas disc $\rm V_{gas}$ and the circular velocity of stars $\rm V_{*}$ is derived 
by modeling the gas and the star disc as thin concentric rings using Groningen Image Processing System (GIPSY) \citep{van1992groningen} task ROTMOD. The dark matter 
rotation curve is modeled using observationally motivated pseudo-isothermal (PIS) dark matter distribution \citep{begeman1991extended} as well as
Navaro Frenk and White (NFW) profile \citep{navarro1997universal} motivated by cold dark matter simulations. The rotation curve due to the cored PIS halo is given by

\begin{uprightmath}
\begin{equation}
 V(R)=\sqrt{4\pi G \rho_{0} R^{2}_{c} \bigg( 1- \frac{R_{c}}{R} arctan(\frac{R}{R_{c}})\bigg)} 
\end{equation}
\end{uprightmath}

where, $\rm \rho_{0}$ is the central density of the halo and $\rm R_{c}$ is the core radius. The asymptotic rotation velocity is defined 
as $\rm V_{\infty}=\sqrt{\bigg(4\pi G R^{2}_{c}\bigg)}$. 

The rotation curve due to the cuspy NFW density distribution is,
\begin{uprightmath}
\begin{equation}
 V(r)=V_{200} \sqrt{\frac{ln(1+cx) -cx/(1+cx)}{x[ln(1+c) -c/(1+c)]}}
\end{equation}
\end{uprightmath}

where, $\rm x=R/R_{200}$, $\rm R_{200}$ is the radius at which the mean density of the dark matter halo is 200 times the critical density. 
$\rm V_{200}$ is the rotation velocity  at $\rm R_{200}$ and $\rm c=R_{200}/R_{s}$. $\rm R_{s}$ is defined as the characteristic scale radius of the NFW density profile. 

We use our rotation curve data to check if baryonic matter alone can explain the observed rotation curve 
in the Modified Newtonian dynamics (MOND) paradigm \citep{{milgrom1983modification}}. The net rotation curve in MOND is given by 
\begin{uprightmath} 
\begin{equation}
  V(r)= \sqrt{\frac{1}{\sqrt{2}}( V^{2}_{gas}+  \gamma^{*}V^{2}_{*})
      \sqrt{1+\sqrt{1+ \bigg( \frac{2Ra}{ V^{2}_{gas}+  \gamma^{*}V^{2}_{stars}} \bigg) }^{2}}} 
\end{equation}
\end{uprightmath}

where $\rm a$ is acceleration and $\rm \gamma^{*}$ is the mass to light ratio.

In order to determine the best-fitting dark matter density profile, we minimize the likelihood function given by 
\begin{uprightmath}
\begin{equation}
 \chi^{2} =\sum _{R} \frac{\bigg(V_{obs}(R) - V_{T}(R) \bigg)^{2} }{V^{2}_{err}}
\end{equation}
\end{uprightmath}

In the above equation, $\rm V_{obs}$ is the observed rotation curve derived using 3D modeling of the data cube 
and $\rm V_{T}$ is the total rotation curve derived by adding the circular velocity of each of the mass components in quadrature.

In order to understand the effect of various mass components on the dark matter distribution, we investigate the following cases.

\begin{itemize}
 \item \textbf{Fixed Mass to Light ratio}\\
 In this case, we derive the mass-to-light ratio $\rm (\gamma)$ using the stellar population synthesis models\citep{bell2001stellar}. We use Kroupa and diet-Salpeter 
 initial mass functions to derive the mass-to-light ratio $\rm \gamma^{*}$. The diet-Salpeter IMF gives maximum stellar mass in a given photometric band.
 
 \item \textbf{Free Mass to light ratio}\\
This model keeps the mass-to-light ratio $\rm (\gamma^{free})$ as a free parameter,
such that constraints on the mass-to-light ratios are obtained from kinematics instead of the photometry

\item \textbf{Maximum disc model}\\
we maximize the contribution of the stellar disc such that the observed rotation curve is entirely due to just the stellar disc. This model allows us to ascertain the lower limits 
on the dark matter content in the galaxy.

\item \textbf{Minimum disc model}\\
We ascertain the upper limits on the dark matter mass by setting the baryonic contribution to zero.

\end{itemize}

\subsection{Neutral gas distribution and rotation curve}
The gas disc is modeled as thin concentric rings for deriving the circular velocity using the GIPSY task ROTMOD \citep{van1992groningen}.
The \HI{} surface density obtained from the tilted ring modeling is used as the input parameter in ROTMOD. The \HI{} surface densities are 
multiplied by 1.4 to correct for the presence of Helium and other metals.

\subsection{Stellar distribution and rotation curve}
We model the stellar distribution to determine the stellar disc's contribution to 
the observed rotation curve. We get the mass-to-light ratio $\rm (\gamma_{*})$ from stellar population synthesis models and the empirical relationships 
in \cite{bell2003optical} and \cite{bell2001stellar}. The relationship between the color-magnitude and $\rm \gamma^{*}$ is given as.

\begin{uprightmath}
\begin{equation}
\gamma^{*}=10^{(a_{\lambda} +b_{\lambda} \times Color )}
\end{equation}
\end{uprightmath}

$\rm \gamma*$ is the mass to light ratio, $\rm a_{\lambda}$ and $\rm b_{\lambda}$ are the intercept and slope of the 
$\rm log_{10}(\gamma_{*})$ versus colour calibration obtained by \cite{bell2001stellar} using stellar population synthesis models.
We compare Salpeter IMF and Kroupa IMF mass models \citep{kroupa2001variation}.
We subtract 0.15 dex from $\rm a_{\lambda}$ to get $\rm \gamma_{*}$ ratios for the Kroupa IMF.
Finally, we derive the stellar rotation curve with the GIPSY task ROTMOD.

\section{Summary}
The topic introduced in this section will serve as a preliminary guide to the topics studied in this thesis. The initial section lays out the historical
foundation of the subject, followed by mathematical techniques and tools needed to study the vertical structure in Newtonian and Alternative Gravity paradigms. In the 
following section, we detail interferometric techniques and image reconstruction methods for observing the neutral \HI{}
distribution in the galaxies. We finally present the 3D modeling 
methods for deriving the structural and kinematic properties of the neutral hydrogen distribution and the techniques to ascertain the dark matter density distribution using mass modeling.

\thispagestyle{empty}

\thispagestyle{empty}
%\chapter[How 'cold' are the stellar discs of superthin galaxies ?]{How 'cold' are the stellar disc of superthin galaxies ?\footnote[1]
%{Adapted from \textbf{K Aditya}, Arunima Banerjee, How $\rm 'cold'$ are the stellar discs of superthin galaxies?, \textit{Monthly Notices of the Royal Astronomical Society, 
%  Volume 502, Issue 4, April 2021, Pages 5049\textendash5064}, \textcolor{blue}{arXiv:2002.09198} } }
%\chaptermark{How 'Cold' are the stellar disc of superthin galaxies ?}

\chapter[How 'cold' are the stellar discs of superthin galaxies ?]{ \fontsize{50}{50}\selectfont Chapter 2 \footnote[1]
{Adapted from \textbf{K Aditya}, Arunima Banerjee, How $\rm 'cold'$ are the stellar discs of superthin galaxies?, \textit{Monthly Notices of the Royal Astronomical Society, 
  Volume 502, Issue 4, April 2021, Pages 5049\textendash5064}, \textcolor{blue}{arXiv:2002.09198} } }
\chaptermark{How 'Cold' are the stellar disc of superthin galaxies ?}

\textbf{\Huge{ How 'Cold' are the stellar disc of superthin galaxies ?}}

\section*{Abstract}
The planar-to-vertical axes ratio $\rm(10 <a/b< 20)$ of superthin galaxies is remarkably large, indicating that an ultra cold 
stellar disc may be present in these galaxies. With the help of the multi-component galactic disc model and the stellar dynamical code AGAMA 
(Action-based Galaxy Modelling Architecture), the vertical velocity dispersion of stars and gas as a function of galactocentric radius 
is determined for a sample of five superthin galaxies (UGC7321, IC5249, FGC1540, IC2233, and UGC00711) using the observed stellar and atomic hydrogen (\HI{}) 
scaleheights as constraints on the models. We find that the central vertical velocity dispersion of the stars in the 
optical band ranges between $\rm \sigma_{0s}$  $\sim$ $10.2 - 18.4$ $\kms$, and the exponential scalelenght between $\rm 2.6$ to $\rm 3.2$ $\rm R_{d}$, 
where $\rm R_{d}$ is the exponential stellar disc scale length. The average value of the stellar dispersions in the 3.6 $\rm \mu$m ranges between $\rm 5.9$ to $\rm 11.8$ $\kms$, 
confirming the existence of "ultracold" stellar discs in superthin galaxies. Superthin galaxies have a higher global median value of the multi-component 
disc dynamical stability parameter $\rm (Q_{N}= 5 \pm 1.5)$ than a previously studied sample of spiral galaxies $\rm (Q_{N}=2.2 \pm 0.6 )$.

\section{Introduction}
There are several possibilities as to why superthin galaxies have thin stellar discs, one of the possibilities is the lack of disc heating perpendicular to 
the galactic plane. In recent years, the stellar velocity dispersion has been accurately measured for face-on or nearly face-on galaxies using Integral Field 
Unit (IFU) surveys (e.g. \citet{Law2015Observing}; \citet{Allen2015Sami}; \citet{Bershady_2010}; \citet{Sanchez2012Califa}).
It is impossible to measure the vertical velocity dispersion of superthin galaxies directly because of their edge-on orientation. 
So, we resort to modeling the vertical velocity dispersion in the superthin galaxies to quantify the excursion of the stars from the midplane.
For five superthin galaxies in the optical and 3.6 $\mu$m, we use observed stellar scaleheight data as a constraint and employ the Markov Chain Monte Carlo (MCMC)
method (see Section 1.3) to constrain the vertical stellar dispersion using the multi-component model of gravitationally-coupled stars and gas in the force field of dark matter halo
\citep{narayan2002vertical}.The details of estimating the posterior probability distribution of the multi-component model is discussed in Section 1.1.3 and 1.3. Using the publicly 
availabe, stellar dynamics code Action-based Galaxy Modelling Architecture (AGAMA)  
\citep{vasiliev2018agama}, we verify the consistency of results from the multi-component model. The action-angle method for setting up equilibrium distribution 
function-based models of superthin galaxies is discussed in Section 1.1.4 in the introduction. We feed the best-fit vertical stellar dispersion from the 
multi-component model into AGAMA to derive the stellar vertical scale heights. The stellar scaleheights obtained from the AGAMA and multi-component model are 
compared with the observations to check the consistency of the results. Finally, we calculate the multi-component disc stability parameters 
proposed by \cite{romeo2011effective} and \cite{romeo2013simple} to verify the dynamical stability of superthin galaxies against axisymmetric instabilities. 
The mathematical formulation of the two-component stability parameter is described in Section 1.1.5.

\section{Sample of superthin galaxies}
A sample of galaxies with observed stellar and \HI{} scale heights is required to determine the vertical velocity dispersion.
As input parameters to the two-component model and for deriving an equilibrium distribution
function in AGAMA, we require the stellar and \HI{} densities as a function of the galactocentric radius and the dark matter density. 
Thus, we selected our sample of superthin galaxies based on the availability of stellar photometry in the optical and 3.6 $\mu$m band,
as well as \HI{} surface densities and mass models from high-resolution \HI{} 21 cm line data.

\begin{table*}
\hspace*{-60cm}
\begin{small}
\setlength{\tabcolsep}{1pt}
%\begin{center}
%\hfill{}
\caption{Input parameters for the model from the observational constraints}
\begin{tabular}{|l|c|c|c|c|c|c|c|c|c|c|c|c|}
\hline
Parameters& $\mu^{\textcolor{red}{1}}_{01}$ &$\Sigma^{\textcolor{red}{2}}_{01}$ & $R^{\textcolor{red}{3}}_{d1}$& $h^{\textcolor{red}{4}}_{z1}$ &$\mu^{\textcolor{red}{5}}_{02}$ &$\Sigma^{\textcolor{red}{6}}_{02}$ & $R^{\textcolor{red}{7}}_{d2}$&$h^{\textcolor{red}{8}}_{z2}$ &$\rho^{\textcolor{red}{9}}_{0}$&$R^{\textcolor{red}{10}}_{c}$   \\
Galaxy          &$mag$ $arcsec^{-2}$&$M_{\odot}$ $pc^{-2}$&kpc&kpc&$mag$ $arcsec^{-2}$&$M_{\odot}$ $pc^{-2}$&kpc&kpc&$M_{\odot}$ $pc^{-3}$&kpc \\
\hline    
\hline
Optical band &   &  &  &   &   & &  &  &  & \\
\hline
UGC 7321 &-&-&-&-&  23.5 &34.7 & 2.1 & 0.150 &0.039&2.99 \\
FGC 1540 &21.67&33.14&1.29&0.675&20.60&88.79&1.29&0.185&0.308&0.64\\  
IC  2233 &  22.90&17.85&2.47&0.332&-&-&-&-&0.0457&1.84\\
UGC 00711&    -   &15.0&1.6&0.317 &-&-&-&-&0.05&2.9 \\
\hline
$3.6\mu$m &   &  &  &   &   & &  &  &  &     \\
\hline
UGC 7321 &21.73&7.165&2.39&0.436&19.9&37.26&1.0&0.134&0.140&1.27\\
IC  5249 & 21.7&5.44&5.24&0.724&20.53&15.97&1.23&0.253&0.026&2.99\\
FGC 1540 &22.23&3.37&1.85&0.43&21.39& 8.167&0.54&0.152&0.319&0.63\\
IC  2233 &21.67&5.59&2.16&0.39&20.53&12.2&0.81&0.08&0.055&1.83\\
UGC 00711& -   &14.6&2.14&0.44&- &-&-&-& 0.033&2.95 \\
\hline
Parameters of \HI{} disc&   &  &  &   &   & &  &  &  &\\
\hline
&$\Sigma^{HI\textcolor{red}{11}}_{01}$& $R^{\textcolor{red}{12}}_{0,1}$&$a^{\textcolor{red}{13}}_{1}$&$\Sigma^{HI \textcolor{red}{14}}_{02}$& $R^{\textcolor{red}{15}}_{0,2}$&$a^{\textcolor{red}{16}}_{2}$ &  &  &  &\\   
          &$M_{\odot}$ $pc^{-2}$& kpc    &   kpc &$M_{\odot}$ $pc^{-2}$& kpc    &   kpc&  &  &  &\\
\hline
 UGC 7321&  4.912&2.85&3.85&2.50&1.51&0.485 &  &  &  &\\
 IC  5249 &  3.669&3.35&5.92& 4.85&4.05&17.06&  &  &  &\\
 FGC 1540&   4.09&5.73&2.48&1.3&5.73&5.08&  &  &  &\\
 IC  2233 &  2.236&1.79&2.52&2.454&1.69&6.14&  &  &  &\\ 
 UGC 00711 & 30.83&3.73&    &     &   &     &  &  &  &\\
 \hline
\end{tabular}
\begin{tablenotes}
\item ($\textcolor{red}1$):  Central surface brightness of disc(1)
\item ($\textcolor{red}{2}$):  Central stellar surface density disc(1)
\item ($\textcolor{red}{3}$):  Exponential scale length for disc(1)
\item ($\textcolor{red}{4}$):  Exponential scale height for disc(1)
\item ($\textcolor{red}{5}$):  Central surface brightness of disc(2) 
\item ($\textcolor{red}{6}$):  Central stellar surface density disc(2)
\item ($\textcolor{red}{7}$):  Exponential scale length for disc(2)
\item ($\textcolor{red}{8}$):  Exponential scale height for disc(2)
\item ($\textcolor{red}{9}$):  Core density of the pseudo-isothermal dark matter halo
\item ($\textcolor{red}{10}$): Core radius of the pseudo-isothermal dark matter halo
\item ($\textcolor{red}{11}$): The central surface density of disc 1 constituting the double gaussian HI profile
\item ($\textcolor{red}{12}$): Scalelength of gaussian disc(1)
\item ($\textcolor{red}{13}$): Offset in the centre of disc(1)
\item ($\textcolor{red}{14}$): The central surface density of disc 2 constituting the double gaussian HI profile
\item ($\textcolor{red}{15}$): Scalelength of gaussian disc(2) 
\item ($\textcolor{red}{16}$): Offset in the centre of disc(2)
\end{tablenotes}
\end{small}
\end{table*}

\subsection{UGC7321} 
UGC 7321 is observed at a distance of $\rm D=10$ Mpc, at an inclination of $\rm i=88^{\circ}$, and has a major-to-minor axes ratio  
$\rm a/b=15.4$\citep{matthews2000h}. It has a sharply rising rotation curve and an asymptotic velocity equal to 112 \kms.
The deprojected B-band central surface brightness is equal to 23.5 mag arcsec$\rm ^{-2}$\citep{matthews1999extraordinary}.
The $\rm M_{dyn}/M_{\HI{}}=31$ and $\rm M_{dyn}/L_{B}=29$, highlights the significance of the  dark matter in the galaxy.
\citep{banerjee2010dark} find $\rm R_{c}=2.9kpc$ and $\rm \rho_{0}=0.039 M_{\odot} pc^{-3}$,  using constraints from measured \HI{} rotation curve and 
the \HI{} vertical scale height. The studies by \cite{banerjee2010dark} find that UGC 7321 hosts a compact dark matter halo. See also \cite{banerjee2016mass} for 
additional information.

\subsection{IC 5249}\rm
IC 5249 is an edge-on galaxy with an axes ratio $\rm a/b =10.2 $, observed at an inclination $\rm i=89^{\circ}$ by the \citep{abe1999observation}. 
The asymptotic rotation velocity is equal to 112 \kms. \cite{yock1999observation} and \cite{van2001kinematics} estimate  
$\rm {M_{dyn}}/{M_{HI}}=9.5$ and  $\rm M_{dyn}/L_{B}=9.5$, respectively. Mass modeling indicates the presence of a compact dark matter 
halo with $\rm R_{c}=2.9 kpc$ and $\rm \rho_{0}=0.026 M_{\odot} pc^{-3}$.

\subsection{FGC 1540}
The superthin galaxy FGC 1540 is located at a distance $\rm D=10$ Mpc. It has an axes ratio $\rm a/b = 7.5$ and is observed at 
an inclination $\rm i=87^{\circ}$. It has an asymptotic rotation velocity equal to 90 \kms, $\rm M_{\HI{}}/L_{B}=4.1$.  
Mass modeling predicts a core radius of the dark matter halo equal to $\rm 0.69$ kpc and a central dark matter 
density  $\rm \rho_{0}=0.262 M_{\odot}/pc^{-3}$  \citep{kurapati2018mass}.

\subsection{IC 2233}
IC 2233 has an axes ratio a/b = 8.9, observed at an inclination equal to $\rm 90^{\circ}$, at a distance equal to 10 Mpc \citep{matthews2007h}.
The asymptotic rotation velocity of IC 2233 is equal to 85 \kms and $ \rm M_{\HI{}}/L_{b} \sim 0.62$ and $\rm M_{dyn}/M_{\HI} \sim 12$, 
indicating abundance of \HI{} gas. Mass models suggest dark matter density $\rm \rho_{0}=0.055 M_{\odot} pc^{-3}$ and a core radius $\rm 1.83 kpc$\citep{banerjee2016mass}.

\subsection{UGC00711}
UGC0711 is a superthin galaxy with an a/b ratio of 15.5 Mpc, observed at a distance equal to 23.4 Mpc and at an inclination equal to $\rm 74^{\circ}$.
The asymptotic rotation velocity is equal to 100 \kms \citep{mendelowitz2000rotation}. Mass modeling indicates a 
dark matter density $\rm \rho_{0}=0.033\, M_{\odot} pc^{-3}$ and a core radius equal to $\rm 2.95$ kpc \citep{banerjee2016mass}.
UGC 00711 is inclined at $\rm 74^{\circ}$ to the line of sight, unlike all the galaxies in our sample, which are perfectly edge-on, 
with an inclination close to $\rm 90^{\circ}$.

\section{Observational constraints}
We model vertical stellar dispersion in optical and 3.6 $\mu$m using observationally determined stellar and \HI{} scaleheights as constraints.
Optical observations of superthin galaxies reveal the young stellar population. 3.6 $\rm \mu$m band, on the other hand, traces 
the older stellar population constitutes the dominant stellar mass component and is free from dust extinction.
All the superthin galaxies in our sample have a single exponential stellar disc in the optical band except FGC1540. 
Further, except for UGC711, all the galaxies in our sample have two exponential stellar discs in the 3.6 $\rm \mu$m band. 
The surface density is given by either,
\begin{uprightmath}
\begin{equation}
{\Sigma}_s(R) = {\Sigma}_{0s} \rm {exp} (-R/R_d) 
\end{equation}
\end{uprightmath}

where $\rm \Sigma_{s0}$ is the central stellar surface density and $\rm R_{d}$ the exponential stellar disc scale length.
or by a double exponential given by
\begin{uprightmath}
\begin{equation}
{\Sigma}_s(R) = {\Sigma}_{01} \rm{exp} (-R/R_{d1})) + {\Sigma}_{02} exp(-R/R_{d2}) 
\end{equation}
\end{uprightmath}

where $\rm \Sigma_{01}$ is central stellar surface density and $\rm R_{d1}$ is the scalelength of the stellar disc 1 and so on. 
We may note here that a recent study of UGC7321 indicated that a double disc is not required to explain the photometric data and  
that the existence of a thick disc is disputable \citep{sarkar2019flaring}. The structural data of the stellar disc for UGC 7321 in $\rm B$-band were taken from \cite{uson2003hi}. 
Stellar photometry for IC 5249 was not available in the literature. \cite{kurapati2018mass} provided the $\rm i$-band stellar photometry 
for FGC1540. For IC2233, \cite{bizyaev2016very} has derived the $\rm r$-band data. The stellar photometry in the $\rm B$-band 
for UGC00711 were taken from \cite{mendelowitz2000rotation}. Table 2.1 summarizes the structural parameters of the stellar disc in the optical band for 
the superthin galaxies in our sample.

All of our sample galaxies in the 3.6 $\rm \mu$m band have both a thick and thin stellar disc, each disc has an exponential surface density and a constant scaleheight.
The structural parameters for the superthin galaxies in this study in the 3.6 $\rm \mu$m  were taken from \cite{salo2015} and presented in Table 2.1. 
To summarize, we find that only three galaxies in our sample have a single exponential disc in the optical band, but the same galaxies have two exponential discs in 
the 3.6 $\rm \mu$m band. Interestingly, the galaxies have optical disc scalelengths that are comparable to those of the thick disc component in 3.6 $\rm \mu$m. 
However, their central surface density and vertical scaleheights are closer to that of the corresponding thin disc component. 
This indicates that the optical disc and the 3.6 $\rm \mu$m stellar disc components for the above galaxies may not originate from the same stellar population.
Curiously, FGC 1540 has a double exponential stellar surface density profile in both optical as well as 3.6$\rm \mu$m .
UGC711, on the other hand, has a single disc in both the optical and the 3.6 $\rm \mu$m bands, and the parameters appear 
comparable, possibly indicating that both the discs represent one \& the same stellar population.

The \HI{} surface density for our sample of superthin galaxies were taken from the following sources: UGC 7321 \citep{uson2003hi}, IC 5249 \citep{van2001kinematics}, 
FGC 1540 \citep{kurapati2018mass}, IC2233 \citep{matthews2007h}, and UGC711 \citep{mendelowitz2000rotation}. Earlier studies suggested that the radial 
profiles of \HI{} surface density might be well-fitted with double-gaussian profiles, possibly indicating two \HI{} discs. 
Commonly, \HI{} surface density peaks away from the center of galaxies, indicating a central 
\HI{} hole. The \HI{} surface density profiles are described by an off-centered double Gaussians given by Equation 1.3 in Section 1.0.1. 
For the gas disc, we only consider atomic hydrogen (\HI{}) surface density because molecular gas in LSBs is negligible, see \cite{banerjee2016mass} for a discussion. 
We use the \HI{} scaleheight for UGC 7321 and IC 5249 from \cite{o2010dark}. The \HI{} scaleheight of FGC 1540 is equal to 0.400 kpc (Kurapati, private communication). 
The \HI{} scaleheight corresponding to IC2232 and UGC711 was calculated using the FWHM versus $\rm \frac{2R}{D_{HI}}$ plot from \cite{o2010dark}:
FWHM= $\rm \frac{2.4}{0.5D_{HI}}R +0.244$, where $\rm D_{\HI{}}$ is the \HI {} diameter. 
The \HI{} scaleheight constrains the \HI{} velocity dispersion but not the dispersion of the stars.
So the changes in the \HI{} scaleheight will not impact the best-fitting stellar vertical velocity dispersion values. 
Table 2.1 summarizes the properties of the \HI{} disc.

For the models using optical photometry, the dark matter profile parameters, i.e., central core density $\rm \rho_{0}$ and core radius $\rm R_{c}$ for UGC 7321, were 
modeled by creating mass models with the $\rm 'rotmas'$ and $\rm 'rotmod'$ tasks in GIPSY \citep{van1992groningen}. The same for FGC1540 was taken from \cite{kurapati2018mass}.
Similarly, the dark matter parameters for IC2233 and UGC711 were determined by constructing mass models using GIPSY. 
For constructing dynamical models of the galaxy in the 3.6 $\mu$m band, the dark matter parameters for UGC 7321, IC5249, and IC2233, 
were taken from \cite{banerjee2016mass}. The dark matter parameters for FGC 1540 were taken from \cite{kurapati2018mass}. 
We have presented the parameters corresponding to the dark matter halos of our sample of superthin galaxies in Table 2.1.
Both optical band and the 3.6 $\rm \mu$m have distinct dark matter parameters. The 3.6 $\rm \mu$m band represents the stellar mass distribution better as it is 
free from dust extinction. To make our dynamical model internally consistent, we employ dark matter properties from mass models created using particular 
photometry as input parameters in the dynamical equations describing the structure and kinematics of the stellar disc in the corresponding photometric band.

\section{Results \& Discussion}
\begin{table}
\hspace{-2.5cm}
\begin{center}
\hfill{}
\caption{Best-fitting parameters of the multi-component models of our sample superthin galaxies.}
\begin{tabular}{|l|c|c|c|c|c|c|}
\hline
Parameters      & $\sigma^{\textcolor{red}{1}}_{0sI}$ &$\alpha^{\textcolor{red}{2}}_{sI}$ & $\sigma^{\textcolor{red}{3}}_{s0II}$& $\alpha^{\textcolor{red}{4}}_{sII}$ &$\sigma^{\textcolor{red}{5}}_{avg}$    \\
Galaxy          &km$s^{-1}$            &              &        km$s^{-1}$    &          &  km$s^{-1}$                  \\
\hline    
\hline
Vertical stellar dispersion:  &  &   &  &    &\\
Optical photometry            &  &   &  &    &\\
\hline 
UGC 7321 & - & - & $10.23\pm 0.64$& $2.58 \pm 0.613$  &\\
FGC 1540 &$36.91\pm 1.14$&$3.72 \pm 0.4$&$13.08\pm 1.17$& $3.32\pm 0.42$&16.78\\

IC  2233 &$14.9\pm 0.57$ & $2.36\pm 0.36$&  &    &\\
UGC 00711&$18.4 \pm 0.87$& $3.21 \pm 0.40$&  &    &\\
\hline
Vertical stellar dispersion: &  &   &  &    &\\
3.6 $\mu$m photometry.       &  &   &  &    &\\
\hline
UGC 7321 &$24.66 \pm 0.88$&$2.15 \pm 0.6$&$9.02 \pm 0.8$& $4.55 \pm 0.68$& 11.58\\ 
IC  5249 &$20.64\pm 0.63$&$2.155 \pm 0.217$&$9.32 \pm 0.39 $&$7.54 \pm 0.23$& 11.08\\
FGC 1540 &$16.20\pm0.87$&$3.77\pm 0.42$&$6.86 \pm 0.57$&$12.1\pm 0.59$&8.63\\
IC  2233 &$15.97 \pm 0.54$&$2.16 \pm 0.42$&$3.9 \pm 0.23$& $6.0 \pm 0.2$& 5.92\\
UGC 00711 &$23.82 \pm 1.45$&$2.42 \pm 0.28$&  &    &\\ 
\hline
Vertical HI dispersion: &  &   &  &    &\\
Optical photometry      &  &   &  &    &\\
\hline
Parameters&$\sigma^{\textcolor{red}{6}}_{0HI}$&$\alpha^{\textcolor{red}{7}}_{HI}$&$\beta^{\textcolor{red}{8}}_{HI}$&  &    \\ 
          &km$s^{-1}$            &  &    & &              \\
\hline

UGC 7321 & $11.06 \pm 0.88 $ & $0.18 \pm 0.07$  &$-0.047 \pm 0.02$ & &\\ 
FGC 1540 & $29.01 \pm 1.16 $ & $4.27 \pm 0.425$ &                  & &\\
IC  2233 & $12.52 \pm 0.515$ & $1.03 \pm 0.14$  &$-0.141 \pm 0.031$& &\\
UGC 00711& $23.10 \pm 1.11 $ & $1.03 \pm 0.145$ &$-0.156 \pm 0.05 $& & \\ 
\hline
Vertical HI dispersion:  &  &   &  &    &\\
3.6 $\mu$m photometry.   &  &   &  &    &\\
\hline
UGC 7321 &$11.19 \pm 0.84$  & $-0.29 \pm0.14$&                   & &\\
IC  5249 &$12.4  \pm 0.53$  & $ -0.99 \pm 0.11$&$0.04 \pm 0.0011$& &\\
FGC 1540 &$17.75 \pm 0.81$  & $6.85 \pm 0.56$&                   & &     \\
IC  2233 &$12.0  \pm  0.56$ & $0.53 \pm 0.23$ &$-0.055 \pm 0.026$& &\\
UGC 00711&$22.03 \pm 1.07$  & $0.92 \pm 0.16$&$-0.1 \pm 0.054$   & &\\
\hline
 \end{tabular}
\hfill{}
\label{table:results}
\end{center}
\begin{tablenotes}
\item (\textcolor{red}{1}): Central vertical stellar velocity dispersion in thick disc 
\item (\textcolor{red}{2}): Scale length of radial fall off of the thick disc stellar dispersion in units of $R_{d1}$ 
\item (\textcolor{red}{3}): Central vertical stellar velocity dispersion in the thin disc
\item (\textcolor{red}{4}): Scale length of radial fall off of the thin disc stellar dispersion in units of $R_{d2}$ 
\item (\textcolor{red}{5}): Average stellar dispersion
\item (\textcolor{red}{6}): Central vertical HI dispersion 
\item (\textcolor{red}{7}): steepness parameter-1 of HI dispersion profile
\item (\textcolor{red}{8}): steepness parameter-2 of HI dispersion profile
\end{tablenotes}
\end{table}

\begin{figure*}
\begin{center}
\begin{tabular}{cc} 
\hspace{-1.5cm}
\resizebox{90mm}{85mm}{\includegraphics{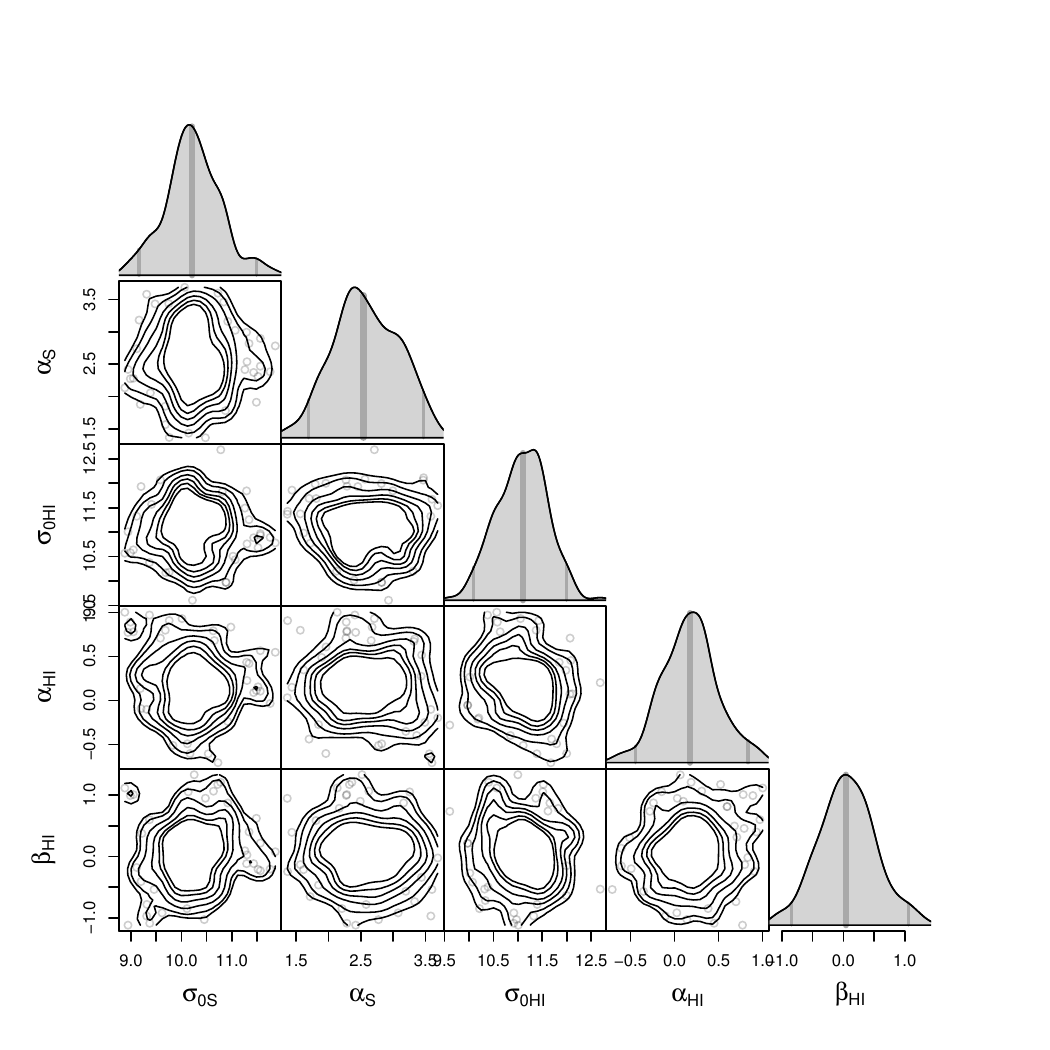}} &
\resizebox{90mm}{85mm}{\includegraphics{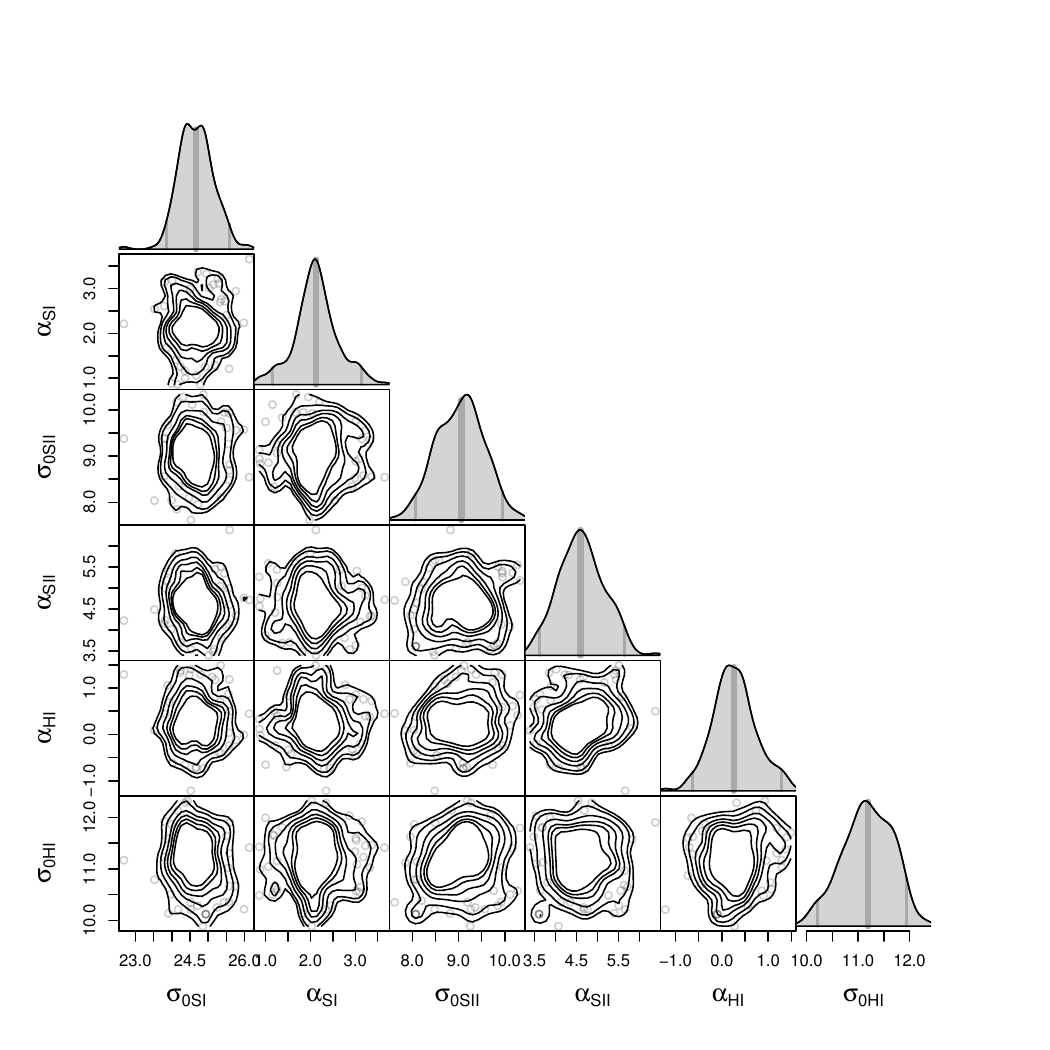}}\\
\end{tabular}
\end{center}
\caption{Posterior probability distribution and covariance plots of the parameters of the multi-component model of the galactic disc of UGC7321 with the stellar component 
modeled by $\rm B$-band [Left Panel] and 3.6$\rm \mu$m photometry [Right Panel]}
\end{figure*}
\begin{figure}
%\begin{center}
\begin{tabular}{c}
\hspace{-1.5cm}
\resizebox{90mm}{85mm}{\includegraphics{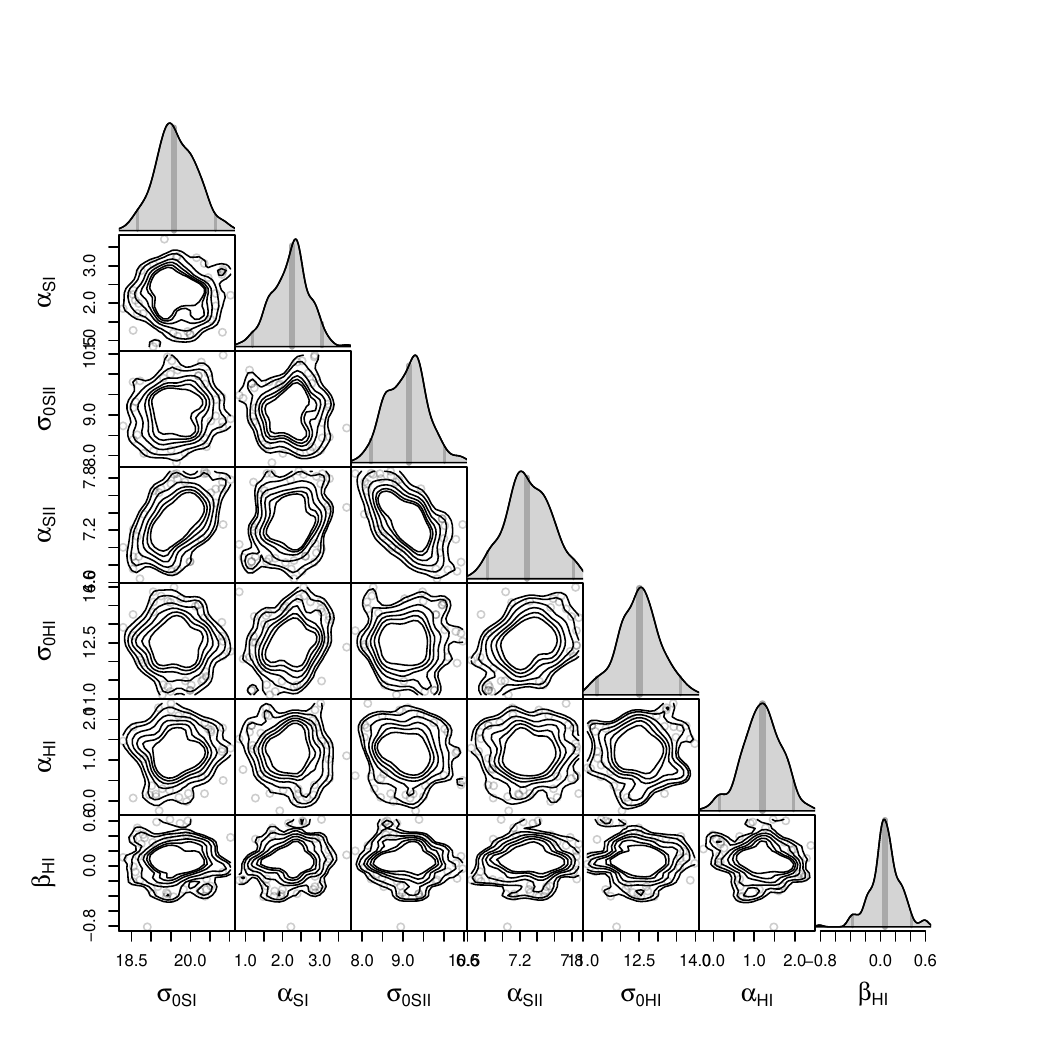}} 
\end{tabular}
%\end{center}
\caption{Posterior probability distribution and covariance plots of the parameters of the multi-component model of the galactic disc of IC5249 with the stellar component
modeled by 3.6$\rm \mu$m photometry}
\end{figure}
\begin{figure*}
\begin{tabular}{cc}
\hspace{-1.5cm}
\resizebox{90mm}{85mm}{\includegraphics{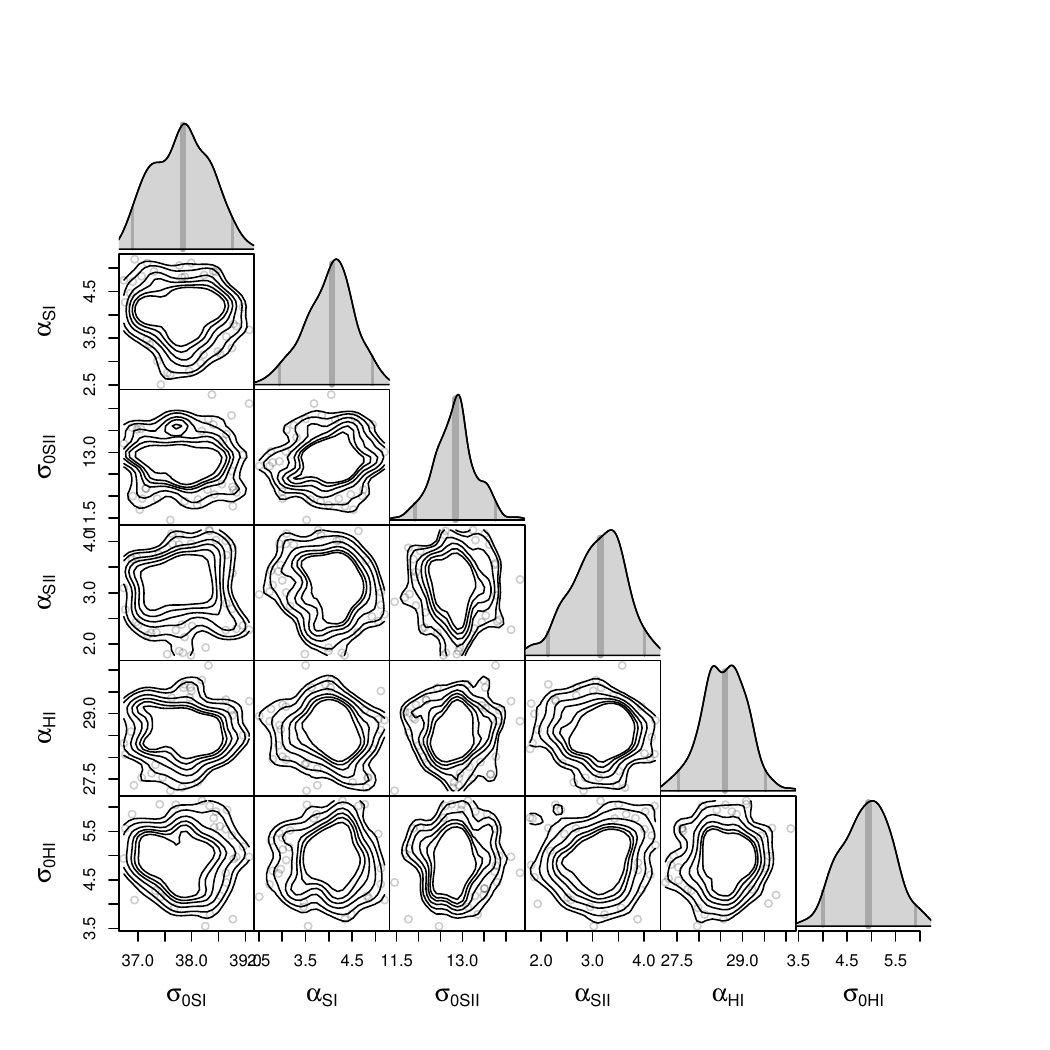}}&
\resizebox{90mm}{85mm}{\includegraphics{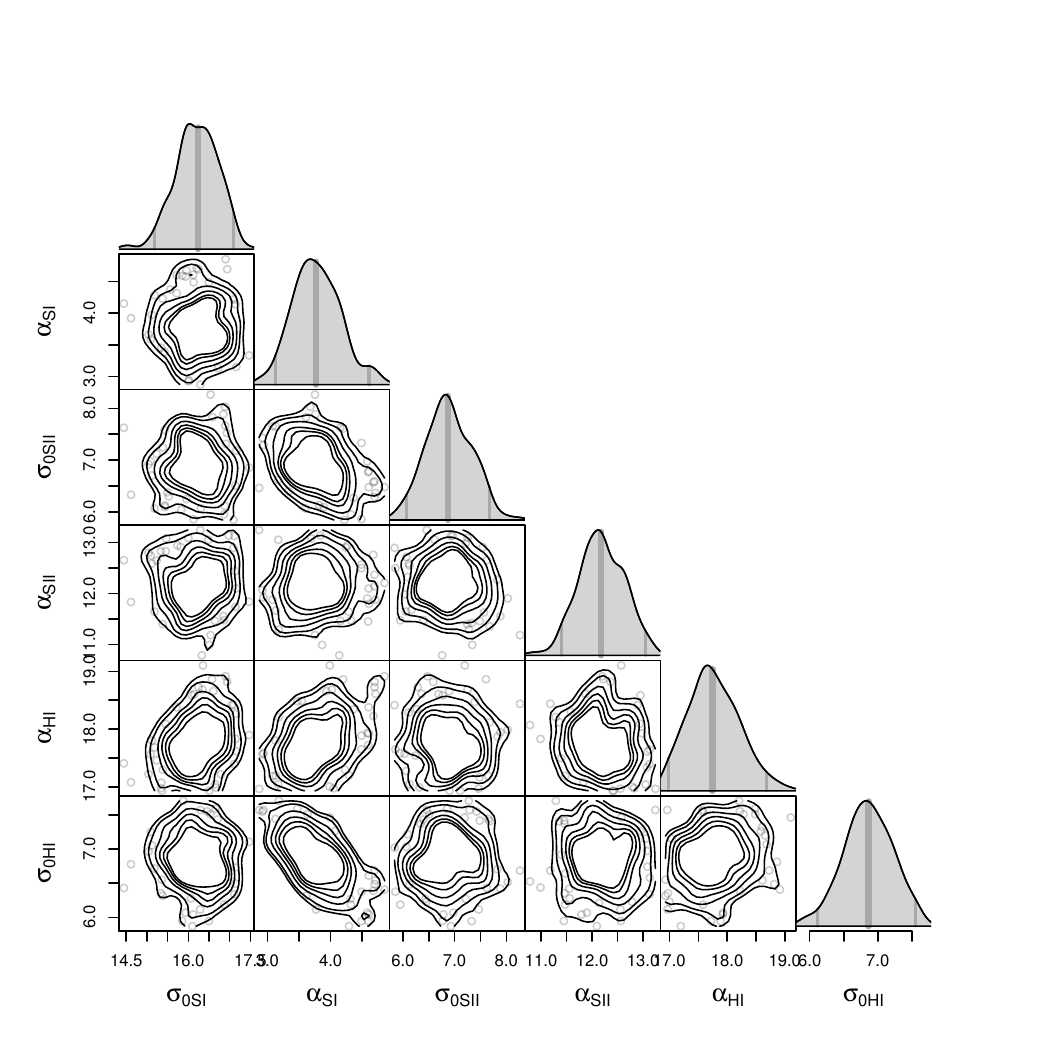}} 
\end{tabular}
\caption{Posterior probability distribution and covariance plots of the parameters of the multi-component model of the galactic disc of FGC1540 with the stellar 
component modelled by $\rm i$-band [Left Panel] and 3.6$\rm \mu$m photometry [Right Panel]}
\end{figure*}
\begin{figure*}
\begin{center}
\begin{tabular}{cc}
\hspace{-1.5cm}
\resizebox{90mm}{85mm}{\includegraphics{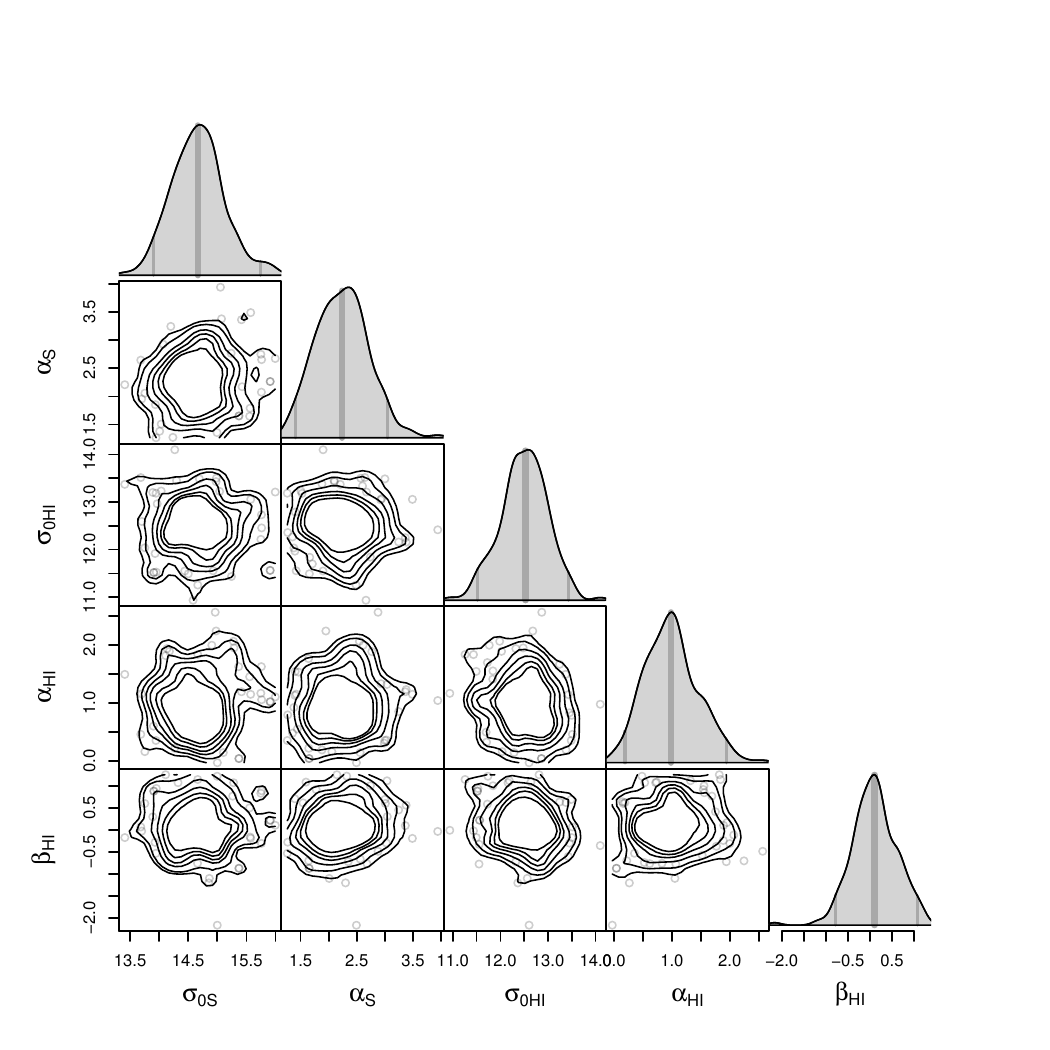}}&
\resizebox{90mm}{85mm}{\includegraphics{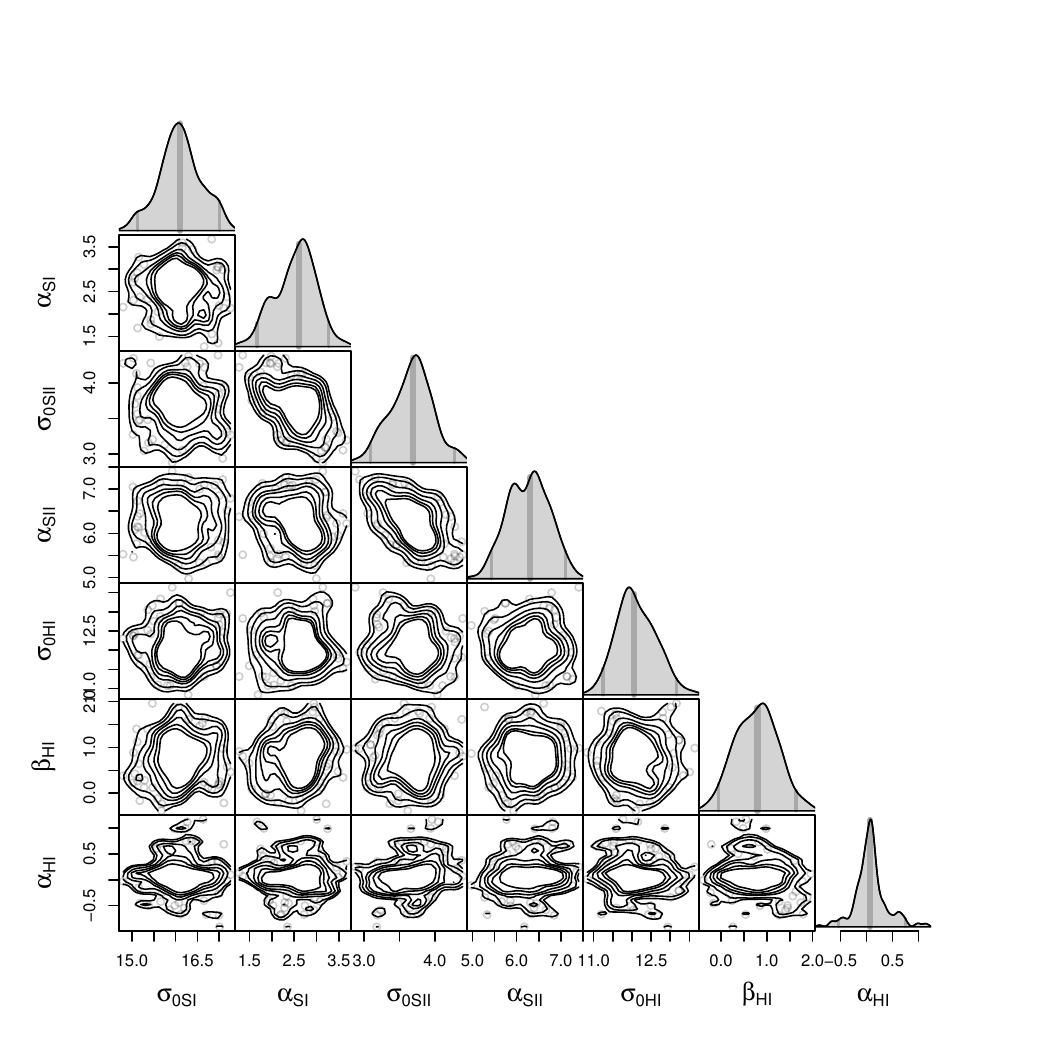}}
\end{tabular}
\end{center}
\caption{Posterior probability distribution and covariance plots of the parameters of the multi-component model of the galactic disc of IC2233 with the stellar component
modeled by $\rm r$-band [Left Panel] and 3.6$\rm \mu$m photometry [Right Panel]}
\end{figure*}
\begin{figure*}
\begin{tabular}{cc}
\hspace{-1.5cm}
\resizebox{90mm}{85mm}{\includegraphics{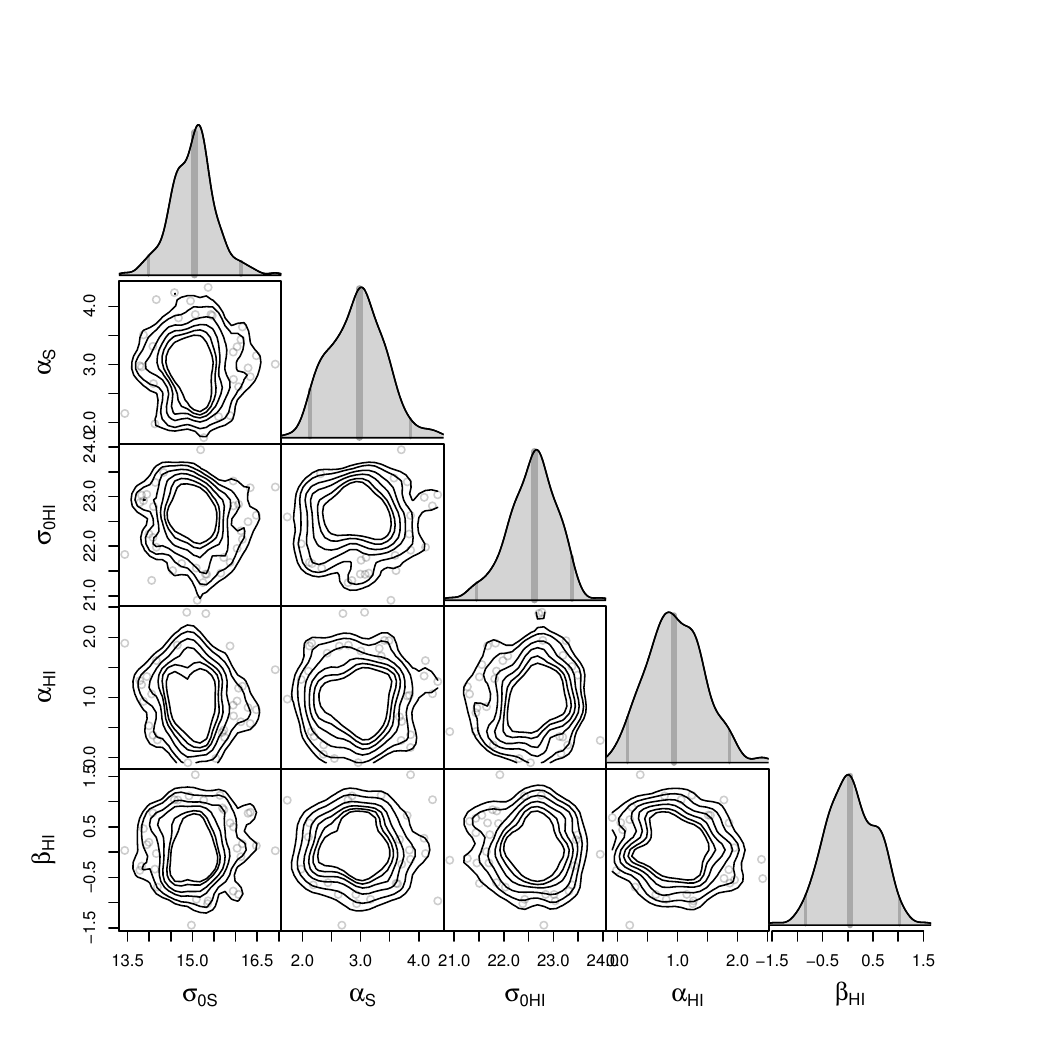}} &
\resizebox{90mm}{85mm}{\includegraphics{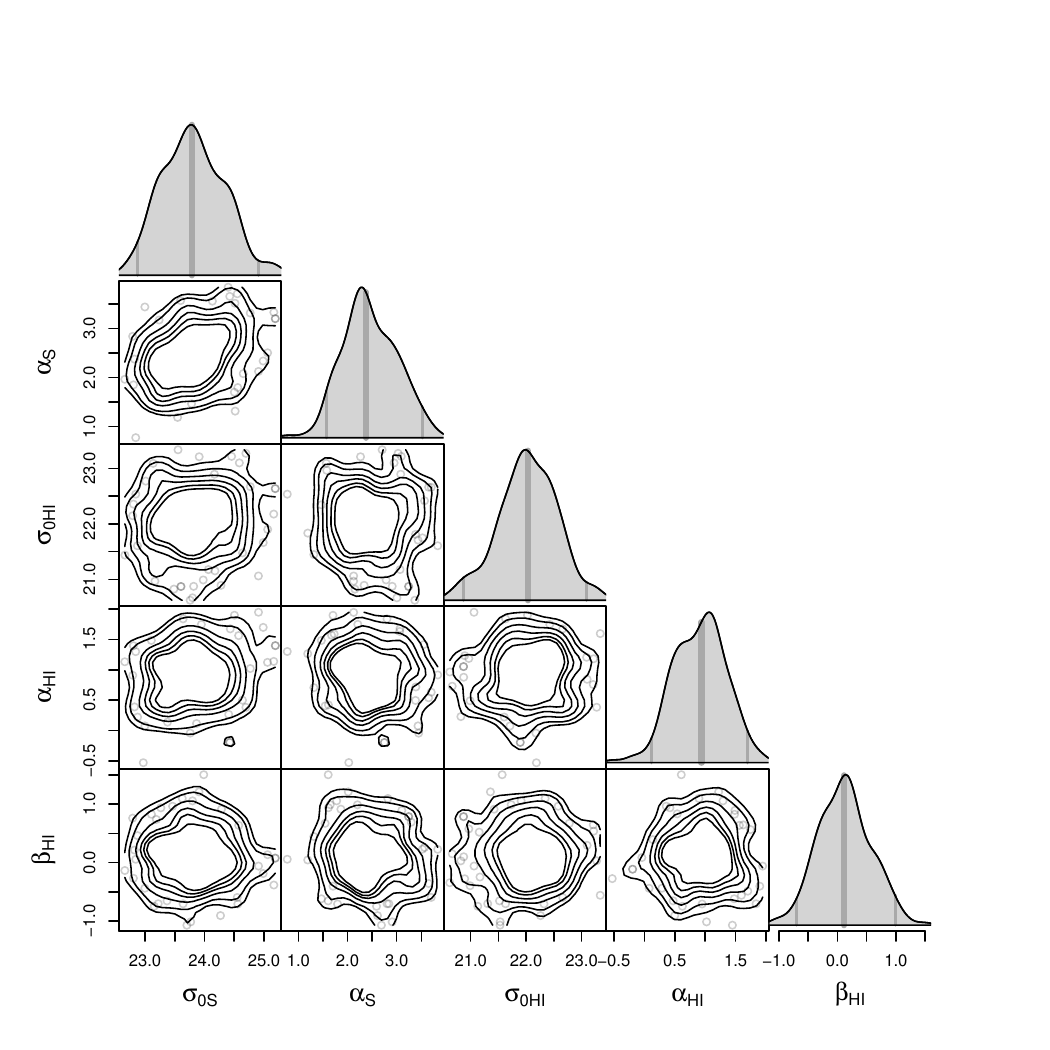}}\\
\end{tabular}
\caption{Posterior probability distribution and covariance plots of the parameters of the multi-component model of the galactic disc of UGC00711 with the stellar 
component modeled by $\rm B$-band [Left Panel] and 3.6$\rm \mu$m photometry [Right Panel]}
\end{figure*}

\subsection{UGC7321}
We present the results obtained from the dynamical modeling of UGC 7321 using the $\rm B$-band stellar photometry and \HI{} 21cm radio-synthesis measurements.
The central stellar vertical velocity dispersion $\rm \sigma_{0s}$ = (10.2 $\pm$ 0.6) \kms and it falls exponentially with (2.6$\pm$0.6)$\rm R_{d}$.
The vertical velocity dispersion of the Milky Way and Andromeda or M31 stellar discs is roughly 53 \kms. This assumes that stellar velocity dispersion 
falls off exponentially with a scale length of 2$\rm R_{d}$, as measured in the Galaxy, and that the vertical-to-radial stellar velocity dispersion ratio is 
0.5 at all radii, equal to its observed value in the solar neighborhood \citep{binney2008princeton}. This confirms that UGC 7321 has an ultracold stellar disc 
with small vertical velocity dispersion.
The \HI{} dispersion is equal to 11.1 $\pm$0.9 \kms with $\rm \alpha_{\HI{}}$=0.2$\pm$0.1 and $\rm \beta_{\HI{}}$=-0.04$\pm$0.02, 
therefore \HI{} dispersion is constant with radius. We use the open-source stellar dynamical code AGAMA to verify the consistency of 
the multi-component model (see Section 1.1.4). We find that the scaleheight predicted by AGAMA agrees with that obtained using the multi-component model 
when using the best-fitting value of the vertical stellar dispersion 
as an input to AGAMA. The multi-component disc dynamical stability parameter $\rm Q_{RW}$ (see Section 1.1.5) is then calculated as a function of $\rm R$ using the value of the radial velocity
dispersion obtained from AGAMA. We find that the minimum value of $\rm Q_{RW}$ is 2.7 at roughly 5$\rm R_{d}$, confirming that UGC 7321, despite having an ultra-cold stellar disc, 
is stable against the growth of axisymmetric perturbations. In the 3.6 $\rm \mu$m, the value of the central vertical velocity dispersion for the thin disc is 9.02$\pm$0.8 \kms. 
It falls off exponentially with a scalelength (4.6$\pm$0.7)$\rm R_{d2}$, where $\rm R_{d2}$ is the scalelength of the thin disc.
It is interesting to note that, within error bars, the vertical velocity dispersion profile of the thin disc in the 3.6 $\rm \mu$m matches the profile from 
the $\rm B$-band component. The value of the central vertical velocity dispersion of the thick disc 
is 24.7$\pm$0.9 \kms, it falls exponentially with disc scalelength (2.2$\pm$0.6)$\rm R_{d1}$, where $\rm R_{d1}$ is the scalelength of the thick disc. 
The density-averaged vertical velocity dispersion reflects the cold, dense, compact, thin disc component. 
The central value of \HI{} velocity dispersion is equal to 
11.2 $\pm$0.8 \kms, together with $\rm \alpha_{\HI{}}$=-0.3$\pm$0.8 and $\rm \beta_{\HI{}}$=-0.04$\pm$0.02, shows that the \HI{} dispersion nearly remains constant 
with $\rm R$. We find similar estimates of the \HI{} vertical velocity dispersion profile, using two different tracers. 
In Figure 2.1, we show the posterior probability distribution of the parameters of the multi-component model of the galactic disc of 
UGC7321 modeled using B-band [Left Panel] and 3.6 $\rm \mu$m photometry [Right Panel]. The vertical velocity dispersion of the stars in UGC 7321 is lower than 
that of the thin disc stars in the Milky Way. Despite the low vertical velocity dispersion of the stars, UGC 7321 is stable against the growth of axisymmetric instabilities.

\subsection{IC5249}
In the 3.6 $\mu$m, the vertical velocity dispersion of the stars in the thick disc has a central value equal to 20.6 $\pm$ 0.6 \kms, 
and it decreases exponentially with scalelength equal to (2.2 $\pm$ 0.2) $\rm R_{d1}$. 
For the thin disc, the central vertical velocity dispersion is equal to 9.3 $\pm$0.4 \kms and falls off exponentially with (7.5 $\pm$0.2) $\rm R_{d2}$. 
The density-averaged vertical velocity dispersion converges to the value of the vertical velocity dispersion profile at $\rm R \geq R_{d1}$ of the hot, 
diffuse, and extended thick disc. The central value of \HI{} dispersion is 12.4 $\pm$0.5 \kms with $\rm \alpha_{\HI}$=-0.9 $\pm$0.1 and $\rm \beta_{\HI{}}$=-0.04$\pm$0.01, 
indicating that the \HI{} dispersion nearly remains constant with $\rm R$. $\rm Q_{N}$ has a minimum value of 1.7 at roughly 3 $\rm R_{d1}$ suggesting that 
IC5249 may be on the brink of dynamical instability. Figure 2.2 shows the posterior probability distribution of the model parameters 
for IC 5249, derived using 3.6$\mu$m photometry. The value of the density averaged vertical velocity dispersion of stars of the composite thick+thin disc system in IC 5249 is lower than the same for Milky Way's
thin disc (20\kms) \citep{sharma2014kinematic}. The $\rm Q_{N}=1.7$ at $3R_{d1}$ indicates that IC 5249 may be susceptible to the growth of axisymmetric instabilities. 

\subsection{FGC1540}
FGC1540 has a thick and thin stellar disc in the optical band, unlike other superthin galaxies in our sample. The vertical velocity dispersion of the thick disc 
is 36.9 $\pm$1.1 \kms and falls off exponentially with 3.7 $\pm$ 0.4 $\rm R_{d1}$. The central dispersion of the thin disc is 13.1 $\pm$1.2 \kms, and 
falls-off exponentially with scalelenght equal to 3.3 $\pm$ 0.4 $\rm R_{d2}$. The density averaged dispersion reflects the thin disc component.
Due to the lack of \HI{} scale height data, we assume the scaleheight equal to 400 $\rm kpc$ at all radii (Kurapati, private communication)
The multi-component stability parameter $\rm Q_{N}$ has a minimum value of 1.9.
In 3.6 $\rm \mu$m band the vertical velocity dispersion of the thick disc is 16.2$\pm$0.9 \kms and falls exponentially with scale length 3.8 $\pm$0.4 $\rm R_{d1}$.
The central vertical velocity dispersion for the thin disc is 6.9 $\pm$ 0.6 \kms and falls off exponentially with scalelength equal to 12.1 $\pm$0.2 $\rm R_{d2}$
At large $\rm R$, density-averaged dispersion converges to thick disc value.
We note that the thin disc's vertical velocity dispersion in the optical band matches the thick disc's velocity dispersion in the 3.6 $\mu$m band.
FGC 1540 can resist axisymmetric instabilities as the $\rm Q_{N}$ is at least equal to 2.9. Figure 2.3 shows the posterior probability distribution of the 
model parameters in the $\rm i$-band [Left Panel] and 3.6$\rm \mu$m photometry [Right Panel].
The vertical velocity dispersion of the thick and the thin disc is comparable to that of Milky Way in the optical band, but the density averaged velocity dispersion converges
to that of the thin disc in the optical and to the thick disc in the 3.6 $\rm \mu m$ band. Further, the $\rm Q_{RW}=2.9$ indicates that the FGC 1440 is stable against
axisymmetric instabilities.

\subsection{IC2233}
The central velocity dispersion for IC 2233 in r-band is equal to 14.9$\pm$0.6 \kms with a scalelength of 2.4$\pm$0.4. $\rm Q_{N}$ has a minimum value $\sim$ 2.2, 
confirming the stability of the galaxy against axisymmetric instabilities. The central vertical velocity dispersion of the thick disc in 3.6 $\rm \mu$m is 
15.9 $\pm$0.5 \kms and falls off exponentially with a scalelength of (2.2 $\pm$ 0.4) $\rm R_{d1}$, and the value of stellar dispersion for the thin disc is 
3.9 $\pm$0.2 \kms and the corresponding scalelenght of dispersion is equal to (6.0 $\pm$0.2) $\rm R_{d2}$. The density averaged vertical velocity dispersion 
does not reflect any component, remaining constant at 6\kms at all radii. 
The disc is robust against axisymmetric instabilities as $\rm Q_{N}$ 
in 3.6 $\rm \mu$m has a minimum value of 5.7. Figure 2.4 shows the posterior probability distribution and covariance plots of the multi-component model of IC2233 
using $\rm r$-band photometry. [Left Panel] the 3.6$\rm \mu$m photometry [Right Panel].
The central vertical velocity dispersion of stars in the optical band and 3.6 $\rm \mu m$ is lower than the Milky Way's 
thin disc ($\rm \sigma_{z}=25\kms$). The values of the vertical velocity dispersion of the thin disc stars in IC 2233 are lower than the typical value (5 \kms) found for 
the molecular gas. The minimum value $\rm Q_{RW}=5.7$ is significantly higher than the marginal stability levels, indicating the stability against axisymmetric instabilities.

\subsection{UGC00711}
The central vertical velocity dispersion in B-band is 18.4$\pm$0.9 \kms, and it decreases exponentially with a scalelength equal to (3.2$\pm$0.4) $\rm R_{d1}$, 
where $\rm R_{d1}$ is the exponential scalelength of the optical disc. The minimum value of $\rm Q_{N}$ is equal to 4.5, which shows that the disc is dynamically stable. 
In 3.6$\rm \mu$m, the central velocity dispersion is 23.8$\pm$1.5 \kms, and it decreases exponentially with scalelength (2.4$\pm$0.3)$\rm R_{d1}$. 
The minimum value for $\rm Q_{N}$ is 4.3. We notice that the optical and the 3.6 $\rm \mu$m discs of UGC00711 have similar profiles 
for vertical velocity dispersion and dynamical disc stability, which could mean that they are the same disc in both the bands. In Figure 2.5, we show 
the posterior probability distribution and covariance plots of the parameters of the multi-component model of the galactic disc of UGC711, with the 
stellar component modeled by $\rm r$-band [left panel] and 3.6$\rm \mu$m photometry [right panel].
The vertical velocity dispersion of stars is comparable to each other in the optical and 3.6 $\rm \mu m$ band. The minimum value of $\rm Q_{N}=4.7$ is significantly
higher than 1, proving that $UGC00711$ is highly stable.

\begin{figure*}
\hspace{-20mm}
\resizebox{185mm}{195mm}{\includegraphics{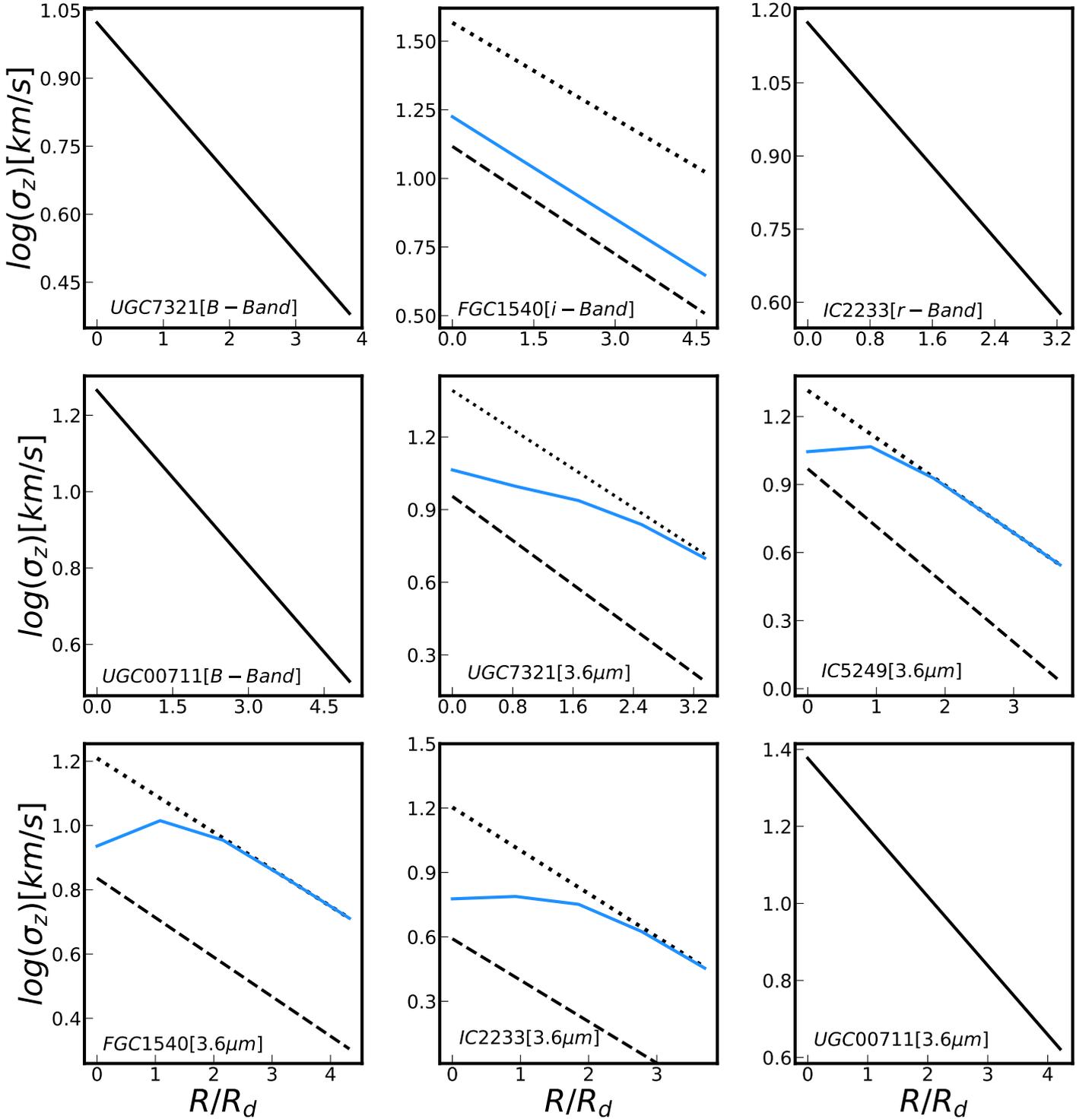}} 
%\vspace{-84}
\caption{Model-predicted stellar vertical velocity dispersion in logarithmic scale for our sample of superthin galaxies as a function of galactocentric 
radius $\rm R$ normalized by exponential stellar disc scalelength $\rm R_{d}$. In double exponential stellar disc, $\rm R_{d}$ is $\rm R_{d1}$, which is the 
thick disc's scale length. In a 1-component stellar disc, vertical velocity dispersion is a single solid line. The dashed line represents the thin disc, the dotted 
line the thick disc, and the solid line the density-averaged value.}
\end{figure*}

\begin{figure*}
\hspace{-20mm}
\resizebox{185mm}{195mm}{\includegraphics{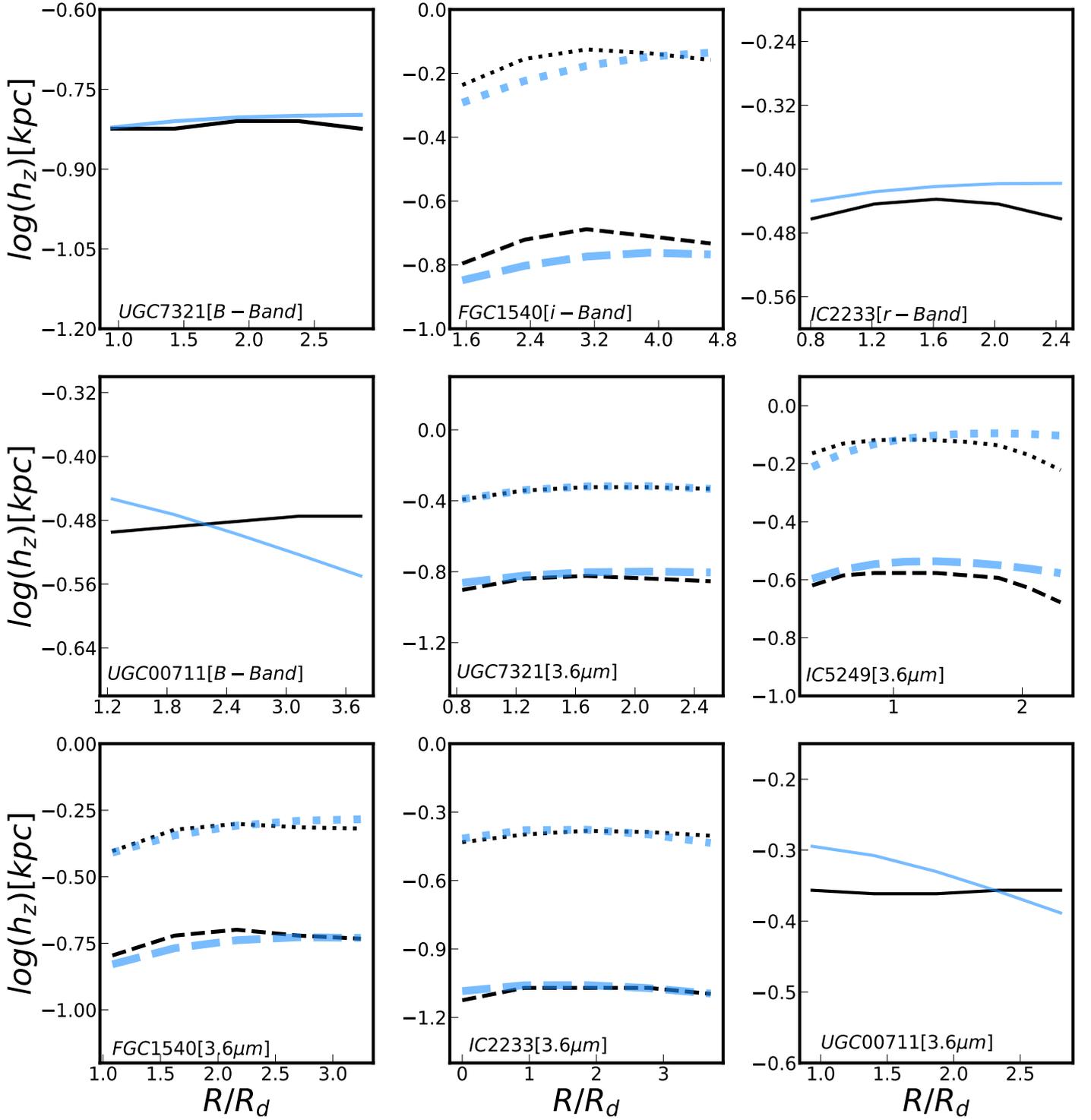}} 
\caption{We plot model-predicted stellar scaleheight in logarithmic scale for our sample superthins as a function of galactocentric radius $\rm R$ normalized by 
exponential stellar disc scalelength $\rm R _{d}$. In the double exponential stellar disc, $\rm R_{d}$ is $\rm R_{d1}$, the thick disc's scale length. In a single 
exponential stellar disc, the vertical scaleheight is a solid line. The dashed line corresponds to the thin disc and the dotted line to the thick disc.
Black lines represent the multi-component model, whereas blue lines represent AGAMA.}
\end{figure*}

\begin{figure*}
\hspace{-20mm}
\resizebox{185mm}{195mm}{\includegraphics{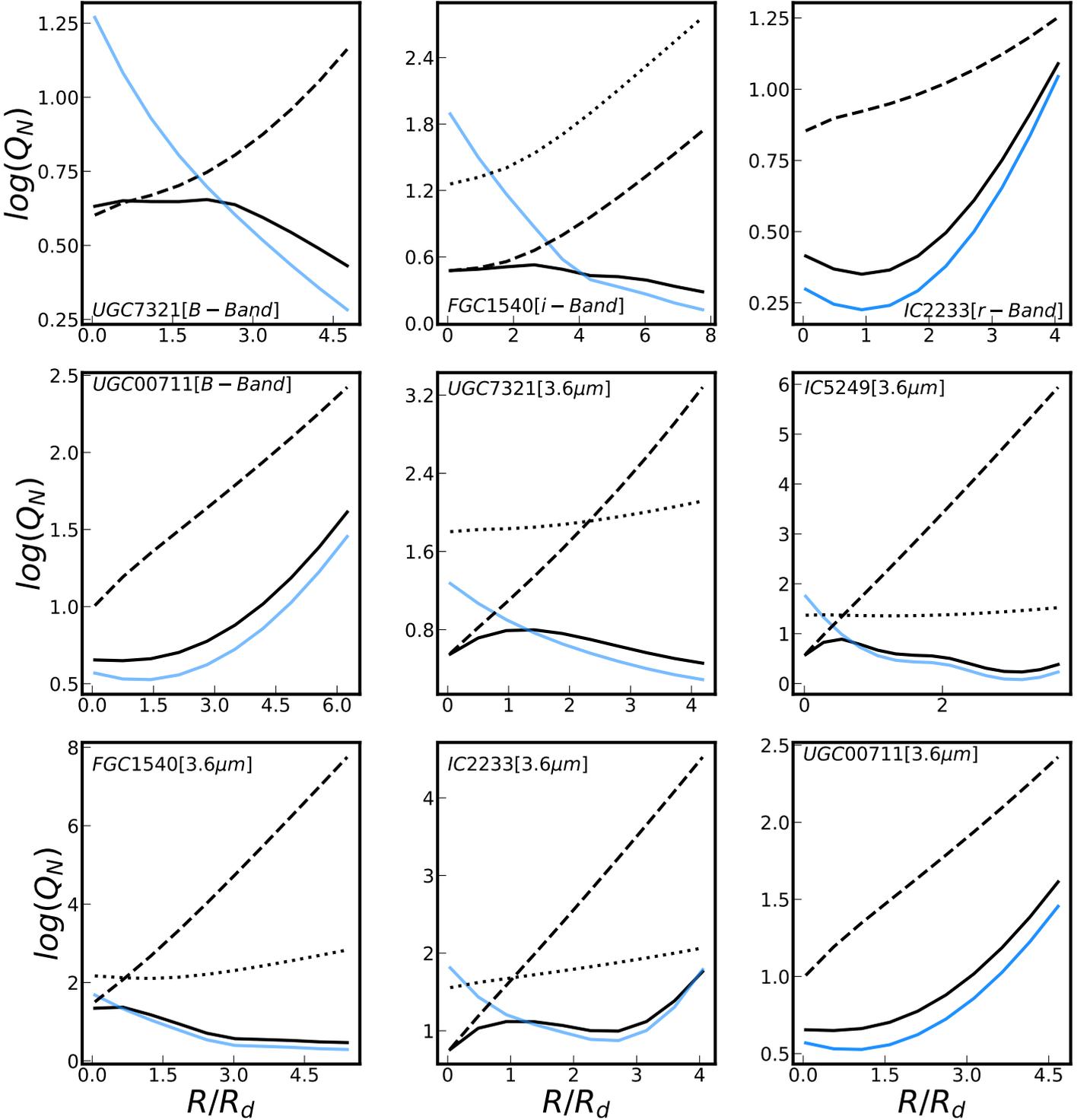}} 
\caption{We plot our sample superthins' stability parameter $\rm Q_{N}$ in logarithmic scale as a function of galactocentric radius $\rm R$ normalized by stellar 
disc scale length $\rm R_{d}$. In double exponential model of stellar discs, $\rm R_{d}$ is $\rm R_{d1}$, the thick disc's scalelength. In a single exponential model of 
stellar disc, the dashed line is Toomre Q of the stellar disc. In a double exponential model of the stellar disc, the dashed line represents the thin disc's Toomre Q, and 
the dotted line represents the Toomre Q of the thick disc. For both cases, the blue line is the Toomre Q of the gas disc, and the solid line is the multi-component disc 
stability parameter $\rm Q_{N}$.}
\end{figure*}

\subsection{Stellar vertical velocity dispersion}
We present dynamical models of superthin galaxies using stellar photometry and \HI{} 21cm radio-synthesis observations.
Figure 2.6 shows the vertical stellar velocity dispersion in logarithmic scale $\rm log(\sigma_{z})$ as a function of $\rm R/R_{d}$ for our sample superthin galaxies.
For each sample galaxy, we explore two instances of the dynamical models, depending on the photometric band: 
i) optical (ii) 3.6 $\rm \mu$m. Except for FGC1540, our sample of superthin galaxies have a single thin stellar disc in the optical band. 
FGC1540 has both thin and thick disc components in the optical band. In the optical band, $\rm \sigma_{0s}$ ranges between $\rm 10.2$\kms and $\rm 18.4$\kms. 
All the superthin galaxies in our sample, except UGC711, have a thin and thick exponential stellar disc in the 3.6 $\rm \mu$m band. UGC711 
has only a thick disc in 3.6 $\rm \mu$m. The vertical stellar velocity dispersion of the thin disc $\rm \sigma_{0s}$ ranges from 3.9 to 9.3 \kms. The vertical 
velocity dispersion of the thick disc lies in between $\rm 15.9$\kms to $\rm 24.7$ \kms. Also, the density-weighted average for thin and thick stellar discs 
lies between $\rm 5.9$\kms $\rm \&$ $\rm 11.6$ \kms. The 3.6 $\rm \mu$m band traces the older stellar population, whereas the optical data traces the young stellar 
population. The thin disc component in UGC 7321, IC 5249, FGC 1540, and IC 2233 exhibit a lower vertical stellar velocity dispersion than the optical disc. This possibly
implies that these galaxies probably underwent a recent star formation event, where the short-lived young stars have entered the red giant phase, emitting in near-infrared. 
Thus, the thin disc in 3.6 $\rm \mu$m may indicate the presence of a cold near-infrared component. Previous studies on the vertical structure of disc galaxies 
assumed $\rm \alpha _{s}=2$ to keep stellar scaleheight constant with radius (see \cite{van1982surface}, \cite{van1988three}, \cite{van2011galaxy}, and 
\cite{sharma2014kinematic}). We keep $\rm \alpha_{s}$ as a free parameter in our model \citep{narayan2002origin}. $\rm \alpha_{s}$ varies between 2.4 and 
3.7 for stellar discs in the optical band. The $\rm \alpha_{s}$ for 3.6 $\rm \mu$m thick disc lies between 2.1-3.7. In comparison, \cite{narayan2002origin}
found $\rm \alpha_{s}$ between 2 and 3 for a sample of ordinary disc galaxies. $\rm \alpha_{s}$ for the thin disc in  3.6 $\rm \mu$m lies in between 4.5 - 12.1.
The value of the central vertical stellar velocity dispersion in the Milky Way and Andromeda (M31) is $\sim$ 53 \kms \citep{lewis1989kinematics} and \citep{tamm2007visible}, 
assuming that the dispersion falls off with a scalelength equal to $\rm 2R_{d}$ and that the ratio of vertical-to-radial stellar velocity dispersion equals 
0.5 \citep{binney2008princeton}. According to kinematic models that were fitted to data from the 
Geneva Copenhagen Survey (GCS) \citep{nordstrom2004geneva} and Radial Velocity Experiment (RAVE) \citep{sharma2014kinematic, steinmetz2006radial}, 
the vertical velocity dispersion of the thin and thick disc of Milky Way stars is given by  $\rm \sigma_{0sII}=25.73^{+0.21}_{-0.21}$km/s and 
$\rm \sigma_{0sI}=34.3^{+0.51}_{-0.57}$\kms, the exponential scalelengths being $\rm 1/(\alpha_{s2}R_{d2})=0.073^{+0.0037}_{-0.003}$ $kpc^{-1}$ 
and $\rm 1/(\alpha_{sI}R_{d1})=0.1328^{+0.005}_{-0.0051}$ 
$\rm kpc^{-1}$ respectively. Central stellar vertical dispersion values of the older stellar population in the two face-on spirals, NGC 628 and NGC 1566, are 60 \kms and 80 \kms, respectively \citep{van1984vertical}, which is substantially greater than the range of 15 \kms- 24 \kms seen in the 3.6 $\rm \mu$m 
thick disc of superthin galaxies in our sample. Compared to our sample of superthin galaxies, the 30 low inclination galaxies in the 
DiskMass survey had a central vertical dispersion between 25.9 \kms to 108.5\kms \citep{martinsson2013diskmass} .
We find that the values of vertical velocity dispersion for our sample of superthin galaxies is very small compared even to the stars in the thin disc of the Milky Way. The 
values of vertical velocity dispersion in superthin is roughly 1/3 of that observed in the two face-on spiral galaxies, NGC 628 (60 \kms) and NGC 1566 (80 \kms).

Multi-component models can constrain the \HI{} dispersion of the galaxies along with the stellar dispersion. However, only UGC 7321 and IC 5249 had 
\HI{} scale height measurements. For the other superthin galaxies, we approximated \HI{} scale height using the relation between FWHM and \HI{} diameter as given 
in \cite{o2010dark}. The \HI{} dispersion for UGC 7321 and IC 5249 is 11.2 \kms and 12.4 \kms, which matches the measured mean dispersion $11.7 \pm 2.3$ \kms 
of nearby galaxies from THINGS \HI{} survey\citep{mogotsi2016hi}. The modeled \HI{} velocity dispersion lies between 11-29 \kms for our sample of superthin galaxies. 
This is comparable to an earlier analysis of spiral galaxies in which \HI{} velocity dispersion ranged from 9\kms to 22\kms with a 
mean value 11.7\kms \citep{mogotsi2016hi}. \cite{tamburro2009driving} found a comparable \HI{} velocity dispersion equal to $10$\kms.
Low \HI{} dispersion estimates are consistent with the velocity dispersion of the cold neutral medium (CNM) 3.4\kms - 14.3 \kms.
\citep{ianjamasimanana2012shapes}. Higher \HI{} dispersion values may indicate a warm neutral medium (WNM) with values between 10.4\kms and 43.2\kms
\citep{ianjamasimanana2012shapes}. In some circumstances, our calculations show that the vertical velocity dispersion of the stars is lower than the \HI{} dispersion.
As stars are collisionless, they cannot dissipate energy through collisions and cannot have smaller dispersion than the gas clouds in which they develop. 
This suggests that thin disc stars were created in very cold low dispersion molecular clouds. Table 2.2 summarizes our results.

\subsection{Model stellar scaleheight: Multi-component model versus AGAMA}
We use the publicly available stellar dynamics code AGAMA to verify the consistency of the multi-component model (see Section 1.1.4). We show the results in Figure 2.7.
We find that the scaleheight predicted by AGAMA agrees with that obtained from the multi-component model when using the best-fitting value of the vertical 
stellar dispersion as an input parameter. As was previously mentioned, AGAMA estimates the radial velocity dispersion of the stars in addition to the 
vertical velocity dispersion, and also the ratio of the same as a function of galactocentric radius.
The ratio of the vertical-to-planar velocity dispersion of the stellar component of the model using optical photometry remains essentially constant at 0.5 within 
3 $\rm R_{d}$. On the other hand, using the 3.6 $\rm \mu$m photometry, we find the following trend:
For the thin disc, the ratio is fixed at 0.5 in UGC7321, IC5249, and UGC711. The ratio of vertical-to-planar dispersion ranges between 0.5 and 0.3 in FGC 1540, 
and is fixed at 0.3 in IC2233. For the thick disc, it varies between 0.4 and 0.3 in UGC7321 and IC2233 but stays constant at 0.5 in IC5249 and UGC711 and is 
equal to 0.3 for FGC1540. This is consistent with the findings of \cite{gerssen2012disc}, who find that the ratio of the 
vertical to the planar stellar velocity dispersion deviates substantially from early to late-type galaxies. The value of the planar to the 
vertical dispersion for Miky Way, which is an Sbc galaxy is equal to $\sim$ 0.5. However, it can be considerably closer than 0.3 for late-type galaxies 
like the superthins (Scd -Sd). Finally, we examine the dynamical stability of our sample of superthin galaxies using the values of the radial stellar dispersion from AGAMA.

\subsection{ Disc dynamical stability}
Figure 2.8 shows the multi-component stability parameters $\rm Q_{N}$ \citep{romeo2013simple}  as a function of $\rm R/R_{d}$ (see Section 1.1.5).
We also plot Toomre's stability parameter for each disc component \citep{toomre1964gravitational}. Except at the innermost galactocentric radii, $\rm Q_{N}$ 
closely tracks the $\rm Q$ value of the gas disc for all our sample galaxies.
\begin{figure*}
%\vspace{-1.2cm}
\hspace{-0.9cm}
\resizebox{130mm}{85mm}{\includegraphics{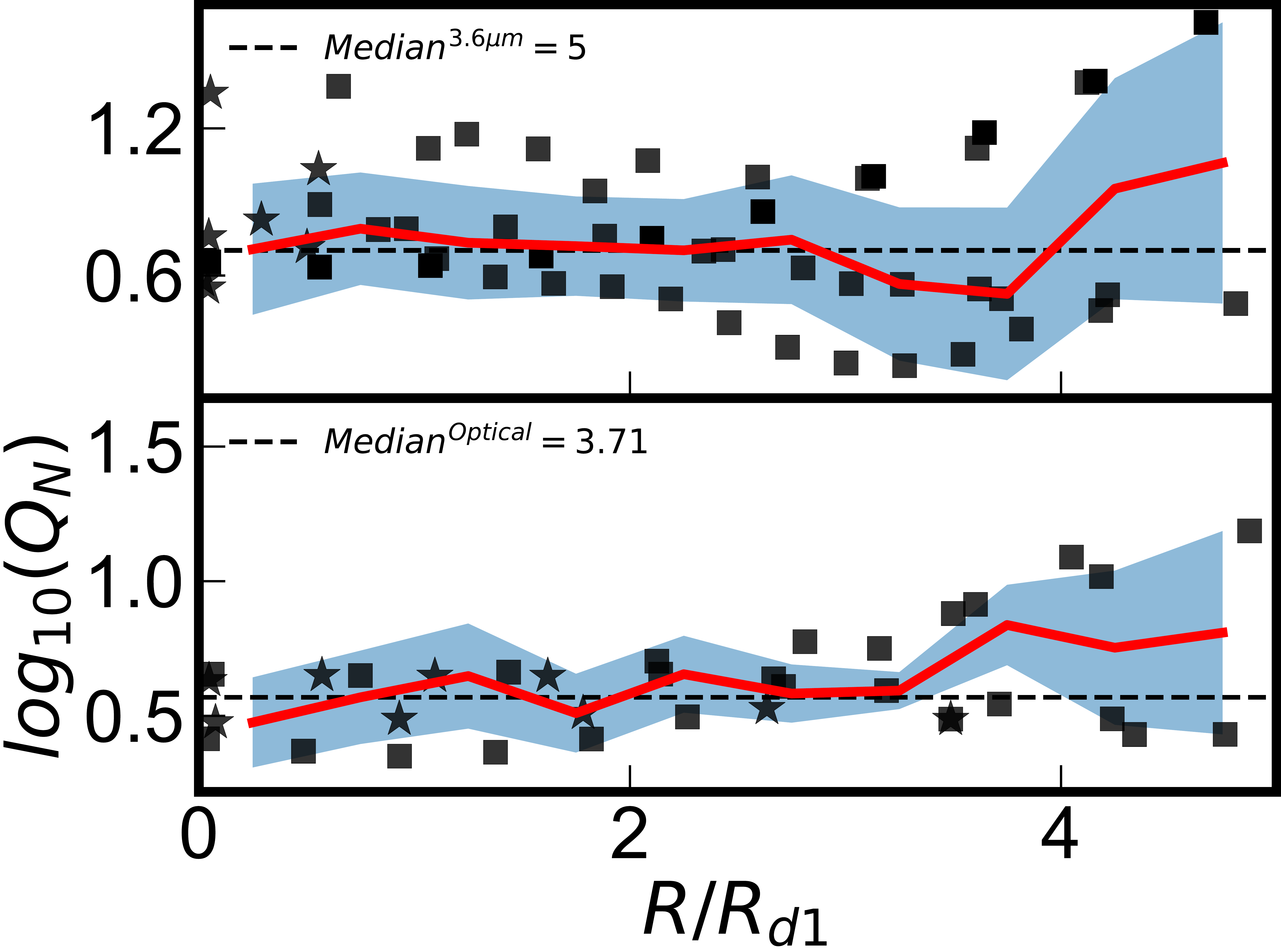}} 
\caption{The figure shows multi-component disc stability parameter $\rm Q_{N}$ of our sample superthins as a function of galactocentric radius $\rm R$ normalized 
by exponential stellar disc scalelength $\rm R_{d}$. In 2-component stellar discs, $\rm R_{d}$ is $\rm R_{d1}$, the thick disc's scale length. 
The upper panel shows the $\rm Q_{N}$ for the stellar disc using 3.6 $\rm \mu$ band, while the lower panel shows the $\rm Q_{N}$  using the optical band. $\rm "Stars"$ indicate 
$\rm Q_{N}$ values driven by the stellar disc, and $"squares"$ show the $\rm Q_{N}$ values driven by the gas disc. The solid red line depicts the sample's median, and the 
blue shaded region shows the 1-sigma scatter. The black dashed line shows the global median.}
\end{figure*}
Figure 2.9 shows Log$\rm Q_{N}$ versus $\rm R/R_{d}$ for the superthin galaxies in our sample. The Upper and lower panels model the stellar disc using 3.6 $\rm \mu$m 
and optical photometry, respectively. The sample's local median (solid lines) and global (dotted lines) are also shown. The blue-shaded region represents one-sigma 
scatter in each radial bin. These values are higher than the median $\rm Q_{N}$ equal to 2.2 $\pm$ 0.6 reported for a sample of nearby star-forming galaxies 
by \cite{romeo2017drives}, who also modeled the galaxies using 3.6 $\rm \mu$m photometry from the SINGS survey \citep{kennicutt2003sings}.
This suggests that the superthin galaxies are more stable than conventional disc galaxies. We also observe that the global median value for superthin galaxies
is larger than $\rm Q_{critical}=2$ \citep{griv2012stability} and $\rm Q_{critical}=2-3$ \citep{elmegreen2011gravitational}, 
the minimum values of $\rm Q_{N}$ for the sample of superthin galaxies ranges between 1.9 to 4.5, with a median equal to 2.5. The same varies between 1.7 to 5.7, with a median equal to 2.9 in the 3.6 $\rm \mu$m band. 
This agrees the median $\rm Q_{N}$ value equal to 2.9 - 3.1 obtained for the low surface brightness galaxies in studies by \cite{garg2017origin}. Therefore, the median stability 
levels of the superthin galaxies are higher than that of the nearby galaxies, suggesting that these galaxies can resist the growth of local axisymmetric
instabilities.

\subsection{How 'cold' are superthin galaxies ?}
\begin{figure*}
\hspace{1.5cm}
\resizebox{130mm}{85mm}{\includegraphics{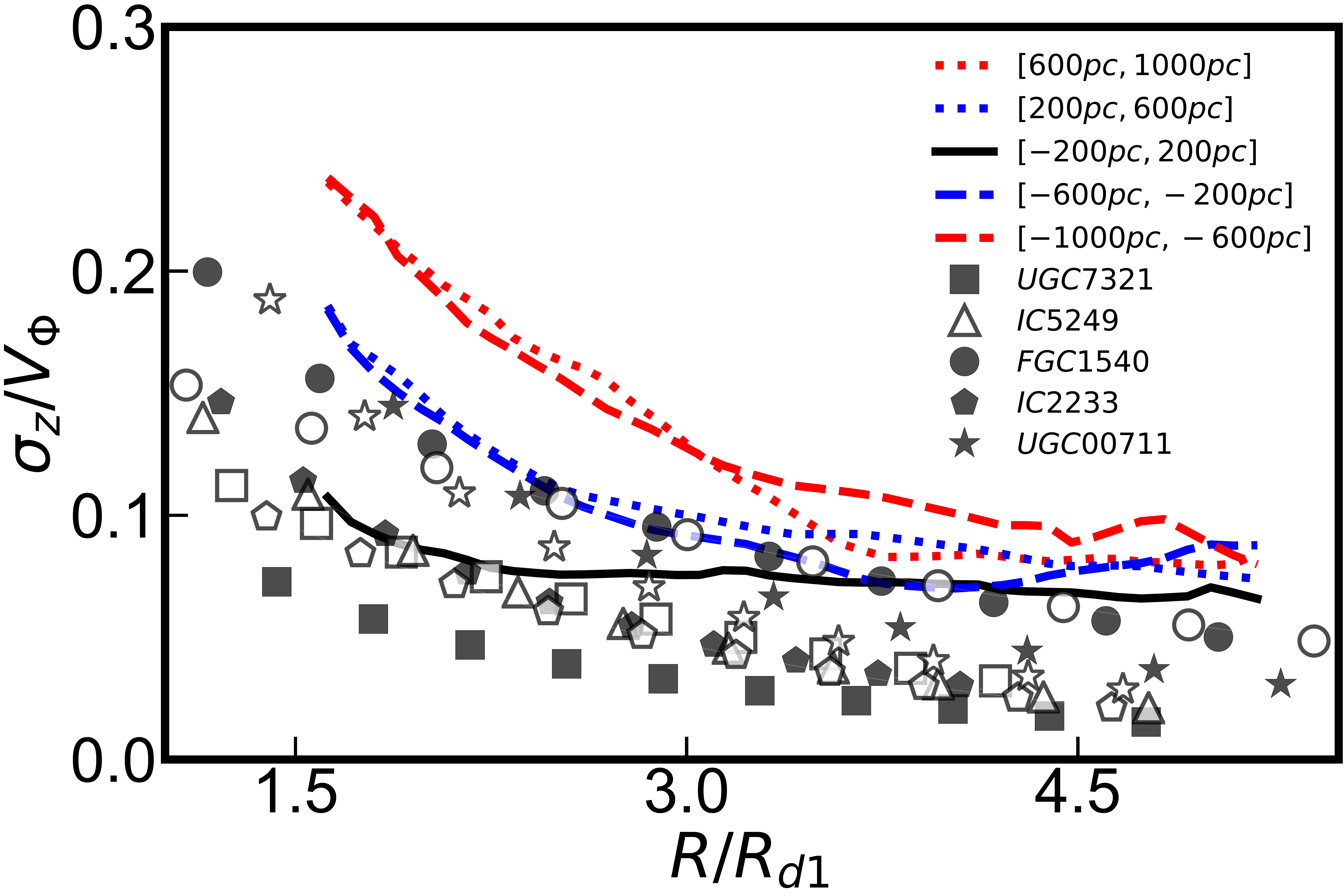}} 
\caption{We depict the ratio of the vertical stellar velocity dispersion to asymptotic rotational velocity $\rm (\sigma_{z}/V_{\phi})$ versus the galactocentric radius $\rm R$ 
normalised by $\rm R_{d}$. $\rm R_{d}$ is the exponential stellar disc scale length for all our sample superthins. $\rm \sigma_{z}$ is the density-averaged vertical velocity 
dispersion of 2-component stellar discs, and $\rm R_{d}$ is the thick disc scale length. Unfilled markers for the same marker-style represent 3.6$\mu$m stellar disc models and the 
filled marker depicts the models derived using optical photometry. The solid and the dotted lines show $\rm \sigma_{z}/V_{\phi}$ versus $\rm R/R_{d}$ for Milky Way stars taken 
from different vertical slices.}
\end{figure*}

Thus far, we have only looked at the absolute values of the vertical velocity dispersion when comparing superthin galaxies with disc galaxies. In terms of absolute 
vertical velocity dispersion, we find that superthin galaxies are colder than spiral galaxies like the Milky Way.
We plot the vertical velocity dispersion ($\rm \sigma_{z}$) of thin stellar discs both in the optical and 3.6 $\rm\mu$m band normalized by their rotation velocity 
$\rm V_{\phi}$ in Figure 2.10. For comparison, we look at the Milky Way stars located at different vertical heights from the galactic mid-plane\citep{katz2018gaia}.
The value of $\rm \sigma_{z}/V_{\phi}$ for all the superthin galaxies is lower than the corresponding value for the Milky Way stars in the -200 pc to 200 pc vertical slice, 
showing that the superthin galaxies are ultra-cold systems. Possible origin of the \emph{cold} stellar discs in superthins\\
Giant molecular clouds (GMCs), satellite galaxies, and spiral arms all contribute to the heating of the galactic disc in important ways.
The spiral arms heat the stellar disc radially \citep{aumer2016age}. On the other hand, GMCs \citep{jenkins1990spiral} heat the disc in both the vertical and radial directions. 
It has been shown that strong bars efficiently heat the galaxy disc, leading to the formation of thick discs \citep{saha2014disc}. Bar strength changes over 
time and correlates with the overall kinetic energy of the stellar particles in the vertical direction, \citep{grand2016vertical}. Thus, weak bars in the superthin galaxies may prevent the disc from heating up vertically, which would keep the superthin structure intact. According to 
\cite{aumer2016age}, satellite galaxies and subhaloes also heat the galactic disc vertically. \textcolor{red}{Further, we find that of all the galaxies in our sample 
only IC 2233 has a blue companion dwarf companion at a distance equal to 46 kpc. This indicates that these galaxies possibly have evolved in isolated
environments therby evading flybys and mergers that play an important role in heating up the stellar disc. See, for example the discussion in 
\cite{kumar2022excitation} on role of flybys and mergers on exciting the vertical breathing modes in galaxies.} Thus, possibly weak bars and a 
sparse environment devoid of satellite galaxies and subhaloes may explain the ultracold superthin structure.

\section{Conclusions}
Superthin galaxies are low surface brightness, bulgeless, disc galaxies with a sharp needle-like appearance in the optical band. Their large planar-to-vertical axes ratios may indicate the presence of an ultra-cold stellar disc, whose dynamical origin is unknown. 
We build dynamical models of superthin galaxies utilizing stellar photometry and \HI{} 21cm radio-synthesis measurements as 
constraints using Markov Chain Monte Carlo approach. We check the consistency of the results obtained with the two-component model using 
AGAMA, i.e., Action-based Galaxy Modelling Architecture \citep{vasiliev2018agama}.

\begin{itemize}
\item The central vertical velocity dispersion for the stellar disc in the optical band lies between $\rm \sigma_{0s}$ $\sim$ $10.2 - 18.4$ \kms  and 
falls off with an exponential scale length of $\rm 2.6$ to $\rm 3.2$ $\rm R_{d}$. In 3.6 $\rm \mu$m, the same, averaged over the two stellar disc components, 
varies between $\rm 5.9$ to $\rm 11.8$ \kms, representing the denser, thinner, and smaller of the two-disc components. 
The massive disc component's dispersion ranges from $\rm 15.9\, to\, 24.7$ \kms with a scalelength equal to $\rm \sim$ 2.2 $\rm R_{d}$. Compared to the value of the 
central vertical stellar velocity dispersion in the Milky Way and Andromeda (M31) is $\sim$ 53 \kms, the absolute value of the vertical velocity dispersion of 
superthins is indeed very small. \emph{Our results indicate that superthin galaxies indeed host ultra-cold stellar discs.}
 
\item Using AGAMA, we set up an equilibrium distribution function for our sample of superthin galaxies for constructing 
the stellar velocity ellipsoid ($\rm \sigma_{z}/\sigma_{R}$). $ \rm\sigma_{z}/\sigma_{R}$ for optical models varies in between 0.48 to 0.59. 
$\rm \sigma_{z}/\sigma_{R}$ for thick disc in 3.6 $\rm \mu$m models lies in between 0.4 - 0.55, while for the thin disc same varies between 0.28 - 0.51.
This is consistent with the findings of \cite{gerssen2012disc}, who find that $ \rm\sigma_{z}/\sigma_{R}\sim 0.3 -0.5$ for late type galaxies Sc -Scd

\item In the 3.6 $\rm \mu$m band, the global median of the multi-component disc dynamical stability parameter $\rm Q_{N}$ 
for superthin galaxies in our sample is 5 $\pm$ 1.5 , which is higher than the global median equal to 2.2 $\pm$ 0.6 obtained for spiral galaxies. The gas component 
drives the dynamical stability of extremely thin galaxies, as demonstrated by $\rm Q_{N}$. 

\end{itemize}

\thispagestyle{empty}

\thispagestyle{empty}
\chapter[Dynamical modeling of vertical disc structure in the superthin galaxy `UGC 7321' in braneworld gravity: An MCMC study]{
\fontsize{50}{50}\selectfont Chapter 3\footnote[1]{
Adapted from \textbf{Aditya Komanduri}, Indrani Banerjee, Arunima Banerjee, Soumitra Sengupta,
Dynamical modelling of disc vertical structure in superthin galaxy UGC 7321 in braneworld gravity: an MCMC study, \textit{Monthly Notices of the Royal Astronomical Society, 
Volume 499, Issue 4, December 2020, Pages 5690\textendash5701}, \textcolor{blue}{arXiv:2004.05627} }}
\chaptermark{Dynamical modeling of UGC 7321 in braneworld gravity}

\textbf{\Huge{ Dynamical modeling of vertical disc structure in the superthin galaxy `UGC 7321' in braneworld gravity: An MCMC study}}
%\vspace {2.5cm}

\section*{Abstract}
Low surface brightness (LSB) are examples of extreme late-type galaxies. The dynamics of the stellar discs in superthin galaxies are 
strongly governed by their dark matter halos. Earlier studies have shown that the higher dimensional Weyl stress term projected onto the 3-brane acts as 
the source of dark matter.
This dark matter is referred to as the \emph{'dark mass'} within the context of the braneworld paradigm.
This model has successfully reproduced the rotation curves of several galaxies with low and high surface brightness. 
Therefore, it is worthwhile to examine the possibility of this model explaining the vertical structure of galaxies, 
which has not been studied in the literature before. Using our 2-component model of gravitationally-coupled stars and gas in 
the external force field of this \emph{'dark mass'}, we use the Markov Chain Monte Carlo method to fit the measured scale heights of the 
stellar and atomic hydrogen (\HI{}) gas of superthin galaxy 'UGC7321'. According to our findings, the observed scaleheights of 'UGC7321' can be successfully 
modeled in the context of the braneworld scenario. In addition, the rotation curve predicted by the model fits the observed one.

\section{Introduction}
Historically, dark matter has been invoked to explain missing mass in spiral galaxies \citep{rubin1979extended} and mass discrepancies in 
galaxy clusters  \citep{zwicky1937masses, zwicky1933rotverschiebung}. Optical tracers or \HI{} 21cm radio-synthesis observations show that disc galaxies have
asymptotically flat rotation curves. On the other hand, the visible matter distribution indicates a Keplerian fall-off beyond the optical disc. The flatness of the 
rotation curve requires the total galactic mass to increase linearly with the galactocentric radius beyond the baryonic disc. This suggests the presence of dark 
matter in spiral galaxies for explaining the mass mismatch and flatness of the rotation curves. "Dark matter" has been hypothesized to explain the rotation curves of 
a wide range of spiral galaxies, from giant HSBs to intermediate-mass LSBs to dwarf irregulars \citep{sofue2001rotation,mcgaugh2001high, kranz2003dark, 
gentile2004cored, de2005halo, de2008high, oh2015high}. Also, "dark matter" resolved galaxy cluster mass discrepancy problem. The mass of a galaxy cluster 
measured by adding together the masses of its member galaxies is substantially smaller than its virial mass determined using the line of sight
velocity dispersion values of its member galaxies \citep{carlberg1997average}. This again suggested the presence of non-luminous materials 
at cluster scales, which may be explained by dark matter.

Despite the widespread application of the dark matter hypothesis for solving a wide range of astrophysical and cosmological problems, 
the fundamental particles that make up dark matter have so far eluded discovery in dark matter search efforts 
(for example, Cryogenic Dark Matter Search (CDMS), \citet{agnese2013silicon}).
Dark matter particles have not been detected, and a number of problems have arisen in the dark-matter particle approach 
\citep{Peebles:2010di}, \citep{kroupa2012dark,Kroupa:2014ria}, \citep{pawlowski2015persistence}. This opens up 
the potential for a gravitational origin of dark matter where Newtonian gravity is adjusted to solve the "missing mass" problem.

Modified Newtonian dynamics (MOND) \citep{milgrom1983modification} is one of the first attempts to modify Newton's laws in order 
to explain the rotation curves of galaxies. This modification has been thoroughly tested in the context of the Milky Way 
\citep{famaey2005modified} as well as on a large sample of spiral galaxies \citep{de1998testing, sanders1998rotation},
Other possibilities for the gravitational origin of dark matter include extra-dimensional models or brane-world models
\citep{Binetruy:1999ut, Csaki:1999jh, Mazumdar:2000gj, Maartens:2000fg, Maartens:2003tw, Koyama:2003be, Haghani:2012zq},
in which the Standard Model particles and fields are contained within the 3-brane while gravity enters the bulk 
\citep{Antoniadis:1990ew,Antoniadis:1998ig,ArkaniHamed:1998rs,Randall:1999vf,Garriga:1999yh,Randall:1999ee,Csaki:1999mp}. 
Several versions of the braneworld do not have the matter fields localized to the brane, e.g. see \cite{Fichet:2019owx}.
It was primarily for this reason that higher dimensional models or brane-worlds were created, which eventually led to string theory 
and M-theory \citep{Kaluza:1921tu,Klein:1926fj,Horava:1995qa,Polchinski:1998rq}. The gauge hierarchy problem in particle physics arises due to the disparity between the Planck scale and the electroweak scale, could be resolved by introducing
extra dimensions \citep{Antoniadis:1990ew,ArkaniHamed:1998rs,Antoniadis:1998ig,Randall:1999ee,Csaki:1999mp}.
The phenomenological and cosmological implications of extra-dimensional models are also interesting, 
for example, see \cite{Davoudiasl:1999tf, Davoudiasl:1999jd, Davoudiasl:2000wi, Hundi:2011dc, Chakraborty:2014zya}.
It is also possible that the deviations from Einstein's gravity in the high energy regime may manifest in the extra dimensions.

\cite{maartens2004brane} explores a single three-brane contained in a five-dimensional bulk.
The bulk Weyl tensor's non-local effects are the primary cause of Einstein's equations being altered. 
Observers on the brane see this Weyl stress term as a fluid with its energy density and pressure, known as the Weyl fluid.
This model has been proven to be in agreement with observed rotation curves of galaxies by \cite{mak2004can}, \cite{harko2006galactic}, 
\cite{boehmer2007galactic}, \cite{rahaman2008galactic} and \cite{gergely2011galactic}. With this study, we will investigate the possibility 
that the braneworld model might explain the observed scaleheight of stars and neutral hydrogen gas (\HI{}) in galaxies. 
We consider UGC7321 which is a prototype of a superthin galaxy with low surface brightness, which was shown to be dominated by dark matter at all radius. 
It has been shown by \cite{banerjee2013some} that the stellar disc's ultrathin vertical structure is strongly regulated by a compact dark matter halo.  
In this study, we will use the observed stellar and \HI{} scale heights of UGC 7321 to constrain the dark matter density profile in 
the braneworld scenario. The density profile of the Weyl fluid in the braneworld scenario resembles the cored dark matter halo profile consistent with mass models 
of low surface brightness galaxies. In the two-component model of gravitationally coupled stars and gas in the dark matter's external force field, 
the Weyl fluid's density profile mimics the effect of the dark matter. The stellar and the \HI{} scaleheights are utilized to constrain 
the vertical velocity dispersion of the stars, gas, and the Weyl parameters. In addition, UGC 7321's measured rotation curve is also 
compared to the rotation curve predicted by the Weyl model. 

\section{Input Parameters}
UGC 7321 is a typical example of a superthin galaxy with a ratio of radial-to-vertical axis ratio of 15.4, 
observed at an inclination equal to $\rm 88^{\circ}$ and at a distance equal to 10 Mpc \citep{matthews2000h,matthews1999extraordinary}. 
The asymptotic rotation velocity equals 110 \kms \citep{uson2003hi}.

\begin{table*}
\hspace{2.2cm}
\begin{minipage}{100mm}
\large
\hfill{}
\caption{Stellar parameters of UGC 7321 in B-band.}
\begin{center}
\centering
\begin{tabular}{|l|c|}
\hline
Parameters& $UGC7321^{B}$   \\
\hline    
\hline
$\mu_{0} (\rm{mag}\, arcsec^{-2})$ \footnote{Central surface brightness of stellar disk} & 23.5   \\
$\Sigma_{0} (M_{\odot}{\rm{pc}}^{-2})$  \footnote{Central surface density of the stellar disk}  & 34.7 \\
$R_{D} (\rm{kpc})$ \footnote{Disc scalelength of the exponential stellar disk} & 2.1  \\
$h_{z} (\rm{kpc})$    \footnote{Scaleheight (HWHM) of the stellar disk}  &  0.105 \\
$M_{stars}$\footnote{Stellar mass calculated using $2\pi\Sigma_{0}R^{2}_{d}$} &$9.6 \times 10^{8} M_\odot$ \\
$M_{HI}$\footnote{\HI{} mass \cite{uson2003hi}}& $1.1\times10^{9} M_\odot$\\
\hline
\end{tabular}
\hfill{}
\end{center}
%\label{table:B-band}
\end{minipage}
\end{table*}

The de-projected B-band centre surface brightness is 23.5 $\rm mag\,arcsec^{-2}$.
UGC 7321 has a high value of dynamical mass-to-light ratio, $\rm M_{dyn}/M_{\HI{}}=31$ and $\rm M_{dyn}/L_{B}=29$, where $\rm {L_{B}}$ is the B-band luminosity and 
$\rm M_{\HI{}}$ is the galaxy's total \HI{} mass. Dark matter dominates the disc dynamics at all radii, as shown by \cite{banerjee2013some}. 
UGC7321 follows an exponential surface density profile in B-band, $\rm \Sigma_{s}(R) = \Sigma_{0}exp (-R/R_{D})$, where $\rm \Sigma_{0}$ 
is the central stellar surface density, and $\rm R_{D}$ is the exponential scalelength of the stellar disc. The parameters for the exponential stellar disc were taken from \cite{uson2003hi}.

\begin{table}
\hspace{2.2cm}
\begin{minipage}{100mm}
\large
%\hfill{}
\centering
\caption{Input parameters for HI}
\begin{center}
\begin{tabular}{|l|c|}
\hline
Parameters& UGC7321 \\ 
\hline    
\hline
$\Sigma_{01} (M_{\odot}{\rm{pc}}^{-2})$ \footnote{Central surface density of the first HI gaussian disk}  &  4.912   \\
$\Sigma_{02} (M_{\odot}{\rm{pc}}^{-2})$ \footnote{Central surface density of the second HI gaussian disk}   & 2.50       \\
$a_{1}$ ({\rm{kpc}}) \footnote{Centre of the first HI gaussian disk}  & 3.85\\
$a_{2}$ ({\rm{kpc}}) \footnote{Centre of the second HI gaussian disk} & 0.485  \\
$r_{01}$ ({\rm{kpc}})\footnote{Scalelength of the first HI gaussian disk } &2.85  \\
$r_{02}$ ({\rm{kpc}})  \footnote{Scalelength of the second HI gaussian disk}  &1.51 \\
\hline
\end{tabular}
\hfill{}
\end{center}
%\label{table:HI parameters}
\end{minipage}
\end{table}

\HI{} surface density and scaleheight of UGC 7321 were taken from \cite{uson2003hi} and \cite{o2010dark}.
Earlier studies suggested that a double-Gaussian profile represents the radial profile of the \HI{} surface density 
(see for example \cite{begum2005dwarf}, \cite{patra2014modelling}) which may indicate presence of two \HI{} discs.
Commonly, \HI{} surface density peaks away from the center of galaxies, indicating a central \HI{} hole.
\HI{} surface density profiles can be fitted with an off-centered double Gaussian defined by 
$$\rm {\Sigma}_{HI} (R) = {\Sigma}_{01} \rm{exp} \Big[ -{\frac{{(r-a_1)}^2}{2 {r_{01}}^2}}\Big] + {\Sigma}_{02} \rm{exp}\Big[-{\frac{{(r-a_2)}^2}{2 {r_{02}}^2}}\Big] $$
where $\rm \Sigma_{01}$ is the central surface density, $\rm a_{1}$ the centre, and $\rm r_{01}$ scalelength of the gas disc 1, etc.
The input parameters for the stars and the gas are summarized in Tables 3.1 and 3.2.

\section{Results \& Discussion }
\begin{figure*}
\hspace*{-2cm}
%\begin{center}
\begin{tabular}{ccc}
\resizebox{57mm}{50mm}{\includegraphics{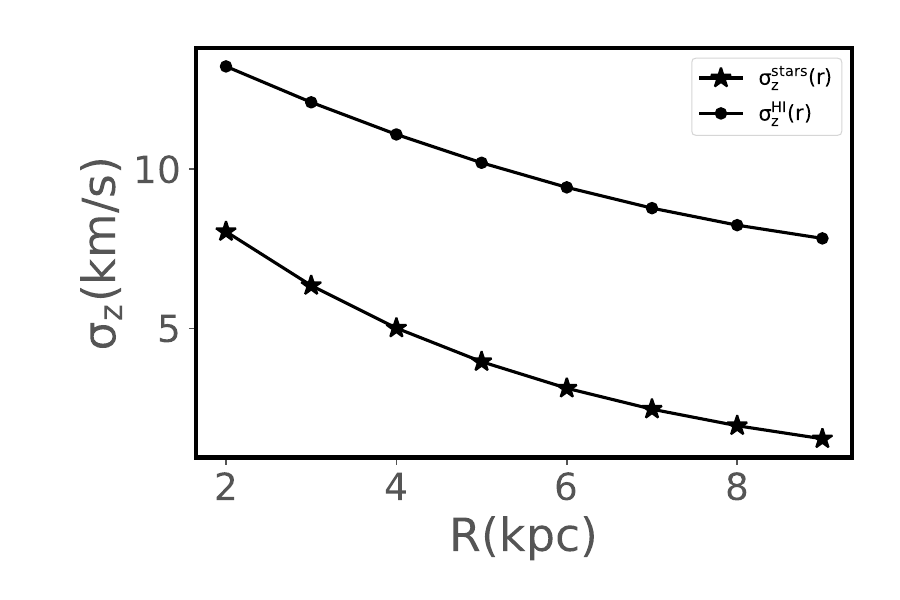}} &
\resizebox{57mm}{50mm}{\includegraphics{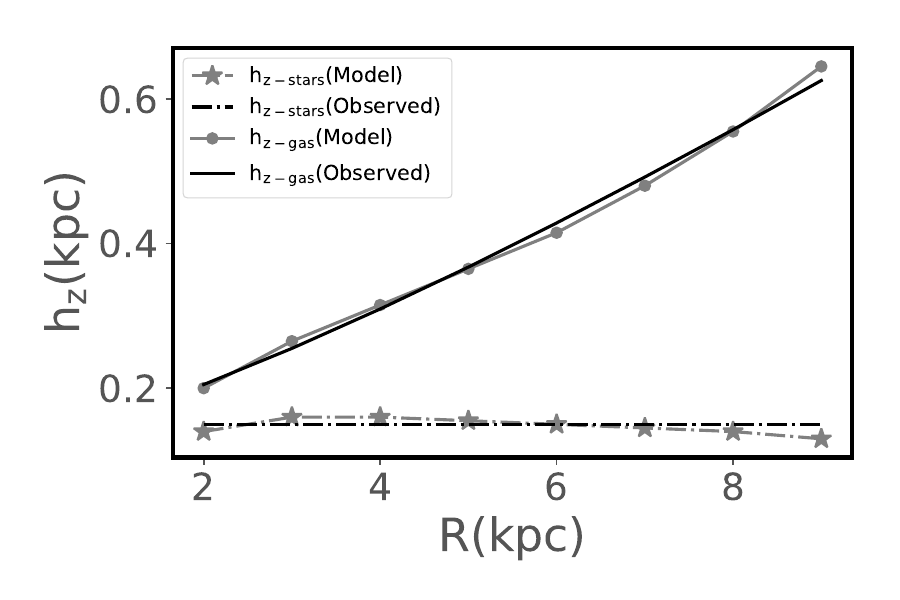}} &
\resizebox{57mm}{52mm}{\includegraphics{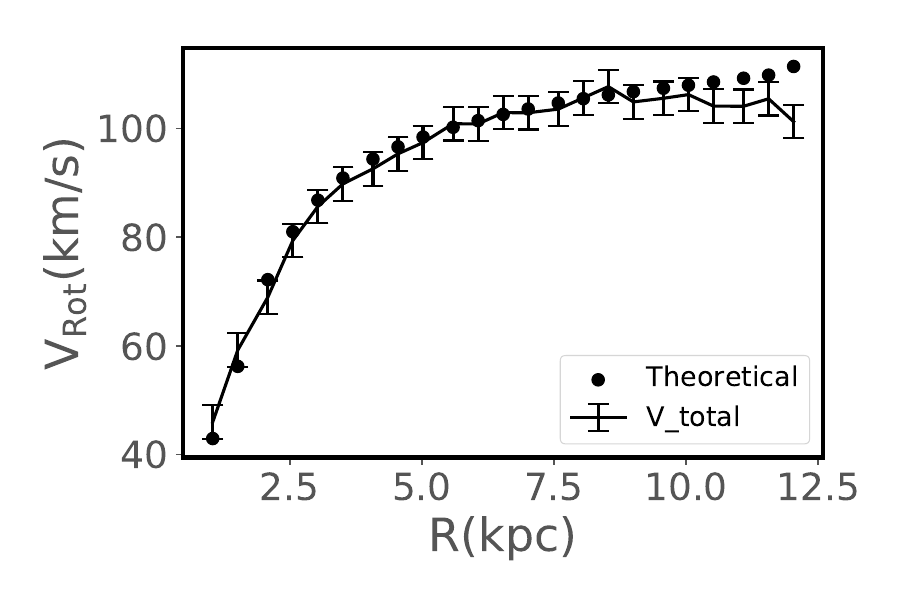}} \\
\end{tabular}
%\end{center}
\caption{In the left panel, we plot the vertical velocity dispersion of the stars and \HI{} as a function of the galactocentric radius $\rm R$
using a model of the baryonic disc in a halo of $\rm dark\, mass$, constrained by the observed stellar and \HI{} scaleheight data. In the middle
panel, the black $\rm 'dash-dot'$ and $\rm 'solid\,lines'$ 
depict the observed stellar and \HI{} scaleheights respectively,  $\rm 'dots'$ and $\rm 'stars'$ in grey indicate the corresponding modeled stellar and \HI{} scaleheights.
The rotation curve of UGC 7321, derived using the best-fit parameters from the braneworld model, is shown in the right panel.}
\end{figure*}

\begin{figure*}
%\begin{center}
%\vspace{-1.5cm}
%\hspace*{-1.6cm}
\resizebox{160mm}{145mm}{\includegraphics{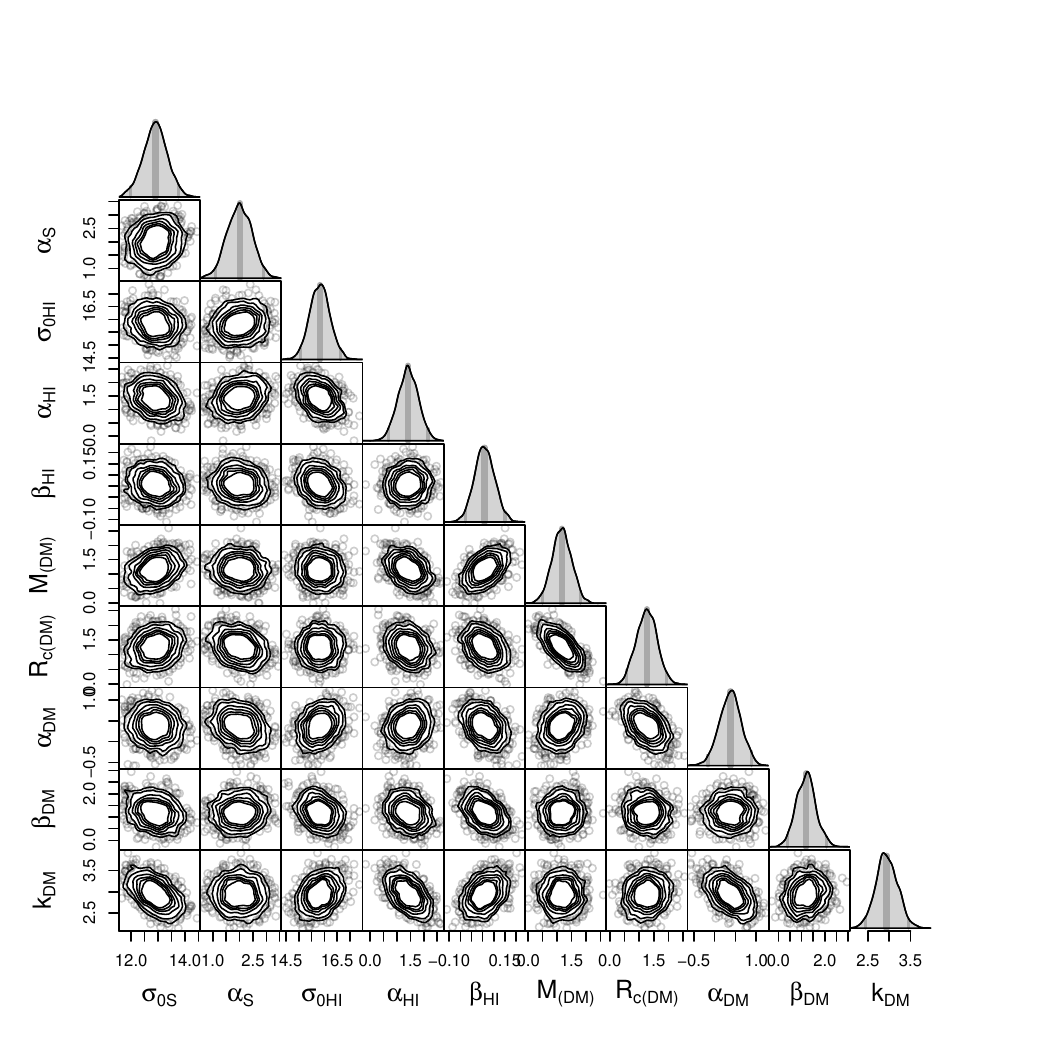}}
%\end{center}
\caption{The posterior probability distribution of the parameters characterizing the dynamical model of UGC 7321 in the braneworld paradigm.}
\end{figure*}

\begin{table}
\hspace{2.2cm}
\begin{minipage}{100mm}
{\small
\hfill{}}
\centering
\caption{Best-fitting parameters found by optimizing the dynamical model of UGC 7321.}
\begin{center}
\begin{tabular}{|l|c|}
\hline
Parameters & UGC7321 \\ 
\hline    
\hline
$\sigma_{0s} $ (\rm{kms}$^{-1}$)\footnote{Central stellar vertical velocity dispersion}  & $13.4 \pm 0.6$      \\
$\alpha_{s} $ $(\rm{kpc}^{-1})$ \footnote{Exponential radial scale length (in units of R$_D$) of the stellar vertical velocity dispersion}   & $2.0 \pm 0.4$        \\
$\sigma_{HI}$ ($\rm{kms}^{-1}$) 
\footnote{Central HI vertical velocity dispersion}  & $15.4 \pm 0.5$  \\
$\alpha_{HI}$ $(\rm{kms}^{-1}\rm{kpc}^{-1})$ \footnote{Radial gradient of HI vertical velocity dispersion} &  $-1.3 \pm 0.2$ \\
$\beta_{HI}$ $(\rm{kms}^{-1}\rm{kpc}^{-2})$ \footnote{Gradient of the radial gradient of HI vertical velocity dispersion} & $0.04 \pm 0.02$ \\
$M_{DM}$ ($M_{\odot}$)  \footnote{Dark mass}  & $(1.3 \pm 0.2) \times10^{9}$\\
$R_{c(DM)}$ (\rm{kpc})  \footnote{Core-radius of the dark mass density profile}  &  $1.3 \pm 0.2$\\
$\alpha_{DM}$  \footnote{Weyl-fluid parameter}  & $-0.4 \pm 0.1$ \\
$\beta_{DM}$   \footnote{Weyl- fluid parameter}  & $(1.4 \pm 0.2) \times 10^{-7}$\\
$k_{DM}$ $(\rm{kpc}^{-1})$  \footnote{Smoothening term associated with dark mass density profile}  & $2.7 \pm 0.3$\\
\hline
\end{tabular}
\end{center}
%\label{table:HI parameters}
\end{minipage}
\end{table}

A total of 10 independent parameters make up our dynamical model of UGC 7321 in the $\rm B$-band. Five free parameters represent the baryonic disc: 
$\rm \sigma_{0s}$ and $\rm {\alpha_{s}}$ for the stellar vertical velocity dispersion profile and $\rm \sigma_{0\HI{}}$,  $\rm \beta_{\HI{}}$ and $\rm \alpha_{\HI{}}$ for 
the vertical velocity dispersion profile of the \HI{} gas. The dark mass profile arising from the braneworld model provides the remaining five free parameters.
The Weyl fluid is described by $\rm M_{DM}$ $\rm R_{DM}$, $\rm \alpha_{DM}$, and $\rm \beta_{DM}$ respectively, while $\rm k_{DM}$ is associated with the 
smoothing function. The braneworld model and the multi-component models are characterized by 10 free parameters, and the effect of the higher dimensional bulk are propagated to the four-dimensional
brane through the parameters $\rm \alpha$ and $\rm \beta$. The formulation of equations of the braneworld theory is presented in Section 1.2. In order to constrain the 10 free parameters listed above, 
we use the observations of stellar and \HI{} scaleheights as the constraints and employ the Markov chain Monte Carlo method (MCMC) (see Section 1.3). Using the 2-component model, 
we calculate the vertical velocity dispersion of stars and \HI{} as a function of galactocentric radius, Figure 3.1 [Left Panel]. The central stellar velocity dispersion 
is equal to (13.4$\pm$0.6) \kms, which falls off exponentially with a scalelength (2.1$\pm$0.4)$\rm R_{D}$, where $\rm R_{D}$ is the exponential stellar disc scale length. 
The vertical velocity dispersion in the central \HI{} regions is approximately (15.4$\pm$0.5) \kms, with $\rm \alpha_{\HI{}}=(-1.3\pm 0.2)$ and $\rm \beta_{\HI{}}=0.04\pm 0.02$,
which indicates that it decreases almost linearly with radius. The stellar velocity dispersion can not be lower than the \HI{} velocity dispersion because stars can not dissipate energy 
through collisions. Therefore, their dispersion can never be lower than the gas clouds from which they form. As a result, it is possible that the thin disc stars formed in an underlying cold component of the gas with low values of velocity dispersion.

For the parameters corresponding to the Weyl fluid we find that $\rm M_{DM}=(1.3\pm 0.2) \times 10^{9} M_{\odot}$, 
core radius $\rm R_{c(DM)}=(1.3\pm 0.2)kpc$, $\rm \alpha_{DM}=-0.4\pm 0.1$, and $\rm \beta_{DM}=(1.4 \pm 0.2) \times 10^{-7}$, 
smoothing function term $\rm k_{DM}$ is equal to $\rm 2.7 \pm 0.3$ $(\rm{kpc}^{-1})$. Table 3.3 summarizes the results. $\rm M_{DM}$ is consistent with the average dark 
matter mass in low-surface-brightness superthin galaxies. The value of $\rm M_{DM}$ is $\rm 10^{9}$, agrees with the dark mass of LSBs published in \citet{gergely2011galactic}. 
$\rm R_{c(DM)}$ is 1.3 kpc ($\sim$0.6 $\rm R_{D}$), showing the Weyl fluid is dense and compact. This confirms a prior study that showed that UGC7321 has a dense and  
compact cored dark matter halo. The values of $\rm \alpha_{DM}$ and $\beta_{DM}$, indicate  the deviations of the spherically symmetric metric from the Schwarzschild 
scenario in general relativity, satisfy $\rm \alpha_{DM}<0$ and $\rm 0<\beta_{DM}<1$. Further, the values of $\rm \alpha_{DM}$ and $\rm \beta_{DM}>0$ are comparable with 
the parameter set obtained by \cite{gergely2011galactic} for a sample of nine LSBs using the observed rotation curve. $\rm k_{DM}$ is 15-150 $\rm kpc^{-1}$ for the 
sample LSBs analysed by \cite{gergely2011galactic}, while for UGC 7321 it's $\rm 2.66 \, kpc^{-1}$. 
UGC7321, like other superthins, has a steeply rising rotation curve compared to LSBs \citep{banerjee2016mass}.

Figure 3.1 [Middle Panel] shows the modeled and the observed scale heights. Our best-fitting model matches the data closely.
In Figure 3.1 [Right Panel], we compare the model rotation curve with the observed rotation curve of UGC 7321. 
The figure shows that the theoretical rotation curve generated using the scaleheight constraint on the two-component model 
conforms with the observed rotation curve within error bars. The rotation curve flattens around $\rm \sim 10$ kpc and remains flat. If we extend the rotation curve of UGC 7321 using the retrieved model parameters (Table 3.3), the curve remains flat and does not show any decline. This asymptotic behavior is compatible with the pseudo-isothermal dark matter density profile used to model galaxy rotation curves \citep{de2001mass,oh2015high,lelli2016sparc}. Figure 3.2 shows the correlation plots and posterior 
distributions from the 2-component MCMC fits. We do not find any strong correlations between the model parameters.

\section{Conclusions}
We study the possibility of higher-dimensional gravity in describing the vertical scaleheight structure of the dark matter-dominated LSB galaxy UGC 7321.
The Einstein's equations in five-dimensional equations are projected onto the 3-brane where our observable universe dwells, such that the effective 
4-dimensional gravitational field equations inherit a source term originating from the bulk. For a brane observer, the bulk acts 
as a fluid (the so-called Weyl fluid) with an energy density and pressure, dark radiation, and dark pressure. Due to the Weyl term, the 4-d effective gravitational 
field equation's static, spherically symmetric, and asymptotically flat solution deviates from Schwarzschild spacetime. 
The parameters connecting the dark radiation and dark pressure terms in the Weyl fluid's equation of state define the deviation from the Schwarzschild scenario. 
The equation of state supplies initial conditions to the off-brane evolution equations (e.g., normalized normals) along extra dimensions. 
In galaxies, $\rm p=(a-2)\mu-B$ corresponds to the Schwarzschild scenario, where $\rm a=3$ and $\rm B=0$. With this equation of 
state and $\rm \tilde{q}\equiv GM/R \approx 10^{-7}<<1$  (which holds 
in the galactic case), one may derive the density profile and rotation curve of LSB galaxies in terms of the parameters corresponding to the equation of state. 
Fitting these with UGC 7321's rotation curve and scale height data constrains the Weyl parameters and stellar and \HI{} vertical velocity dispersion profiles. 
By fitting the rotation curves and vertical scaleheight data, we get distinct Weyl parameter values 
within their physically permissible range. MOND  predicts the rotation curve due to the universality of $\rm g^{\dagger}$. However, the physical origin of the universal 
acceleration scale, the interpolating function between low and high accelerations, and the force law obeyed by particles in the deep-MOND domain are not well understood.
\cite{sanchez2008thickness} examined if the Milky Way's measured \HI{} vertical thickness could be modeled in the MOND scenario.
Using a 2-component galactic disc model of gravitationally-coupled stars and gas, identical to the one investigated in this study, and assuming 
\HI{} vertical velocity dispersion equal to 9 kms$\rm ^{-1}$, \cite{sanchez2008thickness} showed that the model scaleheight fitted the measured scaleheight well 
at $\rm R \geq 17$ kpc, or beyond 5-6 $\rm R_{D}$. At $\rm R \leq 17$ kpc, the model underpredicted the scaleheight by 40$\rm \%$.

In earlier studies Weyl parameters were estimated by fitting the rotation curves of LSB and HSB galaxies.
This motivates us to compare the model stellar and \HI{} vertical scale height obtained using the braneworld model with the observed scaleheights of 
dark matter-dominated LSB galaxies. We use the prototypical low surface brightness superthin galaxies UGC 7321 in this study, which is a well-studied object with a 
dynamical mass of $\rm M_{dyn}/M_{\HI{}}=31$ and $\rm M_{dyn}/M_{L_{B}}=29$. 
Our investigation shows that the Weyl model can explain UGC 7321's vertical scaleheight data within error bars. 
This work opens a new observational path for understanding the role of extra dimensions or alternate gravity models in the galaxy.

\thispagestyle{empty}

\thispagestyle{empty}
\chapter[ \HI{} 21 cm observation and mass models of the extremely thin galaxy FGC 1440]{\fontsize{50}{50}\selectfont Chapter 4
\footnote[1]{Adapted from \textbf{K Aditya}, Peter Kamphuis, Arunima Banerjee, Sviatoslav Borisov, Aleksandr Mosenkov, Aleksandra Antipova, Dmitry Makarov
\HI{} 21 cm observation and mass models of the extremely thin galaxy FGC 1440, \textit{Monthly Notices of the Royal Astronomical Society, 
Volume 509, Issue 3, January 2022, Pages 4071\textendash4093}, \textcolor{blue}{arXiv:2110.15478}}  }
\chaptermark{ \HI{} 21 cm observation and mass models of the extremely thin galaxy FGC 1440}

\textbf{\Huge{\HI{} 21 cm observation and mass models of the extremely thin galaxy FGC 1440 }}

\section*{Abstract}
We report observation of neutral hydrogen (\HI{}) in the extremely thin galaxy FGC 1440 with an optical axial ratio $\rm a/b = 20.4$. 
We observed the galaxy with a spectral resolution of 1.7 \kms and a spatial resolution of  $\rm 15\farcs9\times13\farcs5$. Using GMRT \HI{} 21 cm radio synthesis observations,
we find that the asymptotic rotational velocity is equal to 141.8 \kms. Besides, the \HI{} is found to be confined in a thin disc which is possibly warped 
along the line of sight, although a thick central \HI{} disc cannot be ruled out. We find that the dark matter halo in FGC 1440 is well-represented by a pseudo-isothermal 
(PIS) profile with $\rm R_{c}/ R_{d} \leq 2$, where $\rm R_{c}$ is the core radius of the PIS halo and $\rm R_{d}$ is the scalelength of the stellar disc. 
Despite extra-ordinarily large planar to vertical axis ratio, the ratio of its rotational velocity to the vertical velocity dispersion of stars is comparable 
to other superthin galaxies ($\rm \frac{\sigma_{z}}{V_{Rot}} \sim 0.125 - 0.2$). Further, FGC 1440 complies with the $\rm j_{*} - M_{*}$ relation for normal disc 
galaxies, unlike normal superthin galaxies investigated earlier. The specific angular momentum of the stellar and the gas disc in the FGC 1440 is comparable 
to the previously studied superthin galaxies with the same mass.

\section{Introduction}
Superthin galaxies are edge-on low-surface brightness disc galaxies with an axial ratio $\rm a/b>10$ 
\citep{bothun1997low, mcgaugh1996number}. These late-type disc structures observed at high inclination are some of the least evolved systems in the Universe 
\citep{vorontsov1974specification, kautsch2009edge, uson2003hi} characterized by a high gas mass to blue luminosity ratio, 
$\rm \frac{M_{\rm \HI{}}}{L_{B}} \approx 1 M_{\odot}\,L^{-1}_{\odot}$ \citep{goad1981spectroscopic, uson2003hi}, and low star formation rates 
$\rm \sim 0.01 - 0.05 \, M_{\odot}yr^{-1}$ \citep{wyder2009star, narayanan2022superthin}.

The $\rm \HI{}$ distribution provides insights into the fundamental mechanisms that govern the structure, dynamics, and formation of galaxies.
Recent large $\rm \HI{}$ surveys, such as THINGS (The HI Nearby Galaxy Survey) \citep{walter2008things} and LITTLE THINGS (Local Irregulars That Trace Luminosity Extremes, 
The HI Nearby Galaxy Survey) \citep{hunter2012little}, have mapped the $\rm \HI{}$ distribution in nearby spiral and dwarf galaxies. These \HI{} surveys have given important
insights into the role of the gas in regulating gravitational instabilities and star formation in these galaxies. \citep{leroy2008star, bigiel2008star}. 
The high-resolution \HI{} rotation curves from these surveys have been used to constrain the dark matter mass \citep{de2008high, oh2015high}. 

Measurements of the $\rm \HI{}$ distribution, dispersion, and scaleheight can be used to infer the shape of dark matter haloes
\citep{olling1995usage, olling1996highly, peters2017shape}. It is now well known that a compact dark matter halo 
regulates the superthin structure and stabilizes the galaxy against axisymmetric instabilities \citep{garg2017origin, van2001kinematics, ghosh2014suppression}.
Neutral hydrogen contributes significantly to total potential, and the total potential regulates the vertical structure of
the stellar and $\rm \HI{}$ discs \citep{narayan2002vertical}. \cite{jog1996local, romeo2013simple} study the influence of $\rm \HI{}$ gas 
on stability of disc galaxies. Recent investigations have shown that the low stellar scaleheight is a direct outcome of low vertical velocity dispersion 
of stars and that these superthin galaxies are remarkably stable despite low values of dispersion \cite{10.1093/mnras/stab155}. 
\textcolor{red}{Further, these galaxies possibly have evolved in isolation, because of which the usual process like mergers and galaxy interactions 
which otherwise are effective in thickening the galaxy disc have been ineffective in heating the superthin stellar discs.}

Of all known superthin galaxies studied in the literature, UGC 7321  \citep{uson2003hi, matthews2003high, matthews1999extraordinary, 
banerjee2010dark, sarkar2019flaring, 10.1093/mnras/stab155, komanduri2020dynamical, pohlen2003evidence} has the highest axial ratio $\rm a/b$ equal to 15.4, 
followed by FGC 1540 \citep{kurapati2018mass} which has $a/b$ equal to 15.2.
Other superthin galaxies, IC 2233 \citep{matthews2008corrugations,matthews2007h, gallagher1976surface} and 
IC 5249\citep{abe1999observation, van2001kinematics, byun1998surface,  yock1999observation}, have an a/b of 8.9 and 10.4 respectively.

An axis ratio of 20.36 ($B$-band) makes FGC 1440 one of the flattest known galaxies. We carry of \HI{} 21 cm synthesis observation of FGC 1440 using GMRT
and derive detailed structural properties of the gas distribution using 3D tilted ting methods (see Section 1.4.5). 
The limiting values of scaleheight, velocity dispersion, and inclination are derived by fitting the tilted ring models to the datacubes.
Finally, we infer the dark matter profile of FGC 1440 using the total rotation curve in conjunction with stellar photometry.

Further, we solve the Jean's equation based multi-component model of gravitationally coupled stars + gas in the force field of dark matter halo 
for obtaining the vertical stellar velocity dispersion as a function of the radius. We use the stellar and the \HI{} surface density, as well as the 
dark matter mass models as the input parameters. We constrain the multi-component Jeans equation 
\citep{narayan2002vertical, banerjee2010dark, sarkar2020general, sarkar2019vertical, patra2020theoretical, patra2020h, patra2018molecular} 
using the observed stellar scaleheight and the limits on the \HI{} velocity dispersion and the \HI{} scaleheight.

\section{Target: FGC 1440}
FGC 1440 is an edge-on late type Sd spiral galaxy at 59.6 Mpc \citep{karachentseva2016ultra}. The major and minor axes of FGC 1440 in the 
$\rm B$-band are $\rm 2.24\farcm \times 0.11\farcm$, giving $\rm (a/b)_{B}=20.36$. \cite{hoffman1989hi} report a $\rm \HI{}$ diameter of $\rm 3.8\farcm$ 
and a ratio of dynamical mass to blue luminosity ($\rm M_{dyn}/L_{B}$) of 11.9 $\rm M_{\odot}/L_{\odot}$, with an $\rm M_{\HI{}} /L_{B}>0.86, M_{\odot}/L_{\odot}$.
FGC 1440's $\rm \HI{}$ attributes are delineated in the ALFALFA $\rm \HI{}$ source catalog. It has total single dish flux $\rm F=9.53 \pm 0.09Jy\,\kms$, 
\HI{} mass $\rm log(M_{\HI})=10.08 \pm 0.18 M_{\odot}$, and $\rm W_{50}=298 \pm 2 \kms$, where $\rm W_{50}$ is the 50$\%$ of peak maximum velocity width. 
In their work on the optical photometry of 47 late-type galaxies, \cite{dalcanton2000structural, dalcanton2002structural} discovered that FGC 1440 has 
a small bulge in the center. However, they conclude that it may not be a kinematic bulge but rather an edge-on pseudo-bulge. 
Based on their investigation of vertical color gradients, \cite{dalcanton2002structural} 
find that FGC 1440 may also host dust lanes. In their study of the kinematics of the galaxies with a thick disc, \cite{yoachim2008kinematics} find that 
FGC 1440 does not have a thick disc component based on their measurement of the off-plane rotation curve, 
which is identical to the mid-plane rotation curve. Table 4.1 summarizes the basic properties of FGC 1440.

\begin{table}
%\hspace{4cm}
\begin{minipage}{110mm}
\hfill{}
\caption{Basic properties: FGC 1440}
%\centering
\begin{tabular}{|l|c|}
\hline
\hline
Parameter& Value \\
\hline    
RA(J2000)$^{\textcolor{red}{(a)}}$   \footnote{Right Ascension}       &  $12^{h}28\farcm52.29\farcs$  \\
Dec(J2000)$^{\textcolor{red}{(b)}}$  \footnote{Declination}           & $+04{d}17\farcm35.4\farcs$  \\
P.A$^{\textcolor{red}{(c)}}$         \footnote{Position Angle}        & $53^{\circ}$\\
a/b $^{\textcolor{red}{(d)}}$        \footnote{Major axis to minor axis ratio}                 & 20.4  \\
Hubble type $^{\textcolor{red}{(e)}}$\footnote{Galaxy type}           & Sd \\
$i$ $^{\textcolor{red}{(f)}}$        \footnote{Inclination}           & $90^{\circ}$\\
Distance$^{\textcolor{red}{(g)}}$    \footnote{Distance to the galaxy}&59.6 Mpc\\
log (M$_{\HI{}}/M_{\odot} $)$^{\textcolor{red}{(h)}}$ \footnote{\HI{} mass}            & 10.1\\
W$_{50}$ $^{\textcolor{red}{(i)}}$   \footnote{50$\rm \%$ of the peak maximum velocity widths}& 298 \kms{} \\
D$_{\HI{}}$ $^{\textcolor{red}{(j)}}$   \footnote{\HI{} diameter}        &3.8$\farcm$\\
M$_{\HI{}}$/L$_{B}$ $^{\textcolor{red}{(k)}}$ \footnote{Ratio of \HI{} mass to blue luminosity} & $>$ 0.86\\
M$_{Dyn}$/L$_{B}$ $^{\textcolor{red}{(l)}}$   \footnote{Ratio of dynamical mass to blue luminosity}& 11.8 $M_{\odot}/L_{B}$ \\
\hline
\end{tabular}
\hfill{}
\label{table: table 1}
\end{minipage}
\begin{tablenotes}
\item  $\textcolor{red}{(a, b, c, d)}$:  \cite{karachentsev2003revised} .
\item $\textcolor{red}{(e)}$:       \cite{de1991third}  .
\item $\textcolor{red}{(f)}$:       \cite{makarov2014hyperleda}. 
\item $\textcolor{red}{(g)}$:       \cite{kourkchi2020cosmicflows}. 
\item $\textcolor{red}{(h, i)}$:     \cite{haynes2018arecibo}.
\item $\textcolor{red}{(j, k, l)}$:   \cite{hoffman1989hi}.
\end{tablenotes}
\end{table}

\section{Observations and Data Reduction}
We observed FGC 1440 for 7 hours (including overheads) on August 19, 2019, with 26 antennae. The target FGC 1440 was observed for 5.5 hours in 11 
scans of 30-minute intervals, interspersed by 11 phase calibrator scans of 5-minute duration. 3C286 was observed for 30 minutes at the beginning and at the end of the observations. 
We carried out observations in GSB mode with 512 channels, spectral resolution equal to 8.14 kHz (1.71 \kms), and bandwidth equal to 4.14 MHz was used to observe the central line at 1402.5 MHz. Table 4.2 presents the details of the observations.

\begin{table}
\caption{Summary of FGC 1440 observation}
%\centering
\begin{tabular}{|l|c|}
\hline 
\hline
(a) Observing Setup&      \\
\hline
Parameter& Value \\
\hline    
\hline
Observing Date                     & 19August2019\\
Phase center,$\alpha$(J2000)       & $12^{h}28\farcm52.29\farcs$\\
Phase center,$\delta$(J2000)       & $+04^{d}17\farcm35.4\farcs$\\
Total on-source observation time   & 5 $\frac{1}{2}$ hours\\
Flux  calibrator                   & 3C286  \\
Phase calibrator                   & 1150-003\\
Channel Width                      & 8.14 kHz\\
Velocity separation                & 1.7 \kms\\
Central frequency                  & 1400.5 MHz\\
Total bandwidth                    & 4.14 MHz   \\
\hline
(b) Deconvolved Image Characteristics&          \\
\hline
Weighing                           & Briggs\\
Robustness parameter               & 0\\
Synthesized beam FWHM              & $15.9\farcs \times 13.5\farcs$\\
Synthesized beam position angle    & $23.1^{\circ}$\\
rms noise in channel               & 1.01 mJy/beam\\
\hline
\end{tabular}
\hfill{}
\label{table: table 2}
\end{table}

\subsection{Flagging and Calibration}
We use Common Astronomy Software Applications (CASA) to carry out data analysis of the GMRT  observations \citep{mcmullin2007casa} (see Section 1.4.2 \& 1.4.3). 
We start by flagging the offline antennae E04, E05, E06, and S02. We then visually inspect the data set for RFI, followed by cross-calibration. After cross-calibrating 
and separating the $\rm 'target'$ from the measurement set, we average the visibilities in time to locate the spectral line and flag those channels to generate a 
$\rm 'continuum-only'$ measurement set for self-calibration.

\subsection{Imaging the spectral line}
We manually mask emissions in the dirty image using CASA task $\rm \sc{TCLEAN}$. We lower the cleaning threshold after each $\rm 'phase-only'$ and $\rm 'amplitude-phase'$ 
self-calibration. We employ a total of 4 rounds of $\rm 'phase-only'$ and 3 rounds of $\rm 'amplitude-phase'$ self-calibration, after which the rms saturates.
We do not detect continuum emissions from the center of the galaxy or from its outskirts. We apply the final amplitude-phase self-calibration table to the $\rm 'target-only'$ measurement
set encompassing both spectral line and continuum emission. Then we carry out continuum subtraction using CASA task $\sc{UVCONTSUB}$ with zeroth order 
interpolation excluding the spectral channels. 
We generate a data cube using $\rm \sc{TLCEAN}$ and clean emission within a SoFiA mask up to $\rm 0.5 \sigma$,
and iterate till SoFiA mask is steady. We tested different weighting systems in $\rm \sc{TCLEAN}$ and found that $\sc{briggs}$ scheme
with robust equal to 0 and a $10k\Lambda$ uvtaper delivers the best tradeoff between resolution and sensitivity. We finally carry out Hanning smoothing of the cube 
and find that the final resolution is $\rm 15.9\farcs \times 13.5\farcs$ and the rms noise is 1.01 mJy/beam compared to 1mJy/beam expected theoretically.

\section{Analysis}
\subsection{Global \HI{} profile}
\begin{figure*}
\hspace{4.5cm}
\resizebox{80mm}{60mm}{\includegraphics{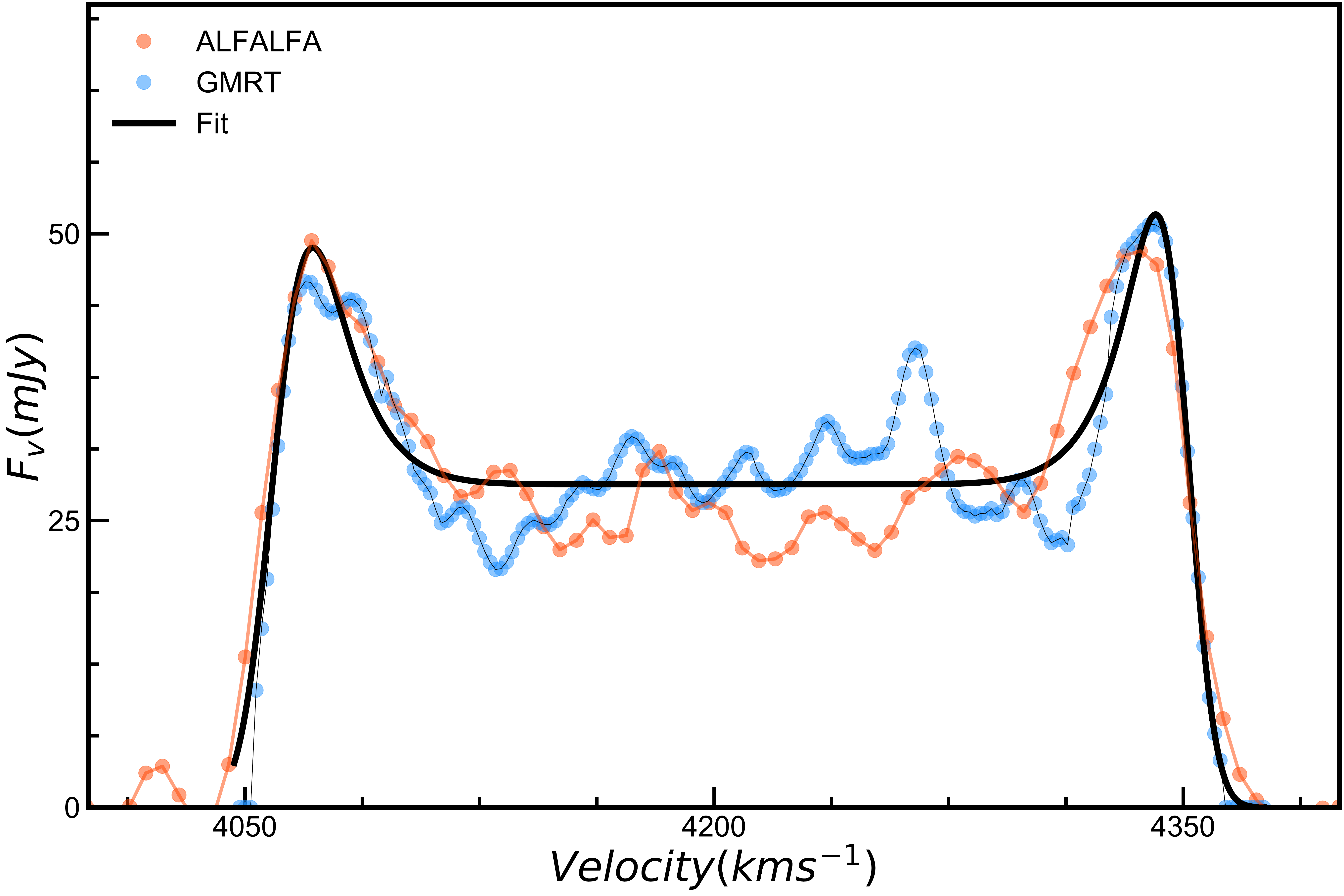}} 
\caption{The fitted busy function overlaid on the spectrum of FGC 1440 derived using GMRT \HI{} 21 cm radio synthesis observations.}	
\end{figure*}

Figure 4.1 shows FGC 1440's global \HI{} profile. The integrated \HI{} flux is equal to $\rm 9.6$ $\rm Jy\kms$ and is comparable to flux obtained by \cite{haynes2018arecibo} using single dish observation. The profile parameters of the observed \HI{} spectrum are 
obtained by fitting the profile with Busy-Function \citep{westmeier2014busy}. 
The 20$\rm \%$ and the 50$\rm \%$ of the peak maximum velocity widths are $\rm W_{50} \sim $ 304.9\kms and $\rm W_{20} \sim $ 293.3 \kms.
Table 4.3 summarizes the results.

\begin{table}
\begin{minipage}{110mm}the
\caption{Best fitting values obtained by fitting busy function.}
\centering
\begin{tabular}{|c|c|c|c|c|}
\hline
\hline
$V^{\textcolor{red}{(a)}}_{0}$   & $W^{\textcolor{red}{(b)}}_{50}$ & $W^{\textcolor{red}{(c)}}_{20}$&$F^{\textcolor{red}{(d)}}_{peak}$& $F^{\textcolor{red}{(e)}}_{int}$ \\
\kms{}&\kms{}& \kms{}& mJy & $Jy\, \kms{}$     \\
\hline
$4206 \pm 1.9$ &$293.3\pm 2.1$&$304.9\pm 2.94 $&$51.7 \pm 3.7$&$9.6 \pm0.2$\\
\hline
\end{tabular}
\hfill{}
\label{table: table 3}
\end{minipage}
\begin{tablenotes}
\item  $\textcolor{red}{(a)}$: Frequency centroid of the \HI{} line.
\item $\textcolor{red}{(b)}$:  Spectral line width at 50$\%$ of the peak flux density.
\item $\textcolor{red}{(c)}$:  Spectral line width at 20$\%$ of the peak flux density. 
\item $\textcolor{red}{(d)}$:  Peak of the \HI{} flux density.
\item $\textcolor{red}{(e)}$:  Integrated \HI{} flux.
\end{tablenotes}
\end{table}

\subsection{Channel and Moment maps}
\begin{figure*}
\resizebox{150mm}{190mm}{\includegraphics{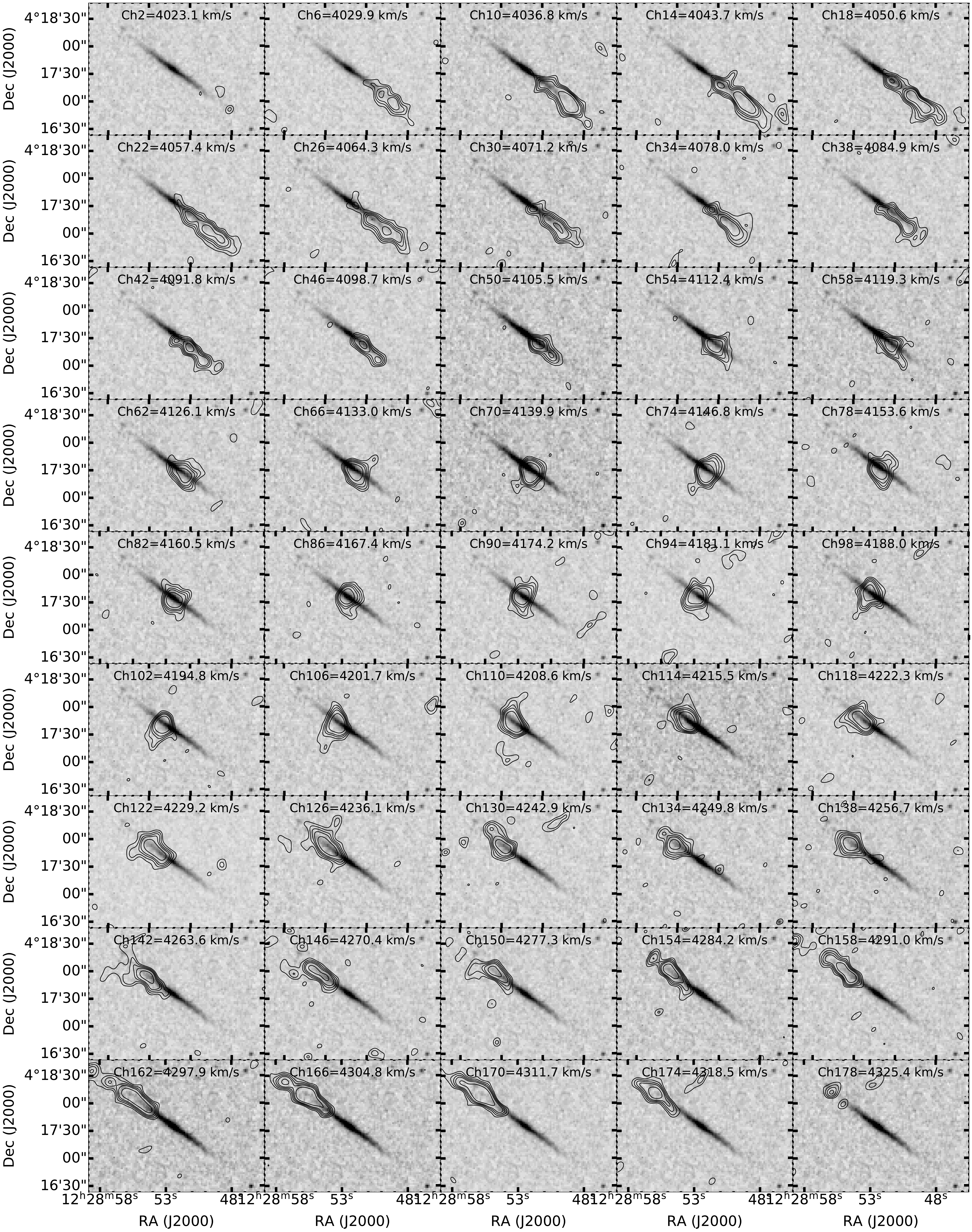}} 
\caption{Channel maps showing \HI{} emission from FGC 1440 overlaid on POSS-II optical image, each panel is separated by four channels. 
The contour levels are at [3, 4, 5 ,6]$\rm \times$ 1.01 mJy beam$\rm ^{-1}$}	
\end{figure*}

Figure 4.2 shows channel maps overlaid on POSS-II optical image of FGC 1440.
In channels with the highest velocity deviation, emission begins at the stellar disc's edge and extends further beyond it.
In the channels close to the systemic velocity (70, 102, 106, 110), the \HI{} emission spreads out of the plane at the center, 
compared to channels further away. For example, the emissions from the central channels 70, 102 are extended away from 
the plane compared to channels 170, 174, or 14 and 18. Figure 4.3 shows Moment 0 and Moment 1 maps. Moment 0 map is overlaid on the POSS-II image of FGC 1440. 
Moment 0 map shows a minor warp in the galaxy's north-east.

\begin{figure}
\hspace{3.5cm}
\resizebox{100mm}{95mm}{\includegraphics{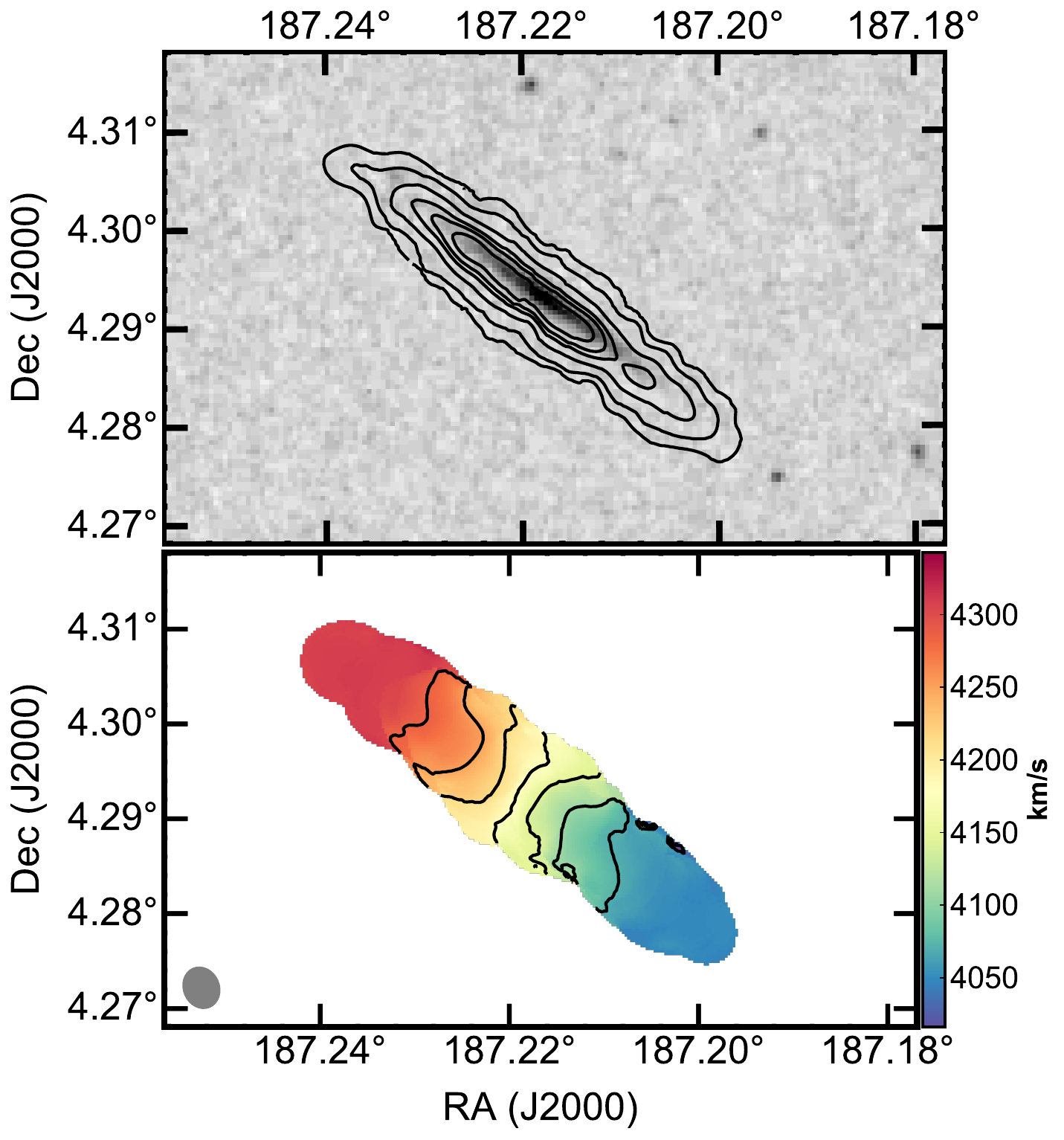}} 
\caption{In the top panel we have plotted the Moment 0 map the contours are at [ 2.5, 5.0, 10, 15, 18, 22]$\rm \times$ 40 mJy beam$\rm ^{-1}$ kms$\rm ^{-1}$. 
In the bottom panel, we have shown the Moment 1 map, and the contours start at 4000\kms increasing at 35 \kms }
\end{figure}

\section{3D - Tilted Ring Modeling (TRM)}
For deriving FGC 1440's kinematic and structural parameters, we employ the tilted ring modeling software TiRiFiC \citep{jozsa2012tirific}.
We model the galaxy as a rotating gas disc with a $\rm sech^{2}$ profile in the vertical direction. Each ring is set to 1.1 times beamwidth. (5.1 kpc).
For details on 3D tilted ring modeling, refer to section 1.4.5.

\subsection{Modeling strategy} 
Fully Automated TiRiFiC \citep[FAT,][]{kamphuis2015fat} estimates the initial fit to the \HI{} observations (see Section 1.4.5). 
We utilize these initial estimates of the fit parameters to compare the model and data by (1) visually inspecting the emissions in individual channels, 
(2) comparing the Moment 0 and Moment 1 maps, and (3) examining the minor axis PV diagrams at various offsets. 
We then manually modify parameters using TiRiFiC, comparing the model and data in each iteration to fine-tune the parameters corresponding to the final model data cube.

\subsection{Automated fit using FAT (Fully Automated TiRiFic)}
We have employed a beta version of FAT, which allows us to vary the Intrinsic velocity dispersion radially.
FAT takes an \HI{} data cube as input and estimates the following parameters;
1)Surface brightness profile, 2)Position angle 3)Inclination 4) Rotation velocity  5) Scaleheight 6)Intrinsic velocity dispersion and the 
7)Central coordinates: Right Ascension, Declination, and Systemic velocity, as a function of the radius. 
FAT fits each parameter ring by ring and smoothens them using a 0, 1, 2, 3, 4, or $\rm 5^{th}$ order polynomial. FAT models the \HI{} disc as 
two halves with 9 semi-rings across each half. FAT fits the surface brightness of the approaching and receding sides independently and uses a 
constant scaleheight equal to $\rm 6.1\farcs$. Table 4.4 shows the model parameters estimated by FAT, and Figure 4.4 depicts the radial profiles of the above parameters. 
Figure 4.5 shows the $\rm (V_{Total})$ derived from the 3d tilted ring model overlaid on a major axis PV diagram.
After estimating the kinematic and structural parameters with FAT, we test the model's sensitivity estimated by FAT
to the variation in parameter by manually running TiRiFiC and comparing the model with the data. The models are degenerate for 
a wide range of inclinations, scale heights, and dispersion values.

\begin{figure}
%\vspace{-5mm}
\hspace{3.5cm}
\resizebox{100mm}{95mm}{\includegraphics{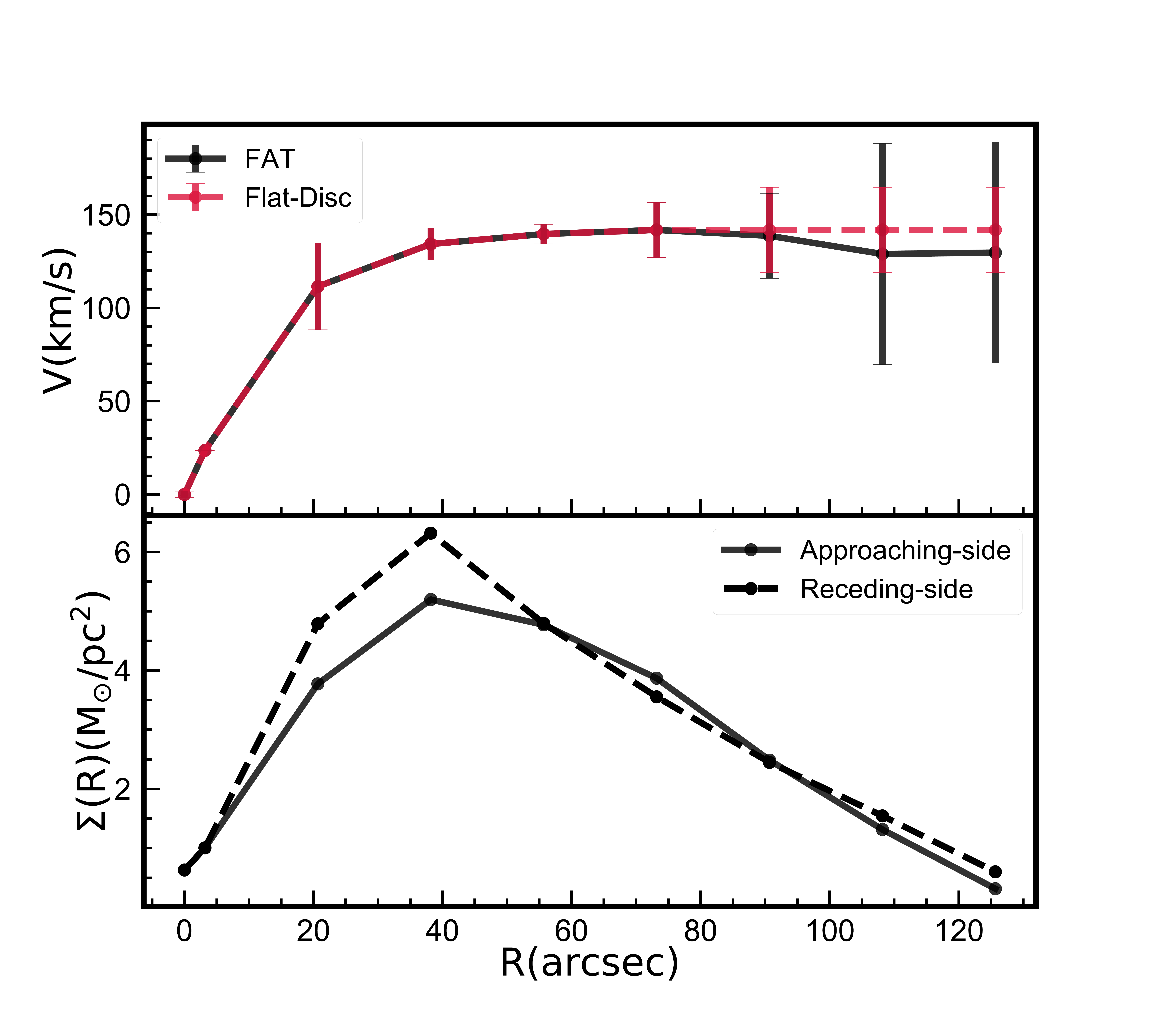}} 
%\vspace*{-15mm}
\caption{The rotational velocity [top panel] and the surface brightness profile [bottom panel] obtained from the tilted ring modeling.
The surface brightness is fitted independently for the approaching and the receding side.}	
\end{figure}

\begin{figure}
\hspace{3.0cm}
\resizebox{120mm}{70mm}{\includegraphics{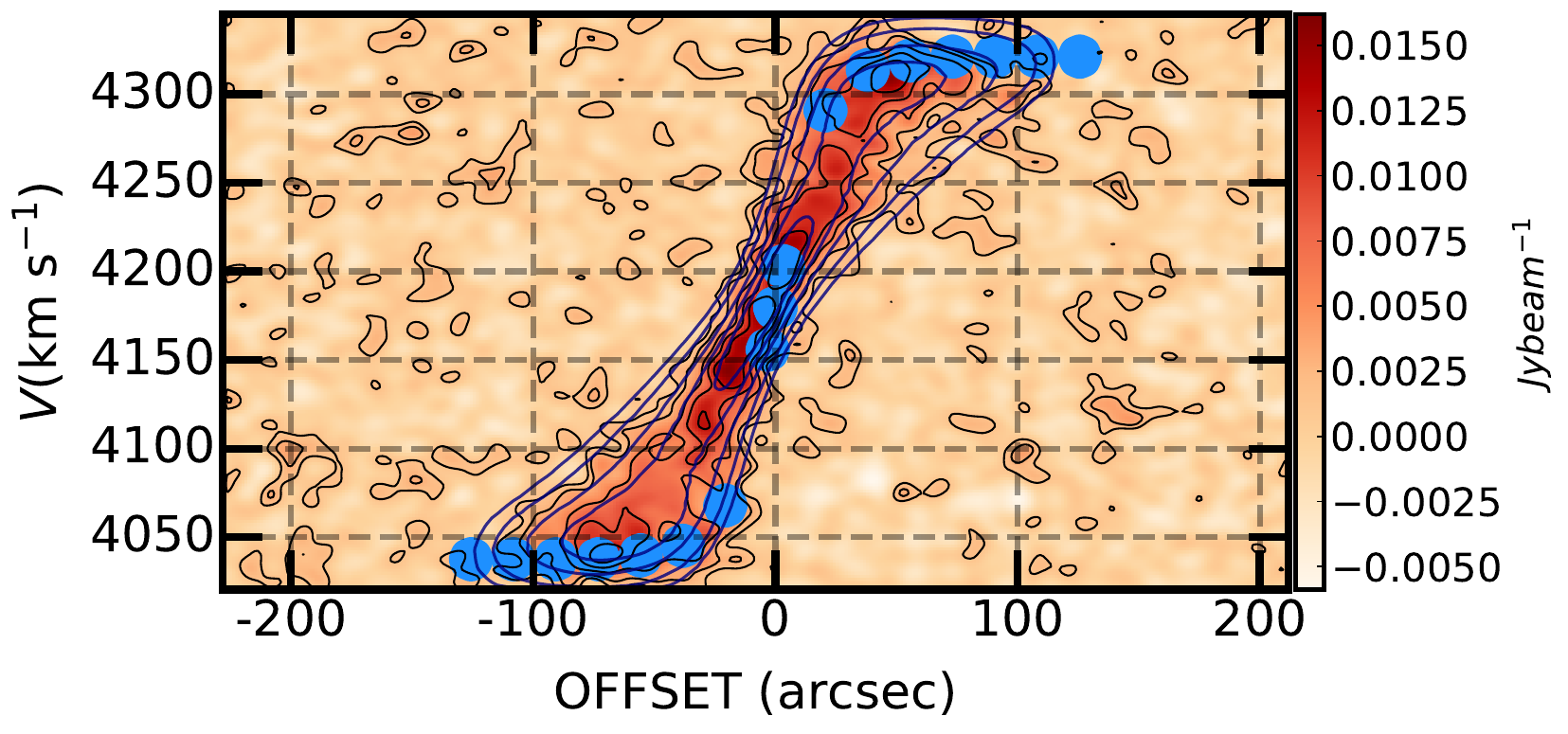}} 
\caption{The major axis PV map with flat disc model overlaid at contours [1.5, 3, 6, 9, 12]$\rm \times$ 1.01 Jy/beam. The blue points represent the \HI{} rotation curve.}
\end{figure}

\begin{table}
\begin{minipage}{110mm}
\hfill{}
\caption{Parameters describing 'FAT' and 'Flat-Disc' models.}
%\centering
\begin{tabular}{|l|c|c|}
\hline
\hline
Parameter&    FAT-Model& Flat-Disc-Model  \\
\hline    
$\rm X_{o}$\footnote{Right ascension}            &    $+187.2$                &   $+187.2$  \\
$\rm Y_{o}$\footnote{Declination}            &    $+4.29$                  &   $+4.29$    \\
$\rm i$ \footnote{Inclination}               &    $85^{\circ}$                  &   $88.5^{\circ}$        \\
$\rm V_{sys}$\footnote{Systemic velocity}          &    4179.45                &   4179.45     \\
PA \footnote{Position angle}                    &    $53.5^{\circ}$         &   $53.5^{\circ}$     \\
Dispersion \footnote{Velocity dispersion}             &    6.6\kms               &   15 \kms\\  
$\rm h_{z}$ \footnote{Scaleheight of the \HI{} disc}           &    $6.10 \farcs$ (1.8 kpc)             &   $0.45 \farcs$ (0.13 kpc)  \\
\hline
\end{tabular}
\hfill{}
\label{table: table 4}
\end{minipage}
\end{table}

\subsection{Manual TRM Models}
We perform multiple rounds of manual fitting with TiRiFiC followed by visual analysis 
of channel maps, PV diagrams at various offsets, and moment maps to estimate $\rm Flat-Disc$ model parameters. 
Iterative visual inspection is used to fine-tune parameters and create the final model data cube. 
(See \cite{allaert2015herschel}, \cite{zschaechner2012halogas}, \cite{gentile2013halogas}, \cite{kamphuis2013halogas}.

Using the FAT model data cube as an initial model, we repeatedly generate model data cubes by modifying the inclination, dispersion, and scaleheight. 
In the $\rm Flat-Disc$ model, we adopt the parameters estimated by FAT for the center (RA, Dec), systemic velocity, surface brightness, and position angle 
and vary the scaleheight, dispersion and inclination angle. We build model data cubes
by altering the parameters (i, $\rm \sigma$, $\rm h _{z}$) individually, in pairs, or simultaneously.
We match each model with the data channel-wise, comparing 3D models of data and the model data cubes using volume rendering software astroslicer \citep{punzo20173d}, and comparing the datacubes by taking slices at various offsets along the major axis (as it preserves the 3D structure of the cube) to arrive at the secondary base model 
called the $\rm 'Flat-Disc'$ model. Figure 4.6 shows the  Moment 0 and Moment 1 maps derived from the 'Flat-disc model' 
superposed on the data. To show the effect of modifying the inclination $\rm (i)$, dispersion $\rm (\sigma)$, and scaleheight $\rm (h_{z})$, 
we plot the minor axis PV diagrams (Figure 4.7 to 4.11) at different offsets, varying each parameter i, $\rm \sigma$, $\rm h_{z}$ one by one, while keeping the 
other two-parameters to those of the $\rm Flat-Disc$ (Table. 4.4). The second row in Figure 4.6 shows how variation in the inclination affects the model. 
Dispersion and the scaleheight are kept constant when the inclination is changed. In the third row, we keep the inclination and scaleheight fixed and only change the dispersion.

In order to understand if the data is sensitive to radial variation of the parameters, we analyze models in which we adjust inclination, dispersion, and 
scaleheight as a function of radius, leaving other parameters constant at the values obtained from the  $\rm (Flat-Disc)$;
\begin{itemize}
\item \textbf{Radially varying inclination $\rm i(R)$:} The inner rings are kept at an inclination equal to 90$\rm ^{\circ}$ and the outer rings at 85$\rm ^{\circ}$.\\
\item \textbf{Radially varying dispersion  $\rm \sigma(R)$:} Dispersion varies from 20 \kms{} in the inner rings to 5 \kms{ at the outer rings in steps of 5 \kms{}}.\\
\item \textbf{Radially varying scaleheight $\rm h_{z}(R)$:} The inner rings are kept at an $\rm h_{z}=0.45\farcs{}$, the central rings are kept at 
$\rm h_{z}=1.97\farcs{}$, and the outer rings at $\rm h_{z}=5.3\farcs$
\end{itemize}

\textbf{Inclination $\rm (i)$.}
For examining the effect of inclination, we fix all parameters in the $\rm 'Flat Disc'$ model and vary only the inclination.
From the minor axis PV diagrams (Figure 4.7 to Figure 4.11), the first immediate observation is that we can rule out the models with an 
$\rm i<85^{\circ}$. For example, the PV plot at an offset equal to 0, see Figure 4.7, the inner model contours at 9$\rm \times$rms are not extended 
sufficiently to describe the emissions, and further models with lower inclination only increase this discrepancy. Comparing PV diagrams at 
different offsets, the inclination is $\rm 85^{\circ}<i<90^{\circ}$, $\rm i=85^{\circ}$ is the lower limit for FGC 1440's inclination. We note that 
the low-inclination models do not represent the data accurately.

\textbf{Dispersion $(\sigma)$.}
By comparing the PV diagrams (Figures 4.7 to 4.11), we discover that the data is not sensitive to the change in dispersion, 
as models with 5 \kms to 15 \kms show little variance at the level of data. However, models with dispersion higher than 
15 \kms deviate significantly from the data. The dispersion for FGC 1440 lies in range $\rm 5\kms<\sigma <15 \kms$.

\textbf{Scaleheight $\rm (h_{z})$.}
Minor axis PV diagram (Figures 4.7 to 4.11) shows that models with higher scaleheight are more spread out spatially than models with lower scaleheight, which are spread 
out along the velocity axis but not spatially. Also, we see that the model contours do not match the data contours in the inner regions, 
at $\rm 9\times rms$, when the scaleheight is higher. In the inner parts of the PV diagrams, the models with lower 
scaleheight follows the data contours, but it is not easy to tell the difference between the models with $\rm h_{z}=0.45\farcs$ and $\rm h_{z}=1.97\farcs$. 
Further, we make a model where the scaleheight varies with radius. This is based on the fact that the data contours in the inner parts of the PV diagram 
are better described by low scaleheight values, while the data contours in the outer parts of the PV diagram are described by 
higher scaleheight values. We find that the models with scaleheight that changes radially are not much different 
from models with $\rm h_{z}=0.45\farcs$ or $\rm h_{z}=1.97\farcs$. This indicates that the data is not sensitive to the changes in the scaleheight.
To better understand the vertical structure of the \HI{} gas, we plotted the normalized 
vertical density profile from the moment 0 maps at different slices and compared it with the major axis FWHM of the synthesized beam in Figure 4.12. We see that 
the synthesized beam is similar to the vertical density profile we get from the data. This indicates that the vertical thickness is hardly resolved in these 
observations. By comparing PV diagrams, we can see that the upper limit on the scaleheight is $\rm 5.3\farcs$.
To summarize, we find that by comparing the model and the data in Figure 4.7 - 4.11, we find the lower and upper limits on the inclination
($\rm 85^{\circ}\leq i \leq 88.5\circ$), the dispersion ($\rm 5\kms \leq \sigma \leq 15\kms$) and the scaleheight ($\rm h_{z} \leq 5.3 \farcs$).

\subsection{Thickness of the \HI{} disc}
Is it a flare, a thick disc or a line of sight warp ?

In Figure 4.2, we can see from the channel maps that the emissions from the channels close to the systemic velocity are higher 
than the emissions from the end channels. This could mean either a thick \HI{} disc or a line of sight warp. We can rule out a 
flaring disc, as a flare would show up as extended emission in channels further away from the systemic velocity. If the \HI{}
disc were flaring then the scaleheight of the \HI{} disc will increase with the radius, which is not the case in FGC 1440.
In order to figure out the origin of thickness of the \HI{} disc 
in FGC 1440, we made Moment 0 maps with only the starting channels (4023\kms - 4100\kms), 
the central channels close to the systemic velocity (4100\kms - 4242\kms), and the ending channels (4242\kms - 4325 \kms).
Then, we compare the data contours for the above velocity ranges with the model in which the scaleheight and inclination 
change radially and the flat disc model. We find that the data contours (see Figure 4.13) are just as thin as the model contours in high-velocity channels. 
This could mean that the \HI{} disc in FGC 1440 is not flaring. More evidence for the absence of flaring comes from the fact that the \HI{} disc 
is not thicker at high-velocity channels but at the center. So, if we see a thick \HI{} disc in the middle that gets thinner 
toward the edges, we only have two options: either a line-of-sight warp or a thick \HI{} disc in the middle.

To better understand this effect, we look at the following models: 1) A model with radially changing inclination, and 2) A thick+ thin \HI{} disc model with a 
lagging thick disc $\rm (\frac{dv}{dz}=-10 \kms\, kpc^{-1},\, h_{z}=5.3\farcs)$, and a thin disc with scaleheight ($\rm h_{z}$)=0.45$\farcs$. In the thick+ thin \HI{} disc model, 
the scaleheight of the thick disc is equal to the upper limit of the scaleheight found in the one-component model. Then, we visually inspect the 
models with the data for different values of $\rm \frac{dv}{dz}$. We find that $\rm \frac{dv}{dz}=-10 \kms\, kpc^{-1}$ matches the data contours better. 
From Figures 4.13 and 4.14, we can see that the thick + thin \HI{} model and the overall model with radially changing inclination are very similar. It 
is almost impossible to tell if the observed central thickness is caused by radially changing inclination or if the galaxy has a thick central \HI{} disc. 
Also, in Figure. 4.12, we compare the synthesized major axis beam to the density profile extracted from different offsets. This shows that we might not be
able to resolve the thick disc in our observations, and we can not say for sure that a thick \HI{} disc does not exist in the center.
We use a model with a line of sight warp because it is simpler than a model with a thick + thin \HI{} disc.

\begin{figure*}
\resizebox{160mm}{60mm}{\includegraphics{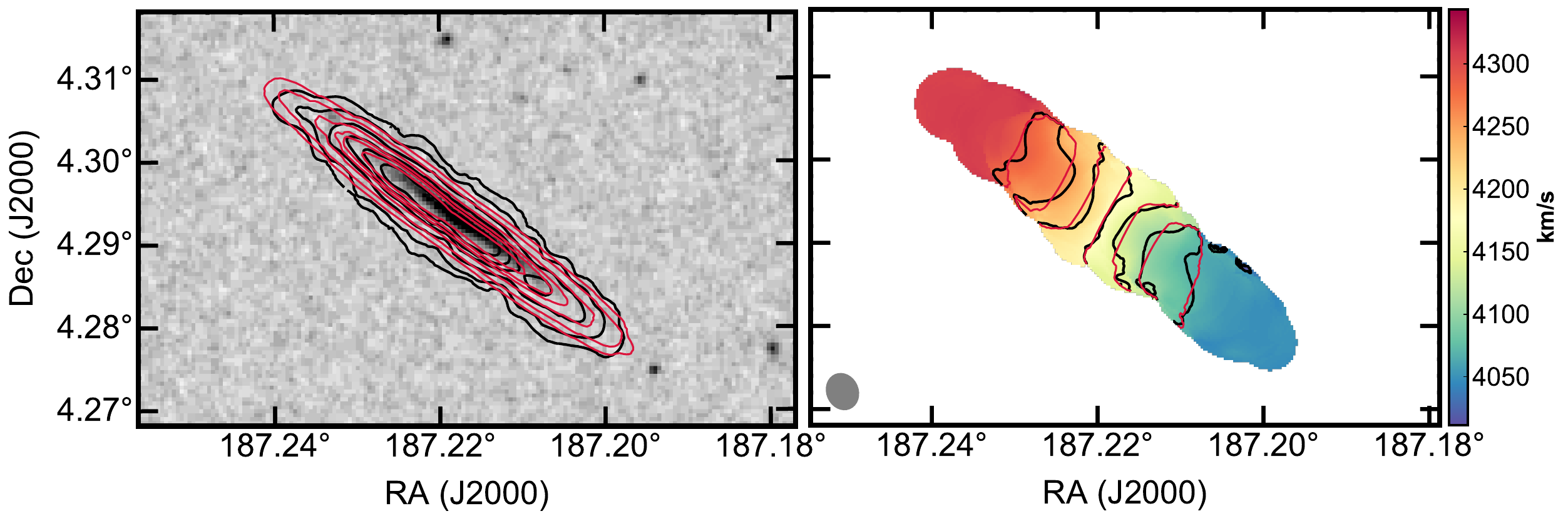}} 
\caption{In the left panel we have plotted the moment 0 (top) and the moment 1 map(bottom)  the contours are at [2.5, 5.0, 10, 15, 18, 22]$\times$ 40 mJy beam$\rm ^{-1}$ kms$\rm ^{-1}$. 
and the contours for moment 1 map start at 4000 kms$\rm ^{-1}$ increasing at 35  km s$\rm^{-1}$ respectively. The contours corresponding to the data are rendered in black 
and the contours of the $\rm Flat-Disc$ Model are in shown in red color.}
\end{figure*}

\begin{figure*}
\resizebox{160mm}{100mm}{\includegraphics{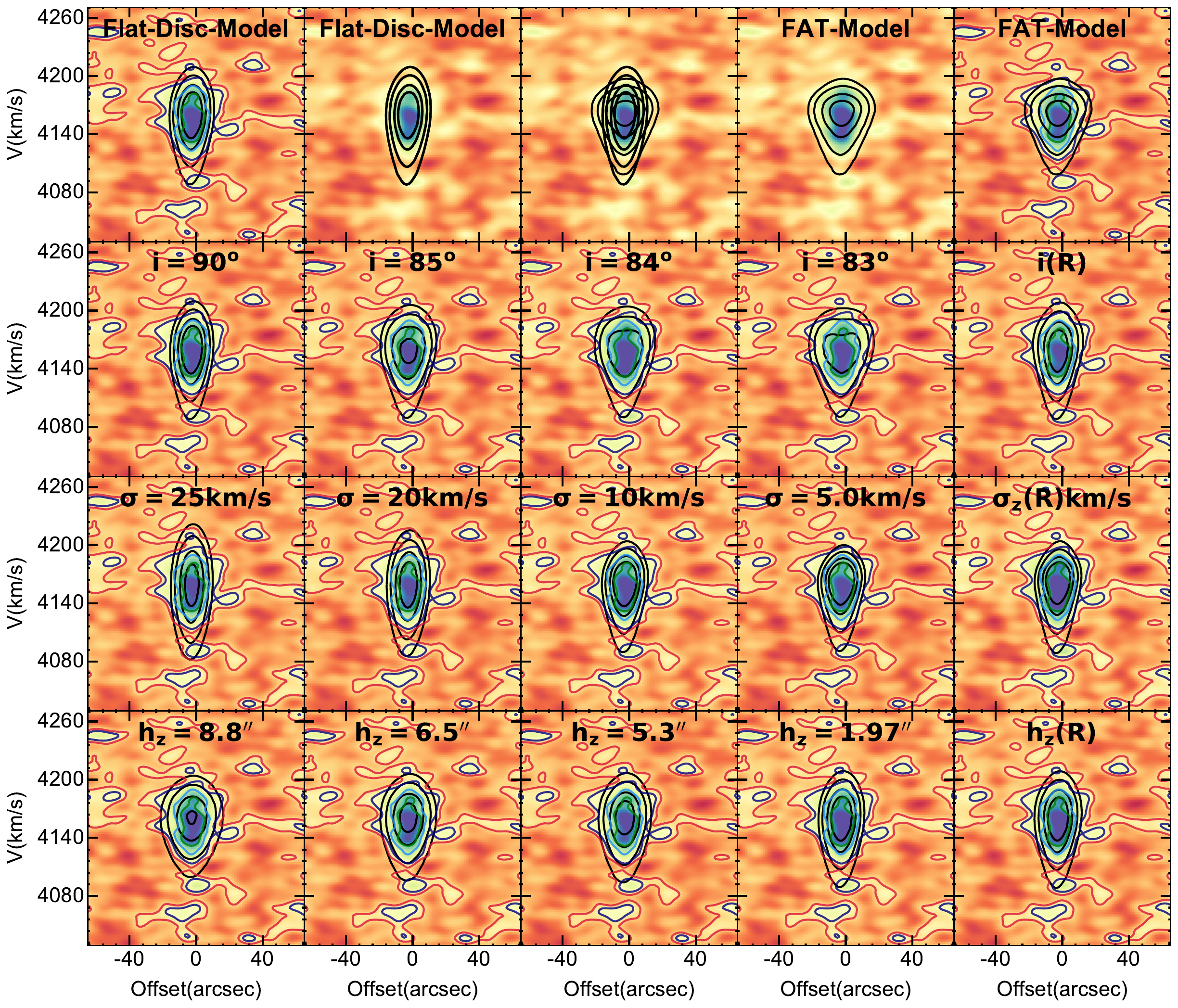}} 
\caption{Position velocity maps parallel to the minor-axis comparing the tilted rings model to data at an offset equal to 0 by varying the model parameters. The contours 
at [1.5, 3, 6, 9]\rm $\times$ 1.01 Jy/beam.}	
\end{figure*}

\begin{figure*}
\resizebox{160mm}{100mm}{\includegraphics{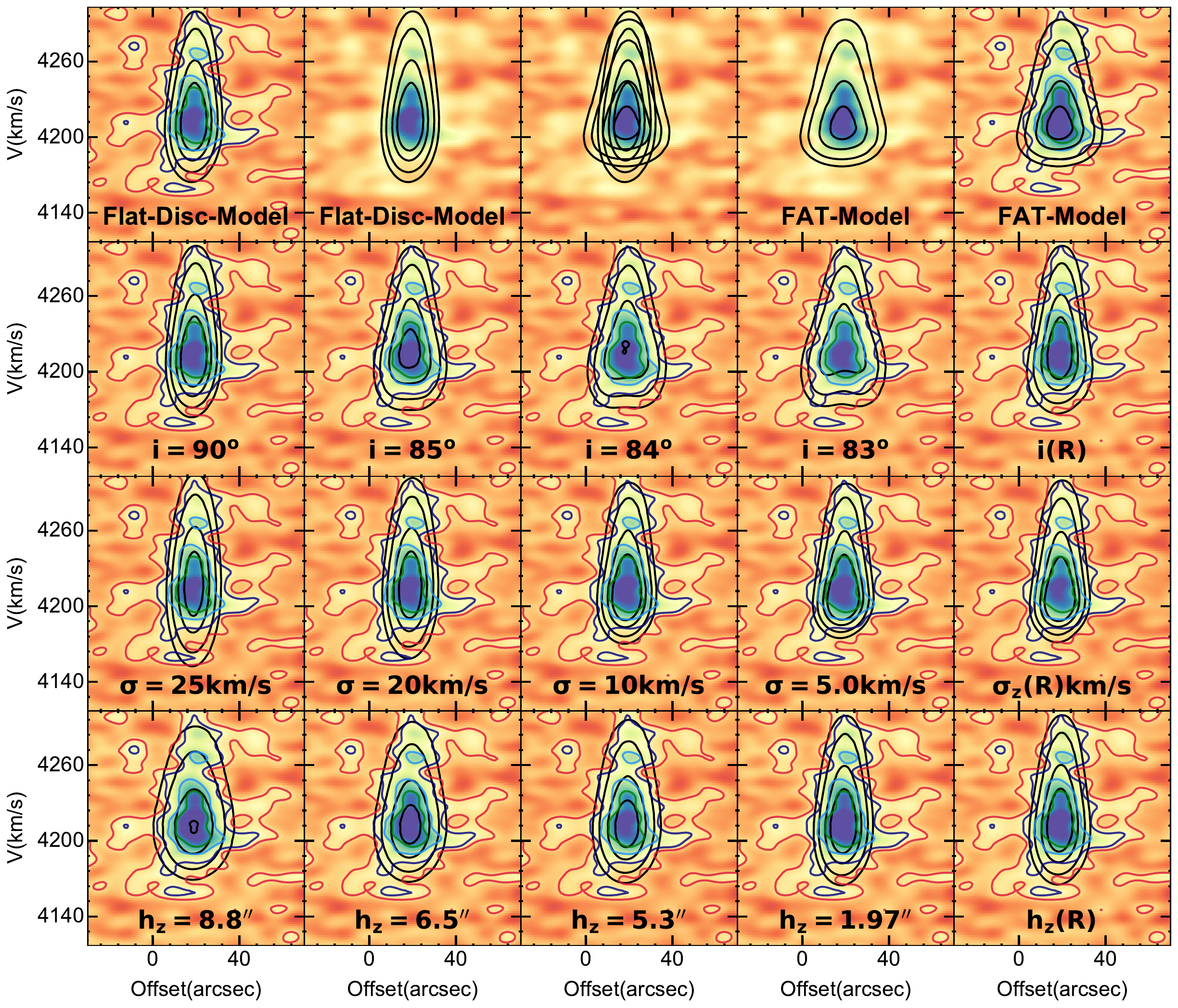}} 
\caption{Same as figure 4.6 but at an offset=20}	
\end{figure*}

\begin{figure*}
\resizebox{160mm}{100mm}{\includegraphics{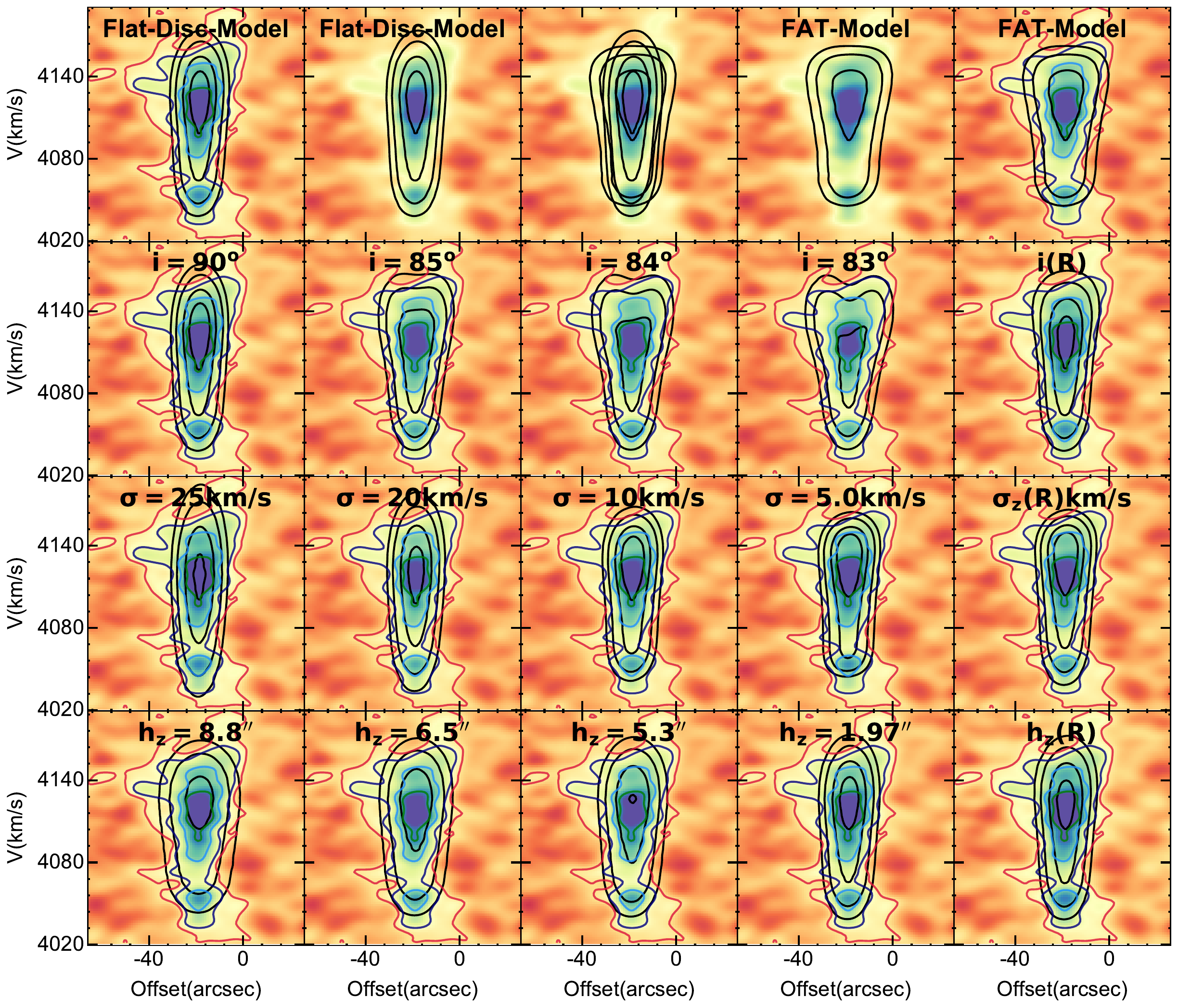}} 
\caption{Same as figure 4.6 but at an offset equal to -20,.}	
\end{figure*}

\begin{figure*}
\resizebox{160mm}{100mm}{\includegraphics{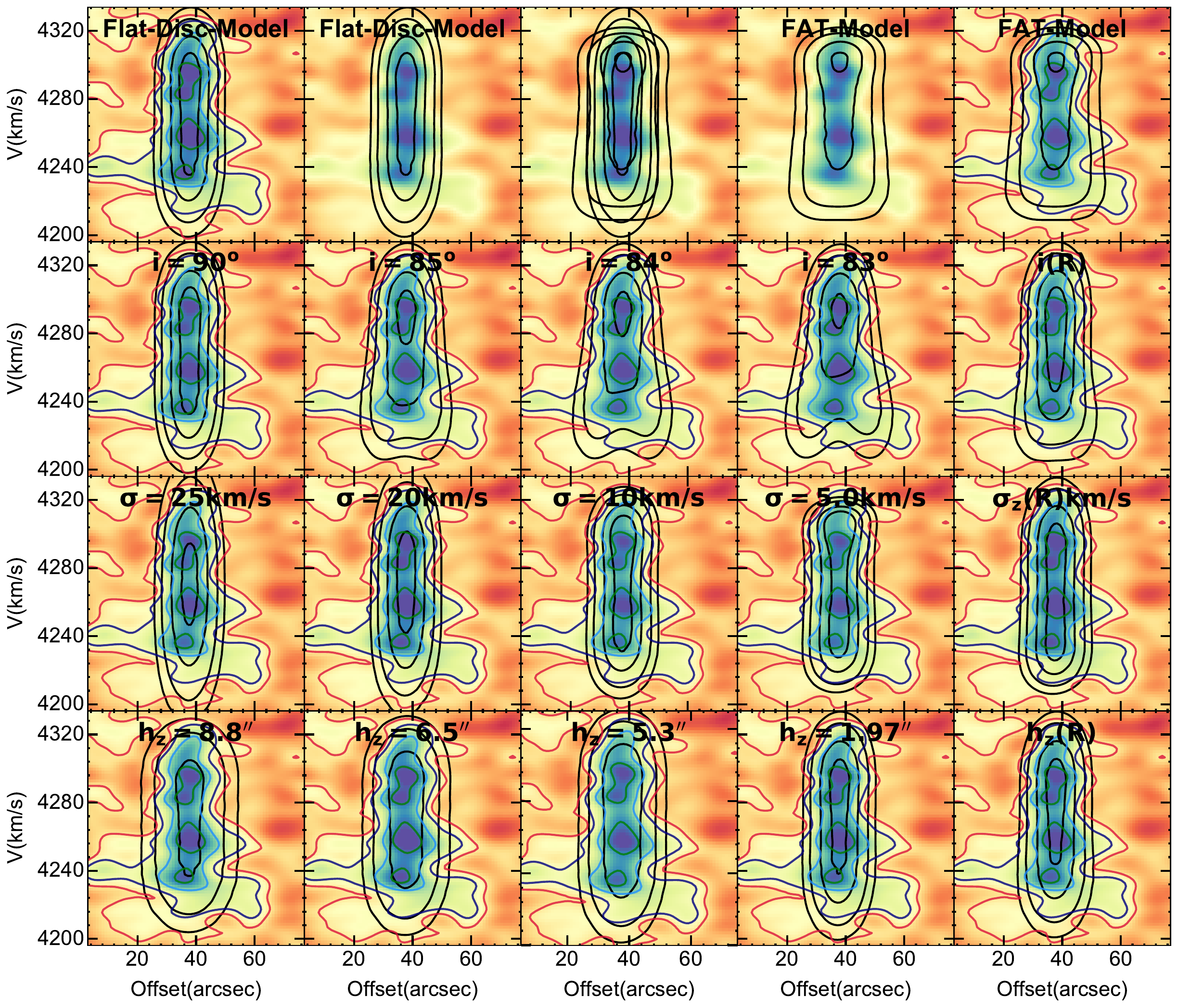}} 
\caption{Same as figure 4.6 but at an offset equal to 40.0 }	
\end{figure*}

\begin{figure*}
\resizebox{160mm}{100mm}{\includegraphics{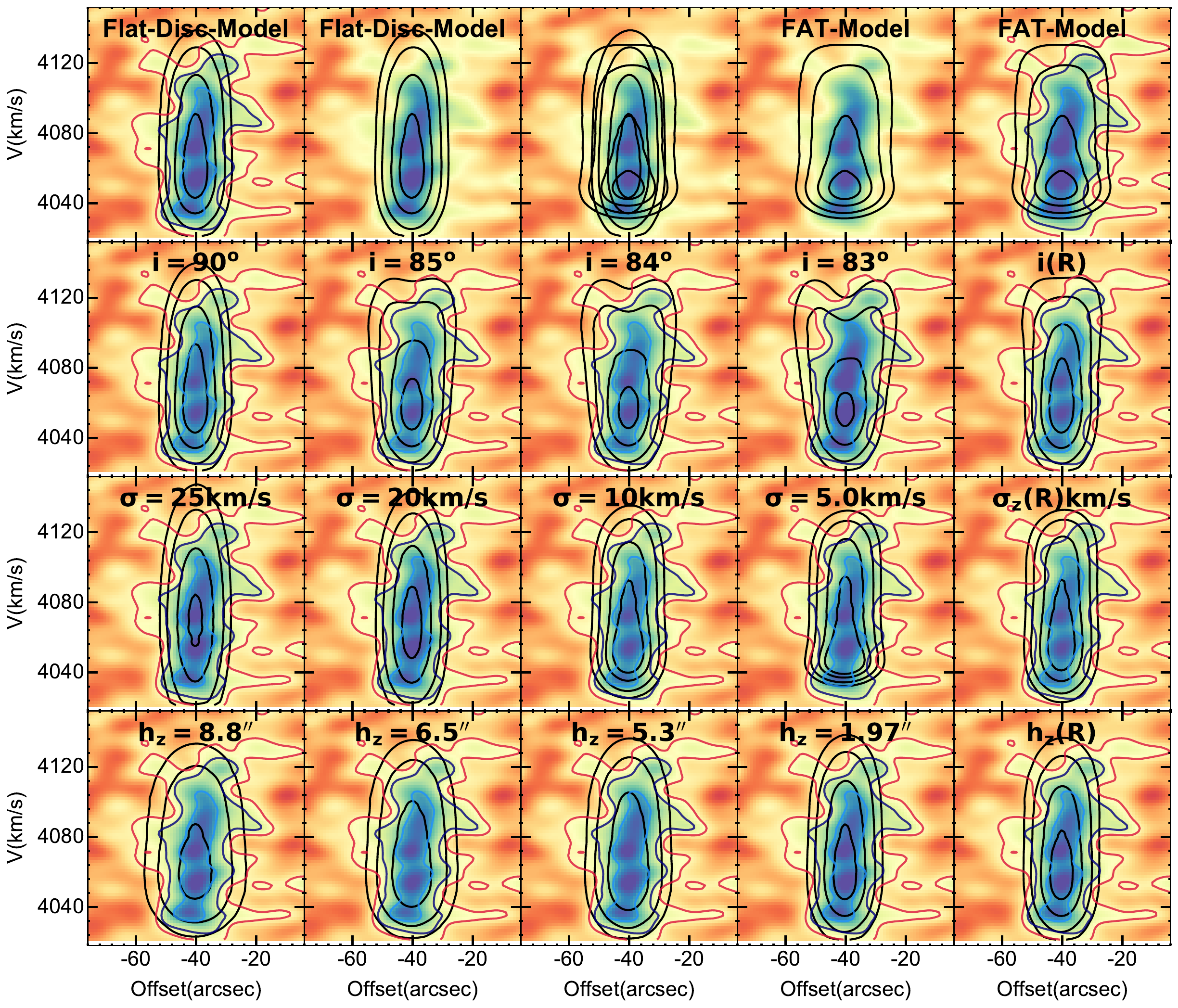}} 
\caption{Same as figure 4.6 but at an offset equal to -40.}	
\end{figure*}

\begin{figure*}
\resizebox{180mm}{50mm}{\includegraphics{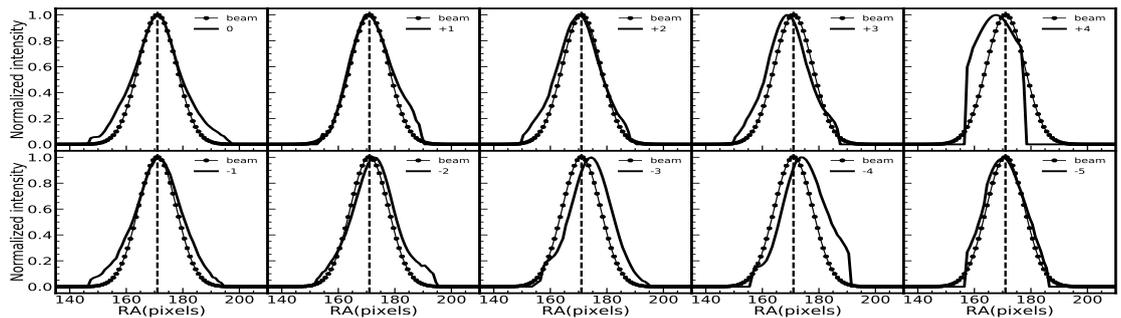}} 
\caption{We have compared the synthesized major axis beam size with the vertical density profile at slices extracted from the moment 0 maps.}	
\end{figure*}

\begin{figure*}
\hspace{-3cm}
\resizebox{220mm}{120mm}{\includegraphics{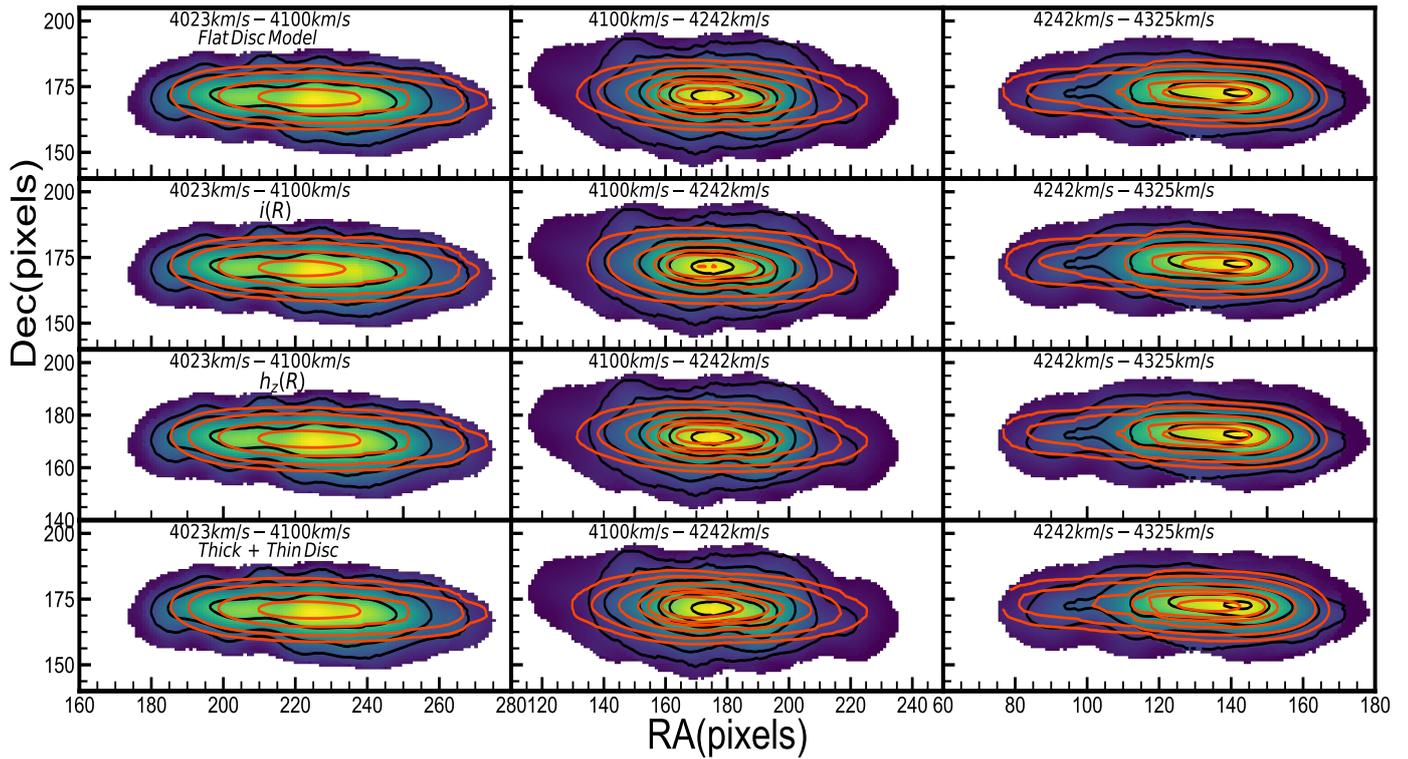}} 
\caption{We compare the moment zero maps derived from the center and edge channels. In the first row we compare the moment maps for the $\rm'Flat-Disc-Model'$
derived at the velocity range 4023 \kms to 4100 \kms in the first column and in the middle panel, moment 0 map for the central velocity range 4100 \kms - 4242 \kms, and in the 
third panel, we show the moment 0 maps for the velocity range 4242 \kms to 4325 \kms. Similarly, in the second, third, and  fourth rows, we plot the contours 
for the models with radially varying inclination, scale height, and the thick + thin disc model. }	
\end{figure*}

\begin{figure*}
\hspace{-2cm}
\resizebox{180mm}{30mm}{\includegraphics{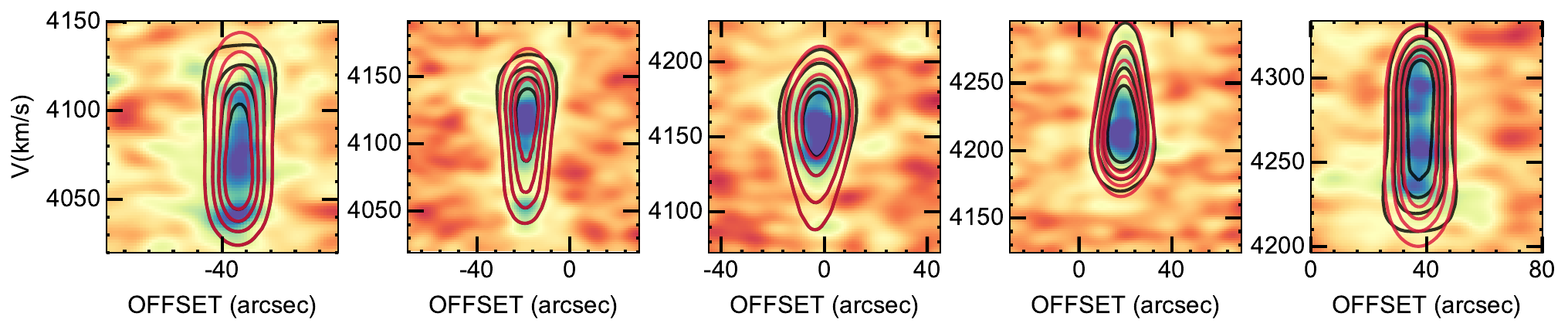}} 
\caption{We show the minor axis PV diagrams comparing the model with radially varying inclination and the thick + thin disc model. Black contours depict a model with 
radially varying inclination, and the red contours show the thick + thin disc model. }	
\end{figure*}

\section{Optical photometry}
We analyze FGC 1440's optical photometry in SDSS g, r, and UKIDSS K bands. First, we mask the surrounding objects and the galaxy itself and 
estimate the center and positional angles (PA) using SExtractor \citep{bertin1996sextractor}. We subtract the background using a two-dimensional 
second-order polynomial and then rotate the frame by the PA. Next, we replace close-by objects with symmetric mid-plane regions.
We integrate light in a rectangle box to measure overall magnitude. The box size was chosen in such a way that the growth curve flattens at the edges. 
The g, r, and K band magnitudes are 15.43, 14.72, and 11.83. We get the galaxy's structural properties using GalFit \citep{peng2011galfit}.
by fitting the intensity profile given as $\rm I \sim R/R_{d} K_{1}(R/R_{d})sech^{2}(z/h_{z})$, where $\rm R_{d}$ is the disc scale radius and $\rm h_{z}$ is the disc scaleheight.
Table 4.5 summarizes parameters obtained from optical photometry. 

\section{Mass Modeling}
This section presents mass models for FGC 1440. By dividing the galaxy's total rotation curve into baryonic (stars+$\rm \HI{}$) and dark matter components, 
we estimate each mass component's contribution to $\rm (V_{Total})$ (see Figure 4.4,top panel). The procedure for carrying out the mass modeling is detailed in Section 1.4.6

\begin{table*}
\begin{minipage}{110mm}
\hfill{}
\caption{Optical photometry}
\centering
%\small\addtolength{\tabcolsep}{-1pt}
\begin{tabular}{|l|c|c|c|c|}
\hline
\hline
Parameter                                 &Value        &Value     &value            &Description \\

                                          &g-band&     r-band      &K-band           &             \\
\hline
Total magnitude                           &15.4&      14.7        &11.8           &             \\
$\mu^{edge-on}_{o}$                       &21.6&      20.8        &16.5           &$\mu^{edge-on}_{0} \, mag/arcsec^{2}$\\
$\mu^{face-on(\textcolor{red}{*})}_{o}$   &23.3&      22.5        &18.6           &$\mu^{face-on}_{0} \, mag/arcsec^{2}$\\
$\Sigma_{o}$                              & 22.7&      31.0        &328.0         &$L_{0}$ $L_{\odot}/pc^2$\\
$R_{d}$                                   & 4.4&      4.2        &2.6             &$R_{d}$kpc\\
$h_{z}$                                   & 0.9&      0.9          &0.4           &$h_{z}$ kpc\\
\hline
Parameters $(\gamma^{*})$&  &    & \\
\hline
$g-r$         &0.7      &        &                     \\
$a_{\lambda}$ & -0.5&  -0.31      &-0.2  &\cite{bell2003optical}\\
$b_{\lambda}$ & 1.5&     1.1       &0.2   &\cite{bell2003optical}\\
$\gamma^{*}$  & 3.8&     3.0        &0.8   &M/L using scaled Salpeter IMF\\
$\gamma^{*}$  & 2.7&     2.1         &0.6   &M/L using Kroupa IMF\\
\hline
\end{tabular}
\hfill{}
\label{table: table 5}
\end{minipage}
\begin{tablenotes}
\item  \textcolor{red}{(*)}: The edge-on surface brightness has been converted to face-on surface brightness using 
$\rm \mu^{face-on}= \mu^{edge-on} + 2.5log( \frac{ R_{d} } {h_{z}})$ \citep{kregel2005structure}.
\end{tablenotes}
\end{table*}

\subsection{H$\alpha$ Rotation curve}
\begin{figure}
\hspace{2cm}
\resizebox{110mm}{85mm}{\includegraphics{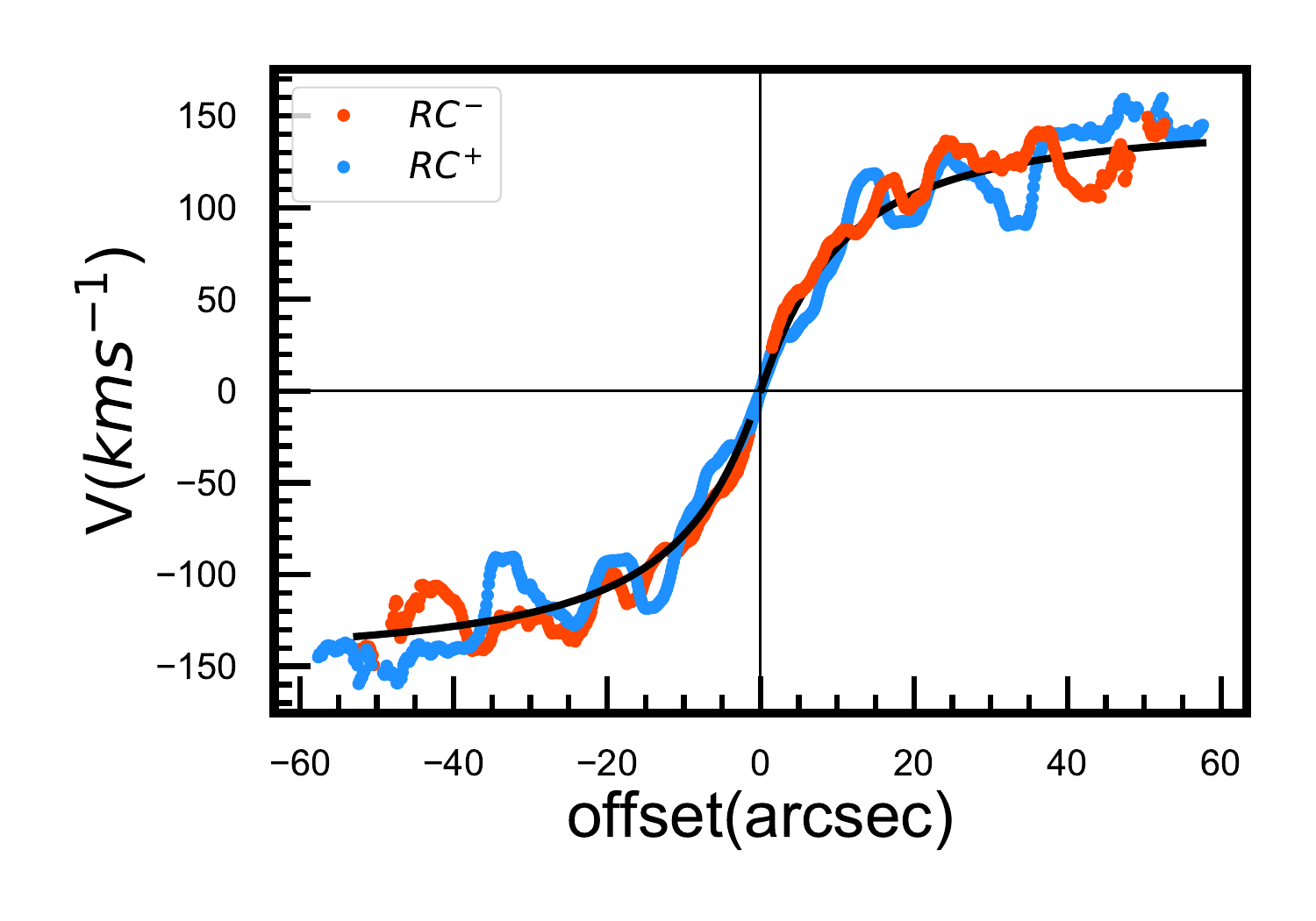}} 
\caption{The blue and red points in the above plot depict the optical rotation curve for the approaching and receding sides respectively. The points have been mirrored along 
the major axis. The smooth model optical rotation curve is shown using the solid black line.}	
\end{figure}

We derive the hybrid rotation curve for FGC 1440, wherein the inner region is composed of points from the $H\alpha$ rotation curve and \HI{} 21 cm 
data define the outer points \citep{de2001high}. The optical rotation curve is taken from \cite{yoachim2008kinematics}.
We apply a simple multi-parameter function to the raw optical rotation curve to get a smooth curve \citep{yoachim2005kinematics,courteau1997optical}.

\begin{uprightmath}
\begin{equation}
 V(r)=V_{0} + \frac{V_{c}}{(1+x^{\gamma})^{\frac{1}{\gamma}}}
\end{equation}
\end{uprightmath}

$\rm V_{0}$ is the galactic center's recession velocity, $\rm V_{c}$ is the asymptotic rotation velocity (the flat part), 
$\rm x$ is defined as $\rm x=r_{t}/R-r_{0}$, where $\rm r_{t}$ is the transition radius between the rising and flat parts of the rotation curve, 
and $\rm r_{0}$ and $\rm \gamma_{*}$ define the center and sharpness of transition. 
Table 4.6 shows the best-fitting parameters for Equation 4.1. Figure 4.15 shows the optical rotation curve and raw data.

\begin{table*}
\begin{minipage}{110mm}
\caption{Best fit values obtained by fitting the multi-parameter model fitted to raw optical curve.}
\centering
\begin{tabular}{|c|c|c|c|c|}
\hline
\hline
$V^{\textcolor{red}{(a)}}_{0}$   & $V^{\textcolor{red}{(b)}}_{c}$ & $r^{\textcolor{red}{(c)}}_{0}$&$r^{\textcolor{red}{(d)}}_{t}$& $\gamma^{\textcolor{red}{(e)}}$ \\
$km s{-1}$&$\kms$&kpc &kpc&    \\
\hline
4253.5&$149.4\pm4.2$&$1.4\pm0.04 $&$12.3 \pm 0.6$&$1.3 \pm0.1$\\
\hline
\end{tabular}
\hfill{}
\label{table: table 6}
\end{minipage}
\begin{tablenotes}
\item $\textcolor{red}{(a)}$: Recession velocity at the galaxy center .
\item $\textcolor{red}{(b)}$:  Asymptotic rotation velocity at the flat part .
\item $\textcolor{red}{(c)}$:  Center of the galaxy. 
\item $\textcolor{red}{(d)}$:  Point of transition between the rising and the flat part.
\item $\textcolor{red}{(e)}$:  Degree of sharpness of transition.
\end{tablenotes}
\end{table*}

We parameterize the dark matter distribution using observationally motivated pseudo isothermal (PIS) 
\citep{begeman1991extended, fuchs1998decomposition} dark matter halo dominated by a constant
density core and the Navarro-Frenk-White (NFW) \cite{navarro1997universal} dark matter 
halo profile obtained in cold dark matter (CDM) simulations. The details of mass modeling are delineated in section 1.3.6.

\subsection{Rotation curve fitting method}
The likelihood function as $\rm exp(-\frac{\chi^{2}}{2})$, where $\chi^{2}$ is given by,

\begin{uprightmath}
\begin{equation}
 \chi^{2} =\sum _{R} \frac{\bigg(V_{obs}(R) - V_{T}(R) \bigg)^{2} }{V^{2}_{err}}
\end{equation}
\end{uprightmath}

The baryonic and dark matter components are added in quadrature, yielding $\rm V_{T}$, which is the total rotation curve, and $\rm V_{err}$ 
is the error bars on the observed rotation curve, the top panel of Figure 4.4 shows $V_{obs}$. The python package LMFIT \citep{newville2016lmfit} 
is used to optimize the likelihood function. Figures 4.16 and 4.17 illustrate the residuals and the reduced chi-square values $\rm \chi^{2}_{red}$.
The inner slope of the dark matter density is critically dependent on the spatial resolution of the \HI{} rotation curve in the inner region, and since we 
only have a limited number of resolution elements in the inner region. The $\rm H\alpha$ rotation curve is used in the inner region when $\rm H\alpha$ data 
is available, while points from the $\HI{}$ rotation curve are used in the outer region. 
We create a smooth representation of the $\rm H\alpha$ rotation curve to account for the scatter between the points. 
We utilize the B-spline method from the python package scipy \citep{virtanen2020scipy} to create a smooth spline approximation of the hybrid rotation curve 
consisting of points from  $\rm H\alpha$ data in the inner region and \HI{} data in the outer region.
We use conservative error bars equal to 10\kms on the data points defining the observed rotation curve.
\cite{de2008mass, mcgaugh2001high}. \cite{mcgaugh2001high} show that the halo parameters are robust against the precise definition of error bars.

\subsection{Results from mass modeling}
In this section, we present mass models generated using the SDSS optical g-band and NIR K-band rotation curves in conjunction with the hybrid rotation curve. 
We fit the cuspy NFW and cored PIS dark matter halo profiles for each photometric band.
\begin{table*}
\hspace{-4cm}
\caption{Dark matter density parameters derived from mass-modeling using the optical g-band and NIR K band photometry using the hybrid (\HI{} + H$\alpha$) rotation curve.}
\scalebox{0.7}{
\begin{tabular}{|l|c|c|c|c|c|c|c|c|c|c|}
\hline
\hline
Model             & $c^{\textcolor{red}{(a)}}$           &$R^{\textcolor{red}{(b)}}_{200}$         & $\gamma^{*\textcolor{red}{(c)}}$ &$\frac{V_{max}}{V_{200}}^{\textcolor{red}{(d)}}$     & $\chi^{2\textcolor{red}{(e)}}_{red}$     & $\rho^{\textcolor{red}{(f)}}_{0}\times10^{-3}$  &$R^{\textcolor{red}{(g)}}_{c}$ & $\gamma^{*\textcolor{red}{(h)}}$& $\frac{R_{c}}{R_{d}}^{\textcolor{red}{(i)}}$     &$\chi^{2\textcolor{red}{(j)}}_{red}$ \\
                  &             &(kpc)             &             &              &                  &  $M_{\odot}/pc^{3}$      &(kpc)   &           &      &                                  \\
g-band            & NFW  profile&                  &             &              &                &PIS profile               &        &             &      & \\
\hline
$'diet'$ Salpeter	&$5.06\pm 0.5	$&$	84.1\pm0.3$	&$	3.8	$&$	2.3$&$	0.1	$& $	56.6\pm0.4	$ &$	2.3\pm0.01	$ &$	3.8	$&$	0.5$&	$0.1$	\\
Kroupa IMF 	&$	5.5 \pm0.06	$& $84.4\pm0.3$	& $2.7$ & $2.3$ & $0.1$ & $	66.1\pm0.6$  & $2.2\pm0.01$ &$	2.7	$&$	0.5	$&$	0.01$	\\
Free $\gamma^{*}$ 	&$3.8\pm0.1	$&$	85.0\pm0.3$	&$	6.4\pm0.2	$&$	2.3	$&$	0.05	$&$	58.0\pm0.61	$&$	2.3\pm0.01	$&$	3.6\pm0.05	$&$	0.5	$&$	0.01$	\\
Maximum Disc 	&$	0.06\pm0.04	$&$	218.4\pm9.6$	&$	14	$&$	0.07	$&$	0.3	$&$	1.4\pm0.05	$&$	19.2\pm0.8	$&$	14	$&$	4.3	$&$	0.2$	\\
Minimum Disc	&$	5.7\pm 0.06	$&$	96.1\pm0.3$	&$	0	$&$	2.0	$&$	0.1	$&$	71.9\pm0.43	$&$	2.4\pm0.0	$&$	0	$&$	0.5	$&$	0.9$	\\
\hline
K-Band          &                    &                  &        &           &                      &   &\\
\hline
$'diet'$ Salpeter	&$2.8\pm0.09	$&$	97.0\pm1.3$	&$	0.5	$&$	2.01	$&$	0.4	$&$	13.9\pm0.4	$&$	4.9\pm0.09	$&$	0.85	$&$	1.9$&$	0.14$	\\
Kroupa IMF 	&$	3.8\pm0.09	$&$	91.6\pm0.8$	&$	0.6	$&$	2.1	$&$	0.3	$&$	24.3\pm0.5	$&$	3.7\pm0.05	$&$	0.6	$&$	1.	$&$	0.08$	\\
Free $\gamma^{*}$ 	&$5.6\pm0.2	$&$	87.9\pm0.6$	&$	0.2\pm0.04	$&$	2.2	$&$	0.2	$&$	38.8\pm2.7	$&$	3.02\pm0.11	$&$	0.4\pm0.03	$&$	1.2	$&$	0.1$	\\
Maximum Disc 	&$	1.6\pm0.10	$&$	128.9\pm3.6	$&$	1.1	$&$	1.5	$&$	0.6	$&$	8.5\pm0.3	$&$	6.7\pm0.2	$&$	1.1	$&$	2.6	$&$	0.3$	\\
Minimum Disc	&$	5.7\pm0.06	$&$	96.1\pm0.3$	&$	0	$&$	2.0	$&$	0.1	$&$	71.8\pm0.4	$&$	2.4\pm0.0	$&$	0	$&$	0.9	$&$	0.1$\\
\hline
MOND                       &                                    &                                        &                &     &    &    &   & & & \\
\hline
        &$a^{\textcolor{red}{(k)}}$ &  $\gamma^{*\textcolor{red}{(l)}}$  &  $\chi^{2\textcolor{red}{(m)}}_{red}$  &                &     &    &    & & & \\
        &$ms^{-2}$  &                                    &                                        &                &     &    &    &  & &\\        
\hline
g-Band              & $0.42 \times 10^{-10}$         & $12.7 \pm 0.1 $&$0.2$  &            &    &    &    &    & &\\    
$g-Band^{a=fixed}$  &$1.2 \times 10^{-10}$                       & $4.1 \pm 0.1$      & 1.8        &    &    &    & & & &    \\
K-Band              &$0.85 \times 10^{-10}$         &1.0                  & 0.5       &    &    &    &    & & & \\             
$K-Band^{a=fixed}$  &$1.2 \times 10^{-10}$ & $0.6 \pm 0.01$    &0.8&                 &            &    &    &    &    &  \\
\hline
\end{tabular} 
}
\label{table: table 7}
\begin{tablenotes}
\item $\textcolor{red}{(a)}$: Concentration parameter of the NFW profile
\item $\textcolor{red}{(b)}$:  Radius at which the mean density equal to 200 times the critical density.
\item $\textcolor{red}{(c)}$:  Mass to light ratio derived using population synthesis models or estimated as a free parameter. 
\item $\textcolor{red}{(d)}$:  Ratio of the asymptotic velocity to the  velocity at $\frac{V_{200}}{\kms}=0.73\frac{R_{200}}{kpc}$\citep{navarro1997universal}
\item $\textcolor{red}{(e)}$:  Reduced chi-square value corresponding to the fit. 
\item $\textcolor{red}{(f)}$:  The central dark matter density of the PIS dark matter halo model
\item $\textcolor{red}{(g)}$:  The core radius of the PIS dark matter halo model
\item $\textcolor{red}{(h)}$:  Mass to light ratio derived using population synthesis models or estimated as a free parameter. 
\item $\textcolor{red}{(i)}$:  Ratio of the core radius and the disc scalelenght.
\item $\textcolor{red}{(j)}$:  Reduced chi-square value corresponding to the fit.
\item $\textcolor{red}{(k)}$:  Acceleration per length in MOND.
\item $\textcolor{red}{(l)}$:  Estimated Mass to light ratio in MOND.
\item $\textcolor{red}{(m)}$:  Reduced chi-square corresponding to the fit.
\end{tablenotes}
\end{table*}

\begin{figure*}
\hspace{-3cm}
\resizebox{210mm}{200mm}{\includegraphics{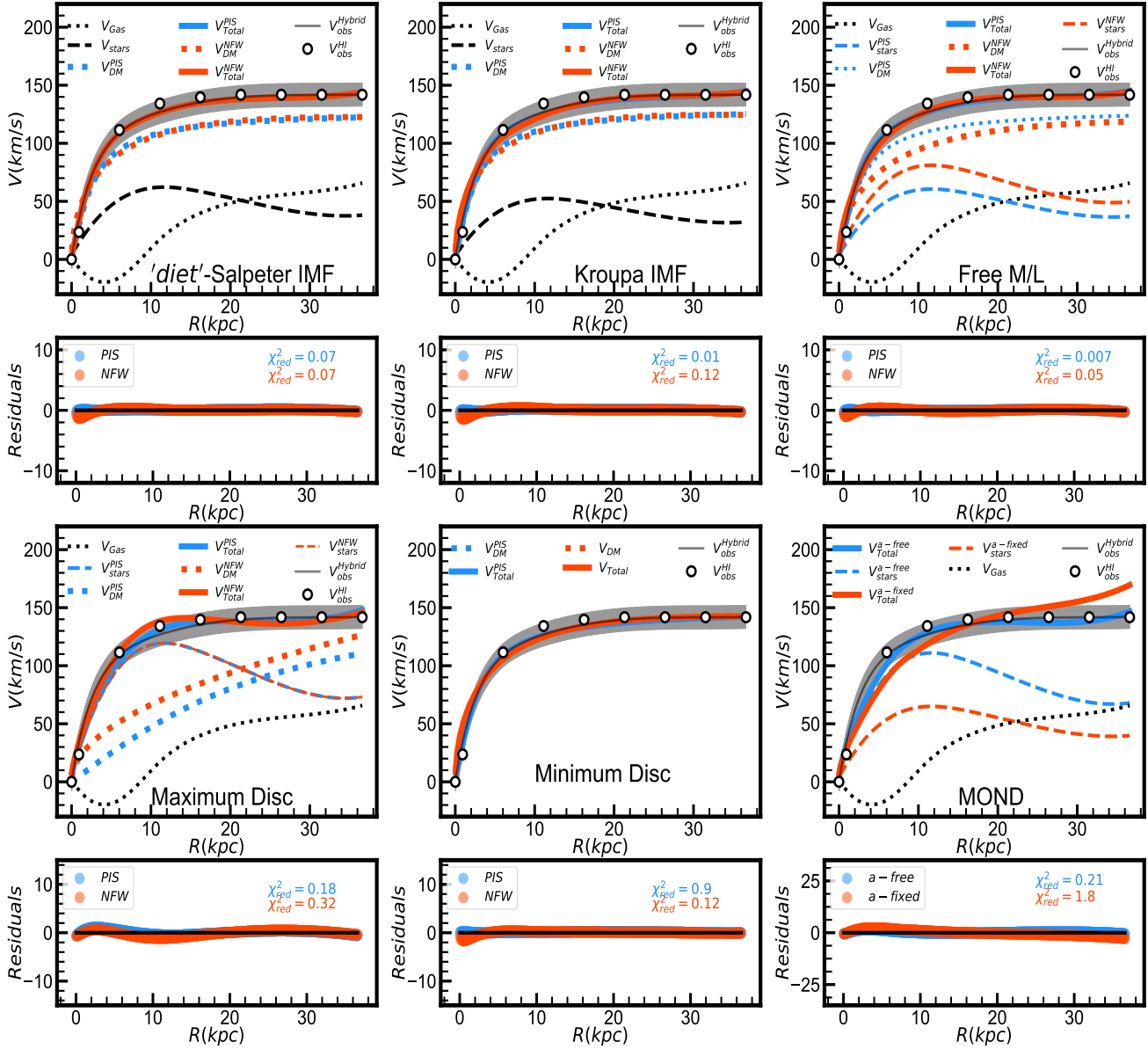}} 
\vspace{-2cm}
\caption{We present the mass model of the galaxy FGC 1440 derived using SDSS g-band photometry. The mass models are constrained using the
hybrid rotation curve.} 
\end{figure*}

\begin{figure*}
\hspace{-3cm}
\resizebox{210mm}{200mm}{\includegraphics{./chap3/MM_K_hybrid_smooth_uniform.pdf}} 
\vspace{-2cm}
\caption{We present the mass model of the galaxy FGC 1440 derived using UKIDSS K-band photometry. The mass models are constrained using the hybrid rotation curve.}
\end{figure*}

Figure 4.16 shows mass models derived using the g-band photometry and the hybrid rotation curves.
The reduced chi square$\rm (\chi^{2}_{red})$ for the cored pseudo-isothermal dark matter halo is lower than the cuspy NFW dark matter halo. 
Only in the model with $\rm 'diet'-Salpeter$ IMF do both halos provide similar $\rm (\chi^{2}_{red})$ values. For mass models employing PIS halo, 
the Kroupa IMF has a smaller $\rm (\chi^{2} _{red})$ than the $'diet'-Salpeter$ IMF, indicating dark matter dominate mass distribution. 
In PIS halo models with a free $\rm \gamma^{*}$, $\rm \gamma^{*}$ values tend to match the $\rm 'diet'-Salpeter$ IMF. 
$\rm \gamma^{*}$ tends to be higher than values predicted by stellar population synthesis models in the case of NFW halo.
In maximum disc models, we scaled the stellar rotation curve by 14 to maximize the stellar disc's contribution to the total mass. 
In minimum disc models, we set the baryonic contribution to zero.
Figure 4.17 shows the mass models using the stellar rotation curve derived using the photometry in K-band.
$\rm \chi^{2}_{red}$ is lower for the mass model with PIS dark matter halo than NFW.
Models with Kroupa IMF have a lower $\chi^{2}_{red}$ value than $\rm 'diet'-Salpeter'$ IMF. In mass models with $\rm \gamma^{*}$ kept as a free parameter, 
$\rm \gamma^{*}$ is lower than $\rm \gamma^{*}$ determined using Kroupa IMF, in case of both PIS halo and NFW halo. 
This indicates that, in the K-band, the mass models prefer IMF, which reduces the contribution of the stellar disc and increases that of the dark matter halo. 

A compact dark matter halo is defined as one in which $\rm R_{c}/R_{d}<2$, \citep{banerjee2013some}. In table 4.7, $\rm R_{c}/R_{d}$ is less than 2 for all 
models except the maximum disc case in K-band. We note that, the compactness parameter for the dark matter halo is thus not independent of the IMF, 
as the maximum disc models in a particular photometric band have a larger core radius and less compact dark matter halo. Constant IMF models lead 
to concentration parameters ranging from 2.8 (diet-Salpeter K-band) to 5.5 (Kroupa IMF in g-band). The IMF models that prefer higher disc masses have 
smaller concentration and compactness parameters. The mass models with a larger value of $\rm \gamma_{*}$ have a higher core radius, indicating that the definition of the compactness parameter is not 
independent of the choice of the IMF.

The scaling relation  between the asymptotic rotation velocity $\rm V_{max}$ and the concentration parameter given by \cite{bottema2015distribution} as 
$\rm c_{exp}=55.5(V max/[km s-1])-0.2933$, yields $\rm c=12.97$. We point out that the concentration parameters derived in this study are significantly 
lower than what is expected from the above scaling relation. The scaling between $\rm V_{max}$ and $R_{200}$ yields $\rm R_{200}=146kpc$, which is closer to the 
$\rm R_{200}$ values obtained for the maximum disc models, where $\rm R_{200}=0.0127(V_{max})^{1.37} c_{exp}$. We compare our results with theoretical predictions 
for the concentration parameter based on $\rm V_{max}/V_{200}$. \cite{dutton2014cold} define the relation between $\rm V_{max}/V_{200}$ and concentration parameter $\rm c$ as $\rm V_{max}/V_{200}= 0.216c_{200}/f(c_{200})$, where $\rm f(c_{200})= ln(1+c) - c/(1+c)$.  We find that the value of $\rm V_{max}/V_{200}$ is close to 1, using the values of the 
concentration parameters from Table 4.7. We use the relationship given by \cite{donato2004cores} $\rm log(R_{c})=(1.05 \pm 0.11) log(R_{d}) + (0.33 \pm 0.04)$ to understand the correlation 
between the size of the dark matter core radius and the scalelength of the disc. With the K-band disc scalelength $\rm R_{d}$ = 2.58 kpc,  we get a core radius 
equal to 5.78 kpc, which is closer to the diet-Salpeter and the maximum disc cases. In the same way, the g-band scalelength gives us a core radius equal to 10.72 kpc.
We compare the parameters $\rm V_{\infty}=\sqrt{4 \pi G \rho_{0}R^{2}_{c}}$ for the PIS halo and $\rm R_{s}=R_{200}/c$ of FGC 1440 with other superthin galaxies in the literature, 
since $\rm R_{s}$ and $\rm V_{\infty}$ make up a single parameter that includes both the fitted parameters. In the study of three superthins, \cite{banerjee2017mass} find 
that $\rm V_{\infty}$ is equal to 110 \kms for UGC7321, 112 \kms for IC5249, and 99 \kms for IC2233. 
In another study, \cite{kurapati2018mass} finds that for FGC 1540, $\rm V_{\infty}$ is equal to 82.7 \kms. For FGC 1440, 
we find that $\rm V_{\infty}=135$\kms. $\rm R_{s}$ is equal to 8.55 and 22.6 for UGC 7321 and IC 5249. \cite{kurapati2018mass} find that $\rm R_{s}$ equal to 5.25 for FGC 1540. 
We find that $\rm R_{s}$ is equal to 24.3 for FGC 1440.

\subsubsection{Mass models in MOND}
The last panels in Figures 4.16 and 4.17 show mass models derived using MOND phenomenology, see Section 1.4.6. Keeping $\rm a$ and $\rm \gamma^{*}$ as free parameters, 
we find $\rm a=0.42 \times 10^{-10} ms^{-2}$ and $\rm \gamma^{*}=12.7$ in g-band. Whereas in K-band $\rm a=0.85 \times 10^{-10} ms^{-2}$ and $\rm \gamma^{*}=0.97$. $\rm \gamma^{*}$ 
tends to maximize the disc mass when both $\rm a$ and $\rm \gamma^{*}$ are kept as free parameters, i.e., closer to the maximum disc case of dark matter models. 
Fixing $\rm a=1.2 \times 10^{-10}ms^{-2}$ and varying $\rm \gamma^{*}$ gives $\rm \gamma^{*}$ equal to 4.14 and 0.65 in g-band and K-band, which are closer to population 
synthesis models.

\section{Vertical structure of FGC 1440}
\begin{figure*}
\hspace{-13.5mm}
\resizebox{190mm}{70mm}{\includegraphics{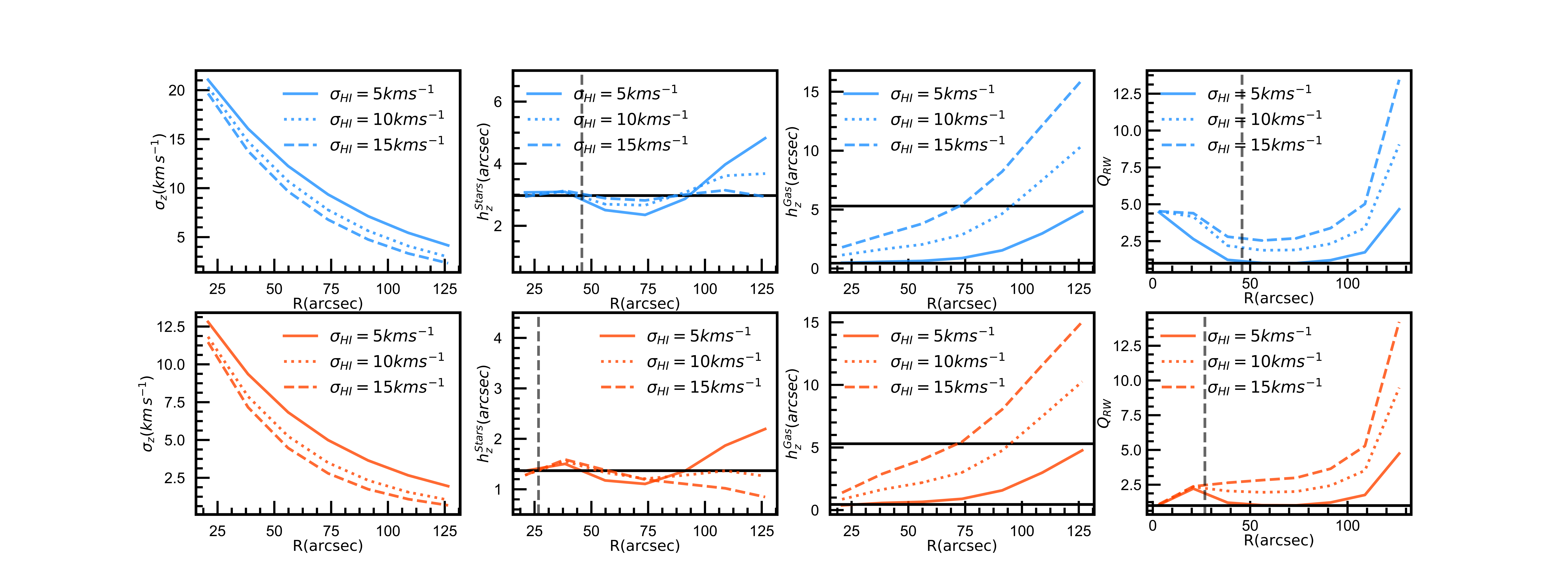}} 
\caption{ The plots show the vertical velocity dispersion $\sigma_{z}$, the modeled stellar $\rm(h^{Stars}_{z})$ and \HI{} scaleheights $\rm(h^{Gas}_{z})$, and 
 the stability parameter $\rm (Q_{RW})$. The top panel in blue color depicts the results for the g-band, and the lower panel in red, shows the results in K-band. The vertical dashed line 
 marks the 3$\rm R_{d}$ in g-band and K-band, respectively. The horizontal black line in the panel depicting $\rm h^{Stars}_{z}$ marks the observed stellar scaleheight. The horizontal
 line in the plot showing $\rm h^{Gas}_{z}$ marks the upper and the lower limit on the \HI{} scaleheight derived using the tilted ring modeling.} 
\end{figure*}

The galaxy disc is modeled as a coplanar and coaxial system of $\rm stars + gas$. The stars and the gas are 
gravitationally coupled under an external dark matter halo force field. The stellar vertical velocity dispersion $\rm (\sigma_{z})$ is determined by
solving the multi-component model using the methods detailed in \cite{10.1093/mnras/stab155} and \cite{komanduri2020dynamical}, see Section 1.1.3. 
The input parameters for our models are the PIS dark matter profile and the stellar surface density scaled by Kroupa IMF along with the \HI{} surface density and 
\HI{} dispersion. Since the \HI{} dispersion is confined between 
$\rm 5\kms \sigma \leq \HI{} \leq 15 \kms$ in the 3-D models of the data cube, we model the 
stellar vertical dispersion by fixing the \HI{} dispersion at 5 \kms, 10 \kms, and 15 \kms at all radii. 
The stellar dispersion is represented using an exponential function $\rm \sigma_{z}(R)=\sigma_{0}e^{\frac{-R}{\alpha R_{d}}}$, where 
$\rm \sigma_{0}$ is the central value of the vertical velocity dispersion, and $\rm \alpha$ is the steepness parameter.
The values of $\rm \sigma_{0}$ and $\alpha$ in g and K-band, estimated by solving the multi-component Jeans equations, are given in Table 4.8.
The central value of the stellar dispersion is not affected by the values of the $\rm \sigma_{\HI}$ in the allowed range, but the steepness parameter 
changes for different values of \HI{} dispersion. Steepness parameter $\rm (\alpha)$ varies from $\rm 3.2 - 4.2$ in the g-band to $\rm 4.2 - 6.3$ in the K band for 
different values of \HI{} dispersion. Scaleheight of stars derived from the two-component model is found to be consistent with the observed scaleheight up to 
$\rm 3R_{d}$. The \HI{} scaleheight obtained from the model is confined between the limits set by tilted ring modeling $\rm (0.45\farcs \leq h_{z} \leq 5.3\farcs$,
shown in panel 3 of Figure 4.18.

\begin{table}
\begin{minipage}{110mm}
\hfill{}
\caption{Modeled values of vertical stellar dispersion.}
%\centering
\begin{tabular}{|l|c|c|c|c|}
\hline
\hline
Parameter                  &           g-Band       &             & K-Band         &            \\
\hline
                           &         $\sigma_{0}$   &  $\alpha$   &  $\sigma_{0}$  &  $\alpha$  \\  
			   &          \kms          &             &   \kms         &            \\
\hline    
$\sigma_{HI}$=5  \kms      &          29.0          &4.2         &18.6            &6.3   \\   
$\sigma_{HI}$=10 \kms      &          29.7          &3.6          &19.1            &4.8  \\
$\sigma_{HI}$=15 \kms      &          29.9          &3.2         &20.1            &4.2   \\ 
\hline
\end{tabular}
\hfill{}
\label{table: table 8}
\end{minipage}
\end{table}

\section{Disc Heating}
Bars, spiral arms, and globular clusters are important disc heating agents. Bars and spiral arms heat the disc radially, while globular clusters heat up the galactic disc 
isotropically in both radial and vertical directions \citep{aumer2016age, jenkins1990spiral, grand2016spiral, saha2014disc}. Superthin galaxies have lower vertical stellar velocity dispersion than the stars in the Milky Way. The ratio of vertical velocity dispersion to total rotation velocity ($\rm \frac{\sigma_{z}}{V_{Rot}}$) for superthin galaxies is comparable to stars in the Milky Way's thin disc, indicating 
that superthin galaxies are dynamically cold \citep{10.1093/mnras/stab155}. To quantify the disc heating, we compare FGC 1440's $\rm \frac{\sigma_{z}}{V_{Rot}}$ to 
that of other superthin galaxies. Using the velocity dispersion constrained using g-band and K-band data, we find that $\rm \frac{\sigma_{z}}{V_{Rot}}$=0.125 for 
FGC 1440, compared to 0.1 for other superthin galaxies, save for IC2233, which has $\rm \frac{\sigma_{z}}{V_{Rot}}=0.07$ and UGC00711 which 
has $\rm \sigma_{z}/V_{Rot}=0.2$. Inspite of having an extra-ordinarily thin stellar disc, the values of $\rm \sigma_{z}/V_{Rot}$ are comparable to 
previous studied normal superthin galaxies

\section{Disc dynamical stability}
\cite{garg2017origin} show that in the absence of dark matter, the superthin disc will be highly unstable and would be subject to 
the growth of axisymmetric instabilities. In \cite{10.1093/mnras/stab155}, it was shown for a sample of superthin galaxies that the median 
stability is higher than the spiral galaxies analyzed by \cite{romeo2017drives}. This section compares 
the stability levels in FGC 1440 to previously studied superthin galaxies. Using the multi-component stability parameter from 
\cite{romeo2011effective}, we compute FGC 1440's dynamical stability using the method detailed in Section 1.1.5 in the introduction.
We compute the stability parameter using $\rm \sigma_{\HI}=5 \kms,10 \kms,15 \kms $ in both
g-band and K-band. $\rm Q^{min}_{RW}$ in g-band is 1.0, 1.9, 2.5, corresponding to $\rm \sigma_{\HI}=5 \kms, 10 \kms, 15\kms$. In g-band $\rm Q_{RW} > 2$ for $\rm R<3R_{d}$, 
indicating that FGC 1440 is stable. However, in the K-band, it is closer to marginal stability. In g-band and K-band, the galaxy disc is 
closer to minimal stability levels for $\rm \sigma_{\HI}=5\kms$ than for $\sigma_{\HI}=10\kms$ or $\sigma_{\HI}=15\kms$ for $R<3R_{d}$. Further, we find that
$\rm Q_{RW}(R<3R_{d})$ is lower than the median value of 5.5 \citep{10.1093/mnras/stab155} for previously examined superthin galaxies and is closer to the median value of 
$\rm Q_{RW}$ for spiral galaxies equal to $\rm 2.2 $ by \citep{romeo2017drives}.

Therefore, despite being one of the flattest galaxies, the minimum stability levels in FGC 1440 are lower than that 
of previously studied superthin galaxies and is closer to the stability levels of the nearby spiral galaxies. 

\begin{figure*}
\hspace{-13.5mm}
\resizebox{180mm}{65mm}{\includegraphics{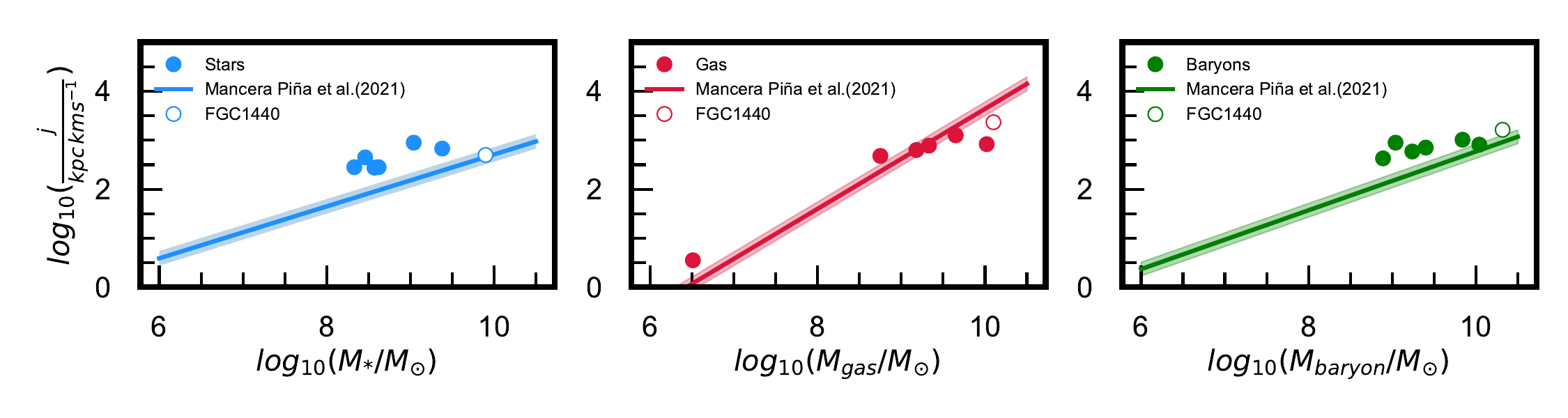}} 
\caption{The plot shows the j - M relation, where 'j' is the specific angular momentum and 'M' is the mass for superthin galaxies (filled-points) and compares it with that of FGC 1440 (open-point). The straight line 
shows the best-fitting Fall relation, and the shaded region indicates the intrinsic scatter obtained by \protect \cite{2021A&A...647A..76M}  for disc galaxies.} 
\end{figure*}
\section{Specific angular momentum of the stellar disc}
The Fall relation relates the mass (M) of the disc galaxy to its specific angular momentum (j). The Fall relation is well established for disc galaxies with diverse
morphologies and masses \citep{posti2019galaxy, 2021A&A...647A..76M, marasco2019angular}. Studies pertaining to superthin galaxies and low surface brightness 
galaxies indicate that low surface brightness galaxies have a higher specific angular momentum than typical 
disc galaxies, suggesting that high specific angular momentum plays a key role in regulating the superthin disc structure. 
We calculate $\rm j-M$ for FGC 1440 to compare its specific angular momentum with a sample of previously studied superthin galaxies and disc galaxies \citep{jadhav2019specific}. 
We compute $\rm j_{*}-M_{*}$, $\rm j_{g}-M_{g}$ and $\rm j_{b}-M_{b}$ relation for the stars  $\rm (*)$, gas $\rm (g)$, and for baryons $\rm (b)$ respectively.
We find that $\rm log_{10}\frac{j_{*}}{kpc \, \kms}$ is equal to 2.7 for a stellar mass of 
$\rm 7.9 \times 10^{9} M_{\odot}$, and $\rm log_{10}\frac{j_{g}}{kpc \, \kms}$ is equal to 3.4 for a gas disc mass of $\rm 1.3 \times 10^{10} M_{\odot}$. 
In Figure 4.19, we compare,
Fall relation obtained for FGC 1440 (open points) with other superthin galaxies (filled points). We also plot $\rm j-M$ relation obtained by \cite{2021A&A...647A..76M} 
$\rm (straight-line)$ for a larger sample of disc galaxies. FGC 1440 closely matches the regression line obtained by \cite{2021A&A...647A..76M} for the 
$\rm log_{10}(j_{*}) - log_{10}(M_{*})$, $log_{10}(j_{b}) - log_{10} (M_{b})$ and the $\rm log_{10}(j_{g}) - log_{10}(M_{g})$ relation. Unlike previously studied 
superthin galaxies, which have a higher stellar specific angular momentum than the ordinary spiral galaxies for a given stellar mass, 
FGC 1440 complies with the $\rm j_{*}-M_{*}$ relation obtained for ordinary spiral galaxies. Hence, the specific angular momentum may not play a fundamental role in driving
the superthin discs.

\section{Conclusions}

\begin{itemize} 
 \item We fit the busy function to the \HI{} spectrum and find the velocity widths are 295 \kms and 306 \kms, respectively. The total flux density is 10.46 $\rm Jy \kms$. 
 FGC 1440 is an intermediate-mass galaxy with mass between dwarfs and spirals like Milky Way.
 
 \item The moment 0 and moment 1 maps of the \HI{} data cube indicate that the $\rm \HI{}$ disc is warped.

 \item We use $\rm \HI{}$ data cube to construct the tilted rings model of the \HI{} emission and estimate kinematic parameters using TiRiFic and FAT. 
 A warped model with 90$\rm ^{\circ}$ for the inner rings and 85$\rm ^{\circ}$ for the outer rings describes the observed \HI{} dis structure of FGC 1440.
 The position angle is equal to 53.6$\rm ^{\circ}$. The rotation curve of FGC 1440 derived from 3D-tilted ring models rises slowly, with an asymptotic velocity equal to 141.8 \kms.
 
 \item By manually comparing PV diagrams at different offsets, we find that the \HI{} velocity dispersion is constrained between $\rm 5\kms \leq \sigma \leq 15 \kms$ and the scaleheight is limited by the resolution of the synthesized beam, $\rm h_{z} \leq 5.3\farcs$.
 
 \item By comparing the data with models having different inclinations and scaleheights, we conclude that \HI{} is confined in a thin disc 
       possibly, warped along the line of sight, although a centrally localized thick \HI{} disc can not be ruled out.
 
 \item We employ stellar photometry and total rotation velocity to derive mass models using SDSS g-band and NIR-K band photometry. 
 The mass models constructed with a cored pseudo-isothermal dark matter halo and Kroupa initial mass function explain the observed rotation curve best. 
 The Kroupa IMF gives lower disc mass than the diet Salpeter IMF, indicating that the mass models predict higher dark matter content in these galaxies.

 \item We also derive mass models in Modified Newtonian paradigm (MOND). In models where acceleration and mass-to-light ratios are kept as free parameters, 
      acceleration is less than universal acceleration scale predicted by MOND $\rm 1.2 \times 10^{-10},ms^{-2}$ and mass-to-light ratio 
      tends to maximize disc mass. $\rm \gamma^{*}$ tends to values predicted by stellar population synthesis models when acceleration is fixed at 
      $\rm 1.2 \times 10^{-10} \,ms^{-2}$.
      
 \item We constrain g-band and K-band vertical velocity dispersion using the measured stellar scaleheight. In g-Band, central vertical velocity dispersion is equal to 
 29.0 \kms, and in K-band the same is equal to 18.6 \kms. Vertical dispersion values are comparable to values obtained for previously 
 studied ordinary superthin galaxies \citep{10.1093/mnras/stab155}, indicating presence of a cold stellar disc     
 
 \item Using the two-component stability criterion derived by \cite{romeo2011effective}, we calculate the disc dynamical stability of FGC 1440. 
 The value of $\rm Q_{RW}>1$ at  $\rm R<3R_{d}$, indicates that FGC 1440 is stable against axisymmetric instabilities. The value of Q for FGC 1440 is lower than the
  median value obtained for superthin galaxies (\citep{10.1093/mnras/stab155} and is closer to the median $\rm Q_{RW}$ for spiral galaxies 
  \citep{romeo2017drives}. Despite extra-ordinarily large axis ratios, $\rm \frac{\sigma_{z}}{V_{Rot}}=0.2$ in g-band and $\rm \frac{\sigma_{z}}{V_{Rot}}=0.125$ 
  in K-band are comparable to previously studied ordinary superthin galaxies. 
 
 \item FGC 1440 follows the $\rm log_{10}(j_{*})- log_{10}(M_{*})$,  $\rm log_{10}(j_{b}) - log_{10}(M_{b})$ and $\rm log_{10}(j_{g}) - log_{10}(M_{g})$ relations 
 akin to ordinary disc galaxies. The specific angular momentum of stars, gas, and baryons in FGC 1440 is comparable to spiral galaxies with similar mass unlike
 previously studied superthin galaxies, which have a higher j values compared to ordinary disc galaxies of the same mass.
 
\end{itemize}

\thispagestyle{empty}

\thispagestyle{empty}
\chapter[ \HI{} 21cm observations and dynamical modeling of the thinnest galaxy: FGC 2366]{\fontsize{50}{50}\selectfont Chapter 5
\footnote[1]{Adapted from \textbf{K Aditya}, Arunima Banerjee, Peter Kamphuis, Aleksandr Mosenkov, Dmitry Makarov, Sviatoslav Borisov
 \HI{} 21cm observations and dynamical modelling of the flattest/thinnest galaxy known: FGC 2366, \textcolor{blue}{under revision} }}
\chaptermark{ \HI{} 21cm observations and dynamical modeling of the thinnest galaxy: FGC 2366}

\textbf{\Huge{ \HI{} 21cm observations and dynamical modeling of the thinnest galaxy: FGC 2366}}

\section*{Abstract}
Superthin galaxies have a major-to-minor axis ratio (a/b) of $\rm 10\leq a/b \leq 22$ and lack a discernable bulge component.
We present GMRT \HI{} 21cm radio-synthesis observations of FGC 2366, the thinnest known galaxy with an a/b ratio of 21.6. We use the 3D tilted-ring model to 
derive the structural and kinematic properties of the \HI{} disc. We find an asymptotic rotation velocity equal to 100 \kms and a total \HI{} mass of $\rm 10^{9} M_{\odot}$. We derive the mass models and self-consistent 
models of FGC2366 using \HI{} data and optical photometry. We find that FGC 2366 has a compact dark matter halo, the ratio of the 
core radius of the pseudo-isothermal dark matter halo to the exponential stellar disc scalelength is $\rm (R_{c}/R_{d})$ = $0.35 \pm 0.03$. 
The minimum of vertical-to-radial stellar velocity dispersion $\rm (\sigma_{z}/\sigma_{R})_{min}$ = $0.42 \pm 0.04$,  
a high value of 2-component (star + gas) disc dynamical stability parameter $\rm Q_{RW} = 7.4 \pm 1.8 $ at 1.5$R_{d}$  and finally 
specific angular momentum $\rm \sim$ $\rm{log}_{10}(j_{*})$ equal to  $\rm  2.67 \pm 0.02 $ for the stellar  
mass $\rm \sim$ $\rm{log}10(M_{*}/M_{\odot})=9.0$. We do a Principal Component Analysis of the following dynamical 
parameters and $a/b$ to find the main physical mechanism driving the vertical structure of superthin galaxies [IC 2233, FGC 1540, UGC 7321, FGC 1440, \& FGC 2366]. 
We see that the first two principal components explain about 80$\%$ of the variation in the data and that $\rm a/b$, $\rm Q_{RW}$, and $\rm V_{\rm rot}/(R_{c}/R_{d})$ 
make a major contribution to the first two principal components. This implies that the superthin stellar discs structure can be explained by high values of the dynamical stability and 
the mass of dark matter halo in the inner galactocentric radii.

\section{Introduction}
Superthin galaxies are bulgeless edge-on disc galaxies with low surface brightness in the $\rm B$-band $\rm \mu_{B}(0)>22.7,mag \, arcsec^{-2}$ and 
have a major-to-minor axes ratio (a/b) $\rm 10 \leq a/b \leq 22$ \citep{bothun1997low, mcgaugh1996number,karachentseva2016ultra}. 
Properties of superthin galaxies have been reviewed in \cite{2000bgfp.conf..107M} and \cite{kautsch2009edge}. The Revised Flat Galaxy Catalogue (RFGC) is the principal catalog of edge-on disc galaxies \citep{1999BSAO...47....5K}. It comprises of 4,444 edge-on galaxies 
with $\rm a/b$ > $\rm 7$, 1150 galaxies with $\rm a/b$ > $10$, and only 6 extremely thin galaxies with $\rm a/b$ > $20$, indicating a paucity of 
thin galaxies. Figure 5.1 shows the distribution of $\rm a/b$ for the galaxies in the RFGC catalog. One can see that extremely thin galaxies form the tail of the distribution of galaxies in RFGC.

\cite{goad1981spectroscopic} coined the term \emph{superthin} for galaxies with $\rm a/b$ between 10 and 20. \cite{karachentseva2016ultra} classified
RFGC galaxies with $\rm a/b >10$ as \emph{ultra-flat} galaxies and uses the term interchangeably with \emph{superthin} galaxy, which is also the nomenclature adopted by 
\cite{kautsch2009edge}. \cite{matthews1999extraordinary} defines \emph{superthin galaxies} as galaxies with an $\rm h_{z}/R_{d} \leq 0.1$,
\citep{o2010dark, banerjee2010dark, peters2017shape, kurapati2018mass, aditya2022h}. We use the term \emph{extremely thin galaxies} for 
RFGC galaxies with $\rm a/b > 20$ and \emph{superthin} for $\rm 10\leq a/b\leq 20 $. 

Galaxy interactions strongly regulate the structure of galaxies, 
including mergers and internal processes like heating by bars and spiral arms \cite{martinez2015contribution, saha2010effect, benson2004heating, grand2016vertical}. 
The role of these physical mechanisms in determining the morphology of the Milky Way has been well-studied \citep{rix2013milky, brook2012thin, minchev2015formation}.
Also, \cite{khoperskov2017disk} shows that normal disc galaxies with a small counter-rotating component can heat up the stellar disc. Observations reveal that
the superthin galaxies and the extremely thin galaxies have not been affected by the mechanisms that heat the discs of ordinary disc galaxies. Even though Figure 1 
shows that superthin galaxies are ubiquitous, the number of superthin and extremely thin galaxies produced in the $\rm Lambda-CDM$ simulations is very small. 
In a recent study, \citep{vogelsberger2014introducing} measured the sky-projected aspect ratio distribution in the $\rm \Lambda-CDM$  
\citep{pillepich2018simulating, 2015MNRAS.446..521S}  simulation and found that these simulations lack galaxies with intrinsically thin discs. 
Also, see \citep{2017MNRAS.467.2879B}.

\begin{figure*} 
\resizebox{110mm}{80mm}{\includegraphics{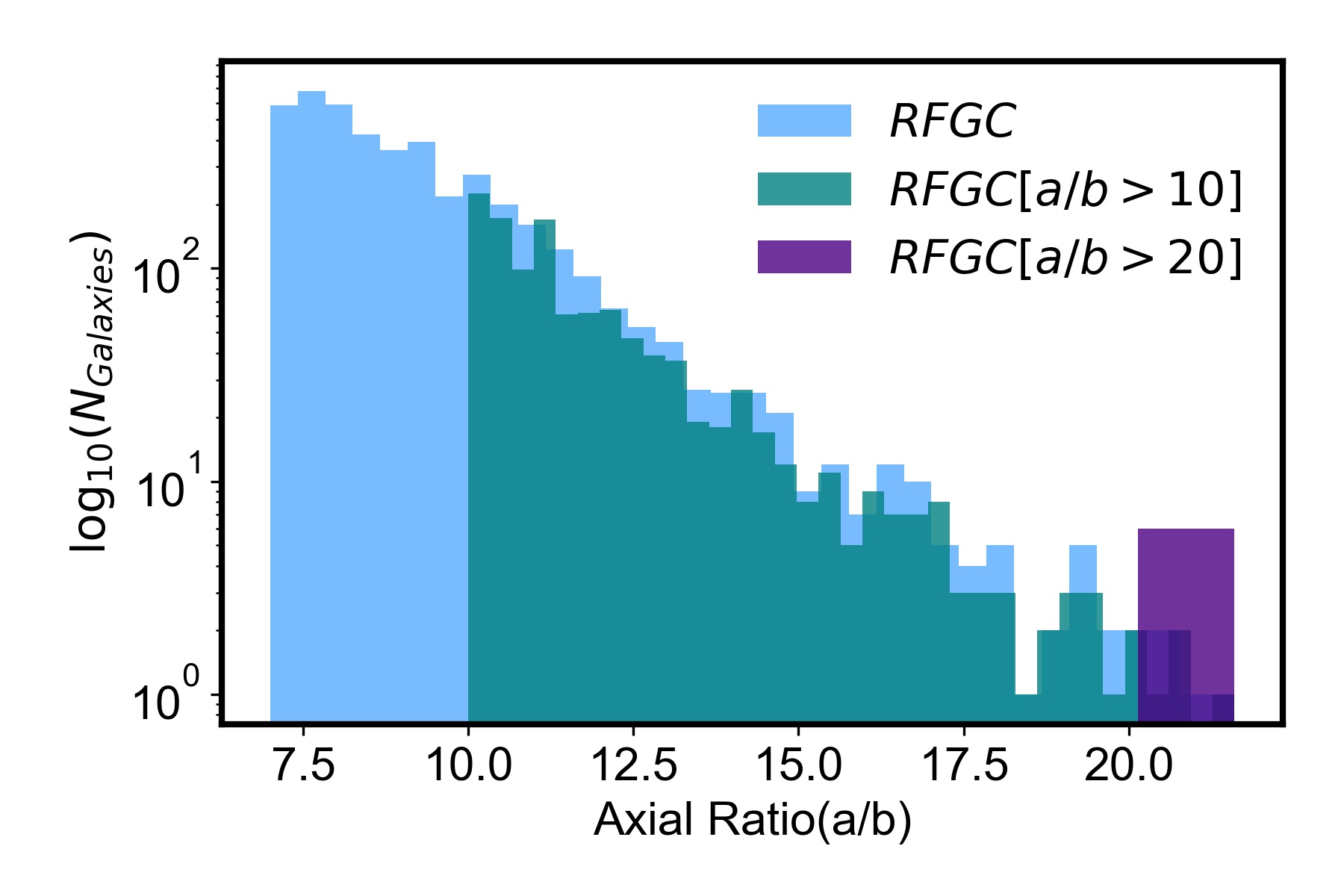}} 
\caption{Histogram showing the distribution of the RFGC galaxies as a function of the major-to-minor axes ratio of their optical discs}   
\end{figure*}

Our understanding of superthin galaxies comes from detailed dynamical modeling of a small sample of superthins for which the stellar photometry 
and \HI{} 21 cm radio-synthesis measurements were available. \HI{} 21 cm observations of edge-on disc galaxies impose a two-fold constraint on the 
net gravitational potential of the galaxy: the rotation curve constrains the radial derivative of the potential, and the vertical 
thickness of the gas constrains the vertical derivative \citep{1995AJ....110..591O,narayan2002origin}.
\cite{zasov1991thickness} in their studies show that a massive dark matter halo stabilizes a superthin disc. \cite{ghosh2014suppression} show that, 
because of its dark-matter halo, the prototypical superthin galaxy UGC7321 can withstand both the axisymmetric and 
non-axisymmetric local perturbations. \cite{banerjee2013some} claimed that UGC 7321 is dominated by the dark matter at all galactocentric radius, 
and \cite{banerjee2017mass} found that superthin galaxies have compact dark matter halos with $\rm R_{c}/R_{d}<2$, where $\rm R_{c}$ is the core radius 
of a pseudo-isothermal dark matter halo, and $\rm R_{d}$ the exponential scale length of the stellar disc. 

According to \cite{2013MNRAS.431..582B}, a compact dark matter halo is critical for regulating the 
thin vertical structure of the stellar disc. Further, the superthin galaxy's 
razor-sharp vertical structure indicates the presence of an ultra-cold stellar disc. Furthermore, the ultra-cold razor-thin vertical structure may be a 
direct consequence of an anisotropic stellar velocity ellipsoid with less heating in the vertical direction than in the radial 
direction  \citep{khoperskov2003minimum, gentile2015disk, grand2016vertical}. 
Due to their edge-on orientation, superthin galaxies cannot be directly studied for the vertical component of 
their stellar velocity dispersion. High-resolution observations of stars in these galaxies impose strict constraints on the vertical structure of stars in these galaxies \citep{seth2005study,abe1999observation, uson2003hi,kregel2004structure, 
matthews2003high, sarkar2019flaring, sarkar2018constraining, sarkar2020vertical, sarkar2020general, sarkar2020vertical, sarkar2020general}. 
We use the observed scaleheight of the stellar disc to determine the stellar vertical velocity dispersion as a function of radius. 
\cite{10.1093/mnras/stab155} derive the vertical velocity dispersion of stars for a sample of five superthin galaxies which have $8 \leq a/b \leq 15$ 
using the observed stellar scaleheight as a constraint. They further showed 
that the ratio of vertical velocity dispersion lies in between 0.3 to 0.5 for superthin galaxies compared to 0.5 for the ordinary disc galaxies.
The ratio of the vertical velocity dispersion to rotational velocity of superthin galaxies is comparable 
to the stars in Milky Way's thin stellar disc. \cite{jadhav2019specific} show that a high planar-to-vertical axis ratio of a stellar disc could be 
due to the high specific angular momentum of the stellar disc. Superthin galaxies have a relatively high specific angular momentum for a given mass compared 
to normal disc galaxies.

In the RFGC catalogue, there are only six galaxies with $\rm a/b > 20$: FGC 2366 (21.6), FGC 2288 (20.8), FGC 1194 (20.6), FGC 1440 (20.4), FGC 1818 (20.2) and FGC 1170 (20.1). 
FGC 1440, which as studied in the earlier chapter and FGC 2366 are the thinnest galaxy known, which have been chosen for the \HI{} 21 cm radio-synthesis observations. We modeled the \HI{} rotation curve and surface 
density using 3D tilted ring modeling (see Section 1.4.5). Further, in conjunction with stellar photometry, we derive the mass models and extensive dynamical models 
of these galaxies. We intend to understand if the structure and kinematics of extremely thin galaxies are governed by the same formation and evolution processes as 
typical superthin galaxies, and if so, to establish the important dynamical factors responsible for their superthin stellar vertical structure. 
This is a follow-up to the analysis done by \cite{aditya2022h} on FGC 1440. Similar to the trend found in previously studied superthins, we found that 
the other thin galaxy FGC 1440 hosts a compact dark matter halo, i.e. the core radius to the disc scalelength $\rm \sim$ $R_{c}/R_{d} \sim 0.5$, 
has low values of the minimum of vertical-to-radial stellar velocity dispersion $\rm (\sigma_{z}/\sigma_{R}) \sim 0.42$, and a two-component stability parameter 
$\rm Q_{RW} \sim 4.0$ at 1.5$\rm R_{d}$. FGC 1440, however, FGC 1440 agrees with the specific angular momentum mass relation for the stars obtained for ordinary disc galaxies.

In this study, we observe the neutral \HI{} distribution in the thinnest known galaxy FGC 2366 using GMRT and derive its structural and kinematics parameters using tilted 
ring modeling software FAT (Fully Automated TiRiFiC) \citep{kamphuis2015fat}.
We focus on four dynamical factors that may drive the high planar-to-vertical axes ratios in these galaxies:\\ 
1) Dark matter density profile \\
2) Vertical and radial stellar velocity dispersion\\
3) Disc stability against axisymmetric perturbations \\
4)Stellar disc's specific angular momentum. \\
We compare the dynamical parameters of FGC 2366 to those of FGC1440, other previously studied superthin galaxies, and normal disc galaxies.

We use the \HI{} data and optical photometry together to derive FGC 2366's mass models to ascertain the dark matter halo's density distribution 
(refer to section 1.4.6 for details about mass modeling). Using measured stellar and \HI{} scaleheights as constraints, we build dynamical models of 
FGC 2366. We model the galaxy as a two-component system of gravitationally coupled stars and gas under the force-field of a dark matter halo to derive the 
vertical velocity dispersion \citep{banerjee2011theoretical, patra2020theoretical,patra2020h}. The details of modeling the vertical structure using the 
multi-component method is given in Section 1.1.3. Next, we determine the radial velocity dispersion using self-consistent iterative modeling 
using a publicly available toolkit called Action-Angle based Galaxy Modeling Architecture (AGAMA) \citep{vasiliev2018agama} (see Section 1.1.4). 
We use the vertical velocity dispersion ($\rm \sigma_{z}$) from the multi-component model as an input in AGAMA and compute $\rm \sigma_{z}/\sigma_{R}$. $\rm \sigma_{z}/\sigma_{R}$ which 
measures the relative relevance of disc heating agents in the vertical and radial directions \citep{2001ASPC..230..221M} and may provide crucial 
clues for understanding FGC 2366's thin disc structure. We calculate the multi-component dynamical stability parameter $\rm Q_{RW}$ suggested by \cite{romeo2011effective} to
assess if the galaxy disc is unstable to local axisymmetric instabilities and thus to local structural distortions. 
In the current galaxy formation and evolution model, proto galaxies acquire angular momentum from nearby galaxies in the local environment
\citep{hoyle1953fragmentation, barnes1987angular}. Thus, the isolated environment may play a key role in regulating the thin disc structure in these galaxies.
For a given stellar mass, superthin galaxies have a higher specific angular momentum than ordinary spiral galaxies, indicating that higher specific angular 
momentum acquired by stellar disc may solve the puzzle of superthin vertical structures in these galaxies \cite{jadhav2019specific}. 

\section{FGC 2366}
FGC 2366 is part of the Revised Flat Galaxy Catalogue (RFGC) \citep{1999BSAO...47....5K}.
The major-to-minor axis ratio of FGC 2366 is 21.6, making it the thinnest or the flattest galaxy.
The Nancy Radio Telescope and the Green Bank Telescope were used to observe 472 late-type edge-on spiral galaxies, including FGC 2366 \citep{matthews2000h}. 
The integrated \HI{} flux density was found to be 5.93 Jy \kms, and the velocity width $\rm (W_{50})$ at 50 percent of peak flux was found to be 188 \kms. Table 5.1 shows the properties of FGC2366.

\begin{table}
\begin{minipage}{110mm}
\hfill{}
\caption{Basic properties: FGC 2366}
%\centering
\begin{tabular}{|l|c|}
\hline
\hline
Parameter& Value \\
\hline    

RA(J2000)$^{\color{red}(1)}$\footnote{Right ascension}   &  $22^h 08^{m} 03^{\farcs} 62$ \\
Dec(J2000)$^{\color{red}(1)}$\footnote{Declination}  & $-10^{\circ} 19^{\farcm} 59^{\farcs}1$ \\
Hubble type$^{\color{red}(1)}$ & Sd \\
$i^{\color{red}(3)}$ \footnote{Inclination}   & $90^{\circ}$\\
Distance (Mpc)$^{\color{red}(4)}$ &32.95 \\
a/b$^{\color{red}(1)}$\footnote{Major axis to minor axis ratio  } & 21.6  \\
log ($M_{HI}/M_{\odot}$)$^{\color{red}(2)}$\footnote{\HI{} mass} & 9.38\\
W$_{50} (\kms)$$^{\color{red}(5)}$\footnote{Spectral line width at 20$\%$ of the peak flux density}  & 204.5\\

\hline
\end{tabular}
\hfill{}
\label{table: table 1}
\end{minipage}
\begin{tablenotes}
\item \textcolor{red}{(1)}: \cite{1999BSAO...47....5K} 
\item \textcolor{red}{(2)}: \cite{matthews2000h} 
\item \textcolor{red}{(3)}: \cite{2014A&A...570A..13M} 
\item \textcolor{red}{(4)}: \cite{kourkchi2020cosmicflows}
\item \textcolor{red}{(5)}: \textcolor{blue}{This Work} 
\end{tablenotes}
\end{table}

\section{\HI{} 21cm radio-synthesis observations}
We observed FGC 2366 with the GMRT for 9 hours, including overheads, using 27 antennae on August 25, 2019.
The source was observed for 7 hours in 14 scans consisting of 30-minute intervals and 14 phase calibrator scans of 5-minute intervals.
Flux calibrators 3C286 and 3C48 were observed for 30 minutes each at the beginning and at the end of the observation. 
We observed the central frequency 1406.9 MHz in GSB mode with 512 channels, bandwidth equal to 4.2 MHz, and resolution equal to 8.14 kHz. 
We use CASA \citep{mcmullin2007casa} to carry out data reduction of our observations. We remove the bad antennae (E04, E05, E06) from our data set, then 
visually analyze the data for radio frequency interference (RFI).
Before separating the target from the measurement set, we carry out cross-calibration. We then create a dirty cube to identify channels 
with spectral line emission and mark them to create a
continuum-only measurement set. Next, we execute 5 rounds of phase-only self-calibration and 4 rounds of amplitude-phase self-calibration. After 5 rounds of 
phase-only self-calibration, the S/N saturates. During amplitude-phase self-calibration, image quality initially degrades but improves after the second round. 
The S/N ratio finally saturates during the third round of amplitude-phase self-calibration. We apply the final amplitude-phase self-calibration table to 
the target-only measurement set and subtract the continuum interpolated with zeroth order polynomial.
Next, we create a dirty cube using the continuum subtracted measurement set and mask the emission using 
the Source Finding Algorithm SoFiA:\citep{serra2015sofia}. The final data cube was cleaned using a mask made with SoFiA. We observe that the 
$\rm 'briggs'$ weighing scheme in CASA task 'tclean' with robustness parameter = 0.5 and uvtaper = 12k$\rm \lambda$ provide the optimum trade-off between the 
sensitivity and resolution. We present the properties of the observations and deconvolved image in Table 5.2. Refer to sections 1.4.2 and 1.4.3 for details of data analysis.

\begin{table}
\caption{Summary of \HI{} observations of FGC2366}

%\centering
\begin{tabular}{|l|c|}	
\hline
\hline
(a) Observing Setup&\\
\hline
Parameter& Value \\
\hline    
\hline
Observing Date                     & 25August2019\\
Phase center,$\alpha$(J2000)       & $22^h 08^m 03^{\farcs}62$\\
Phase center,$\delta$(J2000)       & $-10^{\circ}19^{\farcm}59^{\farcs}1$\\
Total on-source observation time   & 5 $\frac{1}{2}$ hrs\\
Flux  calibrator                   & 3C286, 3C48  \\
Phase calibrator                   & 2136+006\\
Channel Width                      & 8.14 kHz\\
Velocity separation                & 1.7 \kms\\
Central frequency                  & 1406.9 MHz\\
Total bandwidth                    & 4.2 MHz   \\
\hline
(b) Deconvolved Image Characteristics&\\
\hline
Weighing                           & robust\\
Robustness parameter               & 0.5\\
Synthesized beam FWHM              & $13.4\farcs \times 11.7\farcs$\\
Synthesized beam position angle    & $39.4^{\circ}$\\
rms noise in channel               & 0.39 mJy/beam\\
\hline
\end{tabular}
\hfill{}
\label{table: table 2}
\end{table}

\subsection{Observations, Data Reduction $\&$ Analysis}
Figure 5.2 shows FGC 2366's global \HI{} profile. We fit the observed profile with a busy function \citep{westmeier2014busy} and find the integrated flux 
is 5.4 Jy \kms and the peak \HI{} flux is 34 mJy. The integrated flux obtained by fitting busy function is comparable
to that measured directly from the data cube using CASA task \textit{SPECFLUX}. The flux gives us $\rm \log_{10}(M_{\HI{}}/M_{\odot})=9.1$. The 20$\rm \%$ 
$\rm (W_{20})$ and 50$\rm \%$ $(W_{50})$ velocity widths are 214 \kms and 205 \kms, respectively. Table 5.3 shows results obtained by fitting the busy function. 
Figure 5.3 shows the integrated column density (Moment 0) and velocity field (Moment 1) of FGC 2366. We overlay the \HI{} column density map of FGC 2366 on POSS-II (Palomar Optical Sky Survey) 
image. The Moment 0 map is smooth and unperturbed, and there is no enhanced emission from the center or edge of the \HI{} disc. The \HI{} distribution and velocity 
field are similar to other superthin galaxies; see \citep{kurapati2018mass}. Figure 5.4 shows \HI{} channels maps superimposed on the POSS-II optical image. 
Every fourth channel map is shown. In the channels close to the systemic velocity, \HI{} emission lies in a flat plane. The \HI{} emission deviates from the plane only 
in edge channels.

\begin{table}
\centering
\begin{minipage}{110mm}
\caption{Busy function fit to \HI{} global profile}
\begin{tabular}{|c|c|c|c|c|}
\hline
$V^{(\textcolor{red}{a})}_{0}$ & $W^{(\textcolor{red}{b})}_{50}$ & $W^{(\textcolor{red}{c})}_{20}$ & $F^{(\textcolor{red}{d})}_{\text{peak}}$ & $F^{(\textcolor{red}{e})}_{\text{int}}$ \\
\kms{} & \kms{} & \kms{} & mJy & $\rm Jy\, \kms{}$ \\
\hline
$ 2844 \pm 2.4 $ & $204.5 \pm 2.9$ & $214.4 \pm 3.9$ & $33.9 \pm 0.3$ & $5.4 \pm 0.2$ \\
\hline
\end{tabular}
\label{table:table3}
\begin{tablenotes}
\item[\textcolor{red}{(a)}] Frequency centroid of the \HI{} line.
\item[\textcolor{red}{(b)}] Spectral line width at 50\% of the peak flux density.
\item[\textcolor{red}{(c)}] Spectral line width at 20\% of the peak flux density.
\item[\textcolor{red}{(d)}] Peak of the \HI{} flux density.
\item[\textcolor{red}{(e)}] Integrated \HI{} flux.
\end{tablenotes}
\end{minipage}
\end{table}

\begin{figure} 
\hspace{2.2cm}
\resizebox{90mm}{60mm}{\includegraphics{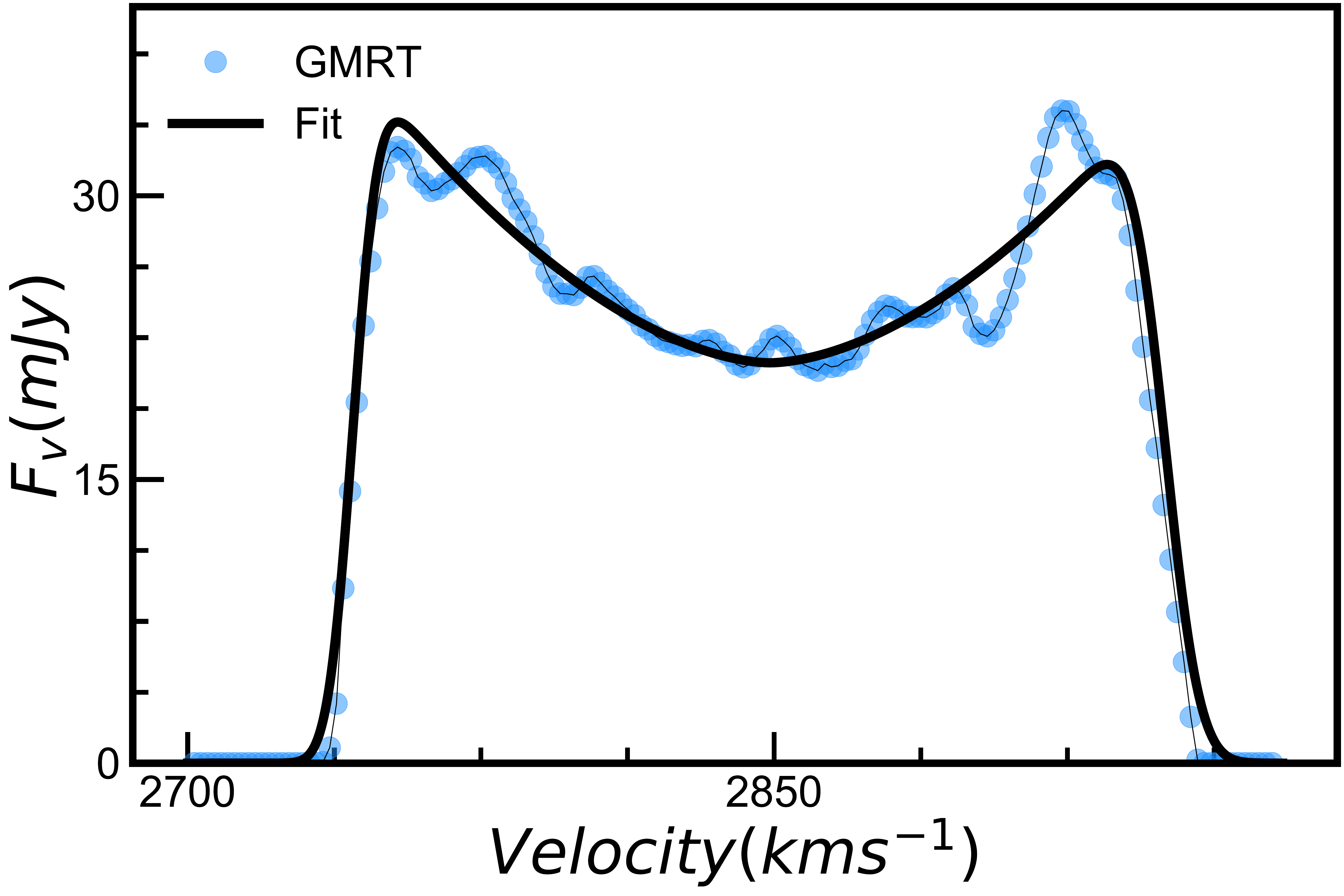}} 
\caption{The global \HI{} profile of FGC 2366 derived from our GMRT observation overplotted with the best-fit busy function.}
\end{figure}

\begin{figure*}
\hspace{-0.5cm}
\resizebox{170mm}{70mm}{\includegraphics{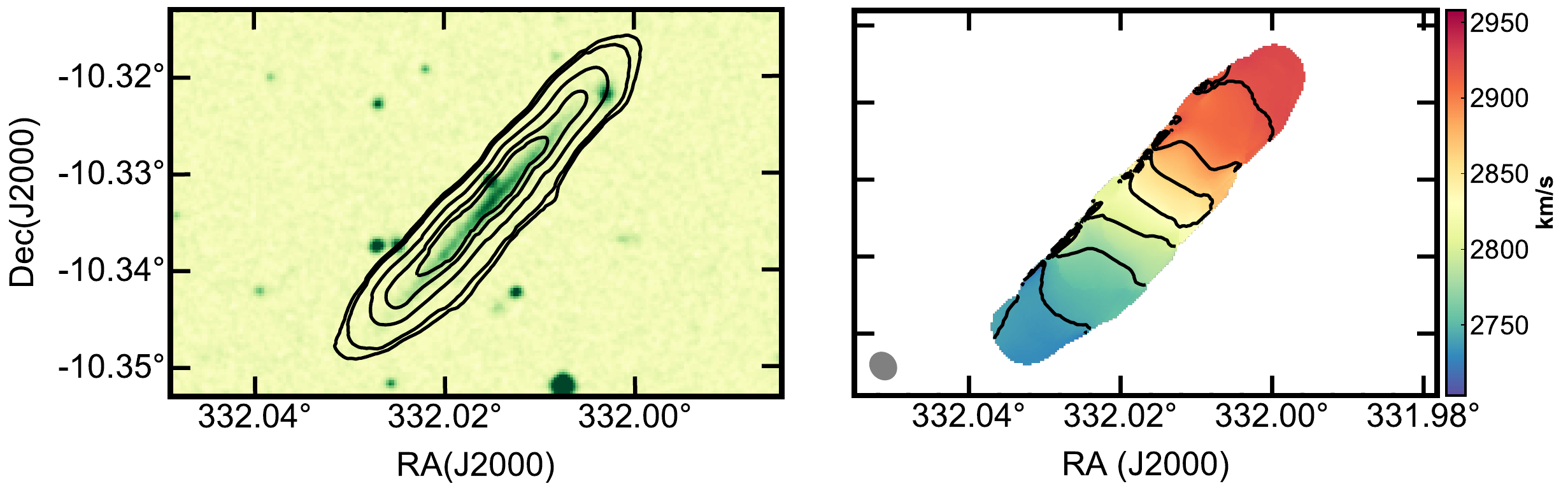}} 
\caption{Moment 0 [Left] and Moment 1 [Right] maps derived from the observed data cube. The data contours are shown in black. The contours levels in 
moment 0 maps are at [ 3.0, 4.0, 8, 12, 14]$\times$ 35 mJy beam$^{-1}$ kms$^{-1}$ and the contour in moment 1 maps start at 2951 \kms and end at 2710 \kms  
increasing by 26 \kms.}
\end{figure*}

\begin{figure*}
\hspace{-2.2cm}
\resizebox{185mm}{195mm}{\includegraphics{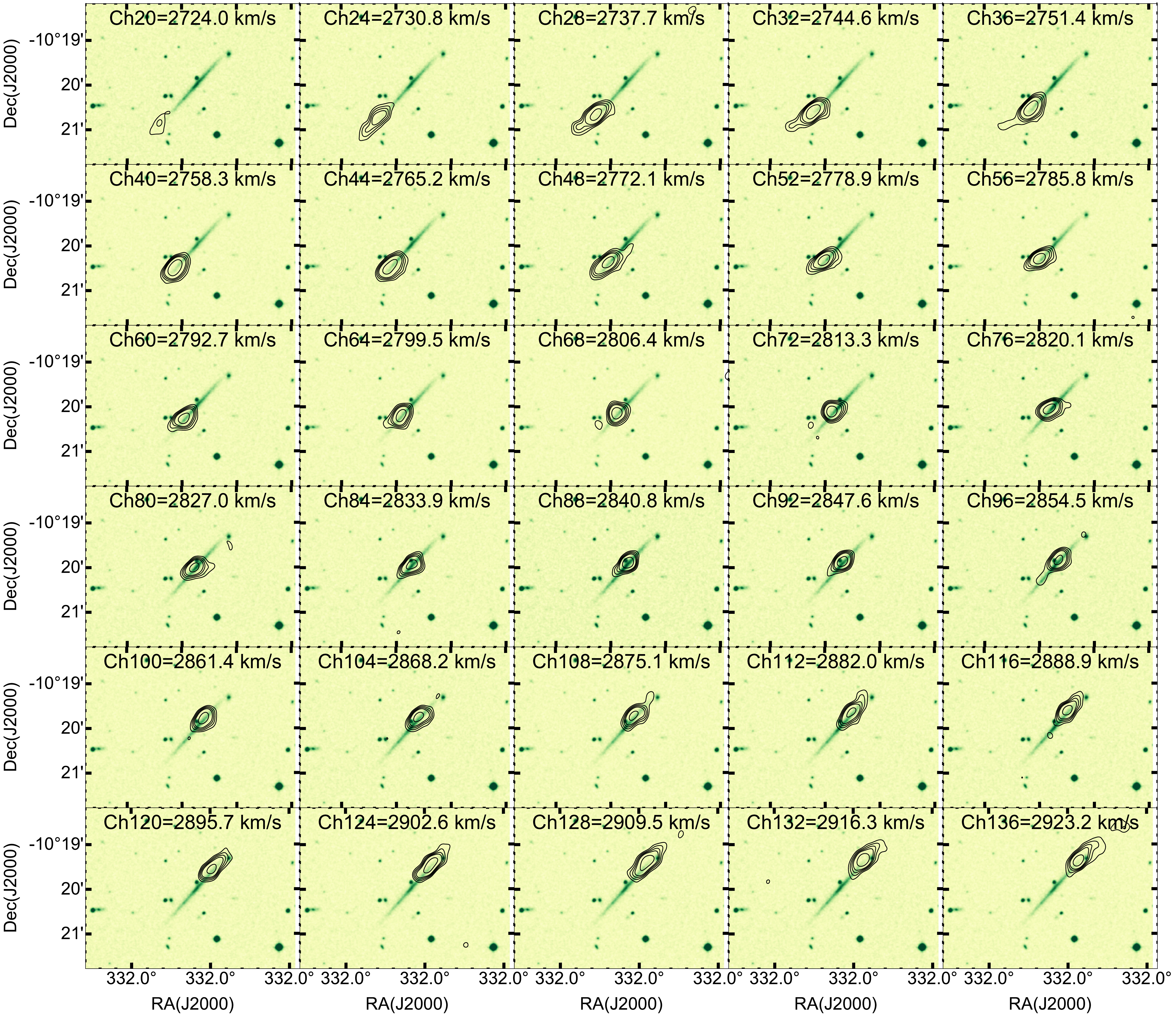}} 
\caption{Channel maps showing \HI{} emission from FGC 2366, each panel is separated by four channels. The \HI{} emissions are overlaid on the POSS-II optical image. 
The contour levels are at [3, 4, 5 ,6, 9]$\times$ 0.8 mJy beam$^{-1}$}	
\end{figure*}

\subsection{3D-Tilted Ring Modeling}
Fully Automated TiRiFiC (FAT) is a GDL wrapper around TiRiFiC, a tilted ring modeling software (see Section 1.4.5 for details). 
TiRiFiC builds model data cubes from tilted ring parametrization of the rotating disc and fits them to observed data cubes. 
Fitting the observed data cubes directly with TiRiFiC allows us to precisely calculate FGC 2366's structural and kinematic properties. 
Conventional 2D approaches rely on the velocity field to derive structural and kinematic parameters and suffer from projection and beam smearing effects. 
Beam smearing impacts the derivation of the rotation curve in the inner regions of the galaxy. On the other hand, projection effects prevent deriving a distinct 
velocity field for edge-on galaxies because the line of sight passes through the disc several times. Instead of fitting to derived data products like 2D-velocity
fields or PV diagrams, FAT directly fits the observed data cube. FAT uses the \HI{} data cube as input and automates the fitting of free parameters, namely
1) Surface brightness, 2)  Position angle 3) Inclination 4) Rotational velocity 5) Scaleheight, 6) Intrinsic velocity dispersion, and 7) Central coordinates: Right Ascension, Declination, 
and Systemic velocity. We divide the \HI{} disc in two halves and fit 19 semi-rings over each half; each ring is 0.6 times the major axis beam size. We model the vertical
gas distribution using the $\rm sech^{2}$ profile. Table 5.4 shows the best-fitting values obtained from 3D models. Figure 5.5 shows the rotation curve, velocity dispersion, and
surface brightness profiles of FGC 2366 as a function of radius. Figure 5.6 shows the rotation curve derived by FAT overlaid over major axis PV diagram. The rotation curve matches the midplane 
emission in the major axis PV diagram. FAT estimates a different central velocity than that obtained by fitting the busy function because it directly fits the 3D data 
cube. Note that we fit the busy function to the 1D spectrum produced from the 3D cube. The \HI{} spectrum in Figure 5.2 exhibits asymmetry, which may 
explain why FAT and busy function fitting produce different central velocities. \\

\textbf{Quality of Fits:}
Figure 5.7 [top panel] compares the observed and model Moment 0 and Moment 1 maps. 
The bottom panel shows the residual maps. We evaluate the residual maps and find that 3D model fitted by FAT reproduces the observed 
surface brightness and velocity field well. Moment maps show surface brightness and velocity information separately, but the observed data cube is three-dimensional. 
In order to preserve the 3D structure of the observed data cube, we compare minor axis PV diagrams. Figure 5.8 [top panel] compares the minor axis PV diagrams 
from the observed and model data cube. Our model data cube reproduces the observed structure of the \HI{} emission. The model and data diverge only at velocities 
away from the galaxy at 3$\rm rms$. Finally, we note that the residuals deviate from the model at values less than 3rms, indicating that the 3d models obtained from FAT 
describes the data well.

\begin{table}
\begin{minipage}{110mm}
\hfill{}
\caption{Best-fitting model derived by FAT }
%\centering
\begin{tabular}{|l|c|}
\hline
Parameter&    Values  \\
\hline    
$\rm X_{o}$    \footnote{Right ascension}            &   $22^h 08^m 03^{\farcs}62$               \\
$\rm Y_{o}$    \footnote{Declination}            &    $-10^{\circ}19^{\farcm}59^{\farcs}1$                 \\
$\rm i$        \footnote{Inclination}            &    $87.1 ^{\circ}\, \pm \, 4^{\circ}$            \\
$\rm V_{sys}$  \footnote{Systemic velocity}         &    2832.5 \kms               \\
PA             \footnote{Position angle}           &    $317.1^{\circ} \, \pm \, 2^{\circ} $         \\
$\rm h_{z}$      \footnote{Scaleheight of the \HI{} disc}         &    $6.7^{\farcs}$            \\
\hline
\end{tabular}
\hfill{}
\label{table: table 4}
\end{minipage}
\end{table}

\begin{figure*}
\hspace{-1.5cm}
\resizebox{180mm}{50mm}{\includegraphics{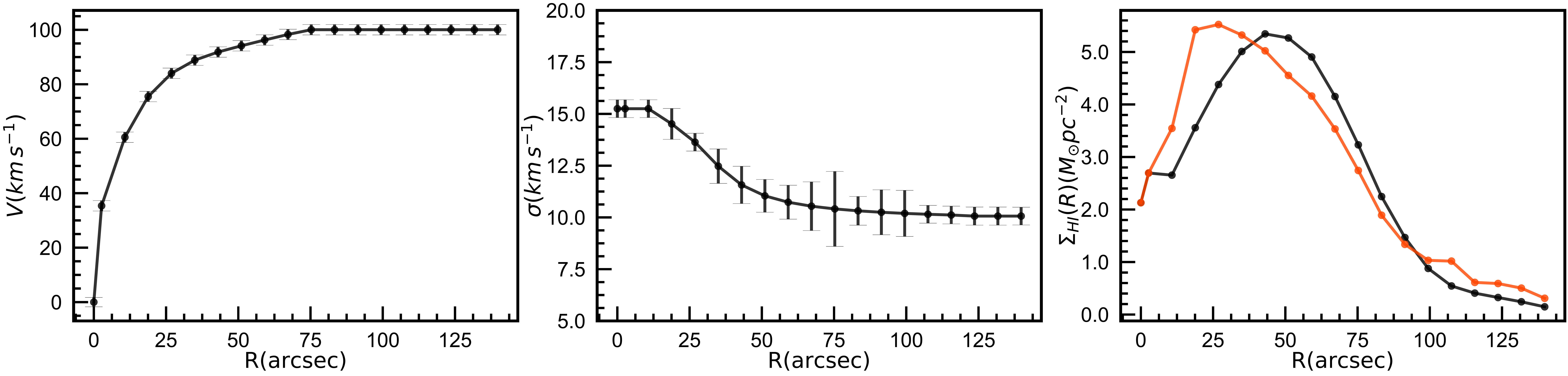}} 
\caption{The best-fitting model derived using the 3D tilted modeling of the observed data cube. The red and the black lines in the third panel indicate the 
surface brightness profile of the approaching and the receding sides. [Left Panel] Rotational velocity [Middle Panel] \HI{} dispersion [Right Panel]} 
\HI{} surface density as a function of galactocentric radii.	
\end{figure*}

\begin{figure}
\hspace{1cm}
\resizebox{140mm}{62mm}{\includegraphics{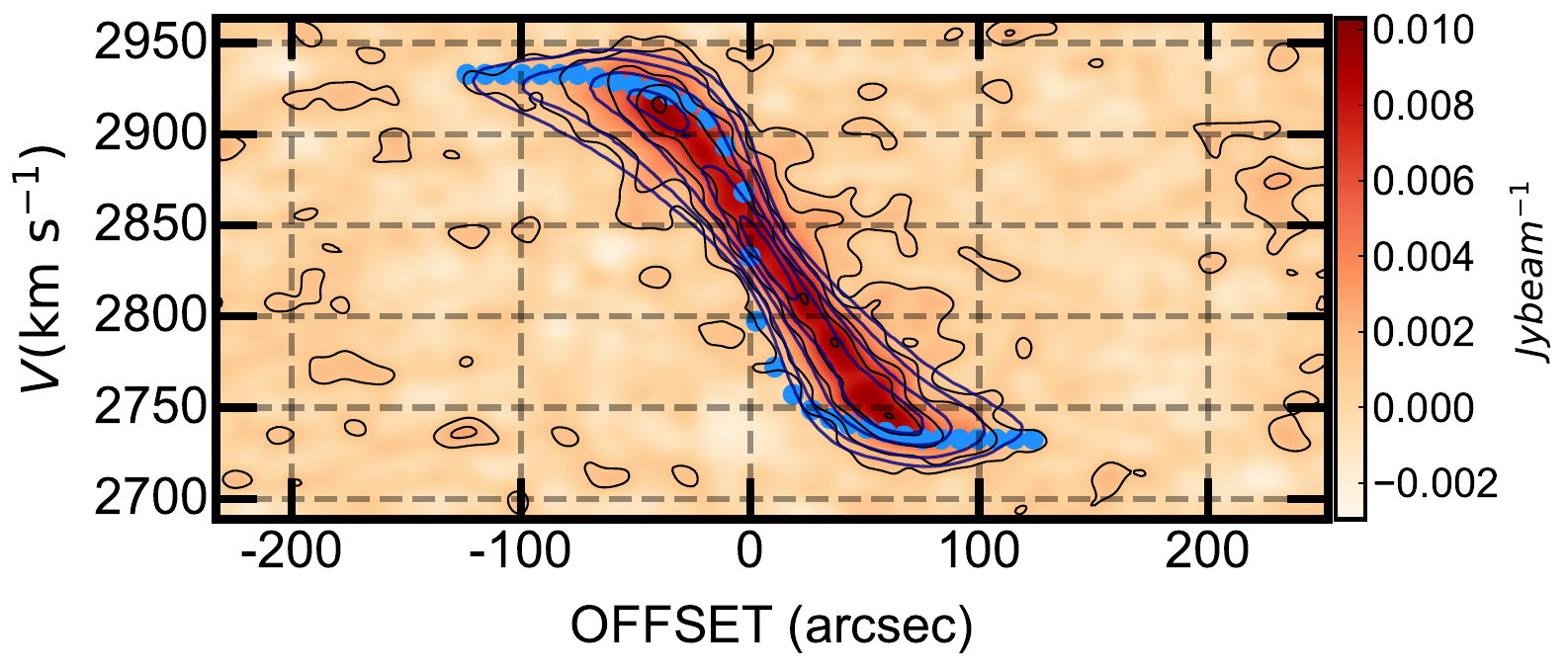}} 
\caption{The PV diagrams of the observed (black) and the model (violet) data cube derived along the major axis. 
The rotation curve derived from the 3D tilted ring modeling (blue points) is overplotted. The contour levels are at  [1.5,3, 6, 9, 12]$\rm \times$ 0.8 mJy beam$\rm ^{-1}$}	
\end{figure}

\begin{figure*}
\begin{center}
\begin{tabular}{cc}
\hspace{-2cm}
\resizebox{160mm}{50mm}{\includegraphics{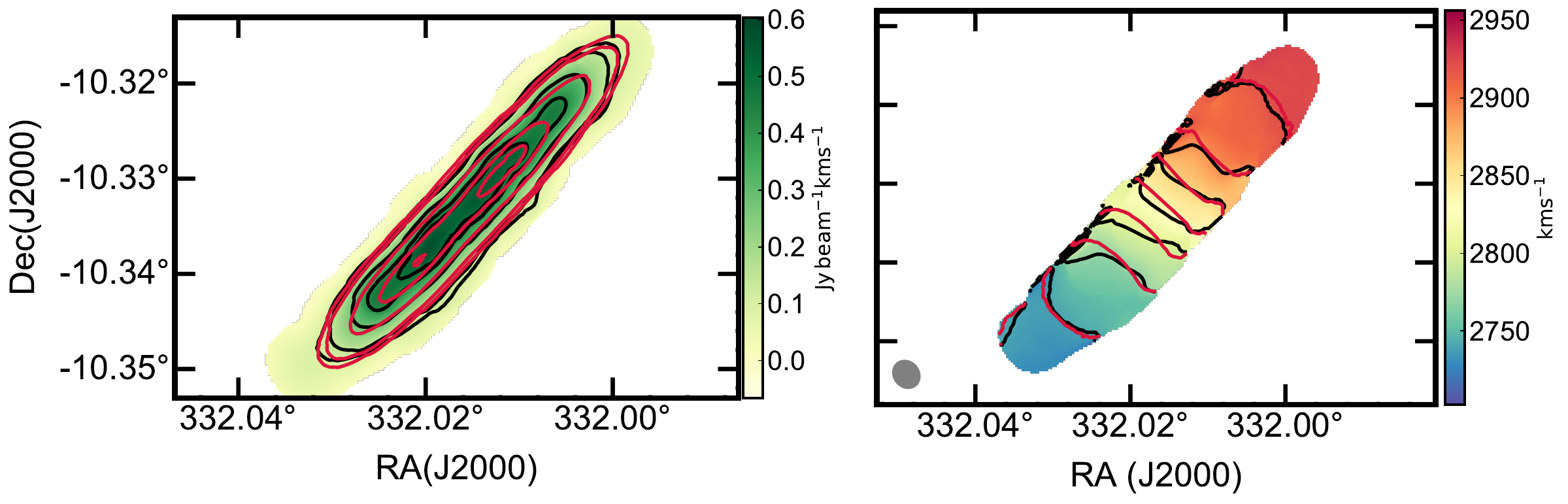}} \\
\hspace{-2cm}
\resizebox{160mm}{50mm}{\includegraphics{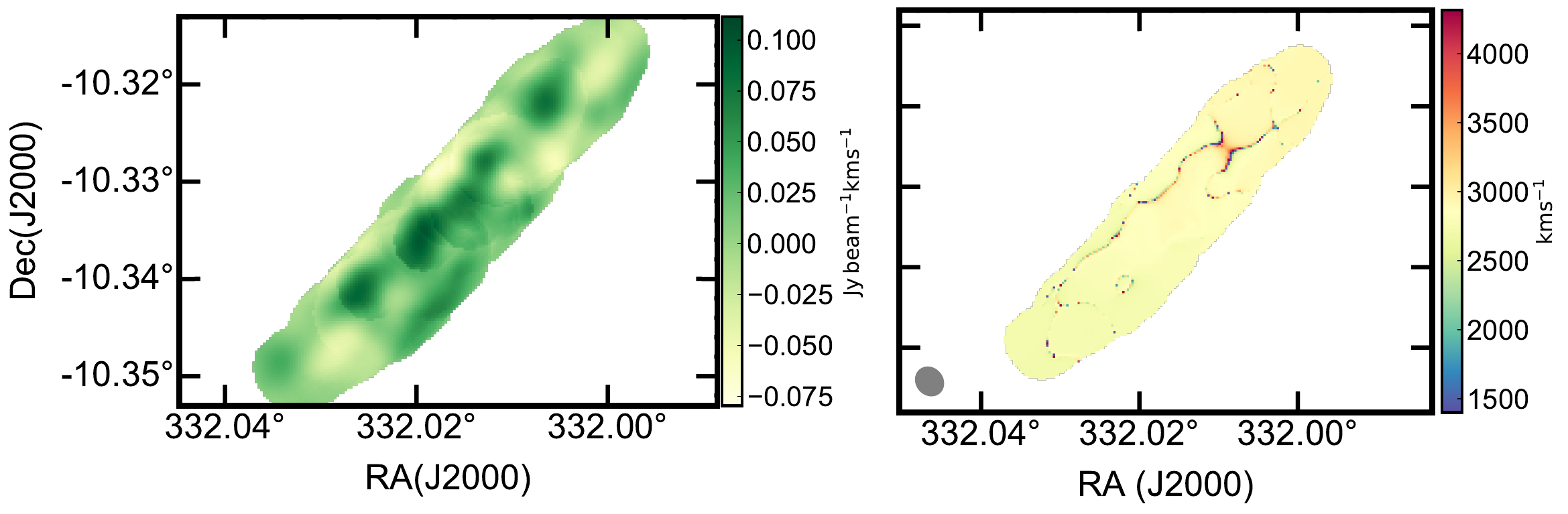}} 
\end{tabular}
\end{center}
\caption{The Moment 0 and Moment 1 maps derived from the observed and the model data cube. The model contours
are shown in crimson, and the data contours are shown in black. The contours levels in moment 0 maps at [ 3.0, 4.0, 8, 12, 14]$\rm \times$ 35 mJy beam$\rm ^{-1}$ kms$\rm ^{-1}$
and the contour in moment 1 maps starts at 2951 \kms and ends at 2710 \kms, increasing 26 \kms. In the lower panel, we have shown the residual maps of the total 
intensity and the velocity field obtained by subtracting the model from the data.}	
\end{figure*}

\begin{figure}

\begin{center}
\begin{tabular}{cc}
\hspace{-2cm}
\resizebox{180mm}{40mm}{\includegraphics{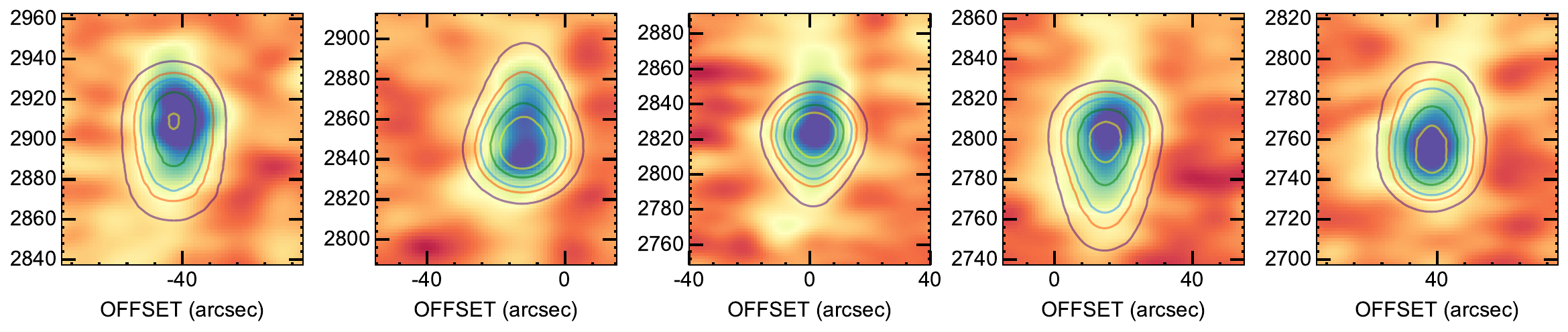}} \\
\hspace{-2cm}
\resizebox{180mm}{40mm}{\includegraphics{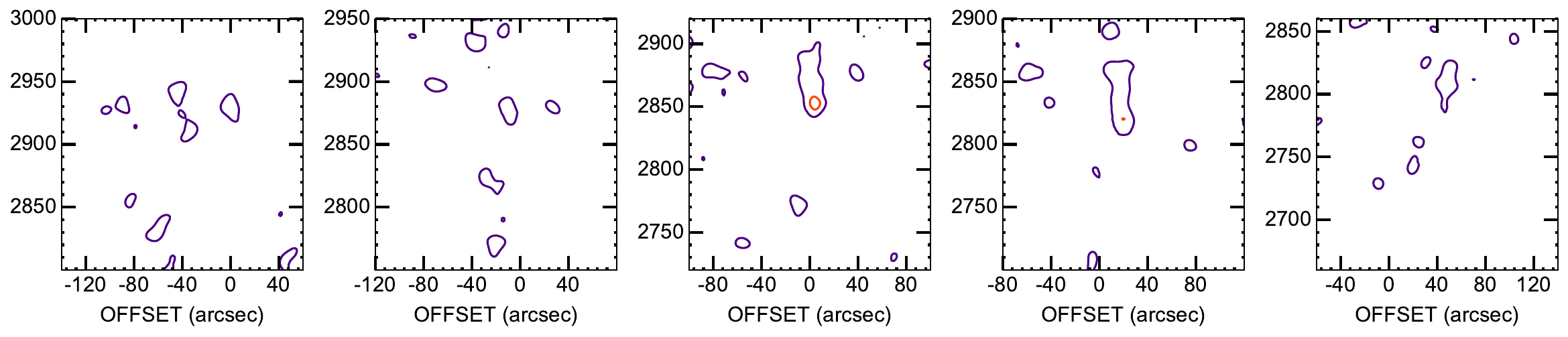}} 
\end{tabular}
\end{center}
\caption{The PV diagrams of the observed data cube and the model data cube along the minor axis. The contour levels are 
at  [1.5, 3, 4, 6, 8]$\rm \times$ 1.0 mJy beam$\rm ^{-1}$. The lower panels show the corresponding residuals of the minor axis PV diagrams 
obtained by subtracting the model from the data.}	
\end{figure}

\section{Optical photometry}
We use optical images from the DESI Legacy Imaging Surveys DR9 to derive the stellar photometry of FGC 2366 in the g, r, and z bands.
Even in the bluest optical bands, there is no sign of a dust lane in this galaxy. So, for a first guess, we can ignore the internal extinction. 
We use the photometric decomposition package IMFIT and use the ExponentialDisk3D function.
We fit the light distribution keeping the central surface brightness, radial scalelength, vertical exponential scaleheight, and the inclination as free parameters.
Table 5.5 shows the results from optical photometry. The total magnitudes of the galaxy model in the $\rm g$, $\rm r$, and $\rm z$ bands, after correcting for the 
Galactic extinction from \cite{schlafly2011measuring} are 15.44, 15.01, and 14.76, respectively. From Table 5.5, we can see that the radial scalelength decreases with wavelength, 
which fits with the inside-out formation scenario. Figure 5.9 shows the model and data profile along the major axis in the z-band. See Section 1.4.8 for details regarding 
the calculation of the mass-to-light ratio.

\begin{figure*}

\begin{center}
\begin{tabular}{ccc}
\hspace{-2cm}
\resizebox{60mm}{50mm}{\includegraphics{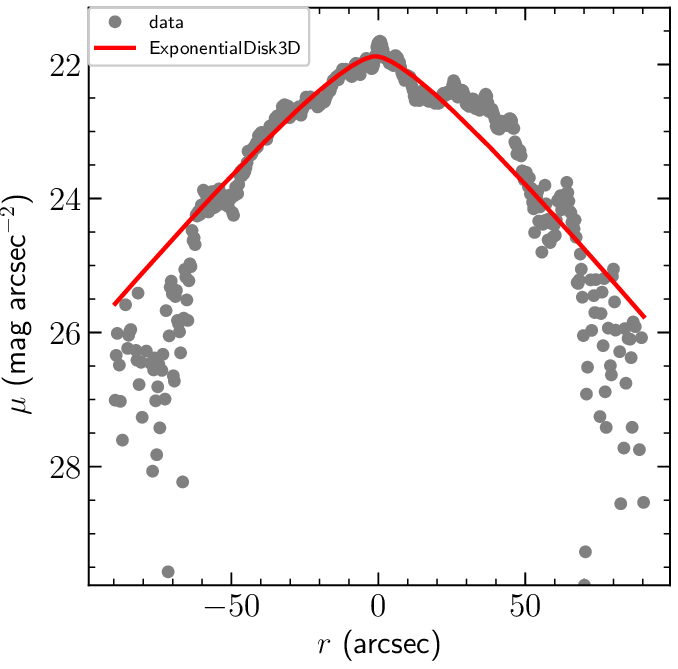}} 
\resizebox{60mm}{50mm}{\includegraphics{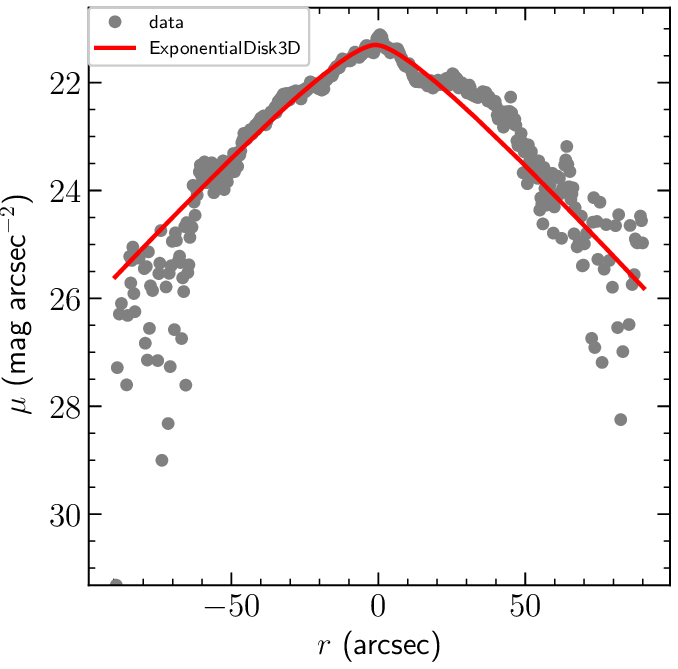}} 
\resizebox{60mm}{50mm}{\includegraphics{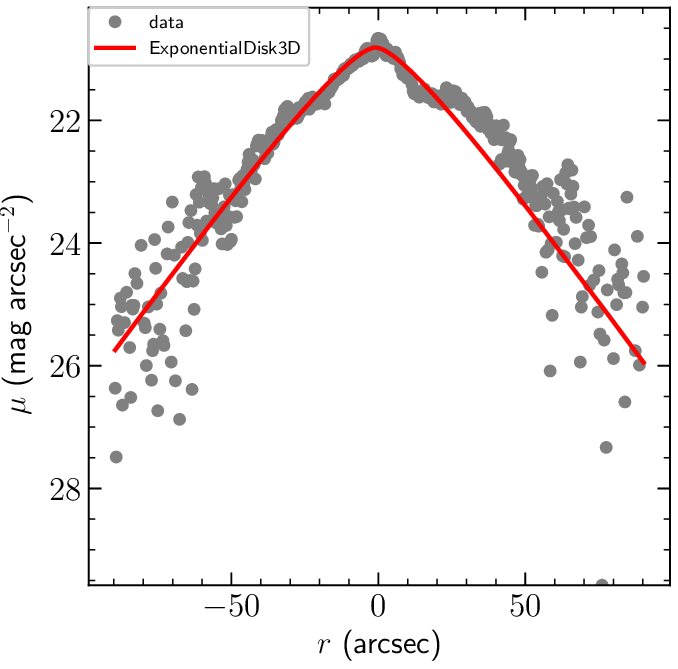}} 
\end{tabular}
\end{center}
\caption{The model and data profile along the major axis of FGC 2366 for the g, r, and z-band optical image. 
The light distribution is modeled using the ExponentialDisk3D function, keeping the surface brightness, scaleheight, 
scalelength, and the inclination as the free parameters.}
\end{figure*}

\begin{table*}
\hspace{-1.5cm}
\begin{minipage}{110mm}
\hfill{}
\caption{Structural parameters from optical photometry of FGC 2366.}
\centering
%\small\addtolength{\tabcolsep}{-1pt}
\begin{tabular}{|l|c|c|c|c|}
\hline
\hline
Parameter                                 &               & 	               &               		&					 Description \\
                                          &g-band 	  &  r-band            &  z-band       		&             				\\
\hline
Total magnitude                           & 15.44         &   15.01            & 14.76          	&   					\\
$\mu^{edge-on}_{o}$                       & $21.58 \pm 0.02$&   $20.97\pm 0.02$    & $20.46 \pm 0.02$	        &	Edge-on surface brightness ($\rm mag/arcsec^{2}$)\\
$R_{d}$                                   & $20.4  \pm 0.2$ &   $17.9 \pm 0.1$     & $16.1  \pm 0.1$	        &	Disc scalelength ($\rm arcsec$)\\
$h_{z}$                                   & $2.0   \pm 0.1$ &   $2.0 \pm 0.1$      & $1.8   \pm 0.1$	       	&       Disc scaleheight ($\rm arcsec$)\\
$i$                                       & $89.3  \pm 0.1$ &   $89.0\pm 0.1$      & $88.6  \pm 0.1$        	&	(\rm degrees).\\
\hline 
\end{tabular}
\hfill{}
\label{table: table 4}
\end{minipage}
\end{table*} 

\section{Dynamical Modelling}
\subsection{Mass Modeling}
By dissecting the total rotation curve into baryonic $\rm (stars + gas)$ and dark matter components, we derive the rotation velocity due to
the net gravitational potential of the dark matter halo. The parameters for deriving the stellar rotation curve are given in Table 5.6.The procedure for carrying out mass modeling is detailed in the introduction in Section 1.4.6
\begin{table*}
\begin{minipage}{110mm}
\hfill{}
\caption{Input parameters for deriving the stellar rotation curve.}
\centering
%\small\addtolength{\tabcolsep}{-1pt}
\scalebox{0.9}{
\begin{tabular}{|l|c|c|}
\hline
\hline
Parameter                                 &Value         &Description \\
                                          &z-band        &            \\
\hline
$\mu^{face-on(*)}_{o}$   &22.8&         Face-on surface brightness ($\rm{mag/arcsec^{2}}$)\\
$\Sigma_{o}$                              &21.5&         Surface density ($L_{\odot}/pc^2$)\\
$R_{d}$                                   &2.6&          Disc scalelength (kpc)\\
$h_{z}$                                   &0.29&         Disc scaleheight (kpc)\\
\hline 
Parameters for deriving  mass to light ratio $(\gamma^{*})$&   & \\
\hline
$g-z$         &  0.7      &                         \\
$a_{\lambda}$ &  $-0.17$  &    \cite{bell2003optical}\\
$b_{\lambda}$ &  0.32     &     \cite{bell2003optical}\\
$M/L$  &  1.12     &     M/L ratio derived using scaled Salpeter IMF\\
$M/L$  &  0.8      &     M/L ratio derived using Kroupa IMF\\
\hline
\end{tabular}}
\hfill{}
\label{table: table 5}
\end{minipage}
\begin{tablenotes}
\item  (*): The edge-on surface brightness has been converted to face-on surface brightness using 
$ \mu^{face-on}= \mu^{edge-on} + 2.5log( \frac{ R_{d} } {h_{z}})$ \citep{kregel2005structure}.
\end{tablenotes}
\end{table*} 

\noindent \textbf{Uncertainty on the rotation curve:}\\
We quantify two key sources of uncertainty on the rotational curve: the difference in rotation velocities between the approaching and receding sides, 
and our observation's spectral resolution, see for example \cite{de2008high, swaters2009rotation}. We define the final uncertainty by adding in quadrature 
spectral resolution and one-fourth of the difference in approaching and receding rotation velocity.

\subsection{Dark Matter Halo}
Table 5.7 and Figure 5.10 show results obtained from mass modeling. Both the cored PIS halo and the cuspy NFW halo fit the observed rotation curve well. 
The Maximum-Disc model fits poorly in both situations, highlighting dark matter dominance at all radii. The Minimum-Disc model fits the observed rotation curve 
better in the case of the NFW halo than a PIS halo. Based on the mass models, it is impossible to distinguish between the Kroupa and the $\rm 'diet'-Salpeter$ IMF.
Mass models produced with PIS halo have a small core radius of 0.9 kpc, indicating that the baryons in FGC 2366 are embedded in a compact dark matter halo 
with $\rm R_{c}/R_{d} <2$ \citep{banerjee2013some}. Mass models using an NFW halo have a concentration parameter equal to 7.9 and $\rm R_{200}$ equal to 58.3 kpc,
which is comparable to the values obtained from mass modeling of the disc galaxies in the THINGS galaxy sample \cite{de2008high} .
We may note here that it is hardly possible to distinguish between an NFW nad PIS halo using mass modeling results.
High-resolution N-body simulations with 5000 haloes show that galaxies with low dark matter concentrations survive in 
isolated environments \citep{Wechsler2002concentrations}. This suggests that the extremely thin galaxies are isolated systems 
formed in underdense regions and have been undisturbed by the tidal forces of adjacent galaxies. \cite{bailin2005internal} showed that 
galaxies with high spin parameters exist in low concentration haloes, which may be vital for explaining exceedingly thin stellar discs. Thus, the 
small concentration parameter obtained from the mass models for extremely thin galaxies suggests that these extremely thin stellar discs have formed in halos with high spin parameters thriving in isolated environments.

In the case of MOND based mass models, the acceleration for FGC 2366 ($\rm 0.5 \times 10^{-10} ms^{-2}$) is lower than expected by MOND ($\rm 1.2 \times 10^{-10} ms^{-2}$) 
and the mass-to-light ratio (= 10.4) is close to values obtained in the maximum disc model. If we fix the acceleration parameter to $\rm (1.2 \times 10^{-10} ms^{-2})$ indicated 
by MOND, the mass-to-light ratio is still three times more than the value predicted by stellar population synthesis models. In a study of the superthin galaxies UGC 7321, 
IC 5249, and IC 2233, \cite{banerjee2016mass} find that the rotation curves of IC 5249 and UGC 7321 are explained by MOND. They find that mass to light ratios obtained 
for UGC 7321 and IC 5249 from MOND agree with values predicted by stellar population synthesis models. The acceleration and mass-to-light ratio predicted by MOND 
for IC 2233 are unrealistic. Finally, the mass-to-light ratio for FGC 2366 obtained from MOND is higher than predicted by population synthesis models.

In Figure 5.11, we compare the mass models of FGC 2366 with FGC 1440 and other superthin galaxies with published mass models i.e. IC 2233, IC 5249, FGC 1540, UGC 7321. 
We plot the ratio of the core radius of the PIS Dark Matter halo to stellar disc scalelength $\rm R_{c}/R_{d}$ [Left Panel], its central core density $\rho_{0}$ [Middle Panel], 
and $\rm V_{\rm rot}/(R_{c}/R_{d})$ [Right Panel] as a function of $\rm a/b$. Regression line fits are superimposed on plots. We note that, $\rm R_{c}/R_{d}$ decreases with $\rm a/b$, 
indicating that galaxies with a larger a/b have a more compact dark matter halo. A compact halo is one in which the core radius is less than twice the disc scalelength. 
All galaxies except UGC7321 have $\rm R_{c}/R_{d}< 1$. \citep{banerjee2013some} argue that a compact dark matter halo may be key to explaining the 
thin disc structure. They show that if the dark matter potential were to be removed from the 2-component model, the stellar disc thickness would start to increase 
from the inner radius itself, and the disc would no longer be super thin. In fact, \cite{2019MNRAS.490.5451D} show that galaxies with larger disc scalelengths have a larger 
core radius. $\rm \rho_{0}$ grows with $\rm a/b$, indicating thinner discs have a denser halo. Further, $\rm V_{\rm rot}/(R_{c}/R_{d})$ 
increases as $\rm a/b$ increases. $\rm V_{\rm rot}$ represents the galaxy's total dynamical mass, which is dominated by dark matter in case of superthins. $\rm R_{c}/R_{d}$ represents the halo's compactness, or how large the dark matter's core is relative to the disc. If $\rm R_{c}/R_{d} \sim 1$, dark matter would dominate the dynamics of the stellar disc.Thus, $\rm V_{\rm rot}/(R_{c}/R_{d})$ indicates the dark matter dominance at inner galactocentric radii. We find that in the galaxies with a larger axis ratio, the dark matter is more centrally concentrated and is characterized by the high value of central density and a small compactness parameter.

\begin{figure*}
\hspace{-2.cm}
\resizebox{180mm}{160mm}{\includegraphics{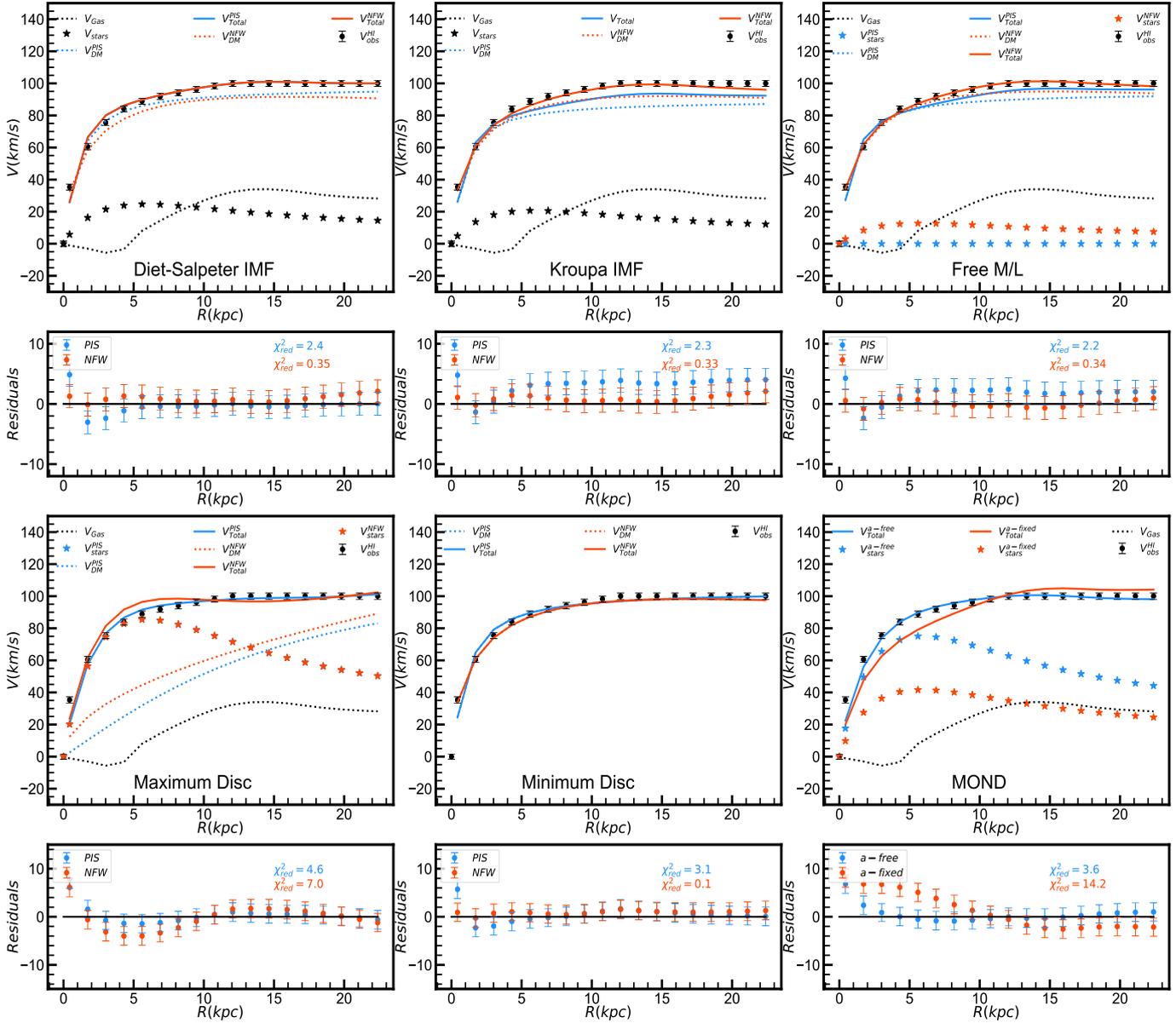}} 
\vspace{1pt}
\caption{Mass-models of the galaxy FGC 2366 derived using SDSS z-band photometry. In the first two panels of 
the first row, we show the results obtained by fixing $\rm M/L$ from population synthesis models, and in the third panel, we derive the mass models by keeping $\rm M/L$ 
as a free parameter. In the second row, we show the residuals obtained by subtracting the data from the model weighed by the 
errors on the observed rotation curve. In the first and second panels of the third row, we derive the mass models keeping 
the disc mass to be maximum and minimum, respectively, and finally, in the last panel, we show the mass models derived using the modified Newtonian (MOND) phenomenology followed by the corresponding residuals in the fourth row.}	
\end{figure*}

\begin{table}
\hspace{-2cm}
\begin{minipage}{110mm}
\hfill{}
\caption{Dark matter density parameters derived from mass modeling using the z-band photometry and \HI{} observations.}
\centering
\begin{tabular}{|l|c|c|c|c|c|c|c|c|}
\hline
\hline
Model                   &$c^{(a)}$ &$R^{(b)}_{200}$&${M/L}^{(c)}$& $\chi^{2 (d)}_{red}$     & $\rho^{(e)}_{0}\times10^{-3}$  &$R^{(f)}_{c}$ & ${M/L}^{(g)}$& $\chi^{2 (h)}_{red}$  \\
                        &                           &(kpc)                           &                                &        &  $M_{\odot}/pc^{3}$                             &(kpc)                          &                                 &        \\
\hline
                        z-band                  & NFW  profile              &                                &                                &        &PIS profile                                      &                               &                                 &              \\
\hline
$'diet'$ Salpeter	&$ 7.9 \pm 0.2 $   &$ 58.3  \pm 0.4 $&$1.1$&$0.35$      		&$217.8\pm40.2$& $0.9  \pm  0.09$ &$1.1$&$2.4$\\
Kroupa IMF 	        &$ 8.1 \pm 0.2 $   &$ 58.3  \pm 0.4 $&$0.8$&$0.33$      		&$230.9\pm42.6$& $0.8  \pm  0.08$ &$0.8$&$2.3$\\
Free $\gamma^{*}$ 	&$ 8.4 \pm 0.4 $   &$ 59.7  \pm 0.4 $&$0.3\pm0.6 $&$0.34$      		&$257.3\pm43.4$& $0.8  \pm  0.07$ &$0.0$&$2.2$\\
Maximum Disc 	        &$  0.0        $   &$ 187   \pm 11.5$&$13.5      $&$7.0 $      		&$2.0  \pm0.4$ & $12.5 \pm 2.9$   &$13.5$&$4.6$\\
Minimum Disc	        &$ 7.9 \pm 0.1 $   &$ 62.5  \pm 0.3$&$0.0       $&$0.1 $      		&$197.3\pm35.7$& $1.0  \pm 0.1$   &$0.0$&$3.1$\\
\hline
MOND                       &                                    &                                        &                &     &    &    & &   \\
\hline
        &$a^{(i)}$ &  ${M/L}^{(j)}$  &  $\chi^{2 (k)}_{red}$    &                &     &    &    & \\
        &$ms^{-2}$  &                                    &                                        &                &     &    &    &\\        
\hline
z-Band              & $( 0.5 \pm 0.06 ) \times 10^{-10}$      & $ 10.4 \pm 1.1$&$   3.6    $        &            &    &    &    &    \\    
z-Band$^{a=fixed}$  &$1.2 \times 10^{-10}$& $  3.2  \pm 0.4 $ &$   14.2  $        &    &    &    &   & \\

\hline
\end{tabular}

\hfill{}
\label{table: table 7}
\end{minipage}
\begin{tablenotes}
\item  $(a)$: Concentration parameter of the NFW profile
\item $(b)$:  Radius at which the mean density equal to 200 times the critical density.
\item $(c)$:  Mass to light ratio derived using population synthesis models or estimated as a free parameter. 
\item $(d)$:  Reduced chi-square value corresponding to the fit. 
\item $(e)$:  The central dark matter density of the PIS dark matter halo model
\item $(f)$:  The core radius of the PIS dark matter halo model
\item $(g)$:  Mass to light ratio derived using population synthesis models or estimated as a free parameter. 
\item $(h)$:  Reduced chi-square value corresponding to the fit.
\item $(i)$:  Acceleration in MOND.
\item $(j)$:  Estimated Mass to light ratio in MOND.
\item $(k)$:  Reduced chi-square value corresponding to the fit.
\end{tablenotes}
\end{table}

\begin{figure*}
\hspace{-2cm}
\resizebox{180mm}{50mm}{\includegraphics{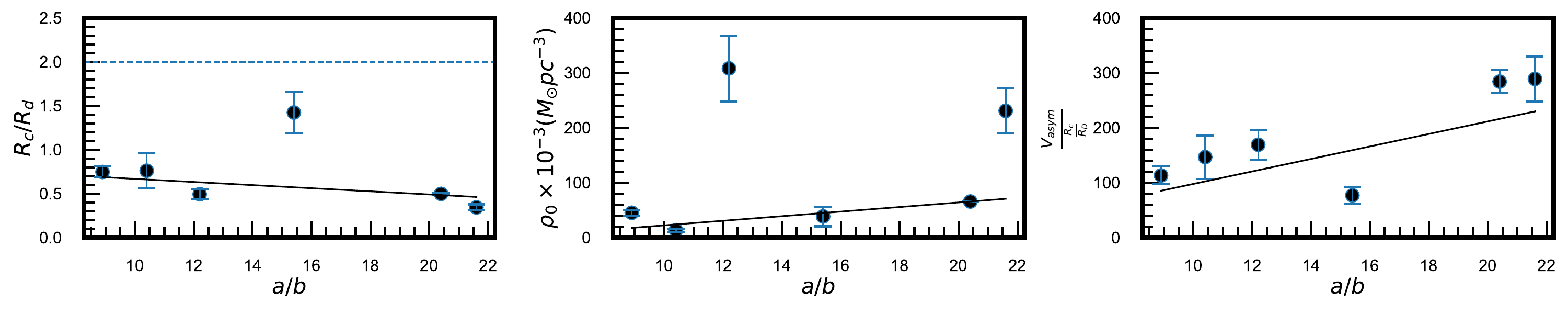}} 
\vspace{-2pt}
\caption{Comparison of the results from mass modeling obtained for extremely thin galaxies FGC 2366 and FGC 1440 with a previously studied sample of 
superthin galaxies for which mass models are available in the literature. In panel 1, we plot the ratio of the core radius to the disc scalelength $\rm R_{c}/R_{d}$, in panel 2 
we plot the central dark matter density $\rm (\rho_{0})$ as a function of the axis ratio. In the third panel, we show the ratio of asymptotic rotation velocity $\rm (V_{asym})$ 
to the compactness parameter $\rm (R_{c}/R_{d})$. Superimposed on the data points in each panel is the best-fitting regression line. }	
\end{figure*}

With the complete mass inventory in place, we construct the dynamical models of FGC 2366 using the multi-component model of gravitationally-coupled 
stars + gas in the dark matter halo's force field. Further, using action-angle formalism in AGAMA, we set up equilibrium distribution function-based 
models of FGC 2366. The multi-component model and the action angle formalism have their own advantages and disadvantages. In the multi-component model, 
the thin disc assumption when solving Poisson's equation and other simplifying assumptions decouple vertical and radial motions. 
The multi-component model can be used to calculate the stellar vertical velocity dispersion using the observed scaleheight as a constraint. 
The SCM approach can be used to derive both the radial dispersion and vertical dispersion from the moments of the distribution function 
once the system's total potential is known. In the SCM model, the gas component is treated as a rigid potential.
Thus, in the SCM model, stars and gas are not treated equally, despite the gas being dynamically as important as the stars. The stellar disc's vertical velocity dispersion is first calculated using the multi-component approach. Next, we use stellar vertical velocity dispersion as initial input and set up 
equilibrium models using AGAMA to determine radial velocity dispersion. The multi-component model and the AGAMA-based distribution functions models are discussed in detail in
Section 1.1.3 and 1.1.4.

\subsection{Multi-component star-gas model of the galaxy} 
FGC 2366 is modeled as a coplanar coaxial equilibrium disc of stars and gas under the influence of the dark matter halo.
The procedure is discussed in \cite{narayan2002vertical,10.1093/mnras/stab155}, and in the context of deriving gas scaleheight in 
\cite{banerjee2011theoretical, patra2020theoretical}. We use the optical photometry, \HI{} surface density, and the \HI{} dispersion in 
conjunction with the mass models as the input parameters and the stellar and the gas scaleheights as constraints on the models. 
We iteratively solve the multi-component equation for stars and gas to determine their vertical velocity ($\rm \sigma_{z}(R)$). (see, Section 1.1.3)

\subsection{Self-consistent model (SCM) of the stellar disc using AGAMA:}
We model FGC 2366 as a system in dynamical equilibrium using the multi-component self-consistent iterative modeling method 
implemented in AGAMA \citep{vasiliev2018agama} to determine the vertical and radial velocity dispersion of the stars. We model FGC 2366 as 
a multi-component system comprising of stars, \HI{} gas, and the dark matter halo. Stellar and dark matter halos are characterized by 
their distribution functions. We use the values of stellar vertical velocity dispersion obtained from the 2-component method to construct 
AGAMA based models. We estimate FGC 2366's total potential using the density of each component estimated using tilted ring modeling, 
optical photometry, and mass models. Then, we generate actions corresponding to this potential using AGAMA's "action-finder" algorithm and compute 
the distribution function and density. We update the potential by solving Poisson's equation for density. We then calculate the actions from 
the updated potential and evaluate the distribution function iteratively till the potentials converge. The self-consistent iterative method is described in 
detail in section 1.1.4 in the introduction.

\subsection{Stellar Velocity dispersion}
Figure 5.12 shows SCM with PIS dark matter halo results. The central value of radial dispersion  $\rm \sigma_{R,0}=63.1\kms$, which falls off exponentially with
scalelength $\rm R_{\sigma_{R}}=6.2 \rm kpc$. $\rm \sigma_{z,0}$ is 21.0 \kms and falls off exponentially with a scalelenght $\rm R_{\sigma_{z}} =4.2 kpc$. In Figure 5.13, 
we show the self-consistent models using NFW halo; the central vertical velocity dispersion of stars is 22.6 \kms and falls off exponentially with 
scalelength $\rm R_{\\sigma_{z}}$ equal to 3.8 kpc. Vertical velocity dispersion in FGC 2366 $\rm \sigma_{z,0}=23\kms$ is close to the values measured for Milky Way's 
thin disc's $\rm \sigma_{z,0}=25\kms$. $\rm \sigma_{R,0}=61 \kms$ for FGC 2366 is higher than $\rm \sigma_{R,0}=40 \kms$ for Milkyway's thin disc obtained by 
\cite{sharma2014kinematic}; see also \cite{bovy2012milky}. We also calculate $\rm \sigma_{z}/\sigma_{R}$ for FGC 2366 and find that the minimum value is 
equal to 0.22 for both the PIS and NFW halos. $\rm \sigma_{z}/\sigma_{R}$ is used to determine the radial and vertical contributions of various disc heating agents.
The  stellar vertical velocity dispersion of FGC 2366 is comparable to that of the thin disc in the Milky Way, and the value of $\rm \frac{\sigma_{z}}\sigma_{R}$
is smaller than the value measured for the stars Milky Way. $\rm \frac{\sigma_{z}}\sigma_{R}$ indicates that the stellar velocity ellipse is dominated by 
anisotropy in the radial direction. Disc thickening is caused by various internal processes, including the presence of bars, spiral arms, giant molecular clouds, and the radial migration of stars.
The effect of disc heating is usually assessed in terms of increased stellar dispersion; i.e., the ratio $\rm \sigma_{z}/\sigma_{R}$, which is an important diagnostic for 
comparing the agents that increase disc heating in the plane to those out of the plane. \citep{pinna2018revisiting}. \cite{jenkins1990spiral} showed that increased
contribution from the spiral arms to the total potential decreases the values $\rm \sigma_{z}/\sigma_{R}$, indicating that the excursion of stars from the plane 
are modest, as shown in FGC 2366 and other superthin galaxies. Due to the edge-on orientation of the extremely thin galaxies, spiral structure can not be detected. 
Thus, it is unclear what suppresses the galaxy's disc heating mechanism. Figure 5.14 shows $\rm Min(\sigma_{z}/\sigma_{R})$ versus $\rm a/b$. 
$\rm Min(\sigma_{z}/\sigma_{R})$ decreases with $\rm a/b$, implying that the stellar velocity ellipsoid's anisotropy increases with increasing planar to vertical axis ratio.

\begin{figure*}
\hspace{-1.7cm}
\resizebox{180mm}{45mm}{\includegraphics{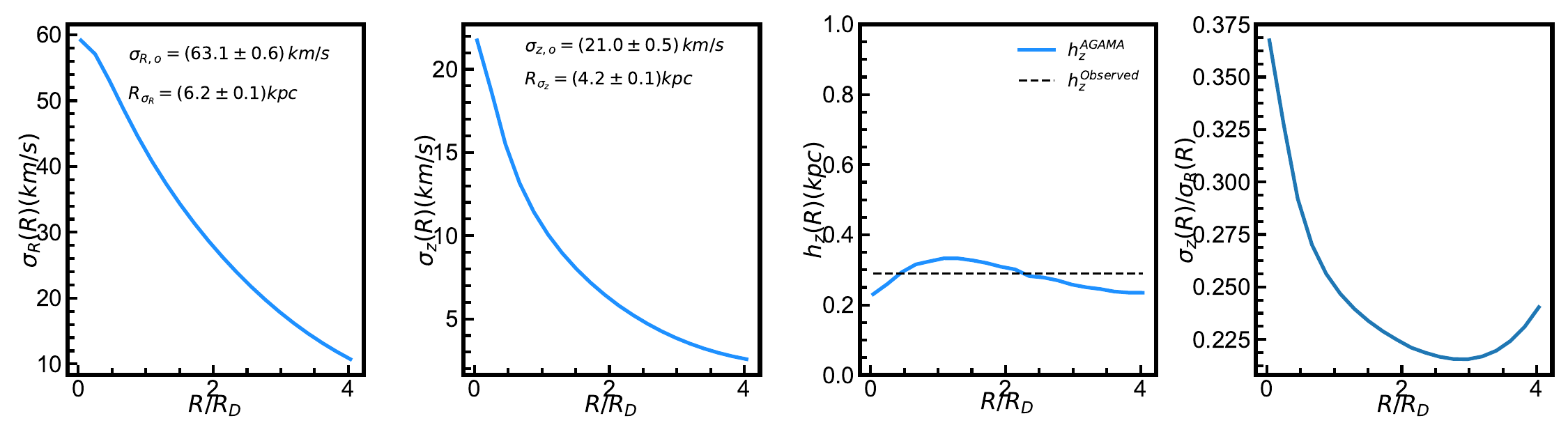}} 
\caption{[Panel 1:]  Radial velocity dispersion 
[Panel 2] Vertical velocity dispersion [Panel 3] Scaleheight [Panel 4] Ratio of vertical-to-stellar velocity dispersion, as a function of galactocentric 
radius normalized by disc scale length, as determined by dynamical modeling with a PIS dark matter halo. In the third panel, the dotted line shows the observed scaleheight.}	
\end{figure*}

\begin{figure*}
\hspace{-1.7cm}
%\vspace{-5cm}
\resizebox{180mm}{45mm}{\includegraphics{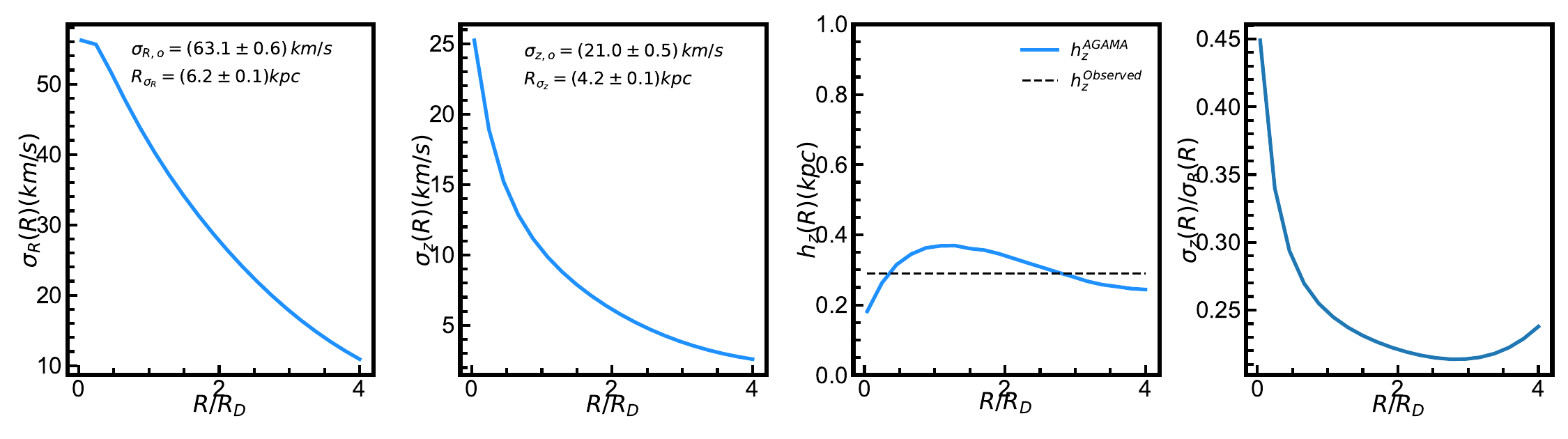}} 
\caption{Same as above, but the dark matter density is modeled using a cuspy NFW profile.}	
\end{figure*}

\begin{figure}
\hspace{3.0cm}
\resizebox{90mm}{70mm}{\includegraphics{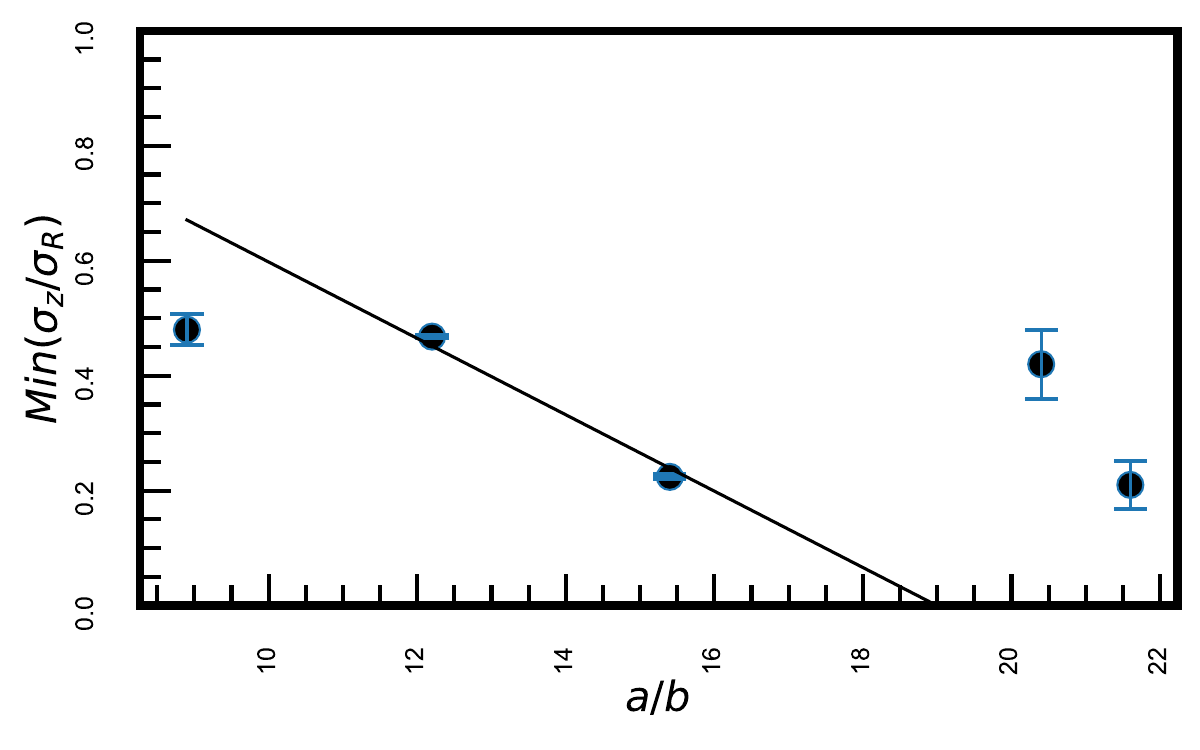}} 
\caption{Ratio of the vertical-to-radial stellar velocity dispersion as a function of the of major-to- minor axes ratio as derived from dynamical models. 
The regression line fit is superposed.}	
\end{figure}

\subsection{Disc stability of FGC 2366}
Figure 5.15 shows $\rm Q_{RW}$ for FGC2366's $\rm star+gas$ disc (see Section 1.1.5). 
The minimum value of $\rm Q_{RW}=3.1$, suggesting the disc is stable against axisymmetric instabilities. $\rm Q_{RW}$ at $\rm 1.5R_{d}$ is higher than the global median 
of spiral galaxies \citep{romeo2017drives} $\rm Q_{RW}2.2$. Figure 5.16 compares the $\rm Q_{RW}$ values at 1.5$\rm R_{d}$ of the extremely thin galaxy FGC 2366 and FGC 1440, 
with a sample of superthin galaxies. $\rm Q_{RW}$ at $\rm 1.5R_{d}$ is shown as function of $\rm a/b$. The regression line indicates that the thin discs have higher dynamical 
stability. Even the minimum value of $Q_{RW}$ is higher than the condition for marginal stability, indicating that the FGC 2366 is stable against the growth of 
axisymmetric instabilities. Further, counterintuitively, thinner discs have higher dynamical stability.

\begin{figure}
\hspace{3.0cm}
\resizebox{90mm}{70mm}{\includegraphics{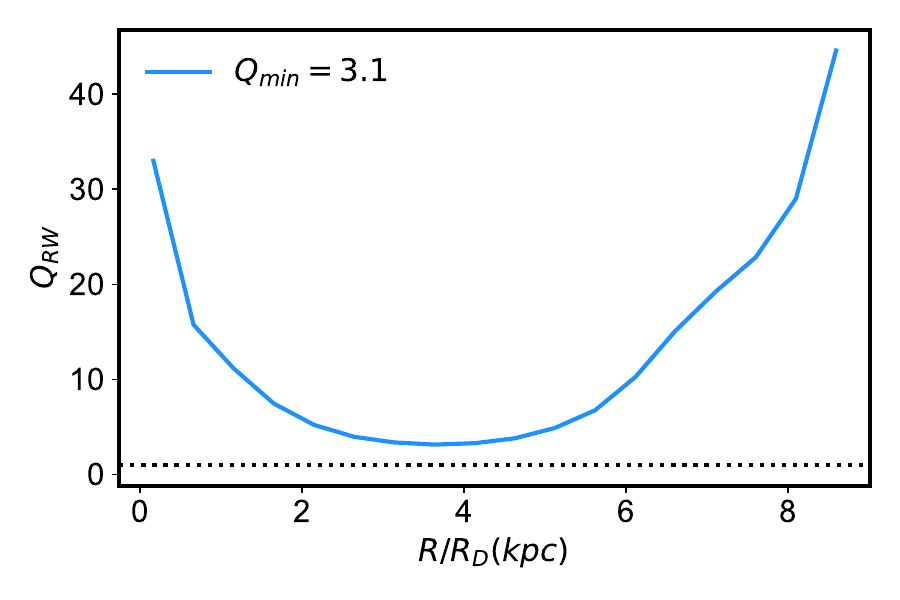}} 
\caption{ 2-component disc dynamical stability parameter $\rm Q_{RW}$ as a function of $\rm R/R_D$ for the FGC 2366.}
\end{figure}

\begin{figure}
\hspace{3.0cm}
\resizebox{90mm}{70mm}{\includegraphics{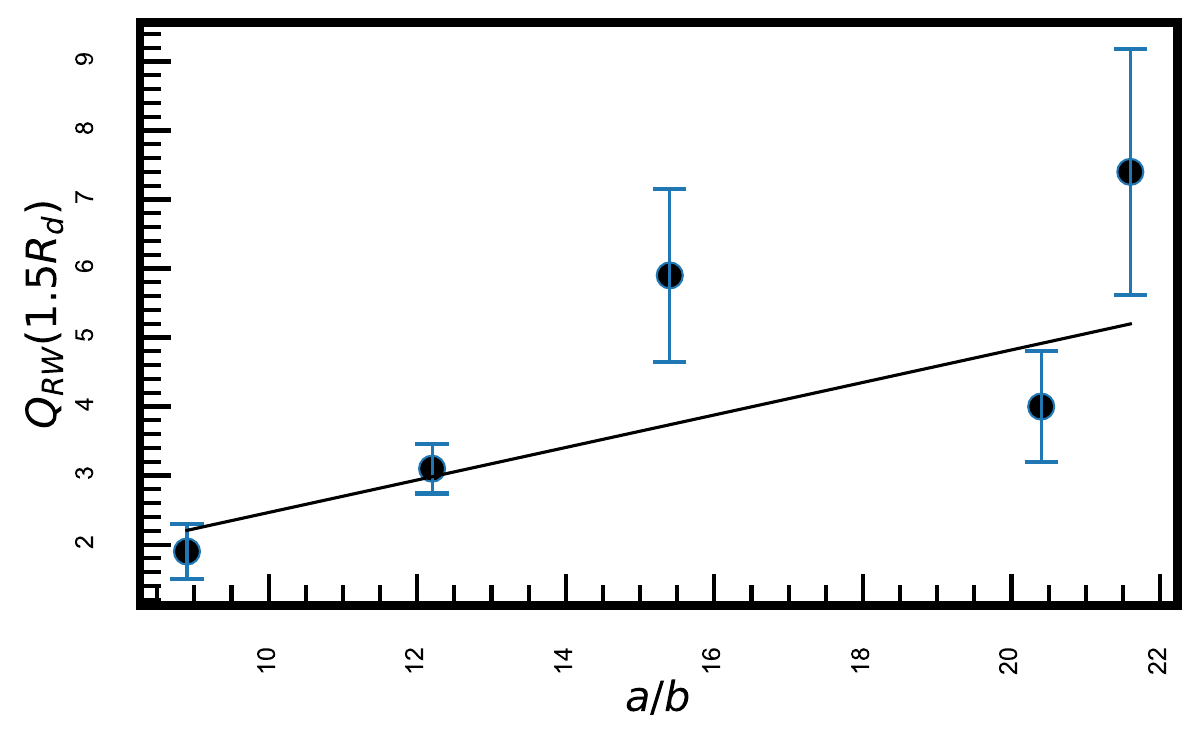}} 
\caption{2-component disc dynamical stability parameter $\rm Q_{RW}$ at $\rm R=1.5 R_d$ as a function of the major-to-minor axes ratio $\rm a/b$}
\end{figure}

\subsection{Specific angular momentum}
Superthin galaxies lie above the 95.4$\rm \%$ confidence interval of $\rm j_{*}-M_{*}$ line obtained for ordinary bulgeless disc galaxies studied \cite{obreschkow2014fundamental}, 
where $\rm j_{*}$ is the specific angular momentum of stars and $M_{*}$ is the corresponding stellar mass.
Thus a high stellar specific angular momentum for a given stellar mass and may provide cues to their origin. In Figure 5.17, we show the specific angular momentum 
for stars [Left Panel], \HI{} disc [Middle Panel], and baryons [Right Panel] further, we superpose the j-M relations for ordinary bulgeless disc galaxies 
\citep{jadhav2019specific}, as well as a sample of disc galaxies spanning a mass range $\rm 6\leq log(M_{gas}/M_{\odot}) \leq11$ \citep{2021A&A...647A..76M}.
We compare it to FGC 1440 and other superthin galaxies. We observe that, unlike FGC1440, FGC 2366 does not follow 
the $\rm j_{*} - M_{*}$ relations. The mismatch between the j - M relations of the stars and gas may be due to the supernovae feedback driving up high angular
momentum gas outwards from the center of the galaxy, as seen by the presence of  
\HI{} hole (see Figure 5.5, Right panel). On the other hand, we not that the FGC 2366 complies with the $\rm j_{g} - M_{g}$ relation like ordinary disc galaxies, 
but deviates from the  $\rm j_{b} - M_{b}$ relations. Next, we compare specific angular momentum obtained for FGC 2366 to that of other superthin galaxies and FGC 1440. We plot 
$\rm j_{*}$ versus $a/b$ in Figure 5.18. Thin discs possibly have a higher specific angular momentum, as $\rm j_{*}$ increases with $\rm a/b$. 
Table 5.8 shows the dynamical parameters that may drive the superthin vertical structure in FGC 2366. The error bars on the parameters are derived by using Monte-Carlo sampling.
FGC 2366 deviates from the j - M relation akin to previously studied superthin galaxies, thus has a higher specific angular momentum corresponding to its stellar mass. Further,
we note that galaxies with larger a/b values have a higher stellar specific angular momentum.

\begin{figure*}
\hspace{-2cm}
\resizebox{180mm}{50mm}{\includegraphics{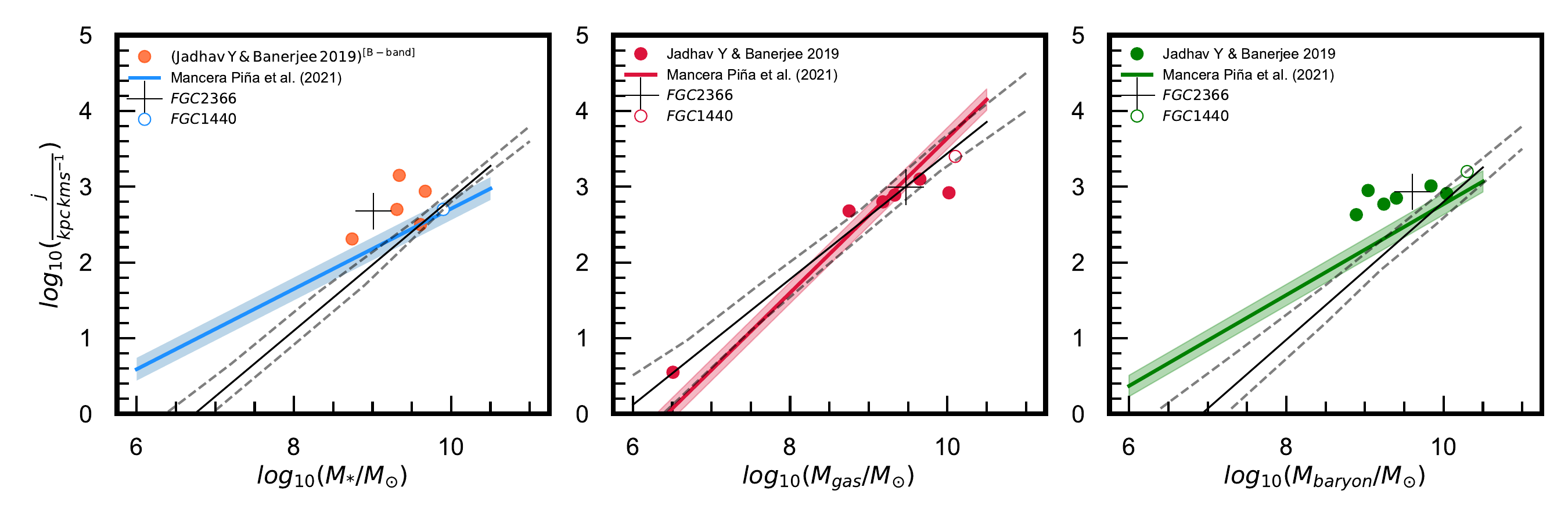}} 
\caption{Comparison of the specific angular momentum of [Left Panel]stars, [Middle Panel] gas, and [Right Panel] baryons [stars + gas] in FGC 2366 with previously 
studied superthin galaxies. Extremely thin galaxies FGC 1440 and FGC 2366 are shown using $\rm 'open-circle'$ and $\rm 'cross'$. The $\rm 'blue'$ regression line is 
for a sample of disc galaxies studied by \cite{2021A&A...647A..76M}. The solid black line shows the regression line obtained by \cite{jadhav2019specific} 
for a sample of six ordinary bulgeless disc galaxies with bulge fraction less than 0.05 from a larger sample of disc galaxies studied by \cite{obreschkow2014fundamental}. 
The dashed line shows the 95.4 $\rm \%$ confidence interval on the regression line obtained by \cite{jadhav2019specific}.}
\end{figure*}

\begin{figure*}
\hspace{3.0cm}
\resizebox{90mm}{70mm}{\includegraphics{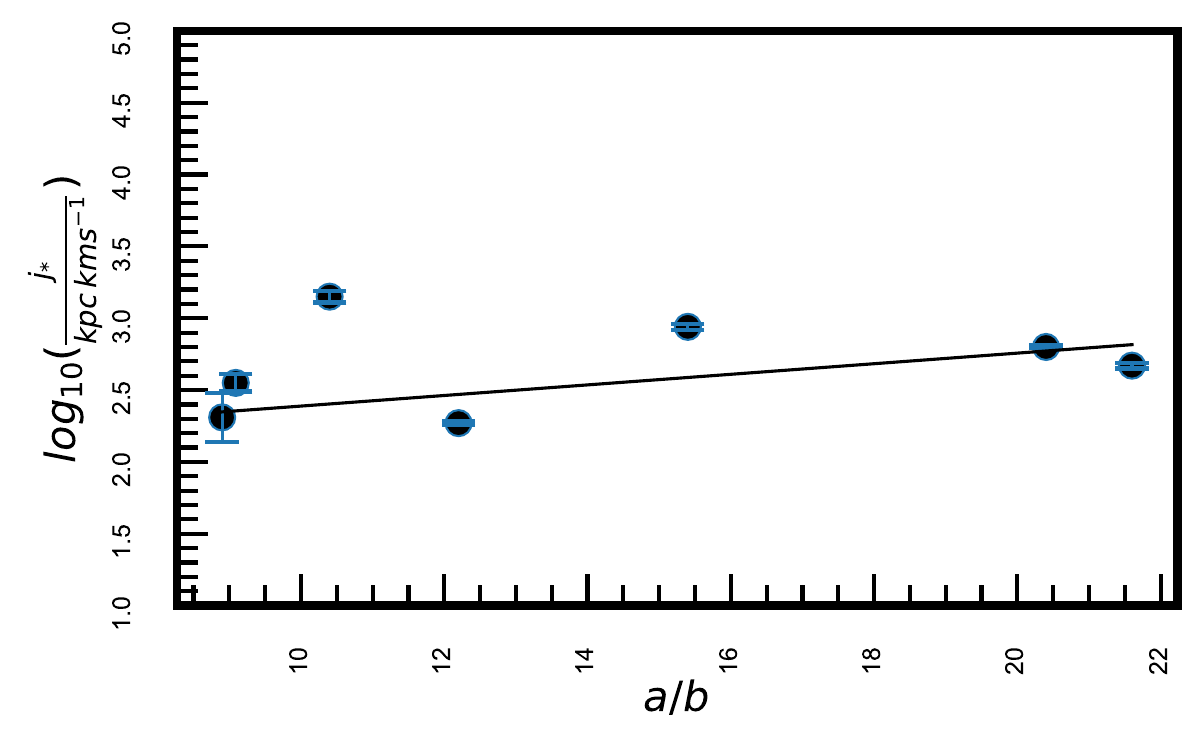}} 
\caption{Stellar specific angular momenta $\rm j_{*}$ of superthin galaxies as a function of the major-to-minor axis ratio $\rm a/b$}
\end{figure*}

\begin{table}
\begin{minipage}{110mm}
\hfill{}
\caption{Best-fitting model dynamical parameters of FGC 2366}
%\centering
\begin{tabular}{|c|l|}
\hline
\hline
Parameter                  & Value         \\
\hline
$R_{c}/R_{d}$ \footnote{Compactness parameter}                       &     $0.35 \pm 0.03$                \\ 
$V_{\rm{rot}}/(R_{c}/R_{d})$   \footnote {Ratio of asymptotic rotational velocity to the compactness parameter}         & $288  \pm 40$           \\
$\sigma_{z}/\sigma_{R}$   \footnote{Ratio of vertical to radial velocity dispersion}           &     $0.2 \pm  0.04$              \\
$Q_{RW}(1.5R_{d})$  \footnote{2-component stability parameter at 1.5 $R_{d}$ }                 &     $7.0  \pm 1.8$               \\
$\rm{log}_{10}( j_{*}/ \rm{kpc} \kms{} )$ \footnote{Specific stellar angular momentum of stars}                   &     $2.67 \pm 0.02$              \\ 
\hline
\end{tabular}
\hfill{}
\label{table: table 8}
\end{minipage}
\end{table}

\subsection{Principal Component Analysis}
To find the most important dynamical component controlling the vertical structure of stellar discs in superthin galaxies, we perform a Principal Component Analysis (PCA) 
of the parameters $\rm j_{*}$, $\rm Q_{RW}(1.5R_{d})$ $\rm Min(\sigma_{z}/\sigma_{R})$, $\rm a/b$, $\rm V_{rot}/(R_{c}/R_{d})$.
We build $\rm \sim$ 100 realizations of each galaxy's 5-D vector of parameter space by randomly sampling from the parameter space of five galaxies: 
IC 2233, FGC 1540, UGC 7321, FGC 1440, \& FGC 2366. We generate a 5-D Gaussian distribution for each parameter using the best-fitting value as the mean and 
the error as the standard deviation. Figure 5.19 shows the results from the principal component analysis of the resulting sample.
The first two components explain 80$\rm \%$ of the variation in data. Our computations show that PC1 and PC2 each have a positive eigenvalue. Figure 5.20 shows 
the parameter breakdown of the principal components. $\rm a/b$ and $\rm Q_{RW}(1.5 R_{d})$ contribute most to PC1, which explains $55\rm \%$ of the variation in data. 
Maximum contribution to PC1 and PC2 are from  $\rm a/b$, $\rm V_{rot}/R_{c}/R_{d}$, and $\rm Q_{RW}$. This suggests that the key drivers of the superthin 
vertical structure are (1) disc dynamical stability as given by $\rm Q_{RW}(1.5 R_{d})$ and (2) dark matter dominance at inner galactocentric radii as 
indicated by $\rm V_{rot}/(R_{c}/R_{d})$.

\emph{Principal component analysis suggests that $\rm V_{rot}/R_{c}/R_{d}$, and $\rm Q_{RW}$  may be driving the superthin vertical structure.}

\begin{figure}
\hspace{3cm}
\resizebox{90mm}{70mm}{\includegraphics{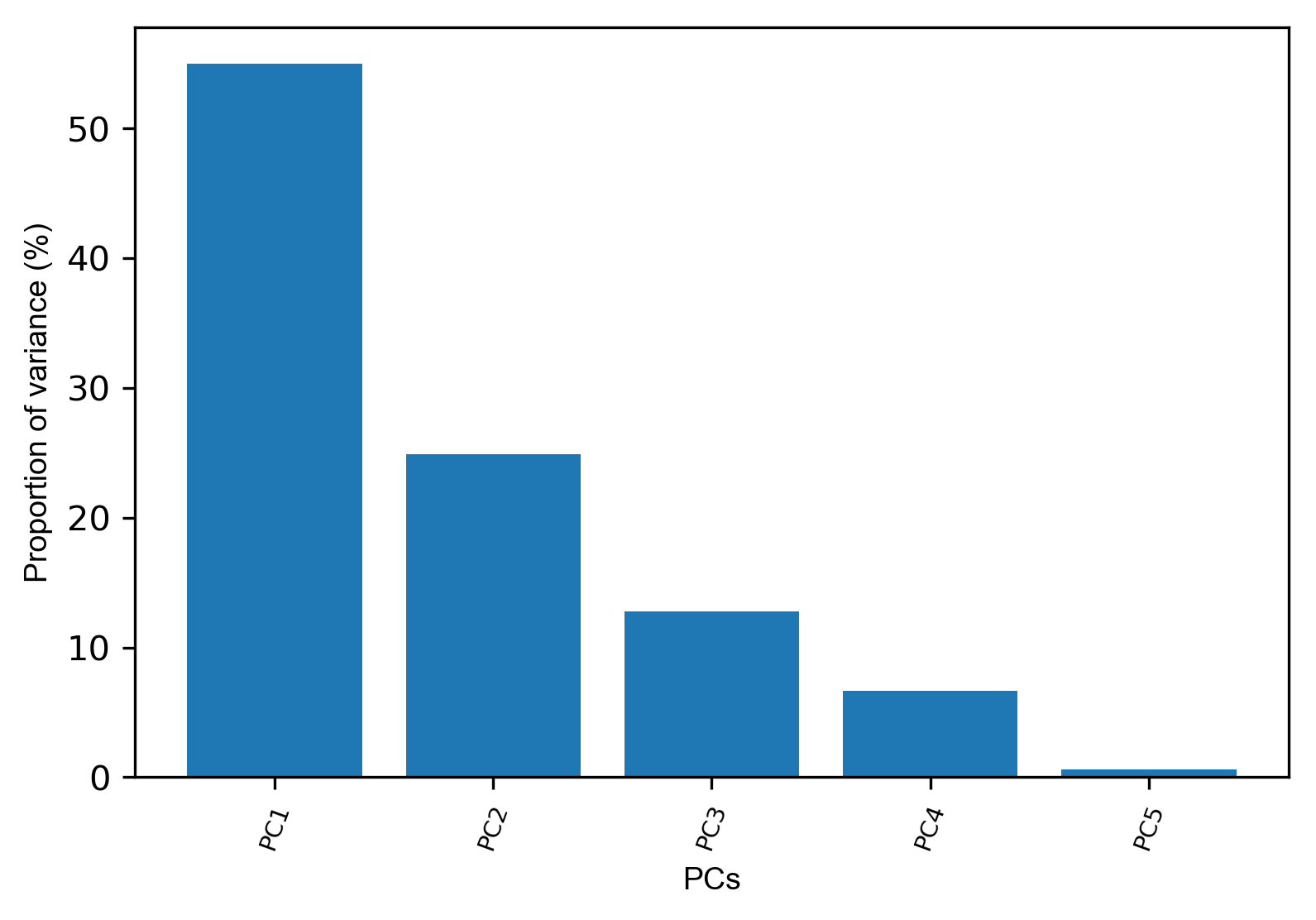}} 
\caption{Scree plot from the principal component analysis of the dynamical parameters responsible for driving the superthin stellar discs. The height indicates the percentage of 
variance in the data each principal component explains.}	
\end{figure}

\begin{figure}
\hspace{1.5cm}
\resizebox{155mm}{140mm}{\includegraphics{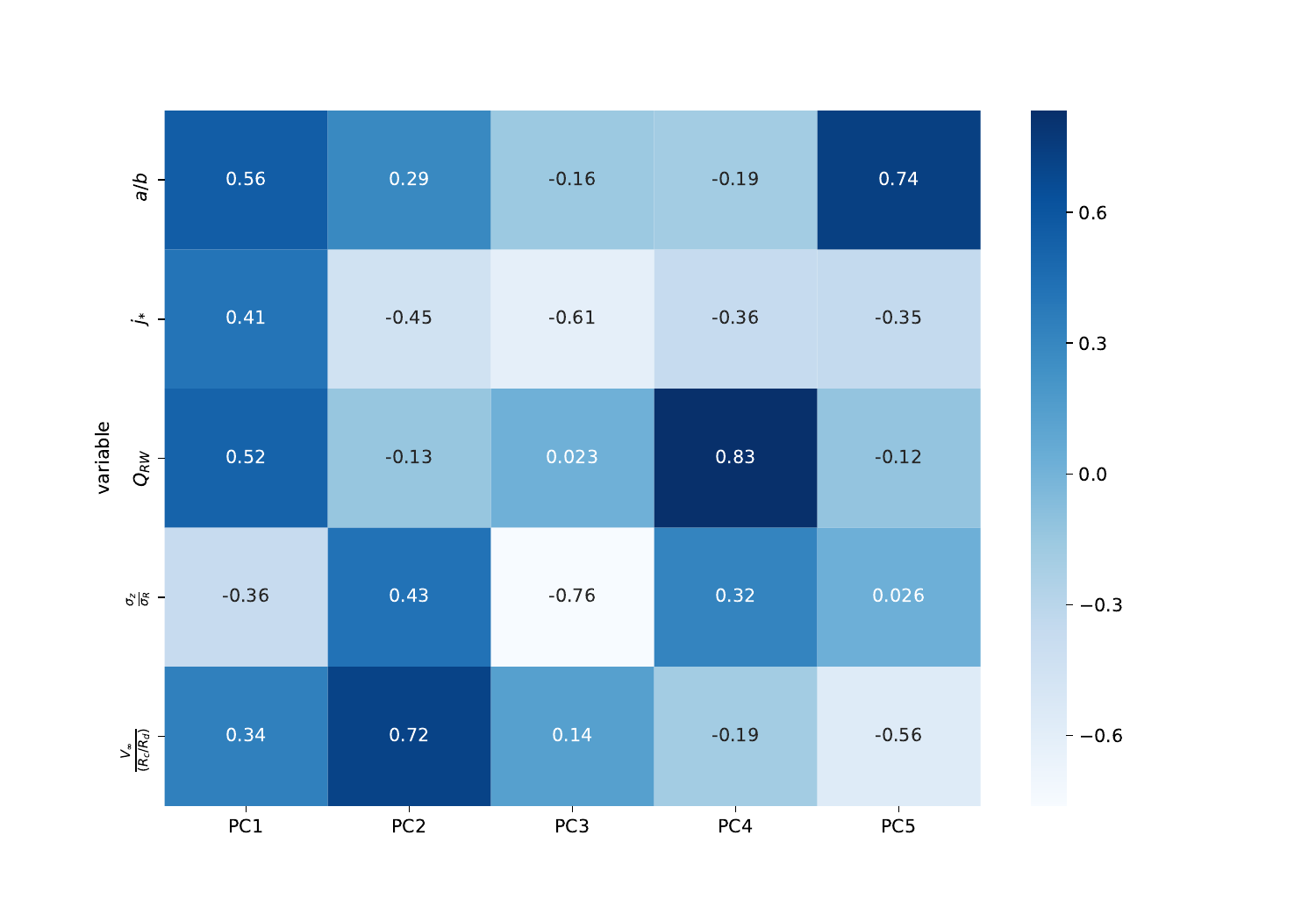}} 
\caption{Loadings of the Principal Component Analysis of the dynamical parameters.}	
\end{figure}

\section{Conclusions}
In this study, we present GMRT \HI{} 21cm radio-synthesis observations of the thinnest known galaxy FGC 2366 (a/b = 21.6) and use
CASA to carry out data reduction and 3D tilted ring modeling to infer the kinematics and structural features of neutral \HI{} distribution. Using the total rotation 
curve and optical photometry, we derive the distribution of the dark matter. We build self-consistent models of FGC 2366 using the 2-component model of 
gravitationally-coupled stars and gas and AGAMA-based self-consistent methods. Finally, we explore FGC 2366's 
structure and dynamics in terms of dark matter density, stellar velocity dispersion, disc dynamical stability, and specific angular momentum.

\begin{itemize}

\item We obtain the global \HI{} profile of FGC 2366 and fit it with the busy function. We find that the total 
      flux is 5.4 Jy \kms and the $\rm log(M_{\HI}/M_{\odot})$ is 9.1.
 
\item From our initial assessment of the \HI{} data cube, moment maps, and position velocity diagrams, we find 
      that \HI{} distribution is uniform and unperturbed, with no warping or flaring.
 
\item FGC 2366 is modeled using 3D tilted rings. FGC 2366 has an asymptotic velocity equal to 100 \kms and an 
      inclination equal to $\rm 87^{\circ}$. FGC 2366's \HI{} central velocity dispersion is 15 \kms at the center and falls to 10 \kms in the outer radius.
 
\item FGC 2366 hosts a compact dark matter halo with $\rm R_{c}/R_{d2}<2$. FGC 2366's dark matter concentration parameter is equal to 7.9.
 
\item Using AGAMA's self-consistent iterative technique, we construct equilibrium models of FGC 2366 to derive radial and vertical velocity dispersion. 
     FGC 2366's radial velocity dispersion is 63.1 \kms and falls off exponentially with scalelength equal to 6.2 kpc. The value of central vertical velocity dispersion 
     is equal to 21 \kms and fall-off with an exponential scalelength equal to 4.2 kpc . $\rm \sigma_{z}/\sigma_{R}$ lies between 0.21 and 0.31,  
     compared to 0.45 for Milkyway stars in the solar neighborhood.
       
\item  We do a Principal Component Analysis of the dynamical parameters i.e. $\rm j_{*}$, $\rm Q_{RW}(1.5R_{d})$, $\rm a/b$, $\rm Min(\sigma_{z}/\sigma_{R})$ and
 $\rm V_{rot}/(R_{c}/R_{d})$ to find out the dominant dynamical component responsible for the superthin vertical structure. We conclude disc dynamical stability, 
 represented by $\rm Q_{RW}(1.5_{R_d})$, and dark matter dominance at inner galactocentric radii, indicated by $\rm V_{rot}/(R_{c}/R_{d})$, are essentially responsible 
 for a superthin stellar vertical structure.      
\end{itemize}

\thispagestyle{empty}

\thispagestyle{empty}
\chapter[Conclusions]{\fontsize{50}{50}\selectfont Chapter 6}
\chaptermark{Conclusions}

\textbf{\Huge{Conclusions}}

Superthin galaxies are a class of low surface brightness, bulgeless, disc galaxies, exhibiting sharp, needle-like images in the optical, implying strikingly high values of planar-to-vertical axes ratios of the stellar disc, which possibly indicates the presence of an ultra-cold stellar disc, the dynamical stability of which continues to be a mystery. In this thesis, we have explored the dynamical origins of the extremely thin galaxies by constructing dynamical models of a sample of superthin galaxies using stellar photometry and HI 21cm radio-synthesis observations as constraints and employing a Markov Chain Monte Carlo method, also checking the consistency of our model results using AGAMA i.e. Action-based Galaxy Modelling Architecture. Our current study of the superthin galaxies is derived from the detailed dynamical modeling of a limited sample of superthin galaxies using stellar photometry  and \HI{} 21cm radio-synthesis observations, already available in the literature ($ 8 < a/b < 16$): UGC 7321, IC5249, IC 2233, FGC 1540, which we refer to as \emph{superthin} galaxies. In addition, we have carried out GMRT HI 21cm radio-synthesis observations of two of the thinnest galaxies known: FGC1440 ($ a/b = 20.4$) and FGC2366 ($a/b = 21.6$), the stellar photometry being available in the literature, which we refer to as \emph{extremely thin} galaxies. The key findings from this thesis may be summarized below:

\begin{itemize}
    \item \textbf{Ultra-cold stellar discs with high disc dynamical stability:}\\ The stellar discs of superthin galaxies are ultra-cold 
    in the vertical direction, with the ratio of the vertical stellar velocity dispersion-to-the rotational velocity comparable to that of 
    the thin disc of the Milky Way. Interestingly, however, inspite of being ultra-cold in the vertical direction, they have high dynamical stability against local, 
    axisymmetric perturbations than ordinary disc galaxies.
    
    \item \textbf{Superthin galaxies in braneworld gravity:}\\ Superthin galaxies are dark matter dominated at all radii and hence are 
    ideal test-beds to constrain dark matter models in alternative theories of gravity. We find that the ultra-cold stellar disc of a 
    superthin galaxy is perfectly consistent with the dark mass derived from the braneworld gravity, also complying with the observational 
    constraints of the rotation curve and the vertical scaleheights of the disc.
    
    \item \textbf{\HI{} Observations and modeling of extremely thin galaxies:}\\ We carry out GMRT HI 21cm radio-synthesis observations of the two  
    thinnest galaxies, FGC1440 $\rm(a/b \sim 20.4)$ and FGC2366 $\rm (a/b \sim 21.6)$, and construct their dynamical models, also using stellar photometry available in the literature. 
    We note that the extremely thin galaxies are not characterized by the extraordinarily high value of any particular dynamical parameters possibly 
    regulating the superthin vertical structure: \\
    \begin{itemize}
    \item Dark matter dominance at inner galactocentric radii given by $V_{\rm{rot}}/{(R_{c}/R_{d})}$,
    \item Disc dynamical stability $Q_{RW}$ 
    \item Ratio of the vertical-to-radial stellar velocity dispersion $(\sigma_{z,s}/\sigma_{R,s})$, and
    \item Specific angular momentum of the stellar disc $j*$.
    \end{itemize}
    
    Therefore we conclude they are governed by the same formation and evolution mechanisms as ordinary superthins.
    
    \item \textbf{Why are the stellar discs of some galaxies superthin?}\\ To identify the physical mechanism primarily responsible for the 
    superthin vertical structure, we carry out a Principal Component Analysis of the above dynamical parameters along with $a/b$ for all 
    superthins and the extremely thin galaxies studied so far. We note that the first two principal components explain $\sim$ 80$\%$ of the variation in the data, 
    and the major contributions are from $a/b$, $Q_{RW}$ and $V_{\rm{rot}}/{(R_{c}/R_{d})}$. This possibly indicates that high values of the disc dynamical 
    stability and dark matter dominance at inner galactocentric radii are fundamentally responsible for the superthin stellar discs.
\end{itemize}
\thispagestyle{empty}

%\thispagestyle{empty}

%\thispagestyle{empty}

%\include{chapter3}

%\thispagestyle{empty}
%\include{chapter4}

%\thispagestyle{empty}
%\include{chapter5}

%\thispagestyle{empty}
%\include{chapter6}

%\thispagestyle{empty}
%\include{chapter7}

%\thispagestyle{empty}
%\include{conclusions}
%\cleardoublepage

%\include{Preliminary}

%\thispagestyle{empty}
%\addcontentsline{toc}{chapter}{Bibliography}

@ARTICLE{1948AnAp...11..247D,
       author = {{de Vaucouleurs}, Gerard},
        title = "{Recherches sur les Nebuleuses Extragalactiques}",
      journal = {Annales d'Astrophysique},
         year = 1948,
        month = jan,
       volume = {11},
        pages = {247},
       adsurl = {https://ui.adsabs.harvard.edu/abs/1948AnAp...11..247D},
      adsnote = {Provided by the SAO/NASA Astrophysics Data System}
}

@article{de1956survey,
  title={Survey of bright galaxies south of-35 deg. declination with the 30-inch Reynolds reflector (1952-1955)},
  author={De Vaucouleurs, G{\'e}rard},
  journal={Canberra: Mount Stromlo},
  year={1956}
}

@article{sersic1968atlas,
  title={Atlas de galaxias australes, ed},
  author={Sersic, JL},
  journal={Sersic, JL},
  volume={2},
  pages={23},
  year={1968}
}

@article{dalcanton2000structural,
  title={A structural and dynamical study of late-type, edge-on galaxies. I. Sample selection and imaging data},
  author={Dalcanton, Julianne J and Bernstein, Rebecca A},
  journal={The Astronomical Journal},
  volume={120},
  number={1},
  pages={203},
  year={2000},
  publisher={IOP Publishing}
}

@article{dalcanton2002structural,
  title={A structural and dynamical study of late-type, edge-on galaxies. II. Vertical color gradients and the detection of ubiquitous thick disks},
  author={Dalcanton, Julianne J and Bernstein, Rebecca A},
  journal={The Astronomical Journal},
  volume={124},
  number={3},
  pages={1328},
  year={2002},
  publisher={IOP Publishing}
}

@article{yoachim2005kinematics,
  title={The kinematics of thick disks in external galaxies},
  author={Yoachim, Peter and Dalcanton, Julianne J},
  journal={The Astrophysical Journal},
  volume={624},
  number={2},
  pages={701},
  year={2005},
  publisher={IOP Publishing}
}

@article{yoachim2008lick,
  title={Lick indices in the thin and thick disks of edge-on disk galaxies},
  author={Yoachim, Peter and Dalcanton, Julianne J},
  journal={The Astrophysical Journal},
  volume={683},
  number={2},
  pages={707},
  year={2008},
  publisher={IOP Publishing}
}

@article{juric2008milky,
  title={The Milky Way tomography with SDSS. I. Stellar number density distribution},
  author={Juri{\'c}, Mario and Ivezi{\'c}, {\v{Z}}eljko and Brooks, Alyson and Lupton, Robert H and Schlegel, David and Finkbeiner, Douglas and Padmanabhan, Nikhil and Bond, Nicholas and Sesar, Branimir and Rockosi, Constance M and others},
  journal={The Astrophysical Journal},
  volume={673},
  number={2},
  pages={864},
  year={2008},
  publisher={IOP Publishing}
}

@article{begum2004kinematics,
  title={Kinematics of two dwarf galaxies in the NGC 6946 group},
  author={Begum, Ayesha and Chengalur, Jayaram N},
  journal={Astronomy \& Astrophysics},
  volume={424},
  number={2},
  pages={509--517},
  year={2004},
  publisher={EDP Sciences}
}

@article{patra2014modelling,
  title={Modelling H i distribution and kinematics in the edge-on dwarf irregular galaxy KK250},
  author={Patra, Narendra Nath and Banerjee, Arunima and Chengalur, Jayaram N and Begum, Ayesha},
  journal={Monthly Notices of the Royal Astronomical Society},
  volume={445},
  number={2},
  pages={1424--1429},
  year={2014},
  publisher={Oxford University Press}
}

@article{agnese2018results,
  title={Results from the super cryogenic dark matter search experiment at Soudan},
  author={Agnese, R and Aramaki, T and Arnquist, IJ and Baker, W and Balakishiyeva, D and Banik, S and Barker, D and Thakur, R Basu and Bauer, DA and Binder, T and others},
  journal={Physical review letters},
  volume={120},
  number={6},
  pages={061802},
  year={2018},
  publisher={APS}
}

@article{springel2005simulations,
  title={Simulations of the formation, evolution and clustering of galaxies and quasars},
  author={Springel, Volker and White, Simon DM and Jenkins, Adrian and Frenk, Carlos S and Yoshida, Naoki and Gao, Liang and Navarro, Julio and Thacker, Robert and Croton, Darren and Helly, John and others},
  journal={nature},
  volume={435},
  number={7042},
  pages={629--636},
  year={2005},
  publisher={Nature Publishing Group}
}

@article{rubin1983dark,
  title={Dark matter in spiral galaxies},
  author={Rubin, Vera C},
  journal={Scientific American},
  volume={248},
  number={6},
  pages={96--109},
  year={1983},
  publisher={JSTOR}
}

@article{zwicky1937masses,
  title={On the Masses of Nebulae and of Clusters of Nebulae},
  author={Zwicky, Fritz},
  journal={The Astrophysical Journal},
  volume={86},
  pages={217},
  year={1937}
}

@article{de2008high,
  title={High-resolution rotation curves and galaxy mass models from THINGS},
  author={De Blok, WJG and Walter, Fabian and Brinks, Elias and Trachternach, C and Oh, SH and Kennicutt, Robert C},
  journal={The Astronomical Journal},
  volume={136},
  number={6},
  pages={2648},
  year={2008},
  publisher={IOP Publishing}
}

@article{oh2015high,
  title={High-resolution mass models of dwarf galaxies from LITTLE THINGS},
  author={Oh, Se-Heon and Hunter, Deidre A and Brinks, Elias and Elmegreen, Bruce G and Schruba, Andreas and Walter, Fabian and Rupen, Michael P and Young, Lisa M and Simpson, Caroline E and Johnson, Megan C and others},
  journal={The Astronomical Journal},
  volume={149},
  number={6},
  pages={180},
  year={2015},
  publisher={IOP Publishing}
}

@article{lelli2016sparc,
  title={SPARC: mass models for 175 disk galaxies with spitzer photometry and accurate rotation curves},
  author={Lelli, Federico and McGaugh, Stacy S and Schombert, James M},
  journal={The Astronomical Journal},
  volume={152},
  pages={157},
  year={2016},
  publisher={IOP Publishing}
}

@article{rizzo2020dynamically,
  title={A dynamically cold disk galaxy in the early Universe},
  author={Rizzo, F and Vegetti, S and Powell, D and Fraternali, F and McKean, JP and Stacey, HR and White, SDM},
  journal={Nature},
  volume={584},
  number={7820},
  pages={201--204},
  year={2020},
  publisher={Nature Publishing Group}
}

@inproceedings{navarro1996structure,
  title={The structure of cold dark matter halos},
  author={Navarro, Julio F},
  booktitle={Symposium-international astronomical union},
  volume={171},
  pages={255--258},
  year={1996},
  organization={Cambridge University Press}
}

@article{navarro1996cores,
  title={The cores of dwarf galaxy haloes},
  author={Navarro, Julio F and Eke, Vincent R and Frenk, Carlos S},
  journal={Monthly Notices of the Royal Astronomical Society},
  volume={283},
  number={3},
  pages={L72--L78},
  year={1996},
  publisher={Blackwell Science Ltd Oxford, UK}

}

@article{gilmore1983new,
  title={New light on faint stars--III. Galactic structure towards the South Pole and the Galactic thick disc},
  author={Gilmore, Gerard and Reid, Neil},
  journal={Monthly Notices of the Royal Astronomical Society},
  volume={202},
  number={4},
  pages={1025--1047},
  year={1983},
  publisher={The Royal Astronomical Society}
}

@article{bovy2012spatial,
  title={The spatial structure of mono-abundance sub-populations of the Milky Way disk},
  author={Bovy, Jo and Rix, Hans-Walter and Liu, Chao and Hogg, David W and Beers, Timothy C and Lee, Young Sun},
  journal={The Astrophysical Journal},
  volume={753},
  number={2},
  pages={148},
  year={2012},
  publisher={IOP Publishing}
}

@article{quinn1993heating,
  title={Heating of galactic disks by mergers},
  author={Quinn, PJ and Hernquist, Lars and Fullagar, DP},
  journal={The Astrophysical Journal},
  volume={403},
  pages={74--93},
  year={1993}
}

@article{villalobos2008simulations,
  title={Simulations of minor mergers--I. General properties of thick discs},
  author={Villalobos, {\'A}lvaro and Helmi, Amina},
  journal={Monthly Notices of the Royal Astronomical Society},
  volume={391},
  number={4},
  pages={1806--1827},
  year={2008},
  publisher={Blackwell Publishing Ltd Oxford, UK}
}

@article{bekki2011origin,
  title={Origin of chemical and dynamical properties of the galactic thick disk},
  author={Bekki, Kenji and Tsujimoto, Takuji},
  journal={The Astrophysical Journal},
  volume={738},
  number={1},
  pages={4},
  year={2011},
  publisher={IOP Publishing}
}

@article{brook2004emergence,
  title={The emergence of the thick disk in a cold dark matter universe},
  author={Brook, Chris B and Kawata, Daisuke and Gibson, Brad K and Freeman, Ken C},
  journal={The Astrophysical Journal},
  volume={612},
  number={2},
  pages={894},
  year={2004},
  publisher={IOP Publishing}
}

@article{agertz2009disc,
  title={Disc formation and the origin of clumpy galaxies at high redshift},
  author={Agertz, Oscar and Teyssier, Romain and Moore, Ben},
  journal={Monthly Notices of the Royal Astronomical Society: Letters},
  volume={397},
  number={1},
  pages={L64--L68},
  year={2009},
  publisher={The Royal Astronomical Society}
}

@article{ceverino2010high,
  title={High-redshift clumpy discs and bulges in cosmological simulations},
  author={Ceverino, Daniel and Dekel, Avishai and Bournaud, Frederic},
  journal={Monthly Notices of the Royal Astronomical Society},
  volume={404},
  number={4},
  pages={2151--2169},
  year={2010},
  publisher={Blackwell Publishing Ltd Oxford, UK}
}

@article{spitzer1951possible,
  title={The Possible Influence of Interstellar Clouds on Stellar Velocities.},
  author={Spitzer Jr, Lyman and Schwarzschild, Martin},
  journal={The Astrophysical Journal},
  volume={114},
  pages={385},
  year={1951}
}

@article{lacey1991tidally,
  title={Tidally triggered galaxy formation. I-Evolution of the galaxy luminosity function.},
  author={Lacey, Cedric and Silk, Joseph},
  journal={Astrophysical journal.},
  volume={381},
  pages={14--32},
  year={1991},
  publisher={Institute of Physics}
}

@article{eggen1962evidence,
  title={Evidence from the motions of old stars that the Galaxy collapsed.},
  author={Eggen, OJ and Lynden-Bell, D and Sandage, AR},
  journal={The Astrophysical Journal},
  volume={136},
  pages={748},
  year={1962}
}

@article{heald2011westerbork,
  title={The Westerbork Hydrogen Accretion in LOcal GAlaxieS (HALOGAS) survey-I. Survey description and pilot observations},
  author={Heald, George and J{\'o}zsa, Gyula and Serra, Paolo and Zschaechner, Laura and Rand, Richard and Fraternali, Filippo and Oosterloo, Tom and Walterbos, Rene and J{\"u}tte, Eva and Gentile, Gianfranco},
  journal={Astronomy \& astrophysics},
  volume={526},
  pages={A118},
  year={2011},
  publisher={EDP Sciences}
}

@article{zibetti2004haloes,
  title={Haloes around edge-on disc galaxies in the Sloan Digital Sky Survey},
  author={Zibetti, Stefano and White, Simon DM and Brinkmann, Jon},
  journal={Monthly Notices of the Royal Astronomical Society},
  volume={347},
  number={2},
  pages={556--568},
  year={2004},
  publisher={Blackwell Science Ltd Oxford, UK}
}

@article{monachesi2016ghosts,
  title={The GHOSTS survey--II. The diversity of halo colour and metallicity profiles of massive disc galaxies},
  author={Monachesi, Antonela and Bell, Eric F and Radburn-Smith, David J and Bailin, Jeremy and de Jong, Roelof S and Holwerda, Benne and Streich, David and Silverstein, Grace},
  journal={Monthly Notices of the Royal Astronomical Society},
  volume={457},
  number={2},
  pages={1419--1446},
  year={2016},
  publisher={Oxford University Press}
}

@article{swaters1997hi,
  title={The HI halo of NGC 891},
  author={Swaters, RA and Sancisi, R and Van Der Hulst, JM},
  journal={The Astrophysical Journal},
  volume={491},
  number={1},
  pages={140},
  year={1997},
  publisher={IOP Publishing}
}

@article{olling1995usage,
  title={On the usage of flaring gas layers to determine the shape of dark matter halos},
  author={Olling, Rob P},
  journal={arXiv preprint astro-ph/9505002},
  year={1995}
}

@article{olling1996highly,
  title={The highly flattened dark matter halo of NGC 4244},
  author={Olling, Rob P},
  journal={arXiv preprint astro-ph/9605111},
  year={1996}
}

@article{banerjee2011progressively,
  title={PROGRESSIVELY MORE PROLATE DARK MATTER HALO IN THE OUTER GALAXY AS TRACED BY FLARING H i GAS},
  author={Banerjee, Arunima and Jog, Chanda J},
  journal={The Astrophysical Journal Letters},
  volume={732},
  number={1},
  pages={L8},
  year={2011},
  publisher={IOP Publishing}
}

@article{banerjee2008flattened,
  title={The flattened dark matter halo of M31 as deduced from the observed HI scale heights},
  author={Banerjee, Arunima and Jog, Chanda J},
  journal={The Astrophysical Journal},
  volume={685},
  number={1},
  pages={254},
  year={2008},
  publisher={IOP Publishing}
}

@article{karachentsev2003revised,
  title={The revised flat galaxy catalogue},
  author={Karachentsev, ID and Karachentseva, VE and Kudrya, Yu N and Sharina, ME and Parnovsky, SL},
  journal={arXiv preprint astro-ph/0305566},
  year={2003}
}

@article{goad1981spectroscopic,
  title={Spectroscopic observations of superthin galaxies},
  author={Goad, JW and Roberts, MS},
  journal={The Astrophysical Journal},
  volume={250},
  pages={79--86},
  year={1981}
}

@article{kautsch2021spectroscopic,
  title={A Spectroscopic Survey of Superthin Galaxies},
  author={Kautsch, Stefan J and Bizyaev, Dmitry and Makarov, Dimitry I and Reshetnikov, Vladimir P and Mosenkov, Alexander V and Antipova, Alexandra V},
  journal={Research Notes of the AAS},
  volume={5},
  number={3},
  pages={43},
  year={2021},
  publisher={IOP Publishing}
}

@article{bizyaev2021spectral,
  title={Spectral Observations of Superthin Galaxies},
  author={Bizyaev, Dmitry and Makarov, DI and Reshetnikov, VP and Mosenkov, AV and Kautsch, SJ and Antipova, AV},
  journal={The Astrophysical Journal},
  volume={914},
  number={2},
  pages={104},
  year={2021},
  publisher={IOP Publishing}
}

@article{bizyaev2020near,
  title={Near-infrared photometry of superthin edge-on galaxies},
  author={Bizyaev, Dmitry and Tatarnikov, Andrey and Shatsky, Nikolai and Najip, Aurik and Birlak, Marina and Voziakova, Olga},
  journal={Astronomische Nachrichten},
  volume={341},
  number={3},
  pages={314--323},
  year={2020},
  publisher={Wiley Online Library}
}

@article{bizyaev2017very,
  title={Very thin disc galaxies in the SDSS catalogue of edge-on galaxies},
  author={Bizyaev, DV and Kautsch, SJ and Sotnikova, N Ya and Reshetnikov, Vladimir P and Mosenkov, Aleksander V},
  journal={Monthly Notices of the Royal Astronomical Society},
  volume={465},
  number={4},
  pages={3784--3792},
  year={2017},
  publisher={Oxford University Press}
}

@article{kautsch2006catalog,
  title={A catalog of edge-on disk galaxies-From galaxies with a bulge to superthin galaxies},
  author={Kautsch, SJ and Grebel, EK and Barazza, FD and Gallagher, JS},
  journal={Astronomy \& Astrophysics},
  volume={445},
  number={2},
  pages={765--778},
  year={2006},
  publisher={EDP Sciences}
}

@article{antipova2021database,
  title={The database for studying edge-on galaxies},
  author={Antipova, AV and Makarov, DI and Savchenko, SS},
  journal={ASTRONOMY AT THE EPOCH OF MULTIMESSENGER STUDIES},
  pages={347},
  year={2021}
}

@article{matthews2000h,
  title={An H I survey of highly flattened, edge-on, pure disk galaxies},
  author={Matthews, LD and Van Driel, W},
  journal={Astronomy and Astrophysics Supplement Series},
  volume={143},
  number={3},
  pages={421--456},
  year={2000},
  publisher={EDP Sciences}
}

@article{huchtmeier2005hi,
  title={HI observations of edge-on spiral galaxies},
  author={Huchtmeier, WK and Karachentsev, ID and Karachentseva, VE and Kudrya, Yu N and Mitronova, SN},
  journal={Astronomy \& Astrophysics},
  volume={435},
  number={2},
  pages={459--463},
  year={2005},
  publisher={EDP Sciences}
}

@article{o2010dark,
  title={The dark matter halo shape of edge-on disk galaxies-III. Modelling the HI observations: results},
  author={O'Brien, Jess Clare and Freeman, KC and van der Kruit, PC},
  journal={Astronomy \& astrophysics},
  volume={515},
  pages={A62},
  year={2010},
  publisher={EDP Sciences}
}

@article{uson2003hi,
  title={HI Imaging observations of superthin galaxies. I. UGC 7321},
  author={Uson, Juan M and Matthews, LD},
  journal={The Astronomical Journal},
  volume={125},
  number={5},
  pages={2455},
  year={2003},
  publisher={IOP Publishing}
}

@article{matthews2007h,
  title={H i Imaging Observations of Superthin Galaxies. II. IC 2233 and the Blue Compact Dwarf NGC 2537},
  author={Matthews, Lynn D and Uson, Juan M},
  journal={The Astronomical Journal},
  volume={135},
  number={1},
  pages={291},
  year={2007},
  publisher={IOP Publishing}
}

@article{abe1999observation,
  title={Observation of the halo of the edge-on galaxy IC 5249},
  author={Abe, F and Bond, IA and Carter, BS and Dodd, RJ and Fujimoto, M and Hearnshaw, JB and Honda, M and Jugaku, J and Kabe, S and Kilmartin, PM and others},
  journal={The Astronomical Journal},
  volume={118},
  number={1},
  pages={261},
  year={1999},
  publisher={IOP Publishing}
}

@article{van2001kinematics,
  title={Kinematics and dynamics of the" superthin" edge-on disk galaxy IC 5249},
  author={Van Der Kruit, PC and Jim{\'e}nez-Vicente, J and Kregel, M and Freeman, KC},
  journal={Astronomy \& Astrophysics},
  volume={379},
  number={2},
  pages={374--383},
  year={2001},
  publisher={EDP Sciences}
}

@article{kurapati2018mass,
  title={Mass modelling of a superthin galaxy, FGC 1540},
  author={Kurapati, Sushma and Banerjee, Arunima and Chengalur, Jayaram N and Makarov, Dmitry and Borisov, Svyatoslav and Afanasiev, Anton and Antipova, Aleksandra},
  journal={Monthly Notices of the Royal Astronomical Society},
  volume={479},
  number={4},
  pages={5686--5695},
  year={2018},
  publisher={Oxford University Press}
}

@article{van1986dark,
  title={Dark matter in spiral galaxies},
  author={Van Albada, TS and Sancisi, R},
  journal={Philosophical Transactions of the Royal Society of London. Series A, Mathematical and Physical Sciences},
  volume={320},
  number={1556},
  pages={447--464},
  year={1986},
  publisher={The Royal Society London}
}

@article{banerjee2017mass,
  title={Mass modelling of superthin galaxies: IC5249, UGC7321 and IC2233},
  author={Banerjee, Arunima and Bapat, Disha},
  journal={Monthly Notices of the Royal Astronomical Society},
  volume={466},
  number={3},
  pages={3753--3761},
  year={2017},
  publisher={Oxford University Press}
}

@article{peters2017shape,
  title={The shape of dark matter haloes--III. Kinematics and structure of the H i disc},
  author={Peters, SPC and van der Kruit, PC and Allen, RJ and Freeman, KC},
  journal={Monthly Notices of the Royal Astronomical Society},
  volume={464},
  number={1},
  pages={32--47},
  year={2017},
  publisher={Oxford University Press}
}

@article{banerjee2010dark,
  title={Dark matter dominance at all radii in the superthin galaxy UGC 7321},
  author={Banerjee, Arunima and Matthews, Lynn D and Jog, Chanda J},
  journal={New Astronomy},
  volume={15},
  number={1},
  pages={89--95},
  year={2010},
  publisher={Elsevier}
}

@article{banerjee2013some,
  title={Why are some galaxy discs extremely thin?},
  author={Banerjee, Arunima and Jog, Chanda J},
  journal={Monthly Notices of the Royal Astronomical Society},
  volume={431},
  number={1},
  pages={582--588},
  year={2013},
  publisher={The Royal Astronomical Society}
}

@article{zasov1991thickness,
  title={Thickness of Thin Stellar Disks and the Mass of the Dark Halo},
  author={Zasov, AV and Makarov, DI and Mikhailova, EA},
  journal={Soviet Astronomy Letters},
  volume={17},
  pages={374},
  year={1991}
}

@article{spitzer1953possible,
  title={The possible influence of interstellar clouds on stellar velocities. II.},
  author={Spitzer Jr, Lyman and Schwarzschild, Martin},
  journal={The Astrophysical Journal},
  volume={118},
  pages={106},
  year={1953}
}

@article{goldreich1965gravitational,
  title={I. Gravitational stability of uniformly rotating disks},
  author={Goldreich, Peter and Lynden-Bell, D},
  journal={Monthly Notices of the Royal Astronomical Society},
  volume={130},
  number={2},
  pages={97--124},
  year={1965},
  publisher={Oxford University Press Oxford, UK}
}

@article{barbanis1967orbits,
  title={Orbits in spiral galaxies and the velocity dispersion of population I stars},
  author={Barbanis, B and Woltjer, L},
  journal={The Astrophysical Journal},
  volume={150},
  pages={461},
  year={1967}
}

@article{toth1992galactic,
  title={Galactic disks, infall, and the global value of Omega},
  author={Toth, G and Ostriker, JP},
  journal={The Astrophysical Journal},
  volume={389},
  pages={5--26},
  year={1992}
}

@article{lacey1985massive,
  title={Massive black holes in galactic halos?},
  author={Lacey, CG and Ostriker, JP},
  journal={Astrophysical journal.},
  volume={299},
  pages={633--652},
  year={1985},
  publisher={IOP}
}

@book{binney2011galactic,
  title={Galactic dynamics},
  author={Binney, James and Tremaine, Scott},
  volume={13},
  year={2011},
  publisher={Princeton university press}
}

@article{rohlfs1977lectures,
  title={Lectures on density wave theory},
  author={Rohlfs, Kristen},
  year={1977}
}

@article{de1988potential,
  title={Potential-density pairs for galaxies},
  author={de Zeeuw, Tim and Pfenniger, Daniel},
  journal={Monthly Notices of the Royal Astronomical Society},
  volume={235},
  pages={949--995},
  year={1988}
}

@article{van1989photometry,
  title={Photometry of disks in galaxies},
  author={Van der Kruit, PC},
  journal={The World of Galaxies},
  pages={256--275},
  year={1989},
  publisher={Springer}
}

@article{hastings1970monte,
  title={Monte Carlo sampling methods using Markov chains and their applications},
  author={Hastings, W Keith},
  year={1970},
  publisher={Oxford University Press}
}

@article{haario2006dram,
  title={DRAM: efficient adaptive MCMC},
  author={Haario, Heikki and Laine, Marko and Mira, Antonietta and Saksman, Eero},
  journal={Statistics and computing},
  volume={16},
  number={4},
  pages={339--354},
  year={2006},
  publisher={Springer}
}

@article{soetaert2010inverse,
  title={Inverse modelling, sensitivity and monte carlo analysis in R using package FME},
  author={Soetaert, Karline and Petzoldt, Thomas and others},
  journal={Journal of Statistical Software},
  volume={33},
  number={3},
  pages={1--28},
  year={2010}
}

@INPROCEEDINGS{2001ASPC..230..221M,
       author = {{Merrifield}, M.~R. and {Gerssen}, J. and {Kuijken}, K.},
        title = "{The Origins of Disk Heating}",
     keywords = {Astrophysics},
    booktitle = {Galaxy Disks and Disk Galaxies},
         year = 2001,
       editor = {{Funes}, Jos{\'e} G. and {Corsini}, Enrico Maria},
       series = {Astronomical Society of the Pacific Conference Series},
       volume = {230},
        month = jan,
        pages = {221-224},
archivePrefix = {arXiv},
       eprint = {astro-ph/0008290},
 primaryClass = {astro-ph},
       adsurl = {https://ui.adsabs.harvard.edu/abs/2001ASPC..230..221M},
      adsnote = {Provided by the SAO/NASA Astrophysics Data System}
}

@article{vasiliev2018agama,
  title={AGAMA: action-based galaxy modelling architecture},
  author={Vasiliev, Eugene},
  journal={Monthly Notices of the Royal Astronomical Society},
  volume={482},
  number={2},
  pages={1525--1544},
  year={2018},
  publisher={Oxford University Press}
}

@article{binney2012actions,
  title={Actions for axisymmetric potentials},
  author={Binney, James},
  journal={Monthly Notices of the Royal Astronomical Society},
  volume={426},
  number={2},
  pages={1324--1327},
  year={2012},
  publisher={Blackwell Science Ltd Oxford, UK}
}

@article{rubin1970rotation,
  title={Rotation of the Andromeda nebula from a spectroscopic survey of emission regions},
  author={Rubin, Vera C and Ford Jr, W Kent},
  journal={The Astrophysical Journal},
  volume={159},
  pages={379},
  year={1970}
}

@phdthesis{bosma1978distribution,
  title={The distribution and kinematics of neutral hydrogen in spiral galaxies of various morphological types},
  author={Bosma, Albert},
  year={1978},
  school={Rijksuniversiteit te Groningen.}
}

@article{agnese2013silicon,
  title={Silicon detector dark matter results from the final exposure of CDMS II},
  author={Agnese, R and Ahmed, Z and Anderson, AJ and Arrenberg, S and Balakishiyeva, D and Thakur, R Basu and Bauer, DA and Billard, J and Borgland, A and Brandt, D and others},
  journal={Physical review letters},
  volume={111},
  number={25},
  pages={251301},
  year={2013},
  publisher={APS}
}

@article{pawlowski2015persistence,
  title={On the persistence of two small-scale problems in $\Lambda$CDM},
  author={Pawlowski, Marcel S and Famaey, Benoit and Merritt, David and Kroupa, Pavel},
  journal={The Astrophysical Journal},
  volume={815},
  number={1},
  pages={19},
  year={2015},
  publisher={IOP Publishing}
}

@article{kroupa2012dark,
  title={The dark matter crisis: falsification of the current standard model of cosmology},
  author={Kroupa, Pavel},
  journal={Publications of the Astronomical Society of Australia},
  volume={29},
  number={4},
  pages={395--433},
  year={2012},
  publisher={Cambridge University Press}
}

@article{Kroupa:2014ria,
    author = "Kroupa, Pavel",
    title = "{Galaxies as simple dynamical systems: observational data disfavor dark matter and stochastic star formation}",
    eprint = "1406.4860",
    archivePrefix = "arXiv",
    primaryClass = "astro-ph.GA",
    doi = "10.1139/cjp-2014-0179",
    journal = "Can. J. Phys.",
    volume = "93",
    number = "2",
    pages = "169--202",
    year = "2015"
}

@article{Peebles:2010di,
    author = "Peebles, P.J.E. and Nusser, Adi",
    title = "{Clues from nearby galaxies to a better theory of cosmic evolution}",
    eprint = "1001.1484",
    archivePrefix = "arXiv",
    primaryClass = "astro-ph.CO",
    doi = "10.1038/nature09101",
    journal = "Nature",
    volume = "465",
    pages = "565--569",
    year = "2010"
}

@article{milgrom1983modification,
  title={A modification of the Newtonian dynamics as a possible alternative to the hidden mass hypothesis},
  author={Milgrom, Mordehai},
  journal={The Astrophysical Journal},
  volume={270},
  pages={365--370},
  year={1983}
}

@article{sanders2007confrontation,
  title={Confrontation of Modified Newtonian Dynamics with the rotation curves of early-type disc galaxies},
  author={Sanders, RH and Noordermeer, E},
  journal={Monthly Notices of the Royal Astronomical Society},
  volume={379},
  number={2},
  pages={702--710},
  year={2007},
  publisher={Blackwell Publishing Ltd Oxford, UK}
}

@article{sanders1998rotation,
  title={Rotation curves of Ursa major galaxies in the context of modified newtonian dynamics},
  author={Sanders, Robert H and Verheijen, MAW},
  journal={The Astrophysical Journal},
  volume={503},
  number={1},
  pages={97},
  year={1998},
  publisher={IOP Publishing}
}

@article{de1998testing,
  title={Testing modified newtonian dynamics with low surface brightness galaxies: rotation curve fits},
  author={de Blok, WJG and McGaugh, SS},
  journal={The Astrophysical Journal},
  volume={508},
  number={1},
  pages={132},
  year={1998},
  publisher={IOP Publishing}
}

@article{binetruy2000brane,
  title={Brane cosmological evolution in a bulk with cosmological constant},
  author={Binetruy, Pierre and Deffayet, Cedric and Ellwanger, Ulrich and Langlois, David},
  journal={Physics Letters B},
  volume={477},
  number={1-3},
  pages={285--291},
  year={2000},
  publisher={Elsevier}
}

@article{Csaki:1999mp,
      author         = "Csaki, Csaba and Graesser, Michael and Randall, Lisa and
                        Terning, John",
      title          = "{Cosmology of brane models with radion stabilization}",
      journal        = "Phys.Rev.",
      volume         = "D62",
      pages          = "045015",
      doi            = "10.1103/PhysRevD.62.045015",
      year           = "2000",
      eprint         = "hep-ph/9911406",
      archivePrefix  = "arXiv",
      primaryClass   = "hep-ph",
      reportNumber   = "SCIPP-99-49, HUTP-A061, NSF-ITP-99-130",
      SLACcitation   = "%%CITATION = HEP-PH/9911406;%%",
}

@inproceedings{Maartens:2001jx,
      author         = "Maartens, Roy",
      title          = "{Geometry and dynamics of the brane world}",
      booktitle      = "{Spanish Relativity Meeting on Reference Frames and
                        Gravitomagnetism (EREs2000) Valladolid, Spain, September
                        6-9, 2000}",
      url            = "http://alice.cern.ch/format/showfull?sysnb=2237527",
      year           = "2001",
      eprint         = "gr-qc/0101059",
      archivePrefix  = "arXiv",
      primaryClass   = "gr-qc",
      SLACcitation   = "%%CITATION = GR-QC/0101059;%%"
}

\%\% Papers On Brane

@article{Horava:1996ma,
      author         = "Horava, Petr and Witten, Edward",
      title          = "{Eleven-dimensional supergravity on a manifold with
                        boundary}",
      journal        = "Nucl.Phys.",
      volume         = "B475",
      pages          = "94-114",
      doi            = "10.1016/0550-3213(96)00308-2",
      year           = "1996",
      eprint         = "hep-th/9603142",
      archivePrefix  = "arXiv",
      primaryClass   = "hep-th",
      reportNumber   = "IASSNS-HEP-96-17, PUPT-1597",
      SLACcitation   = "%%CITATION = HEP-TH/9603142;%%",
}

@article{Horava:1995qa,
      author         = "Horava, Petr and Witten, Edward",
      title          = "{Heterotic and type I string dynamics from
                        eleven-dimensions}",
      journal        = "Nucl.Phys.",
      volume         = "B460",
      pages          = "506-524",
      doi            = "10.1016/0550-3213(95)00621-4",
      year           = "1996",
      eprint         = "hep-th/9510209",
      archivePrefix  = "arXiv",
      primaryClass   = "hep-th",
      reportNumber   = "IASSNS-HEP-95-86, PUPT-1571A",
      SLACcitation   = "%%CITATION = HEP-TH/9510209;%%",
}

@article{Randall:1999ee,
      author         = "Randall, Lisa and Sundrum, Raman",
      title          = "{A Large mass hierarchy from a small extra dimension}",
      journal        = "Phys.Rev.Lett.",
      volume         = "83",
      pages          = "3370-3373",
      doi            = "10.1103/PhysRevLett.83.3370",
      year           = "1999",
      eprint         = "hep-ph/9905221",
      archivePrefix  = "arXiv",
      primaryClass   = "hep-ph",
      reportNumber   = "MIT-CTP-2860, PUPT-1860, BUHEP-99-9",
      SLACcitation   = "%%CITATION = HEP-PH/9905221;%%",
}

@article{Goldberger:1999uk,
      author         = "Goldberger, Walter D. and Wise, Mark B.",
      title          = "{Modulus stabilization with bulk fields}",
      journal        = "Phys.Rev.Lett.",
      volume         = "83",
      pages          = "4922-4925",
      doi            = "10.1103/PhysRevLett.83.4922",
      year           = "1999",
      eprint         = "hep-ph/9907447",
      archivePrefix  = "arXiv",
      primaryClass   = "hep-ph",
      reportNumber   = "CALT-68-2232",
      SLACcitation   = "%%CITATION = HEP-PH/9907447;%%",
}

@article{Randall:1999vf,
      author         = "Randall, Lisa and Sundrum, Raman",
      title          = "{An Alternative to compactification}",
      journal        = "Phys.Rev.Lett.",
      volume         = "83",
      pages          = "4690-4693",
      doi            = "10.1103/PhysRevLett.83.4690",
      year           = "1999",
      eprint         = "hep-th/9906064",
      archivePrefix  = "arXiv",
      primaryClass   = "hep-th",
      reportNumber   = "MIT-CTP-2874, PUPT-1867, BUHEP-99-13",
      SLACcitation   = "%%CITATION = HEP-TH/9906064;%%",
}

@article{Verlinde:1999fy,
      author         = "Verlinde, Herman L.",
      title          = "{Holography and compactification}",
      journal        = "Nucl.Phys.",
      volume         = "B580",
      pages          = "264-274",
      doi            = "10.1016/S0550-3213(00)00224-8",
      year           = "2000",
      eprint         = "hep-th/9906182",
      archivePrefix  = "arXiv",
      primaryClass   = "hep-th",
      reportNumber   = "PUPT-1872, ITFA-99-14",
      SLACcitation   = "%%CITATION = HEP-TH/9906182;%%",
}

@article{Djouadi:2005gi,
      author         = "Djouadi, Abdelhak",
      title          = "{The Anatomy of electro-weak symmetry breaking. I: The
                        Higgs boson in the standard model}",
      journal        = "Phys.Rept.",
      volume         = "457",
      pages          = "1-216",
      doi            = "10.1016/j.physrep.2007.10.004",
      year           = "2008",
      eprint         = "hep-ph/0503172",
      archivePrefix  = "arXiv",
      primaryClass   = "hep-ph",
      reportNumber   = "LPT-ORSAY-05-17",
      SLACcitation   = "%%CITATION = HEP-PH/0503172;%%",
}

@article{Davoudiasl:1999tf,
      author         = "Davoudiasl, H. and Hewett, J.L. and Rizzo, T.G.",
      title          = "{Bulk gauge fields in the Randall-Sundrum model}",
      journal        = "Phys.Lett.",
      volume         = "B473",
      pages          = "43-49",
      doi            = "10.1016/S0370-2693(99)01430-6",
      year           = "2000",
      eprint         = "hep-ph/9911262",
      archivePrefix  = "arXiv",
      primaryClass   = "hep-ph",
      reportNumber   = "SLAC-PUB-8298",
      SLACcitation   = "%%CITATION = HEP-PH/9911262;%%",
}

@article{Davoudiasl:2000wi,
      author         = "Davoudiasl, H. and Hewett, J.L. and Rizzo, T.G.",
      title          = "{Experimental probes of localized gravity: On and off the
                        wall}",
      journal        = "Phys.Rev.",
      volume         = "D63",
      pages          = "075004",
      doi            = "10.1103/PhysRevD.63.075004",
      year           = "2001",
      eprint         = "hep-ph/0006041",
      archivePrefix  = "arXiv",
      primaryClass   = "hep-ph",
      reportNumber   = "SLAC-PUB-8436",
      SLACcitation   = "%%CITATION = HEP-PH/0006041;%%",
}

@article{Hundi:2011dc,
      author         = "Hundi, R.S. and SenGupta, Soumitra",
      title          = "{Fermion mass hierarchy in a multiple warped braneworld
                        model}",
      journal        = "J.Phys.",
      volume         = "G40",
      pages          = "075002",
      doi            = "10.1088/0954-3899/40/7/075002",
      year           = "2013",
      eprint         = "1111.1106",
      archivePrefix  = "arXiv",
      primaryClass   = "hep-th",
      SLACcitation   = "%%CITATION = ARXIV:1111.1106;%%",
}

@article{Mukhopadhyaya:2011gn,
      author         = "Mukhopadhyaya, Biswarup and Sen, Somasri and SenGupta,
                        Soumitra",
      title          = "{Matter-Gravity Interaction in a Multiply Warped
                        Braneworld,}",
      journal        = "J.Phys.",
      volume         = "G40",
      pages          = "015004",
      doi            = "10.1088/0954-3899/40/1/015004",
      year           = "2013",
      eprint         = "1106.1027",
      archivePrefix  = "arXiv",
      primaryClass   = "hep-ph",
      reportNumber   = "HRI-RECAPP-2011-005",
      SLACcitation   = "%%CITATION = ARXIV:1106.1027;%%",
}

@article{Chamblin:1999by,
      author         = "Chamblin, A. and Hawking, S.W. and Reall, H.S.",
      title          = "{Brane world black holes}",
      journal        = "Phys.Rev.",
      volume         = "D61",
      pages          = "065007",
      doi            = "10.1103/PhysRevD.61.065007",
      year           = "2000",
      eprint         = "hep-th/9909205",
      archivePrefix  = "arXiv",
      primaryClass   = "hep-th",
      reportNumber   = "DAMTP-1999-133",
      SLACcitation   = "%%CITATION = HEP-TH/9909205;%%",
}

@article{Garriga:1999yh,
      author         = "Garriga, Jaume and Tanaka, Takahiro",
      title          = "{Gravity in the brane world}",
      journal        = "Phys.Rev.Lett.",
      volume         = "84",
      pages          = "2778-2781",
      doi            = "10.1103/PhysRevLett.84.2778",
      year           = "2000",
      eprint         = "hep-th/9911055",
      archivePrefix  = "arXiv",
      primaryClass   = "hep-th",
      reportNumber   = "UAB-FT-476, OU-TAP-106",
      SLACcitation   = "%%CITATION = HEP-TH/9911055;%%",
}

@article{Lukas:1998qs,
      author         = "Lukas, Andre and Ovrut, Burt A. and Waldram, Daniel",
      title          = "{Cosmological solutions of Horava-Witten theory}",
      journal        = "Phys.Rev.",
      volume         = "D60",
      pages          = "086001",
      doi            = "10.1103/PhysRevD.60.086001",
      year           = "1999",
      eprint         = "hep-th/9806022",
      archivePrefix  = "arXiv",
      primaryClass   = "hep-th",
      reportNumber   = "UPR-805T, PUPT-1795",
      SLACcitation   = "%%CITATION = HEP-TH/9806022;%%",
}

@article{Dadhich:2000am,
      author         = "Dadhich, Naresh and Maartens, Roy and Papadopoulos,
                        Philippos and Rezania, Vahid",
      title          = "{Black holes on the brane}",
      journal        = "Phys.Lett.",
      volume         = "B487",
      pages          = "1-6",
      doi            = "10.1016/S0370-2693(00)00798-X",
      year           = "2000",
      eprint         = "hep-th/0003061",
      archivePrefix  = "arXiv",
      primaryClass   = "hep-th",
      SLACcitation   = "%%CITATION = HEP-TH/0003061;%%",
}

@article{Chakraborty:2007zh,
      author         = "Chakraborty, Subenoy and Bandyopadhyay, Tanwi",
      title          = "{Schwarzschild Solution on the Brane}",
      journal        = "Int.J.Theor.Phys.",
      volume         = "47",
      pages          = "2493-2499",
      doi            = "10.1007/s10773-008-9682-5",
      year           = "2008",
      eprint         = "0707.1182",
      archivePrefix  = "arXiv",
      primaryClass   = "gr-qc",
      SLACcitation   = "%%CITATION = ARXIV:0707.1182;%%",
}

@article{Binetruy:1999ut,
      author         = "Binetruy, Pierre and Deffayet, Cedric and Langlois,
                        David",
      title          = "{Nonconventional cosmology from a brane universe}",
      journal        = "Nucl.Phys.",
      volume         = "B565",
      pages          = "269-287",
      doi            = "10.1016/S0550-3213(99)00696-3",
      year           = "2000",
      eprint         = "hep-th/9905012",
      archivePrefix  = "arXiv",
      primaryClass   = "hep-th",
      reportNumber   = "LPT-ORSAY-99-25",
      SLACcitation   = "%%CITATION = HEP-TH/9905012;%%",
}

@article{Kaloper:1999sm,
      author         = "Kaloper, Nemanja",
      title          = "{Bent domain walls as brane worlds}",
      journal        = "Phys.Rev.",
      volume         = "D60",
      pages          = "123506",
      doi            = "10.1103/PhysRevD.60.123506",
      year           = "1999",
      eprint         = "hep-th/9905210",
      archivePrefix  = "arXiv",
      primaryClass   = "hep-th",
      reportNumber   = "SU-ITP-99-25",
      SLACcitation   = "%%CITATION = HEP-TH/9905210;%%",
}

@article{Csaki:1999jh,
      author         = "Csaki, Csaba and Graesser, Michael and Kolda, Christopher
                        F. and Terning, John",
      title          = "{Cosmology of one extra dimension with localized
                        gravity}",
      journal        = "Phys.Lett.",
      volume         = "B462",
      pages          = "34-40",
      doi            = "10.1016/S0370-2693(99)00896-5",
      year           = "1999",
      eprint         = "hep-ph/9906513",
      archivePrefix  = "arXiv",
      primaryClass   = "hep-ph",
      reportNumber   = "LBNL-43586, UCB-PTH-99-28, LBL-43586",
      SLACcitation   = "%%CITATION = HEP-PH/9906513;%%",
}

@article{Cvetic:1996vr,
      author         = "Cvetic, Mirjam and Soleng, Harald H.",
      title          = "{Supergravity domain walls}",
      journal        = "Phys.Rept.",
      volume         = "282",
      pages          = "159-223",
      doi            = "10.1016/S0370-1573(96)00035-X",
      year           = "1997",
      eprint         = "hep-th/9604090",
      archivePrefix  = "arXiv",
      primaryClass   = "hep-th",
      reportNumber   = "IASSNS-HEP-96-25, CERN-TH-96-97",
      SLACcitation   = "%%CITATION = HEP-TH/9604090;%%",
}

@article{Benakli:1998pe,
      author         = "Benakli, Karim",
      title          = "{Cosmological solution in M theory on S-1 / Z(2)}",
      journal        = "Int.J.Mod.Phys.",
      volume         = "D8",
      pages          = "153-160",
      doi            = "10.1142/S0218271899000134",
      year           = "1999",
      eprint         = "hep-th/9804096",
      archivePrefix  = "arXiv",
      primaryClass   = "hep-th",
      SLACcitation   = "%%CITATION = HEP-TH/9804096;%%",
}

@article{Chacko:1999eb,
      author         = "Chacko, Z. and Nelson, Ann E.",
      title          = "{A Solution to the hierarchy problem with an infinitely
                        large extra dimension and moduli stabilization}",
      journal        = "Phys.Rev.",
      volume         = "D62",
      pages          = "085006",
      doi            = "10.1103/PhysRevD.62.085006",
      year           = "2000",
      eprint         = "hep-th/9912186",
      archivePrefix  = "arXiv",
      primaryClass   = "hep-th",
      reportNumber   = "UW-PTH-28",
      SLACcitation   = "%%CITATION = HEP-TH/9912186;%%",
}

@article{Cohen:1999ia,
      author         = "Cohen, Andrew G. and Kaplan, David B.",
      title          = "{Solving the hierarchy problem with noncompact extra
                        dimensions}",
      journal        = "Phys.Lett.",
      volume         = "B470",
      pages          = "52-58",
      doi            = "10.1016/S0370-2693(99)01331-3",
      year           = "1999",
      eprint         = "hep-th/9910132",
      archivePrefix  = "arXiv",
      primaryClass   = "hep-th",
      reportNumber   = "BUHEP-99-26, DOE-ER-40561-76",
      SLACcitation   = "%%CITATION = HEP-TH/9910132;%%",
}

@article{Gregory:1999gv,
      author         = "Gregory, Ruth",
      title          = "{Nonsingular global string compactifications}",
      journal        = "Phys.Rev.Lett.",
      volume         = "84",
      pages          = "2564-2567",
      doi            = "10.1103/PhysRevLett.84.2564",
      year           = "2000",
      eprint         = "hep-th/9911015",
      archivePrefix  = "arXiv",
      primaryClass   = "hep-th",
      reportNumber   = "DTP-99-77",
      SLACcitation   = "%%CITATION = HEP-TH/9911015;%%",
}

@article{Giovannini:2001hh,
      author         = "Giovannini, Massimo and Meyer, H. and Shaposhnikov,
                        Mikhail E.",
      title          = "{Warped compactification on Abelian vortex in
                        six-dimensions}",
      journal        = "Nucl.Phys.",
      volume         = "B619",
      pages          = "615-645",
      doi            = "10.1016/S0550-3213(01)00520-X",
      year           = "2001",
      eprint         = "hep-th/0104118",
      archivePrefix  = "arXiv",
      primaryClass   = "hep-th",
      reportNumber   = "UNIL-IPT-01-4",
      SLACcitation   = "%%CITATION = HEP-TH/0104118;%%",
}

@article{Choudhury:2006nj,
      author         = "Choudhury, Debajyoti and SenGupta, Soumitra",
      title          = "{Living on the edge in a spacetime with multiple
                        warping}",
      journal        = "Phys.Rev.",
      volume         = "D76",
      pages          = "064030",
      doi            = "10.1103/PhysRevD.76.064030",
      year           = "2007",
      eprint         = "hep-th/0612246",
      archivePrefix  = "arXiv",
      primaryClass   = "hep-th",
      SLACcitation   = "%%CITATION = HEP-TH/0612246;%%",
}

@article{Koley:2006bh,
      author         = "Koley, Ratna and Kar, Sayan",
      title          = "{Braneworlds in six dimensions: New models with bulk
                        scalars}",
      journal        = "Class.Quant.Grav.",
      volume         = "24",
      pages          = "79-94",
      doi            = "10.1088/0264-9381/24/1/004",
      year           = "2007",
      eprint         = "hep-th/0611074",
      archivePrefix  = "arXiv",
      primaryClass   = "hep-th",
      SLACcitation   = "%%CITATION = HEP-TH/0611074;%%",
}

@article{Saharian:2006jv,
      author         = "Saharian, Aram A.",
      title          = "{Surface Casimir densities and induced cosmological
                        constant in higher dimensional braneworlds}",
      journal        = "Phys.Rev.",
      volume         = "D74",
      pages          = "124009",
      doi            = "10.1103/PhysRevD.74.124009",
      year           = "2006",
      eprint         = "hep-th/0608211",
      archivePrefix  = "arXiv",
      primaryClass   = "hep-th",
      SLACcitation   = "%%CITATION = HEP-TH/0608211;%%",
}

@article{Koley:2008dh,
      author         = "Koley, Ratna and Mitra, Joydip and SenGupta, Soumitra",
      title          = "{Chiral fermions in a spacetime with multiple warping}",
      journal        = "Phys.Rev.",
      volume         = "D78",
      pages          = "045005",
      doi            = "10.1103/PhysRevD.78.045005",
      year           = "2008",
      eprint         = "0804.1019",
      archivePrefix  = "arXiv",
      primaryClass   = "hep-th",
      SLACcitation   = "%%CITATION = ARXIV:0804.1019;%%",
}

@article{Das:2011fb,
      author         = "Das, Ashmita and Hundi, R.S. and SenGupta, Soumitra",
      title          = "{Bulk Higgs and Gauge fields in a multiply warped
                        braneworld model}",
      journal        = "Phys.Rev.",
      volume         = "D83",
      pages          = "116003",
      doi            = "10.1103/PhysRevD.83.116003",
      year           = "2011",
      eprint         = "1105.1064",
      archivePrefix  = "arXiv",
      primaryClass   = "hep-ph",
      SLACcitation   = "%%CITATION = ARXIV:1105.1064;%%",
}

@article{Shiromizu:1999wj,
      author         = "Shiromizu, Tetsuya and Maeda, Kei-ichi and Sasaki, Misao",
      title          = "{The Einstein equation on the 3-brane world}",
      journal        = "Phys.Rev.",
      volume         = "D62",
      pages          = "024012",
      doi            = "10.1103/PhysRevD.62.024012",
      year           = "2000",
      eprint         = "gr-qc/9910076",
      archivePrefix  = "arXiv",
      primaryClass   = "gr-qc",
      reportNumber   = "DAMTP-1999-150, OUTAP-103, UTAP-349, RESCEU-40-99",
      SLACcitation   = "%%CITATION = GR-QC/9910076;%%",
}

@article{Harko:2004ui,
      author         = "Harko, T. and Mak, M.K.",
      title          = "{Vacuum solutions of the gravitational field equations in
                        the brane world model}",
      journal        = "Phys.Rev.",
      volume         = "D69",
      pages          = "064020",
      doi            = "10.1103/PhysRevD.69.064020",
      year           = "2004",
      eprint         = "gr-qc/0401049",
      archivePrefix  = "arXiv",
      primaryClass   = "gr-qc",
      SLACcitation   = "%%CITATION = GR-QC/0401049;%%",
}

@article{Vassilevich:2003xt,
      author         = "Vassilevich, D.V.",
      title          = "{Heat kernel expansion: User's manual}",
      journal        = "Phys.Rept.",
      volume         = "388",
      pages          = "279-360",
      doi            = "10.1016/j.physrep.2003.09.002",
      year           = "2003",
      eprint         = "hep-th/0306138",
      archivePrefix  = "arXiv",
      primaryClass   = "hep-th",
      SLACcitation   = "%%CITATION = HEP-TH/0306138;%%",
}

@article{Nojiri:2010wj,
      author         = "Nojiri, Shin'ichi and Odintsov, Sergei D.",
      title          = "{Unified cosmic history in modified gravity: from F(R)
                        theory to Lorentz non-invariant models}",
      journal        = "Phys.Rept.",
      volume         = "505",
      pages          = "59-144",
      doi            = "10.1016/j.physrep.2011.04.001",
      year           = "2011",
      eprint         = "1011.0544",
      archivePrefix  = "arXiv",
      primaryClass   = "gr-qc",
      SLACcitation   = "%%CITATION = ARXIV:1011.0544;%%",
}

@article{Sotiriou:2008rp,
      author         = "Sotiriou, Thomas P. and Faraoni, Valerio",
      title          = "{f(R) Theories Of Gravity}",
      journal        = "Rev.Mod.Phys.",
      volume         = "82",
      pages          = "451-497",
      doi            = "10.1103/RevModPhys.82.451",
      year           = "2010",
      eprint         = "0805.1726",
      archivePrefix  = "arXiv",
      primaryClass   = "gr-qc",
      SLACcitation   = "%%CITATION = ARXIV:0805.1726;%%",
}

@article{Nojiri:2003ft,
      author         = "Nojiri, Shin'ichi and Odintsov, Sergei D.",
      title          = "{Modified gravity with negative and positive powers of
                        the curvature: Unification of the inflation and of the
                        cosmic acceleration}",
      journal        = "Phys.Rev.",
      volume         = "D68",
      pages          = "123512",
      doi            = "10.1103/PhysRevD.68.123512",
      year           = "2003",
      eprint         = "hep-th/0307288",
      archivePrefix  = "arXiv",
      primaryClass   = "hep-th",
      SLACcitation   = "%%CITATION = HEP-TH/0307288;%%",
}

@article{DeFelice:2010aj,
      author         = "De Felice, Antonio and Tsujikawa, Shinji",
      title          = "{f(R) theories}",
      journal        = "Living Rev.Rel.",
      volume         = "13",
      pages          = "3",
      doi            = "10.12942/lrr-2010-3",
      year           = "2010",
      eprint         = "1002.4928",
      archivePrefix  = "arXiv",
      primaryClass   = "gr-qc",
      SLACcitation   = "%%CITATION = ARXIV:1002.4928;%%",
}

@article{Corda:2009re,
      author         = "Corda, Christian",
      title          = "{Interferometric detection of gravitational waves: the
                        definitive test for General Relativity}",
      journal        = "Int.J.Mod.Phys.",
      volume         = "D18",
      pages          = "2275-2282",
      doi            = "10.1142/S0218271809015904",
      year           = "2009",
      eprint         = "0905.2502",
      archivePrefix  = "arXiv",
      primaryClass   = "gr-qc",
      SLACcitation   = "%%CITATION = ARXIV:0905.2502;%%",
}

@article{Aslan:2006qi,
      author         = "Aslan, Onder and Demir, Durmus A.",
      title          = "{Generalized modified gravity in large extra dimensions}",
      journal        = "Phys.Lett.",
      volume         = "B635",
      pages          = "343-349",
      doi            = "10.1016/j.physletb.2006.03.010",
      year           = "2006",
      eprint         = "hep-ph/0603051",
      archivePrefix  = "arXiv",
      primaryClass   = "hep-ph",
      reportNumber   = "IZTECH-P-2006-02",
      SLACcitation   = "%%CITATION = HEP-PH/0603051;%%",
}

@article{Burrington:2012yq,
      author         = "Burrington, Benjamin A. and Peet, Amanda W. and Zadeh,
                        Ida G.",
      title          = "{Operator mixing for string states in the D1-D5 CFT near
                        the orbifold point}",
      journal        = "Phys.Rev.",
      number         = "10",
      volume         = "D87",
      pages          = "106001",
      doi            = "10.1103/PhysRevD.87.106001",
      year           = "2013",
      eprint         = "1211.6699",
      archivePrefix  = "arXiv",
      primaryClass   = "hep-th",
      SLACcitation   = "%%CITATION = ARXIV:1211.6699;%%",
}

@article{Chakraborty:2007ad,
      author         = "Chakraborty, Subenoy and Banerjee, Asit and
                        Bandyopadhyay, Tanwi",
      title          = "{Matter in the bulk and its consequences on the brane: A
                        Possible source of dark energy}",
      year           = "2007",
      eprint         = "0707.0199",
      archivePrefix  = "arXiv",
      primaryClass   = "gr-qc",
      SLACcitation   = "%%CITATION = ARXIV:0707.0199;%%",
}

@article{Polchinski:1998rr,
      author         = "Polchinski, J.",
      title          = "{String theory. Vol. 2: Superstring theory and beyond}",
      year           = "1998",
      SLACcitation   = "%%CITATION = INSPIRE-487241;%%",
}

@article{Polchinski:1998rq,
      author         = "Polchinski, J.",
      title          = "{String theory. Vol. 1: An introduction to the bosonic
                        string}",
      year           = "1998",
      SLACcitation   = "%%CITATION = INSPIRE-487240;%%",
}

@article{ArkaniHamed:1998rs,
      author         = "Arkani-Hamed, Nima and Dimopoulos, Savas and Dvali, G.R.",
      title          = "{The Hierarchy problem and new dimensions at a
                        millimeter}",
      journal        = "Phys.Lett.",
      volume         = "B429",
      pages          = "263-272",
      doi            = "10.1016/S0370-2693(98)00466-3",
      year           = "1998",
      eprint         = "hep-ph/9803315",
      archivePrefix  = "arXiv",
      primaryClass   = "hep-ph",
      reportNumber   = "SLAC-PUB-7769, SU-ITP-98-13",
      SLACcitation   = "%%CITATION = HEP-PH/9803315;%%",
}

@article{Antoniadis:1998ig,
      author         = "Antoniadis, Ignatios and Arkani-Hamed, Nima and
                        Dimopoulos, Savas and Dvali, G.R.",
      title          = "{New dimensions at a millimeter to a Fermi and
                        superstrings at a TeV}",
      journal        = "Phys.Lett.",
      volume         = "B436",
      pages          = "257-263",
      doi            = "10.1016/S0370-2693(98)00860-0",
      year           = "1998",
      eprint         = "hep-ph/9804398",
      archivePrefix  = "arXiv",
      primaryClass   = "hep-ph",
      reportNumber   = "SLAC-PUB-7801, SU-ITP-98-28, CPTH-S608-0498, IC-98-39",
      SLACcitation   = "%%CITATION = HEP-PH/9804398;%%",
}

@article{Antoniadis:1990ew,
      author         = "Antoniadis, Ignatios",
      title          = "{A Possible new dimension at a few TeV}",
      journal        = "Phys.Lett.",
      volume         = "B246",
      pages          = "377-384",
      doi            = "10.1016/0370-2693(90)90617-F",
      year           = "1990",
      reportNumber   = "EP-CPTH-A978-0690",
      SLACcitation   = "%%CITATION = PHLTA,B246,377;%%",
}

@article{Rubakov:1983bz,
      author         = "Rubakov, V.A. and Shaposhnikov, M.E.",
      title          = "{Extra Space-Time Dimensions: Towards a Solution to the
                        Cosmological Constant Problem}",
      journal        = "Phys.Lett.",
      volume         = "B125",
      pages          = "139",
      doi            = "10.1016/0370-2693(83)91254-6",
      year           = "1983",
      reportNumber   = "IC/83/11",
      SLACcitation   = "%%CITATION = PHLTA,B125,139;%%",
}

@article{Rubakov:1983bb,
      author         = "Rubakov, V.A. and Shaposhnikov, M.E.",
      title          = "{Do We Live Inside a Domain Wall?}",
      journal        = "Phys.Lett.",
      volume         = "B125",
      pages          = "136-138",
      doi            = "10.1016/0370-2693(83)91253-4",
      year           = "1983",
      SLACcitation   = "%%CITATION = PHLTA,B125,136;%%",
}

@article{Dadhich:2015lra,
      author         = "Dadhich, Naresh",
      title          = "{A discerning gravitational property for gravitational
                        equation in higher dimensions}",
      year           = "2015",
      eprint         = "1506.08764",
      archivePrefix  = "arXiv",
      primaryClass   = "gr-qc",
      SLACcitation   = "%%CITATION = ARXIV:1506.08764;%%",
}

@article{Ida:1999ui,
      author         = "Ida, Daisuke",
      title          = "{Brane world cosmology}",
      journal        = "JHEP",
      volume         = "0009",
      pages          = "014",
      doi            = "10.1088/1126-6708/2000/09/014",
      year           = "2000",
      eprint         = "gr-qc/9912002",
      archivePrefix  = "arXiv",
      primaryClass   = "gr-qc",
      SLACcitation   = "%%CITATION = GR-QC/9912002;%%",
}

@article{Brax:2004xh,
      author         = "Brax, Philippe and van de Bruck, Carsten and Davis,
                        Anne-Christine",
      title          = "{Brane world cosmology}",
      journal        = "Rept.Prog.Phys.",
      volume         = "67",
      pages          = "2183-2232",
      doi            = "10.1088/0034-4885/67/12/R02",
      year           = "2004",
      eprint         = "hep-th/0404011",
      archivePrefix  = "arXiv",
      primaryClass   = "hep-th",
      reportNumber   = "T04-038",
      SLACcitation   = "%%CITATION = HEP-TH/0404011;%%",
}

@article{Guha:2009pba,
      author         = "Guha, Sarbari and Chakraborty, Subenoy",
      title          = "{Five-dimensional bulk with a time-dependent warp factor
                        and its consequences on brane cosmology}",
      year           = "2009",
      eprint         = "0906.3894",
      archivePrefix  = "arXiv",
      primaryClass   = "gr-qc",
      SLACcitation   = "%%CITATION = ARXIV:0906.3894;%%",
}

@article{Kanno:2002iaa,
      author         = "Kanno, Sugumi and Soda, Jiro",
      title          = "{Brane world effective action at low-energies and AdS /
                        CFT}",
      journal        = "Phys. Rev.",
      volume         = "D66",
      year           = "2002",
      pages          = "043526",
      doi            = "10.1103/PhysRevD.66.043526",
      eprint         = "hep-th/0205188",
      archivePrefix  = "arXiv",
      primaryClass   = "hep-th",
      reportNumber   = "KUCP-0209",
      SLACcitation   = "%%CITATION = HEP-TH/0205188;%%"
}

@article{Kanno:2002kz,
      author         = "Kanno, Sugumi and Soda, Jiro",
      title          = "{Brane world effective action at low-energies}",
      booktitle      = "{Workshop on the Cosmology of Extra Dimensions and
                        Varying Fundamental Constants Porto, Portugal, September
                        3-5, 2002}",
      journal        = "Astrophys. Space Sci.",
      volume         = "283",
      year           = "2003",
      pages          = "481-486",
      doi            = "10.1023/A:1022556701901",
      eprint         = "gr-qc/0209087",
      archivePrefix  = "arXiv",
      primaryClass   = "gr-qc",
      reportNumber   = "KUCP-0219",
      SLACcitation   = "%%CITATION = GR-QC/0209087;%%"
}

@article{Shiromizu:2002qr,
      author         = "Shiromizu, Tetsuya and Koyama, Kazuya",
      title          = "{Low-energy effective theory for two brane systems:
                        Covariant curvature formulation}",
      journal        = "Phys. Rev.",
      volume         = "D67",
      year           = "2003",
      pages          = "084022",
      doi            = "10.1103/PhysRevD.67.084022",
      eprint         = "hep-th/0210066",
      archivePrefix  = "arXiv",
      primaryClass   = "hep-th",
      SLACcitation   = "%%CITATION = HEP-TH/0210066;%%"
}

@article{Kobayashi:2006jw,
      author         = "Kobayashi, Tsutomu and Shiromizu, Tetsuya and Deruelle,
                        Nathalie",
      title          = "{Low energy effective gravitational equations on a
                        Gauss-Bonnet brane}",
      journal        = "Phys. Rev.",
      volume         = "D74",
      year           = "2006",
      pages          = "104031",
      doi            = "10.1103/PhysRevD.74.104031",
      eprint         = "hep-th/0608166",
      archivePrefix  = "arXiv",
      primaryClass   = "hep-th",
      SLACcitation   = "%%CITATION = HEP-TH/0608166;%%"
}

@article{Maeda:2003vq,
      author         = "Maeda, Kei-ichi and Torii, Takashi",
      title          = "{Covariant gravitational equations on brane world with
                        Gauss-Bonnet term}",
      journal        = "Phys. Rev.",
      volume         = "D69",
      year           = "2004",
      pages          = "024002",
      doi            = "10.1103/PhysRevD.69.024002",
      eprint         = "hep-th/0309152",
      archivePrefix  = "arXiv",
      primaryClass   = "hep-th",
      reportNumber   = "WU-AP-173-03",
      SLACcitation   = "%%CITATION = HEP-TH/0309152;%%"
}

@article{Mak:2004hv,
      author         = "Mak, M. K. and Harko, T.",
      title          = "{Can the galactic rotation curves be explained in brane
                        world models?}",
      journal        = "Phys. Rev.",
      volume         = "D70",
      year           = "2004",
      pages          = "024010",
      doi            = "10.1103/PhysRevD.70.024010",
      eprint         = "gr-qc/0404104",
      archivePrefix  = "arXiv",
      primaryClass   = "gr-qc",
      SLACcitation   = "%%CITATION = GR-QC/0404104;%%"
}

@article{Boehmer:2007xh,
      author         = "Boehmer, C. G. and Harko, T.",
      title          = "{Galactic dark matter as a bulk effect on the brane}",
      journal        = "Class. Quant. Grav.",
      volume         = "24",
      year           = "2007",
      pages          = "3191-3210",
      doi            = "10.1088/0264-9381/24/13/004",
      eprint         = "0705.2496",
      archivePrefix  = "arXiv",
      primaryClass   = "gr-qc",
      SLACcitation   = "%%CITATION = ARXIV:0705.2496;%%"
}

@article{Viznyuk:2007ft,
      author         = "Viznyuk, Alexander and Shtanov, Yuri",
      title          = "{Spherically symmetric problem on the brane and galactic
                        rotation curves}",
      journal        = "Phys. Rev.",
      volume         = "D76",
      year           = "2007",
      pages          = "064009",
      doi            = "10.1103/PhysRevD.76.064009",
      eprint         = "0706.0649",
      archivePrefix  = "arXiv",
      primaryClass   = "gr-qc",
      SLACcitation   = "%%CITATION = ARXIV:0706.0649;%%"
}

@article{Boehmer:2007az,
      author         = "Boehmer, C. G. and Harko, T.",
      title          = "{On Einstein clusters as galactic dark matter halos}",
      journal        = "Mon. Not. Roy. Astron. Soc.",
      volume         = "379",
      year           = "2007",
      pages          = "393-398",
      doi            = "10.1111/j.1365-2966.2007.11977.x",
      eprint         = "0705.1756",
      archivePrefix  = "arXiv",
      primaryClass   = "gr-qc",
      SLACcitation   = "%%CITATION = ARXIV:0705.1756;%%"
}

@article{Pal:2004ii,
      author         = "Pal, Supratik and Bharadwaj, Somnath and Kar, Sayan",
      title          = "{Can extra dimensional effects replace dark matter?}",
      journal        = "Phys. Lett.",
      volume         = "B609",
      year           = "2005",
      pages          = "194-199",
      doi            = "10.1016/j.physletb.2005.01.043",
      eprint         = "gr-qc/0409023",
      archivePrefix  = "arXiv",
      primaryClass   = "gr-qc",
      SLACcitation   = "%%CITATION = GR-QC/0409023;%%"
}

@article{Borowiec:2006qr,
      author         = "Borowiec, Andrzej and Godlowski, Wlodzimierz and
                        Szydlowski, Marek",
      title          = "{Dark matter and dark energy as a effects of Modified
                        Gravity}",
      booktitle      = "{Theoretical physics: Current mathematical topics in
                        gravitation and cosmology. Proceedings, 42nd Karpacz
                        Winter School, Ladek, Poland, February 6-11, 2006}",
      journal        = "eConf",
      volume         = "C0602061",
      year           = "2006",
      pages          = "09",
      doi            = "10.1142/S0219887807001898",
      note           = "[Int. J. Geom. Meth. Mod. Phys.4,183(2007)]",
      eprint         = "astro-ph/0607639",
      archivePrefix  = "arXiv",
      primaryClass   = "astro-ph",
      reportNumber   = "KARP-2006-09",
      SLACcitation   = "%%CITATION = ASTRO-PH/0607639;%%"
}

@article{Capozziello:2006uv,
      author         = "Capozziello, Salvatore and Cardone, V. F. and Troisi, A.",
      title          = "{Dark energy and dark matter as curvature effects}",
      journal        = "JCAP",
      volume         = "0608",
      year           = "2006",
      pages          = "001",
      doi            = "10.1088/1475-7516/2006/08/001",
      eprint         = "astro-ph/0602349",
      archivePrefix  = "arXiv",
      primaryClass   = "astro-ph",
      SLACcitation   = "%%CITATION = ASTRO-PH/0602349;%%"
}

@article{Harko:2007yq,
      author         = "Harko, T. and Cheng, K. S.",
      title          = "{The Virial theorem and the dynamics of clusters of
                        galaxies in the brane world models}",
      journal        = "Phys. Rev.",
      volume         = "D76",
      year           = "2007",
      pages          = "044013",
      doi            = "10.1103/PhysRevD.76.044013",
      eprint         = "0707.1128",
      archivePrefix  = "arXiv",
      primaryClass   = "gr-qc",
      SLACcitation   = "%%CITATION = ARXIV:0707.1128;%%"
}

@article{Maia:2001zu,
      author         = "Maia, J. M. F. and Lima, J. A. S.",
      title          = "{Scalar field description of decaying Lambda
                        cosmologies}",
      journal        = "Phys. Rev.",
      volume         = "D65",
      year           = "2002",
      pages          = "083513",
      doi            = "10.1103/PhysRevD.65.083513",
      eprint         = "astro-ph/0112091",
      archivePrefix  = "arXiv",
      primaryClass   = "astro-ph",
      SLACcitation   = "%%CITATION = ASTRO-PH/0112091;%%"
}

@article{Bildhauer:1989dp,
      author         = "Bildhauer, S.",
      title          = "{Transport Equations for Freely Propagating Photons in
                        Curved Space-times: A Derivation by Wigner
                        Transformation}",
      journal        = "Class. Quant. Grav.",
      volume         = "6",
      year           = "1989",
      pages          = "1171-1187",
      doi            = "10.1088/0264-9381/6/8/017",
      SLACcitation   = "%%CITATION = CQGRD,6,1171;%%"
}

@article{Maia:2001gq,
      author         = "Maia, Marcos Duarte and Monte, Edmundo M.",
      title          = "{Geometry of brane worlds}",
      journal        = "Phys. Lett.",
      volume         = "A297",
      year           = "2002",
      pages          = "9-19",
      doi            = "10.1016/S0375-9601(02)00182-2",
      eprint         = "hep-th/0110088",
      archivePrefix  = "arXiv",
      primaryClass   = "hep-th",
      SLACcitation   = "%%CITATION = HEP-TH/0110088;%%"
}

@article{Kar:2015lma,
      author         = "Kar, Sayan and Lahiri, Sayantani and SenGupta, Soumitra",
      title          = "{Can extra dimensional effects allow wormholes without
                        exotic matter?}",
      journal        = "Phys. Lett.",
      volume         = "B750",
      year           = "2015",
      pages          = "319-324",
      doi            = "10.1016/j.physletb.2015.09.039",
      eprint         = "1505.06831",
      archivePrefix  = "arXiv",
      primaryClass   = "gr-qc",
      SLACcitation   = "%%CITATION = ARXIV:1505.06831;%%"
}

@article{Kar:2015fva,
      author         = "Kar, Sayan and Lahiri, Sayantani and SenGupta, Soumitra",
      title          = "{A note on spherically symmetric, static spacetimes in
                        Kanno–Soda on-brane gravity}",
      journal        = "Gen. Rel. Grav.",
      volume         = "47",
      year           = "2015",
      number         = "6",
      pages          = "70",
      doi            = "10.1007/s10714-015-1912-6",
      eprint         = "1501.00686",
      archivePrefix  = "arXiv",
      primaryClass   = "hep-th",
      SLACcitation   = "%%CITATION = ARXIV:1501.00686;%%"
}

@article{Anand:2014vqa,
      author         = "Anand, Sampurn and Choudhury, Debajyoti and Sen, Anjan A.
                        and SenGupta, Soumitra",
      title          = "{A Geometric Approach to Modulus Stabilization}",
      journal        = "Phys. Rev.",
      volume         = "D92",
      year           = "2015",
      number         = "2",
      pages          = "026008",
      doi            = "10.1103/PhysRevD.92.026008",
      eprint         = "1411.5120",
      archivePrefix  = "arXiv",
      primaryClass   = "hep-th",
      SLACcitation   = "%%CITATION = ARXIV:1411.5120;%%"
}

@article{Bazeia:2015dna,
      author         = "Bazeia, D. and Lobao, A. and Losano, L. and Menezes, R.
                        and Petrov, A. {\relax Yu}.",
      title          = "{Note on the Gauss-Bonnet braneworld scenario}",
      journal        = "Phys. Rev.",
      volume         = "D92",
      year           = "2015",
      number         = "6",
      pages          = "064010",
      doi            = "10.1103/PhysRevD.92.064010",
      eprint         = "1502.02564",
      archivePrefix  = "arXiv",
      primaryClass   = "hep-th",
      SLACcitation   = "%%CITATION = ARXIV:1502.02564;%%"
}

@article{Bernardini:2014vba,
      author         = "Bernardini, A. E. and Cavalcanti, R. T. and da Rocha,
                        Roldão",
      title          = "{Spherically Symmetric Thick Branes Cosmological
                        Evolution}",
      journal        = "Gen. Rel. Grav.",
      volume         = "47",
      year           = "2015",
      number         = "1",
      pages          = "1840",
      doi            = "10.1007/s10714-014-1840-x",
      eprint         = "1411.3552",
      archivePrefix  = "arXiv",
      primaryClass   = "gr-qc",
      SLACcitation   = "%%CITATION = ARXIV:1411.3552;%%"
}

@article{Nojiri:2001ae,
      author         = "Nojiri, Shin'ichi and Odintsov, Sergei D. and Ogushi,
                        Sachiko",
      title          = "{Cosmological and black hole brane world universes in
                        higher derivative gravity}",
      journal        = "Phys. Rev.",
      volume         = "D65",
      year           = "2002",
      pages          = "023521",
      doi            = "10.1103/PhysRevD.65.023521",
      eprint         = "hep-th/0108172",
      archivePrefix  = "arXiv",
      primaryClass   = "hep-th",
      reportNumber   = "NDA-FP-97, YITP-01-58",
      SLACcitation   = "%%CITATION = HEP-TH/0108172;%%"
}

@article{Nojiri:2002hz,
      author         = "Nojiri, Shin'ichi and Odintsov, Sergei D. and Ogushi,
                        Sachiko",
      title          = "{Friedmann-Robertson-Walker brane cosmological equations
                        from the five-dimensional bulk (A)dS black hole}",
      journal        = "Int. J. Mod. Phys.",
      volume         = "A17",
      year           = "2002",
      pages          = "4809-4870",
      doi            = "10.1142/S0217751X02012156",
      eprint         = "hep-th/0205187",
      archivePrefix  = "arXiv",
      primaryClass   = "hep-th",
      reportNumber   = "YITP-02-31",
      SLACcitation   = "%%CITATION = HEP-TH/0205187;%%"
}

@article{Davis:2002gn,
      author         = "Davis, Stephen C.",
      title          = "{Generalized Israel junction conditions for a
                        Gauss-Bonnet brane world}",
      journal        = "Phys. Rev.",
      volume         = "D67",
      year           = "2003",
      pages          = "024030",
      doi            = "10.1103/PhysRevD.67.024030",
      eprint         = "hep-th/0208205",
      archivePrefix  = "arXiv",
      primaryClass   = "hep-th",
      SLACcitation   = "%%CITATION = HEP-TH/0208205;%%"
}

@article{Deruelle:2003ur,
      author         = "Deruelle, Nathalie and Germani, Cristiano",
      title          = "{Smooth branes and junction conditions in Einstein
                        Gauss-Bonnet gravity}",
      journal        = "Nuovo Cim.",
      volume         = "B118",
      year           = "2003",
      pages          = "977-988",
      doi            = "10.1393/ncb/i2004-10015-0",
      eprint         = "gr-qc/0306116",
      archivePrefix  = "arXiv",
      primaryClass   = "gr-qc",
      SLACcitation   = "%%CITATION = GR-QC/0306116;%%"
}

@article{Charmousis:2002rc,
      author         = "Charmousis, Christos and Dufaux, Jean-Francois",
      title          = "{General Gauss-Bonnet brane cosmology}",
      journal        = "Class. Quant. Grav.",
      volume         = "19",
      year           = "2002",
      pages          = "4671-4682",
      doi            = "10.1088/0264-9381/19/18/304",
      eprint         = "hep-th/0202107",
      archivePrefix  = "arXiv",
      primaryClass   = "hep-th",
      reportNumber   = "LPT-ORSAY-02-07",
      SLACcitation   = "%%CITATION = HEP-TH/0202107;%%"
}

@article{Ayuso:2014jda,
      author         = "Ayuso, Ismael and Beltrán Jiménez, Jose and de la
                        Cruz-Dombriz, Álvaro",
      title          = "{Consistency of universally nonminimally coupled
                        $f(R,T,R_{μν}T^{μν})$ theories}",
      journal        = "Phys. Rev.",
      volume         = "D91",
      year           = "2015",
      number         = "10",
      pages          = "104003",
      doi            = "10.1103/PhysRevD.91.104003",
      eprint         = "1411.1636",
      archivePrefix  = "arXiv",
      primaryClass   = "hep-th",
      SLACcitation   = "%%CITATION = ARXIV:1411.1636;%%"
}

@article{delaCruz-Dombriz:2013gfa,
      author         = "de la Cruz-Dombriz, Alvaro and Dunsby, Peter K. S. and
                        Busti, Vinicius C. and Kandhai, Sulona",
      title          = "{On tidal forces in f(R) theories of gravity}",
      journal        = "Phys. Rev.",
      volume         = "D89",
      year           = "2014",
      number         = "6",
      pages          = "064029",
      doi            = "10.1103/PhysRevD.89.064029",
      eprint         = "1312.2022",
      archivePrefix  = "arXiv",
      primaryClass   = "gr-qc",
      SLACcitation   = "%%CITATION = ARXIV:1312.2022;%%"
}

@article{Abebe:2013zua,
      author         = "Abebe, Amare and de la Cruz-Dombriz, Alvaro and Dunsby,
                        Peter K. S.",
      title          = "{Large Scale Structure Constraints for a Class of f(R)
                        Theories of Gravity}",
      journal        = "Phys. Rev.",
      volume         = "D88",
      year           = "2013",
      pages          = "044050",
      doi            = "10.1103/PhysRevD.88.044050",
      eprint         = "1304.3462",
      archivePrefix  = "arXiv",
      primaryClass   = "astro-ph.CO",
      SLACcitation   = "%%CITATION = ARXIV:1304.3462;%%"
}

@article{Wiltshire:1985us,
      author         = "Wiltshire, D. L.",
      title          = "{Spherically Symmetric Solutions of Einstein-maxwell
                        Theory With a {Gauss-Bonnet} Term}",
      journal        = "Phys. Lett.",
      volume         = "B169",
      year           = "1986",
      pages          = "36",
      doi            = "10.1016/0370-2693(86)90681-7",
      reportNumber   = "Print-86-0327 (CAMBRIDGE)",
      SLACcitation   = "%%CITATION = PHLTA,B169,36;%%"
}

@article{Wiltshire:1988uq,
      author         = "Wiltshire, David L.",
      title          = "{Black Holes in String Generated Gravity Models}",
      journal        = "Phys. Rev.",
      volume         = "D38",
      year           = "1988",
      pages          = "2445",
      doi            = "10.1103/PhysRevD.38.2445",
      reportNumber   = "IC/88/56",
      SLACcitation   = "%%CITATION = PHRVA,D38,2445;%%"
}

@article{Abdesselam:2001ff,
      author         = "Abdesselam, B. and Mohammedi, N.",
      title          = "{Brane world cosmology with Gauss-Bonnet interaction}",
      journal        = "Phys. Rev.",
      volume         = "D65",
      year           = "2002",
      pages          = "084018",
      doi            = "10.1103/PhysRevD.65.084018",
      eprint         = "hep-th/0110143",
      archivePrefix  = "arXiv",
      primaryClass   = "hep-th",
      SLACcitation   = "%%CITATION = HEP-TH/0110143;%%"
}

@article{Deruelle:2000ge,
      author         = "Deruelle, Nathalie and Dolezel, Tomas",
      title          = "{Brane versus shell cosmologies in Einstein and
                        Einstein-Gauss-Bonnet theories}",
      journal        = "Phys. Rev.",
      volume         = "D62",
      year           = "2000",
      pages          = "103502",
      doi            = "10.1103/PhysRevD.62.103502",
      eprint         = "gr-qc/0004021",
      archivePrefix  = "arXiv",
      primaryClass   = "gr-qc",
      SLACcitation   = "%%CITATION = GR-QC/0004021;%%"
}

@article{Germani:2002pt,
      author         = "Germani, Cristiano and Sopuerta, Carlos F.",
      title          = "{String inspired brane world cosmology}",
      journal        = "Phys. Rev. Lett.",
      volume         = "88",
      year           = "2002",
      pages          = "231101",
      doi            = "10.1103/PhysRevLett.88.231101",
      eprint         = "hep-th/0202060",
      archivePrefix  = "arXiv",
      primaryClass   = "hep-th",
      SLACcitation   = "%%CITATION = HEP-TH/0202060;%%"
}

@article{Gravanis:2002wy,
      author         = "Gravanis, Elias and Willison, Steven",
      title          = "{Israel conditions for the Gauss-Bonnet theory and the
                        Friedmann equation on the brane universe}",
      journal        = "Phys. Lett.",
      volume         = "B562",
      year           = "2003",
      pages          = "118-126",
      doi            = "10.1016/S0370-2693(03)00555-0",
      eprint         = "hep-th/0209076",
      archivePrefix  = "arXiv",
      primaryClass   = "hep-th",
      SLACcitation   = "%%CITATION = HEP-TH/0209076;%%"
}

@article{Mavromatos:2000az,
      author         = "Mavromatos, Nick E. and Rizos, John",
      title          = "{String inspired higher curvature terms and the
                        Randall-Sundrum scenario}",
      journal        = "Phys. Rev.",
      volume         = "D62",
      year           = "2000",
      pages          = "124004",
      doi            = "10.1103/PhysRevD.62.124004",
      eprint         = "hep-th/0008074",
      archivePrefix  = "arXiv",
      primaryClass   = "hep-th",
      reportNumber   = "CERN-TH-2000-234",
      SLACcitation   = "%%CITATION = HEP-TH/0008074;%%"
}

@article{Cho:2001su,
      author         = "Cho, Y. M. and Neupane, Ishwaree P. and Wesson, P. S.",
      title          = "{No ghost state of Gauss-Bonnet interaction in warped
                        background}",
      journal        = "Nucl. Phys.",
      volume         = "B621",
      year           = "2002",
      pages          = "388-412",
      doi            = "10.1016/S0550-3213(01)00579-X",
      eprint         = "hep-th/0104227",
      archivePrefix  = "arXiv",
      primaryClass   = "hep-th",
      reportNumber   = "SNUTP-07-01, WATPHYS-TP-05-01",
      SLACcitation   = "%%CITATION = HEP-TH/0104227;%%"
}

@article{Torii:1996yi,
      author         = "Torii, Takashi and Yajima, Hiroki and Maeda, Kei-ichi",
      title          = "{Dilatonic black holes with Gauss-Bonnet term}",
      journal        = "Phys. Rev.",
      volume         = "D55",
      year           = "1997",
      pages          = "739-753",
      doi            = "10.1103/PhysRevD.55.739",
      eprint         = "gr-qc/9606034",
      archivePrefix  = "arXiv",
      primaryClass   = "gr-qc",
      reportNumber   = "WU-AP-55-96",
      SLACcitation   = "%%CITATION = GR-QC/9606034;%%"
}

@article{Meissner:2000dy,
      author         = "Meissner, Krzysztof A. and Olechowski, Marek",
      title          = "{Domain walls without cosmological constant in higher
                        order gravity}",
      journal        = "Phys. Rev. Lett.",
      volume         = "86",
      year           = "2001",
      pages          = "3708-3711",
      doi            = "10.1103/PhysRevLett.86.3708",
      eprint         = "hep-th/0009122",
      archivePrefix  = "arXiv",
      primaryClass   = "hep-th",
      reportNumber   = "IFT-20-2000",
      SLACcitation   = "%%CITATION = HEP-TH/0009122;%%"
}

@article{Lidsey:2002zw,
      author         = "Lidsey, James E. and Nojiri, Shin'ichi and Odintsov,
                        Sergei D.",
      title          = "{Brane world cosmology in (anti)-de Sitter
                        Einstein-Gauss-Bonnet-Maxwell gravity}",
      journal        = "JHEP",
      volume         = "06",
      year           = "2002",
      pages          = "026",
      doi            = "10.1088/1126-6708/2002/06/026",
      eprint         = "hep-th/0202198",
      archivePrefix  = "arXiv",
      primaryClass   = "hep-th",
      SLACcitation   = "%%CITATION = HEP-TH/0202198;%%"
}

@article{Borzou:2009gn,
      author         = "Borzou, Ahmad and Sepangi, Hamid Reza and Shahidi, Shahab
                        and Yousefi, Razieh",
      title          = "{Brane f(R) gravity}",
      journal        = "Europhys. Lett.",
      volume         = "88",
      year           = "2009",
      pages          = "29001",
      doi            = "10.1209/0295-5075/88/29001",
      eprint         = "0910.1933",
      archivePrefix  = "arXiv",
      primaryClass   = "gr-qc",
      SLACcitation   = "%%CITATION = ARXIV:0910.1933;%%"
}

@article{Haghani:2013oma,
      author         = "Haghani, Zahra and Harko, Tiberiu and Lobo, Francisco S.
                        N. and Sepangi, Hamid Reza and Shahidi, Shahab",
      title          = "{Further matters in space-time geometry:
                        f(R,T,RμνTμν) gravity}",
      journal        = "Phys. Rev.",
      volume         = "D88",
      year           = "2013",
      number         = "4",
      pages          = "044023",
      doi            = "10.1103/PhysRevD.88.044023",
      eprint         = "1304.5957",
      archivePrefix  = "arXiv",
      primaryClass   = "gr-qc",
      SLACcitation   = "%%CITATION = ARXIV:1304.5957;%%"
}

@article{Carames:2012gr,
      author         = "Carames, T. R. P. and Guimaraes, M. E. X. and Hoff da
                        Silva, J. M.",
      title          = "{Effective gravitational equations for $f(R)$ braneworld
                        models}",
      journal        = "Phys. Rev.",
      volume         = "D87",
      year           = "2013",
      number         = "10",
      pages          = "106011",
      doi            = "10.1103/PhysRevD.87.106011",
      eprint         = "1205.4980",
      archivePrefix  = "arXiv",
      primaryClass   = "gr-qc",
      SLACcitation   = "%%CITATION = ARXIV:1205.4980;%%"
}

@article{Haghani:2012zq,
      author         = "Haghani, Zahra and Sepangi, Hamid Reza and Shahidi,
                        Shahab",
      title          = "{Cosmological dynamics of brane f(R) gravity}",
      journal        = "JCAP",
      volume         = "1202",
      year           = "2012",
      pages          = "031",
      doi            = "10.1088/1475-7516/2012/02/031",
      eprint         = "1201.6448",
      archivePrefix  = "arXiv",
      primaryClass   = "gr-qc",
      SLACcitation   = "%%CITATION = ARXIV:1201.6448;%%"
}

@article{Apostolopoulos:2010jg,
      author         = "Apostolopoulos, Pantelis S. and Brouzakis, Nikolaos and
                        Tetradis, Nikolaos",
      title          = "{Effective cosmological equations of induced f(R)
                        gravity}",
      journal        = "JCAP",
      volume         = "1008",
      year           = "2010",
      pages          = "032",
      doi            = "10.1088/1475-7516/2010/08/032",
      eprint         = "1006.4573",
      archivePrefix  = "arXiv",
      primaryClass   = "hep-th",
      SLACcitation   = "%%CITATION = ARXIV:1006.4573;%%"
}

@article{DeWolfe:1999cp,
      author         = "DeWolfe, O. and Freedman, D. Z. and Gubser, S. S. and
                        Karch, A.",
      title          = "{Modeling the fifth-dimension with scalars and gravity}",
      journal        = "Phys. Rev.",
      volume         = "D62",
      year           = "2000",
      pages          = "046008",
      doi            = "10.1103/PhysRevD.62.046008",
      eprint         = "hep-th/9909134",
      archivePrefix  = "arXiv",
      primaryClass   = "hep-th",
      reportNumber   = "HUTP-99-A048, MIT-CTP-2903",
      SLACcitation   = "%%CITATION = HEP-TH/9909134;%%"
}

@article{Lykken:1999nb,
      author         = "Lykken, Joseph D. and Randall, Lisa",
      title          = "{The Shape of gravity}",
      journal        = "JHEP",
      volume         = "06",
      year           = "2000",
      pages          = "014",
      doi            = "10.1088/1126-6708/2000/06/014",
      eprint         = "hep-th/9908076",
      archivePrefix  = "arXiv",
      primaryClass   = "hep-th",
      reportNumber   = "MIT-CTP-2892, PUPT-1883, NSF-ITP-99-092,
                        FERMILAB-PUB-00-267-T",
      SLACcitation   = "%%CITATION = HEP-TH/9908076;%%"
}

@article{Gergely:2011df,
      author         = "Gergely, L. A. and Harko, T. and Dwornik, M. and Kupi, G.
                        and Keresztes, Z.",
      title          = "{Galactic rotation curves in brane world models}",
      journal        = "Mon. Not. Roy. Astron. Soc.",
      volume         = "415",
      year           = "2011",
      pages          = "3275-3290",
      doi            = "10.1111/j.1365-2966.2011.18941.x",
      eprint         = "1105.0159",
      archivePrefix  = "arXiv",
      primaryClass   = "gr-qc",
      SLACcitation   = "%%CITATION = ARXIV:1105.0159;%%"
}

@article{Nojiri:2006gh,
      author         = "Nojiri, Shin'ichi and Odintsov, Sergei D.",
      title          = "{Modified f(R) gravity consistent with realistic
                        cosmology: From matter dominated epoch to dark energy
                        universe}",
      journal        = "Phys. Rev.",
      volume         = "D74",
      year           = "2006",
      pages          = "086005",
      doi            = "10.1103/PhysRevD.74.086005",
      eprint         = "hep-th/0608008",
      archivePrefix  = "arXiv",
      primaryClass   = "hep-th",
      SLACcitation   = "%%CITATION = HEP-TH/0608008;%%"
}

@article{Capozziello:2006dj,
      author         = "Capozziello, Salvatore and Nojiri, S. and Odintsov, S. D.
                        and Troisi, A.",
      title          = "{Cosmological viability of f(R)-gravity as an ideal fluid
                        and its compatibility with a matter dominated phase}",
      journal        = "Phys. Lett.",
      volume         = "B639",
      year           = "2006",
      pages          = "135-143",
      doi            = "10.1016/j.physletb.2006.06.034",
      eprint         = "astro-ph/0604431",
      archivePrefix  = "arXiv",
      primaryClass   = "astro-ph",
      SLACcitation   = "%%CITATION = ASTRO-PH/0604431;%%"
}

@article{Bahamonde:2016wmz,
      author         = "Bahamonde, Sebastian and Odintsov, S. D. and Oikonomou,
                        V. K. and Wright, Matthew",
      title          = "{Correspondence of $F(R)$ Gravity Singularities in Jordan
                        and Einstein Frames}",
      year           = "2016",
      eprint         = "1603.05113",
      archivePrefix  = "arXiv",
      primaryClass   = "gr-qc",
      SLACcitation   = "%%CITATION = ARXIV:1603.05113;%%"
}

@article{Barrow:1988xh,
      author         = "Barrow, John D. and Cotsakis, S.",
      title          = "{Inflation and the Conformal Structure of Higher Order
                        Gravity Theories}",
      journal        = "Phys. Lett.",
      volume         = "B214",
      year           = "1988",
      pages          = "515-518",
      doi            = "10.1016/0370-2693(88)90110-4",
      SLACcitation   = "%%CITATION = PHLTA,B214,515;%%"
}

@article{Maeda:1987xf,
      author         = "Maeda, Kei-ichi",
      title          = "{Inflation as a Transient Attractor in R**2 Cosmology}",
      journal        = "Phys. Rev.",
      volume         = "D37",
      year           = "1988",
      pages          = "858",
      doi            = "10.1103/PhysRevD.37.858",
      reportNumber   = "UTAP-60-87",
      SLACcitation   = "%%CITATION = PHRVA,D37,858;%%"
}

@article{Barrow:1988xi,
      author         = "Barrow, John D.",
      title          = "{The Premature Recollapse Problem in Closed Inflationary
                        Universes}",
      journal        = "Nucl. Phys.",
      volume         = "B296",
      year           = "1988",
      pages          = "697-709",
      doi            = "10.1016/0550-3213(88)90040-5",
      SLACcitation   = "%%CITATION = NUPHA,B296,697;%%"
}

@article{Barrow:2009gx,
      author         = "Barrow, John D. and Hervik, Sigbjorn",
      title          = "{Simple Types of Anisotropic Inflation}",
      journal        = "Phys. Rev.",
      volume         = "D81",
      year           = "2010",
      pages          = "023513",
      doi            = "10.1103/PhysRevD.81.023513",
      eprint         = "0911.3805",
      archivePrefix  = "arXiv",
      primaryClass   = "gr-qc",
      SLACcitation   = "%%CITATION = ARXIV:0911.3805;%%"
}

@article{Barrow:2006xb,
      author         = "Barrow, John D. and Hervik, Sigbjorn",
      title          = "{On the evolution of universes in quadratic theories of
                        gravity}",
      journal        = "Phys. Rev.",
      volume         = "D74",
      year           = "2006",
      pages          = "124017",
      doi            = "10.1103/PhysRevD.74.124017",
      eprint         = "gr-qc/0610013",
      archivePrefix  = "arXiv",
      primaryClass   = "gr-qc",
      SLACcitation   = "%%CITATION = GR-QC/0610013;%%"
}

@article{Paliathanasis:2015aos,
      author         = "Paliathanasis, Andronikos",
      title          = "{$f(R)$-gravity from Killing Tensors}",
      journal        = "Class. Quant. Grav.",
      volume         = "33",
      year           = "2016",
      number         = "7",
      pages          = "075012",
      doi            = "10.1088/0264-9381/33/7/075012",
      eprint         = "1512.03239",
      archivePrefix  = "arXiv",
      primaryClass   = "gr-qc",
      SLACcitation   = "%%CITATION = ARXIV:1512.03239;%%"
}

@article{Basilakos:2011rx,
      author         = "Basilakos, Spyros and Tsamparlis, Michael and
                        Paliathanasis, Andronikos",
      title          = "{Using the Noether symmetry approach to probe the nature
                        of dark energy}",
      journal        = "Phys. Rev.",
      volume         = "D83",
      year           = "2011",
      pages          = "103512",
      doi            = "10.1103/PhysRevD.83.103512",
      eprint         = "1104.2980",
      archivePrefix  = "arXiv",
      primaryClass   = "astro-ph.CO",
      SLACcitation   = "%%CITATION = ARXIV:1104.2980;%%"
}

@article{Liu:2011wi,
      author         = "Liu, Yu-Xiao and Zhong, Yuan and Zhao, Zhen-Hua and Li,
                        Hai-Tao",
      title          = "{Domain wall brane in squared curvature gravity}",
      journal        = "JHEP",
      volume         = "06",
      year           = "2011",
      pages          = "135",
      doi            = "10.1007/JHEP06(2011)135",
      eprint         = "1104.3188",
      archivePrefix  = "arXiv",
      primaryClass   = "hep-th",
      SLACcitation   = "%%CITATION = ARXIV:1104.3188;%%"
}

@article{Zhong:2010ae,
      author         = "Zhong, Yuan and Liu, Yu-Xiao and Yang, Ke",
      title          = "{Tensor perturbations of $f(R)$-branes}",
      journal        = "Phys. Lett.",
      volume         = "B699",
      year           = "2011",
      pages          = "398-402",
      doi            = "10.1016/j.physletb.2011.04.037",
      eprint         = "1010.3478",
      archivePrefix  = "arXiv",
      primaryClass   = "hep-th",
      SLACcitation   = "%%CITATION = ARXIV:1010.3478;%%"
}

@article{Carloni:2010ph,
      author         = "Carloni, Sante and Goswami, Rituparno and Dunsby, Peter
                        K. S.",
      title          = "{A new approach to reconstruction methods in $f(R)$
                        gravity}",
      journal        = "Class. Quant. Grav.",
      volume         = "29",
      year           = "2012",
      pages          = "135012",
      doi            = "10.1088/0264-9381/29/13/135012",
      eprint         = "1005.1840",
      archivePrefix  = "arXiv",
      primaryClass   = "gr-qc",
      SLACcitation   = "%%CITATION = ARXIV:1005.1840;%%"
}

@article{Nojiri:2009xh,
      author         = "Nojiri, Shin'ichi and Odintsov, Sergei D. and Toporensky,
                        Alexey and Tretyakov, Petr",
      title          = "{Reconstruction and deceleration-acceleration transitions
                        in modified gravity}",
      journal        = "Gen. Rel. Grav.",
      volume         = "42",
      year           = "2010",
      pages          = "1997-2008",
      doi            = "10.1007/s10714-010-0977-5",
      eprint         = "0912.2488",
      archivePrefix  = "arXiv",
      primaryClass   = "hep-th",
      SLACcitation   = "%%CITATION = ARXIV:0912.2488;%%"
}

@article{Nojiri:2006be,
      author         = "Nojiri, Shin'ichi and Odintsov, Sergei D.",
      title          = "{Modified gravity and its reconstruction from the
                        universe expansion history}",
      booktitle      = "{Einstein's legacy: From the theoretical paradise to
                        astrophysical observations}",
      journal        = "J. Phys. Conf. Ser.",
      volume         = "66",
      year           = "2007",
      pages          = "012005",
      doi            = "10.1088/1742-6596/66/1/012005",
      eprint         = "hep-th/0611071",
      archivePrefix  = "arXiv",
      primaryClass   = "hep-th",
      SLACcitation   = "%%CITATION = HEP-TH/0611071;%%"
}

@article{Nojiri:2009kx,
      author         = "Nojiri, Shin'ichi and Odintsov, Sergei D. and Saez-Gomez,
                        Diego",
      title          = "{Cosmological reconstruction of realistic modified F(R)
                        gravities}",
      journal        = "Phys. Lett.",
      volume         = "B681",
      year           = "2009",
      pages          = "74-80",
      doi            = "10.1016/j.physletb.2009.09.045",
      eprint         = "0908.1269",
      archivePrefix  = "arXiv",
      primaryClass   = "hep-th",
      SLACcitation   = "%%CITATION = ARXIV:0908.1269;%%"
}

@article{Capozziello:2010sc,
      author         = "Capozziello, S. and Martin-Moruno, P. and Rubano, C.",
      title          = "{Physical non-equivalence of the Jordan and Einstein
                        frames}",
      journal        = "Phys. Lett.",
      volume         = "B689",
      year           = "2010",
      pages          = "117-121",
      doi            = "10.1016/j.physletb.2010.04.058",
      eprint         = "1003.5394",
      archivePrefix  = "arXiv",
      primaryClass   = "gr-qc",
      SLACcitation   = "%%CITATION = ARXIV:1003.5394;%%"
}

@article{Dabrowski:2008kx,
      author         = "Dabrowski, Mariusz P. and Garecki, Janusz and Blaschke,
                        David B.",
      title          = "{Conformal transformations and conformal invariance in
                        gravitation}",
      journal        = "Annalen Phys.",
      volume         = "18",
      year           = "2009",
      pages          = "13-32",
      doi            = "10.1002/andp.200810331",
      eprint         = "0806.2683",
      archivePrefix  = "arXiv",
      primaryClass   = "gr-qc",
      SLACcitation   = "%%CITATION = ARXIV:0806.2683;%%"
}

@article{Iglesias:2007nv,
      author         = "Iglesias, Alberto and Kaloper, Nemanja and Padilla,
                        Antonio and Park, Minjoon",
      title          = "{How (Not) to Palatini}",
      journal        = "Phys. Rev.",
      volume         = "D76",
      year           = "2007",
      pages          = "104001",
      doi            = "10.1103/PhysRevD.76.104001",
      eprint         = "0708.1163",
      archivePrefix  = "arXiv",
      primaryClass   = "astro-ph",
      SLACcitation   = "%%CITATION = ARXIV:0708.1163;%%"
}

@article{Barausse:2007ys,
      author         = "Barausse, E. and Sotiriou, T. P. and Miller, J. C.",
      title          = "{Curvature singularities, tidal forces and the viability
                        of Palatini f(R) gravity}",
      journal        = "Class. Quant. Grav.",
      volume         = "25",
      year           = "2008",
      pages          = "105008",
      doi            = "10.1088/0264-9381/25/10/105008",
      eprint         = "0712.1141",
      archivePrefix  = "arXiv",
      primaryClass   = "gr-qc",
      SLACcitation   = "%%CITATION = ARXIV:0712.1141;%%"
}

@article{Mak:2003kw,
      author         = "Mak, M. K. and Harko, T.",
      title          = "{Quark stars admitting a one parameter group of conformal
                        motions}",
      journal        = "Int. J. Mod. Phys.",
      volume         = "D13",
      year           = "2004",
      pages          = "149-156",
      doi            = "10.1142/S0218271804004451",
      eprint         = "gr-qc/0309069",
      archivePrefix  = "arXiv",
      primaryClass   = "gr-qc",
      SLACcitation   = "%%CITATION = GR-QC/0309069;%%"
}

@article{Maartens:1989ay,
      author         = "Maartens, Roy and Maharaj, M. S.",
      title          = "{CONFORMALLY SYMMETRIC STATIC FLUID SPHERES}",
      journal        = "J. Math. Phys.",
      volume         = "31",
      year           = "1990",
      pages          = "151",
      doi            = "10.1063/1.528853",
      reportNumber   = "CNLS-89-13",
      SLACcitation   = "%%CITATION = JMAPA,31,151;%%"
}

@article{Parry:2005eb,
      author         = "Parry, M. and Pichler, S. and Deeg, D.",
      title          = "{Higher-derivative gravity in brane world models}",
      journal        = "JCAP",
      volume         = "0504",
      year           = "2005",
      pages          = "014",
      doi            = "10.1088/1475-7516/2005/04/014",
      eprint         = "hep-ph/0502048",
      archivePrefix  = "arXiv",
      primaryClass   = "hep-ph",
      reportNumber   = "LMU-ASC-08-05",
      SLACcitation   = "%%CITATION = HEP-PH/0502048;%%"
}

@article{Catena:2006bd,
      author         = "Catena, Riccardo and Pietroni, Massimo and Scarabello,
                        Luca",
      title          = "{Einstein and Jordan reconciled: a frame-invariant
                        approach to scalar-tensor cosmology}",
      journal        = "Phys. Rev.",
      volume         = "D76",
      year           = "2007",
      pages          = "084039",
      doi            = "10.1103/PhysRevD.76.084039",
      eprint         = "astro-ph/0604492",
      archivePrefix  = "arXiv",
      primaryClass   = "astro-ph",
      reportNumber   = "DESY-06-076",
      SLACcitation   = "%%CITATION = ASTRO-PH/0604492;%%"
}

@article{Chiba:2013mha,
      author         = "Chiba, Takeshi and Yamaguchi, Masahide",
      title          = "{Conformal-Frame (In)dependence of Cosmological
                        Observations in Scalar-Tensor Theory}",
      journal        = "JCAP",
      volume         = "1310",
      year           = "2013",
      pages          = "040",
      doi            = "10.1088/1475-7516/2013/10/040",
      eprint         = "1308.1142",
      archivePrefix  = "arXiv",
      primaryClass   = "gr-qc",
      SLACcitation   = "%%CITATION = ARXIV:1308.1142;%%"
}

@article{Capozziello:1996xg,
      author         = "Capozziello, S. and de Ritis, R. and Marino, Alma Angela",
      title          = "{Some aspects of the cosmological conformal equivalence
                        between 'Jordan frame' and 'Einstein frame'}",
      journal        = "Class. Quant. Grav.",
      volume         = "14",
      year           = "1997",
      pages          = "3243-3258",
      doi            = "10.1088/0264-9381/14/12/010",
      eprint         = "gr-qc/9612053",
      archivePrefix  = "arXiv",
      primaryClass   = "gr-qc",
      SLACcitation   = "%%CITATION = GR-QC/9612053;%%"
}

@article{Rizzo:2006wf,
      author         = "Rizzo, Thomas G.",
      title          = "{Higher curvature gravity in TeV-scale extra dimensions}",
      year           = "2006",
      eprint         = "hep-ph/0603242",
      archivePrefix  = "arXiv",
      primaryClass   = "hep-ph",
      reportNumber   = "SLAC-PUB-11666",
      SLACcitation   = "%%CITATION = HEP-PH/0603242;%%"
}

@article{Konoplya:2008ix,
      author         = "Konoplya, R. A. and Zhidenko, A.",
      title          = "{(In)stability of D-dimensional black holes in
                        Gauss-Bonnet theory}",
      journal        = "Phys. Rev.",
      volume         = "D77",
      year           = "2008",
      pages          = "104004",
      doi            = "10.1103/PhysRevD.77.104004",
      eprint         = "0802.0267",
      archivePrefix  = "arXiv",
      primaryClass   = "hep-th",
      SLACcitation   = "%%CITATION = ARXIV:0802.0267;%%"
}

@phdthesis{Brown:2007dd,
      author         = "Brown, Richard A.",
      title          = "{Brane world cosmology with Gauss-Bonnet and induced
                        gravity terms}",
      school         = "Portsmouth U., ICG",
      year           = "2007",
      eprint         = "gr-qc/0701083",
      archivePrefix  = "arXiv",
      primaryClass   = "gr-qc",
      SLACcitation   = "%%CITATION = GR-QC/0701083;%%"
}

@article{Rizzo:2006sb,
      author         = "Rizzo, Thomas G.",
      title          = "{Higher Curvature Effects in ADD and RS Models}",
      booktitle      = "{Linear collider. Proceedings, International Workshop,
                        LCWS06, Bangalore, India, March 9-13, 2006}",
      journal        = "Pramana",
      volume         = "69",
      year           = "2007",
      pages          = "889-894",
      doi            = "10.1007/s12043-007-0200-8",
      eprint         = "hep-ph/0606292",
      archivePrefix  = "arXiv",
      primaryClass   = "hep-ph",
      reportNumber   = "SLAC-PUB-11930",
      SLACcitation   = "%%CITATION = HEP-PH/0606292;%%"
}

@article{Cai:2006pq,
      author         = "Cai, Rong-Gen and Ohta, Nobuyoshi",
      title          = "{Black Holes in Pure Lovelock Gravities}",
      journal        = "Phys. Rev.",
      volume         = "D74",
      year           = "2006",
      pages          = "064001",
      doi            = "10.1103/PhysRevD.74.064001",
      eprint         = "hep-th/0604088",
      archivePrefix  = "arXiv",
      primaryClass   = "hep-th",
      reportNumber   = "KU-TP-001",
      SLACcitation   = "%%CITATION = HEP-TH/0604088;%%"
}

@article{Cembranos:2003mr,
      author         = "Cembranos, J. A. R. and Dobado, A. and Maroto, Antonio
                        Lopez",
      title          = "{Brane world dark matter}",
      journal        = "Phys. Rev. Lett.",
      volume         = "90",
      year           = "2003",
      pages          = "241301",
      doi            = "10.1103/PhysRevLett.90.241301",
      eprint         = "hep-ph/0302041",
      archivePrefix  = "arXiv",
      primaryClass   = "hep-ph",
      SLACcitation   = "%%CITATION = HEP-PH/0302041;%%"
}

@article{Cembranos:2011cm,
      author         = "Cembranos, Jose A. R. and Diaz-Cruz, J. Lorenzo and
                        Prado, Lilian",
      title          = "{Impact of DM direct searches and the LHC analyses on
                        branon phenomenology}",
      journal        = "Phys. Rev.",
      volume         = "D84",
      year           = "2011",
      pages          = "083522",
      doi            = "10.1103/PhysRevD.84.083522",
      eprint         = "1110.0542",
      archivePrefix  = "arXiv",
      primaryClass   = "hep-ph",
      SLACcitation   = "%%CITATION = ARXIV:1110.0542;%%"
}

@article{Cembranos:2016jun,
      author         = "Cembranos, Jose A. R. and Maroto, Antonio L.",
      title          = "{Disformal scalars as dark matter candidates: Branon
                        phenomenology}",
      journal        = "Int. J. Mod. Phys.",
      volume         = "31",
      year           = "2016",
      number         = "14n15",
      pages          = "1630015",
      doi            = "10.1142/S0217751X16300155",
      eprint         = "1602.07270",
      archivePrefix  = "arXiv",
      primaryClass   = "hep-ph",
      SLACcitation   = "%%CITATION = ARXIV:1602.07270;%%"
}

@article{Cembranos:2008kg,
      author         = "Cembranos, J. A. R. and de la Cruz-Dombriz, A. and
                        Dobado, A. and Maroto, Antonio Lopez",
      title          = "{Is the CMB Cold Spot a gate to extra dimensions?}",
      journal        = "JCAP",
      volume         = "0810",
      year           = "2008",
      pages          = "039",
      doi            = "10.1088/1475-7516/2008/10/039",
      eprint         = "0803.0694",
      archivePrefix  = "arXiv",
      primaryClass   = "astro-ph",
      reportNumber   = "FTPI-MINN-08-09, UMN-TH-2640-08",
      SLACcitation   = "%%CITATION = ARXIV:0803.0694;%%"
}

@article{Resco:2016upv,
      author         = "Aparicio Resco, Miguel and de la Cruz-Dombriz, Álvaro
                        and Llanes Estrada, Felipe J. and Zapatero Castrillo,
                        Víctor",
      title          = "{On neutron stars in $f(R)$ theories: Small radii, large
                        masses and large energy emitted in a merger}",
      journal        = "Phys. Dark Univ.",
      volume         = "13",
      year           = "2016",
      pages          = "147-161",
      doi            = "10.1016/j.dark.2016.07.001",
      eprint         = "1602.03880",
      archivePrefix  = "arXiv",
      primaryClass   = "gr-qc",
      SLACcitation   = "%%CITATION = ARXIV:1602.03880;%%"
}

@article{Cembranos:2011sr,
      author         = "Cembranos, J. A. R. and de la Cruz-Dombriz, A. and Jimeno
                        Romero, P.",
      title          = "{Kerr-Newman black holes in $f(R)$ theories}",
      journal        = "Int. J. Geom. Meth. Mod. Phys.",
      volume         = "11",
      year           = "2014",
      pages          = "1450001",
      doi            = "10.1142/S0219887814500017",
      eprint         = "1109.4519",
      archivePrefix  = "arXiv",
      primaryClass   = "gr-qc",
      SLACcitation   = "%%CITATION = ARXIV:1109.4519;%%"
}

@article{delaCruz-Dombriz:2015tye,
      author         = "de la Cruz-Dombriz, Álvaro and Dunsby, Peter K. S. and
                        Kandhai, Sulona and Sáez-Gómez, Diego",
      title          = "{Theoretical and observational constraints of viable f(R)
                        theories of gravity}",
      journal        = "Phys. Rev.",
      volume         = "D93",
      year           = "2016",
      number         = "8",
      pages          = "084016",
      doi            = "10.1103/PhysRevD.93.084016",
      eprint         = "1511.00102",
      archivePrefix  = "arXiv",
      primaryClass   = "gr-qc",
      SLACcitation   = "%%CITATION = ARXIV:1511.00102;%%"
}

@article{delaCruzDombriz:2009et,
      author         = "de la Cruz-Dombriz, A. and Dobado, A. and Maroto, A. L.",
      title          = "{Black Holes in f(R) theories}",
      journal        = "Phys. Rev.",
      volume         = "D80",
      year           = "2009",
      pages          = "124011",
      doi            = "10.1103/PhysRevD.83.029903, 10.1103/PhysRevD.80.124011",
      note           = "[Erratum: Phys. Rev.D83,029903(2011)]",
      eprint         = "0907.3872",
      archivePrefix  = "arXiv",
      primaryClass   = "gr-qc",
      SLACcitation   = "%%CITATION = ARXIV:0907.3872;%%"
}

@article{delaCruzDombriz:2012xy,
      author         = "de la Cruz-Dombriz, A. and Saez-Gomez, D.",
      title          = "{Black holes, cosmological solutions, future
                        singularities, and their thermodynamical properties in
                        modified gravity theories}",
      journal        = "Entropy",
      volume         = "14",
      year           = "2012",
      pages          = "1717-1770",
      doi            = "10.3390/e14091717",
      eprint         = "1207.2663",
      archivePrefix  = "arXiv",
      primaryClass   = "gr-qc",
      SLACcitation   = "%%CITATION = ARXIV:1207.2663;%%"
}

@article{Bronnikov:2002rn,
      author         = "Bronnikov, K. A. and Kim, Sung-Won",
      title          = "{Possible wormholes in a brane world}",
      journal        = "Phys. Rev.",
      volume         = "D67",
      year           = "2003",
      pages          = "064027",
      doi            = "10.1103/PhysRevD.67.064027",
      eprint         = "gr-qc/0212112",
      archivePrefix  = "arXiv",
      primaryClass   = "gr-qc",
      SLACcitation   = "%%CITATION = GR-QC/0212112;%%"
}

@article{Bronnikov:2003gx,
      author         = "Bronnikov, K. A. and Melnikov, V. N. and Dehnen, Heinz",
      title          = "{On a general class of brane world black holes}",
      journal        = "Phys. Rev.",
      volume         = "D68",
      year           = "2003",
      pages          = "024025",
      doi            = "10.1103/PhysRevD.68.024025",
      eprint         = "gr-qc/0304068",
      archivePrefix  = "arXiv",
      primaryClass   = "gr-qc",
      SLACcitation   = "%%CITATION = GR-QC/0304068;%%"
}

@article{Olmo:2011uz,
      author         = "Olmo, Gonzalo J.",
      title          = "{Palatini Approach to Modified Gravity: f(R) Theories and
                        Beyond}",
      journal        = "Int. J. Mod. Phys.",
      volume         = "D20",
      year           = "2011",
      pages          = "413-462",
      doi            = "10.1142/S0218271811018925",
      eprint         = "1101.3864",
      archivePrefix  = "arXiv",
      primaryClass   = "gr-qc",
      SLACcitation   = "%%CITATION = ARXIV:1101.3864;%%"
}

@article{Antoniadis:1993jc,
      author         = "Antoniadis, Ignatios and Rizos, J. and Tamvakis, K.",
      title          = "{Singularity - free cosmological solutions of the
                        superstring effective action}",
      journal        = "Nucl. Phys.",
      volume         = "B415",
      year           = "1994",
      pages          = "497-514",
      doi            = "10.1016/0550-3213(94)90120-1",
      eprint         = "hep-th/9305025",
      archivePrefix  = "arXiv",
      primaryClass   = "hep-th",
      reportNumber   = "CPTH-A239-0593",
      SLACcitation   = "%%CITATION = HEP-TH/9305025;%%"
}

@article{Briscese:2006xu,
      author         = "Briscese, F. and Elizalde, E. and Nojiri, S. and
                        Odintsov, S. D.",
      title          = "{Phantom scalar dark energy as modified gravity:
                        Understanding the origin of the Big Rip singularity}",
      journal        = "Phys. Lett.",
      volume         = "B646",
      year           = "2007",
      pages          = "105-111",
      doi            = "10.1016/j.physletb.2007.01.013",
      eprint         = "hep-th/0612220",
      archivePrefix  = "arXiv",
      primaryClass   = "hep-th",
      SLACcitation   = "%%CITATION = HEP-TH/0612220;%%"
}

@article{Nojiri:2003ft,
      author         = "Nojiri, Shin'ichi and Odintsov, Sergei D.",
      title          = "{Modified gravity with negative and positive powers of
                        the curvature: Unification of the inflation and of the
                        cosmic acceleration}",
      journal        = "Phys. Rev.",
      volume         = "D68",
      year           = "2003",
      pages          = "123512",
      doi            = "10.1103/PhysRevD.68.123512",
      eprint         = "hep-th/0307288",
      archivePrefix  = "arXiv",
      primaryClass   = "hep-th",
      SLACcitation   = "%%CITATION = HEP-TH/0307288;%%"
}

@article{vanderBij:2006pg,
      author         = "van der Bij, J. J. and Dilcher, S.",
      title          = "{A Higher dimensional explanation of the excess of
                        Higgs-like events at CERN LEP}",
      journal        = "Phys. Lett.",
      volume         = "B638",
      year           = "2006",
      pages          = "234-238",
      doi            = "10.1016/j.physletb.2006.05.056",
      eprint         = "hep-ph/0605008",
      archivePrefix  = "arXiv",
      primaryClass   = "hep-ph",
      reportNumber   = "FREIBURG-THEP-06-05",
      SLACcitation   = "%%CITATION = HEP-PH/0605008;%%"
}

@article{Langlois:2002bb,
      author         = "Langlois, David",
      title          = "{Brane cosmology: An Introduction}",
      booktitle      = "{Brane world: New perspective in cosmology. Proceedings,
                        2nd Workshop on relativistic and cosmological aspects of
                        the brane world, Kyoto, Japan, January 15-18, 2002}",
      journal        = "Prog. Theor. Phys. Suppl.",
      volume         = "148",
      year           = "2003",
      pages          = "181-212",
      doi            = "10.1143/PTPS.148.181",
      eprint         = "hep-th/0209261",
      archivePrefix  = "arXiv",
      primaryClass   = "hep-th",
      SLACcitation   = "%%CITATION = HEP-TH/0209261;%%"
}

@article{Goldberger:1999un,
      author         = "Goldberger, Walter D. and Wise, Mark B.",
      title          = "{Phenomenology of a stabilized modulus}",
      journal        = "Phys. Lett.",
      volume         = "B475",
      year           = "2000",
      pages          = "275-279",
      doi            = "10.1016/S0370-2693(00)00099-X",
      eprint         = "hep-ph/9911457",
      archivePrefix  = "arXiv",
      primaryClass   = "hep-ph",
      reportNumber   = "CALT-68-2250",
      SLACcitation   = "%%CITATION = HEP-PH/9911457;%%"
}

@article{Adelberger:2003zx,
      author         = "Adelberger, E. G. and Heckel, Blayne R. and Nelson, A.
                        E.",
      title          = "{Tests of the gravitational inverse square law}",
      journal        = "Ann. Rev. Nucl. Part. Sci.",
      volume         = "53",
      year           = "2003",
      pages          = "77-121",
      doi            = "10.1146/annurev.nucl.53.041002.110503",
      eprint         = "hep-ph/0307284",
      archivePrefix  = "arXiv",
      primaryClass   = "hep-ph",
      SLACcitation   = "%%CITATION = HEP-PH/0307284;%%"
}

@article{Khoury:2003aq,
      author         = "Khoury, Justin and Weltman, Amanda",
      title          = "{Chameleon fields: Awaiting surprises for tests of
                        gravity in space}",
      journal        = "Phys. Rev. Lett.",
      volume         = "93",
      year           = "2004",
      pages          = "171104",
      doi            = "10.1103/PhysRevLett.93.171104",
      eprint         = "astro-ph/0309300",
      archivePrefix  = "arXiv",
      primaryClass   = "astro-ph",
      SLACcitation   = "%%CITATION = ASTRO-PH/0309300;%%"
}

@article{Long:2002wn,
      author         = "Long, Joshua C. and Chan, Hilton W. and Churnside,
                        Allison B. and Gulbis, Eric A. and Varney, Michael C. M.
                        and Price, John C.",
      title          = "{Upper limits to submillimeter-range forces from extra
                        space-time dimensions}",
      year           = "2002",
      doi            = "10.1038/nature01432",
      note           = "[Nature421,922(2003)]",
      eprint         = "hep-ph/0210004",
      archivePrefix  = "arXiv",
      primaryClass   = "hep-ph",
      SLACcitation   = "%%CITATION = HEP-PH/0210004;%%"
}

@article{Damour:1993hw,
      author         = "Damour, Thibault and Esposito-Farese, Gilles",
      title          = "{Nonperturbative strong field effects in tensor - scalar
                        theories of gravitation}",
      journal        = "Phys. Rev. Lett.",
      volume         = "70",
      year           = "1993",
      pages          = "2220-2223",
      doi            = "10.1103/PhysRevLett.70.2220",
      reportNumber   = "IHES-P-93-1, CPT-93-PE-2868",
      SLACcitation   = "%%CITATION = PRLTA,70,2220;%%"
}

@article{Bertotti:2003rm,
      author         = "Bertotti, B. and Iess, L. and Tortora, P.",
      title          = "{A test of general relativity using radio links with the
                        Cassini spacecraft}",
      journal        = "Nature",
      volume         = "425",
      year           = "2003",
      pages          = "374-376",
      doi            = "10.1038/nature01997",
      SLACcitation   = "%%CITATION = NATUA,425,374;%%"
}

@book{Green:1987mn,
      author         = "Green, Michael B. and Schwarz, J. H. and Witten, Edward",
      title          = "{Superstring Theory. Vol. 2: Loop Amplitudes, Anomalies
                        and Phenomenology}",
      year           = "1988",
      url            = "http://www.cambridge.org/us/academic/subjects/physics/theoretical-physics-and-mathematical-physics/superstring-theory-volume-2",
      journal        = "Cambridge, Uk: Univ. Pr. ( 1987) 596 P. ( Cambridge
                        Monographs On Mathematical Physics)",
      ISBN           = "9780521357531",
      SLACcitation   = "%%CITATION = INSPIRE-252419;%%"
}

@article{Kalb:1974yc,
      author         = "Kalb, Michael and Ramond, Pierre",
      title          = "{Classical direct interstring action}",
      journal        = "Phys. Rev.",
      volume         = "D9",
      year           = "1974",
      pages          = "2273-2284",
      doi            = "10.1103/PhysRevD.9.2273",
      SLACcitation   = "%%CITATION = PHRVA,D9,2273;%%"
}

@article{Howe:1996kj,
      author         = "Howe, Paul S. and Papadopoulos, G.",
      title          = "{Twistor spaces for HKT manifolds}",
      journal        = "Phys. Lett.",
      volume         = "B379",
      year           = "1996",
      pages          = "80-86",
      doi            = "10.1016/0370-2693(96)00393-0",
      eprint         = "hep-th/9602108",
      archivePrefix  = "arXiv",
      primaryClass   = "hep-th",
      reportNumber   = "DAMTP-R-96-4, CERN-TH-96-46",
      SLACcitation   = "%%CITATION = HEP-TH/9602108;%%"
}

@article{Kar:2001eb,
      author         = "Kar, Sayan and Majumdar, Parthasarathi and SenGupta,
                        Soumitra and Sur, Saurabh",
      title          = "{Cosmic optical activity from an inhomogeneous
                        Kalb-Ramond field}",
      journal        = "Class. Quant. Grav.",
      volume         = "19",
      year           = "2002",
      pages          = "677-688",
      doi            = "10.1088/0264-9381/19/4/304",
      eprint         = "hep-th/0109135",
      archivePrefix  = "arXiv",
      primaryClass   = "hep-th",
      SLACcitation   = "%%CITATION = HEP-TH/0109135;%%"
}

@article{Letelier:1995ze,
      author         = "Letelier, P. S.",
      title          = "{Spinning strings as torsion line space-time defects}",
      journal        = "Class. Quant. Grav.",
      volume         = "12",
      year           = "1995",
      pages          = "471-478",
      doi            = "10.1088/0264-9381/12/2/016",
      SLACcitation   = "%%CITATION = CQGRD,12,471;%%"
}

@article{Ellis:2013gca,
      author         = "Ellis, John and Mavromatos, Nick E. and Sarkar, Sarben",
      title          = "{Environmental CPT Violation in an Expanding Universe in
                        String Theory}",
      journal        = "Phys. Lett.",
      volume         = "B725",
      year           = "2013",
      pages          = "407-411",
      doi            = "10.1016/j.physletb.2013.07.016",
      eprint         = "1304.5433",
      archivePrefix  = "arXiv",
      primaryClass   = "gr-qc",
      reportNumber   = "KCL-PH-TH-2013-13, LCTS-2013-07, CERN-PH-TH-2013-079",
      SLACcitation   = "%%CITATION = ARXIV:1304.5433;%%"
}

@article{Maity:2004he,
      author         = "Maity, Debaprasad and Majumdar, Parthasarathi and
                        SenGupta, Soumitra",
      title          = "{Parity violating Kalb-Ramond-Maxwell interactions and
                        CMB anisotropy in a brane world}",
      journal        = "JCAP",
      volume         = "0406",
      year           = "2004",
      pages          = "005",
      doi            = "10.1088/1475-7516/2004/06/005",
      eprint         = "hep-th/0401218",
      archivePrefix  = "arXiv",
      primaryClass   = "hep-th",
      SLACcitation   = "%%CITATION = HEP-TH/0401218;%%"
}

@article{Majumdar:1999jd,
      author         = "Majumdar, Parthasarathi and SenGupta, Soumitra",
      title          = "{Parity violating gravitational coupling of
                        electromagnetic fields}",
      journal        = "Class. Quant. Grav.",
      volume         = "16",
      year           = "1999",
      pages          = "L89-L94",
      doi            = "10.1088/0264-9381/16/12/102",
      eprint         = "gr-qc/9906027",
      archivePrefix  = "arXiv",
      primaryClass   = "gr-qc",
      SLACcitation   = "%%CITATION = GR-QC/9906027;%%"
}

@article{Myrzakulov:2010vz,
      author         = "Myrzakulov, Ratbay",
      title          = "{Accelerating universe from F(T) gravity}",
      journal        = "Eur. Phys. J.",
      volume         = "C71",
      year           = "2011",
      pages          = "1752",
      doi            = "10.1140/epjc/s10052-011-1752-9",
      eprint         = "1006.1120",
      archivePrefix  = "arXiv",
      primaryClass   = "gr-qc",
      SLACcitation   = "%%CITATION = ARXIV:1006.1120;%%"
}

@article{Chen:2010va,
      author         = "Chen, Shih-Hung and Dent, James B. and Dutta, Sourish and
                        Saridakis, Emmanuel N.",
      title          = "{Cosmological perturbations in f(T) gravity}",
      journal        = "Phys. Rev.",
      volume         = "D83",
      year           = "2011",
      pages          = "023508",
      doi            = "10.1103/PhysRevD.83.023508",
      eprint         = "1008.1250",
      archivePrefix  = "arXiv",
      primaryClass   = "astro-ph.CO",
      SLACcitation   = "%%CITATION = ARXIV:1008.1250;%%"
}

@article{Dent:2011zz,
      author         = "Dent, James B. and Dutta, Sourish and Saridakis, Emmanuel
                        N.",
      title          = "{f(T) gravity mimicking dynamical dark energy. Background
                        and perturbation analysis}",
      journal        = "JCAP",
      volume         = "1101",
      year           = "2011",
      pages          = "009",
      doi            = "10.1088/1475-7516/2011/01/009",
      eprint         = "1010.2215",
      archivePrefix  = "arXiv",
      primaryClass   = "astro-ph.CO",
      SLACcitation   = "%%CITATION = ARXIV:1010.2215;%%"
}

@article{Bamba:2010wb,
      author         = "Bamba, Kazuharu and Geng, Chao-Qiang and Lee, Chung-Chi
                        and Luo, Ling-Wei",
      title          = "{Equation of state for dark energy in $f(T)$ gravity}",
      journal        = "JCAP",
      volume         = "1101",
      year           = "2011",
      pages          = "021",
      doi            = "10.1088/1475-7516/2011/01/021",
      eprint         = "1011.0508",
      archivePrefix  = "arXiv",
      primaryClass   = "astro-ph.CO",
      SLACcitation   = "%%CITATION = ARXIV:1011.0508;%%"
}

@article{Zhang:2011qp,
      author         = "Zhang, Yi and Li, Hui and Gong, Yungui and Zhu,
                        Zong-Hong",
      title          = "{Notes on $f(T)$ Theories}",
      journal        = "JCAP",
      volume         = "1107",
      year           = "2011",
      pages          = "015",
      doi            = "10.1088/1475-7516/2011/07/015",
      eprint         = "1103.0719",
      archivePrefix  = "arXiv",
      primaryClass   = "astro-ph.CO",
      SLACcitation   = "%%CITATION = ARXIV:1103.0719;%%"
}

@article{Daouda:2011rt,
      author         = "Hamani Daouda, M. and Rodrigues, Manuel E. and Houndjo,
                        M. J. S.",
      title          = "{Static Anisotropic Solutions in f(T) Theory}",
      journal        = "Eur. Phys. J.",
      volume         = "C72",
      year           = "2012",
      pages          = "1890",
      doi            = "10.1140/epjc/s10052-012-1890-8",
      eprint         = "1109.0528",
      archivePrefix  = "arXiv",
      primaryClass   = "physics.gen-ph",
      SLACcitation   = "%%CITATION = ARXIV:1109.0528;%%"
}

@article{Capozziello:2001mq,
      author         = "Capozziello, S. and Lambiase, G. and Stornaiolo, C.",
      title          = "{Geometric classification of the torsion tensor in
                        space-time}",
      journal        = "Annalen Phys.",
      volume         = "10",
      year           = "2001",
      pages          = "713-727",
      doi            = "10.1002/1521-3889(200108)10:8<713::AID-ANDP713>3.0.CO,
                        10.1002/1521-3889(200108)10:8<713::AID-ANDP713>3.0.CO;2-2",
      eprint         = "gr-qc/0101038",
      archivePrefix  = "arXiv",
      primaryClass   = "gr-qc",
      SLACcitation   = "%%CITATION = GR-QC/0101038;%%"
}

@article{Capozziello:2007ec,
      author         = "Capozziello, Salvatore and Francaviglia, Mauro",
      title          = "{Extended Theories of Gravity and their Cosmological and
                        Astrophysical Applications}",
      journal        = "Gen. Rel. Grav.",
      volume         = "40",
      year           = "2008",
      pages          = "357-420",
      doi            = "10.1007/s10714-007-0551-y",
      eprint         = "0706.1146",
      archivePrefix  = "arXiv",
      primaryClass   = "astro-ph",
      SLACcitation   = "%%CITATION = ARXIV:0706.1146;%%"
}

@article{Myrzakulov:2012qp,
      author         = "Myrzakulov, Ratbay",
      title          = "{FRW Cosmology in F(R,T) gravity}",
      journal        = "Eur. Phys. J.",
      volume         = "C72",
      year           = "2012",
      pages          = "2203",
      doi            = "10.1140/epjc/s10052-012-2203-y",
      eprint         = "1207.1039",
      archivePrefix  = "arXiv",
      primaryClass   = "gr-qc",
      SLACcitation   = "%%CITATION = ARXIV:1207.1039;%%"
}

@article{Wanas:2012pu,
      author         = "Wanas, M. I. and Hassan, H. A.",
      title          = "{Torsion and Problems of Standard Cosmology}",
      year           = "2012",
      eprint         = "1209.6218",
      archivePrefix  = "arXiv",
      primaryClass   = "gr-qc",
      SLACcitation   = "%%CITATION = ARXIV:1209.6218;%%"
}

@book{Capozziello:2010zz,
      author         = "Faraoni, Valerio and Capozziello, Salvatore",
      title          = "{Beyond Einstein Gravity}",
      publisher      = "Springer",
      address        = "Dordrecht",
      year           = "2011",
      url            = "http://www.springerlink.com/content/hl1805/#section=801705&page=1",
      journal        = "Fundam. Theor. Phys.",
      volume         = "170. 170",
      series         = "Fundamental Theories of Physics",
      doi            = "10.1007/978-94-007-0165-6",
      ISBN           = "9789400701649, 9789400701656",
      SLACcitation   = "%%CITATION = FTPHD,170,;%%"
}

@article{Ferraro:2011us,
      author         = "Ferraro, Rafael and Fiorini, Franco",
      title          = "{Non trivial frames for f(T) theories of gravity and
                        beyond}",
      journal        = "Phys. Lett.",
      volume         = "B702",
      year           = "2011",
      pages          = "75-80",
      doi            = "10.1016/j.physletb.2011.06.049",
      eprint         = "1103.0824",
      archivePrefix  = "arXiv",
      primaryClass   = "gr-qc",
      SLACcitation   = "%%CITATION = ARXIV:1103.0824;%%"
}

@article{Miao:2011ki,
      author         = "Miao, Rong-Xin and Li, Miao and Miao, Yan-Gang",
      title          = "{Violation of the first law of black hole thermodynamics
                        in $f(T)$ gravity}",
      journal        = "JCAP",
      volume         = "1111",
      year           = "2011",
      pages          = "033",
      doi            = "10.1088/1475-7516/2011/11/033",
      eprint         = "1107.0515",
      archivePrefix  = "arXiv",
      primaryClass   = "hep-th",
      reportNumber   = "USTC-ICTS-11-08",
      SLACcitation   = "%%CITATION = ARXIV:1107.0515;%%"
}

@article{Kar:2002xa,
      author         = "Kar, Sayan and SenGupta, Soumitra and Sur, Saurabh",
      title          = "{Static spherisymmetric solutions, gravitational lensing
                        and perihelion precession in Einstein-Kalb-Ramond theory}",
      journal        = "Phys. Rev.",
      volume         = "D67",
      year           = "2003",
      pages          = "044005",
      doi            = "10.1103/PhysRevD.67.044005",
      eprint         = "hep-th/0210176",
      archivePrefix  = "arXiv",
      primaryClass   = "hep-th",
      SLACcitation   = "%%CITATION = HEP-TH/0210176;%%"
}

@article{SenGupta:2001cs,
      author         = "SenGupta, Soumitra and Sur, Saurabh",
      title          = "{Spherically symmetric solutions of gravitational field
                        equations in Kalb-Ramond background}",
      journal        = "Phys. Lett.",
      volume         = "B521",
      year           = "2001",
      pages          = "350-356",
      doi            = "10.1016/S0370-2693(01)01238-2",
      eprint         = "gr-qc/0102095",
      archivePrefix  = "arXiv",
      primaryClass   = "gr-qc",
      SLACcitation   = "%%CITATION = GR-QC/0102095;%%"
}

@article{Capozziello:2012zj,
      author         = "Capozziello, Salvatore and Gonzalez, P. A. and Saridakis,
                        Emmanuel N. and Vasquez, Yerko",
      title          = "{Exact charged black-hole solutions in D-dimensional f(T)
                        gravity: torsion vs curvature analysis}",
      journal        = "JHEP",
      volume         = "02",
      year           = "2013",
      pages          = "039",
      doi            = "10.1007/JHEP02(2013)039",
      eprint         = "1210.1098",
      archivePrefix  = "arXiv",
      primaryClass   = "hep-th",
      SLACcitation   = "%%CITATION = ARXIV:1210.1098;%%"
}

@article{Ferraro:2006jd,
      author         = "Ferraro, Rafael and Fiorini, Franco",
      title          = "{Modified teleparallel gravity: Inflation without
                        inflaton}",
      journal        = "Phys. Rev.",
      volume         = "D75",
      year           = "2007",
      pages          = "084031",
      doi            = "10.1103/PhysRevD.75.084031",
      eprint         = "gr-qc/0610067",
      archivePrefix  = "arXiv",
      primaryClass   = "gr-qc",
      SLACcitation   = "%%CITATION = GR-QC/0610067;%%"
}

@article{sanchez2008thickness,
  title={The thickness of H i in galactic discs under MOdified Newtonian Dynamics: theory and application to the Galaxy},
  author={S{\'a}nchez-Salcedo, FJ and Saha, K and Narayan, CA},
  journal={Monthly Notices of the Royal Astronomical Society},
  volume={385},
  number={3},
  pages={1585--1596},
  year={2008},
  publisher={Blackwell Publishing Ltd Oxford, UK}
}

@article{Ferraro:2008ey,
      author         = "Ferraro, Rafael and Fiorini, Franco",
      title          = "{On Born-Infeld Gravity in Weitzenbock spacetime}",
      journal        = "Phys. Rev.",
      volume         = "D78",
      year           = "2008",
      pages          = "124019",
      doi            = "10.1103/PhysRevD.78.124019",
      eprint         = "0812.1981",
      archivePrefix  = "arXiv",
      primaryClass   = "gr-qc",
      SLACcitation   = "%%CITATION = ARXIV:0812.1981;%%"
}

@article{Yang:2010ji,
      author         = "Yang, Rong-Jia",
      title          = "{Conformal transformation in $f(T)$ theories}",
      journal        = "Europhys. Lett.",
      volume         = "93",
      year           = "2011",
      pages          = "60001",
      doi            = "10.1209/0295-5075/93/60001",
      eprint         = "1010.1376",
      archivePrefix  = "arXiv",
      primaryClass   = "gr-qc",
      SLACcitation   = "%%CITATION = ARXIV:1010.1376;%%"
}

@article{ArkaniHamed:1999hk,
      author         = "Arkani-Hamed, Nima and Dimopoulos, Savas and Dvali, G. R.
                        and Kaloper, Nemanja",
      title          = "{Infinitely large new dimensions}",
      journal        = "Phys. Rev. Lett.",
      volume         = "84",
      year           = "2000",
      pages          = "586-589",
      doi            = "10.1103/PhysRevLett.84.586",
      eprint         = "hep-th/9907209",
      archivePrefix  = "arXiv",
      primaryClass   = "hep-th",
      reportNumber   = "LBNL-44030, NYU-TH-99-07-02, SU-ITP-99-36, LBL-44030,
                        NYU-TH--99-07-02",
      SLACcitation   = "%%CITATION = HEP-TH/9907209;%%"
}

@article{Chung:1999zs,
      author         = "Chung, Daniel J. H. and Freese, Katherine",
      title          = "{Cosmological challenges in theories with extra
                        dimensions and remarks on the horizon problem}",
      journal        = "Phys. Rev.",
      volume         = "D61",
      year           = "2000",
      pages          = "023511",
      doi            = "10.1103/PhysRevD.61.023511",
      eprint         = "hep-ph/9906542",
      archivePrefix  = "arXiv",
      primaryClass   = "hep-ph",
      SLACcitation   = "%%CITATION = HEP-PH/9906542;%%"
}

@article{Cline:1999ts,
      author         = "Cline, James M. and Grojean, Christophe and Servant,
                        Geraldine",
      title          = "{Cosmological expansion in the presence of extra
                        dimensions}",
      journal        = "Phys. Rev. Lett.",
      volume         = "83",
      year           = "1999",
      pages          = "4245",
      doi            = "10.1103/PhysRevLett.83.4245",
      eprint         = "hep-ph/9906523",
      archivePrefix  = "arXiv",
      primaryClass   = "hep-ph",
      reportNumber   = "MCGILL-99-25, SACLAY-SPH-T-99-065",
      SLACcitation   = "%%CITATION = HEP-PH/9906523;%%"
}

@article{Nihei:1999mt,
      author         = "Nihei, Takeshi",
      title          = "{Inflation in the five-dimensional universe with an
                        orbifold extra dimension}",
      journal        = "Phys. Lett.",
      volume         = "B465",
      year           = "1999",
      pages          = "81-85",
      doi            = "10.1016/S0370-2693(99)01085-0",
      eprint         = "hep-ph/9905487",
      archivePrefix  = "arXiv",
      primaryClass   = "hep-ph",
      reportNumber   = "KEK-TH-630, KEK-PREPRINT-99-32",
      SLACcitation   = "%%CITATION = HEP-PH/9905487;%%"
}

@article{Lipatov:2016ayn,
      author         = "Lipatov, A. V. and Malyshev, M. A.",
      title          = "{On possible small-x effects in the Kaluza-Klein graviton
                        and radion production at high energies}",
      year           = "2016",
      eprint         = "1612.03811",
      archivePrefix  = "arXiv",
      primaryClass   = "hep-ph",
      SLACcitation   = "%%CITATION = ARXIV:1612.03811;%%"
}

@article{Das:2016dtz,
      author         = "Das, Ashmita and Paul, Tanmoy and SenGupta, Soumitra",
      title          = "{Modulus stabilisation in a backreacted warped geometry
                        model via Goldberger-Wise mechanism}",
      year           = "2016",
      eprint         = "1609.07787",
      archivePrefix  = "arXiv",
      primaryClass   = "hep-ph",
      SLACcitation   = "%%CITATION = ARXIV:1609.07787;%%"
}

@article{Arun:2016csq,
      author         = "Arun, Mathew Thomas and Choudhury, Debajyoti",
      title          = "{Stabilization of moduli in spacetime with nested
                        warping}",
      year           = "2016",
      eprint         = "1606.00642",
      archivePrefix  = "arXiv",
      primaryClass   = "hep-th",
      SLACcitation   = "%%CITATION = ARXIV:1606.00642;%%"
}

@article{Bazeia:2014dea,
      author         = "Bazeia, D. and Marques, M. A. and Menezes, R. and
                        Moreira, D. C.",
      title          = "{New braneworld models in the presence of auxiliary
                        fields}",
      journal        = "Annals Phys.",
      volume         = "361",
      year           = "2015",
      pages          = "574-584",
      doi            = "10.1016/j.aop.2015.07.017",
      eprint         = "1412.0135",
      archivePrefix  = "arXiv",
      primaryClass   = "hep-th",
      SLACcitation   = "%%CITATION = ARXIV:1412.0135;%%"
}

@article{Cox:2013rva,
      author         = "Cox, Peter and Medina, Anibal D. and Ray, Tirtha Sankar
                        and Spray, Andrew",
      title          = "{Radion/Dilaton-Higgs Mixing Phenomenology in Light of
                        the LHC}",
      journal        = "JHEP",
      volume         = "02",
      year           = "2014",
      pages          = "032",
      doi            = "10.1007/JHEP02(2014)032",
      eprint         = "1311.3663",
      archivePrefix  = "arXiv",
      primaryClass   = "hep-ph",
      SLACcitation   = "%%CITATION = ARXIV:1311.3663;%%"
}

@article{Pogosian:2007sw,
      author         = "Pogosian, Levon and Silvestri, Alessandra",
      title          = "{The pattern of growth in viable f(R) cosmologies}",
      journal        = "Phys. Rev.",
      volume         = "D77",
      year           = "2008",
      pages          = "023503",
      doi            = "10.1103/PhysRevD.77.023503, 10.1103/PhysRevD.81.049901",
      note           = "[Erratum: Phys. Rev.D81,049901(2010)]",
      eprint         = "0709.0296",
      archivePrefix  = "arXiv",
      primaryClass   = "astro-ph",
      SLACcitation   = "%%CITATION = ARXIV:0709.0296;%%"
}

@article{Mirabelli:1998rt,
      author         = "Mirabelli, Eugene A. and Perelstein, Maxim and Peskin,
                        Michael E.",
      title          = "{Collider signatures of new large space dimensions}",
      journal        = "Phys. Rev. Lett.",
      volume         = "82",
      year           = "1999",
      pages          = "2236-2239",
      doi            = "10.1103/PhysRevLett.82.2236",
      eprint         = "hep-ph/9811337",
      archivePrefix  = "arXiv",
      primaryClass   = "hep-ph",
      reportNumber   = "SLAC-PUB-8002",
      SLACcitation   = "%%CITATION = HEP-PH/9811337;%%"
}

@article{Goldberger:1999wh,
      author         = "Goldberger, Walter D. and Wise, Mark B.",
      title          = "{Bulk fields in the Randall-Sundrum compactification
                        scenario}",
      journal        = "Phys. Rev.",
      volume         = "D60",
      year           = "1999",
      pages          = "107505",
      doi            = "10.1103/PhysRevD.60.107505",
      eprint         = "hep-ph/9907218",
      archivePrefix  = "arXiv",
      primaryClass   = "hep-ph",
      reportNumber   = "CALT-68-2229",
      SLACcitation   = "%%CITATION = HEP-PH/9907218;%%"
}

@article{Dudas:2005vna,
      author         = "Dudas, Emilian and Quiros, Mariano",
      title          = "{Five-dimensional massive vector fields and radion
                        stabilization}",
      journal        = "Nucl. Phys.",
      volume         = "B721",
      year           = "2005",
      pages          = "309-324",
      doi            = "10.1016/j.nuclphysb.2005.05.028",
      eprint         = "hep-th/0503157",
      archivePrefix  = "arXiv",
      primaryClass   = "hep-th",
      reportNumber   = "CPTH-RR-018-0305, LPT-ORSAY-05-20, UAB-FT-580",
      SLACcitation   = "%%CITATION = HEP-TH/0503157;%%"
}

@article{Hewett:1998sn,
      author         = "Hewett, JoAnne L.",
      title          = "{Indirect collider signals for extra dimensions}",
      journal        = "Phys. Rev. Lett.",
      volume         = "82",
      year           = "1999",
      pages          = "4765-4768",
      doi            = "10.1103/PhysRevLett.82.4765",
      eprint         = "hep-ph/9811356",
      archivePrefix  = "arXiv",
      primaryClass   = "hep-ph",
      reportNumber   = "SLAC-PUB-8001",
      SLACcitation   = "%%CITATION = HEP-PH/9811356;%%"
}

@article{Hewett:2016omf,
      author         = "Hewett, JoAnne L. and Rizzo, Thomas G.",
      title          = "{750 GeV Diphoton Resonance in Warped Geometries}",
      year           = "2016",
      eprint         = "1603.08250",
      archivePrefix  = "arXiv",
      primaryClass   = "hep-ph",
      reportNumber   = "SLAC-PUB-16499",
      SLACcitation   = "%%CITATION = ARXIV:1603.08250;%%"
}

@article{Giddings:2016sfr,
      author         = "Giddings, Steven B. and Zhang, Hao",
      title          = "{Kaluza-Klein graviton phenomenology for warped
                        compactifications, and the 750 GeV diphoton excess}",
      journal        = "Phys. Rev.",
      volume         = "D93",
      year           = "2016",
      number         = "11",
      pages          = "115002",
      doi            = "10.1103/PhysRevD.93.115002",
      eprint         = "1602.02793",
      archivePrefix  = "arXiv",
      primaryClass   = "hep-ph",
      SLACcitation   = "%%CITATION = ARXIV:1602.02793;%%"
}

@article{Oliveira:2014kla,
      author         = "Oliveira, Alexandra",
      title          = "{Gravity particles from Warped Extra Dimensions,
                        predictions for LHC}",
      year           = "2014",
      eprint         = "1404.0102",
      archivePrefix  = "arXiv",
      primaryClass   = "hep-ph",
      SLACcitation   = "%%CITATION = ARXIV:1404.0102;%%"
}

@article{Cho:2013mva,
      author         = "Cho, Gi-Chol and Nomura, Daisuke and Ohno, Yoshiko",
      title          = "{Constraints on radion in a warped extra dimension model
                        from Higgs boson searches at the LHC}",
      journal        = "Mod. Phys. Lett.",
      volume         = "A28",
      year           = "2013",
      pages          = "1350148",
      doi            = "10.1142/S0217732313501484",
      eprint         = "1305.4431",
      archivePrefix  = "arXiv",
      primaryClass   = "hep-ph",
      reportNumber   = "OCHA-PP-315, TU-936",
      SLACcitation   = "%%CITATION = ARXIV:1305.4431;%%"
}

@article{Csaki:2000zn,
      author         = "Csaki, Csaba and Graesser, Michael L. and Kribs, Graham
                        D.",
      title          = "{Radion dynamics and electroweak physics}",
      journal        = "Phys. Rev.",
      volume         = "D63",
      year           = "2001",
      pages          = "065002",
      doi            = "10.1103/PhysRevD.63.065002",
      eprint         = "hep-th/0008151",
      archivePrefix  = "arXiv",
      primaryClass   = "hep-th",
      reportNumber   = "SCIPP-00-27",
      SLACcitation   = "%%CITATION = HEP-TH/0008151;%%"
}

@article{bizyaev2016very,
  title={Very thin disc galaxies in the SDSS catalogue of edge-on galaxies},
  author={Bizyaev, DV and Kautsch, SJ and Sotnikova, N Ya and Reshetnikov, Vladimir P and Mosenkov, Aleksander V},
  journal={Monthly Notices of the Royal Astronomical Society},
  volume={465},
  number={4},
  pages={3784--3792},
  year={2016},
  publisher={Oxford University Press}
}

@article{boily2003thickness,
  title={The Thickness of Stellar Disks of Edge-on Galaxies and Their Truncation Radii},
  author={Boily, C and Patsis, P and Portegies, S and Spurzem, R and Theis, C and Zasov, AV and Bizyaev, DV},
  journal={European Astronomical Society Publications Series},
  volume={10},
  pages={121--121},
  year={2003},
  publisher={EDP Sciences}
}

@article{banerjee2016mass,
  title={Mass modelling of superthin galaxies: IC5249, UGC7321 and IC2233},
  author={Banerjee, Arunima and Bapat, Disha},
  journal={Monthly Notices of the Royal Astronomical Society},
  volume={466},
  number={3},
  pages={3753--3761},
  year={2016},
  publisher={Oxford University Press}
}

@article{narayan2002vertical,
  title={Vertical scaleheights in a gravitationally coupled, three-component Galactic disk},
  author={Narayan, Chaitra A and Jog, Chanda J},
  journal={Astronomy \& Astrophysics},
  volume={394},
  number={1},
  pages={89--96},
  year={2002},
  publisher={EDP Sciences}
}

@article{romeo2011effective,
  title={The effective stability parameter for two-component galactic discs},
  author={Romeo, Alessandro B and Wiegert, Joachim},
  journal={Monthly Notices of the Royal Astronomical Society},
  volume={416},
  number={2},
  pages={1191--1196},
  year={2011},
  publisher={The Royal Astronomical Society}
}

@article{rohlfs1977lectures,
  title={Lectures on density wave theory},
  author={Rohlfs, Kristen},
  year={1977}
}

@article{binney2011models,
  title={Models of our Galaxy--II},
  author={Binney, James and McMillan, Paul},
  journal={Monthly Notices of the Royal Astronomical Society},
  volume={413},
  number={3},
  pages={1889--1898},
  year={2011},
  publisher={Blackwell Publishing Ltd Oxford, UK}
}

@article{toomre1964gravitational,
  title={On the gravitational stability of a disk of stars},
  author={Toomre, Alar},
  journal={The Astrophysical Journal},
  volume={139},
  pages={1217--1238},
  year={1964}
}

@article{wang1994gravitational,
  title={Gravitational instability and disk star formation},
  author={Wang, Boqi and Silk, Joseph},
  journal={The Astrophysical Journal},
  volume={427},
  pages={759--769},
  year={1994}
}

@article{romeo2013simple,
  title={A simple and accurate approximation for the Q stability parameter in multicomponent and realistically thick discs},
  author={Romeo, Alessandro B and Falstad, Niklas},
  journal={Monthly Notices of the Royal Astronomical Society},
  volume={433},
  number={2},
  pages={1389--1397},
  year={2013},
  publisher={Oxford University Press}
}

@article{walter2008things,
  title={THINGS: The HI nearby galaxy survey},
  author={Walter, Fabian and Brinks, Elias and de Blok, WJG and Bigiel, Frank and Kennicutt Jr, Robert C and Thornley, Michele D and Leroy, Adam},
  journal={The Astronomical Journal},
  volume={136},
  number={6},
  pages={2563},
  year={2008},
  publisher={IOP Publishing}
}

@article{westfall2014diskmass,
  title={The DiskMass survey. VIII. On the relationship between disk stability and star formation},
  author={Westfall, Kyle B and Andersen, David R and Bershady, Matthew A and Martinsson, Thomas PK and Swaters, Robert A and Verheijen, Marc AW},
  journal={The Astrophysical Journal},
  volume={785},
  number={1},
  pages={43},
  year={2014},
  publisher={IOP Publishing}
}
@article{matthews2000extraordinary,
  title={The Extraordinary “Superthin” Spiral Galaxy UGC 7321. II. The Vertical Disk Structure},
  author={Matthews, LD},
  journal={The Astronomical Journal},
  volume={120},
  number={4},
  pages={1764},
  year={2000},
  publisher={IOP Publishing}
}

@article{matthews1999extraordinary,
  title={The Extraordinary “Superthin” Spiral Galaxy UGC 7321. I. Disk Color Gradients and Global Properties from Multiwavelength Observations},
  author={Matthews, LD and Gallagher III, JS and Van Driel, W},
  journal={The Astronomical Journal},
  volume={118},
  number={6},
  pages={2751},
  year={1999},
  publisher={IOP Publishing}
}

@inproceedings{mendelowitz2000rotation,
  title={Rotation Curve and Mass Decomposition for the Edge-on Spiral Galaxy UGC 711},
  author={Mendelowitz, CM and Matthews, LD and Hibbard, JE and Wilcots, EM},
  booktitle={Bulletin of the American Astronomical Society},
  volume={32},
  pages={1459},
  year={2000}
}

@article{giovanelli1997spectroscopy,
  title={Spectroscopy of edge-on spirals},
  author={Giovanelli, Riccardo and Avera, Eric and Karachentsev, Igor},
  journal={arXiv preprint astro-ph/9704189},
  year={1997}
}

@article{karachentsev1993flat,
  title={Flat galaxy catalogue},
  author={Karachentsev, ID and Karachentseva, VE and Parnovsky, SL},
  journal={Astronomische Nachrichten},
  volume={314},
  number={3},
  pages={97--222},
  year={1993},
  publisher={Wiley Online Library}
}

@article{katz2018gaia,
  title={Gaia Data Release 2-Mapping the Milky Way disc kinematics},
  author={Katz, D and Antoja, T and Romero-G{\'o}mez, Manuel and Drimmel, R and Reyl{\'e}, C and Seabroke, GM and Soubiran, C and Babusiaux, C and Di Matteo, P and Figueras, F and others},
  journal={Astronomy \& astrophysics},
  volume={616},
  pages={A11},
  year={2018},
  publisher={EDP sciences}
}

@article{banerjee2007origin,
  title={The origin of steep vertical stellar distribution in the Galactic disk},
  author={Banerjee, Arunima and Jog, Chanda J},
  journal={The Astrophysical Journal},
  volume={662},
  number={1},
  pages={335},
  year={2007},
  publisher={IOP Publishing}
}

@article{roberts1994physical,
  title={Physical parameters along the Hubble sequence},
  author={Roberts, Morton S and Haynes, Martha P},
  journal={Annual Review of Astronomy and Astrophysics},
  volume={32},
  number={1},
  pages={115--152},
  year={1994},
  publisher={Annual Reviews 4139 El Camino Way, PO Box 10139, Palo Alto, CA 94303-0139, USA}
}

@inproceedings{yock1999observation,
  title={Observation of the Halo of the Edge-On Galaxy IC 5249},
  author={Yock, Philip and Pennycook, Glen and Rattenbury, Nicholas and Koribalski, Baerbel and Muraki, Yasushi and Yanagisawa, Toshi and Jugaku, Jun and Dodd, Richard},
  booktitle={The Third Stromlo Symposium: The Galactic Halo},
  volume={165},
  pages={187},
  year={1999}
}

@article{querejeta2015spitzer,
  title={The Spitzer Survey of Stellar Structure in Galaxies (S4G): precise stellar mass distributions from automated dust correction at 3.6 $\mu$m},
  author={Querejeta, Miguel and Meidt, Sharon E and Schinnerer, Eva and Cisternas, Mauricio and Mu{\~n}oz-Mateos, Juan Carlos and Sheth, Kartik and Knapen, Johan and Van De Ven, Glenn and Norris, Mark A and Peletier, Reynier and others},
  journal={The Astrophysical Journal Supplement Series},
  volume={219},
  number={1},
  pages={5},
  year={2015},
  publisher={IOP Publishing}
}

@inproceedings{van1992groningen,
  title={The Groningen image processing system, GIPSY},
  author={Van der Hulst, JM and Terlouw, JP and Begeman, KG and Zwitser, W and Roelfsema, PR},
  booktitle={Astronomical Data Analysis Software and Systems I},
  volume={25},
  pages={131},
  year={1992}
}

@article{braine2000deep,
  title={Deep search for CO emission in the Low Surface Brightness galaxy Malin 1},
  author={Braine, J and Herpin, F and Radford, SJE},
  journal={Astronomy and Astrophysics},
  volume={358},
  pages={494--498},
  year={2000}
}

@article{pinna2018revisiting,
  title={Revisiting the stellar velocity ellipsoid--Hubble-type relation: observations versus simulations},
  author={Pinna, F and Falc{\'o}n-Barroso, J and Martig, M and Mart{\'\i}nez-Valpuesta, I and M{\'e}ndez-Abreu, J and van de Ven, G and Leaman, R and Lyubenova, M},
  journal={Monthly Notices of the Royal Astronomical Society},
  volume={475},
  number={2},
  pages={2697--2712},
  year={2018},
  publisher={Oxford University Press}
}

@article{aumer2016age,
  title={Age--velocity dispersion relations and heating histories in disc galaxies},
  author={Aumer, Michael and Binney, James and Sch{\"o}nrich, Ralph},
  journal={Monthly Notices of the Royal Astronomical Society},
  volume={462},
  number={2},
  pages={1697--1713},
  year={2016},
  publisher={Oxford University Press}
}

@article{jenkins1990spiral,
  title={Spiral heating of galactic discs},
  author={Jenkins, Adrian and Binney, James},
  journal={Monthly Notices of the Royal Astronomical Society},
  volume={245},
  pages={305--317},
  year={1990}
}

@article{grand2016vertical,
  title={Vertical disc heating in Milky Way-sized galaxies in a cosmological context},
  author={Grand, Robert JJ and Springel, Volker and G{\'o}mez, Facundo A and Marinacci, Federico and Pakmor, R{\"u}diger and Campbell, David JR and Jenkins, Adrian},
  journal={Monthly Notices of the Royal Astronomical Society},
  volume={459},
  number={1},
  pages={199--219},
  year={2016},
  publisher={Oxford University Press}
}

@article{gerssen2012disc,
  title={Disc heating agents across the Hubble sequence},
  author={Gerssen, J and Shapiro Griffin, K},
  journal={Monthly Notices of the Royal Astronomical Society},
  volume={423},
  number={3},
  pages={2726--2735},
  year={2012},
  publisher={Blackwell Publishing Ltd Oxford, UK}
}

@article{saha2014disc,
  title={Disc heating: possible link between weak bars and superthin galaxies},
  author={Saha, Kanak},
  journal={arXiv preprint arXiv:1403.1711},
  year={2014}
}

@article{fernandez2016ggmcmc,
  title={ggmcmc: Analysis of MCMC samples and Bayesian inference},
  author={Fern{\'a}ndez-i-Mar{\i}n, Xavier},
  journal={Journal of Statistical Software},
  volume={70},
  number={9},
  pages={1--20},
  year={2016}
}

@article{hartig2017bayesiantools,
  title={Bayesiantools: General-purpose MCMC and SMC samplers and tools for Bayesian statistics},
  author={Hartig, F and Minunno, F and Paul, S and Cameron, D and Ott, T},
  journal={R package version. R package version 0.1},
  volume={3},
  year={2017}
}

@article{wickham2011ggplot2,
  title={ggplot2},
  author={Wickham, Hadley},
  journal={Wiley Interdisciplinary Reviews: Computational Statistics},
  volume={3},
  number={2},
  pages={180--185},
  year={2011},
  publisher={Wiley Online Library}
}

@misc{meschiari2015latex2exp,
  title={latex2exp: Use LaTeX Expressions in Plots, r package version 0.4. 0},
  author={Meschiari, S},
  year={2015}
}

@article{hunter2007matplotlib,
  title={Matplotlib: A 2D graphics environment},
  author={Hunter, John D},
  journal={Computing in science \& engineering},
  volume={9},
  number={3},
  pages={90--95},
  year={2007},
  publisher={IEEE Computer Society}
}

@article{vaughan2016false,
  title={False periodicities in quasar time-domain surveys},
  author={Vaughan, S and Uttley, P and Markowitz, AG and Huppenkothen, D and Middleton, MJ and Alston, WN and Scargle, JD and Farr, WM},
  journal={Monthly Notices of the Royal Astronomical Society},
  volume={461},
  number={3},
  pages={3145--3152},
  year={2016},
  publisher={Oxford University Press}
}

@article{gergely2011galactic,
  title={Galactic rotation curves in brane world models},
  author={Gergely, L{\'A} and Harko, T and Dwornik, M and Kupi, G and Keresztes, Z},
  journal={Monthly Notices of the Royal Astronomical Society},
  volume={415},
  number={4},
  pages={3275--3290},
  year={2011},
  publisher={Blackwell Publishing Ltd Oxford, UK}
}

@article{garg2017origin,
  title={Origin of low surface brightness galaxies: a dynamical study},
  author={Garg, Prerak and Banerjee, Arunima},
  journal={Monthly Notices of the Royal Astronomical Society},
  volume={472},
  number={1},
  pages={166--173},
  year={2017},
  publisher={Oxford University Press}
}
@article{rahaman2008galactic,
  title={Galactic rotation curves and brane-world models},
  author={Rahaman, F and Kalam, M and DeBenedictis, A and Usmani, AA and Ray, Saibal},
  journal={Monthly Notices of the Royal Astronomical Society},
  volume={389},
  number={1},
  pages={27--33},
  year={2008},
  publisher={The Royal Astronomical Society}
}
@article{de2005halo,
  title={Halo mass profiles and low surface brightness galaxy rotation curves},
  author={de Blok, WJG},
  journal={The Astrophysical Journal},
  volume={634},
  number={1},
  pages={227},
  year={2005},
  publisher={IOP Publishing}
}

@article{Shiromizu:1999wj,
      author         = "Shiromizu, Tetsuya and Maeda, Kei-ichi and Sasaki, Misao",
      title          = "{The Einstein equation on the 3-brane world}",
      journal        = "Phys.Rev.",
      volume         = "D62",
      pages          = "024012",
      doi            = "10.1103/PhysRevD.62.024012",
      year           = "2000",
      eprint         = "gr-qc/9910076",
      archivePrefix  = "arXiv",
      primaryClass   = "gr-qc",
      reportNumber   = "DAMTP-1999-150, OUTAP-103, UTAP-349, RESCEU-40-99",
      SLACcitation   = "%%CITATION = GR-QC/9910076;%%",
}

@article{Harko:2004ui,
      author         = "Harko, T. and Mak, M.K.",
      title          = "{Vacuum solutions of the gravitational field equations in
                        the brane world model}",
      journal        = "Phys.Rev.",
      volume         = "D69",
      pages          = "064020",
      doi            = "10.1103/PhysRevD.69.064020",
      year           = "2004",
      eprint         = "gr-qc/0401049",
      archivePrefix  = "arXiv",
      primaryClass   = "gr-qc",
      SLACcitation   = "%%CITATION = GR-QC/0401049;%%",
}

@article{Keresztes:2009wb,
      author         = "Keresztes, Zoltan and Gergely, Laszlo A.",
      title          = "{Covariant gravitational dynamics in 3+1+1 dimensions}",
      journal        = "Class. Quant. Grav.",
      volume         = "27",
      year           = "2010",
      pages          = "105009",
      doi            = "10.1088/0264-9381/27/10/105009",
      eprint         = "0909.0490",
      archivePrefix  = "arXiv",
      primaryClass   = "gr-qc",
      SLACcitation   = "%%CITATION = ARXIV:0909.0490;%%"
}

@article{Keresztes:2009hy,
      author         = "Keresztes, Zoltan and Gergely, Laszlo A.",
      title          = "{3+1+1 dimensional covariant gravitational dynamics on an
                        asymmetrically embedded brane}",
      booktitle      = "{Proceedings, Grassmannian Conference in Fundamental
                        Cosmology (Grasscosmofun'09): Szczecin, Poland, September
                        14-19, 2009}",
      journal        = "Annalen Phys.",
      volume         = "19",
      year           = "2010",
      pages          = "249-253",
      doi            = "10.1002/andp.201010421, 10.1142/9789814374552_0371",
      note           = "[,1948(2009)]",
      eprint         = "0911.2495",
      archivePrefix  = "arXiv",
      primaryClass   = "gr-qc",
      SLACcitation   = "%%CITATION = ARXIV:0911.2495;%%"
}

@article{Chakraborty:2015zxc,
      author         = "Chakraborty, Sumanta and SenGupta, Soumitra",
      title          = "{Kinematics of Radion field: A possible source of dark
                        matter}",
      journal        = "Eur. Phys. J.",
      volume         = "C76",
      year           = "2016",
      number         = "12",
      pages          = "648",
      doi            = "10.1140/epjc/s10052-016-4512-z",
      eprint         = "1511.00646",
      archivePrefix  = "arXiv",
      primaryClass   = "gr-qc",
      SLACcitation   = "%%CITATION = ARXIV:1511.00646;%%"
}

@article{rubin1979extended,
  title={Extended rotation curves of high-luminosity spiral galaxies. V-NGC 1961, the most massive spiral known},
  author={Rubin, VC and Ford Jr, WK and Roberts, MS},
  journal={The Astrophysical Journal},
  volume={230},
  pages={35--39},
  year={1979}
}

@article{zwicky1933rotverschiebung,
  title={Die rotverschiebung von extragalaktischen nebeln},
  author={Zwicky, Fritz},
  journal={Helvetica physica acta},
  volume={6},
  pages={110--127},
  year={1933}
}

@article{sofue2001rotation,
  title={Rotation curves of spiral galaxies},
  author={Sofue, Yoshiaki and Rubin, Vera},
  journal={Annual Review of Astronomy and Astrophysics},
  volume={39},
  number={1},
  pages={137--174},
  year={2001},
  publisher={Annual Reviews 4139 El Camino Way, PO Box 10139, Palo Alto, CA 94303-0139, USA}
}

@article{kranz2003dark,
  title={Dark matter within high surface brightness spiral galaxies},
  author={Kranz, Thilo and Slyz, Adrianne and Rix, Hans-Walter},
  journal={The Astrophysical Journal},
  volume={586},
  number={1},
  pages={143},
  year={2003},
  publisher={IOP Publishing}
}

@article{gentile2004cored,
  title={The cored distribution of dark matter in spiral galaxies},
  author={Gentile, Gianfranco and Salucci, Paolo and Klein, U and Vergani, D and Kalberla, P},
  journal={Monthly Notices of the Royal Astronomical Society},
  volume={351},
  number={3},
  pages={903--922},
  year={2004},
  publisher={Blackwell Publishing Ltd Oxford, UK}
}

@article{de2001high,
  title={High-resolution rotation curves of low surface brightness galaxies. II. Mass models},
  author={de Blok, WJG and McGaugh, Stacy S and Rubin, Vera C},
  journal={The Astronomical Journal},
  volume={122},
  number={5},
  pages={2396},
  year={2001},
  publisher={IOP Publishing}
}

@article{carlberg1997average,
  title={The average mass profile of galaxy clusters},
  author={Carlberg, RG and Yee, HKC and Ellingson, E and Morris, SL and Abraham, R and Gravel, Pl and Pritchet, CJ and Smecker-Hane, T and Hartwick, FDA and Hesser, JE and others},
  journal={The Astrophysical Journal Letters},
  volume={485},
  number={1},
  pages={L13},
  year={1997},
  publisher={IOP Publishing}
}

@article{mcgaugh2001high,
  title={High-resolution rotation curves of low surface brightness galaxies. I. Data},
  author={McGaugh, Stacy S and Rubin, Vera C and de Blok, WJG},
  journal={The Astronomical Journal},
  volume={122},
  number={5},
  pages={2381},
  year={2001},
  publisher={IOP Publishing}
}

@article{famaey2005modified,
  title={Modified newtonian dynamics in the milky way},
  author={Famaey, Benoit and Binney, James},
  journal={Monthly Notices of the Royal Astronomical Society},
  volume={363},
  number={2},
  pages={603--608},
  year={2005},
  publisher={Blackwell Science Ltd Oxford, UK}
}

@article{boily2003thickness,
  title={The Thickness of Stellar Disks of Edge-on Galaxies and Their Truncation Radii},
  author={Boily, C and Patsis, P and Portegies, S and Spurzem, R and Theis, C and Zasov, AV and Bizyaev, DV},
  journal={European Astronomical Society Publications Series},
  volume={10},
  pages={121--121},
  year={2003},
  publisher={EDP Sciences}
}

@article{binney2011models,
  title={Models of our Galaxy--II},
  author={Binney, James and McMillan, Paul},
  journal={Monthly Notices of the Royal Astronomical Society},
  volume={413},
  number={3},
  pages={1889--1898},
  year={2011},
  publisher={Blackwell Publishing Ltd Oxford, UK}
}

@article{wang1994gravitational,
  title={Gravitational instability and disk star formation},
  author={Wang, Boqi and Silk, Joseph},
  journal={The Astrophysical Journal},
  volume={427},
  pages={759--769},
  year={1994}
}

@article{westfall2014diskmass,
  title={The DiskMass survey. VIII. On the relationship between disk stability and star formation},
  author={Westfall, Kyle B and Andersen, David R and Bershady, Matthew A and Martinsson, Thomas PK and Swaters, Robert A and Verheijen, Marc AW},
  journal={The Astrophysical Journal},
  volume={785},
  number={1},
  pages={43},
  year={2014},
  publisher={IOP Publishing}
}
@article{matthews2000extraordinary,
  title={The Extraordinary “Superthin” Spiral Galaxy UGC 7321. II. The Vertical Disk Structure},
  author={Matthews, LD},
  journal={The Astronomical Journal},
  volume={120},
  number={4},
  pages={1764},
  year={2000},
  publisher={IOP Publishing}
}

@article{giovanelli1997spectroscopy,
  title={Spectroscopy of edge-on spirals},
  author={Giovanelli, Riccardo and Avera, Eric and Karachentsev, Igor},
  journal={arXiv preprint astro-ph/9704189},
  year={1997}
}

@article{karachentsev1993flat,
  title={Flat galaxy catalogue},
  author={Karachentsev, ID and Karachentseva, VE and Parnovsky, SL},
  journal={Astronomische Nachrichten},
  volume={314},
  number={3},
  pages={97--222},
  year={1993},
  publisher={Wiley Online Library}
}

@article{banerjee2007origin,
  title={The origin of steep vertical stellar distribution in the Galactic disk},
  author={Banerjee, Arunima and Jog, Chanda J},
  journal={The Astrophysical Journal},
  volume={662},
  number={1},
  pages={335},
  year={2007},
  publisher={IOP Publishing}
}

@article{roberts1994physical,
  title={Physical parameters along the Hubble sequence},
  author={Roberts, Morton S and Haynes, Martha P},
  journal={Annual Review of Astronomy and Astrophysics},
  volume={32},
  number={1},
  pages={115--152},
  year={1994},
  publisher={Annual Reviews 4139 El Camino Way, PO Box 10139, Palo Alto, CA 94303-0139, USA}
}

@article{querejeta2015spitzer,
  title={The Spitzer Survey of Stellar Structure in Galaxies (S4G): precise stellar mass distributions from automated dust correction at 3.6 $\mu$m},
  author={Querejeta, Miguel and Meidt, Sharon E and Schinnerer, Eva and Cisternas, Mauricio and Mu{\~n}oz-Mateos, Juan Carlos and Sheth, Kartik and Knapen, Johan and Van De Ven, Glenn and Norris, Mark A and Peletier, Reynier and others},
  journal={The Astrophysical Journal Supplement Series},
  volume={219},
  number={1},
  pages={5},
  year={2015},
  publisher={IOP Publishing}
}

@article{braine2000deep,
  title={Deep search for CO emission in the Low Surface Brightness galaxy Malin 1},
  author={Braine, J and Herpin, F and Radford, SJE},
  journal={Astronomy and Astrophysics},
  volume={358},
  pages={494--498},
  year={2000}
}

@article{fernandez2016ggmcmc,
  title={ggmcmc: Analysis of MCMC samples and Bayesian inference},
  author={Fern{\'a}ndez-i-Mar{\i}n, Xavier},
  journal={Journal of Statistical Software},
  volume={70},
  number={9},
  pages={1--20},
  year={2016}
}

@article{hartig2017bayesiantools,
  title={Bayesiantools: General-purpose MCMC and SMC samplers and tools for Bayesian statistics},
  author={Hartig, F and Minunno, F and Paul, S and Cameron, D and Ott, T},
  journal={R package version. R package version 0.1},
  volume={3},
  year={2017}
}

@article{wickham2011ggplot2,
  title={ggplot2},
  author={Wickham, Hadley},
  journal={Wiley Interdisciplinary Reviews: Computational Statistics},
  volume={3},
  number={2},
  pages={180--185},
  year={2011},
  publisher={Wiley Online Library}
}

@misc{meschiari2015latex2exp,
  title={latex2exp: Use LaTeX Expressions in Plots, r package version 0.4. 0},
  author={Meschiari, S},
  year={2015}
}

@article{hunter2007matplotlib,
  title={Matplotlib: A 2D graphics environment},
  author={Hunter, John D},
  journal={Computing in science \& engineering},
  volume={9},
  number={3},
  pages={90--95},
  year={2007},
  publisher={IEEE Computer Society}
}

@article{vaughan2016false,
  title={False periodicities in quasar time-domain surveys},
  author={Vaughan, S and Uttley, P and Markowitz, AG and Huppenkothen, D and Middleton, MJ and Alston, WN and Scargle, JD and Farr, WM},
  journal={Monthly Notices of the Royal Astronomical Society},
  volume={461},
  number={3},
  pages={3145--3152},
  year={2016},
  publisher={Oxford University Press}
}

@article{Shiromizu:1999wj,
      author         = "Shiromizu, Tetsuya and Maeda, Kei-ichi and Sasaki, Misao",
      title          = "{The Einstein equation on the 3-brane world}",
      journal        = "Phys.Rev.",
      volume         = "D62",
      pages          = "024012",
      doi            = "10.1103/PhysRevD.62.024012",
      year           = "2000",
      eprint         = "gr-qc/9910076",
      archivePrefix  = "arXiv",
      primaryClass   = "gr-qc",
      reportNumber   = "DAMTP-1999-150, OUTAP-103, UTAP-349, RESCEU-40-99",
      SLACcitation   = "%%CITATION = GR-QC/9910076;%%",
}

@article{Harko:2004ui,
      author         = "Harko, T. and Mak, M.K.",
      title          = "{Vacuum solutions of the gravitational field equations in
                        the brane world model}",
      journal        = "Phys.Rev.",
      volume         = "D69",
      pages          = "064020",
      doi            = "10.1103/PhysRevD.69.064020",
      year           = "2004",
      eprint         = "gr-qc/0401049",
      archivePrefix  = "arXiv",
      primaryClass   = "gr-qc",
      SLACcitation   = "%%CITATION = GR-QC/0401049;%%",
}

@article{Chakraborty:2015zxc,
      author         = "Chakraborty, Sumanta and SenGupta, Soumitra",
      title          = "{Kinematics of Radion field: A possible source of dark
                        matter}",
      journal        = "Eur. Phys. J.",
      volume         = "C76",
      year           = "2016",
      number         = "12",
      pages          = "648",
      doi            = "10.1140/epjc/s10052-016-4512-z",
      eprint         = "1511.00646",
      archivePrefix  = "arXiv",
      primaryClass   = "gr-qc",
      SLACcitation   = "%%CITATION = ARXIV:1511.00646;%%"
}

@article{boily2003thickness,
  title={The Thickness of Stellar Disks of Edge-on Galaxies and Their Truncation Radii},
  author={Boily, C and Patsis, P and Portegies, S and Spurzem, R and Theis, C and Zasov, AV and Bizyaev, DV},
  journal={European Astronomical Society Publications Series},
  volume={10},
  pages={121--121},
  year={2003},
  publisher={EDP Sciences}
}

@article{binney2011models,
  title={Models of our Galaxy--II},
  author={Binney, James and McMillan, Paul},
  journal={Monthly Notices of the Royal Astronomical Society},
  volume={413},
  number={3},
  pages={1889--1898},
  year={2011},
  publisher={Blackwell Publishing Ltd Oxford, UK}
}

@article{wang1994gravitational,
  title={Gravitational instability and disk star formation},
  author={Wang, Boqi and Silk, Joseph},
  journal={The Astrophysical Journal},
  volume={427},
  pages={759--769},
  year={1994}
}

@article{westfall2014diskmass,
  title={The DiskMass survey. VIII. On the relationship between disk stability and star formation},
  author={Westfall, Kyle B and Andersen, David R and Bershady, Matthew A and Martinsson, Thomas PK and Swaters, Robert A and Verheijen, Marc AW},
  journal={The Astrophysical Journal},
  volume={785},
  number={1},
  pages={43},
  year={2014},
  publisher={IOP Publishing}
}
@article{matthews2000extraordinary,
  title={The Extraordinary “Superthin” Spiral Galaxy UGC 7321. II. The Vertical Disk Structure},
  author={Matthews, LD},
  journal={The Astronomical Journal},
  volume={120},
  number={4},
  pages={1764},
  year={2000},
  publisher={IOP Publishing}
}

@article{giovanelli1997spectroscopy,
  title={Spectroscopy of edge-on spirals},
  author={Giovanelli, Riccardo and Avera, Eric and Karachentsev, Igor},
  journal={arXiv preprint astro-ph/9704189},
  year={1997}
}

@article{karachentsev1993flat,
  title={Flat galaxy catalogue},
  author={Karachentsev, ID and Karachentseva, VE and Parnovsky, SL},
  journal={Astronomische Nachrichten},
  volume={314},
  number={3},
  pages={97--222},
  year={1993},
  publisher={Wiley Online Library}
}

@article{banerjee2007origin,
  title={The origin of steep vertical stellar distribution in the Galactic disk},
  author={Banerjee, Arunima and Jog, Chanda J},
  journal={The Astrophysical Journal},
  volume={662},
  number={1},
  pages={335},
  year={2007},
  publisher={IOP Publishing}
}

@article{roberts1994physical,
  title={Physical parameters along the Hubble sequence},
  author={Roberts, Morton S and Haynes, Martha P},
  journal={Annual Review of Astronomy and Astrophysics},
  volume={32},
  number={1},
  pages={115--152},
  year={1994},
  publisher={Annual Reviews 4139 El Camino Way, PO Box 10139, Palo Alto, CA 94303-0139, USA}
}

@article{querejeta2015spitzer,
  title={The Spitzer Survey of Stellar Structure in Galaxies (S4G): precise stellar mass distributions from automated dust correction at 3.6 $\mu$m},
  author={Querejeta, Miguel and Meidt, Sharon E and Schinnerer, Eva and Cisternas, Mauricio and Mu{\~n}oz-Mateos, Juan Carlos and Sheth, Kartik and Knapen, Johan and Van De Ven, Glenn and Norris, Mark A and Peletier, Reynier and others},
  journal={The Astrophysical Journal Supplement Series},
  volume={219},
  number={1},
  pages={5},
  year={2015},
  publisher={IOP Publishing}
}

@article{braine2000deep,
  title={Deep search for CO emission in the Low Surface Brightness galaxy Malin 1},
  author={Braine, J and Herpin, F and Radford, SJE},
  journal={Astronomy and Astrophysics},
  volume={358},
  pages={494--498},
  year={2000}
}

@article{Koyama:2003be,
    author = "Koyama, Kazuya",
    archivePrefix = "arXiv",
    doi = "10.1103/PhysRevLett.91.221301",
    eprint = "astro-ph/0303108",
    journal = "Phys.\ Rev.\ Lett.",
    pages = "221301",
    reportNumber = "UTAP-438",
    title = "{Cosmic microwave radiation anisotropies in brane worlds}",
    volume = "91",
    year = "2003"
}

@article{Mazumdar:2000gj,
    author = "Mazumdar, Anupam",
    archivePrefix = "arXiv",
    doi = "10.1103/PhysRevD.64.027304",
    eprint = "hep-ph/0007269",
    journal = "Phys.\ Rev.\ D",
    pages = "027304",
    title = "{Interesting consequences of brane cosmology}",
    volume = "64",
    year = "2001"
}

@article{Maartens:2000fg,
    author = "Maartens, Roy",
    archivePrefix = "arXiv",
    doi = "10.1103/PhysRevD.62.084023",
    eprint = "hep-th/0004166",
    journal = "Phys.\ Rev.\ D",
    pages = "084023",
    title = "{Cosmological dynamics on the brane}",
    volume = "62",
    year = "2000"
}

@article{Maartens:2003tw,
    author = "Maartens, Roy",
    archivePrefix = "arXiv",
    doi = "10.12942/lrr-2004-7",
    eprint = "gr-qc/0312059",
    journal = "Living Rev.\ Rel.",
    pages = "7",
    title = "{Brane world gravity}",
    volume = "7",
    year = "2004"
}

@article{Kaluza:1921tu,
    author = "Kaluza, Th.",
    archivePrefix = "arXiv",
    doi = "10.1142/S0218271818700017",
    eprint = "1803.08616",
    journal = "Int.\ J.\ Mod.\ Phys.\ D",
    number = "14",
    pages = "1870001",
    primaryClass = "physics.hist-ph",
    reportNumber = "HUPD-8401",
    title = "{Zum Unitätsproblem der Physik}",
    volume = "27",
    year = "2018"
}

@article{Klein:1926fj,
    author = "Klein, O.",
    doi = "10.1038/118516a0",
    journal = "Nature",
    pages = "516",
    title = "{The Atomicity of Electricity as a Quantum Theory Law}",
    volume = "118",
    year = "1926"
}

@article{ArkaniHamed:1998nn,
    author = "Arkani-Hamed, Nima and Dimopoulos, Savas and Dvali, G.R.",
    archivePrefix = "arXiv",
    doi = "10.1103/PhysRevD.59.086004",
    eprint = "hep-ph/9807344",
    journal = "Phys.\ Rev.\ D",
    pages = "086004",
    reportNumber = "SLAC-PUB-7864, SU-ITP-98-142, IC-98-44",
    title = "{Phenomenology, astrophysics and cosmology of theories with submillimeter dimensions and TeV scale quantum gravity}",
    volume = "59",
    year = "1999"
}

@article{Chakraborty:2014zya,
    author = "Chakraborty, Sumanta and SenGupta, Soumitra",
    archivePrefix = "arXiv",
    doi = "10.1103/PhysRevD.90.047901",
    eprint = "1403.3164",
    journal = "Phys.\ Rev.\ D",
    number = "4",
    pages = "047901",
    primaryClass = "gr-qc",
    title = "{Higher curvature gravity at the LHC}",
    volume = "90",
    year = "2014"
}

@article{Lukas:1999yn,
    author = "Lukas, Andre and Ovrut, Burt A. and Waldram, Daniel",
    archivePrefix = "arXiv",
    doi = "10.1103/PhysRevD.61.023506",
    eprint = "hep-th/9902071",
    journal = "Phys.\ Rev.\ D",
    pages = "023506",
    reportNumber = "OUTP-99-09-P, UPR-831T, PUPT-1837, OUTP-99-09P",
    title = "{Boundary inflation}",
    volume = "61",
    year = "2000"
}
@article{ArkaniHamed:1999gq,
    author = "Arkani-Hamed, Nima and Dimopoulos, Savas and Kaloper, Nemanja and March-Russell, John",
    archivePrefix = "arXiv",
    doi = "10.1016/S0550-3213(99)00667-7",
    eprint = "hep-ph/9903224",
    journal = "Nucl.\ Phys.\ B",
    pages = "189--228",
    reportNumber = "SLAC-PUB-8068, CERN-TH-99-38, SU-ITP-98-68",
    title = "{Rapid asymmetric inflation and early cosmology in theories with submillimeter dimensions}",
    volume = "567",
    year = "2000"
}

@article{Dienes:1998hx,
    author = "Dienes, Keith R. and Dudas, E. and Gherghetta, T. and Riotto, A.",
    archivePrefix = "arXiv",
    doi = "10.1016/S0550-3213(98)00855-4",
    eprint = "hep-ph/9809406",
    journal = "Nucl.\ Phys.\ B",
    pages = "387--422",
    reportNumber = "CERN-TH-98-259, LPTHE-ORSAY-98-56",
    title = "{Cosmological phase transitions and radius stabilization in higher dimensions}",
    volume = "543",
    year = "1999"
}

@article{Mazumdar:2000sw,
    author = "Mazumdar, Anupam and Wang, Jing",
    archivePrefix = "arXiv",
    doi = "10.1016/S0370-2693(00)00741-3",
    eprint = "gr-qc/0004030",
    journal = "Phys.\ Lett.\ B",
    pages = "251--257",
    reportNumber = "FERMILAB-PUB-00-076-T, IMPERIAL-AST-00-4-2",
    title = "{A Note on brane inflation}",
    volume = "490",
    year = "2000"
}

@article{Chakraborty:2013ipa,
    author = "Chakraborty, Sumanta and Sengupta, Soumitra",
    archivePrefix = "arXiv",
    doi = "10.1140/epjc/s10052-014-3045-6",
    eprint = "1306.0805",
    journal = "Eur.\ Phys.\ J.\ C",
    number = "9",
    pages = "3045",
    primaryClass = "gr-qc",
    title = "{Radion cosmology and stabilization}",
    volume = "74",
    year = "2014"
}

@article{Banerjee:2017lxi,
    author = "Banerjee, Narayan and Paul, Tanmoy",
    archivePrefix = "arXiv",
    doi = "10.1140/epjc/s10052-017-5256-0",
    eprint = "1706.05964",
    journal = "Eur.\ Phys.\ J.\ C",
    number = "10",
    pages = "672",
    primaryClass = "hep-th",
    title = "{Inflationary scenario from higher curvature warped spacetime}",
    volume = "77",
    year = "2017"
}

@article{Banerjee:2018kcz,
    author = "Banerjee, Indrani and Chakraborty, Sumanta and SenGupta, Soumitra",
    archivePrefix = "arXiv",
    doi = "10.1103/PhysRevD.99.023515",
    eprint = "1806.11327",
    journal = "Phys.\ Rev.\ D",
    number = "2",
    pages = "023515",
    primaryClass = "hep-th",
    title = "{Radion induced inflation on nonflat brane and modulus stabilization}",
    volume = "99",
    year = "2019"
}

@article{mak2004can,
  title={Can the galactic rotation curves be explained in brane world models?},
  author={Mak, MK and Harko, T},
  journal={Physical Review D},
  volume={70},
  number={2},
  pages={024010},
  year={2004},
  publisher={APS}
}

@article{harko2006galactic,
  title={Galactic metric, dark radiation, dark pressure, and gravitational lensing in brane world models},
  author={Harko, Tiberiu and Cheng, KS},
  journal={The Astrophysical Journal},
  volume={636},
  number={1},
  pages={8},
  year={2006},
  publisher={IOP Publishing}
}

@article{boehmer2007galactic,
  title={Galactic dark matter as a bulk effect on the brane},
  author={Boehmer, CG and Harko, T},
  journal={Classical and Quantum Gravity},
  volume={24},
  number={13},
  pages={3191},
  year={2007},
  publisher={IOP Publishing}
}

@article{Davoudiasl:1999jd,
    author = "Davoudiasl, H. and Hewett, J.L. and Rizzo, T.G.",
    archivePrefix = "arXiv",
    doi = "10.1103/PhysRevLett.84.2080",
    eprint = "hep-ph/9909255",
    journal = "Phys.\ Rev.\ Lett.",
    pages = "2080",
    reportNumber = "SLAC-PUB-8241",
    title = "{Phenomenology of the Randall-Sundrum Gauge Hierarchy Model}",
    volume = "84",
    year = "2000"
}

@article{maartens2004brane,
  title={Brane-world gravity},
  author={Maartens, Roy},
  journal={Living Reviews in Relativity},
  volume={7},
  number={1},
  pages={7},
  year={2004},
  publisher={Springer}
}

@article{narayan2005constraints,
  title={Constraints on the halo density profile using HI flaring in the outer Galaxy},
  author={Narayan, CA and Saha, K and Jog, CJ},
  journal={Astronomy \& Astrophysics},
  volume={440},
  number={2},
  pages={523--530},
  year={2005},
  publisher={EDP Sciences}
}

@article{haario1999adaptive,
  title={Adaptive proposal distribution for random walk Metropolis algorithm},
  author={Haario, Heikki and Saksman, Eero and Tamminen, Johanna},
  journal={Computational Statistics},
  volume={14},
  number={3},
  pages={375--396},
  year={1999},
  publisher={Heidelberg: Physica-Verlag,[1992-}
}

@article{matthews2005detections,
  title={Detections of CO in late-type, low surface brightness spiral galaxies},
  author={Matthews, Lynn D and Gao, Yu and Uson, Juan M and Combes, Fran{\c{c}}oise},
  journal={The Astronomical Journal},
  volume={129},
  number={4},
  pages={1849},
  year={2005},
  publisher={IOP Publishing}
}

@article{begum2005dwarf,
  title={A dwarf galaxy with a giant HI disk},
  author={Begum, Ayesha and Chengalur, Jayaram N and Karachentsev, ID},
  journal={Astronomy \& Astrophysics},
  volume={433},
  number={1},
  pages={L1--L4},
  year={2005},
  publisher={EDP Sciences}
}

@article{McGaugh:2016leg,
    author = "McGaugh, Stacy and Lelli, Federico and Schombert, Jim",
    title = "{Radial Acceleration Relation in Rotationally Supported Galaxies}",
    eprint = "1609.05917",
    archivePrefix = "arXiv",
    primaryClass = "astro-ph.GA",
    doi = "10.1103/PhysRevLett.117.201101",
    journal = "Phys. Rev. Lett.",
    volume = "117",
    number = "20",
    pages = "201101",
    year = "2016"
}

@article{McGaugh:2000sr,
    author = "McGaugh, Stacy S. and Schombert, Jim M. and Bothun, Greg D. and de Blok, W.J.G.",
    title = "{The Baryonic Tully-Fisher relation}",
    eprint = "astro-ph/0003001",
    archivePrefix = "arXiv",
    doi = "10.1086/312628",
    journal = "Astrophys. J. Lett.",
    volume = "533",
    pages = "L99--L102",
    year = "2000"
}

@article{McGaugh:2011ac,
    author = "McGaugh, Stacy",
    title = "{The Baryonic Tully-Fisher Relation of Gas Rich Galaxies as a Test of LCDM and MOND}",
    eprint = "1107.2934",
    archivePrefix = "arXiv",
    primaryClass = "astro-ph.CO",
    doi = "10.1088/0004-6256/143/2/40",
    journal = "Astron. J.",
    volume = "143",
    pages = "40",
    year = "2012"
}

@article{Lelli:2017vgz,
    author = "Lelli, Federico and McGaugh, Stacy S. and Schombert, James M. and Pawlowski, Marcel S.",
    title = "{One Law to Rule Them All: The Radial Acceleration Relation of Galaxies}",
    eprint = "1610.08981",
    archivePrefix = "arXiv",
    primaryClass = "astro-ph.GA",
    doi = "10.3847/1538-4357/836/2/152",
    journal = "Astrophys. J.",
    volume = "836",
    number = "2",
    pages = "152",
    year = "2017"
}

@article{Milgrom:2014usa,
    author = "Milgrom, Mordehai",
    title = "{MOND theory}",
    eprint = "1404.7661",
    archivePrefix = "arXiv",
    primaryClass = "astro-ph.CO",
    doi = "10.1139/cjp-2014-0211",
    journal = "Can. J. Phys.",
    volume = "93",
    number = "2",
    pages = "107--118",
    year = "2015"
}

@article{Fichet:2019owx,
    author = "Fichet, Sylvain",
    title = "{Braneworld effective field theories --- holography, consistency and conformal effects}",
    eprint = "1912.12316",
    archivePrefix = "arXiv",
    primaryClass = "hep-th",
    doi = "10.1007/JHEP04(2020)016",
    journal = "JHEP",
    volume = "04",
    pages = "016",
    year = "2020"
}

@article{Shiromizu:1999wj,
      author         = "Shiromizu, Tetsuya and Maeda, Kei-ichi and Sasaki, Misao",
      title          = "{The Einstein equation on the 3-brane world}",
      journal        = "Phys.Rev.",
      volume         = "D62",
      pages          = "024012",
      doi            = "10.1103/PhysRevD.62.024012",
      year           = "2000",
      eprint         = "gr-qc/9910076",
      archivePrefix  = "arXiv",
      primaryClass   = "gr-qc",
      reportNumber   = "DAMTP-1999-150, OUTAP-103, UTAP-349, RESCEU-40-99",
      SLACcitation   = "%%CITATION = GR-QC/9910076;%%",
}

@article{Harko:2004ui,
      author         = "Harko, T. and Mak, M.K.",
      title          = "{Vacuum solutions of the gravitational field equations in
                        the brane world model}",
      journal        = "Phys.Rev.",
      volume         = "D69",
      pages          = "064020",
      doi            = "10.1103/PhysRevD.69.064020",
      year           = "2004",
      eprint         = "gr-qc/0401049",
      archivePrefix  = "arXiv",
      primaryClass   = "gr-qc",
      SLACcitation   = "%%CITATION = GR-QC/0401049;%%",
}

@article{Chakraborty:2015zxc,
      author         = "Chakraborty, Sumanta and SenGupta, Soumitra",
      title          = "{Kinematics of Radion field: A possible source of dark
                        matter}",
      journal        = "Eur. Phys. J.",
      volume         = "C76",
      year           = "2016",
      number         = "12",
      pages          = "648",
      doi            = "10.1140/epjc/s10052-016-4512-z",
      eprint         = "1511.00646",
      archivePrefix  = "arXiv",
      primaryClass   = "gr-qc",
      SLACcitation   = "%%CITATION = ARXIV:1511.00646;%%"
}

@article{navarro1997universal,
  title={A universal density profile from hierarchical clustering},
  author={Navarro, Julio F and Frenk, Carlos S and White, Simon DM},
  journal={The Astrophysical Journal},
  volume={490},
  number={2},
  pages={493},
  year={1997},
  publisher={IOP Publishing}
}

@book{born2013principles,
  title={Principles of optics: electromagnetic theory of propagation, interference and diffraction of light},
  author={Born, Max and Wolf, Emil},
  year={2013},
  publisher={Elsevier}
}

@inproceedings{mcmullin2007casa,
  title={CASA architecture and applications},
  author={McMullin, Joseph P and Waters, BSDYWGK and Schiebel, Darrell and Young, Wei and Golap, Kumar},
  booktitle={Astronomical data analysis software and systems XVI},
  volume={376},
  pages={127},
  year={2007}
}

@article{offringa2010aoflagger,
  title={AOFlagger: RFI Software},
  author={Offringa, AR},
  journal={Astrophysics Source Code Library},
  pages={ascl--1010},
  year={2010}
}

@article{hogbom1974aperture,
  title={Aperture synthesis with a non-regular distribution of interferometer baselines},
  author={H{\"o}gbom, JA},
  journal={Astronomy and Astrophysics Supplement Series},
  volume={15},
  pages={417},
  year={1974}
}

@article{rogstad1974aperture,
  title={Aperture-synthesis observations of HI in the galaxy M83.},
  author={Rogstad, DH and Lockhart, IA and Wright, MCH},
  journal={The Astrophysical Journal},
  volume={193},
  pages={309--319},
  year={1974}
}

@article{begeman1989hi,
  title={HI rotation curves of spiral galaxies. I-NGC 3198},
  author={Begeman, KG},
  journal={Astronomy and Astrophysics},
  volume={223},
  pages={47--60},
  year={1989}
}

@article{sellwood2015diskfit,
  title={DiskFit: a code to fit simple non-axisymmetric galaxy models either to photometric images or to kinematic maps},
  author={Sellwood, JA and Spekkens, Kristine},
  journal={arXiv preprint arXiv:1509.07120},
  year={2015}
}

@article{mathewson1992southern,
  title={A southern sky survey of the peculiar velocities of 1355 spiral galaxies},
  author={Mathewson, DS and Ford, VL and Buchhorn, M},
  journal={The Astrophysical Journal Supplement Series},
  volume={81},
  pages={413--659},
  year={1992}
}

@article{takamiya2002iteration,
  title={Iteration method to derive exact rotation curves from position-velocity diagrams of spiral galaxies},
  author={Takamiya, Tsutomu and Sofue, Yoshiaki},
  journal={The Astrophysical Journal},
  volume={576},
  number={1},
  pages={L15},
  year={2002},
  publisher={IOP Publishing}
}

@article{swaters1999kinematically,
  title={Kinematically lopsided spiral galaxies},
  author={Swaters, RA and Schoenmakers, RHM and Sancisi, R and Albada, TS van},
  journal={Monthly Notices of the Royal Astronomical Society},
  volume={304},
  number={2},
  pages={330--334},
  year={1999},
  publisher={Blackwell Science Ltd Oxford, UK}
}

@article{jozsa2012tirific,
  title={TiRiFiC: Tilted Ring Fitting Code},
  author={J{\'o}zsa, Gyula IG and Kenn, Franz and Oosterloo, Thomas A and Klein, Ulrich},
  journal={Astrophysics Source Code Library},
  pages={ascl--1208},
  year={2012}
}

@article{kamphuis2015fat,
  title={FAT: Fully Automated TiRiFiC},
  author={Kamphuis, P and J{\'o}zsa, GIG and Oh, S- H and Spekkens, K and Urbancic, N and Serra, P and Koribalski, BS and Dettmar, R-J},
  journal={Astrophysics Source Code Library},
  pages={ascl--1507},
  year={2015}
}

@article{teodoro20153d,
  title={3D BAROLO: a new 3D algorithm to derive rotation curves of galaxies},
  author={Teodoro, EM Di and Fraternali, Filippo},
  journal={Monthly Notices of the Royal Astronomical Society},
  volume={451},
  number={3},
  pages={3021--3033},
  year={2015},
  publisher={Oxford University Press}
}

@article{begeman1991extended,
  title={Extended rotation curves of spiral galaxies: Dark haloes and modified dynamics},
  author={Begeman, KG and Broeils, AH and Sanders, RH},
  journal={Monthly Notices of the Royal Astronomical Society},
  volume={249},
  number={3},
  pages={523--537},
  year={1991},
  publisher={The Royal Astronomical Society}
}

@article{boily2003thickness,
  title={The Thickness of Stellar Disks of Edge-on Galaxies and Their Truncation Radii},
  author={Boily, C and Patsis, P and Portegies, S and Spurzem, R and Theis, C and Zasov, AV and Bizyaev, DV},
  journal={European Astronomical Society Publications Series},
  volume={10},
  pages={121--121},
  year={2003},
  publisher={EDP Sciences}
}

@article{rohlfs1977lectures,
  title={Lectures on density wave theory},
  author={Rohlfs, Kristen},
  year={1977}
}

@article{binney2011models,
  title={Models of our Galaxy--II},
  author={Binney, James and McMillan, Paul},
  journal={Monthly Notices of the Royal Astronomical Society},
  volume={413},
  number={3},
  pages={1889--1898},
  year={2011},
  publisher={Blackwell Publishing Ltd Oxford, UK}
}

@article{wang1994gravitational,
  title={Gravitational instability and disk star formation},
  author={Wang, Boqi and Silk, Joseph},
  journal={The Astrophysical Journal},
  volume={427},
  pages={759--769},
  year={1994}
}

@article{westfall2014diskmass,
  title={The DiskMass survey. VIII. On the relationship between disk stability and star formation},
  author={Westfall, Kyle B and Andersen, David R and Bershady, Matthew A and Martinsson, Thomas PK and Swaters, Robert A and Verheijen, Marc AW},
  journal={The Astrophysical Journal},
  volume={785},
  number={1},
  pages={43},
  year={2014},
  publisher={IOP Publishing}
}
@article{matthews2000extraordinary,
  title={The Extraordinary “Superthin” Spiral Galaxy UGC 7321. II. The Vertical Disk Structure},
  author={Matthews, LD},
  journal={The Astronomical Journal},
  volume={120},
  number={4},
  pages={1764},
  year={2000},
  publisher={IOP Publishing}
}

@article{giovanelli1997spectroscopy,
  title={Spectroscopy of edge-on spirals},
  author={Giovanelli, Riccardo and Avera, Eric and Karachentsev, Igor},
  journal={arXiv preprint astro-ph/9704189},
  year={1997}
}

@article{karachentsev1993flat,
  title={Flat galaxy catalogue},
  author={Karachentsev, ID and Karachentseva, VE and Parnovsky, SL},
  journal={Astronomische Nachrichten},
  volume={314},
  number={3},
  pages={97--222},
  year={1993},
  publisher={Wiley Online Library}
}

@article{banerjee2007origin,
  title={The origin of steep vertical stellar distribution in the Galactic disk},
  author={Banerjee, Arunima and Jog, Chanda J},
  journal={The Astrophysical Journal},
  volume={662},
  number={1},
  pages={335},
  year={2007},
  publisher={IOP Publishing}
}

@article{roberts1994physical,
  title={Physical parameters along the Hubble sequence},
  author={Roberts, Morton S and Haynes, Martha P},
  journal={Annual Review of Astronomy and Astrophysics},
  volume={32},
  number={1},
  pages={115--152},
  year={1994},
  publisher={Annual Reviews 4139 El Camino Way, PO Box 10139, Palo Alto, CA 94303-0139, USA}
}

@article{querejeta2015spitzer,
  title={The Spitzer Survey of Stellar Structure in Galaxies (S4G): precise stellar mass distributions from automated dust correction at 3.6 $\mu$m},
  author={Querejeta, Miguel and Meidt, Sharon E and Schinnerer, Eva and Cisternas, Mauricio and Mu{\~n}oz-Mateos, Juan Carlos and Sheth, Kartik and Knapen, Johan and Van De Ven, Glenn and Norris, Mark A and Peletier, Reynier and others},
  journal={The Astrophysical Journal Supplement Series},
  volume={219},
  number={1},
  pages={5},
  year={2015},
  publisher={IOP Publishing}
}

@article{fernandez2016ggmcmc,
  title={ggmcmc: Analysis of MCMC samples and Bayesian inference},
  author={Fern{\'a}ndez-i-Mar{\i}n, Xavier},
  journal={Journal of Statistical Software},
  volume={70},
  number={9},
  pages={1--20},
  year={2016}
}

@article{hartig2017bayesiantools,
  title={Bayesiantools: General-purpose MCMC and SMC samplers and tools for Bayesian statistics},
  author={Hartig, F and Minunno, F and Paul, S and Cameron, D and Ott, T},
  journal={R package version. R package version 0.1},
  volume={3},
  year={2017}
}

@article{wickham2011ggplot2,
  title={ggplot2},
  author={Wickham, Hadley},
  journal={Wiley Interdisciplinary Reviews: Computational Statistics},
  volume={3},
  number={2},
  pages={180--185},
  year={2011},
  publisher={Wiley Online Library}
}

@misc{meschiari2015latex2exp,
  title={latex2exp: Use LaTeX Expressions in Plots, r package version 0.4. 0},
  author={Meschiari, S},
  year={2015}
}

@article{hunter2007matplotlib,
  title={Matplotlib: A 2D graphics environment},
  author={Hunter, John D},
  journal={Computing in science \& engineering},
  volume={9},
  number={3},
  pages={90--95},
  year={2007},
  publisher={IEEE Computer Society}
}

@article{vaughan2016false,
  title={False periodicities in quasar time-domain surveys},
  author={Vaughan, S and Uttley, P and Markowitz, AG and Huppenkothen, D and Middleton, MJ and Alston, WN and Scargle, JD and Farr, WM},
  journal={Monthly Notices of the Royal Astronomical Society},
  volume={461},
  number={3},
  pages={3145--3152},
  year={2016},
  publisher={Oxford University Press}
}

@article{sarkar2019flaring,
  title={Flaring stellar disk in low surface brightness galaxy UGC 7321},
  author={Sarkar, Suchira and Jog, Chanda J},
  journal={arXiv preprint arXiv:1905.02735},
  year={2019}
}

@article{narayan2002origin,
  title={Origin of radially increasing stellar scaleheight in a galactic disk},
  author={Narayan, Chaitra A and Jog, Chanda J},
  journal={Astronomy \& Astrophysics},
  volume={390},
  number={3},
  pages={L35--L38},
  year={2002},
  publisher={EDP Sciences}
}

@article{sanchez2012califa,
  title={CALIFA, the Calar Alto Legacy Integral Field Area survey-I. Survey presentation},
  author={S{\'a}nchez, SF and Kennicutt, RC and De Paz, A Gil and Van de Ven, G and V{\'\i}lchez, JM and Wisotzki, L and Walcher, CJ and Mast, D and Aguerri, JAL and Albiol-P{\'e}rez, S and others},
  journal={Astronomy \& Astrophysics},
  volume={538},
  pages={A8},
  year={2012},
  publisher={EDP Sciences}
}

@article{Bershady_2010,
   title={THE DISKMASS SURVEY. I. OVERVIEW},
   volume={716},
   ISSN={1538-4357},
   url={http://dx.doi.org/10.1088/0004-637X/716/1/198},
   DOI={10.1088/0004-637x/716/1/198},
   number={1},
   journal={The Astrophysical Journal},
   publisher={IOP Publishing},
   author={Bershady, Matthew A. and Verheijen, Marc A. W. and Swaters, Rob A. and Andersen, David R. and Westfall, Kyle B. and Martinsson, Thomas},
   year={2010},
   month={May},
   pages={198–233}
}

@article{allen2015sami,
  title={The SAMI galaxy survey: early data release},
  author={Allen, JT and Croom, SM and Konstantopoulos, IS and Bryant, JJ and Sharp, R and Cecil, GN and Fogarty, LMR and Foster, C and Green, AW and Ho, I-T and others},
  journal={Monthly Notices of the Royal Astronomical Society},
  volume={446},
  number={2},
  pages={1567--1583},
  year={2015},
  publisher={Oxford University Press}
}

@article{law2015observing,
  title={Observing strategy for the SDSS-IV/MaNGA IFU galaxy survey},
  author={Law, David R and Yan, Renbin and Bershady, Matthew A and Bundy, Kevin and Cherinka, Brian and Drory, Niv and MacDonald, Nicholas and S{\'a}nchez-Gallego, Jos{\'e} R and Wake, David A and Weijmans, Anne-Marie and others},
  journal={The Astronomical Journal},
  volume={150},
  number={1},
  pages={19},
  year={2015},
  publisher={IOP Publishing}
}

@article{young2011atlas3d,
  title={The ATLAS3D project--IV. The molecular gas content of early-type galaxies},
  author={Young, Lisa M and Bureau, Martin and Davis, Timothy A and Combes, Francoise and McDermid, Richard M and Alatalo, Katherine and Blitz, Leo and Bois, Maxime and Bournaud, Fr{\'e}d{\'e}ric and Cappellari, Michele and others},
  journal={Monthly Notices of the Royal Astronomical Society},
  volume={414},
  number={2},
  pages={940--967},
  year={2011},
  publisher={Blackwell Publishing Ltd Oxford, UK}}

@article{binney2008princeton,
  title={Princeton Univ. Press},
  author={Binney, J and Merrifield, M},
  journal={Galactic Astronomy,},
  year={2008}
}

@article{van1982surface,
  title={Surface photometry of edge-on spiral galaxies. III-Properties of the three-dimensional distribution of light and mass in disks of spiral galaxies},
  author={Van der Kruit, PC and Searle, L},
  journal={Astronomy and Astrophysics},
  volume={110},
  pages={61--78},
  year={1982}
}

@article{van1988three,
  title={The three-dimensional distribution of light and mass in disks of spiral galaxies},
  author={Van Der Kruit, PC},
  journal={Astronomy and Astrophysics},
  volume={192},
  pages={117--127},
  year={1988}
}

@article{van2011galaxy,
  title={Galaxy disks},
  author={Van der Kruit, PC and Freeman, KC},
  journal={Annual Review of Astronomy and Astrophysics},
  volume={49},
  pages={301--371},
  year={2011},
  publisher={Annual Reviews}
}

@article{sharma2014kinematic,
  title={Kinematic modeling of the Milky Way using the RAVE and GCS stellar surveys},
  author={Sharma, Sanjib and Bland-Hawthorn, Joss and Binney, J and Freeman, Ken C and Steinmetz, Matthias and Boeche, Corrado and Bienayme, Olivier and Gibson, Brad K and Gilmore, Gerard F and Grebel, Eva K and others},
  journal={The Astrophysical Journal},
  volume={793},
  number={1},
  pages={51},
  year={2014},
  publisher={IOP Publishing}
}

@article{van1984vertical,
  title={The vertical velocity dispersion of the stars in the disks of two spiral galaxies},
  author={Van der Kruit, PC and Freeman, KC},
  journal={The Astrophysical Journal},
  volume={278},
  pages={81--88},
  year={1984}
}

@article{martinsson2013diskmass,
  title={The DiskMass Survey-VI. Gas and stellar kinematics in spiral galaxies from PPak integral-field spectroscopy},
  author={Martinsson, Thomas PK and Verheijen, Marc AW and Westfall, Kyle B and Bershady, Matthew A and Schechtman-Rook, Andrew and Andersen, David R and Swaters, Rob A},
  journal={Astronomy \& Astrophysics},
  volume={557},
  pages={A130},
  year={2013},
  publisher={EDP Sciences}
}

@article{mogotsi2016hi,
  title={HI and CO velocity dispersions in nearby galaxies},
  author={Mogotsi, KM and de Blok, WJG and Cald{\'u}-Primo, A and Walter, F and Ianjamasimanana, R and Leroy, AK},
  journal={The Astronomical Journal},
  volume={151},
  number={1},
  pages={15},
  year={2016},
  publisher={IOP Publishing}
}

@article{nordstrom2004geneva,
  title={The Geneva-Copenhagen survey of the Solar neighbourhood-Ages, metallicities, and kinematic properties of\~{} 14 000 F and G dwarfs},
  author={Nordstr{\"o}m, Birgitta and Mayor, M and Andersen, J and Holmberg, J and Pont, F and J{\o}rgensen, Bjarne Rosenkilde and Olsen, EH and Udry, S and Mowlavi, N},
  journal={Astronomy \& Astrophysics},
  volume={418},
  number={3},
  pages={989--1019},
  year={2004},
  publisher={EDP Sciences}
}

@article{steinmetz2006radial,
  title={The radial velocity experiment (RAVE): first data release},
  author={Steinmetz, Matthias and Zwitter, Toma{\v{z}} and Siebert, Arnaud and Watson, Fred G and Freeman, Kenneth C and Munari, Ulisse and Campbell, Rachel and Williams, Mary and Seabroke, George M and Wyse, Rosemary FG and others},
  journal={The Astronomical Journal},
  volume={132},
  number={4},
  pages={1645},
  year={2006},
  publisher={IOP Publishing}
}

@article{romeo2017drives,
  title={What drives gravitational instability in nearby star-forming spirals? The impact of CO and H i velocity dispersions},
  author={Romeo, Alessandro B and Mogotsi, Keoikantse Moses},
  journal={Monthly Notices of the Royal Astronomical Society},
  volume={469},
  number={1},
  pages={286--294},
  year={2017},
  publisher={Oxford University Press}
}

@article{leroy2008star,
  title={The star formation efficiency in nearby galaxies: measuring where gas forms stars effectively},
  author={Leroy, Adam K and Walter, Fabian and Brinks, Elias and Bigiel, Frank and De Blok, WJG and Madore, Barry and Thornley, MD},
  journal={The Astronomical Journal},
  volume={136},
  number={6},
  pages={2782},
  year={2008},
  publisher={IOP Publishing}
}

@article{wyder2009star,
  title={The star formation law at low surface density},
  author={Wyder, Ted K and Martin, D Christopher and Barlow, Tom A and Foster, Karl and Friedman, Peter G and Morrissey, Patrick and Neff, Susan G and Neill, James D and Schiminovich, David and Seibert, Mark and others},
  journal={The Astrophysical Journal},
  volume={696},
  number={2},
  pages={1834},
  year={2009},
  publisher={IOP Publishing}
}

@article{elmegreen2011gravitational,
  title={Gravitational instabilities in two-component galaxy disks with gas dissipation},
  author={Elmegreen, Bruce G},
  journal={The Astrophysical Journal},
  volume={737},
  number={1},
  pages={10},
  year={2011},
  publisher={IOP Publishing}
}

@article{griv2012stability,
  title={Stability of galactic discs: finite arm-inclination and finite-thickness effects},
  author={Griv, Evgeny and Gedalin, Michael},
  journal={Monthly Notices of the Royal Astronomical Society},
  volume={422},
  number={1},
  pages={600--609},
  year={2012},
  publisher={The Royal Astronomical Society}
}

@article{bottema1993stellar,
  title={The stellar kinematics of galactic disks},
  author={Bottema, Roelof},
  journal={Astronomy and Astrophysics},
  volume={275},
  pages={16},
  year={1993}
}

@article{romeo2020massive,
  title={From massive spirals to dwarf irregulars: a new set of tight scaling relations for cold gas and stars driven by disc gravitational instability},
  author={Romeo, Alessandro B},
  journal={Monthly Notices of the Royal Astronomical Society},
  volume={491},
  number={4},
  pages={4843--4851},
  year={2020},
  publisher={Oxford University Press}
}

@article{romeo2020lenticulars,
  title={From lenticulars to blue compact dwarfs: the stellar mass fraction is regulated by disc gravitational instability},
  author={Romeo, Alessandro B and Agertz, Oscar and Renaud, Florent},
  journal={arXiv preprint arXiv:2006.09159},
  year={2020}
}

@article{tamburro2009driving,
  title={What is driving the H i velocity dispersion?},
  author={Tamburro, D and Rix, H-W and Leroy, AK and Mac Low, M-M and Walter, F and Kennicutt, RC and Brinks, E and De Blok, WJG},
  journal={The Astronomical Journal},
  volume={137},
  number={5},
  pages={4424},
  year={2009},
  publisher={IOP Publishing}
}

@article{ianjamasimanana2012shapes,
  title={The shapes of the HI velocity profiles of the THINGS galaxies},
  author={Ianjamasimanana, R and De Blok, WJG and Walter, Fabian and Heald, George H},
  journal={The Astronomical Journal},
  volume={144},
  number={4},
  pages={96},
  year={2012},
  publisher={IOP Publishing}
}

@article{10.1093/mnras/stv1220,
    author = {Romeo, Alessandro B. and Fathi, Kambiz},
    title = "{A double molecular disc in the triple-barred starburst galaxy NGC 6946: structure and stability}",
    journal = {Monthly Notices of the Royal Astronomical Society},
    volume = {451},
    number = {3},
    pages = {3107-3116},
    year = {2015},
    month = {06},
    abstract = "{The late-type spiral galaxy NGC 6946 is a prime example of molecular gas dynamics driven by ‘bars within bars’. Here, we use data from the BIMA Survey of Nearby Galaxies and HERA CO-Line Extragalactic Survey to analyse the structure and stability of its molecular disc. Our radial profiles exhibit a clear transition at distance R ∼ 1 kpc from the galaxy centre. In particular, the surface density profile breaks at R ≈ 0.8 kpc and is well fitted by a double exponential distribution with scalelengths R1 ≈ 200 pc and R2 ≈ 3 kpc, while the 1D velocity dispersion σ decreases steeply in the central kpc and is approximately constant at larger radii. The fact that we derive and use the full radial profile of σ rather than a constant value is perhaps the most novel feature of our stability analysis. We show that the profile of the Q stability parameter traced by CO emission is remarkably flat and well above unity, while the characteristic instability wavelength exhibits clear signatures of the nuclear starburst and inner bar within bar. We also show that CO-dark molecular gas, stars and other factors can play a significant role in the stability scenario of NGC 6946. Our results provide strong evidence that gravitational instability, radial inflow and disc heating have driven the formation of the inner structures and the dynamics of molecular gas in the central kpc.}",
    issn = {0035-8711},
    doi = {10.1093/mnras/stv1220},
    url = {https://doi.org/10.1093/mnras/stv1220},
    eprint = {https://academic.oup.com/mnras/article-pdf/451/3/3107/4031552/stv1220.pdf},
}

@article{boily2003thickness,
  title={The Thickness of Stellar Disks of Edge-on Galaxies and Their Truncation Radii},
  author={Boily, C and Patsis, P and Portegies, S and Spurzem, R and Theis, C and Zasov, AV and Bizyaev, DV},
  journal={European Astronomical Society Publications Series},
  volume={10},
  pages={121--121},
  year={2003},
  publisher={EDP Sciences}
}

@article{rohlfs1977lectures,
  title={Lectures on density wave theory},
  author={Rohlfs, Kristen},
  year={1977}
}

@article{binney2011models,
  title={Models of our Galaxy--II},
  author={Binney, James and McMillan, Paul},
  journal={Monthly Notices of the Royal Astronomical Society},
  volume={413},
  number={3},
  pages={1889--1898},
  year={2011},
  publisher={Blackwell Publishing Ltd Oxford, UK}
}

@article{wang1994gravitational,
  title={Gravitational instability and disk star formation},
  author={Wang, Boqi and Silk, Joseph},
  journal={The Astrophysical Journal},
  volume={427},
  pages={759--769},
  year={1994}
}

@article{westfall2014diskmass,
  title={The DiskMass survey. VIII. On the relationship between disk stability and star formation},
  author={Westfall, Kyle B and Andersen, David R and Bershady, Matthew A and Martinsson, Thomas PK and Swaters, Robert A and Verheijen, Marc AW},
  journal={The Astrophysical Journal},
  volume={785},
  number={1},
  pages={43},
  year={2014},
  publisher={IOP Publishing}
}
@article{matthews2000extraordinary,
  title={The Extraordinary “Superthin” Spiral Galaxy UGC 7321. II. The Vertical Disk Structure},
  author={Matthews, LD},
  journal={The Astronomical Journal},
  volume={120},
  number={4},
  pages={1764},
  year={2000},
  publisher={IOP Publishing}
}

@article{giovanelli1997spectroscopy,
  title={Spectroscopy of edge-on spirals},
  author={Giovanelli, Riccardo and Avera, Eric and Karachentsev, Igor},
  journal={arXiv preprint astro-ph/9704189},
  year={1997}
}

@article{karachentsev1993flat,
  title={Flat galaxy catalogue},
  author={Karachentsev, ID and Karachentseva, VE and Parnovsky, SL},
  journal={Astronomische Nachrichten},
  volume={314},
  number={3},
  pages={97--222},
  year={1993},
  publisher={Wiley Online Library}
}

@article{banerjee2007origin,
  title={The origin of steep vertical stellar distribution in the Galactic disk},
  author={Banerjee, Arunima and Jog, Chanda J},
  journal={The Astrophysical Journal},
  volume={662},
  number={1},
  pages={335},
  year={2007},
  publisher={IOP Publishing}
}

@article{roberts1994physical,
  title={Physical parameters along the Hubble sequence},
  author={Roberts, Morton S and Haynes, Martha P},
  journal={Annual Review of Astronomy and Astrophysics},
  volume={32},
  number={1},
  pages={115--152},
  year={1994},
  publisher={Annual Reviews 4139 El Camino Way, PO Box 10139, Palo Alto, CA 94303-0139, USA}
}

@article{querejeta2015spitzer,
  title={The Spitzer Survey of Stellar Structure in Galaxies (S4G): precise stellar mass distributions from automated dust correction at 3.6 $\mu$m},
  author={Querejeta, Miguel and Meidt, Sharon E and Schinnerer, Eva and Cisternas, Mauricio and Mu{\~n}oz-Mateos, Juan Carlos and Sheth, Kartik and Knapen, Johan and Van De Ven, Glenn and Norris, Mark A and Peletier, Reynier and others},
  journal={The Astrophysical Journal Supplement Series},
  volume={219},
  number={1},
  pages={5},
  year={2015},
  publisher={IOP Publishing}
}

@article{braine2000deep,
  title={Deep search for CO emission in the Low Surface Brightness galaxy Malin 1},
  author={Braine, J and Herpin, F and Radford, SJE},
  journal={Astronomy and Astrophysics},
  volume={358},
  pages={494--498},
  year={2000}
}

@article{fernandez2016ggmcmc,
  title={ggmcmc: Analysis of MCMC samples and Bayesian inference},
  author={Fern{\'a}ndez-i-Mar{\i}n, Xavier},
  journal={Journal of Statistical Software},
  volume={70},
  number={9},
  pages={1--20},
  year={2016}
}

@article{hartig2017bayesiantools,
  title={Bayesiantools: General-purpose MCMC and SMC samplers and tools for Bayesian statistics},
  author={Hartig, F and Minunno, F and Paul, S and Cameron, D and Ott, T},
  journal={R package version. R package version 0.1},
  volume={3},
  year={2017}
}

@article{wickham2011ggplot2,
  title={ggplot2},
  author={Wickham, Hadley},
  journal={Wiley Interdisciplinary Reviews: Computational Statistics},
  volume={3},
  number={2},
  pages={180--185},
  year={2011},
  publisher={Wiley Online Library}
}

@misc{meschiari2015latex2exp,
  title={latex2exp: Use LaTeX Expressions in Plots, r package version 0.4. 0},
  author={Meschiari, S},
  year={2015}
}

@article{hunter2007matplotlib,
  title={Matplotlib: A 2D graphics environment},
  author={Hunter, John D},
  journal={Computing in science \& engineering},
  volume={9},
  number={3},
  pages={90--95},
  year={2007},
  publisher={IEEE Computer Society}
}

@article{vaughan2016false,
  title={False periodicities in quasar time-domain surveys},
  author={Vaughan, S and Uttley, P and Markowitz, AG and Huppenkothen, D and Middleton, MJ and Alston, WN and Scargle, JD and Farr, WM},
  journal={Monthly Notices of the Royal Astronomical Society},
  volume={461},
  number={3},
  pages={3145--3152},
  year={2016},
  publisher={Oxford University Press}
}

@article{karachentsev1993flat,
  title={Flat galaxy catalogue},
  author={Karachentsev, ID and Karachentseva, VE and Parnovsky, SL},
  journal={Astronomische Nachrichten},
  volume={314},
  number={3},
  pages={97--222},
  year={1993},
  publisher={Wiley Online Library}
}

@article{karachentseva2016ultra,
  title={Ultra-flat galaxies selected from RFGC catalog. I. The sample properties},
  author={Karachentseva, VE and Kudrya, Yu N and Karachentsev, ID and Makarov, DI and Melnyk, OV},
  journal={Astrophysical Bulletin},
  volume={71},
  number={1},
  pages={1--13},
  year={2016},
  publisher={Springer}
}

@misc{de1991third,
  title={Third Reference Catalogue of Bright Galaxies, Version 3.9. Springer, New York, NY},
  author={De Vaucouleurs, G and De Vaucouleurs, A and Corwin, JHG and Buta, RJ and Paturel, G and Fouque, P},
  year={1991}
}

@article{nilson1973uppsala,
  title={Uppsala General Catalogue of Galaxies, 1973, Acta Universitatis Upsalienis, Nova Regiae Societatis Upsaliensis, Series v: a Vol.},
  author={Nilson, P},
  journal={UGC},
  pages={0},
  year={1973}
}

@article{karachentsev2001list,
  title={A list of peculiar velocities of RFGC galaxies},
  author={Karachentsev, ID and Karachentseva, VE and Kudrya, Yu N and Makarov, DI and Parnovsky, SL},
  journal={arXiv preprint astro-ph/0107058},
  year={2001}
}

@article{binggeli1985studies,
  title={Studies of the Virgo Cluster. II-A catalog of 2096 galaxies in the Virgo Cluster area.},
  author={Binggeli, B and Sandage, A and Tammann, GA},
  journal={The Astronomical Journal},
  volume={90},
  pages={1681--1759},
  year={1985}
}

@article{hoffman1989hi,
  title={HI observations in the Virgo cluster area. III-All'member'spirals},
  author={Hoffman, G Lyle and Williams, BM and Lewis, BM and Helou, George and Salpeter, EE},
  journal={The Astrophysical Journal Supplement Series},
  volume={69},
  pages={65--98},
  year={1989}
}

@article{haynes2018arecibo,
  title={The Arecibo Legacy Fast ALFA Survey: The ALFALFA Extragalactic H i Source Catalog},
  author={Haynes, Martha P and Giovanelli, Riccardo and Kent, Brian R and Adams, Elizabeth AK and Balonek, Thomas J and Craig, David W and Fertig, Derek and Finn, Rose and Giovanardi, Carlo and Hallenbeck, Gregory and others},
  journal={The Astrophysical Journal},
  volume={861},
  number={1},
  pages={49},
  year={2018},
  publisher={IOP Publishing}
}

@article{dalcanton2004formation,
  title={The formation of dust lanes: Implications for galaxy evolution},
  author={Dalcanton, Julianne J and Yoachim, Peter and Bernstein, Rebecca A},
  journal={The Astrophysical Journal},
  volume={608},
  number={1},
  pages={189},
  year={2004},
  publisher={IOP Publishing}
}

@article{yoachim2006structural,
  title={Structural parameters of thin and thick disks in edge-on disk galaxies},
  author={Yoachim, Peter and Dalcanton, Julianne J},
  journal={The Astronomical Journal},
  volume={131},
  number={1},
  pages={226},
  year={2006},
  publisher={IOP Publishing}
}

@article{yoachim2008kinematics,
  title={The kinematics of thick disks in nine external galaxies},
  author={Yoachim, Peter and Dalcanton, Julianne J},
  journal={The Astrophysical Journal},
  volume={682},
  number={2},
  pages={1004},
  year={2008},
  publisher={IOP Publishing}
}

@article{serra2015sofia,
  title={SOFIA: a flexible source finder for 3D spectral line data},
  author={Serra, Paolo and Westmeier, Tobias and Giese, Nadine and Jurek, Russell and Fl{\"o}er, Lars and Popping, Attila and Winkel, Benjamin and van der Hulst, Thijs and Meyer, Martin and Koribalski, B{\"a}rbel S and others},
  journal={Monthly Notices of the Royal Astronomical Society},
  volume={448},
  number={2},
  pages={1922--1929},
  year={2015},
  publisher={Oxford University Press}
}

@inproceedings{jaeger2008common,
  title={The common astronomy software application (casa)},
  author={Jaeger, S},
  booktitle={Astronomical Data Analysis Software and Systems XVII},
  volume={394},
  pages={623},
  year={2008}
}
@article{westmeier2014busy,
  title={The busy function: a new analytic function for describing the integrated 21-cm spectral profile of galaxies},
  author={Westmeier, Tobias and Jurek, Russell and Obreschkow, Danail and Koribalski, B{\"a}rbel S and Staveley-Smith, Lister},
  journal={Monthly Notices of the Royal Astronomical Society},
  volume={438},
  number={2},
  pages={1176--1190},
  year={2014},
  publisher={The Royal Astronomical Society}}

@article{bell2001stellar,
  title={Stellar mass-to-light ratios and the Tully-Fisher relation},
  author={Bell, Eric F and de Jong, Roelof S},
  journal={The Astrophysical Journal},
  volume={550},
  number={1},
  pages={212},
  year={2001},
  publisher={IOP Publishing}
}

@article{bell2003optical,
  title={The optical and near-infrared properties of galaxies. I. Luminosity and stellar mass functions},
  author={Bell, Eric F and McIntosh, Daniel H and Katz, Neal and Weinberg, Martin D},
  journal={The Astrophysical Journal Supplement Series},
  volume={149},
  number={2},
  pages={289},
  year={2003},
  publisher={IOP Publishing}
}

@techreport{fukugita1996sloan,
  title={The Sloan digital sky survey photometric system},
  author={Fukugita, M and Shimasaku, K and Ichikawa, T and Gunn, JE and others},
  year={1996},
  institution={SCAN-9601313}
}

@article{kroupa2001variation,
  title={On the variation of the initial mass function},
  author={Kroupa, Pavel},
  journal={Monthly Notices of the Royal Astronomical Society},
  volume={322},
  number={2},
  pages={231--246},
  year={2001},
  publisher={Blackwell Science Ltd Oxford, UK}
}

@article{fuchs1998decomposition,
  title={Decomposition of the rotation curves of distant field galaxies},
  author={Fuchs, B and M{\"o}llenhoff, C and Heidt, J},
  journal={arXiv preprint astro-ph/9806117},
  year={1998}
}

@article{portinari2004mass,
  title={On the mass-to-light ratio and the initial mass function in disc galaxies},
  author={Portinari, Laura and Sommer-Larsen, J and Tantalo, R},
  journal={Monthly Notices of the Royal Astronomical Society},
  volume={347},
  number={3},
  pages={691--719},
  year={2004},
  publisher={Blackwell Science Ltd Oxford, UK}
}

@article{palunas2000maximum,
  title={Maximum disk mass models for spiral galaxies},
  author={Palunas, Povilas and Williams, TB},
  journal={The Astronomical Journal},
  volume={120},
  number={6},
  pages={2884},
  year={2000},
  publisher={IOP Publishing}
}

@article{mcgaugh2014color,
  title={Color-mass-to-light-ratio relations for disk galaxies},
  author={McGaugh, Stacy S and Schombert, James M},
  journal={The Astronomical Journal},
  volume={148},
  number={5},
  pages={77},
  year={2014},
  publisher={IOP Publishing}
}

@article{newville2016lmfit,
  title={LMFIT: Non-linear least-square minimization and curve-fitting for Python},
  author={Newville, Matthew and Stensitzki, Till and Allen, Daniel B and Rawlik, Michal and Ingargiola, Antonino and Nelson, Andrew},
  journal={Astrophysics Source Code Library},
  pages={ascl--1606},
  year={2016}
}

@article{foreman2013emcee,
  title={emcee: the MCMC hammer},
  author={Foreman-Mackey, Daniel and Hogg, David W and Lang, Dustin and Goodman, Jonathan},
  journal={Publications of the Astronomical Society of the Pacific},
  volume={125},
  number={925},
  pages={306},
  year={2013},
  publisher={IOP Publishing}
}

@article{bothun1997low,
  title={Low-surface-brightness galaxies: hidden galaxies revealed},
  author={Bothun, Greg and Impey, Chris and McGaugh, Stacy},
  journal={Publications of the Astronomical Society of the Pacific},
  volume={109},
  number={737},
  pages={745},
  year={1997},
  publisher={IOP Publishing}
}

@article{mcgaugh1996number,
  title={The number, luminosity and mass density of spiral galaxies as a function of surface brightness},
  author={McGaugh, Stacy S},
  journal={Monthly Notices of the Royal Astronomical Society},
  volume={280},
  number={2},
  pages={337--354},
  year={1996},
  publisher={Blackwell Science Ltd Oxford, UK}
}

@article{de2002high,
  title={High-resolution rotation curves of low surface brightness galaxies},
  author={De Blok, WJG and Bosma, A},
  journal={Astronomy \& Astrophysics},
  volume={385},
  number={3},
  pages={816--846},
  year={2002},
  publisher={EDP Sciences}
}

@article{monnier2003search,
  title={A search for Low Surface Brightness galaxies in the near-infrared. III. Nan{\c{c}}ay HI line observations},
  author={Monnier Ragaigne, D and van Driel, Wim and Schneider, SE and Balkowski, C and Jarrett, TH},
  journal={Astronomy and Astrophysics},
  volume={408},
  number={2},
  pages={465--477},
  year={2003},
  publisher={EDP Sciences}
}

@article{yoachim2006structural,
  title={Structural parameters of thin and thick disks in edge-on disk galaxies},
  author={Yoachim, Peter and Dalcanton, Julianne J},
  journal={The Astronomical Journal},
  volume={131},
  number={1},
  pages={226},
  year={2006},
  publisher={IOP Publishing}
}

@article{dalcanton1996chain,
  title={“Chain” Galaxies are Edge-On Low Surface Brightness Galaxies},
  author={Dalcanton, Julianne J and Shectman, Stephen A},
  journal={The Astrophysical Journal Letters},
  volume={465},
  number={1},
  pages={L9},
  year={1996},
  publisher={IOP Publishing}
}

@article{kregel2005structure,
  title={Structure and kinematics of edge-on galaxy discs--V. The dynamics of stellar discs},
  author={Kregel, M and Van Der Kruit, PC and Freeman, KC},
  journal={Monthly Notices of the Royal Astronomical Society},
  volume={358},
  number={2},
  pages={503--520},
  year={2005},
  publisher={The Royal Astronomical Society}
}

@article{de2002high,
  title={High-resolution rotation curves of low surface brightness galaxies},
  author={De Blok, WJG and Bosma, A},
  journal={Astronomy \& Astrophysics},
  volume={385},
  number={3},
  pages={816--846},
  year={2002},
  publisher={EDP Sciences}
}

@article{bottema2015distribution,
  title={The distribution of dark and luminous matter inferred from extended rotation curves},
  author={Bottema, Roelof and Pestana, Jos{\'e} Luis G},
  journal={Monthly Notices of the Royal Astronomical Society},
  volume={448},
  number={3},
  pages={2566--2593},
  year={2015},
  publisher={Oxford University Press}
}

@article{donato2004cores,
  title={Cores of dark matter haloes correlate with stellar scalelengths},
  author={Donato, Fiorenza and Gentile, Gianfranco and Salucci, Paolo},
  journal={Monthly Notices of the Royal Astronomical Society},
  volume={353},
  number={2},
  pages={L17--L22},
  year={2004},
  publisher={Blackwell Science Ltd}
}

@article{mcgaugh2000baryonic,
  title={The baryonic tully-fisher relation},
  author={McGaugh, Stacy S and Schombert, Jim M and Bothun, Greg D and De Blok, WJG},
  journal={The Astrophysical Journal Letters},
  volume={533},
  number={2},
  pages={L99},
  year={2000},
  publisher={IOP Publishing}
}

@article{mcgaugh2012baryonic,
  title={The baryonic Tully-Fisher relation of gas-rich galaxies as a test of $\Lambda$CDM and MOND},
  author={McGaugh, Stacy S},
  journal={The Astronomical Journal},
  volume={143},
  number={2},
  pages={40},
  year={2012},
  publisher={IOP Publishing}
}

@article{mcgaugh2000baryonic,
  title={The baryonic tully-fisher relation},
  author={McGaugh, Stacy S and Schombert, Jim M and Bothun, Greg D and De Blok, WJG},
  journal={The Astrophysical Journal Letters},
  volume={533},
  number={2},
  pages={L99},
  year={2000},
  publisher={IOP Publishing}
}

@article{mcgaugh2005baryonic,
  title={The Baryonic Tully-Fisher relation of galaxies with extended rotation curves and the stellar mass of rotating galaxies},
  author={McGaugh, Stacy S},
  journal={The Astrophysical Journal},
  volume={632},
  number={2},
  pages={859},
  year={2005},
  publisher={IOP Publishing}
}

@misc{ss2016radial,
  title={The radial acceleration relation in rotationally supported galaxies},
  author={SS, Lelli F McGaugh and Schombert, JM},
  year={2016}
}
@article{lelli2017one,
  title={One law to rule them all: the radial acceleration relation of galaxies},
  author={Lelli, Federico and McGaugh, Stacy S and Schombert, James M and Pawlowski, Marcel S},
  journal={The Astrophysical Journal},
  volume={836},
  number={2},
  pages={152},
  year={2017},
  publisher={IOP Publishing}
}

@misc{wolfe1967modern,
  title={Modern Astrophysics, ed. M. Hack},
  author={Wolfe, RH and Horak, HG and Storer, NW},
  year={1967},
  publisher={New York: Gordon \& Breach}
}

@article{kautsch2009edge,
  title={The Edge-On Perspective of Bulgeless, Simple Disk Galaxies},
  author={Kautsch, Stefan J},
  journal={Publications of the Astronomical Society of the Pacific},
  volume={121},
  number={886},
  pages={1297},
  year={2009},
  publisher={IOP Publishing}
}

@article{vorontsov1974specification,
  title={Specification of the apparent flattening of spiral galaxies},
  author={Vorontsov-Vel'yaminov, BA},
  journal={Soviet Astronomy},
  volume={17},
  pages={452},
  year={1974}
}

@article{10.1093/mnras/stab155,
    author = {Aditya, K and Banerjee, Arunima},
    title = "{How “cold” are the stellar discs of superthin galaxies?}",
    journal = {Monthly Notices of the Royal Astronomical Society},
    year = {2021},
    month = {01},
    abstract = "{Superthin galaxies are a class of bulgeless, low surface brightness galaxies with strikingly high values of planar-to-vertical axes ratio \\$\\rm (b/a\\&gt; 10 - 20)\\$, possibly indicating the presence of an ultra-cold stellar disc. Using the multi-component galactic disc model of gravitationally-coupled stars and gas in the force field of the dark matter halo as well as the stellar dynamical code AGAMA (Action-based Galaxy Modelling Architecture), we determine the vertical velocity dispersion of stars and gas as a function of galacto-centric radius for five superthin galaxies (UGC 7321, IC 5249, FGC 1540, IC2233 and UGC00711) using observed stellar and atomic hydrogen (HI) scale heights as constraints, using a Markov Chain Monte Carlo Method. We find that the central vertical velocity dispersion for the stellar disc in the optical band varies between σ0s ∼ 10.2 − 18.4 \\$\\rm \\{kms\\}^\\{-1\\}\\$ and falls off with an exponential scale length of 2.6 to 3.2 Rd where Rd is the exponential stellar disc scale length. Interestingly, in the 3.6 μm, the same, averaged over the two components of the stellar disc, varies between 5.9 to 11.8 \\$\\rm \\{kms\\}^\\{-1\\}\\$, both of which confirm the presence of ”ultra-cold” stellar discs in superthin galaxies. Interestingly, the global median of the multi-component disc dynamical stability parameter QN of our sample superthins is found to be 5 ± 1.5, which higher than the global median value of 2.2 ± 0.6 for a sample}",
    issn = {0035-8711},
    doi = {10.1093/mnras/stab155},
    url = {https://doi.org/10.1093/mnras/stab155},
    note = {stab155},
    eprint = {https://academic.oup.com/mnras/advance-article-pdf/doi/10.1093/mnras/stab155/35931348/stab155.pdf},
}

@article{hunter2012little,
  title={Little things},
  author={Hunter, Deidre A and Ficut-Vicas, Dana and Ashley, Trisha and Brinks, Elias and Cigan, Phil and Elmegreen, Bruce G and Heesen, Volker and Herrmann, Kimberly A and Johnson, Megan and Oh, Se-Heon and others},
  journal={The Astronomical Journal},
  volume={144},
  number={5},
  pages={134},
  year={2012},
  publisher={IOP Publishing}
}

@article{keller2017lambdacdm,
  title={$\Lambda$CDM is consistent with SPARC radial acceleration relation},
  author={Keller, BW and Wadsley, JW},
  journal={The Astrophysical Journal Letters},
  volume={835},
  number={1},
  pages={L17},
  year={2017},
  publisher={IOP Publishing}
}

@article{li2018fitting,
  title={Fitting the radial acceleration relation to individual SPARC galaxies},
  author={Li, Pengfei and Lelli, Federico and McGaugh, Stacy and Schombert, James},
  journal={Astronomy \& Astrophysics},
  volume={615},
  pages={A3},
  year={2018},
  publisher={EDP Sciences}
}

@article{li2018fitting,
  title={Fitting the radial acceleration relation to individual SPARC galaxies},
  author={Li, Pengfei and Lelli, Federico and McGaugh, Stacy and Schombert, James},
  journal={Astronomy \& Astrophysics},
  volume={615},
  pages={A3},
  year={2018},
  publisher={EDP Sciences}
}

@article{ghari2019dark,
  title={Dark matter--baryon scaling relations from Einasto halo fits to SPARC galaxy rotation curves},
  author={Ghari, Amir and Famaey, Benoit and Laporte, Chervin and Haghi, Hosein},
  journal={Astronomy \& Astrophysics},
  volume={623},
  pages={A123},
  year={2019},
  publisher={EDP Sciences}
}

@article{naik2019constraints,
  title={Constraints on chameleon f (R)-gravity from galaxy rotation curves of the SPARC sample},
  author={Naik, Aneesh P and Puchwein, Ewald and Davis, Anne-Christine and Sijacki, Debora and Desmond, Harry},
  journal={Monthly Notices of the Royal Astronomical Society},
  volume={489},
  number={1},
  pages={771--787},
  year={2019},
  publisher={Oxford University Press}
}

@article{li2019constant,
  title={A constant characteristic volume density of dark matter haloes from SPARC rotation curve fits},
  author={Li, Pengfei and Lelli, Federico and McGaugh, Stacy S and Starkman, Nathaniel and Schombert, James M},
  journal={Monthly Notices of the Royal Astronomical Society},
  volume={482},
  number={4},
  pages={5106--5124},
  year={2019},
  publisher={Oxford University Press}
}

@article{chan2018testing,
  title={Testing the Cubic Galileon Gravity model by the Milky Way rotation curve and SPARC data},
  author={Chan, Man Ho and Hui, Hon Ka},
  journal={The Astrophysical Journal},
  volume={856},
  number={2},
  pages={177},
  year={2018},
  publisher={IOP Publishing}
}

@article{keller2017lambdacdm,
  title={$\Lambda$CDM is consistent with SPARC radial acceleration relation},
  author={Keller, BW and Wadsley, JW},
  journal={The Astrophysical Journal Letters},
  volume={835},
  number={1},
  pages={L17},
  year={2017},
  publisher={IOP Publishing}
}

@article{kamphuis2013halogas,
  title={HALOGAS observations of NGC 5023 and UGC 2082: modelling of non-cylindrically symmetric gas distributions in edge-on galaxies},
  author={Kamphuis, P and Rand, RJ and J{\'o}zsa, GIG and Zschaechner, LK and Heald, GH and Patterson, MT and Gentile, Gianfranco and Walterbos, RAM and Serra, P and de Blok, WJG},
  journal={Monthly Notices of the Royal Astronomical Society},
  volume={434},
  number={3},
  pages={2069--2093},
  year={2013},
  publisher={The Royal Astronomical Society}
}

@article{reynolds2020h,
  title={H i asymmetries in LVHIS, VIVA, and HALOGAS galaxies},
  author={Reynolds, TN and Westmeier, T and Staveley-Smith, L and Chauhan, G and Lagos, CDP},
  journal={Monthly Notices of the Royal Astronomical Society},
  volume={493},
  number={4},
  pages={5089--5106},
  year={2020},
  publisher={Oxford University Press}
}

@article{banerjee2013slow,
  title={A slow bar in the dwarf irregular galaxy NGC 3741},
  author={Banerjee, Arunima and Patra, Narendra Nath and Chengalur, Jayaram N and Begum, Ayesha},
  journal={Monthly Notices of the Royal Astronomical Society},
  volume={434},
  number={2},
  pages={1257--1263},
  year={2013},
  publisher={The Royal Astronomical Society}
}

@article{patra2019detection,
  title={Detection of a slow H i bar in the dwarf irregular galaxy DDO 168},
  author={Patra, Narendra Nath and Jog, Chanda J},
  journal={Monthly Notices of the Royal Astronomical Society},
  volume={488},
  number={4},
  pages={4942--4951},
  year={2019},
  publisher={Oxford University Press}
}

@article{jozsa2007kinematic,
  title={Kinematic modelling of disk galaxies-II. A case-study of symmetrically warped galaxy disks},
  author={J{\'o}zsa, GIG},
  journal={Astronomy \& Astrophysics},
  volume={468},
  number={3},
  pages={903--917},
  year={2007},
  publisher={EDP Sciences}
}

@article{allaert2015herschel,
  title={HERschel Observations of Edge-on Spirals (HEROES)-II. Tilted-ring modelling of the atomic gas disks},
  author={Allaert, Flor and Gentile, Gianfranco and Baes, Maarten and De Geyter, Gert and Hughes, TM and Lewis, F and Bianchi, SIMONE and De Looze, Ilse and Fritz, Jacopo and Holwerda, Benne W and others},
  journal={Astronomy \& Astrophysics},
  volume={582},
  pages={A18},
  year={2015},
  publisher={EDP Sciences}
}

@article{bolatto2017edge,
  title={The EDGE-CALIFA Survey: Interferometric Observations of 126 Galaxies with CARMA},
  author={Bolatto, Alberto D and Wong, Tony and Utomo, Dyas and Blitz, Leo and Vogel, Stuart N and S{\'a}nchez, Sebasti{\'a}n F and Barrera-Ballesteros, Jorge and Cao, Yixian and Colombo, Dario and Dannerbauer, Helmut and others},
  journal={The Astrophysical Journal},
  volume={846},
  number={2},
  pages={159},
  year={2017},
  publisher={IOP Publishing}
}

@article{narayanvertical,
  title={Vertical Scale Height of Stars and Gas in Disk Galaxies},
  author={Narayan, CA and Jog, CJ},
  publisher={Citeseer}
}

@article{jog1996local,
  title={Local stability criterion for stars and gas in a galactic disc},
  author={Jog, Chanda J},
  journal={Monthly Notices of the Royal Astronomical Society},
  volume={278},
  number={1},
  pages={209--218},
  year={1996},
  publisher={Blackwell Science Ltd}
}

@article{phookun1993ngc,
  title={NGC 4254: a spiral galaxy with an m= 1 mode and infalling gas},
  author={Phookun, Bikram and Vogel, Stuart N and Mundy, Lee G},
  journal={The Astrophysical Journal},
  volume={418},
  pages={113},
  year={1993}
}

@article{courteau1997optical,
  title={Optical Rotation Curves and Linewidths for Tully-Fisher Applications},
  author={Courteau, Stephane},
  journal={The Astronomical Journal},
  volume={114},
  pages={2402},
  year={1997}
}

@article{gentile2011things,
  title={THINGS about MoND},
  author={Gentile, Gianfranco and Famaey, B and de Blok, WJG},
  journal={Astronomy \& Astrophysics},
  volume={527},
  pages={A76},
  year={2011},
  publisher={EDP Sciences}
}

@ARTICLE{2014A&A...570A..13M,
   author = {{Makarov}, D. and {Prugniel}, P. and {Terekhova}, N. and {Courtois}, H. and
{Vauglin}, I.},
    title = "{HyperLEDA. III. The catalogue of extragalactic distances}",
  journal = {Astronomy \& Astrophysics},
 keywords = {astronomical databases: miscellaneous, catalogs, galaxies: distances, and redshifts},
     year = 2014,
    month = oct,
   volume = 570,
      eid = {A13},
    pages = {A13},
      doi = {10.1051/0004-6361/201423496},
   adsurl = {http://adsabs.harvard.edu/abs/2014A%26A...570A..13M},
  adsnote = {Provided by the SAO/NASA Astrophysics Data System}
}

@article{bertin1996sextractor,
  title={SExtractor: Software for source extraction},
  author={Bertin, Emmanuel and Arnouts, Stephane},
  journal={Astronomy and astrophysics supplement series},
  volume={117},
  number={2},
  pages={393--404},
  year={1996},
  publisher={EDP Sciences}
}

@article{peng2011galfit,
  title={GALFIT: Detailed Structural Decomposition of Galaxy Images},
  author={Peng, Chien Y and Ho, Luis C and Impey, Chris D and Rix, Hans-Walter},
  journal={Astrophysics Source Code Library},
  pages={ascl--1104},
  year={2011}
}

@article{kourkchi2020cosmicflows,
  title={Cosmicflows-3: Two Distance--Velocity Calculators},
  author={Kourkchi, Ehsan and Courtois, H{\'e}l{\`e}ne M and Graziani, Romain and Hoffman, Yehuda and Pomar{\`e}de, Daniel and Shaya, Edward J and Tully, R Brent},
  journal={The Astronomical Journal},
  volume={159},
  number={2},
  pages={67},
  year={2020},
  publisher={IOP Publishing}
}

@article{vasiliev2019agama,
  title={AGAMA: action-based galaxy modelling architecture},
  author={Vasiliev, Eugene},
  journal={Monthly Notices of the Royal Astronomical Society},
  volume={482},
  number={2},
  pages={1525--1544},
  year={2019},
  publisher={Oxford University Press}
}

@article{binney2014self,
  title={Self-consistent flattened isochrones},
  author={Binney, James},
  journal={Monthly Notices of the Royal Astronomical Society},
  volume={440},
  number={1},
  pages={787--798},
  year={2014},
  publisher={Oxford University Press}
}

@article{piffl2015bringing,
  title={Bringing the Galaxy's dark halo to life},
  author={Piffl, T and Penoyre, Z and Binney, J},
  journal={Monthly Notices of the Royal Astronomical Society},
  volume={451},
  number={1},
  pages={639--650},
  year={2015},
  publisher={Oxford University Press}
}

@article{bosma198121,
  title={21-cm line studies of spiral galaxies. I-Observations of the galaxies NGC 5033, 3198, 5055, 2841, and 7331},
  author={Bosma, Albert},
  journal={The Astronomical Journal},
  volume={86},
  pages={1791--1824},
  year={1981}
}

@article{bosma198121,
  title={21-cm line studies of spiral galaxies. II. The distribution and kinematics of neutral hydrogen in spiral galaxies of various morphological types.},
  author={Bosma, Astrori},
  journal={The Astronomical Journal},
  volume={86},
  pages={1825--1846},
  year={1981}
}

@article{kent1987dark,
  title={Dark matter in spiral galaxies. II-Galaxies with HI rotation curves},
  author={Kent, Stephen M},
  journal={The Astronomical Journal},
  volume={93},
  pages={816--832},
  year={1987}
}

@incollection{bosma1990frequency,
  title={The frequency of warped HI discs},
  author={Bosma, A and Athanassoula, E},
  booktitle={Dynamics and Interactions of Galaxies},
  pages={356--357},
  year={1990},
  publisher={Springer}
}

@article{sancisi2008cold,
  title={Cold gas accretion in galaxies},
  author={Sancisi, Renzo and Fraternali, Filippo and Oosterloo, Tom and Van Der Hulst, Thijs},
  journal={The Astronomy and Astrophysics Review},
  volume={15},
  number={3},
  pages={189--223},
  year={2008},
  publisher={Springer}
}

@article{marinacci2010mode,
  title={The mode of gas accretion on to star-forming galaxies},
  author={Marinacci, Federico and Binney, James and Fraternali, Filippo and Nipoti, Carlo and Ciotti, Luca and Londrillo, Pasquale},
  journal={Monthly Notices of the Royal Astronomical Society},
  volume={404},
  number={3},
  pages={1464--1474},
  year={2010},
  publisher={The Royal Astronomical Society}
}

@article{boomsma2005extra,
  title={Extra-planar HI in the starburst galaxy NGC 253},
  author={Boomsma, Rense and Oosterloo, TA and Fraternali, Filippo and van der Hulst, JM and Sancisi, Renzo},
  journal={Astronomy \& Astrophysics},
  volume={431},
  number={1},
  pages={65--72},
  year={2005},
  publisher={EDP Sciences}
}

@article{dettmar1992extraplanar,
  title={Extraplanar diffuse ionized gas and the disk-halo connection in spiral galaxies},
  author={Dettmar, Ralf-J{\"u}rgen},
  journal={Fundamentals of Cosmic Physics},
  volume={15},
  pages={143--208},
  year={1992}
}

@article{fraternali2008new,
  title={New evidence for halo gas accretion onto disk galaxies},
  author={Fraternali, Filippo},
  journal={Proceedings of the International Astronomical Union},
  volume={4},
  number={S254},
  pages={255--262},
  year={2008},
  publisher={Cambridge University Press}
}

@inproceedings{schoenmakers1999kinematic,
  title={Kinematic Lopsidedness in Spiral Galaxies},
  author={Schoenmakers, RHM and Swaters, RA},
  booktitle={Galaxy Dynamics-A Rutgers Symposium},
  volume={182},
  year={1999}
}

@article{matthews1999properties,
  title={Properties ofSuperthin'Galaxies},
  author={Matthews, LD and Van Driel, W and Gallagher, JS},
  journal={arXiv preprint astro-ph/9911022},
  year={1999}
}

@INPROCEEDINGS{2000bgfp.conf..107M,
       author = {{Matthew}, L.~D. and {van Driel}, W. and {Gallagher}, J.~S.},
        title = "{Properties of ``superthin'' galaxies}",
    booktitle = {Building Galaxies; from the Primordial Universe to the Present},
         year = 2000,
       editor = {{Hammer}, F. and {Thuan}, T.~X. and {Cayatte}, V. and {Guiderdoni}, B. and {Thanh Van}, J.~T.},
        month = jan,
        pages = {107},
       adsurl = {https://ui.adsabs.harvard.edu/abs/2000bgfp.conf..107M},
      adsnote = {Provided by the SAO/NASA Astrophysics Data System}
}

@article{alam2015eleventh,
  title={The eleventh and twelfth data releases of the Sloan Digital Sky Survey: final data from SDSS-III},
  author={Alam, Shadab and Albareti, Franco D and Prieto, Carlos Allende and Anders, Friedrich and Anderson, Scott F and Anderton, Timothy and Andrews, Brett H and Armengaud, Eric and Aubourg, {\'E}ric and Bailey, Stephen and others},
  journal={The Astrophysical Journal Supplement Series},
  volume={219},
  number={1},
  pages={12},
  year={2015},
  publisher={IOP Publishing}
}

@article{erwin2015imfit,
  title={IMFIT: a fast, flexible new program for astronomical image fitting},
  author={Erwin, Peter},
  journal={The Astrophysical Journal},
  volume={799},
  number={2},
  pages={226},
  year={2015},
  publisher={IOP Publishing}
}

@article{schlafly2011measuring,
  title={Measuring reddening with Sloan Digital Sky Survey stellar spectra and recalibrating SFD},
  author={Schlafly, Edward F and Finkbeiner, Douglas P},
  journal={The Astrophysical Journal},
  volume={737},
  number={2},
  pages={103},
  year={2011},
  publisher={IOP Publishing}
}

@article{gentile2008mond,
  title={MOND and the universal rotation curve: similar phenomenologies},
  author={Gentile, Gianfranco},
  journal={The Astrophysical Journal},
  volume={684},
  number={2},
  pages={1018},
  year={2008},
  publisher={IOP Publishing}
}

@article{gentile2011things,
  title={THINGS about MoND},
  author={Gentile, Gianfranco and Famaey, B and de Blok, WJG},
  journal={Astronomy \& Astrophysics},
  volume={527},
  pages={A76},
  year={2011},
  publisher={EDP Sciences}
}

@article{matthews2003high,
  title={High-latitude HI in the low surface brightness galaxy UGC 7321},
  author={Matthews, Lynn D and Wood, Kenneth},
  journal={The Astrophysical Journal},
  volume={593},
  number={2},
  pages={721},
  year={2003},
  publisher={IOP Publishing}
}

@article{elmegreen2020observing,
  title={Observing the Earliest Stages of Star Formation in Galaxies: 8 $\mu$m Cores in Three Edge-on Disks},
  author={Elmegreen, Bruce G and Elmegreen, Debra Meloy},
  journal={The Astrophysical Journal},
  volume={895},
  number={1},
  pages={71},
  year={2020},
  publisher={IOP Publishing}
}

@article{kregel2004structure,
  title={Structure and kinematics of edge-on galaxy discs--II. Observations of the neutral hydrogen},
  author={Kregel, M and Van Der Kruit, PC and De Blok, WJG},
  journal={Monthly Notices of the Royal Astronomical Society},
  volume={352},
  number={3},
  pages={768--786},
  year={2004},
  publisher={Blackwell Science Ltd Oxford, UK}
}

@article{krumm1979neutral,
  title={Neutral hydrogen observations of 14 nearly edge-on spiral galaxies},
  author={Krumm, N and Salpeter, EE},
  journal={The Astronomical Journal},
  volume={84},
  pages={1138--1148},
  year={1979}
}

@article{bottema1995prodigious,
  title={The prodigious warp of NGC 4013. II. Detailed observations of the neutral hydrogen gas.},
  author={Bottema, R},
  journal={Astronomy and Astrophysics},
  volume={295},
  pages={605},
  year={1995}
}

@article{seth2005study,
  title={A study of edge-on galaxies with the Hubble Space Telescope Advanced Camera for Surveys. I. Initial results},
  author={Seth, Anil C and Dalcanton, Julianne J and de Jong, Roelof S},
  journal={The Astronomical Journal},
  volume={129},
  number={3},
  pages={1331},
  year={2005},
  publisher={IOP Publishing}
}

@inproceedings{fraternali2005extra,
  title={The extra-planar neutral gas in the edge-on spiral galaxy NGC 891},
  author={Fraternali, F and Oosterloo, TA and Sancisi, R and Swaters, R},
  booktitle={Extra-Planar Gas},
  volume={331},
  pages={239},
  year={2005}
}

@article{lehner2011reservoir,
  title={A reservoir of ionized gas in the galactic halo to sustain star formation in the Milky Way},
  author={Lehner, Nicolas and Howk, J Christopher},
  journal={Science},
  volume={334},
  number={6058},
  pages={955--958},
  year={2011},
  publisher={American Association for the Advancement of Science}
}

@article{khoperskov2003minimum,
  title={Minimum velocity dispersion in stable stellar disks. Numerical simulations},
  author={Khoperskov, AV and Zasov, AV and Tyurina, NV},
  journal={Astronomy Reports},
  volume={47},
  number={5},
  pages={357--376},
  year={2003},
  publisher={Springer}
}

@article{gentile2015disk,
  title={Disk mass and disk heating in the spiral galaxy NGC 3223},
  author={Gentile, Gianfranco and Tydtgat, C and Baes, Maarten and De Geyter, Gert and Koleva, Mina and Angus, GW and De Blok, WJG and Saftly, Waad and Viaene, S{\'e}bastien},
  journal={Astronomy \& Astrophysics},
  volume={576},
  pages={A57},
  year={2015},
  publisher={EDP Sciences}
}

@article{swaters2009rotation,
  title={The rotation curves shapes of late-type dwarf galaxies},
  author={Swaters, RA and Sancisi, R and Van Albada, TS and Van Der Hulst, JM},
  journal={Astronomy \& Astrophysics},
  volume={493},
  number={3},
  pages={871--892},
  year={2009},
  publisher={EDP Sciences}
}

@ARTICLE{2019A&A...628A..58S,
       author = {{Sarkar}, S. and {Jog}, C.~J.},
        title = "{Flaring stellar disk in the low surface brightness galaxy UGC 7321}",
      journal = {Astronomy \& Astrophysics},
     keywords = {galaxies: halos, galaxies: ISM, galaxies: individual: UGC 7321, galaxies: kinematics and dynamics, galaxies: spiral, galaxies: structure, Astrophysics - Astrophysics of Galaxies},
         year = 2019,
        month = aug,
       volume = {628},
          eid = {A58},
        pages = {A58},
          doi = {10.1051/0004-6361/201935430},
archivePrefix = {arXiv},
       eprint = {1905.02735},
 primaryClass = {astro-ph.GA},
       adsurl = {https://ui.adsabs.harvard.edu/abs/2019A&A...628A..58S},
      adsnote = {Provided by the SAO/NASA Astrophysics Data System}
}

@ARTICLE{1999BSAO...47....5K,
       author = {{Karachentsev}, I.~D. and {Karachentseva}, V.~E. and {Kudrya}, Yu. N. and {Sharina}, M.~E. and {Parnovskij}, S.~L.},
        title = "{The revised Flat Galaxy Catalogue.}",
      journal = {Bulletin of the Special Astrophysics Observatory},
     keywords = {Catalogues: Galaxies, Catalogues: Spiral Galaxies, Astrophysics},
         year = 1999,
        month = jan,
       volume = {47},
        pages = {5-185},
archivePrefix = {arXiv},
       eprint = {astro-ph/0305566},
 primaryClass = {astro-ph},
       adsurl = {https://ui.adsabs.harvard.edu/abs/1999BSAO...47....5K},
      adsnote = {Provided by the SAO/NASA Astrophysics Data System}
}

@article{boily2003thickness,
  title={The Thickness of Stellar Disks of Edge-on Galaxies and Their Truncation Radii},
  author={Boily, C and Patsis, P and Portegies, S and Spurzem, R and Theis, C and Zasov, AV and Bizyaev, DV},
  journal={European Astronomical Society Publications Series},
  volume={10},
  pages={121--121},
  year={2003},
  publisher={EDP Sciences}
}

@article{rohlfs1977lectures,
  title={Lectures on density wave theory},
  author={Rohlfs, Kristen},
  year={1977}
}

@article{binney2011models,
  title={Models of our Galaxy--II},
  author={Binney, James and McMillan, Paul},
  journal={Monthly Notices of the Royal Astronomical Society},
  volume={413},
  number={3},
  pages={1889--1898},
  year={2011},
  publisher={Blackwell Publishing Ltd Oxford, UK}
}

@article{wang1994gravitational,
  title={Gravitational instability and disk star formation},
  author={Wang, Boqi and Silk, Joseph},
  journal={The Astrophysical Journal},
  volume={427},
  pages={759--769},
  year={1994}
}

@article{westfall2014diskmass,
  title={The DiskMass survey. VIII. On the relationship between disk stability and star formation},
  author={Westfall, Kyle B and Andersen, David R and Bershady, Matthew A and Martinsson, Thomas PK and Swaters, Robert A and Verheijen, Marc AW},
  journal={The Astrophysical Journal},
  volume={785},
  number={1},
  pages={43},
  year={2014},
  publisher={IOP Publishing}
}
@article{matthews2000extraordinary,
  title={The Extraordinary “Superthin” Spiral Galaxy UGC 7321. II. The Vertical Disk Structure},
  author={Matthews, LD},
  journal={The Astronomical Journal},
  volume={120},
  number={4},
  pages={1764},
  year={2000},
  publisher={IOP Publishing}
}

@article{giovanelli1997spectroscopy,
  title={Spectroscopy of edge-on spirals},
  author={Giovanelli, Riccardo and Avera, Eric and Karachentsev, Igor},
  journal={arXiv preprint astro-ph/9704189},
  year={1997}
}

@article{karachentsev1993flat,
  title={Flat galaxy catalogue},
  author={Karachentsev, ID and Karachentseva, VE and Parnovsky, SL},
  journal={Astronomische Nachrichten},
  volume={314},
  number={3},
  pages={97--222},
  year={1993},
  publisher={Wiley Online Library}
}

@article{banerjee2007origin,
  title={The origin of steep vertical stellar distribution in the Galactic disk},
  author={Banerjee, Arunima and Jog, Chanda J},
  journal={The Astrophysical Journal},
  volume={662},
  number={1},
  pages={335},
  year={2007},
  publisher={IOP Publishing}
}

@article{roberts1994physical,
  title={Physical parameters along the Hubble sequence},
  author={Roberts, Morton S and Haynes, Martha P},
  journal={Annual Review of Astronomy and Astrophysics},
  volume={32},
  number={1},
  pages={115--152},
  year={1994},
  publisher={Annual Reviews 4139 El Camino Way, PO Box 10139, Palo Alto, CA 94303-0139, USA}
}

@article{querejeta2015spitzer,
  title={The Spitzer Survey of Stellar Structure in Galaxies (S4G): precise stellar mass distributions from automated dust correction at 3.6 $\mu$m},
  author={Querejeta, Miguel and Meidt, Sharon E and Schinnerer, Eva and Cisternas, Mauricio and Mu{\~n}oz-Mateos, Juan Carlos and Sheth, Kartik and Knapen, Johan and Van De Ven, Glenn and Norris, Mark A and Peletier, Reynier and others},
  journal={The Astrophysical Journal Supplement Series},
  volume={219},
  number={1},
  pages={5},
  year={2015},
  publisher={IOP Publishing}
}

@article{braine2000deep,
  title={Deep search for CO emission in the Low Surface Brightness galaxy Malin 1},
  author={Braine, J and Herpin, F and Radford, SJE},
  journal={Astronomy and Astrophysics},
  volume={358},
  pages={494--498},
  year={2000}
}

@article{fernandez2016ggmcmc,
  title={ggmcmc: Analysis of MCMC samples and Bayesian inference},
  author={Fern{\'a}ndez-i-Mar{\i}n, Xavier},
  journal={Journal of Statistical Software},
  volume={70},
  number={9},
  pages={1--20},
  year={2016}
}

@article{hartig2017bayesiantools,
  title={Bayesiantools: General-purpose MCMC and SMC samplers and tools for Bayesian statistics},
  author={Hartig, F and Minunno, F and Paul, S and Cameron, D and Ott, T},
  journal={R package version. R package version 0.1},
  volume={3},
  year={2017}
}

@article{wickham2011ggplot2,
  title={ggplot2},
  author={Wickham, Hadley},
  journal={Wiley Interdisciplinary Reviews: Computational Statistics},
  volume={3},
  number={2},
  pages={180--185},
  year={2011},
  publisher={Wiley Online Library}
}

@misc{meschiari2015latex2exp,
  title={latex2exp: Use LaTeX Expressions in Plots, r package version 0.4. 0},
  author={Meschiari, S},
  year={2015}
}

@article{hunter2007matplotlib,
  title={Matplotlib: A 2D graphics environment},
  author={Hunter, John D},
  journal={Computing in science \& engineering},
  volume={9},
  number={3},
  pages={90--95},
  year={2007},
  publisher={IEEE Computer Society}
}

@article{vaughan2016false,
  title={False periodicities in quasar time-domain surveys},
  author={Vaughan, S and Uttley, P and Markowitz, AG and Huppenkothen, D and Middleton, MJ and Alston, WN and Scargle, JD and Farr, WM},
  journal={Monthly Notices of the Royal Astronomical Society},
  volume={461},
  number={3},
  pages={3145--3152},
  year={2016},
  publisher={Oxford University Press}
}

@article{nilson1973uppsala,
  title={Uppsala General Catalogue of Galaxies, 1973, Acta Universitatis Upsalienis, Nova Regiae Societatis Upsaliensis, Series v: a Vol.},
  author={Nilson, P},
  journal={UGC},
  pages={0},
  year={1973}
}

@article{karachentsev2001list,
  title={A list of peculiar velocities of RFGC galaxies},
  author={Karachentsev, ID and Karachentseva, VE and Kudrya, Yu N and Makarov, DI and Parnovsky, SL},
  journal={arXiv preprint astro-ph/0107058},
  year={2001}
}

@article{binggeli1985studies,
  title={Studies of the Virgo Cluster. II-A catalog of 2096 galaxies in the Virgo Cluster area.},
  author={Binggeli, B and Sandage, A and Tammann, GA},
  journal={The Astronomical Journal},
  volume={90},
  pages={1681--1759},
  year={1985}
}

@article{dalcanton2004formation,
  title={The formation of dust lanes: Implications for galaxy evolution},
  author={Dalcanton, Julianne J and Yoachim, Peter and Bernstein, Rebecca A},
  journal={The Astrophysical Journal},
  volume={608},
  number={1},
  pages={189},
  year={2004},
  publisher={IOP Publishing}
}

@article{yoachim2006structural,
  title={Structural parameters of thin and thick disks in edge-on disk galaxies},
  author={Yoachim, Peter and Dalcanton, Julianne J},
  journal={The Astronomical Journal},
  volume={131},
  number={1},
  pages={226},
  year={2006},
  publisher={IOP Publishing}
}

@inproceedings{jaeger2008common,
  title={The common astronomy software application (casa)},
  author={Jaeger, S},
  booktitle={Astronomical Data Analysis Software and Systems XVII},
  volume={394},
  pages={623},
  year={2008}
}

@techreport{fukugita1996sloan,
  title={The Sloan digital sky survey photometric system},
  author={Fukugita, M and Shimasaku, K and Ichikawa, T and Gunn, JE and others},
  year={1996},
  institution={SCAN-9601313}
}

@ARTICLE{1998A&A...336..878F,
       author = {{Fuchs}, B. and {M{\"o}llenhoff}, C. and {Heidt}, H.},
        title = "{Decomposition of the rotation curves of distant field galaxies}",
      journal = {Astronomy \& Astrophysics},
     keywords = {Astrophysics},
         year = 1998,
        month = aug,
       volume = {336},
        pages = {878},
archivePrefix = {arXiv},
       eprint = {astro-ph/9806117},
 primaryClass = {astro-ph},
       adsurl = {https://ui.adsabs.harvard.edu/abs/1998A&A...336..878F},
      adsnote = {Provided by the SAO/NASA Astrophysics Data System}
}

@article{portinari2004mass,
  title={On the mass-to-light ratio and the initial mass function in disc galaxies},
  author={Portinari, Laura and Sommer-Larsen, J and Tantalo, R},
  journal={Monthly Notices of the Royal Astronomical Society},
  volume={347},
  number={3},
  pages={691--719},
  year={2004},
  publisher={Blackwell Science Ltd Oxford, UK}
}

@article{palunas2000maximum,
  title={Maximum disk mass models for spiral galaxies},
  author={Palunas, Povilas and Williams, TB},
  journal={The Astronomical Journal},
  volume={120},
  number={6},
  pages={2884},
  year={2000},
  publisher={IOP Publishing}
}

@article{mcgaugh2014color,
  title={Color-mass-to-light-ratio relations for disk galaxies},
  author={McGaugh, Stacy S and Schombert, James M},
  journal={The Astronomical Journal},
  volume={148},
  number={5},
  pages={77},
  year={2014},
  publisher={IOP Publishing}
}

@article{de2002high,
  title={High-resolution rotation curves of low surface brightness galaxies},
  author={De Blok, WJG and Bosma, A},
  journal={Astronomy \& Astrophysics},
  volume={385},
  number={3},
  pages={816--846},
  year={2002},
  publisher={EDP Sciences}
}

@article{monnier2003search,
  title={A search for Low Surface Brightness galaxies in the near-infrared. III. Nan{\c{c}}ay HI line observations},
  author={Monnier Ragaigne, D and van Driel, Wim and Schneider, SE and Balkowski, C and Jarrett, TH},
  journal={Astronomy and Astrophysics},
  volume={408},
  number={2},
  pages={465--477},
  year={2003},
  publisher={EDP Sciences}
}

@article{dalcanton1996chain,
  title={“Chain” Galaxies are Edge-On Low Surface Brightness Galaxies},
  author={Dalcanton, Julianne J and Shectman, Stephen A},
  journal={The Astrophysical Journal Letters},
  volume={465},
  number={1},
  pages={L9},
  year={1996},
  publisher={IOP Publishing}
}

@article{mcgaugh2000baryonic,
  title={The baryonic tully-fisher relation},
  author={McGaugh, Stacy S and Schombert, Jim M and Bothun, Greg D and De Blok, WJG},
  journal={The Astrophysical Journal Letters},
  volume={533},
  number={2},
  pages={L99},
  year={2000},
  publisher={IOP Publishing}
}

@article{mcgaugh2012baryonic,
  title={The baryonic Tully-Fisher relation of gas-rich galaxies as a test of $\Lambda$CDM and MOND},
  author={McGaugh, Stacy S},
  journal={The Astronomical Journal},
  volume={143},
  number={2},
  pages={40},
  year={2012},
  publisher={IOP Publishing}
}

@article{mcgaugh2005baryonic,
  title={The Baryonic Tully-Fisher relation of galaxies with extended rotation curves and the stellar mass of rotating galaxies},
  author={McGaugh, Stacy S},
  journal={The Astrophysical Journal},
  volume={632},
  number={2},
  pages={859},
  year={2005},
  publisher={IOP Publishing}
}

@misc{ss2016radial,
  title={The radial acceleration relation in rotationally supported galaxies},
  author={SS, Lelli F McGaugh and Schombert, JM},
  year={2016}
}
@article{lelli2017one,
  title={One law to rule them all: the radial acceleration relation of galaxies},
  author={Lelli, Federico and McGaugh, Stacy S and Schombert, James M and Pawlowski, Marcel S},
  journal={The Astrophysical Journal},
  volume={836},
  number={2},
  pages={152},
  year={2017},
  publisher={IOP Publishing}
}

@misc{wolfe1967modern,
  title={Modern Astrophysics, ed. M. Hack},
  author={Wolfe, RH and Horak, HG and Storer, NW},
  year={1967},
  publisher={New York: Gordon \& Breach}
}

@article{keller2017lambdacdm,
  title={$\Lambda$CDM is consistent with SPARC radial acceleration relation},
  author={Keller, BW and Wadsley, JW},
  journal={The Astrophysical Journal Letters},
  volume={835},
  number={1},
  pages={L17},
  year={2017},
  publisher={IOP Publishing}
}

@article{li2018fitting,
  title={Fitting the radial acceleration relation to individual SPARC galaxies},
  author={Li, Pengfei and Lelli, Federico and McGaugh, Stacy and Schombert, James},
  journal={Astronomy \& Astrophysics},
  volume={615},
  pages={A3},
  year={2018},
  publisher={EDP Sciences}
}

@article{ghari2019dark,
  title={Dark matter--baryon scaling relations from Einasto halo fits to SPARC galaxy rotation curves},
  author={Ghari, Amir and Famaey, Benoit and Laporte, Chervin and Haghi, Hosein},
  journal={Astronomy \& Astrophysics},
  volume={623},
  pages={A123},
  year={2019},
  publisher={EDP Sciences}
}

@article{naik2019constraints,
  title={Constraints on chameleon f (R)-gravity from galaxy rotation curves of the SPARC sample},
  author={Naik, Aneesh P and Puchwein, Ewald and Davis, Anne-Christine and Sijacki, Debora and Desmond, Harry},
  journal={Monthly Notices of the Royal Astronomical Society},
  volume={489},
  number={1},
  pages={771--787},
  year={2019},
  publisher={Oxford University Press}
}

@article{li2019constant,
  title={A constant characteristic volume density of dark matter haloes from SPARC rotation curve fits},
  author={Li, Pengfei and Lelli, Federico and McGaugh, Stacy S and Starkman, Nathaniel and Schombert, James M},
  journal={Monthly Notices of the Royal Astronomical Society},
  volume={482},
  number={4},
  pages={5106--5124},
  year={2019},
  publisher={Oxford University Press}
}

@article{chan2018testing,
  title={Testing the Cubic Galileon Gravity model by the Milky Way rotation curve and SPARC data},
  author={Chan, Man Ho and Hui, Hon Ka},
  journal={The Astrophysical Journal},
  volume={856},
  number={2},
  pages={177},
  year={2018},
  publisher={IOP Publishing}
}

@ARTICLE{1995AJ....110..591O,
       author = {{Olling}, Rob P.},
        title = "{On the Usage of Flaring Gas Layers to Determine the Shape of Dark Matter Halos}",
      journal = {The Astronomical Journal},
     keywords = {GALAXIES: SPIRAL, DARK MATTER, GALAXIES: ISM, Astrophysics},
         year = 1995,
        month = aug,
       volume = {110},
        pages = {591},
          doi = {10.1086/117545},
archivePrefix = {arXiv},
       eprint = {astro-ph/9505002},
 primaryClass = {astro-ph},
       adsurl = {https://ui.adsabs.harvard.edu/abs/1995AJ....110..591O},
      adsnote = {Provided by the SAO/NASA Astrophysics Data System}
}

@ARTICLE{1996AJ....112..481O,
       author = {{Olling}, Rob P.},
        title = "{The Highly Flattened Dark Matter Halo of NGC 4244}",
      journal = {The Astronomical Journal},
     keywords = {GALAXIES: INDIVIDUAL: NGC 4244, GALAXIES: SPIRAL, GALAXIES: ISM, Astrophysics},
         year = 1996,
        month = aug,
       volume = {112},
        pages = {481},
          doi = {10.1086/118029},
archivePrefix = {arXiv},
       eprint = {astro-ph/9605111},
 primaryClass = {astro-ph},
       adsurl = {https://ui.adsabs.harvard.edu/abs/1996AJ....112..481O},
      adsnote = {Provided by the SAO/NASA Astrophysics Data System}
}

@article{reynolds2020h,
  title={H i asymmetries in LVHIS, VIVA, and HALOGAS galaxies},
  author={Reynolds, TN and Westmeier, T and Staveley-Smith, L and Chauhan, G and Lagos, CDP},
  journal={Monthly Notices of the Royal Astronomical Society},
  volume={493},
  number={4},
  pages={5089--5106},
  year={2020},
  publisher={Oxford University Press}
}

@article{banerjee2013slow,
  title={A slow bar in the dwarf irregular galaxy NGC 3741},
  author={Banerjee, Arunima and Patra, Narendra Nath and Chengalur, Jayaram N and Begum, Ayesha},
  journal={Monthly Notices of the Royal Astronomical Society},
  volume={434},
  number={2},
  pages={1257--1263},
  year={2013},
  publisher={The Royal Astronomical Society}
}

@article{patra2019detection,
  title={Detection of a slow H i bar in the dwarf irregular galaxy DDO 168},
  author={Patra, Narendra Nath and Jog, Chanda J},
  journal={Monthly Notices of the Royal Astronomical Society},
  volume={488},
  number={4},
  pages={4942--4951},
  year={2019},
  publisher={Oxford University Press}
}

@article{jozsa2007kinematic,
  title={Kinematic modelling of disk galaxies-II. A case-study of symmetrically warped galaxy disks},
  author={J{\'o}zsa, GIG},
  journal={Astronomy \& Astrophysics},
  volume={468},
  number={3},
  pages={903--917},
  year={2007},
  publisher={EDP Sciences}
}

@article{bolatto2017edge,
  title={The EDGE-CALIFA Survey: Interferometric Observations of 126 Galaxies with CARMA},
  author={Bolatto, Alberto D and Wong, Tony and Utomo, Dyas and Blitz, Leo and Vogel, Stuart N and S{\'a}nchez, Sebasti{\'a}n F and Barrera-Ballesteros, Jorge and Cao, Yixian and Colombo, Dario and Dannerbauer, Helmut and others},
  journal={The Astrophysical Journal},
  volume={846},
  number={2},
  pages={159},
  year={2017},
  publisher={IOP Publishing}
}

@article{phookun1993ngc,
  title={NGC 4254: a spiral galaxy with an m= 1 mode and infalling gas},
  author={Phookun, Bikram and Vogel, Stuart N and Mundy, Lee G},
  journal={The Astrophysical Journal},
  volume={418},
  pages={113},
  year={1993}
}

@article{gentile2011things,
  title={THINGS about MoND},
  author={Gentile, Gianfranco and Famaey, B and de Blok, WJG},
  journal={Astronomy \& Astrophysics},
  volume={527},
  pages={A76},
  year={2011},
  publisher={EDP Sciences}
}

@article{makarov2014hyperleda,
  title={HyperLEDA. III. The catalogue of extragalactic distances},
  author={Makarov, Dmitry and Prugniel, Philippe and Terekhova, Nataliya and Courtois, H{\'e}l{\`e}ne and Vauglin, Isabelle},
  journal={Astronomy \& Astrophysics},
  volume={570},
  pages={A13},
  year={2014},
  publisher={EDP Sciences}
}

@ARTICLE{2000BSAO...50....5K,
       author = {{Karachentsev}, I.~D. and {Karachentseva}, V.~E. and {Kudrya}, Yu. N. and {Makarov}, D.~I. and {Parnovsky}, S.~L.},
        title = "{A list of peculiar velocities of RFGC galaxies}",
      journal = {Bulletin of the Special Astrophysics Observatory},
     keywords = {Galaxies: Observations, Galaxies: Kinematics and Dynamics, RFGC Catalogue, Astrophysics},
         year = 2000,
        month = jan,
       volume = {50},
        pages = {5-38},
archivePrefix = {arXiv},
       eprint = {astro-ph/0107058},
 primaryClass = {astro-ph},
       adsurl = {https://ui.adsabs.harvard.edu/abs/2000BSAO...50....5K},
      adsnote = {Provided by the SAO/NASA Astrophysics Data System}
}

@article{jadhav2019specific,
  title={The specific angular momenta of superthin galaxies: Cue to their origin?},
  author={Jadhav Y, Vikas and Banerjee, Arunima},
  journal={Monthly Notices of the Royal Astronomical Society},
  volume={488},
  number={1},
  pages={547--558},
  year={2019},
  publisher={Oxford University Press}
}

@article{narayanan2021star,
  title={Star Formation in Superthin Galaxies},
  author={Narayanan, Ganesh and Banerjee, Arunima},
  journal={arXiv preprint arXiv:2104.04216},
  year={2021}
}

@article{bigiel2008star,
  title={The star formation law in nearby galaxies on sub-kpc scales},
  author={Bigiel, Frank and Leroy, Adam and Walter, Fabian and Brinks, Elias and De Blok, WJG and Madore, Barry and Thornley, Michele D},
  journal={The Astronomical Journal},
  volume={136},
  number={6},
  pages={2846},
  year={2008},
  publisher={IOP Publishing}
}

@article{sarkar2020general,
  title={General model of vertical distribution of stars in the Milky Way using complete Jeans equations},
  author={Sarkar, Suchira and Jog, Chanda J},
  journal={Monthly Notices of the Royal Astronomical Society},
  volume={492},
  number={1},
  pages={628--633},
  year={2020},
  publisher={Oxford University Press}
}

@article{sarkar2019vertical,
  title={Vertical distribution of stars and flaring in the Milky Way},
  author={Sarkar, Suchira and Jog, Chanda J},
  journal={Proceedings of the International Astronomical Union},
  volume={14},
  number={S353},
  pages={13--15},
  year={2019},
  publisher={Cambridge University Press}
}

@article{fall1980formation,
  title={Formation and rotation of disc galaxies with haloes},
  author={Fall, S Michael and Efstathiou, George},
  journal={Monthly Notices of the Royal Astronomical Society},
  volume={193},
  number={2},
  pages={189--206},
  year={1980},
  publisher={Oxford University Press Oxford, UK}
}

@article{romanowsky2012angular,
  title={Angular momentum and galaxy formation revisited},
  author={Romanowsky, Aaron J and Fall, S Michael},
  journal={The Astrophysical Journal Supplement Series},
  volume={203},
  number={2},
  pages={17},
  year={2012},
  publisher={IOP Publishing}
}

@article{posti2018angular,
  title={The angular momentum-mass relation: a fundamental law from dwarf irregulars to massive spirals},
  author={Posti, Lorenzo and Fraternali, Filippo and Di Teodoro, Enrico M and Pezzulli, Gabriele},
  journal={Astronomy \& Astrophysics},
  volume={612},
  pages={L6},
  year={2018},
  publisher={EDP Sciences}
}

@article{patra2020theoretical,
  title={Theoretical modelling of two-component molecular discs in spiral galaxies},
  author={Patra, Narendra Nath},
  journal={Astronomy \& Astrophysics},
  volume={638},
  pages={A66},
  year={2020},
  publisher={EDP Sciences}
}

@article{patra2020h,
  title={H i scale height in spiral galaxies},
  author={Patra, Narendra Nath},
  journal={Monthly Notices of the Royal Astronomical Society},
  volume={499},
  number={2},
  pages={2063--2075},
  year={2020},
  publisher={Oxford University Press}
}

@article{patra2018molecular,
  title={Molecular scale height in NGC 7331},
  author={Patra, Narendra Nath},
  journal={Monthly Notices of the Royal Astronomical Society},
  volume={478},
  number={4},
  pages={4931--4938},
  year={2018},
  publisher={Oxford University Press}
}

@article{ghosh2014suppression,
  title={Suppression of gravitational instabilities by dominant dark matter halo in low-surface-brightness galaxies},
  author={Ghosh, Soumavo and Jog, Chanda J},
  journal={Monthly Notices of the Royal Astronomical Society},
  volume={439},
  number={1},
  pages={929--935},
  year={2014},
  publisher={Oxford University Press}
}

@phdthesis{punzo20173d,
  title={3D visualization and analysis of HI in and around galaxies},
  author={Punzo, Davide},
  year={2017},
  school={Rijksuniversiteit Groningen}
}

@article{zschaechner2012halogas,
  title={HALOGAS: H I OBSERVATIONS AND MODELING OF THE NEARBY EDGE-ON SPIRAL GALAXY NGC 4565},
  author={Zschaechner, Laura K and Rand, Richard J and Heald, George H and Gentile, Gianfranco and J{\'o}zsa, Gyula},
  journal={The Astrophysical Journal},
  volume={760},
  number={1},
  pages={37},
  year={2012},
  publisher={IOP Publishing}
}

@article{gentile2013halogas,
  title={HALOGAS: Extraplanar gas in NGC 3198},
  author={Gentile, Gianfranco and J{\'o}zsa, GIG and Serra, P and Heald, GH and de Blok, WJG and Fraternali, F and Patterson, MT and Walterbos, RAM and Oosterloo, T},
  journal={Astronomy \& Astrophysics},
  volume={554},
  pages={A125},
  year={2013},
  publisher={EDP Sciences}
}

@article{posti2019galaxy,
  title={Galaxy disc scaling relations: A tight linear galaxy--halo connection challenges abundance matching},
  author={Posti, Lorenzo and Marasco, Antonino and Fraternali, Filippo and Famaey, Benoit},
  journal={Astronomy \& Astrophysics},
  volume={629},
  pages={A59},
  year={2019},
  publisher={EDP Sciences}
}

@ARTICLE{2021A&A...647A..76M,
author = {{Mancera Pi{\~n}a}, Pavel E. and {Posti}, Lorenzo and {Fraternali}, Filippo and {Adams}, Elizabeth A.~K. and {Oosterloo}, Tom},
title = "{The baryonic specific angular momentum of disc galaxies}",
journal = {Astronomy \& Astrophysics},
keywords = {galaxies: kinematics and dynamics, galaxies: formation, galaxies: fundamental parameters, galaxies: evolution, galaxies: dwarf, galaxies: spiral, Astrophysics - Astrophysics of Galaxies, Astrophysics - Cosmology and Nongalactic Astrophysics},
year = 2021,
month = mar,
volume = {647},
eid = {A76},
pages = {A76},
doi = {10.1051/0004-6361/202039340},
archivePrefix = {arXiv},
eprint = {2009.06645},
primaryClass = {astro-ph.GA},
adsurl = {https://ui.adsabs.harvard.edu/abs/2021A&A...647A..76M},
adsnote = {Provided by the SAO/NASA Astrophysics Data System}
}

@article{marasco2019angular,
  title={The angular momentum of disc galaxies at z= 1},
  author={Marasco, A and Fraternali, F and Posti, L and Ijtsma, M and Di Teodoro, EM and Oosterloo, T},
  journal={Astronomy \& Astrophysics},
  volume={621},
  pages={L6},
  year={2019},
  publisher={EDP Sciences}
}

@article{posti2018galaxy,
  title={Galaxy spin as a formation probe: the stellar-to-halo specific angular momentum relation},
  author={Posti, Lorenzo and Pezzulli, Gabriele and Fraternali, Filippo and Di Teodoro, Enrico M},
  journal={Monthly Notices of the Royal Astronomical Society},
  volume={475},
  number={1},
  pages={232--243},
  year={2018},
  publisher={Oxford University Press}
}

@article{dutton2014cold,
  title={Cold dark matter haloes in the Planck era: evolution of structural parameters for Einasto and NFW profiles},
  author={Dutton, Aaron A and Maccio, Andrea V},
  journal={Monthly Notices of the Royal Astronomical Society},
  volume={441},
  number={4},
  pages={3359--3374},
  year={2014},
  publisher={Oxford University Press}
}

@article{virtanen2020scipy,
  title={SciPy 1.0: fundamental algorithms for scientific computing in Python},
  author={Virtanen, Pauli and Gommers, Ralf and Oliphant, Travis E and Haberland, Matt and Reddy, Tyler and Cournapeau, David and Burovski, Evgeni and Peterson, Pearu and Weckesser, Warren and Bright, Jonathan and others},
  journal={Nature methods},
  volume={17},
  number={3},
  pages={261--272},
  year={2020},
  publisher={Nature Publishing Group}
}

@article{grand2016spiral,
  title={Spiral-induced velocity and metallicity patterns in a cosmological zoom simulation of a Milky Way-sized galaxy},
  author={Grand, Robert JJ and Springel, Volker and Kawata, Daisuke and Minchev, Ivan and S{\'a}nchez-Bl{\'a}zquez, Patricia and G{\'o}mez, Facundo A and Marinacci, Federico and Pakmor, R{\"u}diger and Campbell, David JR},
  journal={Monthly Notices of the Royal Astronomical Society: Letters},
  volume={460},
  number={1},
  pages={L94--L98},
  year={2016},
  publisher={The Royal Astronomical Society}
}

@article{kurapati2018angular,
  title={Angular momentum of dwarf galaxies},
  author={Kurapati, Sushma and Chengalur, Jayaram N and Pustilnik, Simon and Kamphuis, Peter},
  journal={Monthly Notices of the Royal Astronomical Society},
  volume={479},
  number={1},
  pages={228--239},
  year={2018},
  publisher={Oxford University Press}}

@article{bovy2012milky,
  title={The Milky Way's Circular-velocity Curve between 4 and 14 kpc from APOGEE data},
  author={Bovy, Jo and Prieto, Carlos Allende and Beers, Timothy C and Bizyaev, Dmitry and Da Costa, Luiz N and Cunha, Katia and Ebelke, Garrett L and Eisenstein, Daniel J and Frinchaboy, Peter M and P{\'e}rez, Ana Elia Garc{\'\i}a and others},
  journal={The Astrophysical Journal},
  volume={759},
  number={2},
  pages={131},
  year={2012},
  publisher={IOP Publishing}
}

@article{dehnen1998local,
  title={Local stellar kinematics from Hipparcos data},
  author={Dehnen, Walter and Binney, James J},
  journal={Monthly Notices of the Royal Astronomical Society},
  volume={298},
  number={2},
  pages={387--394},
  year={1998},
  publisher={Blackwell Science Ltd Oxford, UK and Cambridge, USA}
}

@article{butler2016angular,
  title={Angular Momentum of Dwarf Galaxies},
  author={Butler, Kirsty M and Obreschkow, Danail and Oh, Se-Heon},
  journal={The Astrophysical Journal Letters},
  volume={834},
  number={1},
  pages={L4},
  year={2016},
  publisher={IOP Publishing}
}

@article{kurapati2018angular,
  title={Angular momentum of dwarf galaxies},
  author={Kurapati, Sushma and Chengalur, Jayaram N and Pustilnik, Simon and Kamphuis, Peter},
  journal={Monthly Notices of the Royal Astronomical Society},
  volume={479},
  number={1},
  pages={228--239},
  year={2018},
  publisher={Oxford University Press}
}

@article{10.1093/mnras/stab3143,
    author = {Aditya, K and Kamphuis, Peter and Banerjee, Arunima and Borisov, Sviatoslav and Mosenkov, Aleksandr and Antipova, Aleksandra and Makarov, Dmitry},
    title = "{H i 21 cm observation and mass models of the extremely thin galaxy FGC 1440}",
    journal = {Monthly Notices of the Royal Astronomical Society},
    year = {2021},
    month = {11},
    abstract = "{We present observations and models of the kinematics and distribution of neutral hydrogen (H i) in the superthin galaxy FGC 1440 with an optical axial ratio a/b = 20.4. Using the Giant Meterwave Radio telescope (GMRT), we imaged the galaxy with a spectral resolution of 1.7 km s−1 and a spatial resolution of 15\\$\\{^\\{\\prime \\prime \\}\_\\{.\\}\\}\\$9 × 13\\$\\{^\\{\\prime \\prime \\}\_\\{.\\}\\}\\$5. We find that FGC 1440 has an asymptotic rotational velocity of 141.8 km s−1. The structure of the H i disc in FGC 1440 is that of a typical thin disc warped along the line of sight, but we can\\$\\{^\\{\\prime \\prime \\}\_\\{.\\}\\}\\$ not rule out the presence of a central thick H i disc. We find that the dark matter halo in FGC 1440 could be modeled by a pseudo- isothermal (PIS) profile with \\$\\rm R\_\\{c\\}/ R\_\\{d\\} \\&lt;2\\$, where Rc is the core radius of the PIS halo and Rd the exponential stellar disc scale length. We note that in spite of the unusually large axial ratio of FGC 1440, the ratio of the rotational velocity to stellar vertical velocity dispersion, \\$\\frac\\{V\_\\{Rot\\}\\}\\{\\sigma \_\\{z\\}\\} \\sim 5 - 8\\$, which is comparable to other superthins. Interestingly, unlike previously studied superthin galaxies which are outliers in the log10(j*) − log10(M*) relation for ordinary bulgeless disc galaxies, FGC 1440 is found to comply with the same. The values of j for the stars, gas and the baryons in FGC 1440 are consistent with those of normal spiral galaxies with similar mass.}",
    issn = {0035-8711},
    doi = {10.1093/mnras/stab3143},
    url = {https://doi.org/10.1093/mnras/stab3143},
    note = {stab3143},
    eprint = {https://academic.oup.com/mnras/advance-article-pdf/doi/10.1093/mnras/stab3143/41035100/stab3143.pdf},
}

@article{rix2013milky,
  title={The Milky Way’s stellar disk},
  author={Rix, Hans-Walter and Bovy, Jo},
  journal={The Astronomy and Astrophysics Review},
  volume={21},
  number={1},
  pages={1--58},
  year={2013},
  publisher={Springer}
}

@article{martinez2015contribution,
  title={The Contribution of Spiral Arms to the Thick Disk Along the Hubble Sequence},
  author={Martinez-Medina, LA and Pichardo, B and P{\'e}rez-Villegas, A and Moreno, E},
  journal={The Astrophysical Journal},
  volume={802},
  number={2},
  pages={109},
  year={2015},
  publisher={IOP Publishing}
}

@article{saha2010effect,
  title={The effect of bars and transient spirals on the vertical heating in disk galaxies},
  author={Saha, Kanak and Tseng, Yao-Huan and Taam, Ronald E},
  journal={The Astrophysical Journal},
  volume={721},
  number={2},
  pages={1878},
  year={2010},
  publisher={IOP Publishing}
}

@article{benson2004heating,
  title={Heating of galactic discs by infalling satellites},
  author={Benson, AJ and Lacey, CG and Frenk, CS and Baugh, CM and Cole, S},
  journal={Monthly Notices of the Royal Astronomical Society},
  volume={351},
  number={4},
  pages={1215--1236},
  year={2004},
  publisher={Blackwell Science Ltd Oxford, UK}
}

@article{brook2012thin,
  title={Thin disc, thick disc and halo in a simulated galaxy},
  author={Brook, Chris B and Stinson, GS and Gibson, Brad K and Kawata, Daisuke and House, Elisa L and Miranda, Marco S and Macci{\`o}, Andrea V and Pilkington, Kate and Ro{\v{s}}kar, R and Wadsley, J and others},
  journal={Monthly Notices of the Royal Astronomical Society},
  volume={426},
  number={1},
  pages={690--700},
  year={2012},
  publisher={Blackwell Science Ltd Oxford, UK}
}

@article{minchev2015formation,
  title={On the formation of galactic thick disks},
  author={Minchev, I and Martig, M and Streich, D and Scannapieco, C and De Jong, RS and Steinmetz, M},
  journal={The Astrophysical Journal Letters},
  volume={804},
  number={1},
  pages={L9},
  year={2015},
  publisher={IOP Publishing}
}

@article{haslbauer2022high,
  title={The High Fraction of Thin Disk Galaxies Continues to Challenge $\Lambda$CDM Cosmology},
  author={Haslbauer, Moritz and Banik, Indranil and Kroupa, Pavel and Wittenburg, Nils and Javanmardi, Behnam},
  journal={The Astrophysical Journal},
  volume={925},
  number={2},
  pages={183},
  year={2022},
  publisher={IOP Publishing}
}

@ARTICLE{2015MNRAS.446..521S,
       author = {{Schaye}, Joop and {Crain}, Robert A. and {Bower}, Richard G. and {Furlong}, Michelle and {Schaller}, Matthieu and {Theuns}, Tom and {Dalla Vecchia}, Claudio and {Frenk}, Carlos S. and {McCarthy}, I.~G. and {Helly}, John C. and {Jenkins}, Adrian and {Rosas-Guevara}, Y.~M. and {White}, Simon D.~M. and {Baes}, Maarten and {Booth}, C.~M. and {Camps}, Peter and {Navarro}, Julio F. and {Qu}, Yan and {Rahmati}, Alireza and {Sawala}, Till and {Thomas}, Peter A. and {Trayford}, James},
        title = "{The EAGLE project: simulating the evolution and assembly of galaxies and their environments}",
      journal = {Monthly Notices of the Royal Astronomical Society},
     keywords = {methods: numerical, galaxies: evolution, galaxies: formation, cosmology: theory, Astrophysics - Astrophysics of Galaxies, Astrophysics - Cosmology and Nongalactic Astrophysics},
         year = 2015,
        month = jan,
       volume = {446},
       number = {1},
        pages = {521-554},
          doi = {10.1093/mnras/stu2058},
archivePrefix = {arXiv},
       eprint = {1407.7040},
 primaryClass = {astro-ph.GA},
       adsurl = {https://ui.adsabs.harvard.edu/abs/2015MNRAS.446..521S},
      adsnote = {Provided by the SAO/NASA Astrophysics Data System}
}

@article{vogelsberger2014introducing,
  title={Introducing the Illustris Project: simulating the coevolution of dark and visible matter in the Universe},
  author={Vogelsberger, Mark and Genel, Shy and Springel, Volker and Torrey, Paul and Sijacki, Debora and Xu, Dandan and Snyder, Greg and Nelson, Dylan and Hernquist, Lars},
  journal={Monthly Notices of the Royal Astronomical Society},
  volume={444},
  number={2},
  pages={1518--1547},
  year={2014},
  publisher={Oxford University Press}
}

@article{pillepich2018simulating,
  title={Simulating galaxy formation with the IllustrisTNG model},
  author={Pillepich, Annalisa and Springel, Volker and Nelson, Dylan and Genel, Shy and Naiman, Jill and Pakmor, R{\"u}diger and Hernquist, Lars and Torrey, Paul and Vogelsberger, Mark and Weinberger, Rainer and others},
  journal={Monthly Notices of the Royal Astronomical Society},
  volume={473},
  number={3},
  pages={4077--4106},
  year={2018},
  publisher={Oxford University Press}
}

@ARTICLE{2013MNRAS.431..582B,
       author = {{Banerjee}, Arunima and {Jog}, Chanda J.},
        title = "{Why are some galaxy discs extremely thin?}",
      journal = {Monthly Notices of the Royal Astronomical Society},
     keywords = {galaxies: haloes, galaxies: individual: UGC 7321, galaxies: individual: Holmberg II, galaxies: ISM, galaxies: kinematics and dynamics, galaxies: structure, Astrophysics - Cosmology and Nongalactic Astrophysics},
         year = 2013,
        month = may,
       volume = {431},
       number = {1},
        pages = {582-588},
          doi = {10.1093/mnras/stt186},
archivePrefix = {arXiv},
       eprint = {1210.8244},
 primaryClass = {astro-ph.CO},
       adsurl = {https://ui.adsabs.harvard.edu/abs/2013MNRAS.431..582B},
      adsnote = {Provided by the SAO/NASA Astrophysics Data System}
}

@article{romanowsky2012angular,
  title={Angular momentum and galaxy formation revisited},
  author={Romanowsky, Aaron J and Fall, S Michael},
  journal={The Astrophysical Journal Supplement Series},
  volume={203},
  number={2},
  pages={17},
  year={2012},
  publisher={IOP Publishing}
}

@article{sarkar2018constraining,
  title={The constraining effect of gas and the dark matter halo on the vertical stellar distribution of the Milky Way},
  author={Sarkar, Suchira and Jog, Chanda J},
  journal={Astronomy \& Astrophysics},
  volume={617},
  pages={A142},
  year={2018},
  publisher={EDP Sciences}
}

@article{banerjee2011theoretical,
  title={Theoretical determination of H i vertical scale heights in the dwarf galaxies DDO 154, Ho II, IC 2574 and NGC 2366},
  author={Banerjee, Arunima and Jog, Chanda J and Brinks, Elias and Bagetakos, Ioannis},
  journal={Monthly Notices of the Royal Astronomical Society},
  volume={415},
  number={1},
  pages={687--694},
  year={2011},
  publisher={Blackwell Publishing Ltd Oxford, UK}
}

@article{hoyle1953fragmentation,
  title={On the Fragmentation of Gas Clouds Into Galaxies and Stars.},
  author={Hoyle, Fred},
  journal={The Astrophysical Journal},
  volume={118},
  pages={513},
  year={1953}
}

@article{barnes1987angular,
  title={Angular momentum from tidal torques},
  author={Barnes, Joshua and Efstathiou, George},
  journal={The Astrophysical Journal},
  volume={319},
  pages={575--600},
  year={1987}
}

@ARTICLE{2017MNRAS.467.2879B,
       author = {{Bottrell}, Connor and {Torrey}, Paul and {Simard}, Luc and {Ellison}, Sara L.},
        title = "{Galaxies in the Illustris simulation as seen by the Sloan Digital Sky Survey - II. Size-luminosity relations and the deficit of bulge-dominated galaxies in Illustris at low mass}",
      journal = {Monthly Notices of the Royal Astronomical Society},
     keywords = {hydrodynamics, surveys, galaxies: structure, Astrophysics - Astrophysics of Galaxies},
         year = 2017,
        month = may,
       volume = {467},
       number = {3},
        pages = {2879-2895},
          doi = {10.1093/mnras/stx276},
archivePrefix = {arXiv},
       eprint = {1701.08206},
 primaryClass = {astro-ph.GA},
       adsurl = {https://ui.adsabs.harvard.edu/abs/2017MNRAS.467.2879B},
      adsnote = {Provided by the SAO/NASA Astrophysics Data System}
}

@ARTICLE{2019MNRAS.490.5451D,
       author = {{Di Paolo}, Chiara and {Salucci}, Paolo and {Erkurt}, Adnan},
        title = "{The universal rotation curve of low surface brightness galaxies - IV. The interrelation between dark and luminous matter}",
      journal = {Monthly Notices of the Royal Astronomical Society},
     keywords = {galaxies: fundamental parameters, galaxies: kinematics and dynamics, dark matter, Astrophysics - Astrophysics of Galaxies},
         year = 2019,
        month = dec,
       volume = {490},
       number = {4},
        pages = {5451-5477},
          doi = {10.1093/mnras/stz2700},
archivePrefix = {arXiv},
       eprint = {1805.07165},
 primaryClass = {astro-ph.GA},
       adsurl = {https://ui.adsabs.harvard.edu/abs/2019MNRAS.490.5451D},
      adsnote = {Provided by the SAO/NASA Astrophysics Data System}
}

@article{bullock2001profiles,
  title={Profiles of dark haloes: evolution, scatter and environment},
  author={Bullock, James S and Kolatt, Tsafrir S and Sigad, Yair and Somerville, Rachel S and Kravtsov, Andrey V and Klypin, Anatoly A and Primack, Joel R and Dekel, Avishai},
  journal={Monthly Notices of the Royal Astronomical Society},
  volume={321},
  number={3},
  pages={559--575},
  year={2001},
  publisher={Blackwell Science Ltd Oxford, UK}
}

@article{rovskar2013effects,
  title={The effects of radial migration on the vertical structure of Galactic discs},
  author={Ro{\v{s}}kar, Rok and Debattista, Victor P and Loebman, Sarah R},
  journal={Monthly Notices of the Royal Astronomical Society},
  volume={433},
  number={2},
  pages={976--985},
  year={2013},
  publisher={Oxford University Press}
}

@article{wechsler2002concentrations,
  title={Concentrations of dark halos from their assembly histories},
  author={Wechsler, Risa H and Bullock, James S and Primack, Joel R and Kravtsov, Andrey V and Dekel, Avishai},
  journal={The Astrophysical Journal},
  volume={568},
  number={1},
  pages={52},
  year={2002},
  publisher={IOP Publishing}
}

@article{bailin2005internal,
  title={Internal and external alignment of the shapes and angular momenta of $\Lambda$CDM halos},
  author={Bailin, Jeremy and Steinmetz, Matthias},
  journal={The Astrophysical Journal},
  volume={627},
  number={2},
  pages={647},
  year={2005},
  publisher={IOP Publishing}
}

@article{salo2015,
  title={The Spitzer Survey of Stellar Structure in Galaxies (S4G): Multi-component Decomposition Strategies and Data Release},
  author={Salo. H, Laurikainen. E and others},
  journal={The Astrophysical Journal Supplement Series},
  volume={219},
  pages={4},
  year={2015},
  publisher={ The American Astronomical Society}
}

@book{kamphuis2008structure,
  title={The structure and kinematics of halos in disk galaxies},
  author={Kamphuis, Peter},
  year={2008},
  publisher={University Library Groningen][Host]}
}

@article{voigtlander2013kinematics,
  title={The kinematics of the diffuse ionized gas in NGC 4666},
  author={Voigtl{\"a}nder, Pierre and Kamphuis, Peter and Marcelin, Michel and Bomans, Dominik J and Dettmar, R-J},
  journal={Astronomy \& Astrophysics},
  volume={554},
  pages={A133},
  year={2013},
  publisher={EDP Sciences}
}

@article{barrera2021edge,
  title={EDGE-CALIFA survey: Self-regulation of Star formation at kpc scales},
  author={Barrera-Ballesteros, JK and S{\'a}nchez, SF and Heckman, T and Wong, T and Bolatto, A and Ostriker, E and Rosolowsky, E and Carigi, L and Vogel, S and Levy, RC and others},
  journal={arXiv preprint arXiv:2101.04683},
  year={2021}
}

@article{rubin1980rotational,
  title={Rotational properties of 21 SC galaxies with a large range of luminosities and radii, from NGC 4605/R= 4kpc/to UGC 2885/R= 122 kpc},
  author={Rubin, Vera C and Ford Jr, W Kent and Thonnard, Norbert},
  journal={The Astrophysical Journal},
  volume={238},
  pages={471--487},
  year={1980}
}
@article{rubin1985rotation,
  title={Rotation velocities of 16 SA galaxies and a comparison of Sa, Sb, and SC rotation properties},
  author={Rubin, Vera C and Burstein, David and Ford Jr, W Kent and Thonnard, Norbert},
  journal={The Astrophysical Journal},
  volume={289},
  pages={81--98},
  year={1985}
}

@article{de2001mass,
  title={Mass density profiles of low surface brightness galaxies},
  author={De Blok, WJG and McGaugh, Stacy S and Bosma, Albert and Rubin, Vera C},
  journal={The Astrophysical Journal Letters},
  volume={552},
  number={1},
  pages={L23},
  year={2001},
  publisher={IOP Publishing}
}

@article{komanduri2020dynamical,
  title={Dynamical modelling of disc vertical structure in superthin galaxy ‘UGC 7321’in braneworld gravity: an MCMC study},
  author={Komanduri, Aditya and Banerjee, Indrani and Banerjee, Arunima and Sengupta, Soumitra},
  journal={Monthly Notices of the Royal Astronomical Society},
  volume={499},
  number={4},
  pages={5690--5701},
  year={2020},
  publisher={Oxford University Press}
}

@article{pohlen2003evidence,
  title={Evidence for a large stellar bar in the Low Surface Brightness galaxy UGC 7321},
  author={Pohlen, M and Balcells, M and L{\"u}tticke, R and Dettmar, R-J},
  journal={Astronomy \& Astrophysics},
  volume={409},
  number={2},
  pages={485--490},
  year={2003},
  publisher={EDP Sciences}
}

@article{byun1998surface,
  title={Surface Photometry of Edge-on Galaxies: IC 5249 and ES0 404-G18},
  author={Byun, Yong-Ik},
  journal={Chinese Journal of Physics},
  volume={36},
  number={5},
  pages={677--692},
  year={1998},
  publisher={台灣物理學會}
}

@article{matthews2008corrugations,
  title={Corrugations in the disk of the edge-on spiral galaxy IC 2233},
  author={Matthews, LD and Uson, Juan M},
  journal={The Astrophysical Journal},
  volume={688},
  number={1},
  pages={237},
  year={2008},
  publisher={IOP Publishing}
}

@article{gallagher1976surface,
  title={Surface photometry of the spiral galaxy IC 2233 and the existence of massive halos},
  author={Gallagher, JS and Hudson, HS},
  journal={The Astrophysical Journal},
  volume={209},
  pages={389--391},
  year={1976}
}

@article{de2008mass,
  title={Mass models for low surface brightness galaxies with high-resolution optical velocity fields},
  author={De Naray, Rachel Kuzio and McGaugh, Stacy S and De Blok, WJG},
  journal={The Astrophysical Journal},
  volume={676},
  number={2},
  pages={920},
  year={2008},
  publisher={IOP Publishing}
}

@article{mcgaugh2014color,
  title={Color-mass-to-light-ratio relations for disk galaxies},
  author={McGaugh, Stacy S and Schombert, James M},
  journal={The Astronomical Journal},
  volume={148},
  number={5},
  pages={77},
  year={2014},
  publisher={IOP Publishing}
}

@ARTICLE{2003ApJS..149..289B,
       author = {{Bell}, Eric F. and {McIntosh}, Daniel H. and {Katz}, Neal and {Weinberg}, Martin D.},
        title = "{The Optical and Near-Infrared Properties of Galaxies. I. Luminosity and Stellar Mass Functions}",
      journal = {\apjs},
     keywords = {Galaxies: Evolution, Galaxies: General, Galaxies: Luminosity Function, Mass Function, Galaxies: Stellar Content, Astrophysics},
         year = 2003,
        month = dec,
       volume = {149},
       number = {2},
        pages = {289-312},
          doi = {10.1086/378847},
archivePrefix = {arXiv},
       eprint = {astro-ph/0302543},
 primaryClass = {astro-ph},
       adsurl = {https://ui.adsabs.harvard.edu/abs/2003ApJS..149..289B},
      adsnote = {Provided by the SAO/NASA Astrophysics Data System}
}

@article{boily2003thickness,
  title={The Thickness of Stellar Disks of Edge-on Galaxies and Their Truncation Radii},
  author={Boily, C and Patsis, P and Portegies, S and Spurzem, R and Theis, C and Zasov, AV and Bizyaev, DV},
  journal={European Astronomical Society Publications Series},
  volume={10},
  pages={121--121},
  year={2003},
  publisher={EDP Sciences}
}

@article{rohlfs1977lectures,
  title={Lectures on density wave theory},
  author={Rohlfs, Kristen},
  year={1977}
}

@article{binney2011models,
  title={Models of our Galaxy--II},
  author={Binney, James and McMillan, Paul},
  journal={Monthly Notices of the Royal Astronomical Society},
  volume={413},
  number={3},
  pages={1889--1898},
  year={2011},
  publisher={Blackwell Publishing Ltd Oxford, UK}
}

@article{wang1994gravitational,
  title={Gravitational instability and disk star formation},
  author={Wang, Boqi and Silk, Joseph},
  journal={The Astrophysical Journal},
  volume={427},
  pages={759--769},
  year={1994}
}

@article{westfall2014diskmass,
  title={The DiskMass survey. VIII. On the relationship between disk stability and star formation},
  author={Westfall, Kyle B and Andersen, David R and Bershady, Matthew A and Martinsson, Thomas PK and Swaters, Robert A and Verheijen, Marc AW},
  journal={The Astrophysical Journal},
  volume={785},
  number={1},
  pages={43},
  year={2014},
  publisher={IOP Publishing}
}
@article{matthews2000extraordinary,
  title={The Extraordinary “Superthin” Spiral Galaxy UGC 7321. II. The Vertical Disk Structure},
  author={Matthews, LD},
  journal={The Astronomical Journal},
  volume={120},
  number={4},
  pages={1764},
  year={2000},
  publisher={IOP Publishing}
}

@article{giovanelli1997spectroscopy,
  title={Spectroscopy of edge-on spirals},
  author={Giovanelli, Riccardo and Avera, Eric and Karachentsev, Igor},
  journal={arXiv preprint astro-ph/9704189},
  year={1997}
}

@article{karachentsev1993flat,
  title={Flat galaxy catalogue},
  author={Karachentsev, ID and Karachentseva, VE and Parnovsky, SL},
  journal={Astronomische Nachrichten},
  volume={314},
  number={3},
  pages={97--222},
  year={1993},
  publisher={Wiley Online Library}
}

@article{banerjee2007origin,
  title={The origin of steep vertical stellar distribution in the Galactic disk},
  author={Banerjee, Arunima and Jog, Chanda J},
  journal={The Astrophysical Journal},
  volume={662},
  number={1},
  pages={335},
  year={2007},
  publisher={IOP Publishing}
}

@article{roberts1994physical,
  title={Physical parameters along the Hubble sequence},
  author={Roberts, Morton S and Haynes, Martha P},
  journal={Annual Review of Astronomy and Astrophysics},
  volume={32},
  number={1},
  pages={115--152},
  year={1994},
  publisher={Annual Reviews 4139 El Camino Way, PO Box 10139, Palo Alto, CA 94303-0139, USA}
}

@article{querejeta2015spitzer,
  title={The Spitzer Survey of Stellar Structure in Galaxies (S4G): precise stellar mass distributions from automated dust correction at 3.6 $\mu$m},
  author={Querejeta, Miguel and Meidt, Sharon E and Schinnerer, Eva and Cisternas, Mauricio and Mu{\~n}oz-Mateos, Juan Carlos and Sheth, Kartik and Knapen, Johan and Van De Ven, Glenn and Norris, Mark A and Peletier, Reynier and others},
  journal={The Astrophysical Journal Supplement Series},
  volume={219},
  number={1},
  pages={5},
  year={2015},
  publisher={IOP Publishing}
}

@article{braine2000deep,
  title={Deep search for CO emission in the Low Surface Brightness galaxy Malin 1},
  author={Braine, J and Herpin, F and Radford, SJE},
  journal={Astronomy and Astrophysics},
  volume={358},
  pages={494--498},
  year={2000}
}

@article{fernandez2016ggmcmc,
  title={ggmcmc: Analysis of MCMC samples and Bayesian inference},
  author={Fern{\'a}ndez-i-Mar{\i}n, Xavier},
  journal={Journal of Statistical Software},
  volume={70},
  number={9},
  pages={1--20},
  year={2016}
}

@article{hartig2017bayesiantools,
  title={Bayesiantools: General-purpose MCMC and SMC samplers and tools for Bayesian statistics},
  author={Hartig, F and Minunno, F and Paul, S and Cameron, D and Ott, T},
  journal={R package version. R package version 0.1},
  volume={3},
  year={2017}
}

@article{wickham2011ggplot2,
  title={ggplot2},
  author={Wickham, Hadley},
  journal={Wiley Interdisciplinary Reviews: Computational Statistics},
  volume={3},
  number={2},
  pages={180--185},
  year={2011},
  publisher={Wiley Online Library}
}

@misc{meschiari2015latex2exp,
  title={latex2exp: Use LaTeX Expressions in Plots, r package version 0.4. 0},
  author={Meschiari, S},
  year={2015}
}

@article{hunter2007matplotlib,
  title={Matplotlib: A 2D graphics environment},
  author={Hunter, John D},
  journal={Computing in science \& engineering},
  volume={9},
  number={3},
  pages={90--95},
  year={2007},
  publisher={IEEE Computer Society}
}

@article{vaughan2016false,
  title={False periodicities in quasar time-domain surveys},
  author={Vaughan, S and Uttley, P and Markowitz, AG and Huppenkothen, D and Middleton, MJ and Alston, WN and Scargle, JD and Farr, WM},
  journal={Monthly Notices of the Royal Astronomical Society},
  volume={461},
  number={3},
  pages={3145--3152},
  year={2016},
  publisher={Oxford University Press}
}

@article{karachentsev1993flat,
  title={Flat galaxy catalogue},
  author={Karachentsev, ID and Karachentseva, VE and Parnovsky, SL},
  journal={Astronomische Nachrichten},
  volume={314},
  number={3},
  pages={97--222},
  year={1993},
  publisher={Wiley Online Library}
}

@article{nilson1973uppsala,
  title={Uppsala General Catalogue of Galaxies, 1973, Acta Universitatis Upsalienis, Nova Regiae Societatis Upsaliensis, Series v: a Vol.},
  author={Nilson, P},
  journal={UGC},
  pages={0},
  year={1973}
}

@article{karachentsev2001list,
  title={A list of peculiar velocities of RFGC galaxies},
  author={Karachentsev, ID and Karachentseva, VE and Kudrya, Yu N and Makarov, DI and Parnovsky, SL},
  journal={arXiv preprint astro-ph/0107058},
  year={2001}
}

@article{binggeli1985studies,
  title={Studies of the Virgo Cluster. II-A catalog of 2096 galaxies in the Virgo Cluster area.},
  author={Binggeli, B and Sandage, A and Tammann, GA},
  journal={The Astronomical Journal},
  volume={90},
  pages={1681--1759},
  year={1985}
}

@article{dalcanton2004formation,
  title={The formation of dust lanes: Implications for galaxy evolution},
  author={Dalcanton, Julianne J and Yoachim, Peter and Bernstein, Rebecca A},
  journal={The Astrophysical Journal},
  volume={608},
  number={1},
  pages={189},
  year={2004},
  publisher={IOP Publishing}
}

@article{yoachim2006structural,
  title={Structural parameters of thin and thick disks in edge-on disk galaxies},
  author={Yoachim, Peter and Dalcanton, Julianne J},
  journal={The Astronomical Journal},
  volume={131},
  number={1},
  pages={226},
  year={2006},
  publisher={IOP Publishing}
}

@inproceedings{jaeger2008common,
  title={The common astronomy software application (casa)},
  author={Jaeger, S},
  booktitle={Astronomical Data Analysis Software and Systems XVII},
  volume={394},
  pages={623},
  year={2008}
}

@techreport{fukugita1996sloan,
  title={The Sloan digital sky survey photometric system},
  author={Fukugita, M and Shimasaku, K and Ichikawa, T and Gunn, JE and others},
  year={1996},
  institution={SCAN-9601313}
}

@article{portinari2004mass,
  title={On the mass-to-light ratio and the initial mass function in disc galaxies},
  author={Portinari, Laura and Sommer-Larsen, J and Tantalo, R},
  journal={Monthly Notices of the Royal Astronomical Society},
  volume={347},
  number={3},
  pages={691--719},
  year={2004},
  publisher={Blackwell Science Ltd Oxford, UK}
}

@article{palunas2000maximum,
  title={Maximum disk mass models for spiral galaxies},
  author={Palunas, Povilas and Williams, TB},
  journal={The Astronomical Journal},
  volume={120},
  number={6},
  pages={2884},
  year={2000},
  publisher={IOP Publishing}
}

@article{mcgaugh2014color,
  title={Color-mass-to-light-ratio relations for disk galaxies},
  author={McGaugh, Stacy S and Schombert, James M},
  journal={The Astronomical Journal},
  volume={148},
  number={5},
  pages={77},
  year={2014},
  publisher={IOP Publishing}
}

@article{de2002high,
  title={High-resolution rotation curves of low surface brightness galaxies},
  author={De Blok, WJG and Bosma, A},
  journal={Astronomy \& Astrophysics},
  volume={385},
  number={3},
  pages={816--846},
  year={2002},
  publisher={EDP Sciences}
}

@article{monnier2003search,
  title={A search for Low Surface Brightness galaxies in the near-infrared. III. Nan{\c{c}}ay HI line observations},
  author={Monnier Ragaigne, D and van Driel, Wim and Schneider, SE and Balkowski, C and Jarrett, TH},
  journal={Astronomy and Astrophysics},
  volume={408},
  number={2},
  pages={465--477},
  year={2003},
  publisher={EDP Sciences}
}

@article{yoachim2006structural,
  title={Structural parameters of thin and thick disks in edge-on disk galaxies},
  author={Yoachim, Peter and Dalcanton, Julianne J},
  journal={The Astronomical Journal},
  volume={131},
  number={1},
  pages={226},
  year={2006},
  publisher={IOP Publishing}
}

@article{dalcanton1996chain,
  title={“Chain” Galaxies are Edge-On Low Surface Brightness Galaxies},
  author={Dalcanton, Julianne J and Shectman, Stephen A},
  journal={The Astrophysical Journal Letters},
  volume={465},
  number={1},
  pages={L9},
  year={1996},
  publisher={IOP Publishing}
}

@article{de2002high,
  title={High-resolution rotation curves of low surface brightness galaxies},
  author={De Blok, WJG and Bosma, A},
  journal={Astronomy \& Astrophysics},
  volume={385},
  number={3},
  pages={816--846},
  year={2002},
  publisher={EDP Sciences}
}

@article{mcgaugh2000baryonic,
  title={The baryonic tully-fisher relation},
  author={McGaugh, Stacy S and Schombert, Jim M and Bothun, Greg D and De Blok, WJG},
  journal={The Astrophysical Journal Letters},
  volume={533},
  number={2},
  pages={L99},
  year={2000},
  publisher={IOP Publishing}
}

@article{mcgaugh2012baryonic,
  title={The baryonic Tully-Fisher relation of gas-rich galaxies as a test of $\Lambda$CDM and MOND},
  author={McGaugh, Stacy S},
  journal={The Astronomical Journal},
  volume={143},
  number={2},
  pages={40},
  year={2012},
  publisher={IOP Publishing}
}

@article{mcgaugh2000baryonic,
  title={The baryonic tully-fisher relation},
  author={McGaugh, Stacy S and Schombert, Jim M and Bothun, Greg D and De Blok, WJG},
  journal={The Astrophysical Journal Letters},
  volume={533},
  number={2},
  pages={L99},
  year={2000},
  publisher={IOP Publishing}
}

@article{mcgaugh2005baryonic,
  title={The Baryonic Tully-Fisher relation of galaxies with extended rotation curves and the stellar mass of rotating galaxies},
  author={McGaugh, Stacy S},
  journal={The Astrophysical Journal},
  volume={632},
  number={2},
  pages={859},
  year={2005},
  publisher={IOP Publishing}
}

@misc{ss2016radial,
  title={The radial acceleration relation in rotationally supported galaxies},
  author={SS, Lelli F McGaugh and Schombert, JM},
  year={2016}
}
@article{lelli2017one,
  title={One law to rule them all: the radial acceleration relation of galaxies},
  author={Lelli, Federico and McGaugh, Stacy S and Schombert, James M and Pawlowski, Marcel S},
  journal={The Astrophysical Journal},
  volume={836},
  number={2},
  pages={152},
  year={2017},
  publisher={IOP Publishing}
}

@misc{wolfe1967modern,
  title={Modern Astrophysics, ed. M. Hack},
  author={Wolfe, RH and Horak, HG and Storer, NW},
  year={1967},
  publisher={New York: Gordon \& Breach}
}

@article{keller2017lambdacdm,
  title={$\Lambda$CDM is consistent with SPARC radial acceleration relation},
  author={Keller, BW and Wadsley, JW},
  journal={The Astrophysical Journal Letters},
  volume={835},
  number={1},
  pages={L17},
  year={2017},
  publisher={IOP Publishing}
}

@article{li2018fitting,
  title={Fitting the radial acceleration relation to individual SPARC galaxies},
  author={Li, Pengfei and Lelli, Federico and McGaugh, Stacy and Schombert, James},
  journal={Astronomy \& Astrophysics},
  volume={615},
  pages={A3},
  year={2018},
  publisher={EDP Sciences}
}

@article{li2018fitting,
  title={Fitting the radial acceleration relation to individual SPARC galaxies},
  author={Li, Pengfei and Lelli, Federico and McGaugh, Stacy and Schombert, James},
  journal={Astronomy \& Astrophysics},
  volume={615},
  pages={A3},
  year={2018},
  publisher={EDP Sciences}
}

@article{ghari2019dark,
  title={Dark matter--baryon scaling relations from Einasto halo fits to SPARC galaxy rotation curves},
  author={Ghari, Amir and Famaey, Benoit and Laporte, Chervin and Haghi, Hosein},
  journal={Astronomy \& Astrophysics},
  volume={623},
  pages={A123},
  year={2019},
  publisher={EDP Sciences}
}

@article{naik2019constraints,
  title={Constraints on chameleon f (R)-gravity from galaxy rotation curves of the SPARC sample},
  author={Naik, Aneesh P and Puchwein, Ewald and Davis, Anne-Christine and Sijacki, Debora and Desmond, Harry},
  journal={Monthly Notices of the Royal Astronomical Society},
  volume={489},
  number={1},
  pages={771--787},
  year={2019},
  publisher={Oxford University Press}
}

@article{li2019constant,
  title={A constant characteristic volume density of dark matter haloes from SPARC rotation curve fits},
  author={Li, Pengfei and Lelli, Federico and McGaugh, Stacy S and Starkman, Nathaniel and Schombert, James M},
  journal={Monthly Notices of the Royal Astronomical Society},
  volume={482},
  number={4},
  pages={5106--5124},
  year={2019},
  publisher={Oxford University Press}
}

@article{chan2018testing,
  title={Testing the Cubic Galileon Gravity model by the Milky Way rotation curve and SPARC data},
  author={Chan, Man Ho and Hui, Hon Ka},
  journal={The Astrophysical Journal},
  volume={856},
  number={2},
  pages={177},
  year={2018},
  publisher={IOP Publishing}
}

@article{keller2017lambdacdm,
  title={$\Lambda$CDM is consistent with SPARC radial acceleration relation},
  author={Keller, BW and Wadsley, JW},
  journal={The Astrophysical Journal Letters},
  volume={835},
  number={1},
  pages={L17},
  year={2017},
  publisher={IOP Publishing}
}

@article{reynolds2020h,
  title={H i asymmetries in LVHIS, VIVA, and HALOGAS galaxies},
  author={Reynolds, TN and Westmeier, T and Staveley-Smith, L and Chauhan, G and Lagos, CDP},
  journal={Monthly Notices of the Royal Astronomical Society},
  volume={493},
  number={4},
  pages={5089--5106},
  year={2020},
  publisher={Oxford University Press}
}

@article{banerjee2013slow,
  title={A slow bar in the dwarf irregular galaxy NGC 3741},
  author={Banerjee, Arunima and Patra, Narendra Nath and Chengalur, Jayaram N and Begum, Ayesha},
  journal={Monthly Notices of the Royal Astronomical Society},
  volume={434},
  number={2},
  pages={1257--1263},
  year={2013},
  publisher={The Royal Astronomical Society}
}

@article{patra2019detection,
  title={Detection of a slow H i bar in the dwarf irregular galaxy DDO 168},
  author={Patra, Narendra Nath and Jog, Chanda J},
  journal={Monthly Notices of the Royal Astronomical Society},
  volume={488},
  number={4},
  pages={4942--4951},
  year={2019},
  publisher={Oxford University Press}
}

@article{jozsa2007kinematic,
  title={Kinematic modelling of disk galaxies-II. A case-study of symmetrically warped galaxy disks},
  author={J{\'o}zsa, GIG},
  journal={Astronomy \& Astrophysics},
  volume={468},
  number={3},
  pages={903--917},
  year={2007},
  publisher={EDP Sciences}
}

@article{bolatto2017edge,
  title={The EDGE-CALIFA Survey: Interferometric Observations of 126 Galaxies with CARMA},
  author={Bolatto, Alberto D and Wong, Tony and Utomo, Dyas and Blitz, Leo and Vogel, Stuart N and S{\'a}nchez, Sebasti{\'a}n F and Barrera-Ballesteros, Jorge and Cao, Yixian and Colombo, Dario and Dannerbauer, Helmut and others},
  journal={The Astrophysical Journal},
  volume={846},
  number={2},
  pages={159},
  year={2017},
  publisher={IOP Publishing}
}

@article{narayanvertical,
  title={Vertical Scale Height of Stars and Gas in Disk Galaxies},
  author={Narayan, CA and Jog, CJ},
  publisher={Citeseer}
}

@article{phookun1993ngc,
  title={NGC 4254: a spiral galaxy with an m= 1 mode and infalling gas},
  author={Phookun, Bikram and Vogel, Stuart N and Mundy, Lee G},
  journal={The Astrophysical Journal},
  volume={418},
  pages={113},
  year={1993}
}

@article{gentile2011things,
  title={THINGS about MoND},
  author={Gentile, Gianfranco and Famaey, B and de Blok, WJG},
  journal={Astronomy \& Astrophysics},
  volume={527},
  pages={A76},
  year={2011},
  publisher={EDP Sciences}
}

@article{vasiliev2019agama,
  title={AGAMA: action-based galaxy modelling architecture},
  author={Vasiliev, Eugene},
  journal={Monthly Notices of the Royal Astronomical Society},
  volume={482},
  number={2},
  pages={1525--1544},
  year={2019},
  publisher={Oxford University Press}
}

@article{binney2014self,
  title={Self-consistent flattened isochrones},
  author={Binney, James},
  journal={Monthly Notices of the Royal Astronomical Society},
  volume={440},
  number={1},
  pages={787--798},
  year={2014},
  publisher={Oxford University Press}
}

@article{piffl2015bringing,
  title={Bringing the Galaxy's dark halo to life},
  author={Piffl, T and Penoyre, Z and Binney, J},
  journal={Monthly Notices of the Royal Astronomical Society},
  volume={451},
  number={1},
  pages={639--650},
  year={2015},
  publisher={Oxford University Press}
}

@article{bosma198121,
  title={21-cm line studies of spiral galaxies. I-Observations of the galaxies NGC 5033, 3198, 5055, 2841, and 7331},
  author={Bosma, Albert},
  journal={The Astronomical Journal},
  volume={86},
  pages={1791--1824},
  year={1981}
}

@article{bosma198121,
  title={21-cm line studies of spiral galaxies. II. The distribution and kinematics of neutral hydrogen in spiral galaxies of various morphological types.},
  author={Bosma, Astrori},
  journal={The Astronomical Journal},
  volume={86},
  pages={1825--1846},
  year={1981}
}

@article{kent1987dark,
  title={Dark matter in spiral galaxies. II-Galaxies with HI rotation curves},
  author={Kent, Stephen M},
  journal={The Astronomical Journal},
  volume={93},
  pages={816--832},
  year={1987}
}

@incollection{bosma1990frequency,
  title={The frequency of warped HI discs},
  author={Bosma, A and Athanassoula, E},
  booktitle={Dynamics and Interactions of Galaxies},
  pages={356--357},
  year={1990},
  publisher={Springer}
}

@article{sancisi2008cold,
  title={Cold gas accretion in galaxies},
  author={Sancisi, Renzo and Fraternali, Filippo and Oosterloo, Tom and Van Der Hulst, Thijs},
  journal={The Astronomy and Astrophysics Review},
  volume={15},
  number={3},
  pages={189--223},
  year={2008},
  publisher={Springer}
}

@article{marinacci2010mode,
  title={The mode of gas accretion on to star-forming galaxies},
  author={Marinacci, Federico and Binney, James and Fraternali, Filippo and Nipoti, Carlo and Ciotti, Luca and Londrillo, Pasquale},
  journal={Monthly Notices of the Royal Astronomical Society},
  volume={404},
  number={3},
  pages={1464--1474},
  year={2010},
  publisher={The Royal Astronomical Society}
}

@article{boomsma2005extra,
  title={Extra-planar HI in the starburst galaxy NGC 253},
  author={Boomsma, Rense and Oosterloo, TA and Fraternali, Filippo and van der Hulst, JM and Sancisi, Renzo},
  journal={Astronomy \& Astrophysics},
  volume={431},
  number={1},
  pages={65--72},
  year={2005},
  publisher={EDP Sciences}
}

@article{dettmar1992extraplanar,
  title={Extraplanar diffuse ionized gas and the disk-halo connection in spiral galaxies},
  author={Dettmar, Ralf-J{\"u}rgen},
  journal={Fundamentals of Cosmic Physics},
  volume={15},
  pages={143--208},
  year={1992}
}

@article{fraternali2008new,
  title={New evidence for halo gas accretion onto disk galaxies},
  author={Fraternali, Filippo},
  journal={Proceedings of the International Astronomical Union},
  volume={4},
  number={S254},
  pages={255--262},
  year={2008},
  publisher={Cambridge University Press}
}

@inproceedings{schoenmakers1999kinematic,
  title={Kinematic Lopsidedness in Spiral Galaxies},
  author={Schoenmakers, RHM and Swaters, RA},
  booktitle={Galaxy Dynamics-A Rutgers Symposium},
  volume={182},
  year={1999}
}

@article{matthews1999properties,
  title={Properties ofSuperthin'Galaxies},
  author={Matthews, LD and Van Driel, W and Gallagher, JS},
  journal={arXiv preprint astro-ph/9911022},
  year={1999}
}

@article{alam2015eleventh,
  title={The eleventh and twelfth data releases of the Sloan Digital Sky Survey: final data from SDSS-III},
  author={Alam, Shadab and Albareti, Franco D and Prieto, Carlos Allende and Anders, Friedrich and Anderson, Scott F and Anderton, Timothy and Andrews, Brett H and Armengaud, Eric and Aubourg, {\'E}ric and Bailey, Stephen and others},
  journal={The Astrophysical Journal Supplement Series},
  volume={219},
  number={1},
  pages={12},
  year={2015},
  publisher={IOP Publishing}
}

@article{erwin2015imfit,
  title={IMFIT: a fast, flexible new program for astronomical image fitting},
  author={Erwin, Peter},
  journal={The Astrophysical Journal},
  volume={799},
  number={2},
  pages={226},
  year={2015},
  publisher={IOP Publishing}
}

@article{gentile2008mond,
  title={MOND and the universal rotation curve: similar phenomenologies},
  author={Gentile, Gianfranco},
  journal={The Astrophysical Journal},
  volume={684},
  number={2},
  pages={1018},
  year={2008},
  publisher={IOP Publishing}
}

@article{gentile2011things,
  title={THINGS about MoND},
  author={Gentile, Gianfranco and Famaey, B and de Blok, WJG},
  journal={Astronomy \& Astrophysics},
  volume={527},
  pages={A76},
  year={2011},
  publisher={EDP Sciences}
}

@article{elmegreen2020observing,
  title={Observing the Earliest Stages of Star Formation in Galaxies: 8 $\mu$m Cores in Three Edge-on Disks},
  author={Elmegreen, Bruce G and Elmegreen, Debra Meloy},
  journal={The Astrophysical Journal},
  volume={895},
  number={1},
  pages={71},
  year={2020},
  publisher={IOP Publishing}
}

@article{krumm1979neutral,
  title={Neutral hydrogen observations of 14 nearly edge-on spiral galaxies},
  author={Krumm, N and Salpeter, EE},
  journal={The Astronomical Journal},
  volume={84},
  pages={1138--1148},
  year={1979}
}

@article{bottema1995prodigious,
  title={The prodigious warp of NGC 4013. II. Detailed observations of the neutral hydrogen gas.},
  author={Bottema, R},
  journal={Astronomy and Astrophysics},
  volume={295},
  pages={605},
  year={1995}
}

@inproceedings{fraternali2005extra,
  title={The extra-planar neutral gas in the edge-on spiral galaxy NGC 891},
  author={Fraternali, F and Oosterloo, TA and Sancisi, R and Swaters, R},
  booktitle={Extra-Planar Gas},
  volume={331},
  pages={239},
  year={2005}
}

@article{lehner2011reservoir,
  title={A reservoir of ionized gas in the galactic halo to sustain star formation in the Milky Way},
  author={Lehner, Nicolas and Howk, J Christopher},
  journal={Science},
  volume={334},
  number={6058},
  pages={955--958},
  year={2011},
  publisher={American Association for the Advancement of Science}
}

@article{aditya2022h,
  title={H i 21 cm observation and mass models of the extremely thin galaxy FGC 1440},
  author={Aditya, K and Kamphuis, Peter and Banerjee, Arunima and Borisov, Sviatoslav and Mosenkov, Aleksandr and Antipova, Aleksandra and Makarov, Dmitry},
  journal={Monthly Notices of the Royal Astronomical Society},
  volume={509},
  number={3},
  pages={4071--4093},
  year={2022},
  publisher={Oxford University Press}
}

@article{boily2003thickness,
  title={The Thickness of Stellar Disks of Edge-on Galaxies and Their Truncation Radii},
  author={Boily, C and Patsis, P and Portegies, S and Spurzem, R and Theis, C and Zasov, AV and Bizyaev, DV},
  journal={European Astronomical Society Publications Series},
  volume={10},
  pages={121--121},
  year={2003},
  publisher={EDP Sciences}
}

@article{rohlfs1977lectures,
  title={Lectures on density wave theory},
  author={Rohlfs, Kristen},
  year={1977}
}

@article{binney2011models,
  title={Models of our Galaxy--II},
  author={Binney, James and McMillan, Paul},
  journal={Monthly Notices of the Royal Astronomical Society},
  volume={413},
  number={3},
  pages={1889--1898},
  year={2011},
  publisher={Blackwell Publishing Ltd Oxford, UK}
}

@article{wang1994gravitational,
  title={Gravitational instability and disk star formation},
  author={Wang, Boqi and Silk, Joseph},
  journal={The Astrophysical Journal},
  volume={427},
  pages={759--769},
  year={1994}
}

@article{westfall2014diskmass,
  title={The DiskMass survey. VIII. On the relationship between disk stability and star formation},
  author={Westfall, Kyle B and Andersen, David R and Bershady, Matthew A and Martinsson, Thomas PK and Swaters, Robert A and Verheijen, Marc AW},
  journal={The Astrophysical Journal},
  volume={785},
  number={1},
  pages={43},
  year={2014},
  publisher={IOP Publishing}
}
@article{matthews2000extraordinary,
  title={The Extraordinary “Superthin” Spiral Galaxy UGC 7321. II. The Vertical Disk Structure},
  author={Matthews, LD},
  journal={The Astronomical Journal},
  volume={120},
  number={4},
  pages={1764},
  year={2000},
  publisher={IOP Publishing}
}

@article{giovanelli1997spectroscopy,
  title={Spectroscopy of edge-on spirals},
  author={Giovanelli, Riccardo and Avera, Eric and Karachentsev, Igor},
  journal={arXiv preprint astro-ph/9704189},
  year={1997}
}

@article{karachentsev1993flat,
  title={Flat galaxy catalogue},
  author={Karachentsev, ID and Karachentseva, VE and Parnovsky, SL},
  journal={Astronomische Nachrichten},
  volume={314},
  number={3},
  pages={97--222},
  year={1993},
  publisher={Wiley Online Library}
}

@article{banerjee2007origin,
  title={The origin of steep vertical stellar distribution in the Galactic disk},
  author={Banerjee, Arunima and Jog, Chanda J},
  journal={The Astrophysical Journal},
  volume={662},
  number={1},
  pages={335},
  year={2007},
  publisher={IOP Publishing}
}

@article{roberts1994physical,
  title={Physical parameters along the Hubble sequence},
  author={Roberts, Morton S and Haynes, Martha P},
  journal={Annual Review of Astronomy and Astrophysics},
  volume={32},
  number={1},
  pages={115--152},
  year={1994},
  publisher={Annual Reviews 4139 El Camino Way, PO Box 10139, Palo Alto, CA 94303-0139, USA}
}

@article{querejeta2015spitzer,
  title={The Spitzer Survey of Stellar Structure in Galaxies (S4G): precise stellar mass distributions from automated dust correction at 3.6 $\mu$m},
  author={Querejeta, Miguel and Meidt, Sharon E and Schinnerer, Eva and Cisternas, Mauricio and Mu{\~n}oz-Mateos, Juan Carlos and Sheth, Kartik and Knapen, Johan and Van De Ven, Glenn and Norris, Mark A and Peletier, Reynier and others},
  journal={The Astrophysical Journal Supplement Series},
  volume={219},
  number={1},
  pages={5},
  year={2015},
  publisher={IOP Publishing}
}

@article{braine2000deep,
  title={Deep search for CO emission in the Low Surface Brightness galaxy Malin 1},
  author={Braine, J and Herpin, F and Radford, SJE},
  journal={Astronomy and Astrophysics},
  volume={358},
  pages={494--498},
  year={2000}
}

@article{fernandez2016ggmcmc,
  title={ggmcmc: Analysis of MCMC samples and Bayesian inference},
  author={Fern{\'a}ndez-i-Mar{\i}n, Xavier},
  journal={Journal of Statistical Software},
  volume={70},
  number={9},
  pages={1--20},
  year={2016}
}

@article{hartig2017bayesiantools,
  title={Bayesiantools: General-purpose MCMC and SMC samplers and tools for Bayesian statistics},
  author={Hartig, F and Minunno, F and Paul, S and Cameron, D and Ott, T},
  journal={R package version. R package version 0.1},
  volume={3},
  year={2017}
}

@article{wickham2011ggplot2,
  title={ggplot2},
  author={Wickham, Hadley},
  journal={Wiley Interdisciplinary Reviews: Computational Statistics},
  volume={3},
  number={2},
  pages={180--185},
  year={2011},
  publisher={Wiley Online Library}
}

@misc{meschiari2015latex2exp,
  title={latex2exp: Use LaTeX Expressions in Plots, r package version 0.4. 0},
  author={Meschiari, S},
  year={2015}
}

@article{hunter2007matplotlib,
  title={Matplotlib: A 2D graphics environment},
  author={Hunter, John D},
  journal={Computing in science \& engineering},
  volume={9},
  number={3},
  pages={90--95},
  year={2007},
  publisher={IEEE Computer Society}
}

@article{vaughan2016false,
  title={False periodicities in quasar time-domain surveys},
  author={Vaughan, S and Uttley, P and Markowitz, AG and Huppenkothen, D and Middleton, MJ and Alston, WN and Scargle, JD and Farr, WM},
  journal={Monthly Notices of the Royal Astronomical Society},
  volume={461},
  number={3},
  pages={3145--3152},
  year={2016},
  publisher={Oxford University Press}
}

@article{nilson1973uppsala,
  title={Uppsala General Catalogue of Galaxies, 1973, Acta Universitatis Upsalienis, Nova Regiae Societatis Upsaliensis, Series v: a Vol.},
  author={Nilson, P},
  journal={UGC},
  pages={0},
  year={1973}
}

@article{karachentsev2001list,
  title={A list of peculiar velocities of RFGC galaxies},
  author={Karachentsev, ID and Karachentseva, VE and Kudrya, Yu N and Makarov, DI and Parnovsky, SL},
  journal={arXiv preprint astro-ph/0107058},
  year={2001}
}

@article{binggeli1985studies,
  title={Studies of the Virgo Cluster. II-A catalog of 2096 galaxies in the Virgo Cluster area.},
  author={Binggeli, B and Sandage, A and Tammann, GA},
  journal={The Astronomical Journal},
  volume={90},
  pages={1681--1759},
  year={1985}
}

@article{dalcanton2004formation,
  title={The formation of dust lanes: Implications for galaxy evolution},
  author={Dalcanton, Julianne J and Yoachim, Peter and Bernstein, Rebecca A},
  journal={The Astrophysical Journal},
  volume={608},
  number={1},
  pages={189},
  year={2004},
  publisher={IOP Publishing}
}

@article{yoachim2006structural,
  title={Structural parameters of thin and thick disks in edge-on disk galaxies},
  author={Yoachim, Peter and Dalcanton, Julianne J},
  journal={The Astronomical Journal},
  volume={131},
  number={1},
  pages={226},
  year={2006},
  publisher={IOP Publishing}
}

@inproceedings{jaeger2008common,
  title={The common astronomy software application (casa)},
  author={Jaeger, S},
  booktitle={Astronomical Data Analysis Software and Systems XVII},
  volume={394},
  pages={623},
  year={2008}
}

@techreport{fukugita1996sloan,
  title={The Sloan digital sky survey photometric system},
  author={Fukugita, M and Shimasaku, K and Ichikawa, T and Gunn, JE and others},
  year={1996},
  institution={SCAN-9601313}
}

@ARTICLE{1998A&A...336..878F,
       author = {{Fuchs}, B. and {M{\"o}llenhoff}, C. and {Heidt}, H.},
        title = "{Decomposition of the rotation curves of distant field galaxies}",
      journal = {Astronomy \& Astrophysics},
     keywords = {Astrophysics},
         year = 1998,
        month = aug,
       volume = {336},
        pages = {878},
archivePrefix = {arXiv},
       eprint = {astro-ph/9806117},
 primaryClass = {astro-ph},
       adsurl = {https://ui.adsabs.harvard.edu/abs/1998A&A...336..878F},
      adsnote = {Provided by the SAO/NASA Astrophysics Data System}
}

@article{portinari2004mass,
  title={On the mass-to-light ratio and the initial mass function in disc galaxies},
  author={Portinari, Laura and Sommer-Larsen, J and Tantalo, R},
  journal={Monthly Notices of the Royal Astronomical Society},
  volume={347},
  number={3},
  pages={691--719},
  year={2004},
  publisher={Blackwell Science Ltd Oxford, UK}
}

@article{palunas2000maximum,
  title={Maximum disk mass models for spiral galaxies},
  author={Palunas, Povilas and Williams, TB},
  journal={The Astronomical Journal},
  volume={120},
  number={6},
  pages={2884},
  year={2000},
  publisher={IOP Publishing}
}

@article{mcgaugh2014color,
  title={Color-mass-to-light-ratio relations for disk galaxies},
  author={McGaugh, Stacy S and Schombert, James M},
  journal={The Astronomical Journal},
  volume={148},
  number={5},
  pages={77},
  year={2014},
  publisher={IOP Publishing}
}

@article{de2002high,
  title={High-resolution rotation curves of low surface brightness galaxies},
  author={De Blok, WJG and Bosma, A},
  journal={Astronomy \& Astrophysics},
  volume={385},
  number={3},
  pages={816--846},
  year={2002},
  publisher={EDP Sciences}
}

@article{monnier2003search,
  title={A search for Low Surface Brightness galaxies in the near-infrared. III. Nan{\c{c}}ay HI line observations},
  author={Monnier Ragaigne, D and van Driel, Wim and Schneider, SE and Balkowski, C and Jarrett, TH},
  journal={Astronomy and Astrophysics},
  volume={408},
  number={2},
  pages={465--477},
  year={2003},
  publisher={EDP Sciences}
}

@article{dalcanton1996chain,
  title={“Chain” Galaxies are Edge-On Low Surface Brightness Galaxies},
  author={Dalcanton, Julianne J and Shectman, Stephen A},
  journal={The Astrophysical Journal Letters},
  volume={465},
  number={1},
  pages={L9},
  year={1996},
  publisher={IOP Publishing}
}

@article{mcgaugh2000baryonic,
  title={The baryonic tully-fisher relation},
  author={McGaugh, Stacy S and Schombert, Jim M and Bothun, Greg D and De Blok, WJG},
  journal={The Astrophysical Journal Letters},
  volume={533},
  number={2},
  pages={L99},
  year={2000},
  publisher={IOP Publishing}
}

@article{mcgaugh2012baryonic,
  title={The baryonic Tully-Fisher relation of gas-rich galaxies as a test of $\Lambda$CDM and MOND},
  author={McGaugh, Stacy S},
  journal={The Astronomical Journal},
  volume={143},
  number={2},
  pages={40},
  year={2012},
  publisher={IOP Publishing}
}

@article{mcgaugh2005baryonic,
  title={The Baryonic Tully-Fisher relation of galaxies with extended rotation curves and the stellar mass of rotating galaxies},
  author={McGaugh, Stacy S},
  journal={The Astrophysical Journal},
  volume={632},
  number={2},
  pages={859},
  year={2005},
  publisher={IOP Publishing}
}

@misc{ss2016radial,
  title={The radial acceleration relation in rotationally supported galaxies},
  author={SS, Lelli F McGaugh and Schombert, JM},
  year={2016}
}
@article{lelli2017one,
  title={One law to rule them all: the radial acceleration relation of galaxies},
  author={Lelli, Federico and McGaugh, Stacy S and Schombert, James M and Pawlowski, Marcel S},
  journal={The Astrophysical Journal},
  volume={836},
  number={2},
  pages={152},
  year={2017},
  publisher={IOP Publishing}
}

@misc{wolfe1967modern,
  title={Modern Astrophysics, ed. M. Hack},
  author={Wolfe, RH and Horak, HG and Storer, NW},
  year={1967},
  publisher={New York: Gordon \& Breach}
}

@article{keller2017lambdacdm,
  title={$\Lambda$CDM is consistent with SPARC radial acceleration relation},
  author={Keller, BW and Wadsley, JW},
  journal={The Astrophysical Journal Letters},
  volume={835},
  number={1},
  pages={L17},
  year={2017},
  publisher={IOP Publishing}
}

@article{li2018fitting,
  title={Fitting the radial acceleration relation to individual SPARC galaxies},
  author={Li, Pengfei and Lelli, Federico and McGaugh, Stacy and Schombert, James},
  journal={Astronomy \& Astrophysics},
  volume={615},
  pages={A3},
  year={2018},
  publisher={EDP Sciences}
}

@article{ghari2019dark,
  title={Dark matter--baryon scaling relations from Einasto halo fits to SPARC galaxy rotation curves},
  author={Ghari, Amir and Famaey, Benoit and Laporte, Chervin and Haghi, Hosein},
  journal={Astronomy \& Astrophysics},
  volume={623},
  pages={A123},
  year={2019},
  publisher={EDP Sciences}
}

@article{naik2019constraints,
  title={Constraints on chameleon f (R)-gravity from galaxy rotation curves of the SPARC sample},
  author={Naik, Aneesh P and Puchwein, Ewald and Davis, Anne-Christine and Sijacki, Debora and Desmond, Harry},
  journal={Monthly Notices of the Royal Astronomical Society},
  volume={489},
  number={1},
  pages={771--787},
  year={2019},
  publisher={Oxford University Press}
}

@article{li2019constant,
  title={A constant characteristic volume density of dark matter haloes from SPARC rotation curve fits},
  author={Li, Pengfei and Lelli, Federico and McGaugh, Stacy S and Starkman, Nathaniel and Schombert, James M},
  journal={Monthly Notices of the Royal Astronomical Society},
  volume={482},
  number={4},
  pages={5106--5124},
  year={2019},
  publisher={Oxford University Press}
}

@article{chan2018testing,
  title={Testing the Cubic Galileon Gravity model by the Milky Way rotation curve and SPARC data},
  author={Chan, Man Ho and Hui, Hon Ka},
  journal={The Astrophysical Journal},
  volume={856},
  number={2},
  pages={177},
  year={2018},
  publisher={IOP Publishing}
}

@ARTICLE{1996AJ....112..481O,
       author = {{Olling}, Rob P.},
        title = "{The Highly Flattened Dark Matter Halo of NGC 4244}",
      journal = {The Astronomical Journal},
     keywords = {GALAXIES: INDIVIDUAL: NGC 4244, GALAXIES: SPIRAL, GALAXIES: ISM, Astrophysics},
         year = 1996,
        month = aug,
       volume = {112},
        pages = {481},
          doi = {10.1086/118029},
archivePrefix = {arXiv},
       eprint = {astro-ph/9605111},
 primaryClass = {astro-ph},
       adsurl = {https://ui.adsabs.harvard.edu/abs/1996AJ....112..481O},
      adsnote = {Provided by the SAO/NASA Astrophysics Data System}
}

@article{reynolds2020h,
  title={H i asymmetries in LVHIS, VIVA, and HALOGAS galaxies},
  author={Reynolds, TN and Westmeier, T and Staveley-Smith, L and Chauhan, G and Lagos, CDP},
  journal={Monthly Notices of the Royal Astronomical Society},
  volume={493},
  number={4},
  pages={5089--5106},
  year={2020},
  publisher={Oxford University Press}
}

@article{banerjee2013slow,
  title={A slow bar in the dwarf irregular galaxy NGC 3741},
  author={Banerjee, Arunima and Patra, Narendra Nath and Chengalur, Jayaram N and Begum, Ayesha},
  journal={Monthly Notices of the Royal Astronomical Society},
  volume={434},
  number={2},
  pages={1257--1263},
  year={2013},
  publisher={The Royal Astronomical Society}
}

@article{patra2019detection,
  title={Detection of a slow H i bar in the dwarf irregular galaxy DDO 168},
  author={Patra, Narendra Nath and Jog, Chanda J},
  journal={Monthly Notices of the Royal Astronomical Society},
  volume={488},
  number={4},
  pages={4942--4951},
  year={2019},
  publisher={Oxford University Press}
}

@article{jozsa2007kinematic,
  title={Kinematic modelling of disk galaxies-II. A case-study of symmetrically warped galaxy disks},
  author={J{\'o}zsa, GIG},
  journal={Astronomy \& Astrophysics},
  volume={468},
  number={3},
  pages={903--917},
  year={2007},
  publisher={EDP Sciences}
}

@article{bolatto2017edge,
  title={The EDGE-CALIFA Survey: Interferometric Observations of 126 Galaxies with CARMA},
  author={Bolatto, Alberto D and Wong, Tony and Utomo, Dyas and Blitz, Leo and Vogel, Stuart N and S{\'a}nchez, Sebasti{\'a}n F and Barrera-Ballesteros, Jorge and Cao, Yixian and Colombo, Dario and Dannerbauer, Helmut and others},
  journal={The Astrophysical Journal},
  volume={846},
  number={2},
  pages={159},
  year={2017},
  publisher={IOP Publishing}
}

@article{phookun1993ngc,
  title={NGC 4254: a spiral galaxy with an m= 1 mode and infalling gas},
  author={Phookun, Bikram and Vogel, Stuart N and Mundy, Lee G},
  journal={The Astrophysical Journal},
  volume={418},
  pages={113},
  year={1993}
}

@article{gentile2011things,
  title={THINGS about MoND},
  author={Gentile, Gianfranco and Famaey, B and de Blok, WJG},
  journal={Astronomy \& Astrophysics},
  volume={527},
  pages={A76},
  year={2011},
  publisher={EDP Sciences}
}

@ARTICLE{2000BSAO...50....5K,
       author = {{Karachentsev}, I.~D. and {Karachentseva}, V.~E. and {Kudrya}, Yu. N. and {Makarov}, D.~I. and {Parnovsky}, S.~L.},
        title = "{A list of peculiar velocities of RFGC galaxies}",
      journal = {Bulletin of the Special Astrophysics Observatory},
     keywords = {Galaxies: Observations, Galaxies: Kinematics and Dynamics, RFGC Catalogue, Astrophysics},
         year = 2000,
        month = jan,
       volume = {50},
        pages = {5-38},
archivePrefix = {arXiv},
       eprint = {astro-ph/0107058},
 primaryClass = {astro-ph},
       adsurl = {https://ui.adsabs.harvard.edu/abs/2000BSAO...50....5K},
      adsnote = {Provided by the SAO/NASA Astrophysics Data System}
}

@article{narayanan2021star,
  title={Star Formation in Superthin Galaxies},
  author={Narayanan, Ganesh and Banerjee, Arunima},
  journal={arXiv preprint arXiv:2104.04216},
  year={2021}
}

@article{fall1980formation,
  title={Formation and rotation of disc galaxies with haloes},
  author={Fall, S Michael and Efstathiou, George},
  journal={Monthly Notices of the Royal Astronomical Society},
  volume={193},
  number={2},
  pages={189--206},
  year={1980},
  publisher={Oxford University Press Oxford, UK}
}

@article{romanowsky2012angular,
  title={Angular momentum and galaxy formation revisited},
  author={Romanowsky, Aaron J and Fall, S Michael},
  journal={The Astrophysical Journal Supplement Series},
  volume={203},
  number={2},
  pages={17},
  year={2012},
  publisher={IOP Publishing}
}

@article{posti2018angular,
  title={The angular momentum-mass relation: a fundamental law from dwarf irregulars to massive spirals},
  author={Posti, Lorenzo and Fraternali, Filippo and Di Teodoro, Enrico M and Pezzulli, Gabriele},
  journal={Astronomy \& Astrophysics},
  volume={612},
  pages={L6},
  year={2018},
  publisher={EDP Sciences}
}

@article{posti2018galaxy,
  title={Galaxy spin as a formation probe: the stellar-to-halo specific angular momentum relation},
  author={Posti, Lorenzo and Pezzulli, Gabriele and Fraternali, Filippo and Di Teodoro, Enrico M},
  journal={Monthly Notices of the Royal Astronomical Society},
  volume={475},
  number={1},
  pages={232--243},
  year={2018},
  publisher={Oxford University Press}
}

@article{kurapati2018angular,
  title={Angular momentum of dwarf galaxies},
  author={Kurapati, Sushma and Chengalur, Jayaram N and Pustilnik, Simon and Kamphuis, Peter},
  journal={Monthly Notices of the Royal Astronomical Society},
  volume={479},
  number={1},
  pages={228--239},
  year={2018},
  publisher={Oxford University Press}}

@article{dehnen1998local,
  title={Local stellar kinematics from Hipparcos data},
  author={Dehnen, Walter and Binney, James J},
  journal={Monthly Notices of the Royal Astronomical Society},
  volume={298},
  number={2},
  pages={387--394},
  year={1998},
  publisher={Blackwell Science Ltd Oxford, UK and Cambridge, USA}
}

@article{butler2016angular,
  title={Angular Momentum of Dwarf Galaxies},
  author={Butler, Kirsty M and Obreschkow, Danail and Oh, Se-Heon},
  journal={The Astrophysical Journal Letters},
  volume={834},
  number={1},
  pages={L4},
  year={2016},
  publisher={IOP Publishing}
}

@article{kurapati2018angular,
  title={Angular momentum of dwarf galaxies},
  author={Kurapati, Sushma and Chengalur, Jayaram N and Pustilnik, Simon and Kamphuis, Peter},
  journal={Monthly Notices of the Royal Astronomical Society},
  volume={479},
  number={1},
  pages={228--239},
  year={2018},
  publisher={Oxford University Press}
}

@article{10.1093/mnras/stab3143,
    author = {Aditya, K and Kamphuis, Peter and Banerjee, Arunima and Borisov, Sviatoslav and Mosenkov, Aleksandr and Antipova, Aleksandra and Makarov, Dmitry},
    title = "{H i 21 cm observation and mass models of the extremely thin galaxy FGC 1440}",
    journal = {Monthly Notices of the Royal Astronomical Society},
    year = {2021},
    month = {11},
    abstract = "{We present observations and models of the kinematics and distribution of neutral hydrogen (H i) in the superthin galaxy FGC 1440 with an optical axial ratio a/b = 20.4. Using the Giant Meterwave Radio telescope (GMRT), we imaged the galaxy with a spectral resolution of 1.7 km s−1 and a spatial resolution of 15\\$\\{^\\{\\prime \\prime \\}\_\\{.\\}\\}\\$9 × 13\\$\\{^\\{\\prime \\prime \\}\_\\{.\\}\\}\\$5. We find that FGC 1440 has an asymptotic rotational velocity of 141.8 km s−1. The structure of the H i disc in FGC 1440 is that of a typical thin disc warped along the line of sight, but we can\\$\\{^\\{\\prime \\prime \\}\_\\{.\\}\\}\\$ not rule out the presence of a central thick H i disc. We find that the dark matter halo in FGC 1440 could be modeled by a pseudo- isothermal (PIS) profile with \\$\\rm R\_\\{c\\}/ R\_\\{d\\} \\&lt;2\\$, where Rc is the core radius of the PIS halo and Rd the exponential stellar disc scale length. We note that in spite of the unusually large axial ratio of FGC 1440, the ratio of the rotational velocity to stellar vertical velocity dispersion, \\$\\frac\\{V\_\\{Rot\\}\\}\\{\\sigma \_\\{z\\}\\} \\sim 5 - 8\\$, which is comparable to other superthins. Interestingly, unlike previously studied superthin galaxies which are outliers in the log10(j*) − log10(M*) relation for ordinary bulgeless disc galaxies, FGC 1440 is found to comply with the same. The values of j for the stars, gas and the baryons in FGC 1440 are consistent with those of normal spiral galaxies with similar mass.}",
    issn = {0035-8711},
    doi = {10.1093/mnras/stab3143},
    url = {https://doi.org/10.1093/mnras/stab3143},
    note = {stab3143},
    eprint = {https://academic.oup.com/mnras/advance-article-pdf/doi/10.1093/mnras/stab3143/41035100/stab3143.pdf},
}

@article{haslbauer2022high,
  title={The High Fraction of Thin Disk Galaxies Continues to Challenge $\Lambda$CDM Cosmology},
  author={Haslbauer, Moritz and Banik, Indranil and Kroupa, Pavel and Wittenburg, Nils and Javanmardi, Behnam},
  journal={The Astrophysical Journal},
  volume={925},
  number={2},
  pages={183},
  year={2022},
  publisher={IOP Publishing}
}

@article{romanowsky2012angular,
  title={Angular momentum and galaxy formation revisited},
  author={Romanowsky, Aaron J and Fall, S Michael},
  journal={The Astrophysical Journal Supplement Series},
  volume={203},
  number={2},
  pages={17},
  year={2012},
  publisher={IOP Publishing}
}

@article{bullock2001profiles,
  title={Profiles of dark haloes: evolution, scatter and environment},
  author={Bullock, James S and Kolatt, Tsafrir S and Sigad, Yair and Somerville, Rachel S and Kravtsov, Andrey V and Klypin, Anatoly A and Primack, Joel R and Dekel, Avishai},
  journal={Monthly Notices of the Royal Astronomical Society},
  volume={321},
  number={3},
  pages={559--575},
  year={2001},
  publisher={Blackwell Science Ltd Oxford, UK}
}

@article{rovskar2013effects,
  title={The effects of radial migration on the vertical structure of Galactic discs},
  author={Ro{\v{s}}kar, Rok and Debattista, Victor P and Loebman, Sarah R},
  journal={Monthly Notices of the Royal Astronomical Society},
  volume={433},
  number={2},
  pages={976--985},
  year={2013},
  publisher={Oxford University Press}
}

@book{kamphuis2008structure,
  title={The structure and kinematics of halos in disk galaxies},
  author={Kamphuis, Peter},
  year={2008},
  publisher={University Library Groningen][Host]}
}

@article{voigtlander2013kinematics,
  title={The kinematics of the diffuse ionized gas in NGC 4666},
  author={Voigtl{\"a}nder, Pierre and Kamphuis, Peter and Marcelin, Michel and Bomans, Dominik J and Dettmar, R-J},
  journal={Astronomy \& Astrophysics},
  volume={554},
  pages={A133},
  year={2013},
  publisher={EDP Sciences}
}

@article{barrera2021edge,
  title={EDGE-CALIFA survey: Self-regulation of Star formation at kpc scales},
  author={Barrera-Ballesteros, JK and S{\'a}nchez, SF and Heckman, T and Wong, T and Bolatto, A and Ostriker, E and Rosolowsky, E and Carigi, L and Vogel, S and Levy, RC and others},
  journal={arXiv preprint arXiv:2101.04683},
  year={2021}
}

@article{rubin1980rotational,
  title={Rotational properties of 21 SC galaxies with a large range of luminosities and radii, from NGC 4605/R= 4kpc/to UGC 2885/R= 122 kpc},
  author={Rubin, Vera C and Ford Jr, W Kent and Thonnard, Norbert},
  journal={The Astrophysical Journal},
  volume={238},
  pages={471--487},
  year={1980}
}
@article{rubin1985rotation,
  title={Rotation velocities of 16 SA galaxies and a comparison of Sa, Sb, and SC rotation properties},
  author={Rubin, Vera C and Burstein, David and Ford Jr, W Kent and Thonnard, Norbert},
  journal={The Astrophysical Journal},
  volume={289},
  pages={81--98},
  year={1985}
}

@article{mcgaugh2014color,
  title={Color-mass-to-light-ratio relations for disk galaxies},
  author={McGaugh, Stacy S and Schombert, James M},
  journal={The Astronomical Journal},
  volume={148},
  number={5},
  pages={77},
  year={2014},
  publisher={IOP Publishing}
}

@ARTICLE{2003ApJS..149..289B,
       author = {{Bell}, Eric F. and {McIntosh}, Daniel H. and {Katz}, Neal and {Weinberg}, Martin D.},
        title = "{The Optical and Near-Infrared Properties of Galaxies. I. Luminosity and Stellar Mass Functions}",
      journal = {\apjs},
     keywords = {Galaxies: Evolution, Galaxies: General, Galaxies: Luminosity Function, Mass Function, Galaxies: Stellar Content, Astrophysics},
         year = 2003,
        month = dec,
       volume = {149},
       number = {2},
        pages = {289-312},
          doi = {10.1086/378847},
archivePrefix = {arXiv},
       eprint = {astro-ph/0302543},
 primaryClass = {astro-ph},
       adsurl = {https://ui.adsabs.harvard.edu/abs/2003ApJS..149..289B},
      adsnote = {Provided by the SAO/NASA Astrophysics Data System}
}

@article{khoperskov2017disk,
  title={Disk heating and bending instability in galaxies with counterrotation},
  author={Khoperskov, Sergey and Bertin, Giuseppe},
  journal={Astronomy \& Astrophysics},
  volume={597},
  pages={A103},
  year={2017},
  publisher={EDP Sciences}
}

@ARTICLE{2021ApJ...914..104B,
       author = {{Bizyaev}, Dmitry and {Makarov}, D.~I. and {Reshetnikov}, V.~P. and {Mosenkov}, A.~V. and {Kautsch}, S.~J. and {Antipova}, A.~V.},
        title = "{Spectral Observations of Superthin Galaxies}",
      journal = {\apj},
     keywords = {Galactic and extragalactic astronomy, Disk galaxies, Galaxy dark matter halos, 563, 391, 1880, Astrophysics - Astrophysics of Galaxies},
         year = 2021,
        month = jun,
       volume = {914},
       number = {2},
          eid = {104},
        pages = {104},
          doi = {10.3847/1538-4357/abfb03},
archivePrefix = {arXiv},
       eprint = {2105.11855},
 primaryClass = {astro-ph.GA},
       adsurl = {https://ui.adsabs.harvard.edu/abs/2021ApJ...914..104B},
      adsnote = {Provided by the SAO/NASA Astrophysics Data System}
}

@article{matthews2000extraordinary,
  title={The Extraordinary “Superthin” Spiral Galaxy UGC 7321. II. The Vertical Disk Structure},
  author={Matthews, LD},
  journal={The Astronomical Journal},
  volume={120},
  number={4},
  pages={1764},
  year={2000},
  publisher={IOP Publishing}
}

@article{matthews2001modeling,
  title={Modeling the interstellar medium of low surface brightness galaxies: Constraining internal extinction, disk color gradients, and intrinsic rotation curve shapes},
  author={Matthews, LD and Wood, Kenneth},
  journal={The Astrophysical Journal},
  volume={548},
  number={1},
  pages={150},
  year={2001},
  publisher={IOP Publishing}
}

@article{jc2010dark,
  title={The dark matter halo shape of edge-on disk galaxies-IV. UGC 7321},
  author={JC O, Brien},
  journal={Astronomy \& Astrophysics},
  volume={515},
  number={1},
  pages={A63},
  year={2010},
  publisher={EDP Sciences}
}

@article{zasov2002relationship,
  title={Relationship between the thickness of stellar disks and the relative mass of a dark galactic halo},
  author={Zasov, AV and Bizyaev, DV and Makarov, DI and Tyurina, NV},
  journal={Astronomy Letters},
  volume={28},
  number={8},
  pages={527--535},
  year={2002},
  publisher={Springer}
}

@article{de1996structure,
  title={Structure analysis of edge-on spiral galaxies},
  author={De Grijs, R and Van der Kruit, PC},
  journal={Astronomy and Astrophysics Supplement Series},
  volume={117},
  number={1},
  pages={429--445},
  year={1996},
  publisher={EDP Sciences}
}

@article{sarkar2020vertical,
  title={Vertical stellar density distribution in a non-isothermal galactic disc},
  author={Sarkar, Suchira and Jog, Chanda J},
  journal={Monthly Notices of the Royal Astronomical Society},
  volume={499},
  number={2},
  pages={2523--2533},
  year={2020},
  publisher={Oxford University Press}
}

@article{sarkar2021flaring,
  title={Flaring stellar disk in the low surface brightness galaxy UGC 7321 (Corrigendum)},
  author={Sarkar, S and Jog, CJ},
  journal={Astronomy \& Astrophysics},
  volume={652},
  pages={C1},
  year={2021},
  publisher={EDP Sciences}
}

@article{obreschkow2014fundamental,
  title={FUNDAMENTAL MASS--SPIN--MORPHOLOGY RELATION OF SPIRAL GALAXIES},
  author={Obreschkow, Danail and Glazebrook, Karl},
  journal={The Astrophysical Journal},
  volume={784},
  number={1},
  pages={26},
  year={2014},
  publisher={IOP Publishing}
}

@article{hogg2018data,
  title={Data analysis recipes: Using markov chain monte carlo},
  author={Hogg, David W and Foreman-Mackey, Daniel},
  journal={The Astrophysical Journal Supplement Series},
  volume={236},
  number={1},
  pages={11},
  year={2018},
  publisher={IOP Publishing}
}

@book{draine2010physics,
  title={Physics of the interstellar and intergalactic medium},
  author={Draine, Bruce T},
  volume={19},
  year={2010},
  publisher={Princeton University Press}
}

@book{thompson2017interferometry,
  title={Interferometry and synthesis in radio astronomy},
  author={Thompson, A Richard and Moran, James M and Swenson, George W},
  year={2017},
  publisher={Springer Nature}
}

@misc{chengalur2007low,
  title={Low frequency radio astronomy 3rd edition},
  author={Chengalur, Jayaram N and Gupta, Yashwant and Dwarkanath, KS},
  year={2007},
  publisher={NCRA-TIFR}
}

@misc{lal2013gmrt,
  title={GMRT Observer’s Manual},
  author={Lal, Dharam Vir},
  year={2013},
  publisher={National Centre for Radio Astrophysics, Ganeshkhind, Pune, India}
}

@article{kennicutt2003sings,
  title={SINGS: The SIRTF nearby galaxies survey},
  author={Kennicutt, Robert C and Armus, Lee and Bendo, George and Calzetti, Daniela and Dale, Daniel A and Draine, Bruce T and Engelbracht, Charles W and Gordon, Karl D and Grauer, Albert D and Helou, George and others},
  journal={Publications of the Astronomical Society of the Pacific},
  volume={115},
  number={810},
  pages={928},
  year={2003},
  publisher={IOP Publishing}
}

@article{narayanan2022superthin,
  title={Are superthin galaxies low-surface-brightness galaxies seen edge-on? The star formation probe},
  author={Narayanan, Ganesh and Banerjee, Arunima},
  journal={Monthly Notices of the Royal Astronomical Society},
  volume={514},
  number={4},
  pages={5126--5140},
  year={2022},
  publisher={Oxford University Press}
}

@article{kumar2022excitation,
  title={Excitation of vertical breathing motion in disc galaxies by tidally-induced spirals in fly-by interactions},
  author={Kumar, Ankit and Ghosh, Soumavo and Kataria, Sandeep Kumar and Das, Mousumi and Debattista, Victor P},
  journal={Monthly Notices of the Royal Astronomical Society},
  volume={516},
  number={1},
  pages={1114--1126},
  year={2022},
  publisher={Oxford University Press}
}

@article{reshetnikov1997tidally,
  title={Tidally-triggered disk thickening. II. Results and interpretations},
  author={Reshetnikov, Vladimir and Combes, Fran{\c{c}}oise},
  journal={arXiv preprint astro-ph/9702023},
  year={1997}
}

\bibliographystyle{unsrtnat}
\bibliography{Thesis.bib}

\begin{thebibliography}{284}
\providecommand{\natexlab}[1]{#1}
\providecommand{\url}[1]{\texttt{#1}}
\expandafter\ifx\csname urlstyle\endcsname\relax
  \providecommand{\doi}[1]{doi: #1}\else
  \providecommand{\doi}{doi: \begingroup \urlstyle{rm}\Url}\fi

\bibitem[Matthews and Wood(2003)]{matthews2003high}
Lynn~D Matthews and Kenneth Wood.
\newblock High-latitude hi in the low surface brightness galaxy ugc 7321.
\newblock \emph{The Astrophysical Journal}, 593\penalty0 (2):\penalty0 721,
  2003.

\bibitem[Banerjee et~al.(2010)Banerjee, Matthews, and Jog]{banerjee2010dark}
Arunima Banerjee, Lynn~D Matthews, and Chanda~J Jog.
\newblock Dark matter dominance at all radii in the superthin galaxy ugc 7321.
\newblock \emph{New Astronomy}, 15\penalty0 (1):\penalty0 89--95, 2010.

\bibitem[Goad and Roberts(1981)]{goad1981spectroscopic}
JW~Goad and MS~Roberts.
\newblock Spectroscopic observations of superthin galaxies.
\newblock \emph{The Astrophysical Journal}, 250:\penalty0 79--86, 1981.

\bibitem[{Karachentsev} et~al.(1999){Karachentsev}, {Karachentseva}, {Kudrya},
  {Sharina}, and {Parnovskij}]{1999BSAO...47....5K}
I.~D. {Karachentsev}, V.~E. {Karachentseva}, Yu.~N. {Kudrya}, M.~E. {Sharina},
  and S.~L. {Parnovskij}.
\newblock {The revised Flat Galaxy Catalogue.}
\newblock \emph{Bulletin of the Special Astrophysics Observatory}, 47:\penalty0
  5--185, January 1999.

\bibitem[Pillepich et~al.(2018)Pillepich, Springel, Nelson, Genel, Naiman,
  Pakmor, Hernquist, Torrey, Vogelsberger, Weinberger,
  et~al.]{pillepich2018simulating}
Annalisa Pillepich, Volker Springel, Dylan Nelson, Shy Genel, Jill Naiman,
  R{\"u}diger Pakmor, Lars Hernquist, Paul Torrey, Mark Vogelsberger, Rainer
  Weinberger, et~al.
\newblock Simulating galaxy formation with the illustristng model.
\newblock \emph{Monthly Notices of the Royal Astronomical Society},
  473\penalty0 (3):\penalty0 4077--4106, 2018.

\bibitem[Khoperskov and Bertin(2017)]{khoperskov2017disk}
Sergey Khoperskov and Giuseppe Bertin.
\newblock Disk heating and bending instability in galaxies with
  counterrotation.
\newblock \emph{Astronomy \& Astrophysics}, 597:\penalty0 A103, 2017.

\bibitem[Zasov et~al.(1991)Zasov, Makarov, and Mikhailova]{zasov1991thickness}
AV~Zasov, DI~Makarov, and EA~Mikhailova.
\newblock Thickness of thin stellar disks and the mass of the dark halo.
\newblock \emph{Soviet Astronomy Letters}, 17:\penalty0 374, 1991.

\bibitem[O'Brien et~al.(2010)O'Brien, Freeman, and van~der Kruit]{o2010dark}
Jess~Clare O'Brien, KC~Freeman, and PC~van~der Kruit.
\newblock The dark matter halo shape of edge-on disk galaxies-iii. modelling
  the hi observations: results.
\newblock \emph{Astronomy \& astrophysics}, 515:\penalty0 A62, 2010.

\bibitem[Banerjee and Jog(2013)]{banerjee2013some}
Arunima Banerjee and Chanda~J Jog.
\newblock Why are some galaxy discs extremely thin?
\newblock \emph{Monthly Notices of the Royal Astronomical Society},
  431\penalty0 (1):\penalty0 582--588, 2013.

\bibitem[Jadhav~Y and Banerjee(2019)]{jadhav2019specific}
Vikas Jadhav~Y and Arunima Banerjee.
\newblock The specific angular momenta of superthin galaxies: Cue to their
  origin?
\newblock \emph{Monthly Notices of the Royal Astronomical Society},
  488\penalty0 (1):\penalty0 547--558, 2019.

\bibitem[{de Vaucouleurs}(1948)]{1948AnAp...11..247D}
Gerard {de Vaucouleurs}.
\newblock {Recherches sur les Nebuleuses Extragalactiques}.
\newblock \emph{Annales d'Astrophysique}, 11:\penalty0 247, January 1948.

\bibitem[De~Vaucouleurs(1956)]{de1956survey}
G{\'e}rard De~Vaucouleurs.
\newblock Survey of bright galaxies south of-35 deg. declination with the
  30-inch reynolds reflector (1952-1955).
\newblock \emph{Canberra: Mount Stromlo}, 1956.

\bibitem[Sersic(1968)]{sersic1968atlas}
JL~Sersic.
\newblock Atlas de galaxias australes, ed.
\newblock \emph{Sersic, JL}, 2:\penalty0 23, 1968.

\bibitem[Dalcanton and Bernstein(2000)]{dalcanton2000structural}
Julianne~J Dalcanton and Rebecca~A Bernstein.
\newblock A structural and dynamical study of late-type, edge-on galaxies. i.
  sample selection and imaging data.
\newblock \emph{The Astronomical Journal}, 120\penalty0 (1):\penalty0 203,
  2000.

\bibitem[Dalcanton and Bernstein(2002)]{dalcanton2002structural}
Julianne~J Dalcanton and Rebecca~A Bernstein.
\newblock A structural and dynamical study of late-type, edge-on galaxies. ii.
  vertical color gradients and the detection of ubiquitous thick disks.
\newblock \emph{The Astronomical Journal}, 124\penalty0 (3):\penalty0 1328,
  2002.

\bibitem[Yoachim and Dalcanton(2005)]{yoachim2005kinematics}
Peter Yoachim and Julianne~J Dalcanton.
\newblock The kinematics of thick disks in external galaxies.
\newblock \emph{The Astrophysical Journal}, 624\penalty0 (2):\penalty0 701,
  2005.

\bibitem[Yoachim and Dalcanton(2008{\natexlab{a}})]{yoachim2008lick}
Peter Yoachim and Julianne~J Dalcanton.
\newblock Lick indices in the thin and thick disks of edge-on disk galaxies.
\newblock \emph{The Astrophysical Journal}, 683\penalty0 (2):\penalty0 707,
  2008{\natexlab{a}}.

\bibitem[Juri{\'c} et~al.(2008)Juri{\'c}, Ivezi{\'c}, Brooks, Lupton, Schlegel,
  Finkbeiner, Padmanabhan, Bond, Sesar, Rockosi, et~al.]{juric2008milky}
Mario Juri{\'c}, {\v{Z}}eljko Ivezi{\'c}, Alyson Brooks, Robert~H Lupton, David
  Schlegel, Douglas Finkbeiner, Nikhil Padmanabhan, Nicholas Bond, Branimir
  Sesar, Constance~M Rockosi, et~al.
\newblock The milky way tomography with sdss. i. stellar number density
  distribution.
\newblock \emph{The Astrophysical Journal}, 673\penalty0 (2):\penalty0 864,
  2008.

\bibitem[Begum and Chengalur(2004)]{begum2004kinematics}
Ayesha Begum and Jayaram~N Chengalur.
\newblock Kinematics of two dwarf galaxies in the ngc 6946 group.
\newblock \emph{Astronomy \& Astrophysics}, 424\penalty0 (2):\penalty0
  509--517, 2004.

\bibitem[Patra et~al.(2014)Patra, Banerjee, Chengalur, and
  Begum]{patra2014modelling}
Narendra~Nath Patra, Arunima Banerjee, Jayaram~N Chengalur, and Ayesha Begum.
\newblock Modelling h i distribution and kinematics in the edge-on dwarf
  irregular galaxy kk250.
\newblock \emph{Monthly Notices of the Royal Astronomical Society},
  445\penalty0 (2):\penalty0 1424--1429, 2014.

\bibitem[Agnese et~al.(2018)Agnese, Aramaki, Arnquist, Baker, Balakishiyeva,
  Banik, Barker, Thakur, Bauer, Binder, et~al.]{agnese2018results}
R~Agnese, T~Aramaki, IJ~Arnquist, W~Baker, D~Balakishiyeva, S~Banik, D~Barker,
  R~Basu Thakur, DA~Bauer, T~Binder, et~al.
\newblock Results from the super cryogenic dark matter search experiment at
  soudan.
\newblock \emph{Physical review letters}, 120\penalty0 (6):\penalty0 061802,
  2018.

\bibitem[Springel et~al.(2005)Springel, White, Jenkins, Frenk, Yoshida, Gao,
  Navarro, Thacker, Croton, Helly, et~al.]{springel2005simulations}
Volker Springel, Simon~DM White, Adrian Jenkins, Carlos~S Frenk, Naoki Yoshida,
  Liang Gao, Julio Navarro, Robert Thacker, Darren Croton, John Helly, et~al.
\newblock Simulations of the formation, evolution and clustering of galaxies
  and quasars.
\newblock \emph{nature}, 435\penalty0 (7042):\penalty0 629--636, 2005.

\bibitem[Rubin(1983)]{rubin1983dark}
Vera~C Rubin.
\newblock Dark matter in spiral galaxies.
\newblock \emph{Scientific American}, 248\penalty0 (6):\penalty0 96--109, 1983.

\bibitem[Zwicky(1937)]{zwicky1937masses}
Fritz Zwicky.
\newblock On the masses of nebulae and of clusters of nebulae.
\newblock \emph{The Astrophysical Journal}, 86:\penalty0 217, 1937.

\bibitem[De~Blok et~al.(2008)De~Blok, Walter, Brinks, Trachternach, Oh, and
  Kennicutt]{de2008high}
WJG De~Blok, Fabian Walter, Elias Brinks, C~Trachternach, SH~Oh, and Robert~C
  Kennicutt.
\newblock High-resolution rotation curves and galaxy mass models from things.
\newblock \emph{The Astronomical Journal}, 136\penalty0 (6):\penalty0 2648,
  2008.

\bibitem[Oh et~al.(2015)Oh, Hunter, Brinks, Elmegreen, Schruba, Walter, Rupen,
  Young, Simpson, Johnson, et~al.]{oh2015high}
Se-Heon Oh, Deidre~A Hunter, Elias Brinks, Bruce~G Elmegreen, Andreas Schruba,
  Fabian Walter, Michael~P Rupen, Lisa~M Young, Caroline~E Simpson, Megan~C
  Johnson, et~al.
\newblock High-resolution mass models of dwarf galaxies from little things.
\newblock \emph{The Astronomical Journal}, 149\penalty0 (6):\penalty0 180,
  2015.

\bibitem[Lelli et~al.(2016)Lelli, McGaugh, and Schombert]{lelli2016sparc}
Federico Lelli, Stacy~S McGaugh, and James~M Schombert.
\newblock Sparc: mass models for 175 disk galaxies with spitzer photometry and
  accurate rotation curves.
\newblock \emph{The Astronomical Journal}, 152:\penalty0 157, 2016.

\bibitem[Navarro(1996)]{navarro1996structure}
Julio~F Navarro.
\newblock The structure of cold dark matter halos.
\newblock In \emph{Symposium-international astronomical union}, volume 171,
  pages 255--258. Cambridge University Press, 1996.

\bibitem[Gilmore and Reid(1983)]{gilmore1983new}
Gerard Gilmore and Neil Reid.
\newblock New light on faint stars--iii. galactic structure towards the south
  pole and the galactic thick disc.
\newblock \emph{Monthly Notices of the Royal Astronomical Society},
  202\penalty0 (4):\penalty0 1025--1047, 1983.

\bibitem[Bovy et~al.(2012{\natexlab{a}})Bovy, Rix, Liu, Hogg, Beers, and
  Lee]{bovy2012spatial}
Jo~Bovy, Hans-Walter Rix, Chao Liu, David~W Hogg, Timothy~C Beers, and
  Young~Sun Lee.
\newblock The spatial structure of mono-abundance sub-populations of the milky
  way disk.
\newblock \emph{The Astrophysical Journal}, 753\penalty0 (2):\penalty0 148,
  2012{\natexlab{a}}.

\bibitem[Quinn et~al.(1993)Quinn, Hernquist, and Fullagar]{quinn1993heating}
PJ~Quinn, Lars Hernquist, and DP~Fullagar.
\newblock Heating of galactic disks by mergers.
\newblock \emph{The Astrophysical Journal}, 403:\penalty0 74--93, 1993.

\bibitem[Villalobos and Helmi(2008)]{villalobos2008simulations}
{\'A}lvaro Villalobos and Amina Helmi.
\newblock Simulations of minor mergers--i. general properties of thick discs.
\newblock \emph{Monthly Notices of the Royal Astronomical Society},
  391\penalty0 (4):\penalty0 1806--1827, 2008.

\bibitem[Bekki and Tsujimoto(2011)]{bekki2011origin}
Kenji Bekki and Takuji Tsujimoto.
\newblock Origin of chemical and dynamical properties of the galactic thick
  disk.
\newblock \emph{The Astrophysical Journal}, 738\penalty0 (1):\penalty0 4, 2011.

\bibitem[Agertz et~al.(2009)Agertz, Teyssier, and Moore]{agertz2009disc}
Oscar Agertz, Romain Teyssier, and Ben Moore.
\newblock Disc formation and the origin of clumpy galaxies at high redshift.
\newblock \emph{Monthly Notices of the Royal Astronomical Society: Letters},
  397\penalty0 (1):\penalty0 L64--L68, 2009.

\bibitem[Ceverino et~al.(2010)Ceverino, Dekel, and Bournaud]{ceverino2010high}
Daniel Ceverino, Avishai Dekel, and Frederic Bournaud.
\newblock High-redshift clumpy discs and bulges in cosmological simulations.
\newblock \emph{Monthly Notices of the Royal Astronomical Society},
  404\penalty0 (4):\penalty0 2151--2169, 2010.

\bibitem[Brook et~al.(2004)Brook, Kawata, Gibson, and
  Freeman]{brook2004emergence}
Chris~B Brook, Daisuke Kawata, Brad~K Gibson, and Ken~C Freeman.
\newblock The emergence of the thick disk in a cold dark matter universe.
\newblock \emph{The Astrophysical Journal}, 612\penalty0 (2):\penalty0 894,
  2004.

\bibitem[Spitzer~Jr and Schwarzschild(1951)]{spitzer1951possible}
Lyman Spitzer~Jr and Martin Schwarzschild.
\newblock The possible influence of interstellar clouds on stellar velocities.
\newblock \emph{The Astrophysical Journal}, 114:\penalty0 385, 1951.

\bibitem[Lacey and Silk(1991)]{lacey1991tidally}
Cedric Lacey and Joseph Silk.
\newblock Tidally triggered galaxy formation. i-evolution of the galaxy
  luminosity function.
\newblock \emph{Astrophysical journal.}, 381:\penalty0 14--32, 1991.

\bibitem[Eggen et~al.(1962)Eggen, Lynden-Bell, and Sandage]{eggen1962evidence}
OJ~Eggen, D~Lynden-Bell, and AR~Sandage.
\newblock Evidence from the motions of old stars that the galaxy collapsed.
\newblock \emph{The Astrophysical Journal}, 136:\penalty0 748, 1962.

\bibitem[Karachentsev et~al.(2003)Karachentsev, Karachentseva, Kudrya, Sharina,
  and Parnovsky]{karachentsev2003revised}
ID~Karachentsev, VE~Karachentseva, Yu~N Kudrya, ME~Sharina, and SL~Parnovsky.
\newblock The revised flat galaxy catalogue.
\newblock \emph{arXiv preprint astro-ph/0305566}, 2003.

\bibitem[Swaters et~al.(1997)Swaters, Sancisi, and Van
  Der~Hulst]{swaters1997hi}
RA~Swaters, R~Sancisi, and JM~Van Der~Hulst.
\newblock The hi halo of ngc 891.
\newblock \emph{The Astrophysical Journal}, 491\penalty0 (1):\penalty0 140,
  1997.

\bibitem[Heald et~al.(2011)Heald, J{\'o}zsa, Serra, Zschaechner, Rand,
  Fraternali, Oosterloo, Walterbos, J{\"u}tte, and
  Gentile]{heald2011westerbork}
George Heald, Gyula J{\'o}zsa, Paolo Serra, Laura Zschaechner, Richard Rand,
  Filippo Fraternali, Tom Oosterloo, Rene Walterbos, Eva J{\"u}tte, and
  Gianfranco Gentile.
\newblock The westerbork hydrogen accretion in local galaxies (halogas)
  survey-i. survey description and pilot observations.
\newblock \emph{Astronomy \& astrophysics}, 526:\penalty0 A118, 2011.

\bibitem[Zibetti et~al.(2004)Zibetti, White, and Brinkmann]{zibetti2004haloes}
Stefano Zibetti, Simon~DM White, and Jon Brinkmann.
\newblock Haloes around edge-on disc galaxies in the sloan digital sky survey.
\newblock \emph{Monthly Notices of the Royal Astronomical Society},
  347\penalty0 (2):\penalty0 556--568, 2004.

\bibitem[Monachesi et~al.(2016)Monachesi, Bell, Radburn-Smith, Bailin, de~Jong,
  Holwerda, Streich, and Silverstein]{monachesi2016ghosts}
Antonela Monachesi, Eric~F Bell, David~J Radburn-Smith, Jeremy Bailin, Roelof~S
  de~Jong, Benne Holwerda, David Streich, and Grace Silverstein.
\newblock The ghosts survey--ii. the diversity of halo colour and metallicity
  profiles of massive disc galaxies.
\newblock \emph{Monthly Notices of the Royal Astronomical Society},
  457\penalty0 (2):\penalty0 1419--1446, 2016.

\bibitem[Olling(1995)]{olling1995usage}
Rob~P Olling.
\newblock On the usage of flaring gas layers to determine the shape of dark
  matter halos.
\newblock \emph{arXiv preprint astro-ph/9505002}, 1995.

\bibitem[Olling(1996)]{olling1996highly}
Rob~P Olling.
\newblock The highly flattened dark matter halo of ngc 4244.
\newblock \emph{arXiv preprint astro-ph/9605111}, 1996.

\bibitem[Banerjee and Jog(2011)]{banerjee2011progressively}
Arunima Banerjee and Chanda~J Jog.
\newblock Progressively more prolate dark matter halo in the outer galaxy as
  traced by flaring h i gas.
\newblock \emph{The Astrophysical Journal Letters}, 732\penalty0 (1):\penalty0
  L8, 2011.

\bibitem[Banerjee and Jog(2008)]{banerjee2008flattened}
Arunima Banerjee and Chanda~J Jog.
\newblock The flattened dark matter halo of m31 as deduced from the observed hi
  scale heights.
\newblock \emph{The Astrophysical Journal}, 685\penalty0 (1):\penalty0 254,
  2008.

\bibitem[Matthews and Van~Driel(2000)]{matthews2000h}
LD~Matthews and W~Van~Driel.
\newblock An h i survey of highly flattened, edge-on, pure disk galaxies.
\newblock \emph{Astronomy and Astrophysics Supplement Series}, 143\penalty0
  (3):\penalty0 421--456, 2000.

\bibitem[Huchtmeier et~al.(2005)Huchtmeier, Karachentsev, Karachentseva,
  Kudrya, and Mitronova]{huchtmeier2005hi}
WK~Huchtmeier, ID~Karachentsev, VE~Karachentseva, Yu~N Kudrya, and
  SN~Mitronova.
\newblock Hi observations of edge-on spiral galaxies.
\newblock \emph{Astronomy \& Astrophysics}, 435\penalty0 (2):\penalty0
  459--463, 2005.

\bibitem[Kautsch et~al.(2006)Kautsch, Grebel, Barazza, and
  Gallagher]{kautsch2006catalog}
SJ~Kautsch, EK~Grebel, FD~Barazza, and JS~Gallagher.
\newblock A catalog of edge-on disk galaxies-from galaxies with a bulge to
  superthin galaxies.
\newblock \emph{Astronomy \& Astrophysics}, 445\penalty0 (2):\penalty0
  765--778, 2006.

\bibitem[Bizyaev et~al.(2017)Bizyaev, Kautsch, Sotnikova, Reshetnikov, and
  Mosenkov]{bizyaev2017very}
DV~Bizyaev, SJ~Kautsch, N~Ya Sotnikova, Vladimir~P Reshetnikov, and
  Aleksander~V Mosenkov.
\newblock Very thin disc galaxies in the sdss catalogue of edge-on galaxies.
\newblock \emph{Monthly Notices of the Royal Astronomical Society},
  465\penalty0 (4):\penalty0 3784--3792, 2017.

\bibitem[Kautsch et~al.(2021)Kautsch, Bizyaev, Makarov, Reshetnikov, Mosenkov,
  and Antipova]{kautsch2021spectroscopic}
Stefan~J Kautsch, Dmitry Bizyaev, Dimitry~I Makarov, Vladimir~P Reshetnikov,
  Alexander~V Mosenkov, and Alexandra~V Antipova.
\newblock A spectroscopic survey of superthin galaxies.
\newblock \emph{Research Notes of the AAS}, 5\penalty0 (3):\penalty0 43, 2021.

\bibitem[Bizyaev et~al.(2021)Bizyaev, Makarov, Reshetnikov, Mosenkov, Kautsch,
  and Antipova]{bizyaev2021spectral}
Dmitry Bizyaev, DI~Makarov, VP~Reshetnikov, AV~Mosenkov, SJ~Kautsch, and
  AV~Antipova.
\newblock Spectral observations of superthin galaxies.
\newblock \emph{The Astrophysical Journal}, 914\penalty0 (2):\penalty0 104,
  2021.

\bibitem[Bizyaev et~al.(2020)Bizyaev, Tatarnikov, Shatsky, Najip, Birlak, and
  Voziakova]{bizyaev2020near}
Dmitry Bizyaev, Andrey Tatarnikov, Nikolai Shatsky, Aurik Najip, Marina Birlak,
  and Olga Voziakova.
\newblock Near-infrared photometry of superthin edge-on galaxies.
\newblock \emph{Astronomische Nachrichten}, 341\penalty0 (3):\penalty0
  314--323, 2020.

\bibitem[Antipova et~al.(2021)Antipova, Makarov, and
  Savchenko]{antipova2021database}
AV~Antipova, DI~Makarov, and SS~Savchenko.
\newblock The database for studying edge-on galaxies.
\newblock \emph{ASTRONOMY AT THE EPOCH OF MULTIMESSENGER STUDIES}, page 347,
  2021.

\bibitem[Abe et~al.(1999)Abe, Bond, Carter, Dodd, Fujimoto, Hearnshaw, Honda,
  Jugaku, Kabe, Kilmartin, et~al.]{abe1999observation}
F~Abe, IA~Bond, BS~Carter, RJ~Dodd, M~Fujimoto, JB~Hearnshaw, M~Honda,
  J~Jugaku, S~Kabe, PM~Kilmartin, et~al.
\newblock Observation of the halo of the edge-on galaxy ic 5249.
\newblock \emph{The Astronomical Journal}, 118\penalty0 (1):\penalty0 261,
  1999.

\bibitem[Van Der~Kruit et~al.(2001)Van Der~Kruit, Jim{\'e}nez-Vicente, Kregel,
  and Freeman]{van2001kinematics}
PC~Van Der~Kruit, J~Jim{\'e}nez-Vicente, M~Kregel, and KC~Freeman.
\newblock Kinematics and dynamics of the" superthin" edge-on disk galaxy ic
  5249.
\newblock \emph{Astronomy \& Astrophysics}, 379\penalty0 (2):\penalty0
  374--383, 2001.

\bibitem[Uson and Matthews(2003)]{uson2003hi}
Juan~M Uson and LD~Matthews.
\newblock Hi imaging observations of superthin galaxies. i. ugc 7321.
\newblock \emph{The Astronomical Journal}, 125\penalty0 (5):\penalty0 2455,
  2003.

\bibitem[Matthews and Uson(2007)]{matthews2007h}
Lynn~D Matthews and Juan~M Uson.
\newblock H i imaging observations of superthin galaxies. ii. ic 2233 and the
  blue compact dwarf ngc 2537.
\newblock \emph{The Astronomical Journal}, 135\penalty0 (1):\penalty0 291,
  2007.

\bibitem[Kurapati et~al.(2018)Kurapati, Banerjee, Chengalur, Makarov, Borisov,
  Afanasiev, and Antipova]{kurapati2018mass}
Sushma Kurapati, Arunima Banerjee, Jayaram~N Chengalur, Dmitry Makarov,
  Svyatoslav Borisov, Anton Afanasiev, and Aleksandra Antipova.
\newblock Mass modelling of a superthin galaxy, fgc 1540.
\newblock \emph{Monthly Notices of the Royal Astronomical Society},
  479\penalty0 (4):\penalty0 5686--5695, 2018.

\bibitem[Van~Albada and Sancisi(1986)]{van1986dark}
TS~Van~Albada and R~Sancisi.
\newblock Dark matter in spiral galaxies.
\newblock \emph{Philosophical Transactions of the Royal Society of London.
  Series A, Mathematical and Physical Sciences}, 320\penalty0 (1556):\penalty0
  447--464, 1986.

\bibitem[Banerjee and Bapat(2017)]{banerjee2017mass}
Arunima Banerjee and Disha Bapat.
\newblock Mass modelling of superthin galaxies: Ic5249, ugc7321 and ic2233.
\newblock \emph{Monthly Notices of the Royal Astronomical Society},
  466\penalty0 (3):\penalty0 3753--3761, 2017.

\bibitem[Peters et~al.(2017)Peters, van~der Kruit, Allen, and
  Freeman]{peters2017shape}
SPC Peters, PC~van~der Kruit, RJ~Allen, and KC~Freeman.
\newblock The shape of dark matter haloes--iii. kinematics and structure of the
  h i disc.
\newblock \emph{Monthly Notices of the Royal Astronomical Society},
  464\penalty0 (1):\penalty0 32--47, 2017.

\bibitem[Banerjee and Bapat(2016)]{banerjee2016mass}
Arunima Banerjee and Disha Bapat.
\newblock Mass modelling of superthin galaxies: Ic5249, ugc7321 and ic2233.
\newblock \emph{Monthly Notices of the Royal Astronomical Society},
  466\penalty0 (3):\penalty0 3753--3761, 2016.

\bibitem[Garg and Banerjee(2017)]{garg2017origin}
Prerak Garg and Arunima Banerjee.
\newblock Origin of low surface brightness galaxies: a dynamical study.
\newblock \emph{Monthly Notices of the Royal Astronomical Society},
  472\penalty0 (1):\penalty0 166--173, 2017.

\bibitem[Romeo and Wiegert(2011)]{romeo2011effective}
Alessandro~B Romeo and Joachim Wiegert.
\newblock The effective stability parameter for two-component galactic discs.
\newblock \emph{Monthly Notices of the Royal Astronomical Society},
  416\penalty0 (2):\penalty0 1191--1196, 2011.

\bibitem[Spitzer~Jr and Schwarzschild(1953)]{spitzer1953possible}
Lyman Spitzer~Jr and Martin Schwarzschild.
\newblock The possible influence of interstellar clouds on stellar velocities.
  ii.
\newblock \emph{The Astrophysical Journal}, 118:\penalty0 106, 1953.

\bibitem[Goldreich and Lynden-Bell(1965)]{goldreich1965gravitational}
Peter Goldreich and D~Lynden-Bell.
\newblock I. gravitational stability of uniformly rotating disks.
\newblock \emph{Monthly Notices of the Royal Astronomical Society},
  130\penalty0 (2):\penalty0 97--124, 1965.

\bibitem[Barbanis and Woltjer(1967)]{barbanis1967orbits}
B~Barbanis and L~Woltjer.
\newblock Orbits in spiral galaxies and the velocity dispersion of population i
  stars.
\newblock \emph{The Astrophysical Journal}, 150:\penalty0 461, 1967.

\bibitem[Binney and Tremaine(2011)]{binney2011galactic}
James Binney and Scott Tremaine.
\newblock \emph{Galactic dynamics}, volume~13.
\newblock Princeton university press, 2011.

\bibitem[Toth and Ostriker(1992)]{toth1992galactic}
G~Toth and JP~Ostriker.
\newblock Galactic disks, infall, and the global value of omega.
\newblock \emph{The Astrophysical Journal}, 389:\penalty0 5--26, 1992.

\bibitem[Lacey and Ostriker(1985)]{lacey1985massive}
CG~Lacey and JP~Ostriker.
\newblock Massive black holes in galactic halos?
\newblock \emph{Astrophysical journal.}, 299:\penalty0 633--652, 1985.

\bibitem[Narayan and Jog(2002{\natexlab{a}})]{narayan2002vertical}
Chaitra~A Narayan and Chanda~J Jog.
\newblock Vertical scaleheights in a gravitationally coupled, three-component
  galactic disk.
\newblock \emph{Astronomy \& Astrophysics}, 394\penalty0 (1):\penalty0 89--96,
  2002{\natexlab{a}}.

\bibitem[de~Zeeuw and Pfenniger(1988)]{de1988potential}
Tim de~Zeeuw and Daniel Pfenniger.
\newblock Potential-density pairs for galaxies.
\newblock \emph{Monthly Notices of the Royal Astronomical Society},
  235:\penalty0 949--995, 1988.

\bibitem[Van~der Kruit(1989)]{van1989photometry}
PC~Van~der Kruit.
\newblock Photometry of disks in galaxies.
\newblock \emph{The World of Galaxies}, pages 256--275, 1989.

\bibitem[{Merrifield} et~al.(2001){Merrifield}, {Gerssen}, and
  {Kuijken}]{2001ASPC..230..221M}
M.~R. {Merrifield}, J.~{Gerssen}, and K.~{Kuijken}.
\newblock {The Origins of Disk Heating}.
\newblock In Jos{\'e}~G. {Funes} and Enrico~Maria {Corsini}, editors,
  \emph{Galaxy Disks and Disk Galaxies}, volume 230 of \emph{Astronomical
  Society of the Pacific Conference Series}, pages 221--224, January 2001.

\bibitem[Vasiliev(2018)]{vasiliev2018agama}
Eugene Vasiliev.
\newblock Agama: action-based galaxy modelling architecture.
\newblock \emph{Monthly Notices of the Royal Astronomical Society},
  482\penalty0 (2):\penalty0 1525--1544, 2018.

\bibitem[Aditya and Banerjee(2021)]{10.1093/mnras/stab155}
K~Aditya and Arunima Banerjee.
\newblock {How “cold” are the stellar discs of superthin galaxies?}
\newblock \emph{Monthly Notices of the Royal Astronomical Society}, 01 2021.
\newblock ISSN 0035-8711.
\newblock \doi{10.1093/mnras/stab155}.
\newblock URL \url{https://doi.org/10.1093/mnras/stab155}.
\newblock stab155.

\bibitem[Binney(2012)]{binney2012actions}
James Binney.
\newblock Actions for axisymmetric potentials.
\newblock \emph{Monthly Notices of the Royal Astronomical Society},
  426\penalty0 (2):\penalty0 1324--1327, 2012.

\bibitem[Toomre(1964)]{toomre1964gravitational}
Alar Toomre.
\newblock On the gravitational stability of a disk of stars.
\newblock \emph{The Astrophysical Journal}, 139:\penalty0 1217--1238, 1964.

\bibitem[Posti et~al.(2019)Posti, Marasco, Fraternali, and
  Famaey]{posti2019galaxy}
Lorenzo Posti, Antonino Marasco, Filippo Fraternali, and Benoit Famaey.
\newblock Galaxy disc scaling relations: A tight linear galaxy--halo connection
  challenges abundance matching.
\newblock \emph{Astronomy \& Astrophysics}, 629:\penalty0 A59, 2019.

\bibitem[{Mancera Pi{\~n}a} et~al.(2021){Mancera Pi{\~n}a}, {Posti},
  {Fraternali}, {Adams}, and {Oosterloo}]{2021A&A...647A..76M}
Pavel~E. {Mancera Pi{\~n}a}, Lorenzo {Posti}, Filippo {Fraternali}, Elizabeth
  A.~K. {Adams}, and Tom {Oosterloo}.
\newblock {The baryonic specific angular momentum of disc galaxies}.
\newblock \emph{Astronomy \& Astrophysics}, 647:\penalty0 A76, March 2021.
\newblock \doi{10.1051/0004-6361/202039340}.

\bibitem[Marasco et~al.(2019)Marasco, Fraternali, Posti, Ijtsma, Di~Teodoro,
  and Oosterloo]{marasco2019angular}
A~Marasco, F~Fraternali, L~Posti, M~Ijtsma, EM~Di~Teodoro, and T~Oosterloo.
\newblock The angular momentum of disc galaxies at z= 1.
\newblock \emph{Astronomy \& Astrophysics}, 621:\penalty0 L6, 2019.

\bibitem[Rubin and Ford~Jr(1970)]{rubin1970rotation}
Vera~C Rubin and W~Kent Ford~Jr.
\newblock Rotation of the andromeda nebula from a spectroscopic survey of
  emission regions.
\newblock \emph{The Astrophysical Journal}, 159:\penalty0 379, 1970.

\bibitem[Bosma(1978)]{bosma1978distribution}
Albert Bosma.
\newblock \emph{The distribution and kinematics of neutral hydrogen in spiral
  galaxies of various morphological types}.
\newblock PhD thesis, Rijksuniversiteit te Groningen., 1978.

\bibitem[Agnese et~al.(2013)Agnese, Ahmed, Anderson, Arrenberg, Balakishiyeva,
  Thakur, Bauer, Billard, Borgland, Brandt, et~al.]{agnese2013silicon}
R~Agnese, Z~Ahmed, AJ~Anderson, S~Arrenberg, D~Balakishiyeva, R~Basu Thakur,
  DA~Bauer, J~Billard, A~Borgland, D~Brandt, et~al.
\newblock Silicon detector dark matter results from the final exposure of cdms
  ii.
\newblock \emph{Physical review letters}, 111\penalty0 (25):\penalty0 251301,
  2013.

\bibitem[Pawlowski et~al.(2015)Pawlowski, Famaey, Merritt, and
  Kroupa]{pawlowski2015persistence}
Marcel~S Pawlowski, Benoit Famaey, David Merritt, and Pavel Kroupa.
\newblock On the persistence of two small-scale problems in $\lambda$cdm.
\newblock \emph{The Astrophysical Journal}, 815\penalty0 (1):\penalty0 19,
  2015.

\bibitem[Kroupa(2015)]{Kroupa:2014ria}
Pavel Kroupa.
\newblock {Galaxies as simple dynamical systems: observational data disfavor
  dark matter and stochastic star formation}.
\newblock \emph{Can. J. Phys.}, 93\penalty0 (2):\penalty0 169--202, 2015.
\newblock \doi{10.1139/cjp-2014-0179}.

\bibitem[Peebles and Nusser(2010)]{Peebles:2010di}
P.J.E. Peebles and Adi Nusser.
\newblock {Clues from nearby galaxies to a better theory of cosmic evolution}.
\newblock \emph{Nature}, 465:\penalty0 565--569, 2010.
\newblock \doi{10.1038/nature09101}.

\bibitem[Milgrom(1983)]{milgrom1983modification}
Mordehai Milgrom.
\newblock A modification of the newtonian dynamics as a possible alternative to
  the hidden mass hypothesis.
\newblock \emph{The Astrophysical Journal}, 270:\penalty0 365--370, 1983.

\bibitem[Sanders and Verheijen(1998)]{sanders1998rotation}
Robert~H Sanders and MAW Verheijen.
\newblock Rotation curves of ursa major galaxies in the context of modified
  newtonian dynamics.
\newblock \emph{The Astrophysical Journal}, 503\penalty0 (1):\penalty0 97,
  1998.

\bibitem[de~Blok and McGaugh(1998)]{de1998testing}
WJG de~Blok and SS~McGaugh.
\newblock Testing modified newtonian dynamics with low surface brightness
  galaxies: rotation curve fits.
\newblock \emph{The Astrophysical Journal}, 508\penalty0 (1):\penalty0 132,
  1998.

\bibitem[Sanders and Noordermeer(2007)]{sanders2007confrontation}
RH~Sanders and E~Noordermeer.
\newblock Confrontation of modified newtonian dynamics with the rotation curves
  of early-type disc galaxies.
\newblock \emph{Monthly Notices of the Royal Astronomical Society},
  379\penalty0 (2):\penalty0 702--710, 2007.

\bibitem[Csaki et~al.(2000)Csaki, Graesser, Randall, and Terning]{Csaki:1999mp}
Csaba Csaki, Michael Graesser, Lisa Randall, and John Terning.
\newblock {Cosmology of brane models with radion stabilization}.
\newblock \emph{Phys.Rev.}, D62:\penalty0 045015, 2000.
\newblock \doi{10.1103/PhysRevD.62.045015}.

\bibitem[Binetruy et~al.(2000{\natexlab{a}})Binetruy, Deffayet, Ellwanger, and
  Langlois]{binetruy2000brane}
Pierre Binetruy, Cedric Deffayet, Ulrich Ellwanger, and David Langlois.
\newblock Brane cosmological evolution in a bulk with cosmological constant.
\newblock \emph{Physics Letters B}, 477\penalty0 (1-3):\penalty0 285--291,
  2000{\natexlab{a}}.

\bibitem[Maartens(2001)]{Maartens:2001jx}
Roy Maartens.
\newblock {Geometry and dynamics of the brane world}.
\newblock In \emph{{Spanish Relativity Meeting on Reference Frames and
  Gravitomagnetism (EREs2000) Valladolid, Spain, September 6-9, 2000}}, 2001.
\newblock URL \url{http://alice.cern.ch/format/showfull?sysnb=2237527}.

\bibitem[Antoniadis(1990)]{Antoniadis:1990ew}
Ignatios Antoniadis.
\newblock {A Possible new dimension at a few TeV}.
\newblock \emph{Phys.Lett.}, B246:\penalty0 377--384, 1990.
\newblock \doi{10.1016/0370-2693(90)90617-F}.

\bibitem[Antoniadis et~al.(1998)Antoniadis, Arkani-Hamed, Dimopoulos, and
  Dvali]{Antoniadis:1998ig}
Ignatios Antoniadis, Nima Arkani-Hamed, Savas Dimopoulos, and G.R. Dvali.
\newblock {New dimensions at a millimeter to a Fermi and superstrings at a
  TeV}.
\newblock \emph{Phys.Lett.}, B436:\penalty0 257--263, 1998.
\newblock \doi{10.1016/S0370-2693(98)00860-0}.

\bibitem[Arkani-Hamed et~al.(1998)Arkani-Hamed, Dimopoulos, and
  Dvali]{ArkaniHamed:1998rs}
Nima Arkani-Hamed, Savas Dimopoulos, and G.R. Dvali.
\newblock {The Hierarchy problem and new dimensions at a millimeter}.
\newblock \emph{Phys.Lett.}, B429:\penalty0 263--272, 1998.
\newblock \doi{10.1016/S0370-2693(98)00466-3}.

\bibitem[Randall and Sundrum(1999{\natexlab{a}})]{Randall:1999vf}
Lisa Randall and Raman Sundrum.
\newblock {An Alternative to compactification}.
\newblock \emph{Phys.Rev.Lett.}, 83:\penalty0 4690--4693, 1999{\natexlab{a}}.
\newblock \doi{10.1103/PhysRevLett.83.4690}.

\bibitem[Randall and Sundrum(1999{\natexlab{b}})]{Randall:1999ee}
Lisa Randall and Raman Sundrum.
\newblock {A Large mass hierarchy from a small extra dimension}.
\newblock \emph{Phys.Rev.Lett.}, 83:\penalty0 3370--3373, 1999{\natexlab{b}}.
\newblock \doi{10.1103/PhysRevLett.83.3370}.

\bibitem[Garriga and Tanaka(2000)]{Garriga:1999yh}
Jaume Garriga and Takahiro Tanaka.
\newblock {Gravity in the brane world}.
\newblock \emph{Phys.Rev.Lett.}, 84:\penalty0 2778--2781, 2000.
\newblock \doi{10.1103/PhysRevLett.84.2778}.

\bibitem[Maartens(2004{\natexlab{a}})]{maartens2004brane}
Roy Maartens.
\newblock Brane-world gravity.
\newblock \emph{Living Reviews in Relativity}, 7\penalty0 (1):\penalty0 7,
  2004{\natexlab{a}}.

\bibitem[Mak and Harko(2004)]{mak2004can}
MK~Mak and T~Harko.
\newblock Can the galactic rotation curves be explained in brane world models?
\newblock \emph{Physical Review D}, 70\penalty0 (2):\penalty0 024010, 2004.

\bibitem[Harko and Cheng(2006)]{harko2006galactic}
Tiberiu Harko and KS~Cheng.
\newblock Galactic metric, dark radiation, dark pressure, and gravitational
  lensing in brane world models.
\newblock \emph{The Astrophysical Journal}, 636\penalty0 (1):\penalty0 8, 2006.

\bibitem[Boehmer and Harko(2007)]{boehmer2007galactic}
CG~Boehmer and T~Harko.
\newblock Galactic dark matter as a bulk effect on the brane.
\newblock \emph{Classical and Quantum Gravity}, 24\penalty0 (13):\penalty0
  3191, 2007.

\bibitem[Rahaman et~al.(2008)Rahaman, Kalam, DeBenedictis, Usmani, and
  Ray]{rahaman2008galactic}
F~Rahaman, M~Kalam, A~DeBenedictis, AA~Usmani, and Saibal Ray.
\newblock Galactic rotation curves and brane-world models.
\newblock \emph{Monthly Notices of the Royal Astronomical Society},
  389\penalty0 (1):\penalty0 27--33, 2008.

\bibitem[Gergely et~al.(2011)Gergely, Harko, Dwornik, Kupi, and
  Keresztes]{gergely2011galactic}
L{\'A}~Gergely, T~Harko, M~Dwornik, G~Kupi, and Z~Keresztes.
\newblock Galactic rotation curves in brane world models.
\newblock \emph{Monthly Notices of the Royal Astronomical Society},
  415\penalty0 (4):\penalty0 3275--3290, 2011.

\bibitem[Navarro et~al.(1997)Navarro, Frenk, and White]{navarro1997universal}
Julio~F Navarro, Carlos~S Frenk, and Simon~DM White.
\newblock A universal density profile from hierarchical clustering.
\newblock \emph{The Astrophysical Journal}, 490\penalty0 (2):\penalty0 493,
  1997.

\bibitem[Keresztes and Gergely(2010{\natexlab{a}})]{Keresztes:2009hy}
Zoltan Keresztes and Laszlo~A. Gergely.
\newblock {3+1+1 dimensional covariant gravitational dynamics on an
  asymmetrically embedded brane}.
\newblock \emph{Annalen Phys.}, 19:\penalty0 249--253, 2010{\natexlab{a}}.
\newblock \doi{10.1002/andp.201010421, 10.1142/9789814374552_0371}.
\newblock [,1948(2009)].

\bibitem[Keresztes and Gergely(2010{\natexlab{b}})]{Keresztes:2009wb}
Zoltan Keresztes and Laszlo~A. Gergely.
\newblock {Covariant gravitational dynamics in 3+1+1 dimensions}.
\newblock \emph{Class. Quant. Grav.}, 27:\penalty0 105009, 2010{\natexlab{b}}.
\newblock \doi{10.1088/0264-9381/27/10/105009}.

\bibitem[de~Blok(2005)]{de2005halo}
WJG de~Blok.
\newblock Halo mass profiles and low surface brightness galaxy rotation curves.
\newblock \emph{The Astrophysical Journal}, 634\penalty0 (1):\penalty0 227,
  2005.

\bibitem[Komanduri et~al.(2020)Komanduri, Banerjee, Banerjee, and
  Sengupta]{komanduri2020dynamical}
Aditya Komanduri, Indrani Banerjee, Arunima Banerjee, and Soumitra Sengupta.
\newblock Dynamical modelling of disc vertical structure in superthin galaxy
  ‘ugc 7321’in braneworld gravity: an mcmc study.
\newblock \emph{Monthly Notices of the Royal Astronomical Society},
  499\penalty0 (4):\penalty0 5690--5701, 2020.

\bibitem[Foreman-Mackey et~al.(2013)Foreman-Mackey, Hogg, Lang, and
  Goodman]{foreman2013emcee}
Daniel Foreman-Mackey, David~W Hogg, Dustin Lang, and Jonathan Goodman.
\newblock emcee: the mcmc hammer.
\newblock \emph{Publications of the Astronomical Society of the Pacific},
  125\penalty0 (925):\penalty0 306, 2013.

\bibitem[Hogg and Foreman-Mackey(2018)]{hogg2018data}
David~W Hogg and Daniel Foreman-Mackey.
\newblock Data analysis recipes: Using markov chain monte carlo.
\newblock \emph{The Astrophysical Journal Supplement Series}, 236\penalty0
  (1):\penalty0 11, 2018.

\bibitem[Hastings(1970)]{hastings1970monte}
W~Keith Hastings.
\newblock Monte carlo sampling methods using markov chains and their
  applications.
\newblock 1970.

\bibitem[Soetaert et~al.(2010)Soetaert, Petzoldt, et~al.]{soetaert2010inverse}
Karline Soetaert, Thomas Petzoldt, et~al.
\newblock Inverse modelling, sensitivity and monte carlo analysis in r using
  package fme.
\newblock \emph{Journal of Statistical Software}, 33\penalty0 (3):\penalty0
  1--28, 2010.

\bibitem[Haario et~al.(2006)Haario, Laine, Mira, and Saksman]{haario2006dram}
Heikki Haario, Marko Laine, Antonietta Mira, and Eero Saksman.
\newblock Dram: efficient adaptive mcmc.
\newblock \emph{Statistics and computing}, 16\penalty0 (4):\penalty0 339--354,
  2006.

\bibitem[Draine(2010)]{draine2010physics}
Bruce~T Draine.
\newblock \emph{Physics of the interstellar and intergalactic medium},
  volume~19.
\newblock Princeton University Press, 2010.

\bibitem[Thompson et~al.(2017)Thompson, Moran, and
  Swenson]{thompson2017interferometry}
A~Richard Thompson, James~M Moran, and George~W Swenson.
\newblock \emph{Interferometry and synthesis in radio astronomy}.
\newblock Springer Nature, 2017.

\bibitem[Chengalur et~al.(2007)Chengalur, Gupta, and
  Dwarkanath]{chengalur2007low}
Jayaram~N Chengalur, Yashwant Gupta, and KS~Dwarkanath.
\newblock Low frequency radio astronomy 3rd edition, 2007.

\bibitem[Born and Wolf(2013)]{born2013principles}
Max Born and Emil Wolf.
\newblock \emph{Principles of optics: electromagnetic theory of propagation,
  interference and diffraction of light}.
\newblock Elsevier, 2013.

\bibitem[Lal(2013)]{lal2013gmrt}
Dharam~Vir Lal.
\newblock Gmrt observer’s manual, 2013.

\bibitem[Offringa(2010)]{offringa2010aoflagger}
AR~Offringa.
\newblock Aoflagger: Rfi software.
\newblock \emph{Astrophysics Source Code Library}, pages ascl--1010, 2010.

\bibitem[H{\"o}gbom(1974)]{hogbom1974aperture}
JA~H{\"o}gbom.
\newblock Aperture synthesis with a non-regular distribution of interferometer
  baselines.
\newblock \emph{Astronomy and Astrophysics Supplement Series}, 15:\penalty0
  417, 1974.

\bibitem[Rogstad et~al.(1974)Rogstad, Lockhart, and
  Wright]{rogstad1974aperture}
DH~Rogstad, IA~Lockhart, and MCH Wright.
\newblock Aperture-synthesis observations of hi in the galaxy m83.
\newblock \emph{The Astrophysical Journal}, 193:\penalty0 309--319, 1974.

\bibitem[Begeman(1989)]{begeman1989hi}
KG~Begeman.
\newblock Hi rotation curves of spiral galaxies. i-ngc 3198.
\newblock \emph{Astronomy and Astrophysics}, 223:\penalty0 47--60, 1989.

\bibitem[Sellwood and Spekkens(2015)]{sellwood2015diskfit}
JA~Sellwood and Kristine Spekkens.
\newblock Diskfit: a code to fit simple non-axisymmetric galaxy models either
  to photometric images or to kinematic maps.
\newblock \emph{arXiv preprint arXiv:1509.07120}, 2015.

\bibitem[Mathewson et~al.(1992)Mathewson, Ford, and
  Buchhorn]{mathewson1992southern}
DS~Mathewson, VL~Ford, and M~Buchhorn.
\newblock A southern sky survey of the peculiar velocities of 1355 spiral
  galaxies.
\newblock \emph{The Astrophysical Journal Supplement Series}, 81:\penalty0
  413--659, 1992.

\bibitem[Takamiya and Sofue(2002)]{takamiya2002iteration}
Tsutomu Takamiya and Yoshiaki Sofue.
\newblock Iteration method to derive exact rotation curves from
  position-velocity diagrams of spiral galaxies.
\newblock \emph{The Astrophysical Journal}, 576\penalty0 (1):\penalty0 L15,
  2002.

\bibitem[J{\'o}zsa et~al.(2012)J{\'o}zsa, Kenn, Oosterloo, and
  Klein]{jozsa2012tirific}
Gyula~IG J{\'o}zsa, Franz Kenn, Thomas~A Oosterloo, and Ulrich Klein.
\newblock Tirific: Tilted ring fitting code.
\newblock \emph{Astrophysics Source Code Library}, pages ascl--1208, 2012.

\bibitem[Van~der Hulst et~al.(1992)Van~der Hulst, Terlouw, Begeman, Zwitser,
  and Roelfsema]{van1992groningen}
JM~Van~der Hulst, JP~Terlouw, KG~Begeman, W~Zwitser, and PR~Roelfsema.
\newblock The groningen image processing system, gipsy.
\newblock In \emph{Astronomical Data Analysis Software and Systems I},
  volume~25, page 131, 1992.

\bibitem[Swaters et~al.(1999)Swaters, Schoenmakers, Sancisi, and
  Albada]{swaters1999kinematically}
RA~Swaters, RHM Schoenmakers, R~Sancisi, and TS~van Albada.
\newblock Kinematically lopsided spiral galaxies.
\newblock \emph{Monthly Notices of the Royal Astronomical Society},
  304\penalty0 (2):\penalty0 330--334, 1999.

\bibitem[Kamphuis et~al.(2015)Kamphuis, J{\'o}zsa, Oh, Spekkens, Urbancic,
  Serra, Koribalski, and Dettmar]{kamphuis2015fat}
P~Kamphuis, GIG J{\'o}zsa, S-H Oh, K~Spekkens, N~Urbancic, P~Serra,
  BS~Koribalski, and R-J Dettmar.
\newblock Fat: Fully automated tirific.
\newblock \emph{Astrophysics Source Code Library}, pages ascl--1507, 2015.

\bibitem[Teodoro and Fraternali(2015)]{teodoro20153d}
EM~Di Teodoro and Filippo Fraternali.
\newblock 3d barolo: a new 3d algorithm to derive rotation curves of galaxies.
\newblock \emph{Monthly Notices of the Royal Astronomical Society},
  451\penalty0 (3):\penalty0 3021--3033, 2015.

\bibitem[Begeman et~al.(1991)Begeman, Broeils, and
  Sanders]{begeman1991extended}
KG~Begeman, AH~Broeils, and RH~Sanders.
\newblock Extended rotation curves of spiral galaxies: Dark haloes and modified
  dynamics.
\newblock \emph{Monthly Notices of the Royal Astronomical Society},
  249\penalty0 (3):\penalty0 523--537, 1991.

\bibitem[Bell and de~Jong(2001)]{bell2001stellar}
Eric~F Bell and Roelof~S de~Jong.
\newblock Stellar mass-to-light ratios and the tully-fisher relation.
\newblock \emph{The Astrophysical Journal}, 550\penalty0 (1):\penalty0 212,
  2001.

\bibitem[Bell et~al.(2003)Bell, McIntosh, Katz, and Weinberg]{bell2003optical}
Eric~F Bell, Daniel~H McIntosh, Neal Katz, and Martin~D Weinberg.
\newblock The optical and near-infrared properties of galaxies. i. luminosity
  and stellar mass functions.
\newblock \emph{The Astrophysical Journal Supplement Series}, 149\penalty0
  (2):\penalty0 289, 2003.

\bibitem[Kroupa(2001)]{kroupa2001variation}
Pavel Kroupa.
\newblock On the variation of the initial mass function.
\newblock \emph{Monthly Notices of the Royal Astronomical Society},
  322\penalty0 (2):\penalty0 231--246, 2001.

\bibitem[Law et~al.(2015)Law, Yan, Bershady, Bundy, Cherinka, Drory, MacDonald,
  S{\'a}nchez-Gallego, Wake, Weijmans, et~al.]{Law2015Observing}
David~R Law, Renbin Yan, Matthew~A Bershady, Kevin Bundy, Brian Cherinka, Niv
  Drory, Nicholas MacDonald, Jos{\'e}~R S{\'a}nchez-Gallego, David~A Wake,
  Anne-Marie Weijmans, et~al.
\newblock Observing strategy for the sdss-iv/manga ifu galaxy survey.
\newblock \emph{The Astronomical Journal}, 150\penalty0 (1):\penalty0 19, 2015.

\bibitem[Allen et~al.(2015)Allen, Croom, Konstantopoulos, Bryant, Sharp, Cecil,
  Fogarty, Foster, Green, Ho, et~al.]{Allen2015Sami}
JT~Allen, SM~Croom, IS~Konstantopoulos, JJ~Bryant, R~Sharp, GN~Cecil, LMR
  Fogarty, C~Foster, AW~Green, I-T Ho, et~al.
\newblock The sami galaxy survey: early data release.
\newblock \emph{Monthly Notices of the Royal Astronomical Society},
  446\penalty0 (2):\penalty0 1567--1583, 2015.

\bibitem[Bershady et~al.(2010)Bershady, Verheijen, Swaters, Andersen, Westfall,
  and Martinsson]{Bershady_2010}
Matthew~A. Bershady, Marc A.~W. Verheijen, Rob~A. Swaters, David~R. Andersen,
  Kyle~B. Westfall, and Thomas Martinsson.
\newblock The diskmass survey. i. overview.
\newblock \emph{The Astrophysical Journal}, 716\penalty0 (1):\penalty0
  198–233, May 2010.
\newblock ISSN 1538-4357.
\newblock \doi{10.1088/0004-637x/716/1/198}.
\newblock URL \url{http://dx.doi.org/10.1088/0004-637X/716/1/198}.

\bibitem[S{\'a}nchez et~al.(2012)S{\'a}nchez, Kennicutt, De~Paz, Van~de Ven,
  V{\'\i}lchez, Wisotzki, Walcher, Mast, Aguerri, Albiol-P{\'e}rez,
  et~al.]{Sanchez2012Califa}
SF~S{\'a}nchez, RC~Kennicutt, A~Gil De~Paz, G~Van~de Ven, JM~V{\'\i}lchez,
  L~Wisotzki, CJ~Walcher, D~Mast, JAL Aguerri, S~Albiol-P{\'e}rez, et~al.
\newblock Califa, the calar alto legacy integral field area survey-i. survey
  presentation.
\newblock \emph{Astronomy \& Astrophysics}, 538:\penalty0 A8, 2012.

\bibitem[Romeo and Falstad(2013)]{romeo2013simple}
Alessandro~B Romeo and Niklas Falstad.
\newblock A simple and accurate approximation for the q stability parameter in
  multicomponent and realistically thick discs.
\newblock \emph{Monthly Notices of the Royal Astronomical Society},
  433\penalty0 (2):\penalty0 1389--1397, 2013.

\bibitem[Matthews et~al.(1999)Matthews, Gallagher~III, and
  Van~Driel]{matthews1999extraordinary}
LD~Matthews, JS~Gallagher~III, and W~Van~Driel.
\newblock The extraordinary “superthin” spiral galaxy ugc 7321. i. disk
  color gradients and global properties from multiwavelength observations.
\newblock \emph{The Astronomical Journal}, 118\penalty0 (6):\penalty0 2751,
  1999.

\bibitem[Yock et~al.(1999)Yock, Pennycook, Rattenbury, Koribalski, Muraki,
  Yanagisawa, Jugaku, and Dodd]{yock1999observation}
Philip Yock, Glen Pennycook, Nicholas Rattenbury, Baerbel Koribalski, Yasushi
  Muraki, Toshi Yanagisawa, Jun Jugaku, and Richard Dodd.
\newblock Observation of the halo of the edge-on galaxy ic 5249.
\newblock In \emph{The Third Stromlo Symposium: The Galactic Halo}, volume 165,
  page 187, 1999.

\bibitem[Mendelowitz et~al.(2000)Mendelowitz, Matthews, Hibbard, and
  Wilcots]{mendelowitz2000rotation}
CM~Mendelowitz, LD~Matthews, JE~Hibbard, and EM~Wilcots.
\newblock Rotation curve and mass decomposition for the edge-on spiral galaxy
  ugc 711.
\newblock In \emph{Bulletin of the American Astronomical Society}, volume~32,
  page 1459, 2000.

\bibitem[Sarkar and Jog(2019{\natexlab{a}})]{sarkar2019flaring}
Suchira Sarkar and Chanda~J Jog.
\newblock Flaring stellar disk in low surface brightness galaxy ugc 7321.
\newblock \emph{arXiv preprint arXiv:1905.02735}, 2019{\natexlab{a}}.

\bibitem[Bizyaev et~al.(2016)Bizyaev, Kautsch, Sotnikova, Reshetnikov, and
  Mosenkov]{bizyaev2016very}
DV~Bizyaev, SJ~Kautsch, N~Ya Sotnikova, Vladimir~P Reshetnikov, and
  Aleksander~V Mosenkov.
\newblock Very thin disc galaxies in the sdss catalogue of edge-on galaxies.
\newblock \emph{Monthly Notices of the Royal Astronomical Society},
  465\penalty0 (4):\penalty0 3784--3792, 2016.

\bibitem[Salo.~H et~al.(2015)]{salo2015}
Laurikainen.~E Salo.~H et~al.
\newblock The spitzer survey of stellar structure in galaxies (s4g):
  Multi-component decomposition strategies and data release.
\newblock \emph{The Astrophysical Journal Supplement Series}, 219:\penalty0 4,
  2015.

\bibitem[Binney and Merrifield(2008)]{binney2008princeton}
J~Binney and M~Merrifield.
\newblock Princeton univ. press.
\newblock \emph{Galactic Astronomy,}, 2008.

\bibitem[Sharma et~al.(2014)Sharma, Bland-Hawthorn, Binney, Freeman, Steinmetz,
  Boeche, Bienayme, Gibson, Gilmore, Grebel, et~al.]{sharma2014kinematic}
Sanjib Sharma, Joss Bland-Hawthorn, J~Binney, Ken~C Freeman, Matthias
  Steinmetz, Corrado Boeche, Olivier Bienayme, Brad~K Gibson, Gerard~F Gilmore,
  Eva~K Grebel, et~al.
\newblock Kinematic modeling of the milky way using the rave and gcs stellar
  surveys.
\newblock \emph{The Astrophysical Journal}, 793\penalty0 (1):\penalty0 51,
  2014.

\bibitem[Van~der Kruit and Searle(1982)]{van1982surface}
PC~Van~der Kruit and L~Searle.
\newblock Surface photometry of edge-on spiral galaxies. iii-properties of the
  three-dimensional distribution of light and mass in disks of spiral galaxies.
\newblock \emph{Astronomy and Astrophysics}, 110:\penalty0 61--78, 1982.

\bibitem[Van Der~Kruit(1988)]{van1988three}
PC~Van Der~Kruit.
\newblock The three-dimensional distribution of light and mass in disks of
  spiral galaxies.
\newblock \emph{Astronomy and Astrophysics}, 192:\penalty0 117--127, 1988.

\bibitem[Van~der Kruit and Freeman(2011)]{van2011galaxy}
PC~Van~der Kruit and KC~Freeman.
\newblock Galaxy disks.
\newblock \emph{Annual Review of Astronomy and Astrophysics}, 49:\penalty0
  301--371, 2011.

\bibitem[Narayan and Jog(2002{\natexlab{b}})]{narayan2002origin}
Chaitra~A Narayan and Chanda~J Jog.
\newblock Origin of radially increasing stellar scaleheight in a galactic disk.
\newblock \emph{Astronomy \& Astrophysics}, 390\penalty0 (3):\penalty0
  L35--L38, 2002{\natexlab{b}}.

\bibitem[Nordstr{\"o}m et~al.(2004)Nordstr{\"o}m, Mayor, Andersen, Holmberg,
  Pont, J{\o}rgensen, Olsen, Udry, and Mowlavi]{nordstrom2004geneva}
Birgitta Nordstr{\"o}m, M~Mayor, J~Andersen, J~Holmberg, F~Pont,
  Bjarne~Rosenkilde J{\o}rgensen, EH~Olsen, S~Udry, and N~Mowlavi.
\newblock The geneva-copenhagen survey of the solar neighbourhood-ages,
  metallicities, and kinematic properties of\~{} 14 000 f and g dwarfs.
\newblock \emph{Astronomy \& Astrophysics}, 418\penalty0 (3):\penalty0
  989--1019, 2004.

\bibitem[Steinmetz et~al.(2006)Steinmetz, Zwitter, Siebert, Watson, Freeman,
  Munari, Campbell, Williams, Seabroke, Wyse, et~al.]{steinmetz2006radial}
Matthias Steinmetz, Toma{\v{z}} Zwitter, Arnaud Siebert, Fred~G Watson,
  Kenneth~C Freeman, Ulisse Munari, Rachel Campbell, Mary Williams, George~M
  Seabroke, Rosemary~FG Wyse, et~al.
\newblock The radial velocity experiment (rave): first data release.
\newblock \emph{The Astronomical Journal}, 132\penalty0 (4):\penalty0 1645,
  2006.

\bibitem[Van~der Kruit and Freeman(1984)]{van1984vertical}
PC~Van~der Kruit and KC~Freeman.
\newblock The vertical velocity dispersion of the stars in the disks of two
  spiral galaxies.
\newblock \emph{The Astrophysical Journal}, 278:\penalty0 81--88, 1984.

\bibitem[Martinsson et~al.(2013)Martinsson, Verheijen, Westfall, Bershady,
  Schechtman-Rook, Andersen, and Swaters]{martinsson2013diskmass}
Thomas~PK Martinsson, Marc~AW Verheijen, Kyle~B Westfall, Matthew~A Bershady,
  Andrew Schechtman-Rook, David~R Andersen, and Rob~A Swaters.
\newblock The diskmass survey-vi. gas and stellar kinematics in spiral galaxies
  from ppak integral-field spectroscopy.
\newblock \emph{Astronomy \& Astrophysics}, 557:\penalty0 A130, 2013.

\bibitem[Mogotsi et~al.(2016)Mogotsi, de~Blok, Cald{\'u}-Primo, Walter,
  Ianjamasimanana, and Leroy]{mogotsi2016hi}
KM~Mogotsi, WJG de~Blok, A~Cald{\'u}-Primo, F~Walter, R~Ianjamasimanana, and
  AK~Leroy.
\newblock Hi and co velocity dispersions in nearby galaxies.
\newblock \emph{The Astronomical Journal}, 151\penalty0 (1):\penalty0 15, 2016.

\bibitem[Tamburro et~al.(2009)Tamburro, Rix, Leroy, Mac~Low, Walter, Kennicutt,
  Brinks, and De~Blok]{tamburro2009driving}
D~Tamburro, H-W Rix, AK~Leroy, M-M Mac~Low, F~Walter, RC~Kennicutt, E~Brinks,
  and WJG De~Blok.
\newblock What is driving the h i velocity dispersion?
\newblock \emph{The Astronomical Journal}, 137\penalty0 (5):\penalty0 4424,
  2009.

\bibitem[Ianjamasimanana et~al.(2012)Ianjamasimanana, De~Blok, Walter, and
  Heald]{ianjamasimanana2012shapes}
R~Ianjamasimanana, WJG De~Blok, Fabian Walter, and George~H Heald.
\newblock The shapes of the hi velocity profiles of the things galaxies.
\newblock \emph{The Astronomical Journal}, 144\penalty0 (4):\penalty0 96, 2012.

\bibitem[Gerssen and Shapiro~Griffin(2012)]{gerssen2012disc}
J~Gerssen and K~Shapiro~Griffin.
\newblock Disc heating agents across the hubble sequence.
\newblock \emph{Monthly Notices of the Royal Astronomical Society},
  423\penalty0 (3):\penalty0 2726--2735, 2012.

\bibitem[Romeo and Mogotsi(2017)]{romeo2017drives}
Alessandro~B Romeo and Keoikantse~Moses Mogotsi.
\newblock What drives gravitational instability in nearby star-forming spirals?
  the impact of co and h i velocity dispersions.
\newblock \emph{Monthly Notices of the Royal Astronomical Society},
  469\penalty0 (1):\penalty0 286--294, 2017.

\bibitem[Kennicutt et~al.(2003)Kennicutt, Armus, Bendo, Calzetti, Dale, Draine,
  Engelbracht, Gordon, Grauer, Helou, et~al.]{kennicutt2003sings}
Robert~C Kennicutt, Lee Armus, George Bendo, Daniela Calzetti, Daniel~A Dale,
  Bruce~T Draine, Charles~W Engelbracht, Karl~D Gordon, Albert~D Grauer, George
  Helou, et~al.
\newblock Sings: The sirtf nearby galaxies survey.
\newblock \emph{Publications of the Astronomical Society of the Pacific},
  115\penalty0 (810):\penalty0 928, 2003.

\bibitem[Griv and Gedalin(2012)]{griv2012stability}
Evgeny Griv and Michael Gedalin.
\newblock Stability of galactic discs: finite arm-inclination and
  finite-thickness effects.
\newblock \emph{Monthly Notices of the Royal Astronomical Society},
  422\penalty0 (1):\penalty0 600--609, 2012.

\bibitem[Elmegreen(2011)]{elmegreen2011gravitational}
Bruce~G Elmegreen.
\newblock Gravitational instabilities in two-component galaxy disks with gas
  dissipation.
\newblock \emph{The Astrophysical Journal}, 737\penalty0 (1):\penalty0 10,
  2011.

\bibitem[Katz et~al.(2018)Katz, Antoja, Romero-G{\'o}mez, Drimmel, Reyl{\'e},
  Seabroke, Soubiran, Babusiaux, Di~Matteo, Figueras, et~al.]{katz2018gaia}
D~Katz, T~Antoja, Manuel Romero-G{\'o}mez, R~Drimmel, C~Reyl{\'e}, GM~Seabroke,
  C~Soubiran, C~Babusiaux, P~Di~Matteo, F~Figueras, et~al.
\newblock Gaia data release 2-mapping the milky way disc kinematics.
\newblock \emph{Astronomy \& astrophysics}, 616:\penalty0 A11, 2018.

\bibitem[Aumer et~al.(2016)Aumer, Binney, and Sch{\"o}nrich]{aumer2016age}
Michael Aumer, James Binney, and Ralph Sch{\"o}nrich.
\newblock Age--velocity dispersion relations and heating histories in disc
  galaxies.
\newblock \emph{Monthly Notices of the Royal Astronomical Society},
  462\penalty0 (2):\penalty0 1697--1713, 2016.

\bibitem[Jenkins and Binney(1990)]{jenkins1990spiral}
Adrian Jenkins and James Binney.
\newblock Spiral heating of galactic discs.
\newblock \emph{Monthly Notices of the Royal Astronomical Society},
  245:\penalty0 305--317, 1990.

\bibitem[Saha(2014)]{saha2014disc}
Kanak Saha.
\newblock Disc heating: possible link between weak bars and superthin galaxies.
\newblock \emph{arXiv preprint arXiv:1403.1711}, 2014.

\bibitem[Grand et~al.(2016{\natexlab{a}})Grand, Springel, G{\'o}mez, Marinacci,
  Pakmor, Campbell, and Jenkins]{grand2016vertical}
Robert~JJ Grand, Volker Springel, Facundo~A G{\'o}mez, Federico Marinacci,
  R{\"u}diger Pakmor, David~JR Campbell, and Adrian Jenkins.
\newblock Vertical disc heating in milky way-sized galaxies in a cosmological
  context.
\newblock \emph{Monthly Notices of the Royal Astronomical Society},
  459\penalty0 (1):\penalty0 199--219, 2016{\natexlab{a}}.

\bibitem[Kumar et~al.(2022)Kumar, Ghosh, Kataria, Das, and
  Debattista]{kumar2022excitation}
Ankit Kumar, Soumavo Ghosh, Sandeep~Kumar Kataria, Mousumi Das, and Victor~P
  Debattista.
\newblock Excitation of vertical breathing motion in disc galaxies by
  tidally-induced spirals in fly-by interactions.
\newblock \emph{Monthly Notices of the Royal Astronomical Society},
  516\penalty0 (1):\penalty0 1114--1126, 2022.

\bibitem[Rubin et~al.(1979)Rubin, Ford~Jr, and Roberts]{rubin1979extended}
VC~Rubin, WK~Ford~Jr, and MS~Roberts.
\newblock Extended rotation curves of high-luminosity spiral galaxies. v-ngc
  1961, the most massive spiral known.
\newblock \emph{The Astrophysical Journal}, 230:\penalty0 35--39, 1979.

\bibitem[Zwicky(1933)]{zwicky1933rotverschiebung}
Fritz Zwicky.
\newblock Die rotverschiebung von extragalaktischen nebeln.
\newblock \emph{Helvetica physica acta}, 6:\penalty0 110--127, 1933.

\bibitem[Sofue and Rubin(2001)]{sofue2001rotation}
Yoshiaki Sofue and Vera Rubin.
\newblock Rotation curves of spiral galaxies.
\newblock \emph{Annual Review of Astronomy and Astrophysics}, 39\penalty0
  (1):\penalty0 137--174, 2001.

\bibitem[McGaugh et~al.(2001)McGaugh, Rubin, and de~Blok]{mcgaugh2001high}
Stacy~S McGaugh, Vera~C Rubin, and WJG de~Blok.
\newblock High-resolution rotation curves of low surface brightness galaxies.
  i. data.
\newblock \emph{The Astronomical Journal}, 122\penalty0 (5):\penalty0 2381,
  2001.

\bibitem[Kranz et~al.(2003)Kranz, Slyz, and Rix]{kranz2003dark}
Thilo Kranz, Adrianne Slyz, and Hans-Walter Rix.
\newblock Dark matter within high surface brightness spiral galaxies.
\newblock \emph{The Astrophysical Journal}, 586\penalty0 (1):\penalty0 143,
  2003.

\bibitem[Gentile et~al.(2004)Gentile, Salucci, Klein, Vergani, and
  Kalberla]{gentile2004cored}
Gianfranco Gentile, Paolo Salucci, U~Klein, D~Vergani, and P~Kalberla.
\newblock The cored distribution of dark matter in spiral galaxies.
\newblock \emph{Monthly Notices of the Royal Astronomical Society},
  351\penalty0 (3):\penalty0 903--922, 2004.

\bibitem[Carlberg et~al.(1997)Carlberg, Yee, Ellingson, Morris, Abraham,
  Gravel, Pritchet, Smecker-Hane, Hartwick, Hesser,
  et~al.]{carlberg1997average}
RG~Carlberg, HKC Yee, E~Ellingson, SL~Morris, R~Abraham, Pl~Gravel,
  CJ~Pritchet, T~Smecker-Hane, FDA Hartwick, JE~Hesser, et~al.
\newblock The average mass profile of galaxy clusters.
\newblock \emph{The Astrophysical Journal Letters}, 485\penalty0 (1):\penalty0
  L13, 1997.

\bibitem[Kroupa(2012)]{kroupa2012dark}
Pavel Kroupa.
\newblock The dark matter crisis: falsification of the current standard model
  of cosmology.
\newblock \emph{Publications of the Astronomical Society of Australia},
  29\penalty0 (4):\penalty0 395--433, 2012.

\bibitem[Famaey and Binney(2005)]{famaey2005modified}
Benoit Famaey and James Binney.
\newblock Modified newtonian dynamics in the milky way.
\newblock \emph{Monthly Notices of the Royal Astronomical Society},
  363\penalty0 (2):\penalty0 603--608, 2005.

\bibitem[Binetruy et~al.(2000{\natexlab{b}})Binetruy, Deffayet, and
  Langlois]{Binetruy:1999ut}
Pierre Binetruy, Cedric Deffayet, and David Langlois.
\newblock {Nonconventional cosmology from a brane universe}.
\newblock \emph{Nucl.Phys.}, B565:\penalty0 269--287, 2000{\natexlab{b}}.
\newblock \doi{10.1016/S0550-3213(99)00696-3}.

\bibitem[Csaki et~al.(1999)Csaki, Graesser, Kolda, and Terning]{Csaki:1999jh}
Csaba Csaki, Michael Graesser, Christopher~F. Kolda, and John Terning.
\newblock {Cosmology of one extra dimension with localized gravity}.
\newblock \emph{Phys.Lett.}, B462:\penalty0 34--40, 1999.
\newblock \doi{10.1016/S0370-2693(99)00896-5}.

\bibitem[Mazumdar(2001)]{Mazumdar:2000gj}
Anupam Mazumdar.
\newblock {Interesting consequences of brane cosmology}.
\newblock \emph{Phys.\ Rev.\ D}, 64:\penalty0 027304, 2001.
\newblock \doi{10.1103/PhysRevD.64.027304}.

\bibitem[Maartens(2000)]{Maartens:2000fg}
Roy Maartens.
\newblock {Cosmological dynamics on the brane}.
\newblock \emph{Phys.\ Rev.\ D}, 62:\penalty0 084023, 2000.
\newblock \doi{10.1103/PhysRevD.62.084023}.

\bibitem[Maartens(2004{\natexlab{b}})]{Maartens:2003tw}
Roy Maartens.
\newblock {Brane world gravity}.
\newblock \emph{Living Rev.\ Rel.}, 7:\penalty0 7, 2004{\natexlab{b}}.
\newblock \doi{10.12942/lrr-2004-7}.

\bibitem[Koyama(2003)]{Koyama:2003be}
Kazuya Koyama.
\newblock {Cosmic microwave radiation anisotropies in brane worlds}.
\newblock \emph{Phys.\ Rev.\ Lett.}, 91:\penalty0 221301, 2003.
\newblock \doi{10.1103/PhysRevLett.91.221301}.

\bibitem[Haghani et~al.(2012)Haghani, Sepangi, and Shahidi]{Haghani:2012zq}
Zahra Haghani, Hamid~Reza Sepangi, and Shahab Shahidi.
\newblock {Cosmological dynamics of brane f(R) gravity}.
\newblock \emph{JCAP}, 1202:\penalty0 031, 2012.
\newblock \doi{10.1088/1475-7516/2012/02/031}.

\bibitem[Fichet(2020)]{Fichet:2019owx}
Sylvain Fichet.
\newblock {Braneworld effective field theories --- holography, consistency and
  conformal effects}.
\newblock \emph{JHEP}, 04:\penalty0 016, 2020.
\newblock \doi{10.1007/JHEP04(2020)016}.

\bibitem[Kaluza(2018)]{Kaluza:1921tu}
Th. Kaluza.
\newblock {Zum Unitätsproblem der Physik}.
\newblock \emph{Int.\ J.\ Mod.\ Phys.\ D}, 27\penalty0 (14):\penalty0 1870001,
  2018.
\newblock \doi{10.1142/S0218271818700017}.

\bibitem[Klein(1926)]{Klein:1926fj}
O.~Klein.
\newblock {The Atomicity of Electricity as a Quantum Theory Law}.
\newblock \emph{Nature}, 118:\penalty0 516, 1926.
\newblock \doi{10.1038/118516a0}.

\bibitem[Horava and Witten(1996)]{Horava:1995qa}
Petr Horava and Edward Witten.
\newblock {Heterotic and type I string dynamics from eleven-dimensions}.
\newblock \emph{Nucl.Phys.}, B460:\penalty0 506--524, 1996.
\newblock \doi{10.1016/0550-3213(95)00621-4}.

\bibitem[Polchinski(1998)]{Polchinski:1998rq}
J.~Polchinski.
\newblock {String theory. Vol. 1: An introduction to the bosonic string}.
\newblock 1998.

\bibitem[Davoudiasl et~al.(2000{\natexlab{a}})Davoudiasl, Hewett, and
  Rizzo]{Davoudiasl:1999tf}
H.~Davoudiasl, J.L. Hewett, and T.G. Rizzo.
\newblock {Bulk gauge fields in the Randall-Sundrum model}.
\newblock \emph{Phys.Lett.}, B473:\penalty0 43--49, 2000{\natexlab{a}}.
\newblock \doi{10.1016/S0370-2693(99)01430-6}.

\bibitem[Davoudiasl et~al.(2000{\natexlab{b}})Davoudiasl, Hewett, and
  Rizzo]{Davoudiasl:1999jd}
H.~Davoudiasl, J.L. Hewett, and T.G. Rizzo.
\newblock {Phenomenology of the Randall-Sundrum Gauge Hierarchy Model}.
\newblock \emph{Phys.\ Rev.\ Lett.}, 84:\penalty0 2080, 2000{\natexlab{b}}.
\newblock \doi{10.1103/PhysRevLett.84.2080}.

\bibitem[Davoudiasl et~al.(2001)Davoudiasl, Hewett, and
  Rizzo]{Davoudiasl:2000wi}
H.~Davoudiasl, J.L. Hewett, and T.G. Rizzo.
\newblock {Experimental probes of localized gravity: On and off the wall}.
\newblock \emph{Phys.Rev.}, D63:\penalty0 075004, 2001.
\newblock \doi{10.1103/PhysRevD.63.075004}.

\bibitem[Hundi and SenGupta(2013)]{Hundi:2011dc}
R.S. Hundi and Soumitra SenGupta.
\newblock {Fermion mass hierarchy in a multiple warped braneworld model}.
\newblock \emph{J.Phys.}, G40:\penalty0 075002, 2013.
\newblock \doi{10.1088/0954-3899/40/7/075002}.

\bibitem[Chakraborty and SenGupta(2014)]{Chakraborty:2014zya}
Sumanta Chakraborty and Soumitra SenGupta.
\newblock {Higher curvature gravity at the LHC}.
\newblock \emph{Phys.\ Rev.\ D}, 90\penalty0 (4):\penalty0 047901, 2014.
\newblock \doi{10.1103/PhysRevD.90.047901}.

\bibitem[Begum et~al.(2005)Begum, Chengalur, and Karachentsev]{begum2005dwarf}
Ayesha Begum, Jayaram~N Chengalur, and ID~Karachentsev.
\newblock A dwarf galaxy with a giant hi disk.
\newblock \emph{Astronomy \& Astrophysics}, 433\penalty0 (1):\penalty0 L1--L4,
  2005.

\bibitem[De~Blok et~al.(2001)De~Blok, McGaugh, Bosma, and Rubin]{de2001mass}
WJG De~Blok, Stacy~S McGaugh, Albert Bosma, and Vera~C Rubin.
\newblock Mass density profiles of low surface brightness galaxies.
\newblock \emph{The Astrophysical Journal Letters}, 552\penalty0 (1):\penalty0
  L23, 2001.

\bibitem[S{\'a}nchez-Salcedo et~al.(2008)S{\'a}nchez-Salcedo, Saha, and
  Narayan]{sanchez2008thickness}
FJ~S{\'a}nchez-Salcedo, K~Saha, and CA~Narayan.
\newblock The thickness of h i in galactic discs under modified newtonian
  dynamics: theory and application to the galaxy.
\newblock \emph{Monthly Notices of the Royal Astronomical Society},
  385\penalty0 (3):\penalty0 1585--1596, 2008.

\bibitem[Bothun et~al.(1997)Bothun, Impey, and McGaugh]{bothun1997low}
Greg Bothun, Chris Impey, and Stacy McGaugh.
\newblock Low-surface-brightness galaxies: hidden galaxies revealed.
\newblock \emph{Publications of the Astronomical Society of the Pacific},
  109\penalty0 (737):\penalty0 745, 1997.

\bibitem[McGaugh(1996)]{mcgaugh1996number}
Stacy~S McGaugh.
\newblock The number, luminosity and mass density of spiral galaxies as a
  function of surface brightness.
\newblock \emph{Monthly Notices of the Royal Astronomical Society},
  280\penalty0 (2):\penalty0 337--354, 1996.

\bibitem[Vorontsov-Vel'yaminov(1974)]{vorontsov1974specification}
BA~Vorontsov-Vel'yaminov.
\newblock Specification of the apparent flattening of spiral galaxies.
\newblock \emph{Soviet Astronomy}, 17:\penalty0 452, 1974.

\bibitem[Kautsch(2009)]{kautsch2009edge}
Stefan~J Kautsch.
\newblock The edge-on perspective of bulgeless, simple disk galaxies.
\newblock \emph{Publications of the Astronomical Society of the Pacific},
  121\penalty0 (886):\penalty0 1297, 2009.

\bibitem[Wyder et~al.(2009)Wyder, Martin, Barlow, Foster, Friedman, Morrissey,
  Neff, Neill, Schiminovich, Seibert, et~al.]{wyder2009star}
Ted~K Wyder, D~Christopher Martin, Tom~A Barlow, Karl Foster, Peter~G Friedman,
  Patrick Morrissey, Susan~G Neff, James~D Neill, David Schiminovich, Mark
  Seibert, et~al.
\newblock The star formation law at low surface density.
\newblock \emph{The Astrophysical Journal}, 696\penalty0 (2):\penalty0 1834,
  2009.

\bibitem[Narayanan and Banerjee(2022)]{narayanan2022superthin}
Ganesh Narayanan and Arunima Banerjee.
\newblock Are superthin galaxies low-surface-brightness galaxies seen edge-on?
  the star formation probe.
\newblock \emph{Monthly Notices of the Royal Astronomical Society},
  514\penalty0 (4):\penalty0 5126--5140, 2022.

\bibitem[Walter et~al.(2008)Walter, Brinks, de~Blok, Bigiel, Kennicutt~Jr,
  Thornley, and Leroy]{walter2008things}
Fabian Walter, Elias Brinks, WJG de~Blok, Frank Bigiel, Robert~C Kennicutt~Jr,
  Michele~D Thornley, and Adam Leroy.
\newblock Things: The hi nearby galaxy survey.
\newblock \emph{The Astronomical Journal}, 136\penalty0 (6):\penalty0 2563,
  2008.

\bibitem[Hunter et~al.(2012)Hunter, Ficut-Vicas, Ashley, Brinks, Cigan,
  Elmegreen, Heesen, Herrmann, Johnson, Oh, et~al.]{hunter2012little}
Deidre~A Hunter, Dana Ficut-Vicas, Trisha Ashley, Elias Brinks, Phil Cigan,
  Bruce~G Elmegreen, Volker Heesen, Kimberly~A Herrmann, Megan Johnson, Se-Heon
  Oh, et~al.
\newblock Little things.
\newblock \emph{The Astronomical Journal}, 144\penalty0 (5):\penalty0 134,
  2012.

\bibitem[Leroy et~al.(2008)Leroy, Walter, Brinks, Bigiel, De~Blok, Madore, and
  Thornley]{leroy2008star}
Adam~K Leroy, Fabian Walter, Elias Brinks, Frank Bigiel, WJG De~Blok, Barry
  Madore, and MD~Thornley.
\newblock The star formation efficiency in nearby galaxies: measuring where gas
  forms stars effectively.
\newblock \emph{The Astronomical Journal}, 136\penalty0 (6):\penalty0 2782,
  2008.

\bibitem[Bigiel et~al.(2008)Bigiel, Leroy, Walter, Brinks, De~Blok, Madore, and
  Thornley]{bigiel2008star}
Frank Bigiel, Adam Leroy, Fabian Walter, Elias Brinks, WJG De~Blok, Barry
  Madore, and Michele~D Thornley.
\newblock The star formation law in nearby galaxies on sub-kpc scales.
\newblock \emph{The Astronomical Journal}, 136\penalty0 (6):\penalty0 2846,
  2008.

\bibitem[Ghosh and Jog(2014)]{ghosh2014suppression}
Soumavo Ghosh and Chanda~J Jog.
\newblock Suppression of gravitational instabilities by dominant dark matter
  halo in low-surface-brightness galaxies.
\newblock \emph{Monthly Notices of the Royal Astronomical Society},
  439\penalty0 (1):\penalty0 929--935, 2014.

\bibitem[Jog(1996)]{jog1996local}
Chanda~J Jog.
\newblock Local stability criterion for stars and gas in a galactic disc.
\newblock \emph{Monthly Notices of the Royal Astronomical Society},
  278\penalty0 (1):\penalty0 209--218, 1996.

\bibitem[Pohlen et~al.(2003)Pohlen, Balcells, L{\"u}tticke, and
  Dettmar]{pohlen2003evidence}
M~Pohlen, M~Balcells, R~L{\"u}tticke, and R-J Dettmar.
\newblock Evidence for a large stellar bar in the low surface brightness galaxy
  ugc 7321.
\newblock \emph{Astronomy \& Astrophysics}, 409\penalty0 (2):\penalty0
  485--490, 2003.

\bibitem[Matthews and Uson(2008)]{matthews2008corrugations}
LD~Matthews and Juan~M Uson.
\newblock Corrugations in the disk of the edge-on spiral galaxy ic 2233.
\newblock \emph{The Astrophysical Journal}, 688\penalty0 (1):\penalty0 237,
  2008.

\bibitem[Gallagher and Hudson(1976)]{gallagher1976surface}
JS~Gallagher and HS~Hudson.
\newblock Surface photometry of the spiral galaxy ic 2233 and the existence of
  massive halos.
\newblock \emph{The Astrophysical Journal}, 209:\penalty0 389--391, 1976.

\bibitem[Byun(1998)]{byun1998surface}
Yong-Ik Byun.
\newblock Surface photometry of edge-on galaxies: Ic 5249 and es0 404-g18.
\newblock \emph{Chinese Journal of Physics}, 36\penalty0 (5):\penalty0
  677--692, 1998.

\bibitem[Sarkar and Jog(2020{\natexlab{a}})]{sarkar2020general}
Suchira Sarkar and Chanda~J Jog.
\newblock General model of vertical distribution of stars in the milky way
  using complete jeans equations.
\newblock \emph{Monthly Notices of the Royal Astronomical Society},
  492\penalty0 (1):\penalty0 628--633, 2020{\natexlab{a}}.

\bibitem[Sarkar and Jog(2019{\natexlab{b}})]{sarkar2019vertical}
Suchira Sarkar and Chanda~J Jog.
\newblock Vertical distribution of stars and flaring in the milky way.
\newblock \emph{Proceedings of the International Astronomical Union},
  14\penalty0 (S353):\penalty0 13--15, 2019{\natexlab{b}}.

\bibitem[Patra(2020{\natexlab{a}})]{patra2020theoretical}
Narendra~Nath Patra.
\newblock Theoretical modelling of two-component molecular discs in spiral
  galaxies.
\newblock \emph{Astronomy \& Astrophysics}, 638:\penalty0 A66,
  2020{\natexlab{a}}.

\bibitem[Patra(2020{\natexlab{b}})]{patra2020h}
Narendra~Nath Patra.
\newblock H i scale height in spiral galaxies.
\newblock \emph{Monthly Notices of the Royal Astronomical Society},
  499\penalty0 (2):\penalty0 2063--2075, 2020{\natexlab{b}}.

\bibitem[Patra(2018)]{patra2018molecular}
Narendra~Nath Patra.
\newblock Molecular scale height in ngc 7331.
\newblock \emph{Monthly Notices of the Royal Astronomical Society},
  478\penalty0 (4):\penalty0 4931--4938, 2018.

\bibitem[Karachentseva et~al.(2016)Karachentseva, Kudrya, Karachentsev,
  Makarov, and Melnyk]{karachentseva2016ultra}
VE~Karachentseva, Yu~N Kudrya, ID~Karachentsev, DI~Makarov, and OV~Melnyk.
\newblock Ultra-flat galaxies selected from rfgc catalog. i. the sample
  properties.
\newblock \emph{Astrophysical Bulletin}, 71\penalty0 (1):\penalty0 1--13, 2016.

\bibitem[Hoffman et~al.(1989)Hoffman, Williams, Lewis, Helou, and
  Salpeter]{hoffman1989hi}
G~Lyle Hoffman, BM~Williams, BM~Lewis, George Helou, and EE~Salpeter.
\newblock Hi observations in the virgo cluster area. iii-all'member'spirals.
\newblock \emph{The Astrophysical Journal Supplement Series}, 69:\penalty0
  65--98, 1989.

\bibitem[Yoachim and Dalcanton(2008{\natexlab{b}})]{yoachim2008kinematics}
Peter Yoachim and Julianne~J Dalcanton.
\newblock The kinematics of thick disks in nine external galaxies.
\newblock \emph{The Astrophysical Journal}, 682\penalty0 (2):\penalty0 1004,
  2008{\natexlab{b}}.

\bibitem[De~Vaucouleurs et~al.(1991)De~Vaucouleurs, De~Vaucouleurs, Corwin,
  Buta, Paturel, and Fouque]{de1991third}
G~De~Vaucouleurs, A~De~Vaucouleurs, JHG Corwin, RJ~Buta, G~Paturel, and
  P~Fouque.
\newblock Third reference catalogue of bright galaxies, version 3.9. springer,
  new york, ny, 1991.

\bibitem[Makarov et~al.(2014)Makarov, Prugniel, Terekhova, Courtois, and
  Vauglin]{makarov2014hyperleda}
Dmitry Makarov, Philippe Prugniel, Nataliya Terekhova, H{\'e}l{\`e}ne Courtois,
  and Isabelle Vauglin.
\newblock Hyperleda. iii. the catalogue of extragalactic distances.
\newblock \emph{Astronomy \& Astrophysics}, 570:\penalty0 A13, 2014.

\bibitem[Kourkchi et~al.(2020)Kourkchi, Courtois, Graziani, Hoffman,
  Pomar{\`e}de, Shaya, and Tully]{kourkchi2020cosmicflows}
Ehsan Kourkchi, H{\'e}l{\`e}ne~M Courtois, Romain Graziani, Yehuda Hoffman,
  Daniel Pomar{\`e}de, Edward~J Shaya, and R~Brent Tully.
\newblock Cosmicflows-3: Two distance--velocity calculators.
\newblock \emph{The Astronomical Journal}, 159\penalty0 (2):\penalty0 67, 2020.

\bibitem[Haynes et~al.(2018)Haynes, Giovanelli, Kent, Adams, Balonek, Craig,
  Fertig, Finn, Giovanardi, Hallenbeck, et~al.]{haynes2018arecibo}
Martha~P Haynes, Riccardo Giovanelli, Brian~R Kent, Elizabeth~AK Adams,
  Thomas~J Balonek, David~W Craig, Derek Fertig, Rose Finn, Carlo Giovanardi,
  Gregory Hallenbeck, et~al.
\newblock The arecibo legacy fast alfa survey: The alfalfa extragalactic h i
  source catalog.
\newblock \emph{The Astrophysical Journal}, 861\penalty0 (1):\penalty0 49,
  2018.

\bibitem[McMullin et~al.(2007)McMullin, Waters, Schiebel, Young, and
  Golap]{mcmullin2007casa}
Joseph~P McMullin, BSDYWGK Waters, Darrell Schiebel, Wei Young, and Kumar
  Golap.
\newblock Casa architecture and applications.
\newblock In \emph{Astronomical data analysis software and systems XVI}, volume
  376, page 127, 2007.

\bibitem[Westmeier et~al.(2014)Westmeier, Jurek, Obreschkow, Koribalski, and
  Staveley-Smith]{westmeier2014busy}
Tobias Westmeier, Russell Jurek, Danail Obreschkow, B{\"a}rbel~S Koribalski,
  and Lister Staveley-Smith.
\newblock The busy function: a new analytic function for describing the
  integrated 21-cm spectral profile of galaxies.
\newblock \emph{Monthly Notices of the Royal Astronomical Society},
  438\penalty0 (2):\penalty0 1176--1190, 2014.

\bibitem[Allaert et~al.(2015)Allaert, Gentile, Baes, De~Geyter, Hughes, Lewis,
  Bianchi, De~Looze, Fritz, Holwerda, et~al.]{allaert2015herschel}
Flor Allaert, Gianfranco Gentile, Maarten Baes, Gert De~Geyter, TM~Hughes,
  F~Lewis, SIMONE Bianchi, Ilse De~Looze, Jacopo Fritz, Benne~W Holwerda,
  et~al.
\newblock Herschel observations of edge-on spirals (heroes)-ii. tilted-ring
  modelling of the atomic gas disks.
\newblock \emph{Astronomy \& Astrophysics}, 582:\penalty0 A18, 2015.

\bibitem[Zschaechner et~al.(2012)Zschaechner, Rand, Heald, Gentile, and
  J{\'o}zsa]{zschaechner2012halogas}
Laura~K Zschaechner, Richard~J Rand, George~H Heald, Gianfranco Gentile, and
  Gyula J{\'o}zsa.
\newblock Halogas: H i observations and modeling of the nearby edge-on spiral
  galaxy ngc 4565.
\newblock \emph{The Astrophysical Journal}, 760\penalty0 (1):\penalty0 37,
  2012.

\bibitem[Gentile et~al.(2013)Gentile, J{\'o}zsa, Serra, Heald, de~Blok,
  Fraternali, Patterson, Walterbos, and Oosterloo]{gentile2013halogas}
Gianfranco Gentile, GIG J{\'o}zsa, P~Serra, GH~Heald, WJG de~Blok,
  F~Fraternali, MT~Patterson, RAM Walterbos, and T~Oosterloo.
\newblock Halogas: Extraplanar gas in ngc 3198.
\newblock \emph{Astronomy \& Astrophysics}, 554:\penalty0 A125, 2013.

\bibitem[Kamphuis et~al.(2013)Kamphuis, Rand, J{\'o}zsa, Zschaechner, Heald,
  Patterson, Gentile, Walterbos, Serra, and de~Blok]{kamphuis2013halogas}
P~Kamphuis, RJ~Rand, GIG J{\'o}zsa, LK~Zschaechner, GH~Heald, MT~Patterson,
  Gianfranco Gentile, RAM Walterbos, P~Serra, and WJG de~Blok.
\newblock Halogas observations of ngc 5023 and ugc 2082: modelling of
  non-cylindrically symmetric gas distributions in edge-on galaxies.
\newblock \emph{Monthly Notices of the Royal Astronomical Society},
  434\penalty0 (3):\penalty0 2069--2093, 2013.

\bibitem[Punzo(2017)]{punzo20173d}
Davide Punzo.
\newblock \emph{3D visualization and analysis of HI in and around galaxies}.
\newblock PhD thesis, Rijksuniversiteit Groningen, 2017.

\bibitem[Bertin and Arnouts(1996)]{bertin1996sextractor}
Emmanuel Bertin and Stephane Arnouts.
\newblock Sextractor: Software for source extraction.
\newblock \emph{Astronomy and astrophysics supplement series}, 117\penalty0
  (2):\penalty0 393--404, 1996.

\bibitem[Peng et~al.(2011)Peng, Ho, Impey, and Rix]{peng2011galfit}
Chien~Y Peng, Luis~C Ho, Chris~D Impey, and Hans-Walter Rix.
\newblock Galfit: Detailed structural decomposition of galaxy images.
\newblock \emph{Astrophysics Source Code Library}, pages ascl--1104, 2011.

\bibitem[Kregel et~al.(2005)Kregel, Van Der~Kruit, and
  Freeman]{kregel2005structure}
M~Kregel, PC~Van Der~Kruit, and KC~Freeman.
\newblock Structure and kinematics of edge-on galaxy discs--v. the dynamics of
  stellar discs.
\newblock \emph{Monthly Notices of the Royal Astronomical Society},
  358\penalty0 (2):\penalty0 503--520, 2005.

\bibitem[de~Blok et~al.(2001)de~Blok, McGaugh, and Rubin]{de2001high}
WJG de~Blok, Stacy~S McGaugh, and Vera~C Rubin.
\newblock High-resolution rotation curves of low surface brightness galaxies.
  ii. mass models.
\newblock \emph{The Astronomical Journal}, 122\penalty0 (5):\penalty0 2396,
  2001.

\bibitem[Courteau(1997)]{courteau1997optical}
Stephane Courteau.
\newblock Optical rotation curves and linewidths for tully-fisher applications.
\newblock \emph{The Astronomical Journal}, 114:\penalty0 2402, 1997.

\bibitem[Fuchs et~al.(1998)Fuchs, M{\"o}llenhoff, and
  Heidt]{fuchs1998decomposition}
B~Fuchs, C~M{\"o}llenhoff, and J~Heidt.
\newblock Decomposition of the rotation curves of distant field galaxies.
\newblock \emph{arXiv preprint astro-ph/9806117}, 1998.

\bibitem[Newville et~al.(2016)Newville, Stensitzki, Allen, Rawlik, Ingargiola,
  and Nelson]{newville2016lmfit}
Matthew Newville, Till Stensitzki, Daniel~B Allen, Michal Rawlik, Antonino
  Ingargiola, and Andrew Nelson.
\newblock Lmfit: Non-linear least-square minimization and curve-fitting for
  python.
\newblock \emph{Astrophysics Source Code Library}, pages ascl--1606, 2016.

\bibitem[Virtanen et~al.(2020)Virtanen, Gommers, Oliphant, Haberland, Reddy,
  Cournapeau, Burovski, Peterson, Weckesser, Bright, et~al.]{virtanen2020scipy}
Pauli Virtanen, Ralf Gommers, Travis~E Oliphant, Matt Haberland, Tyler Reddy,
  David Cournapeau, Evgeni Burovski, Pearu Peterson, Warren Weckesser, Jonathan
  Bright, et~al.
\newblock Scipy 1.0: fundamental algorithms for scientific computing in python.
\newblock \emph{Nature methods}, 17\penalty0 (3):\penalty0 261--272, 2020.

\bibitem[De~Naray et~al.(2008)De~Naray, McGaugh, and De~Blok]{de2008mass}
Rachel~Kuzio De~Naray, Stacy~S McGaugh, and WJG De~Blok.
\newblock Mass models for low surface brightness galaxies with high-resolution
  optical velocity fields.
\newblock \emph{The Astrophysical Journal}, 676\penalty0 (2):\penalty0 920,
  2008.

\bibitem[Bottema and Pestana(2015)]{bottema2015distribution}
Roelof Bottema and Jos{\'e} Luis~G Pestana.
\newblock The distribution of dark and luminous matter inferred from extended
  rotation curves.
\newblock \emph{Monthly Notices of the Royal Astronomical Society},
  448\penalty0 (3):\penalty0 2566--2593, 2015.

\bibitem[Dutton and Maccio(2014)]{dutton2014cold}
Aaron~A Dutton and Andrea~V Maccio.
\newblock Cold dark matter haloes in the planck era: evolution of structural
  parameters for einasto and nfw profiles.
\newblock \emph{Monthly Notices of the Royal Astronomical Society},
  441\penalty0 (4):\penalty0 3359--3374, 2014.

\bibitem[Donato et~al.(2004)Donato, Gentile, and Salucci]{donato2004cores}
Fiorenza Donato, Gianfranco Gentile, and Paolo Salucci.
\newblock Cores of dark matter haloes correlate with stellar scalelengths.
\newblock \emph{Monthly Notices of the Royal Astronomical Society},
  353\penalty0 (2):\penalty0 L17--L22, 2004.

\bibitem[Grand et~al.(2016{\natexlab{b}})Grand, Springel, Kawata, Minchev,
  S{\'a}nchez-Bl{\'a}zquez, G{\'o}mez, Marinacci, Pakmor, and
  Campbell]{grand2016spiral}
Robert~JJ Grand, Volker Springel, Daisuke Kawata, Ivan Minchev, Patricia
  S{\'a}nchez-Bl{\'a}zquez, Facundo~A G{\'o}mez, Federico Marinacci,
  R{\"u}diger Pakmor, and David~JR Campbell.
\newblock Spiral-induced velocity and metallicity patterns in a cosmological
  zoom simulation of a milky way-sized galaxy.
\newblock \emph{Monthly Notices of the Royal Astronomical Society: Letters},
  460\penalty0 (1):\penalty0 L94--L98, 2016{\natexlab{b}}.

\bibitem[{Matthew} et~al.(2000){Matthew}, {van Driel}, and
  {Gallagher}]{2000bgfp.conf..107M}
L.~D. {Matthew}, W.~{van Driel}, and J.~S. {Gallagher}.
\newblock {Properties of ``superthin'' galaxies}.
\newblock In F.~{Hammer}, T.~X. {Thuan}, V.~{Cayatte}, B.~{Guiderdoni}, and
  J.~T. {Thanh Van}, editors, \emph{Building Galaxies; from the Primordial
  Universe to the Present}, page 107, January 2000.

\bibitem[Aditya et~al.(2022)Aditya, Kamphuis, Banerjee, Borisov, Mosenkov,
  Antipova, and Makarov]{aditya2022h}
K~Aditya, Peter Kamphuis, Arunima Banerjee, Sviatoslav Borisov, Aleksandr
  Mosenkov, Aleksandra Antipova, and Dmitry Makarov.
\newblock H i 21 cm observation and mass models of the extremely thin galaxy
  fgc 1440.
\newblock \emph{Monthly Notices of the Royal Astronomical Society},
  509\penalty0 (3):\penalty0 4071--4093, 2022.

\bibitem[Martinez-Medina et~al.(2015)Martinez-Medina, Pichardo,
  P{\'e}rez-Villegas, and Moreno]{martinez2015contribution}
LA~Martinez-Medina, B~Pichardo, A~P{\'e}rez-Villegas, and E~Moreno.
\newblock The contribution of spiral arms to the thick disk along the hubble
  sequence.
\newblock \emph{The Astrophysical Journal}, 802\penalty0 (2):\penalty0 109,
  2015.

\bibitem[Saha et~al.(2010)Saha, Tseng, and Taam]{saha2010effect}
Kanak Saha, Yao-Huan Tseng, and Ronald~E Taam.
\newblock The effect of bars and transient spirals on the vertical heating in
  disk galaxies.
\newblock \emph{The Astrophysical Journal}, 721\penalty0 (2):\penalty0 1878,
  2010.

\bibitem[Benson et~al.(2004)Benson, Lacey, Frenk, Baugh, and
  Cole]{benson2004heating}
AJ~Benson, CG~Lacey, CS~Frenk, CM~Baugh, and S~Cole.
\newblock Heating of galactic discs by infalling satellites.
\newblock \emph{Monthly Notices of the Royal Astronomical Society},
  351\penalty0 (4):\penalty0 1215--1236, 2004.

\bibitem[Rix and Bovy(2013)]{rix2013milky}
Hans-Walter Rix and Jo~Bovy.
\newblock The milky way’s stellar disk.
\newblock \emph{The Astronomy and Astrophysics Review}, 21\penalty0
  (1):\penalty0 1--58, 2013.

\bibitem[Brook et~al.(2012)Brook, Stinson, Gibson, Kawata, House, Miranda,
  Macci{\`o}, Pilkington, Ro{\v{s}}kar, Wadsley, et~al.]{brook2012thin}
Chris~B Brook, GS~Stinson, Brad~K Gibson, Daisuke Kawata, Elisa~L House,
  Marco~S Miranda, Andrea~V Macci{\`o}, Kate Pilkington, R~Ro{\v{s}}kar,
  J~Wadsley, et~al.
\newblock Thin disc, thick disc and halo in a simulated galaxy.
\newblock \emph{Monthly Notices of the Royal Astronomical Society},
  426\penalty0 (1):\penalty0 690--700, 2012.

\bibitem[Minchev et~al.(2015)Minchev, Martig, Streich, Scannapieco, De~Jong,
  and Steinmetz]{minchev2015formation}
I~Minchev, M~Martig, D~Streich, C~Scannapieco, RS~De~Jong, and M~Steinmetz.
\newblock On the formation of galactic thick disks.
\newblock \emph{The Astrophysical Journal Letters}, 804\penalty0 (1):\penalty0
  L9, 2015.

\bibitem[Vogelsberger et~al.(2014)Vogelsberger, Genel, Springel, Torrey,
  Sijacki, Xu, Snyder, Nelson, and Hernquist]{vogelsberger2014introducing}
Mark Vogelsberger, Shy Genel, Volker Springel, Paul Torrey, Debora Sijacki,
  Dandan Xu, Greg Snyder, Dylan Nelson, and Lars Hernquist.
\newblock Introducing the illustris project: simulating the coevolution of dark
  and visible matter in the universe.
\newblock \emph{Monthly Notices of the Royal Astronomical Society},
  444\penalty0 (2):\penalty0 1518--1547, 2014.

\bibitem[{Schaye} et~al.(2015){Schaye}, {Crain}, {Bower}, {Furlong},
  {Schaller}, {Theuns}, {Dalla Vecchia}, {Frenk}, {McCarthy}, {Helly},
  {Jenkins}, {Rosas-Guevara}, {White}, {Baes}, {Booth}, {Camps}, {Navarro},
  {Qu}, {Rahmati}, {Sawala}, {Thomas}, and {Trayford}]{2015MNRAS.446..521S}
Joop {Schaye}, Robert~A. {Crain}, Richard~G. {Bower}, Michelle {Furlong},
  Matthieu {Schaller}, Tom {Theuns}, Claudio {Dalla Vecchia}, Carlos~S.
  {Frenk}, I.~G. {McCarthy}, John~C. {Helly}, Adrian {Jenkins}, Y.~M.
  {Rosas-Guevara}, Simon D.~M. {White}, Maarten {Baes}, C.~M. {Booth}, Peter
  {Camps}, Julio~F. {Navarro}, Yan {Qu}, Alireza {Rahmati}, Till {Sawala},
  Peter~A. {Thomas}, and James {Trayford}.
\newblock {The EAGLE project: simulating the evolution and assembly of galaxies
  and their environments}.
\newblock \emph{Monthly Notices of the Royal Astronomical Society},
  446\penalty0 (1):\penalty0 521--554, January 2015.
\newblock \doi{10.1093/mnras/stu2058}.

\bibitem[{Bottrell} et~al.(2017){Bottrell}, {Torrey}, {Simard}, and
  {Ellison}]{2017MNRAS.467.2879B}
Connor {Bottrell}, Paul {Torrey}, Luc {Simard}, and Sara~L. {Ellison}.
\newblock {Galaxies in the Illustris simulation as seen by the Sloan Digital
  Sky Survey - II. Size-luminosity relations and the deficit of bulge-dominated
  galaxies in Illustris at low mass}.
\newblock \emph{Monthly Notices of the Royal Astronomical Society},
  467\penalty0 (3):\penalty0 2879--2895, May 2017.
\newblock \doi{10.1093/mnras/stx276}.

\bibitem[{Olling}(1995)]{1995AJ....110..591O}
Rob~P. {Olling}.
\newblock {On the Usage of Flaring Gas Layers to Determine the Shape of Dark
  Matter Halos}.
\newblock \emph{The Astronomical Journal}, 110:\penalty0 591, August 1995.
\newblock \doi{10.1086/117545}.

\bibitem[{Banerjee} and {Jog}(2013)]{2013MNRAS.431..582B}
Arunima {Banerjee} and Chanda~J. {Jog}.
\newblock {Why are some galaxy discs extremely thin?}
\newblock \emph{Monthly Notices of the Royal Astronomical Society},
  431\penalty0 (1):\penalty0 582--588, May 2013.
\newblock \doi{10.1093/mnras/stt186}.

\bibitem[Khoperskov et~al.(2003)Khoperskov, Zasov, and
  Tyurina]{khoperskov2003minimum}
AV~Khoperskov, AV~Zasov, and NV~Tyurina.
\newblock Minimum velocity dispersion in stable stellar disks. numerical
  simulations.
\newblock \emph{Astronomy Reports}, 47\penalty0 (5):\penalty0 357--376, 2003.

\bibitem[Gentile et~al.(2015)Gentile, Tydtgat, Baes, De~Geyter, Koleva, Angus,
  De~Blok, Saftly, and Viaene]{gentile2015disk}
Gianfranco Gentile, C~Tydtgat, Maarten Baes, Gert De~Geyter, Mina Koleva,
  GW~Angus, WJG De~Blok, Waad Saftly, and S{\'e}bastien Viaene.
\newblock Disk mass and disk heating in the spiral galaxy ngc 3223.
\newblock \emph{Astronomy \& Astrophysics}, 576:\penalty0 A57, 2015.

\bibitem[Seth et~al.(2005)Seth, Dalcanton, and de~Jong]{seth2005study}
Anil~C Seth, Julianne~J Dalcanton, and Roelof~S de~Jong.
\newblock A study of edge-on galaxies with the hubble space telescope advanced
  camera for surveys. i. initial results.
\newblock \emph{The Astronomical Journal}, 129\penalty0 (3):\penalty0 1331,
  2005.

\bibitem[Kregel et~al.(2004)Kregel, Van Der~Kruit, and
  De~Blok]{kregel2004structure}
M~Kregel, PC~Van Der~Kruit, and WJG De~Blok.
\newblock Structure and kinematics of edge-on galaxy discs--ii. observations of
  the neutral hydrogen.
\newblock \emph{Monthly Notices of the Royal Astronomical Society},
  352\penalty0 (3):\penalty0 768--786, 2004.

\bibitem[Sarkar and Jog(2018)]{sarkar2018constraining}
Suchira Sarkar and Chanda~J Jog.
\newblock The constraining effect of gas and the dark matter halo on the
  vertical stellar distribution of the milky way.
\newblock \emph{Astronomy \& Astrophysics}, 617:\penalty0 A142, 2018.

\bibitem[Sarkar and Jog(2020{\natexlab{b}})]{sarkar2020vertical}
Suchira Sarkar and Chanda~J Jog.
\newblock Vertical stellar density distribution in a non-isothermal galactic
  disc.
\newblock \emph{Monthly Notices of the Royal Astronomical Society},
  499\penalty0 (2):\penalty0 2523--2533, 2020{\natexlab{b}}.

\bibitem[Banerjee et~al.(2011)Banerjee, Jog, Brinks, and
  Bagetakos]{banerjee2011theoretical}
Arunima Banerjee, Chanda~J Jog, Elias Brinks, and Ioannis Bagetakos.
\newblock Theoretical determination of h i vertical scale heights in the dwarf
  galaxies ddo 154, ho ii, ic 2574 and ngc 2366.
\newblock \emph{Monthly Notices of the Royal Astronomical Society},
  415\penalty0 (1):\penalty0 687--694, 2011.

\bibitem[Hoyle(1953)]{hoyle1953fragmentation}
Fred Hoyle.
\newblock On the fragmentation of gas clouds into galaxies and stars.
\newblock \emph{The Astrophysical Journal}, 118:\penalty0 513, 1953.

\bibitem[Barnes and Efstathiou(1987)]{barnes1987angular}
Joshua Barnes and George Efstathiou.
\newblock Angular momentum from tidal torques.
\newblock \emph{The Astrophysical Journal}, 319:\penalty0 575--600, 1987.

\bibitem[{Makarov} et~al.(2014){Makarov}, {Prugniel}, {Terekhova}, {Courtois},
  and {Vauglin}]{2014A&A...570A..13M}
D.~{Makarov}, P.~{Prugniel}, N.~{Terekhova}, H.~{Courtois}, and I.~{Vauglin}.
\newblock {HyperLEDA. III. The catalogue of extragalactic distances}.
\newblock \emph{Astronomy \& Astrophysics}, 570:\penalty0 A13, October 2014.
\newblock \doi{10.1051/0004-6361/201423496}.

\bibitem[Serra et~al.(2015)Serra, Westmeier, Giese, Jurek, Fl{\"o}er, Popping,
  Winkel, van~der Hulst, Meyer, Koribalski, et~al.]{serra2015sofia}
Paolo Serra, Tobias Westmeier, Nadine Giese, Russell Jurek, Lars Fl{\"o}er,
  Attila Popping, Benjamin Winkel, Thijs van~der Hulst, Martin Meyer,
  B{\"a}rbel~S Koribalski, et~al.
\newblock Sofia: a flexible source finder for 3d spectral line data.
\newblock \emph{Monthly Notices of the Royal Astronomical Society},
  448\penalty0 (2):\penalty0 1922--1929, 2015.

\bibitem[Schlafly and Finkbeiner(2011)]{schlafly2011measuring}
Edward~F Schlafly and Douglas~P Finkbeiner.
\newblock Measuring reddening with sloan digital sky survey stellar spectra and
  recalibrating sfd.
\newblock \emph{The Astrophysical Journal}, 737\penalty0 (2):\penalty0 103,
  2011.

\bibitem[Swaters et~al.(2009)Swaters, Sancisi, Van~Albada, and Van
  Der~Hulst]{swaters2009rotation}
RA~Swaters, R~Sancisi, TS~Van~Albada, and JM~Van Der~Hulst.
\newblock The rotation curves shapes of late-type dwarf galaxies.
\newblock \emph{Astronomy \& Astrophysics}, 493\penalty0 (3):\penalty0
  871--892, 2009.

\bibitem[Wechsler et~al.(2002)Wechsler, Bullock, Primack, Kravtsov, and
  Dekel]{Wechsler2002concentrations}
Risa~H Wechsler, James~S Bullock, Joel~R Primack, Andrey~V Kravtsov, and
  Avishai Dekel.
\newblock Concentrations of dark halos from their assembly histories.
\newblock \emph{The Astrophysical Journal}, 568\penalty0 (1):\penalty0 52,
  2002.

\bibitem[Bailin and Steinmetz(2005)]{bailin2005internal}
Jeremy Bailin and Matthias Steinmetz.
\newblock Internal and external alignment of the shapes and angular momenta of
  $\lambda$cdm halos.
\newblock \emph{The Astrophysical Journal}, 627\penalty0 (2):\penalty0 647,
  2005.

\bibitem[{Di Paolo} et~al.(2019){Di Paolo}, {Salucci}, and
  {Erkurt}]{2019MNRAS.490.5451D}
Chiara {Di Paolo}, Paolo {Salucci}, and Adnan {Erkurt}.
\newblock {The universal rotation curve of low surface brightness galaxies -
  IV. The interrelation between dark and luminous matter}.
\newblock \emph{Monthly Notices of the Royal Astronomical Society},
  490\penalty0 (4):\penalty0 5451--5477, December 2019.
\newblock \doi{10.1093/mnras/stz2700}.

\bibitem[Bovy et~al.(2012{\natexlab{b}})Bovy, Prieto, Beers, Bizyaev, Da~Costa,
  Cunha, Ebelke, Eisenstein, Frinchaboy, P{\'e}rez, et~al.]{bovy2012milky}
Jo~Bovy, Carlos~Allende Prieto, Timothy~C Beers, Dmitry Bizyaev, Luiz~N
  Da~Costa, Katia Cunha, Garrett~L Ebelke, Daniel~J Eisenstein, Peter~M
  Frinchaboy, Ana Elia~Garc{\'\i}a P{\'e}rez, et~al.
\newblock The milky way's circular-velocity curve between 4 and 14 kpc from
  apogee data.
\newblock \emph{The Astrophysical Journal}, 759\penalty0 (2):\penalty0 131,
  2012{\natexlab{b}}.

\bibitem[Pinna et~al.(2018)Pinna, Falc{\'o}n-Barroso, Martig,
  Mart{\'\i}nez-Valpuesta, M{\'e}ndez-Abreu, van~de Ven, Leaman, and
  Lyubenova]{pinna2018revisiting}
F~Pinna, J~Falc{\'o}n-Barroso, M~Martig, I~Mart{\'\i}nez-Valpuesta,
  J~M{\'e}ndez-Abreu, G~van~de Ven, R~Leaman, and M~Lyubenova.
\newblock Revisiting the stellar velocity ellipsoid--hubble-type relation:
  observations versus simulations.
\newblock \emph{Monthly Notices of the Royal Astronomical Society},
  475\penalty0 (2):\penalty0 2697--2712, 2018.

\bibitem[Obreschkow and Glazebrook(2014)]{obreschkow2014fundamental}
Danail Obreschkow and Karl Glazebrook.
\newblock Fundamental mass--spin--morphology relation of spiral galaxies.
\newblock \emph{The Astrophysical Journal}, 784\penalty0 (1):\penalty0 26,
  2014.

\end{thebibliography}

\end{document}